\documentclass[11pt,twoside]{ucthesism}

\newcommand{\arxivonlyshrink}{\vspace{-10pt}}

\usepackage{hyperref}
\usepackage{bookmark}  
\hypersetup{
  pdftitle={Structure of collective modes in transitional and deformed nuclei},
  pdfauthor={Mark Anthony Caprio},
  bookmarksnumbered,
  hypertexnames=false
}
\makeatletter
\newcommand{\toclevel@partstar}{-1}
\makeatother

\usepackage{vmargin}
\usepackage{xspace}
\usepackage{xperiodspace}

\usepackage[centertags]{amsmath}
\usepackage{amssymb}
\usepackage[mathscr]{eucal}
\usepackage{color}

\usepackage{graphicx}

\usepackage{cite}

\newenvironment{doubleindent}[1][0.375in]
  { \begin{list}{}{\setlength{\topsep}{0pt}\setlength{\partopsep}{0pt}
  \setlength{\leftmargin}{#1}\setlength{\rightmargin}{#1}} \item[] } {
  \end{list} }

\newcounter{enumeratelistctr}
\newenvironment{enumeratelist}[2][0.375in]
  {
    \begin{list}
          {\arabic{enumeratelistctr}.}
          {\setlength{\topsep}{0pt}\setlength{\partopsep}{0pt}
           \setlength{\itemsep}{0pt}\setlength{\parsep}{#2}
           \setlength{\leftmargin}{#1}\setlength{\rightmargin}{0pt}
           \setlength{\labelsep}{0.5em}
           \setlength{\itemindent}{0pt}
           \setlength{\labelwidth}{#1}\addtolength{\labelwidth}{-\labelsep}
           \usecounter{enumeratelistctr}
          }
  }
  {
    \end{list}
  }

\newcommand{\pseudoruledtabular}{\hline\hline}

\usepackage{afterpage}
\usepackage{longtable}
\setlongtables
\usepackage{dcolumn}
\newcolumntype{.}{@{\extracolsep{0pt}}}
\newcolumntype{,}{@{\extracolsep{3pt}}}
\newcolumntype{_}{@{\extracolsep{3ptplus1fil}}}
\newcolumntype{=}{@{\extracolsep{15ptplus1fil}}}
\newcommand{\multi}[1]{\multicolumn{2}{c}{#1}}
\newcommand{\multisplit}[2]{\multicolumn{2}{l}{\ensuremath{\begin{array}[t]{l}\left\lbrace\begin{array}{@{}l@{}}\text{#1}\\\text{#2}\end{array}\right.\end{array}}}}
\newcommand{\multisplitthree}[3]{\multicolumn{2}{l}{\ensuremath{\begin{array}[t]{l}\left\lbrace\begin{array}{@{}l@{}}\text{#1}\\\text{#2}\\\text{#3}\end{array}\right.\end{array}}}}
\usepackage{float}
\floatstyle{plain}
\restylefloat{table}

\newcommand{\colrule}{\hline}

\newcommand{\scedilla}{\c{s}}
\newcommand{\tcedilla}{\c{t}}
\newcommand{\abreve}{\u{a}}

\newcommand{\captionopensquare}{\resizebox{0.7em}{!}{$\square$}}
\newcommand{\captionsolidcircle}{\raisebox{-0.3ex}{\resizebox{0.9em}{!}{$\bullet$}}}
\newcommand{\captionopencircle}{\raisebox{-0.3ex}{\resizebox{0.9em}{!}{$\circ$}}}

\newcommand{\captionopenuptriangle}{\resizebox{0.8em}{!}{$\vartriangle$}}

\newcommand{\captionopendowntriangle}{\resizebox{0.8em}{!}{$\triangledown$}}
\newcommand{\captionsoliddiamond}{\resizebox{0.8em}{!}{\rotatebox{45}{$\blacksquare$}}}

\newcommand{\identity}[1]{#1}  
\newcommand{\mathbox}[1]{\mbox{\ensuremath{#1}}}  

\newcommand{\ie}{\textit{i.e}\xperiodspace}
\newcommand{\eg}{\textit{e.g}\xperiodspace}
\newcommand{\vs}{\textit{vs}\xperiodspace}
\newcommand{\etc}{\textit{etc}\xperiodspace}
\newcommand{\etal}{\textit{et al}\xperiodspace}

\newcommand{\Wu}{\ensuremath{\mathrm{W.u}}\xperiodspace}
\newcommand{\timesSunits}{\ensuremath{\times10^{42}\,\mathrm{MeV}^{-2}\mathrm{s}^{-2}}\xspace}
\newcommand{\timesSgammaunits}{\ensuremath{\times10^{-42}\,\mathrm{MeV}^2\mathrm{s}^2}\xspace}

\newcommand{\lt}{\ensuremath{<}}
\newcommand{\gt}{\ensuremath{>}}
\newcommand{\sci}[2]{#1\ensuremath{\times}10\ensuremath{^{#2}}\xspace}
\DeclareMathOperator{\Real}{Re}  
\DeclareMathOperator{\Imag}{Im}

\newenvironment{tablenotes}[1][0.375in]
  { \ssp
    \footnotesize
    \begin{list}{}{\setlength{\topsep}{-0.5\baselineskip}\setlength{\partopsep}{0pt}
                   \setlength{\leftmargin}{#1}\setlength{\rightmargin}{#1}}
    \item[]
  }
  {
    \vspace{0.1in}
    \end{list}
  }
\newenvironment{widetablenotes}[1][0in]
  { \ssp
    \footnotesize
    \begin{list}{}{\setlength{\topsep}{-0.5\baselineskip}\setlength{\partopsep}{0pt}
                   \setlength{\leftmargin}{#1}\setlength{\rightmargin}{#1}}
    \item[]
  }
  {
    \end{list}
  }

\newenvironment{dissenumeratelist}
  {
    \vspace{0.37\baselineskip} 
    \begin{doubleindent}[\parindent]
    \ssp
    \begin{enumeratelist}[0pt]{0.37\baselineskip}
  }
  {
    \end{enumeratelist}
    \dsp
    \end{doubleindent}
  }       

\newenvironment{disscases}
  {\left\lbrace\begin{array}{l@{\hspace{.2in}}l}}
  {\end{array}\right.}
\newenvironment{disscasesclosed}
  {\left\lbrace\begin{array}{l@{\hspace{.2in}}l}}
  {\end{array}\right\rbrace}

\renewcommand{\floatsep}{24pt}  

\dsp 

\begin{document}

\setpapersize{USletter}
\setmarginsrb{1.5in}{1in}{1in}{0.5in}{0pt}{0pt}{11pt}{61pt}

\title{Structure of collective modes\\in transitional and deformed nuclei}
\abstracttitle{Structure of collective modes in transitional and deformed nuclei}
\author{Mark Anthony Caprio}
\degreeyear{2003}
\degreesemester{May}
\degree{Doctor of Philosophy}
\chair{Professor Richard F. Casten}

\begin{frontmatter}
\maketitle
\copyrightpage
\begin{abstract}
\enlargethispage{10pt}

The collective structure of atomic nuclei intermediate between
spherical and quadrupole deformed structure presents challenges to
theoretical understanding.  However, models have recently been
proposed in terms of potentials which are largely independent of the
quadrupole deformation $\beta$ [beta].  These models, E(5) and X(5),
describe a transitional nucleus either free to undergo deviations from
axial symmetry ($\gamma$[gamma]-soft) or confined to axial symmetry,
respectively.

To test these models, information is needed on low-spin states of
transitional nuclei.  The present work involves measurement of
electromagnetic decay properties of low-spin states for nuclei in the
$A$$\approx$100 ($\gamma$-soft) and $N$$\approx$90 (axially symmetric)
transition regions.  Population in $\beta$-decay and thermal neutron
capture are used, and measurements are carried out using $\gamma$-ray
coincidence spectroscopy, fast electronic scintillation timing, and
$\gamma$-ray induced Doppler broadening techniques.  Results are
obtained for $^{102}$Pd, $^{152}$Sm, $^{154}$Dy, $^{156}$Dy,
$^{162}$Yb, and $^{162}$Er, from experiments at the Yale University
Wright Nuclear Structure Laboratory, the TRIUMF ISAC radioactive beam
facility (Vancouver), and the Institut Laue-Langevin high-flux reactor
(Grenoble).

The present data allow detailed comparison of transitional nuclei to
model predictions.  Many characteristics of the $N$=90 transitional
nuclei (Nd, Sm, Gd, Dy) are found to be reproduced by X(5) and similar
$\beta$-soft axially symmetric descriptions.  The nucleus $^{162}$Er,
bordering the $N$$\approx$90 region, appears to support a low-lying
$\beta$-vibrational excitation, indicating comparatively $\beta$-soft
structure.  And $^{102}$Pd is found to match descriptions involving a
$\beta$-soft, nearly $\gamma$-independent, potential.

To facilitate interpretation of these nuclei, a new approach is
developed that simplifies the application of the geometric collective
model (GCM) by use of scaling properties.  Solutions are also obtained
for the E(5) Hamiltonian for finite well depths.

These results demonstrate the relevance of $\beta$-soft potentials,
and in particular the new E(5) and X(5) models, to transitional
nuclei.  They suggest that such nuclei, historically among the most
difficult to describe theoretically, are amenable to descriptions of
comparable simplicity to those used for spherical and well-deformed
nuclei.

\end{abstract}

\tableofcontents
\listoffigures
\listoftables

\chapter*{Acknowledgements}
\addcontentsline{toc}{chapter}{Acknowledgements}

I would like to acknowledge the many people who collaborated and
otherwise contributed throughout the course of this work.

I express my gratitude to Prof.~Richard Casten and Dr.~Victor Zamfir
for being so continuously supportive of this work, and for their
extensive involvement and feedback throughout the process.  I also
especially thank Prof.~Cornelius Beausang and Prof.~Reiner Kr\"ucken
for their contributions to the research presented here and to my
general experience at the Yale Wright Nuclear Structure Laboratory.
The members of the Yale Nuclear Structure Group over the past several
years have made this work both possible and enjoyable, and I thank
Charles Barton, Daeg Brenner, Libby McCutchan, John Ai, Hanan Amro,
Zvi Berant, Gheorghe C{\abreve}ta-Danil, Wen-Tsae Chou, Jeff Cooper,
Ron Gill, G\"ulhan G\"urdal, Adam Hecht, Carsten Hutter, Benyuan Liu,
Deseree Meyer, Heather Newman, John Novak, Norbert Pietralla, Paddy
Regan, Jo Ressler, Zhijun Wang, Alex Wolf, Jing-Ye Zhang, and Lissa
Zyromski for their involvement in this work.  I am also grateful to
the operations and support staff at WNSL~--- Jeff Ashenfelter, John
Baris, Tom Barker, Phil Clarkin, Karen DeFelice, Salvatore
DeFrancesco, Sam Ezeokoli, Paula Fox, Walter Garnett, Bob McGrath,
Craig Miller, Cheryl Millman, Rusty Reeder, Mary Anne Schulz, Jennifer
Tenedine, and Dick Wagner~--- and to Prof.~Peter Parker.

I thank Prof.~Francesco Iachello for his suggestions regarding and
contributions to the model analysis portions of this work.

I also would like to acknowledge those who contributed to the
experiments outside of Yale, including Hans B\"orner, Milan
Krti\v{c}ka, and Paolo Mutti at the Institut Laue-Langevin and Gordon
Ball, Peter Jackson, Pierre Amaudruz, John D'Auria, Marik Dombsky,
Paul Schmor, and Jean-Charles Thomas at TRIUMF.

I thank Prof.~Charles Baltay, Prof.~Beausang, Prof.~Casten,
Prof.~Iachello, and Prof.~Zamfir for serving on the dissertation
committee and Prof.~David Warner for acting as outside reader.

Support was provided during this time by the US Department of Energy,
under grant DE-FG02-91ER-40609, and by Yale University.

\end{frontmatter}


\part{Introduction}
\chapter{Investigation of collective nuclear structure}
\label{chapintro}

\section{The nuclear structure problem}

The internal structure of the atomic nucleus varies greatly and often
suddenly with the number of constituent protons and neutrons.  These
changes in structure are associated with corresponding changes in the
nuclear excitation spectrum and in the decay properties of the excited
states.  The predominant undertaking of the field of nuclear structure
physics is to extract from observed properties of the ground and
excited states of the nucleus an understanding of the physical
structure of these states and to develop a comprehensive theoretical
description of the nuclear system.

The study of nuclei with more than a few
constituent nucleons is inherently the problem of interpreting a
``complex system'', a many-body system with too many constituents to
treated by simple few-body techniques and with too few to
be adequately treated by pure thermodynamic techniques.  Fundamentally,
the behavior of the aggregate nuclear system is entirely determined by
the interaction of its constituents, but it is largely impossible to
deduce even the most basic structural behavior of the system directly
from the intrinsic properties of these constituents.  This
limitation arises in part since the underlying interactions of protons and
neutrons in the nuclear medium are not entirely understood, but, more
importantly, the computational problem of describing a system of tens
or hundreds of interacting protons and neutrons is intractible without
the benefit of some additional simplifications.  Consequently, there
is a need for ``phenomenological'' models of nuclear structure, which
require some degree of empirical input regarding the properties of the
nuclear system in order to make predictions of further properties.
These models serve at the least to provide a rough conceptual
understanding of the properties of nuclei and ideally can allow
detailed quantitative descriptions to be obtained.

\section{Collectivity in nuclear structure}
\label{seccollective}

Specific phenomenological models of nuclear structure are applicable
only to certain classes of nuclei, and the relevant characteristic
for categorization of nuclei is, broadly, the number of nucleons which
``participate'' in the structure, actively contributing to the ground
state properties and excitation modes.  The motivation for this
classification may be understood in terms of a few elementary
properties of the nuclear system.

The nucleus is a system consisting of a fixed number of spin-1/2
particles, the nucleons, and so the space of possible states for the nucleus is the
direct product of the single-particle spaces for the individual
constituents (\eg, Refs.~\cite{dirac1958:qm,messiah1999:qm}).  This space is
spanned by the direct product states
\begin{equation}\label{eqncartesianbasis}
\left\lbrace 
        \left | \mathbf{x}^{(1)}m_s^{(1)} \right \rangle 
        \left | \mathbf{x}^{(2)}m_s^{(2)} \right \rangle 
        \cdots
        \left | \mathbf{x}^{(A)}m_s^{(A)} \right \rangle 
        _\text{AS},
        ~\cdots
\right\rbrace,
\end{equation}
where $A$ is the total number of nucleons, $\mathbf{x}^{(i)}$ and
$m_s^{(i)}$ are the coordinates and spin projection of nucleon $i$,
and the subscript $\text{AS}$ indicates antisymmetrization over like
nucleons.  However, an alternative set of basis states turns out to be
more useful for the present purposes.  The solutions
$\varphi_{nlm_l}(\mathbf{x})$ of the Schr\"odinger equation for a
central potential $V(r)$,
\begin{equation}\label{eqncentralse}
\left [ -\frac{\hbar^2}{2m} \nabla^2 + V(r) \right ]
\varphi_{nlm_l}(\mathbf{x})
= \varepsilon_{nlm_l}\varphi_{nlm_l}(\mathbf{x}),
\end{equation}
where $n$, $l$, and $m_l$ are radial, orbital angular momentum, and
orbital angular momentum projection quantum
numbers~\cite{dirac1958:qm,messiah1999:qm}, form a complete set of
functions on three-dimensional coordinate space.  Consequently, the
direct products of the single particle states constructed from these
functions constitute a basis for states of the nucleus as a whole,
\begin{equation}\label{centralbasis}
\left\lbrace 
        \left | nlm_lm_s^{(1)} \right \rangle 
        \left | nlm_lm_s^{(2)} \right \rangle 
        \cdots
        \left | nlm_lm_s^{(A)} \right \rangle 
        _\text{AS},
        ~\cdots
\right\rbrace.
\end{equation}
This basis has the virtue that, if the Hamiltonian for the nucleus
happens to have the special form
\begin{equation}\label{eqnmf}
\hat{H}_0= \left [ \sum_i \frac{\hat{p}_i^2}{2m_i}+V(\hat{r}_i)\right],
\end{equation}
where $\hat{p}_i$ and $m_i$ are the momentum and mass of the $i$th nucleon,
then the nuclear eigenstates will be these very basis states.  If the
nuclear Hamiltonian is only \textit{approximately} of this form, \ie,
\begin{equation}\label{eqnmfres}
\hat{H}= \hat{H}_0 + \hat{V}_\text{res},
\end{equation}
where the difference is denoted as the ``residual'' interaction
$\hat{V}_\text{res}$, then the nuclear eigenstates will still
\textit{predominantly} be individual direct product states, but with  
admixtures of the other basis states.  Extensive empirical
evidence~\cite{heyde1990:sm} suggests that such a Hamiltonian
successfully describes a wide range of nuclear phenomena.

\begin{figure}%
\begin{center}%
\includegraphics*[width=0.65\textwidth]{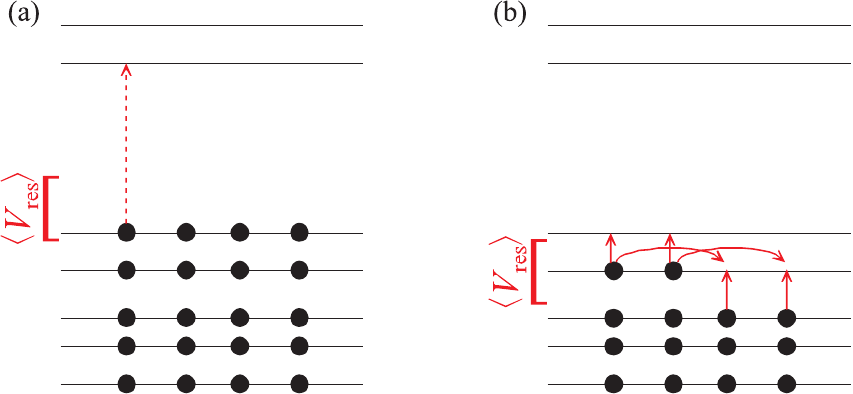}%
\end{center}%
\caption%
[Filling of single-particle levels in closed-shell and open-shell
nuclei.] {Filling of single-particle levels for a single species of
nucleon in (a) closed-shell and (b) open-shell nuclei.  In an
open-shell nucleus, the residual excitation can introduce amplitudes
for states with nucleons excited to nearby single-particle levels
and/or to different $m$ substates of the same level (spin recoupling).
\label{figfilling}
}%
\end{figure}

The spacing between the single-particle energy levels
$\varepsilon_{nlm_lm_s}$, and whether or not sizeable gaps exist
between successive energy levels, is a central factor determining the
structure of a nucleus (\eg, Ref.~\cite{heyde1990:sm}).  In the
absence of a residual interaction, the nuclear ground state would
simply be that direct product state, degenerate in some $m_l$ and
$m_s$, in which nucleons occupy, or ``fill'', exactly the
lowest-energy single-particle energy levels $\varepsilon_{nlm_lm_s}$,
as shown schematically in Fig.~\ref{figfilling}(a).  The residual
interaction can introduce significant amplitudes for occupation of
higher-lying single-particle levels only if the spacing between these
levels is not much larger than the residual interaction strength.
Consequently, depending upon whether or not nucleons have filled all
the energy levels below a sizeable gap in the single-particle level
spectrum [compare Fig.~\ref{figfilling}(a) and (b)], the ground state
and excited spectra of the nucleus will have markedly different
properties.%

The presence of amplitudes for many direct product states in a single
nuclear eigenstate allows for the existence of so-called
``collective'' phenomena, resulting from the addition of these
components with coherent phases.  Collective structure corresponds
semi-classically to coherent motion of nucleons, including possible
bulk motion of nuclear matter.  Features identified with collectivity
include deformation of the nucleus from spherical shape, smaller spacing
between energies of the nuclear states than would arise from the
single-particle energy scale, and enhanced electromagnetic transition
rates between collective levels.  A wide
variety of collective phenomena are possible, ranging from excitations
involving small numbers of nucleons to bulk hydrodynamic flow of
matter throughout the entire
nucleus~\cite{bohr1998:v2,eisenberg1987:v1}.

\begin{figure}
\begin{center}
\includegraphics*[width=0.65\textwidth]{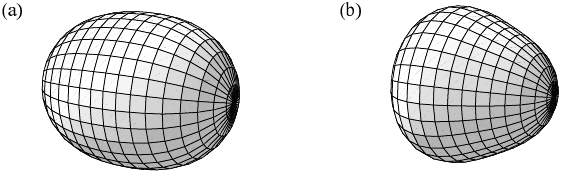}
\end{center}
\caption
[
Liquid drop shapes with quadrupole and octupole deformation.
]
{
Liquid drop shapes with (a) quadrupole and (b) octupole deformation.
\label{figmultipole}
}
\end{figure}

The collective structure of nuclear states at low excitation energy is
dominated by surface deformations, in which a surface layer of
nucleons takes on nonspherical shapes.  If the surface shape is
described by a multipole expansion of the radius as a function of
angle relative to the center of the nucleus,
\begin{equation}
\label{eqnmultipoleshape}
R(\theta,\varphi) = R_0 \left [1+ \frac{1}{\sqrt{4\pi}}\alpha_{00}
+\sum_\mu \alpha_{2\mu} Y_{2\mu}^*(\theta,\varphi)
+\sum_\mu \alpha_{3\mu} Y_{3\mu}^*(\theta,\varphi)
+\cdots\right] ,
\end{equation}
then usually the lowest order terms~--- monopole, quadrupole, and
sometimes octupole~--- are sufficent to describe the phenomena of
interest.  (The dipole term produces center of mass translation.)
Fig.~\ref{figmultipole} illustrates quadrupole-deformed and
octupole-deformed shapes.  The nucleus does not exist in any one shape
but rather in a quantum superposition of shapes: the quantities
$\alpha_{2\mu}$ are dynamical variables, for which Schr\"odinger-like
equations can be devised (Chapter~\ref{chapgcm}) and which have
probability distributions associated with them.  Loosely, however, we
may distinguish between ``undeformed'' nuclei, in which the shape
fluctuations are centered about a spherical shape, and ``deformed''
nuclei, in which the fluctuations are centered about a deformed,
usually quadrupole-deformed, shape.

\begin{figure}
\begin{center}
\includegraphics*[width=1.0\textwidth]{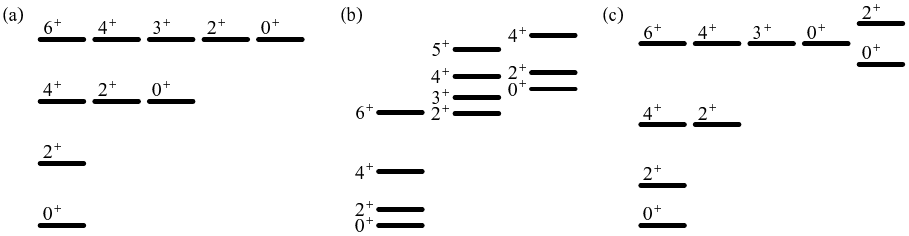}
\end{center}
\caption
[ Level schemes for quantized liquid drops.]  { Level schemes for quantized
liquid drops: (a) spherical oscillator, (b) deformed rotor, and (c)
deformed $\gamma$-soft.  Total angular momentum (``spin'') and parity
quantum numbers for the levels are indicated.
\label{figbasicschemes}
}
\end{figure}

\begin{figure}
\begin{center}
\includegraphics*[width=0.8\textwidth]{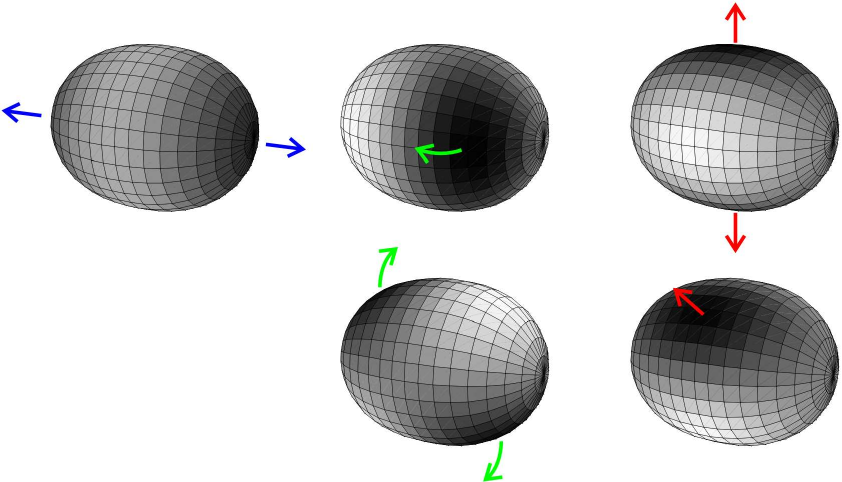}
\end{center}
\arxivonlyshrink
\caption
[Classical quadrupole modes of a deformed liquid drop.]  {Classical
quadrupole modes of a liquid drop with deformed equilibrium shape.
Dark portions of the surface indicate outward radial motion, and light
portions indicate inward radial motion (or vice versa, as an
oscillatory function of time).  The modes shown are those due to
$Y_{20}$ (left), $Y_{2\pm1}$ (middle), and $Y_{2\pm2}$ (right),
defined relative to the symmetry axis as the polar axis.  The
resulting motions are, respectively, oscillations in which the poles and equator
alternately expand, rotation of the surface shape, and standing or
travelling waves about the equator.
\label{figquadrupolemodes}
}
\end{figure}

Classically, the modes of excitation available to a nucleus undergoing
quadrupole surface deformation are the quadrupole normal modes of a
liquid drop~\cite{bohr1998:v2}, and the quantized system shows
features recognizably related to these classical modes.  For a liquid
drop with a spherical shape in equilibrium, the five vibrational modes
available, one for each $Y_{2\mu}$, are degenerate, resulting in just
one fundamental frequency $\omega$ of oscillation.  The corresponding
quantized energy spectrum consists of a series of phonon excitations,
with excitation energies $E=n\hbar\omega$ spaced according to that
fundamental frequency [Fig.~\ref{figbasicschemes}(a)].
If a liquid drop instead has a deformed equilibrium shape,
the degeneracy of its modes is broken.  Fig.~\ref{figquadrupolemodes}
depicts the surface motion induced by each of the $Y_{2\mu}$ for the
case of an axially-symmetric deformed equilibrium shape: $Y_{20}$
induces an oscillation in which the poles and equator alternately
expand, $Y_{2\pm1}$ simply induce a rotation of the surface shape, and
$Y_{2\pm2}$ produce standing and travelling waves about the equator,
causing oscillations about axial symmetry.  The quantized spectrum for
an axially-symmetric deformed liquid drop exhibits ``bands'' of
rotationally excited levels with energies following the spacing
dictated by quantized angular momentum
[$E=(2\mathcal{J})^{-1}|\mathbf{J}|^2=(2\mathcal{J})^{-1}\hbar^2J(J+1)$,
where $\mathcal{J}$ is a moment of inertia] superposed upon the ground
state and upon excitations corresponding to the vibrational modes
[Fig.~\ref{figbasicschemes}(b)].  A special case occurs classically
for a liquid drop which energetically prefers a deformed shape but
which is insensitive to deviations from axial symmetry, since such an
object does not support equatorial vibrations.  The quantum treatment
of such a system (termed ``$\gamma$-soft'' for reasons explained in
Appendix~\ref{appbench}) is especially simple mathematically and is
directly relevant to the description of certain nuclei, so its
excitation spectrum is also shown [Fig.~\ref{figbasicschemes}(c)].

The forms of quadrupole collective structure just described each thus
posess a distinctive, recognizable characteristic pattern of level
energies, and there are also specific selection rules and strength
patterns followed by the electromagnetic transitions between levels.
(The quantitative details are important to discussions throughout the
present work and so are summarized in Appendix~\ref{appbench}.)
Several examples exist of nuclei which exhibit, to varying degrees,
the properties of a harmonic oscillator, an axially-symmetric rotor,
or a deformed $\gamma$-soft system (see, \eg,
Ref.~\cite{casten2000:ns}).  These structures are, however, extreme
cases, only found when the deformation potential for the nucleus takes
on idealized forms, stated more precisely in Appendix~\ref{appbench}.
A description of collective nuclear quadrupole structure must address
the phenomena encountered in the ``transitional'' nuclei, those with
structures intermediate between the ideal limits.

\begin{figure}
\begin{center}
\includegraphics*[width=0.45\hsize]{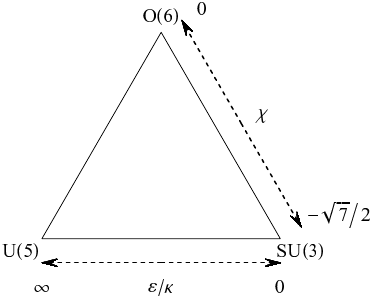}
\end{center}
\caption[Symmetry triangle illustrating the parameter space for the IBM.]
{Symmetry triangle illustrating the parameter space for the IBM with a
quadrupole Hamiltonian.  The U(5), SU(3), and O(6) limits correspond
to an oscillator, rotor, and deformed $\gamma$-soft structure, respectively.
\label{figibmtriangle}
}
\end{figure}

Approaches exist which allow treatment of these three extreme forms of
structure together with a continuum of transitional cases using a
single parametrized model Hamiltonian.  Great success in the
description of the low-lying levels of a large number of collective
nuclei has been achieved using the interacting boson model
(IBM)~\cite{iachello1987:ibm}.  This is an algebraic model in which
the Hamiltonian is built from the Casimir operators of subalgebras of
U(6).  These operators act upon a bosonic basis for a representation of U(6)
consisting of states defined by their occupation numbers for two
species of boson: a spin-0 boson with creation operator $s^\dagger$
and a spin-2 boson with creation operator $d^\dagger$.  Such a description is
appropriate for a system with five coordinate
degrees of freedom and thus for a liquid drop undergoing quadrupole
deformation~\cite{iachello1987:ibm}.  The three limiting
structures just discussed correspond to three dynamical symmetries~--- U(5), O(6),
and SU(3)~--- of the model (Appendix~\ref{appbench}).  A simple
``quadrupole'' form of the IBM
Hamiltonian~\cite{warner1983:cqf,lipas1985:ecqf}
\begin{equation}
H=\varepsilon \hat{n}_d-\kappa \hat{Q}^\chi\cdot\hat{Q}^\chi, 
\end{equation}
where $\hat{n}_d$$\equiv$$(d^\dag\cdot\tilde{d})$ and
$\hat{Q}^\chi$$\equiv$$[d^\dag\times\tilde{s}+s^\dag\times\tilde{d}]^{(2)}+\chi[d^\dag\times\tilde{d}]^{(2)}$,
reproduces all three limiting structures using only
two parameters, $\varepsilon/\kappa$ and $\chi$.  The parameter space
is commonly illustrated using a symmetry triangle
diagram~\cite{casten1978:triangle}, as shown in
Fig.~\ref{figibmtriangle}.  In the limit of large boson number for the
U(6) representation basis, a geometric Hamiltonian is recovered from
the IBM~\cite{ginocchio1980:ibm-classical,dieperink1980:ibm-classical}.  Approaches
using a parametrized geometric Hamiltonian are discussed in
Chapter~\ref{chapgcm}.

\section{Transitional phenomena}
\label{sectrans}

Some basic aspects of the transition between spherical and deformed
shape may be understood schematically from the simple picture already
presented.  Recent work, outlined in this section, provides components
of a more complete understanding.  The shape transition occurs over a
series of a few ``neighboring'', \ie, adjacent on the nuclear chart,
nuclides.  The degree of deformation of a particular nucleus can
usually be quite well estimated simply on the basis of the number of
protons and neutrons present outside closed shells (``valence
nucleons'')~\cite{casten1985:npnn,casten1996:evolution}, though the
onset of deformation does also depend upon the particular orbitals
involved~\cite{federman1977:onset,federman1978:onset,wolf1987:n90onset}.

\begin{figure}
\begin{center}
\includegraphics*[width=1.0\textwidth]{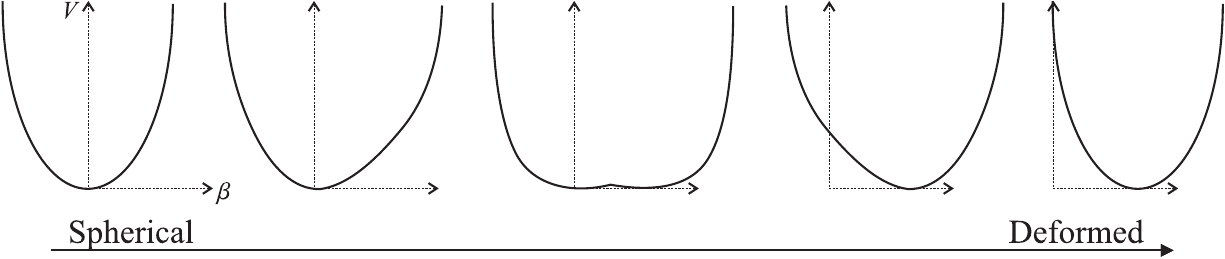}
\end{center}
\caption
[Schematic view of the evolution from undeformed to deformed
structure.]  {Schematic view of the evolution from undeformed to
deformed structure, proceeding through anharmonicity, strongly
transitional structure, and then softly deformed structure.  The
potential is shown here simply as a function the deformation
coordinate $\beta$, though the actual quadrupole shape potential is a
function of two variables (see text).
\label{figpotlevoln}
}
\end{figure}

A schematic illustration showing how a shape transition can be
expected to evolve is given in Fig.~\ref{figpotlevoln}.  For each of
several steps along the evolution from spherical to deformed
structure, Fig.~\ref{figpotlevoln} shows the potential energy as a
function of a deformation coordinate $\beta$.  (The quadrupole shape
deformation can be reexpressed in terms of a variable $\beta$ giving
the overall magnitude of deformation and a variable $\gamma$
describing the deviation from axial symmetry, as defined in
Appendix~\ref{appbench}.)

\begin{figure}[tb]
\begin{center}
\includegraphics*[width=0.45\hsize]{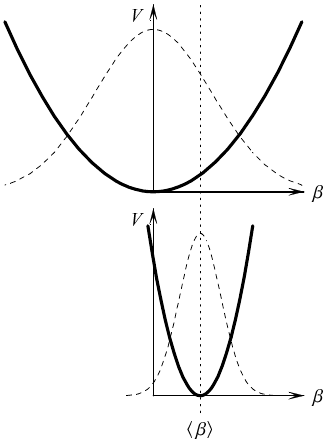}
\end{center}
\caption[Illustration of distinction between deformation magnitude and structure.]  
{Illustration of the distinction between deformation magnitude and
structure.  A structurally ``undeformed''
nucleus can have the same average deformation (dotted line) as a
structurally ``deformed'' nucleus.  Schematic deformation potentials
(solid lines) and wave functions (dashed lines) are shown.  [The
deformation $\beta$ is a radial variable (Appendix~\ref{appbench}) and
assumes only positive values, hence the nonzero mean in the upper
panel.]
\label{figfatosc}
}
\end{figure}

Two distinct but coupled aspects of the nuclear deformation vary over
the course of a shape transition~--- the magnitude of the deformation
and the range of deformation (softness or stiffness) accessible to the
nucleus~--- and it is essential to distinguish these.  From a
structural point of view, a ``spherical'' nucleus is one for which a
deformation $\beta$=0 is strongly energetically preferred, a
``deformed'' nucleus is one in which a specific deformation
$\beta$$\neq$0 is strongly energetically preferred, and a
``transitional'' nucleus is one for which $\beta$=0 and $\beta$$\neq$0
are both accessible without significant energetic preference for one
over the other.  It is this range of accessible $\beta$ values which
determines the qualitative nature of the energy spectrum as being
oscillator-like, rotor-like, \etc; a nucleus with a harmonic potential
centered on $\beta$=0 will posess a ``spherical'' oscillator-like
spectrum even if the expectation value for its deformation is large,
and a nucleus with a stiff nonzero deformation will behave like a
``deformed'' rotor even if this deformation is small (see
Fig.~\ref{figfatosc}).  On the other hand, once rotor-like behavior is
established, the magnitude of deformation directly controls the moment
of inertia of the nucleus and hence the energy scale of rotational
excitations.  This relationship holds since, for a collective nucleus,
only the deformed surface layer of the nucleus is free to participate
in rotation.  Thus, the moment of inertia is much more directly
related to deformation than it is for rigid body rotation and vanishes
at zero deformation.

For nuclei with a structure near one of the ideal limits, \ie, near
one end of the transition in Fig.~\ref{figpotlevoln}, a perturbative
approach is fruitful.  The eigenstates for the ideal limit are taken
as a zeroth-order approximation to the states of the system, and then
the actual perturbed states are obtained as the result of weak mixing between
the original unperturbed states.  This description has been very
successful for relatively well-deformed rotor nuclei
(Section~\ref{secbenchrotor}).  

In the main part of the transition region, the eigenstates of the
transitional nucleus can still be decomposed in terms of the
eigenstates of one of the ideal structural limits but will in general
have significant amplitudes for a large number of basis states,
precluding any simple perturbative approach.  Description of the
nuclear excitation spectrum must usually be based upon numerical
solution of some model Hamiltonian.  In transitional
nuclei, a given nuclear eigenstate may have a probability distribution
spanning a range of deformations.  Also, among the spectrum of
eigenstates, there may be ``coexistence''
of states of large deformation with those of small deformation.

\begin{figure}
\begin{center}
\includegraphics*[width=0.48\hsize]{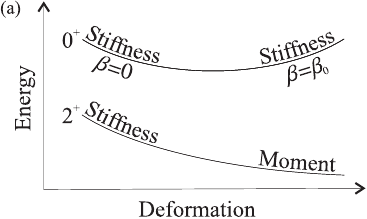}
\hfill
\includegraphics*[width=0.48\hsize]{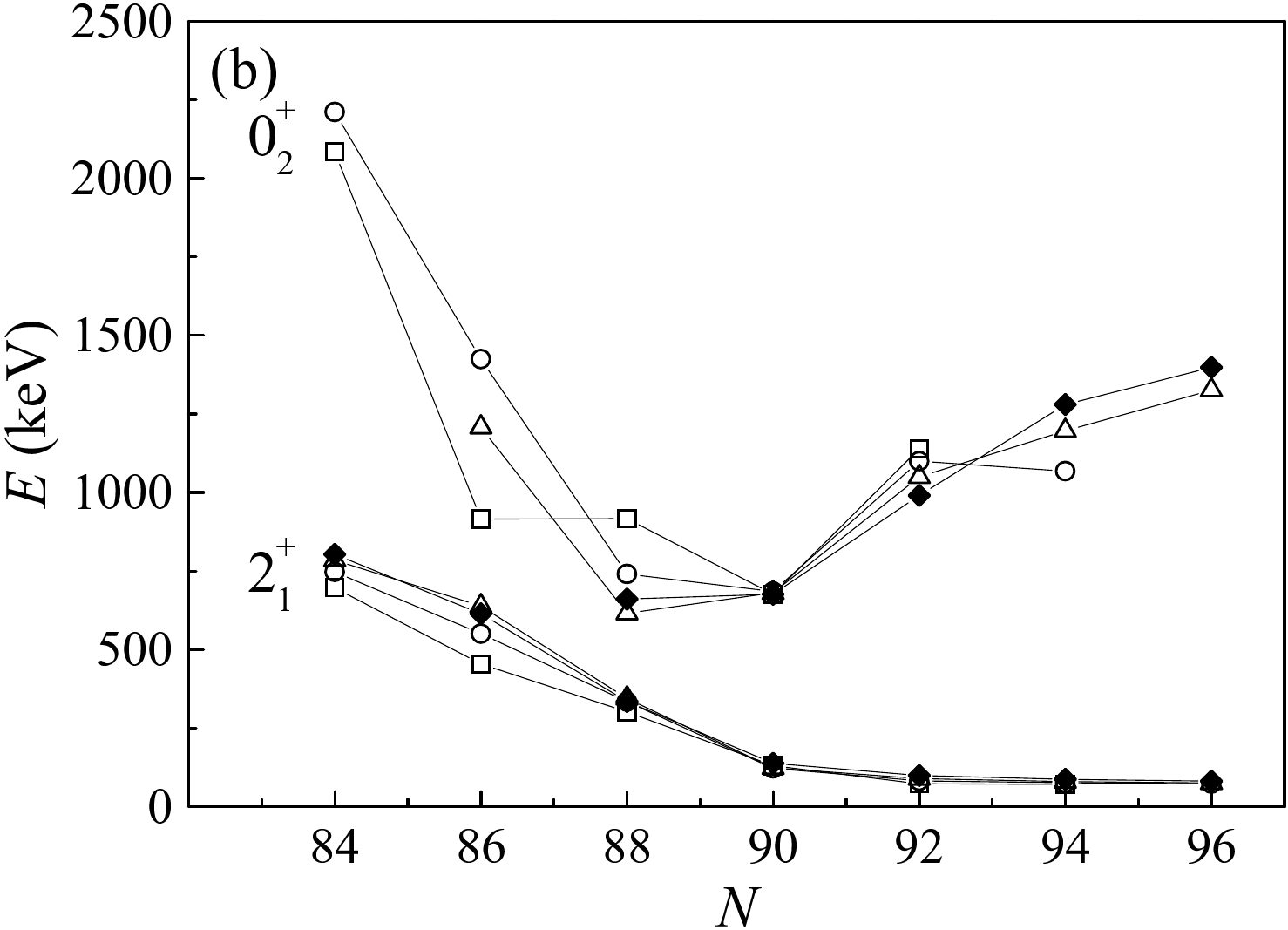}
\end{center}
\caption[Evolution of the $2^+_1$ and $0^+_2$ states across the transition.]  {Evolution of the $2^+_1$ and
$0^+_2$ states across the spherical to axially-symmetric deformed
transition (a) in a schematic picture (see text) and (b) for
Nd~(\captionopensquare), Sm~(\captionopencircle),
Gd~(\captionopenuptriangle), and Dy~(\captionsoliddiamond) isotopic
chains in the $N$$\approx$90 region.  Data are from
Refs.~\cite{nds1989:144,nds1997:146,nds2000:148,nds1995:150,nds1996:152,nds1998:154,ndsboth:156,nds1996:158,nds1996:160,nds1999:162}.
(The first excited $0^+$ state might not be correctly known in all
cases and is not necessarily collective.)
\label{fig2vs0}
}
\end{figure}

Different nuclear observables are sensitive to different aspects of
the onset of nuclear deformation.  Consider the behavior qualitatively
expected for the first excited $2^+$ and $0^+$ states over the course
of the evolution portrayed in Fig.~\ref{figpotlevoln}.  For the
``stiff'' oscillator, starting at the left side of the figure, the
$0^+$ energy is twice the $2^+$ energy, and both energies are
relatively high.  As the potential well becomes wider or ``softer'',
both energies settle lower, and some anharmonicity is likely to be
introduced as well.  Proceeding further to the right in the figure,
rotational behavior sets in, and the first $2^+$ state changes in
nature from being a vibrational excitation, with an energy determined
by the well stiffness, to being a rotational band member, hence with
an energy determined by the nuclear moment of inertia.  The $2^+$
level energy therefore continues to decrease as the deformation, and
thus moment of inertia, increase.  The $0^+$ state, however, retains
its vibrational nature, evolving into the $\beta$ vibrational bandhead
(Section~\ref{secbenchrotor}).  As the potential becomes increasingly
stiff, now about a deformed $\beta$ value, the $0^+$ level again rises
in energy.  This schematic picture of the roles of the $2^+$ and $0^+$
excitations is summarized in Fig.~\ref{fig2vs0}(a).  Such a general
trend in behavior is illustrated with actual data from the
$N$$\approx$90 transition region in Fig.~\ref{fig2vs0}(b).  This
description should not be viewed as either complete or universally
valid~--- it is qualitative in nature and certainly does not contain
the full complexities of a solution of the quadrupole collective
Hamiltonian, and, moreover, the lowest $0^+$ excitation often lies
outside the collective model space~--- but it does indicate which
levels may be sensitive to different aspects of the nuclear structure.

Over the past several years, renewed experimental attention has been
directed towards transitional nuclei.  The experimental
investigation benefits greatly from the development of high-efficiency
multi-detector arrays of high-resolution Ge detectors for $\gamma$-ray
spectroscopy.  Greatly improved spectroscopy (Chapter~\ref{chapspec})
removes a considerable ``clutter'' of misinformation previously
hindering the interpretation of transitional nuclei and also makes
possible intensity measurements for weak transitions sensitive to the
structure of these nuclei.  Measurements of one such transition, the
$2^+_3\rightarrow0^+_2$ transition in
$^{152}$Sm~(Refs.~\cite{casten1998:152sm-beta,zamfir1999:152sm-beta}
and Chapter~\ref{chap152sm}), and the discovery of its unexpectedly
low strength, significantly constrained model descriptions and
encouraged much of the recent theoretical and experimental work.  More
generally, model analysis benefits from having as full and accurate a
set of data available as possible.

Several theoretical ideas have been advanced regarding structural
evolution in the transition region.  Studies involving the IBM
revealed that the transition from spherical to deformed structure has
characteristics of a first or second order phase
transition~\cite{dieperink1980:ibm-classical,iachello1998:phasecoexistence,jolie1999:ibm-phase},
depending upon the path taken through parameter space.  (The second
order phase transition occurs only for the special case in which
complete $\gamma$ softness is maintained along the transition.)
Following a very different approach, empirical examination of basic
observables, \eg, level energies and separation energies, for large
sets of nuclei shows sharp discontinuities in
behavior~\cite{casten1993:anharm,wolf1994:difflobs} to occur at a
specific point along the transition [specifically, where
$E(2^+_1)$$\approx$130--145\,keV], again closely resembling the
behavior expected for a first order phase transition.

With the new experimental data available, model Hamiltonian parameter
values relevant to specific transitional nuclei can be identified much
more closely, and several examples have been found of nuclei near the
first order phase transition in the IBM. In the immediate region of
parameter space surrounding the first order phase transition,
individual states can be decomposed as superpositions of two
well-defined components~--- one oscillator-like and one rotor-like~---
with only a small residual remaining~\cite{zamfir1999:152sm-beta}.
Among the eigenstates, some with predominantly oscillator-like IBM
wave functions are found to coexist with others with predominantly
rotor-like wave
functions~\cite{iachello1998:phasecoexistence,jolie1999:ibm-phase}.
These observations have led to a treatment, in
Ref.~\cite{zamfir2002:ibm-twostate}, of the evolution of structure
across the transition region in the IBM as approximately a two-state
mixing problem, in which a ``spherical'' $0^+$ state and ``deformed''
$0^+$ state undergo an avoided crossing at the critical point of the
transition.  The improved data also constrain the appropriate
parameter values in geometric models (Chapter~\ref{chapphenom}).

\begin{figure}
\begin{center}
\includegraphics*[width=0.85\hsize]{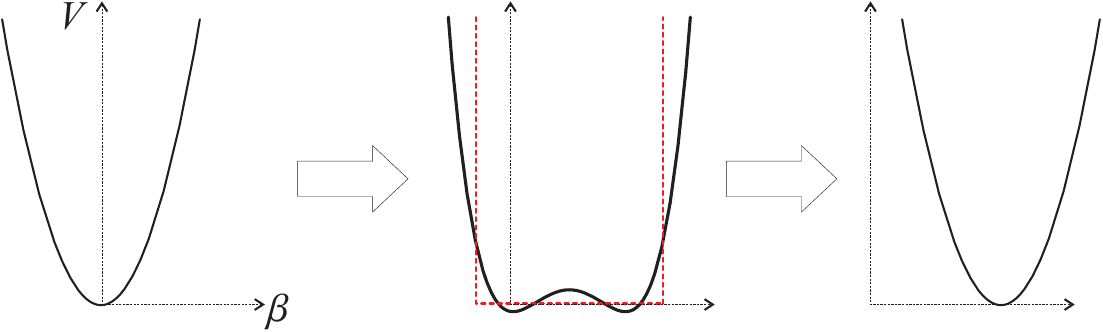}
\end{center}
\caption[The square well as a transitional potential.]
{The square well as an approximation for the transitional potential.
Schematic potentials for the transition from spherical to
deformed structure are shown.
\label{figflatbottom}
}
\end{figure}

An especially simple geometric description for nuclei near the
critical point of the phase transition has been proposed by
Iachello~\cite{iachello2000:e5,iachello2001:x5}.  For such nuclei, the
deformation potential is expected to be relatively ``flat'' in
$\beta$, allowing the nucleus to assume either a spherical or deformed
shape with minimal energy penalty.  A flat-bottomed potential can
reasonably be approximated by an infinite square well potential, as
illustrated in Fig.~\ref{figflatbottom}.
For a square well potential, the solution wave functions are analytic
in form (spherical Bessel functions) and the eigenvalues are simply
given in terms of the zeros of the Bessel functions.  A simple square
well in the deformation variable $\beta$~--- termed E(5)~--- is
appropriate to the $\gamma$-soft transition.  A square well in $\beta$
with an additional term providing stabilization in the $\gamma$ degree
of freedom~--- termed X(5)~--- is used for the description of
axially-symmetric transitional nuclei.  The E(5) and X(5)
models are summarized in Section~\ref{secbenche5x5}.  Empirical
examples of nuclei described by the E(5) and X(5) Hamiltonians have
been identified~(see
Refs.~\cite{casten2000:134ba-e5,casten2001:152sm-x5,kruecken2002:150nd-rdm,arias2001:134ba-e5,frank2001:104ru-e5}
and Chapters~\ref{chap152sm}, \ref{chap156dy}, and~\ref{chap102pd}).
The E(5) Hamiltonian has the special property that it exhibits a
dynamical symmetry, that of the five-dimensional Euclidan algebra, and
is thus the first recognized example a dynamical symmetry based upon a
nonsemisimple Lie algebra~\cite{iachello2001:x5,iachello2001:dynsym}.

\begin{figure}
\begin{center}
\includegraphics*[width=\textwidth]{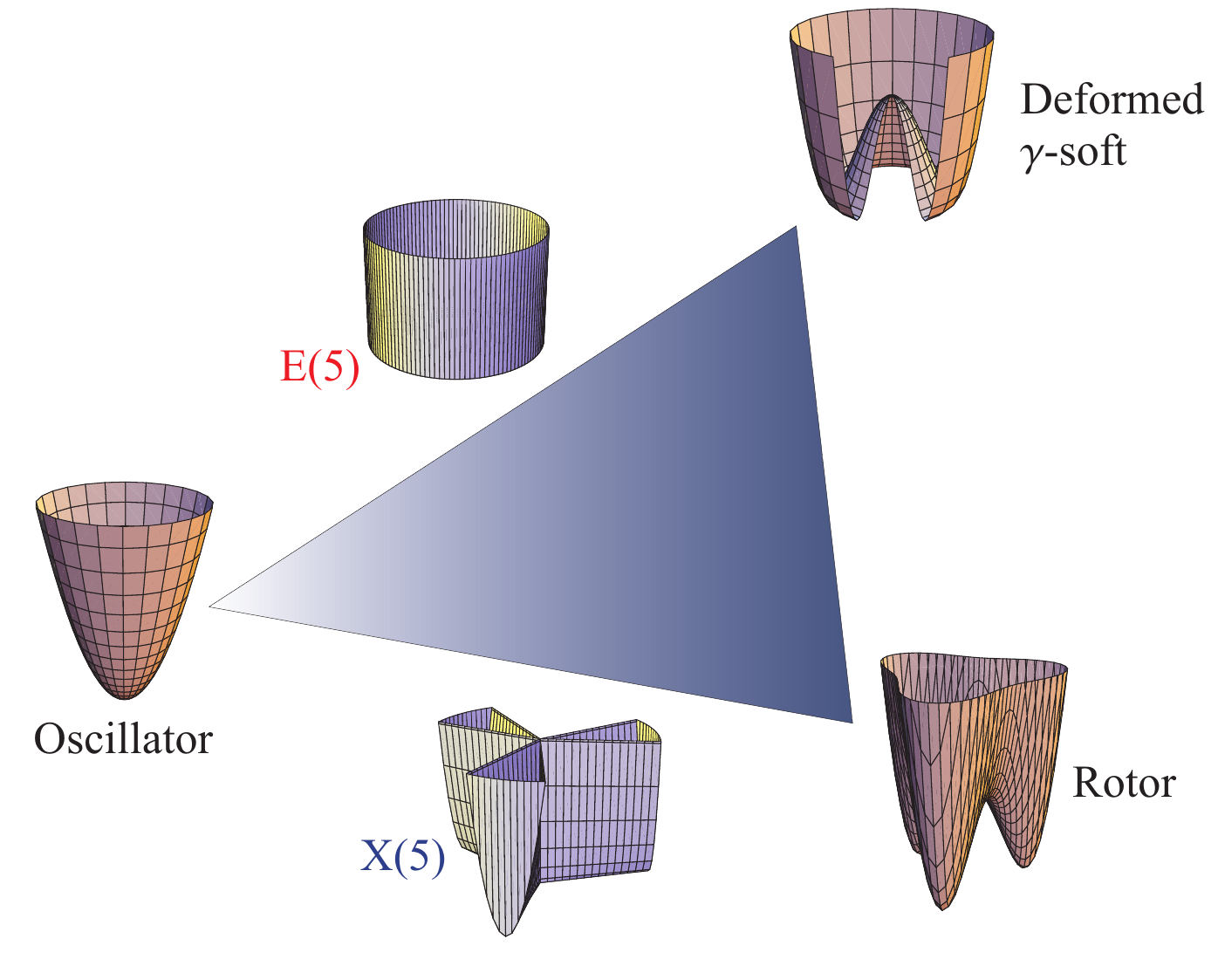}
\end{center}
\caption
[Potential energy surfaces, including the E(5) and X(5) potentials.]
{Potential energy surfaces producing spherical oscillator,
deformed $\gamma$-soft, and rotor structures along with the recently-proposed
E(5) and X(5) potentials for transitional cases.  The radial
coordinate for the potential plots is the deformation $\beta$,
and the angular coordinate is the deviation $\gamma$ from axial
symmetry.  (Figure from Ref.~\cite{caprio2002:critex}.)
\label{figflyingtriangle}
}
\end{figure}

The E(5) and X(5) models provide analytical ``benchmarks'' for the
interpretation of transitional nuclei much as the structural limits
introduced in Section~\ref{seccollective} provide benchmarks for the
interpretation of undeformed or well-deformed nuclei.  The E(5) and
X(5) potentials, as functions of the deformations variables $\beta$
and $\gamma$, are shown along with the potentials for the structural
limits in Fig.~\ref{figflyingtriangle}.

\section{Overview of the present work}
\label{secoverview}

\begin{figure}
\begin{center}
\includegraphics*[width=\textwidth]{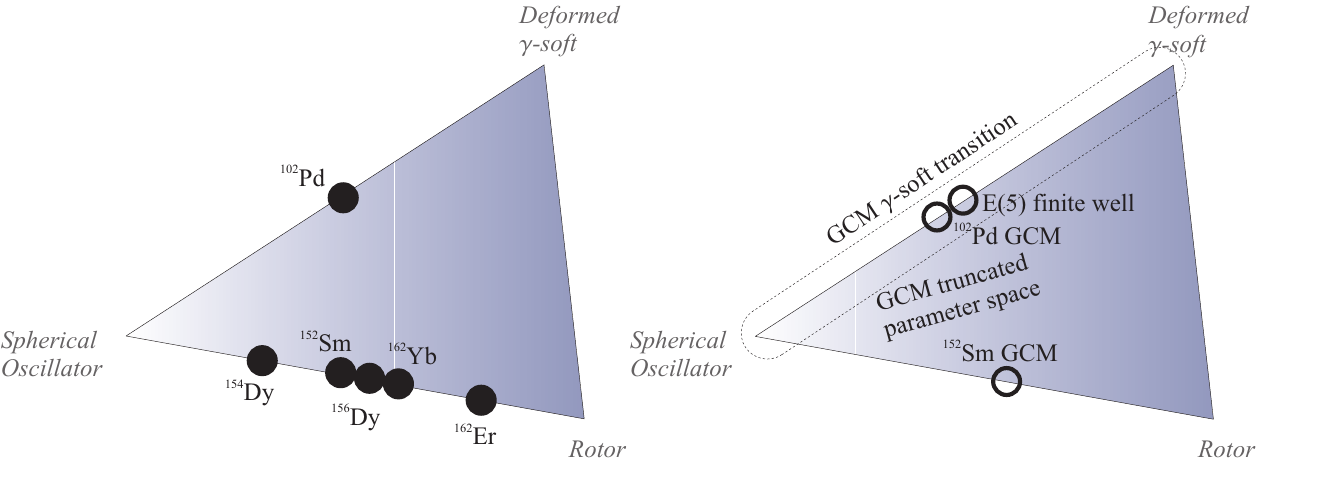}
\end{center}
\caption
[Structural investigations discussed in the following chapters.]
{Structural investigations discussed in the following chapters,
including both experimental work (left) and theoretical
studies (right), shown schematically in the context of the
axially-symmetric and $\gamma$-soft transitions.  
\label{figtrianglemap}
}
\end{figure}

The following chapters present experimental and theoretical
investigations of the nuclear quadrupole shape transitions, involving
both axially-symmetric and $\gamma$-soft structures.  The structural
domain covered by these studies is plotted in
Fig.~\ref{figtrianglemap}, which shows where each individual nucleus
or model investigation falls on a schematic version of the
symmetry triangle diagram.

\begin{figure}[t]
\begin{center}
\includegraphics*[width=0.9\textwidth]{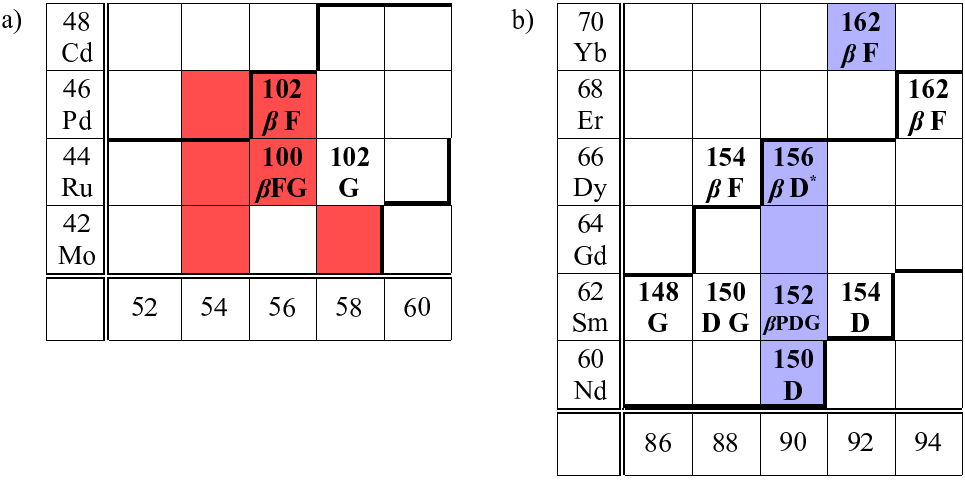}
\end{center}
\caption[Recent experimental work in the
$A$$\approx$100 and $N$$\approx$90 regions.] {Recent
experimental work by the Yale nuclear structure group and collaborators
investigating low-energy collective modes in the (a) $A$$\approx$100 and
(b) $N$$\approx$90 ($A$$\approx$150--160) transitional regions.
Nuclei studied are shown on a standard $Z$ \vs $N$ nuclear
chart. Shaded squares in parts (a) and (b) indicate a $4^+$ to $2^+$
level energy ratio approximately consistent with the E(5) or X(5)
model predictions, respectively.  Experimental techniques indicated
are $\gamma\gamma$ or $\gamma\gamma(\theta)$ spectroscopy in $\beta$
decay~($\beta$), FEST lifetime measurement in $\beta$ decay~(F),
Compton polarimetry in $\beta$ decay~(P), Doppler lifetime measurement
following population by Coulomb excitation (D), Doppler lifetime
measurement following population by heavy-ion
fusion-evaporation~(D$^*$), and GRID lifetime measurment and
eV-resolution $\gamma$-ray spectroscopy following neutron capture~(G)
(see Part~\ref{partexpttechniques}).
\label{figyaleexpts}
}
\end{figure}

The experimental work described consists of spectroscopic studies and
related measurements making use of $\gamma$-ray detection.
Experiments concentrated on the $N$$\approx$90 ($A$$\approx$150--160)
axially-symmetric transition region and the $A$$\approx$100 (Pd, Ru)
$\gamma$-soft transition region.  A summary of recent experiments
involving the Yale nuclear structure group investigating nuclei in
these regions, including the work discussed here, is given in
Fig.~\ref{figyaleexpts}.  Much of the work described in the following
chapters was carried out at Yale University's Wright Nuclear Structure
Laboratory (WNSL), which houses the 21\,MV Extended Stretched
Trans-Uranic (ESTU) tandem Van de Graaff
accelerator~\cite{hyder1988:estu,ashenfelter2002:estu}.  The research
presented also includes one of the first experiments to be carried out
at the TRIUMF Isotope Separator and Accelerator (ISAC)
facility~\cite{schmor1999:isac} in Vancouver, 
Canada, and measurements performed at the 58\,MW high-flux reactor of
the Institut Laue-Langevin in Grenoble, France.

The experimental goals (Chapter~\ref{chapexpt}) and the methods
applied are outlined in Part~\ref{partexpttechniques}.  Experiments
were carried out making use of $\beta$ decay 
at stable and radioactive beam facilities (Chapter~\ref{chapbeta}) and of neutron
capture (Chapter~\ref{chapgrid}).  The availability of high-efficiency
multidetector arrays for $\gamma$-ray detection requires new
approaches to be taken to spectroscopic measurements in decay studies
(Chapter~\ref{chapspec}).

Experimental results for specific nuclei are detailed in
Part~\ref{partexptresults}.  The nuclei discussed include the 
transitional nuclei $^{152}$Sm (Chapter~\ref{chap152sm}), $^{156}$Dy
(Chapter~\ref{chap156dy}), $^{162}$Yb (Chapter~\ref{chap162yb}),
$^{154}$Dy (Chapter~\ref{chap154dy}), and $^{102}$Pd
(Chapter~\ref{chap102pd}), as well as the relatively well-deformed
rotational nucleus $^{162}$Er (Chapter~\ref{chap162er}) bordering the
$N$$\approx$90 transitional region.%

Model investigations are discussed in Part~\ref{partmodelanalysis}.
Analytic relations are used to simplify the use of the geometric
collective model (GCM) in Chapter~\ref{chapgcm}.  These results are
used to survey structural evolution throughout the GCM model space.
Several special topics~--- the $\gamma$-soft transition in the GCM,
and E(5)-like and X(5)-like structure in the GCM~--- are also
considered (Chapter~\ref{chapgcmspecial}).  The E(5) model is solved
for finite well depth, addressing concerns which had been raised about
the application of the E(5) model to actual
nuclei~\cite{iachello2000:e5}, and the effects of finite depth on
observables are studied (Chapter~\ref{chapfwell}).  The phenomenology
of the $N$$\approx$90 transition region is summarized in the context
of the X(5) model and GCM predictions, and an analysis of $^{102}$Pd
using the GCM is carried out (Chapter~\ref{chapphenom}).

\part{Experimental techniques}
\label{partexpttechniques}
\chapter{Gamma-ray spectroscopy: Aims and methods}
\label{chapexpt}

Understanding of nuclear structure is largely obtained through
comparison of measured nuclear properties with model predictions.  The
experimental task is to measure those properties which can most
effectively be used to distinguish between different model
descriptions or to refine the interpretation using a given model (\eg,
find the most suitable set of parameter values).  The collective
models discussed in Chapter~\ref{chapintro}
provide predictions for such quantities as the energies of low-lying
levels and for electromagnetic matrix elements between these levels,
and the study of these is the focus of the present experimental work.

\begin{figure}
\begin{center}
\includegraphics*[width=0.7\textwidth]{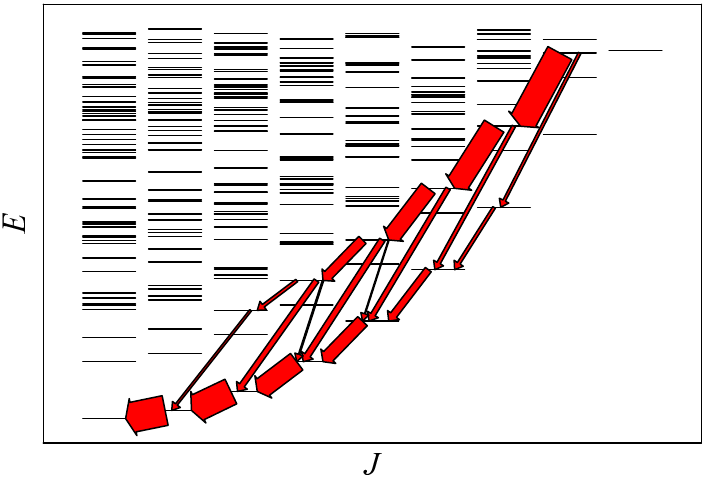}
\end{center}
\caption
[Population patterns in heavy-ion
fusion-evaporation.] {Population patterns
in heavy-ion fusion-evaporation.  Electromagnetic quadrupole
transitions can induce a change in spin of at most two units, and
larger-energy transitions are preferred, leading to a population
pattern in which intensity flow migrates to the yrast states.
\label{figyrastpop}
}
\end{figure}

Much successful work has been performed over the past decades
populating states through heavy-ion fusion-evaporation
reactions~\cite{newton1974:higamma}.  Such reactions tend to initially
produce nuclear states at high spin and high energy, which are usually
either ``yrast'' states (\ie, the lowest-energy states of their given
spin) or nearly yrast.  These states then deexcite by emission of a
series low-multipolarity (dipole and quadrupole) $\gamma$ rays.  Each
low-multipolarity $\gamma$ ray emission can induce at most a small
change in spin, and higher-energy transitions are enhanced due to the
greater available phase space for the radiation.  Consequently, at
each step in the decay the sequence of levels populated tends ever
closer to being yrast, as shown in Fig.~\ref{figyrastpop}.  Population
of near-yrast states at high spin thus leads almost exclusively to the
population of yrast states at low energy.

The useful predictions of the collective models pertain, though, to
the whole set of low-energy collective states, not just the yrast
states.  Mechanisms other than direct population in fusion-evaporation
reactions are necessary for the effective population of non-yrast
states.  Several such mechanisms exist, including $\beta$ decay from a
low-spin parent nucleus (Chapter~\ref{chapbeta}), neutron capture
(Chapter~\ref{chapgrid}), and Coulomb excitation, as well as various
forms of transfer reaction and inelastic scattering.

Once the levels of interest are populated, these levels must be
observed and their properties, especially electromagnetic decay
properties, determined.  The probability per unit time of deexcitation
of one state to another by $\gamma$-ray or conversion electron
emission (an experimentally accessible quantity) is directly related
to electromagnetic multipole operator matrix elements between the two
states (quantities predicted by models).  Experimentally, several
pieces of data are needed to deduce a single matrix element.  A given
level may be depopulated by several different possible transitions or
``branches'', including both $\gamma$-ray and conversion electron
transitions, and the partial transition probabilities for all these
branches combine to produce an aggregate decay rate
\begin{equation}
\label{eqntsum}
T=\sum_i T^\gamma_i + \sum_i T^\text{c.e.}_i,
\end{equation}
and hence lifetime $\tau=T^{-1}$, for the level.  The problem of
measuring the decay rates for individual transitions is broken into
two parts: a lifetime measurement and a ``spectroscopic'' study, in
which the relative intensities of the various branches are determined.
Each individual $\gamma$-ray transition may also proceed by more than
one multipolarity, as described by a mixing ratio $\delta$
(Appendix~\ref{appgamma}), and this $\delta$ must be determined
experimentally in order for the contributions from the different
multipole matrix elements to be separated.  These matters are reviewed
in more detail in Appendix~\ref{appgamma}.

Directly or indirectly, a broad variety of information can be obtained
through detection of the emitted electromagnetic radiation.  The
techniques discussed in the following chapters all rely at least
partially upon $\gamma$-ray detection.  Energies of observed
$\gamma$-ray transitions, and the coincidence relations between these
transitions (Section~\ref{seccoin}), allow identification of the
levels.  Intensities of the emitted $\gamma$ rays
(Section~\ref{seccoin}) provide information on relative
electromagnetic transition strengths, and when combined with
conversion electron data they can also yield information on
$\gamma$-ray multipolarities~\cite{hager1968:convcoeff}.  Angular
corelations and polarizations (Section~\ref{secangpol}) of the emitted
$\gamma$ rays provide information on level spins and $\gamma$-ray
multipolarities.  Detection of the $\gamma$-rays is also integral to
many techniques for the determination of matrix elements in Coulomb
excitation~\cite{mcgowan1974:coulex,cline1986:coulex} and for the measurement of
level lifetimes, including electronic timing methods
(Section~\ref{secfest}) and recoil Doppler shift methods
(Ref.~\cite{nolan1979:lifetime} and Chapter~\ref{chapgrid}).

\chapter{Beta decay and tape collector techniques}
\label{chapbeta}

\section{Beta decay as a population mechanism}
\label{secbetapop}

The process of $\beta$ decay can directly provide certain nuclear
physics information, for instance, on weak interaction coupling
strengths, nuclear ground-state mass
differences, or nuclear spin assignments, but
in the following discussion it is considered primarily as a mechanism
for populating states to be studied by their subsequent
electromagnetic interaction.  ``Beta decay'' is used as a generic term
to include the processes of $\beta^-$ and $\beta^+$ emission and
electron capture (or $\varepsilon$ decay).

Beta decay is a highly spin-selective population mechanism.  The beta
emission wave function may be decomposed (\eg,
Refs.~\cite{heyde1994:nucl,eisenberg1988:v2}) into multipoles of the
orbital angular momentum $L$ of the emitted particles, much as for
electromagnetic radiation~(\eg, Ref.~\cite{moskowski1965:multipole}).
The parity change induced by each multipole is $(-)^L$.  The angular
momentum difference between the parent and daughter nuclei must be
carried off by some combination of this orbital angular momentum and
the combined spin $S$ of the $\beta$ and $\nu$ ($S$=0,1).  The
transition probability for each higher multipole is suppressed by a
factor of approximately $(R_\text{nucl}/\lambda_\beta)^2$, where
$R_\text{nucl}$ is the nuclear radius and $\lambda_\beta$ is the
$\beta$-particle de~Broglie wavelength, which is typically
$\sim$$10^4$.  Consequently, most decays from the parent will proceed
to daughter states which are accessible by $L$=0 decay, thus with
$\Delta J$=0,$\pm$1 and no change in parity.

\begin{figure}
\begin{center}
\includegraphics*[width=0.7\textwidth]{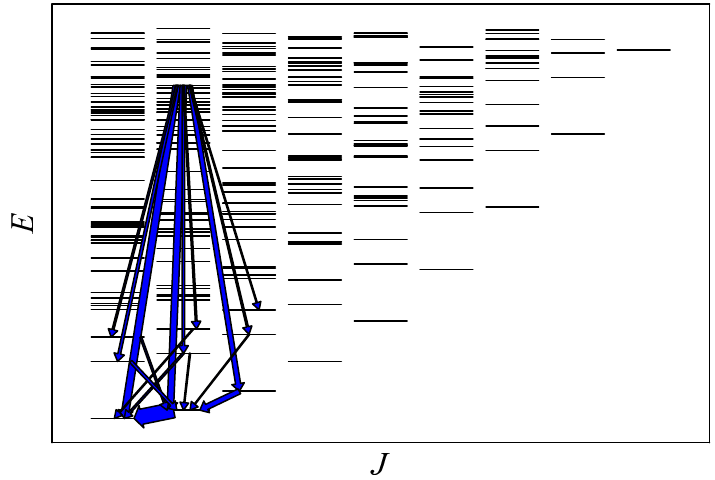}
\end{center}
\caption
[Population patterns in a $\beta$-decay daughter nucleus.] {Population
patterns in a $\beta$ decay daughter nucleus from a low-spin parent.
Transitions from one representative high-lying, low-spin level
directly fed by the $\beta$ decay are shown, along with subsequent
transitions among the levels populated.  
\label{figbetapop}
}
\end{figure}

Beta decay from a low-spin parent can therefore populate low-energy,
low-spin states in the daughter directly.  It can also populate
higher-energy, low-spin levels, which then decay primarily by dipole
or quadrupole radiation to the lower-energy levels of similar spin;
although decays directly to the lowest-energy levels are energetically
preferred, decays to intermediate-energy levels may also be
considerable.  Fig.~\ref{figbetapop} illustrates the resulting
$\gamma$-decay patterns in the daughter nucleus.  These should be
compared to the population patterns shown for fusion-evaporation in
Fig.~\ref{figyrastpop}.

Both $\beta^+$ decay and electron capture proceed from the same parent
nucleus to the same daughter nucleus but do so producing different population
patterns of states in the daughter nucleus.  In $\beta^+$ emission,
some of the energy provided by the mass difference between the parent
and daugther nucleus must be used to create the $\beta^+$ particle;
whereas, in electron capture, the mass energy of an atomic electron is
liberated.  The energy available for excitation of the daughter
nucleus in each case is
\begin{align}
Q_{\beta}&=M_P^\text{nucl}c^2-M_D^\text{nucl}c^2-m_e c^2 \\
\intertext{and}
Q_{\varepsilon}&=M_P^\text{nucl}c^2-M_D^\text{nucl}c^2+m_e c^2, 
\end{align}
where $M_P^\text{nucl}$ is the parent nuclear mass and
$M_D^\text{nucl}$ is the daughter nuclear mass, excluding minor
contributions from atomic binding effects.  Consequently, electron
capture can populate higher-lying states in the daughter nucleus than
can $\beta^+$ decay and can even proceed in some cases for which
$\beta^+$ decay is energetically forbidden
(0$<$$Q_\varepsilon$$<$$2m_e c^2$).  Where both $\beta^+$ decay and
$\varepsilon$ decay to a given level in the daughter nucleus are
possible, the relative probability of these depends upon the maximum
$\beta$ energy for that transition: electron capture dominates for
small transition energy (high-lying levels), while $\beta^+$ decay
dominates for large transition energy~\cite{zweifelboth:betaplusec}.
The distinction between $\beta^+$ decay and electron capture is
relevant to several applications discussed in the following chapters:
radiation from $\beta^+$ annihilation can be a significant source of
background in coincidence spectroscopy experiments, $\beta^+$
particles may be used for timing in lifetime measurements
(Section~\ref{secfest}), and $\beta$ tagging may be used to separate
$\beta^\pm$ decay processes from electron capture processes in
contaminated radioactive beam experiments.

\section{Beta decay in an accelerator environment: The Yale MTC}
\label{secmtc}

Electrostatic heavy-ion accelerators are effective tools for the
production of proton-rich or near-stable nuclei through a variety of
possible reactions, including fusion-evaporation and transfer
reactions.  Fusion-evaporation \textit{per se} is not an efficient
population mechanism for the low-spin, non-yrast states we wish to
study, as discussed in Chapter~\ref{chapexpt}.  However,
fusion-evaporation may be used as a \textit{production} mechanism for
$\beta$-decay parent nuclei, which then populate such states of
interest in the daughter nucleus.  Fusion-evaporation is flexible in
the nuclei it can create, due to the wide variety of available
combinations of beam and target nuclei (which control the compound
nucleus produced) and beam energy (which influences the following
statistical decay process)~\cite{gavron1980:fusevap}, and it can
provide relatively high reaction cross-sections of hundreds of mb.

In the $A$$\approx$100--170 region of interest in this work, the
particle evaporation following compound nuclear formation is mainly of
neutrons, although proton and alpha emission are significant towards
the low end of this mass range.  For beam energies within a few MeV
above the threshold energy for fusion, relatively few
particles ($\sim$3--4) are evaporated, and the evaporation process is
dominated by a single channel.  At energies further above threshold,
the total fusion cross section increases until saturation at a
geometric limit.  The average number of evaporated
particles increases with beam energy as well (about one additional
particle emitted per few-MeV increase in beam energy), and fluctuation
in the number of emitted particles leads to decreased dominance of the
main evaporation channel and hence decreased production selectivity.
At higher energies still (tens of MeV above threshold), compound
nuclear fission begins to occur instead of particle evaporation.  
As an example of a clean, near threshold fusion-evaporation reaction,
one experiment in the present work used
$^{148}$Sm($^{12}$C,4$n$)$^{156}$Er at 73\,MeV beam energy, or
$\sim$20\,MeV above the Coulomb barrier.  Of a total $\sim$990\,mb cross
section for fusion in this reaction, $\sim$74$\%$ results in
$^{156}$Er production, with the neighboring 3$n$ and 5$n$ channels
accounting for most ($\sim$3/4) of the remaining cross
section~\cite{gavron1980:fusevap,gavron:pace}.  

\begin{figure}
\begin{center}
\includegraphics*[width=0.7\textwidth]{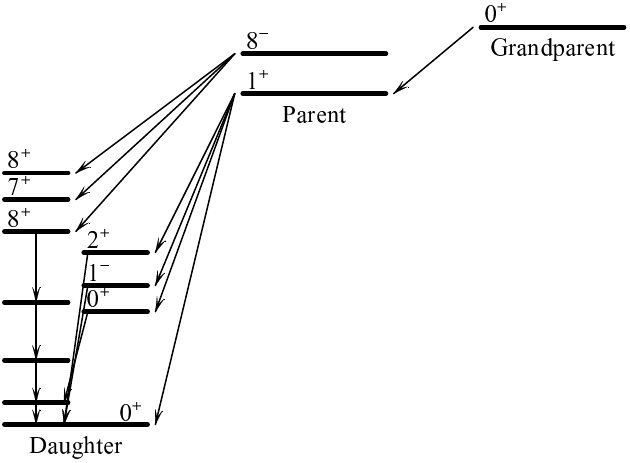}
\end{center}
\caption
[Population of an even-even daughter by two-step $\beta$ decay. ]
{Population of an even-even daughter by two-step $\beta$ decay to
avoid $\beta$ decay from a high-spin isomer in the odd-$A$ parent.
\label{figtwostep}
}
\end{figure}

Beta decay following fusion-evaporation is only effective at
populating low-spin, non-yrast states if the parent nucleus, which is
created at high-spin, $\gamma$-decays to low spin before
$\beta$-decaying.  This is not guaranteed to occur, since the study of
even-even daughter nuclei requires the use of odd-odd parent nuclei.
These nuclei commonly posess yrast high-spin isomeric states (``yrast
traps''), which are strongly populated in fusion-evaporation, and
$\gamma$ decay from these states is often so suppressed that their
preferred decay mode is $\beta$ decay.  Alternatively, odd-odd nuclei
may have a high-spin ground state strongly populated in
fusion-evaporation, with low-spin $\beta$-decaying isomers.  In either
case, the $\beta$-decaying parent nucleus state populated in
fusion-evaporation has a higher spin than desired.  A means of
circumventing such difficulties is to use fusion-evaporation to
produce the $\beta$-decay
\textit{grandparent} of the nucleus of interest, which, as an
even-even nucleus, will $\gamma$ decay to its ground state before
$\beta$ decaying.  It will therefore predominantly populate the
low-spin $\beta$-decaying states in the odd-odd parent, be they the
ground state or isomeric, and therefore lead to the population of the
states of interest in the daughter.  The differing population patterns
for one-step and two-step $\beta$ decay are compared in Fig.~\ref{figtwostep}.

\begin{figure}[p]
\begin{center}
\includegraphics*[width=0.50\textwidth]{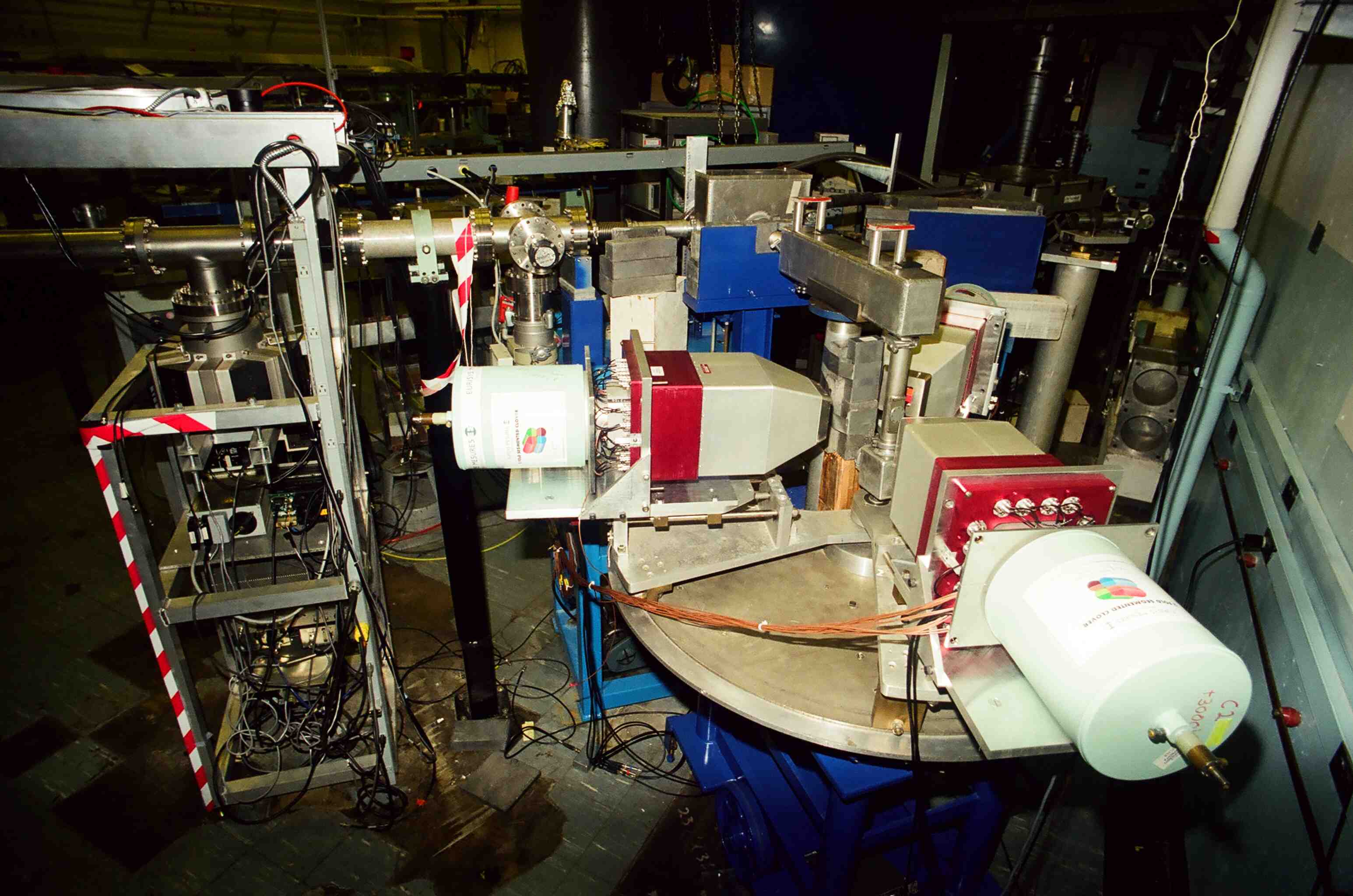}
\hfill
\includegraphics*[width=0.46\textwidth]{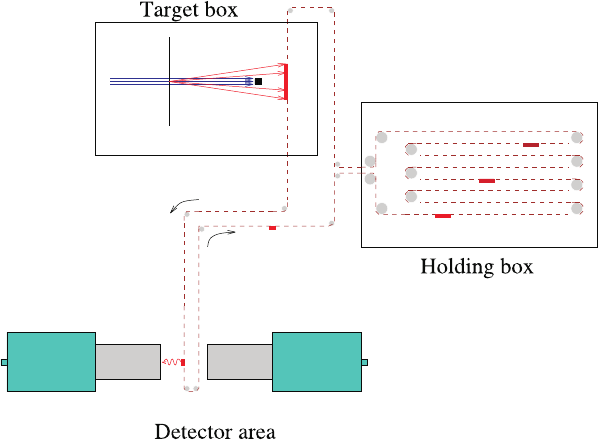}
\end{center}
\caption
[Photograph and schematic diagram of the Yale Moving Tape Collector.]
{Photograph and schematic diagram of the Yale Moving Tape Collector
with an array of clover detectors. (Figure from
Ref.~\cite{caprio2001:156dy-beta-rjp}.)
\label{figmtc}
}
\end{figure}

The Yale Moving Tape Collector (MTC)~\cite{zamfir2000:mtc}, at the
Wright Nuclear Structure Laboratory, is a dedicated setup for the
study of $\beta$-decay daughters of nuclei produced using beam from
the Yale ESTU accelerator.  Reaction product nuclei are collected on a
tape and transported to a detector area, where
their daughter nuclei can be studied using an array of $\gamma$-ray
and $\beta$-particle detectors.  A photograph and schematic diagram of
the Yale MTC are shown in Fig.~\ref{figmtc}.

\begin{figure}[p]
\begin{center}
\includegraphics*[width=0.48\textwidth]{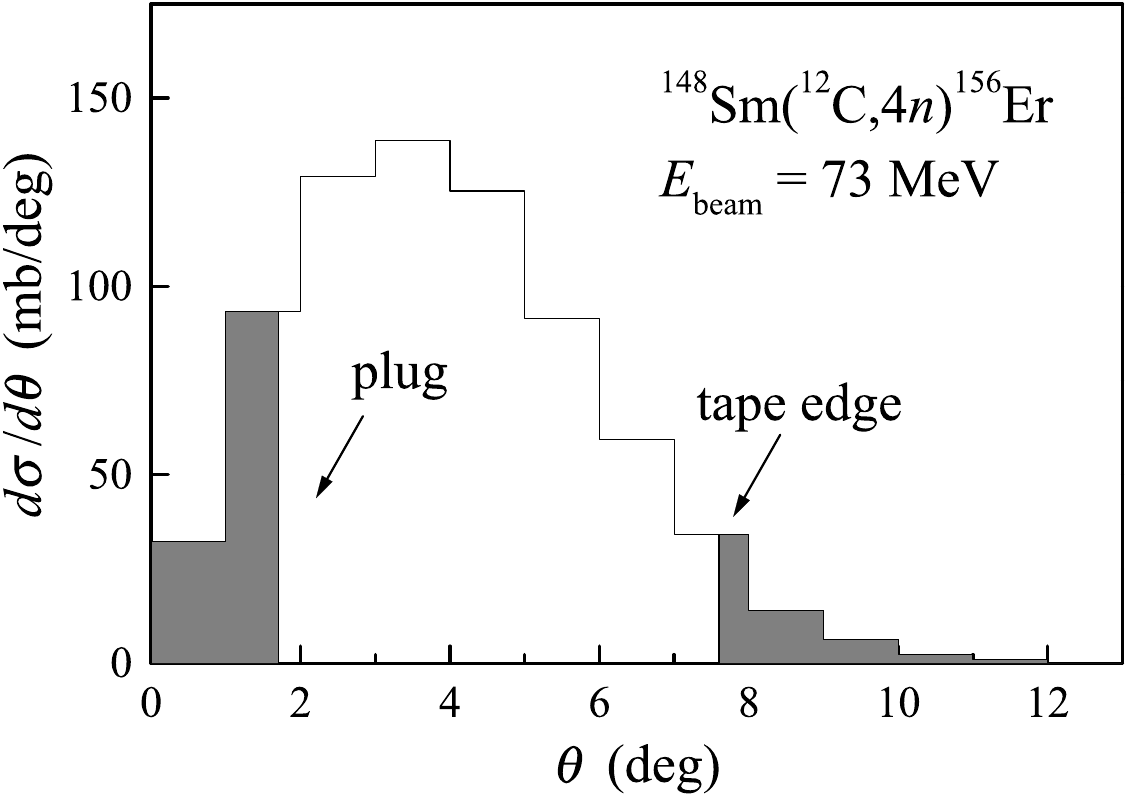}
\hfill
\raisebox{20pt}{
\includegraphics*[width=0.45\textwidth]{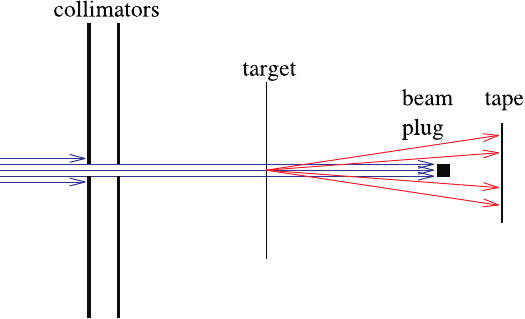}
}
\end{center}
\caption
[Kinematic separation of beam and product nuclei in the Yale MTC.]
{Kinematic separation of beam and product nuclei in the Yale MTC: 
calculated~\cite{gavron:pace} angular distribution of $^{156}$Er
evaporation residue nuclei from $^{148}$Sm($^{12}$C,4$n$)$^{156}$Er at
73\,MeV beam energy (left) and the target chamber configuration (right).
\label{figplug}
}
\end{figure}

The $\beta$-decay parent nuclei are produced in a target and recoil
out from the downstream side of the target.  It is necessary to
separate these product nuclei from unreacted beam nuclei which have
passed through the target, since direct beam rapidly vaporizes the
tape material at the $\sim$1--100\,nA beam currents used.  The beam
spot at the target location is focused to $\lesssim$1\,mm diameter by
a quadrupole magnet upstream of the target chamber and also passes
through 3\,mm diameter collimation apertures before reaching the
target.  The unreacted primary beam nuclei pass largely undeflected
through the target and are stopped by a 3\,mm diameter gold plug
[Fig.~\ref{figplug}(a)].
In contrast, the fusion-evaporation product nuclei are emitted from
the target with an angular distribution spanning several degrees
[Fig.~\ref{figplug}(b)], largely bypass the plug, and are embedded
into a 16\,mm wide Kapton tape 1.5\,cm further downstream.  The
target-plug separation, typically $\sim$4--8\,cm, is chosen on the
basis of the angular distribution for each specific reaction: it must
be sufficient to allow the nuclei to bypass the plug, but small enough
that significant numbers of nuclei do not completely bypass the tape.

The tape transports the collected activity $\sim$3\,m to a shielded
detector area at regular intervals, chosen according to the parent
nucleus half-lives (Appendix~\ref{appmtc}).  The time required for the
tape to be advanced the necessary few meters, starting from rest, is
$\lesssim$1\,s.  The tape is a continuous loop of 16\,mm Kapton motion
picture film leader, of $\sim$50\,m total length, which after
advancing from the target to the detector area continues on to a
storage box, where it passes back and forth among a series of rollers.
It is strung on plain unsprocketed rollers and driven by a single
sprocketed wheel on a stepper motor with adjustable acceleration.

The Yale MTC is equipped with a flexible detector array for
$\gamma$-$\gamma$ and $\beta$-$\gamma$ coincidence spectroscopy.  Its
usual configuration consists of three Compton-suppressed YRAST Ball
clover segmented HPGe
detectors~\cite{duchene1999:clover,beausang2000:yrastball} and one
low-energy photon spectrometer (LEPS) planar Ge detector on an angular
correlation table.  The clover detectors, which are large-volume
composite high-purity germanium detectors, provide high-efficiency
$\gamma$-ray detection at energies above $\sim$50\,keV, giving an
array efficiency of $\sim$1.1$\%$ at 1.3\,MeV.  They may be used as
Compton polarimeters (Section~\ref{secangpol}) as well.  The clover
detectors are placed inside bismuth germanate (BGO) Compton
suppression detectors, with heavy-metal front collimators, to allow
the identification and rejection of clover detector hits in which the
incident $\gamma$ ray is Compton scattered, leaving only a portion of
its energy in the clover.

Auxiliary detectors are regularly used with the array to
extend its capabilities.  Variations upon the standard configuration
have included:
\begin{enumerate}
\item Three Compton-suppressed clover detectors and a LEPS detector,
in combination with a thin plastic scintillation detector and BaF$_2$
scintillation detector, for subnanosecond lifetime measurements by
electronic timing (described in Section~\ref{secfest}).
\item Three Compton-suppressed clover detectors and a LEPS detector, 
in combination with a thick plastic scintillation detector, for
$\gamma$-ray gated $\beta$ energy endpoint
measurements~\cite{tomlin2001:mass,brenner2002:80y-mass}.
\item Four unshielded clover detectors, for $\gamma$-$\gamma$ angular
correlations between individual clover elements and $\gamma$-$\gamma$
Compton polarimetry~\cite{wolf2002:128ba-beta}.
\item Four unshielded clover detectors and a thin plastic
scintillation detector, for $\beta$-$\gamma$ Compton
polarimetry~\cite{pietrallaUNP}.
\end{enumerate}

Electronic pulse processing and data acquisition for the Yale MTC are
accomplished using the YRAST Ball electronics
suite~\cite{beausang2000:yrastball}.  The acquisition hardware and
sorting software are described in Appendix~\ref{appacq}.
  
\section{Beta decay at an ISOL facility: The ISAC GPS}
\label{secgps}

In an isotope separation on-line (ISOL) ion source, fission or
spallation is induced by a proton or light ion beam in a thick
production target.  Reaction products, including unstable species,
escape the target and are ionized, electrostatically extracted as a
low energy (tens of keV) beam, and electromagnetically mass separated.
Although ISOL radioactive ion sources have existed for several
decades~\cite{dauria1995:rib-facilities}, a new, much more powerful,
generation of facilities is currently beginning operation (\eg,
Refs.~\cite{habs1997:rex-isolde,alton1998:hribf,schmor1999:isac}) with
the central purpose of providing high-energy ``postaccelerated'' beams
of unstable nuclei.  The availability of high-purity, high-intensity
$\beta$-decay activities from these new ISOL radioactive beam
production sources promises to provide a major improvement in the
quality of $\gamma$-ray spectroscopy data on collective excitations,
allowing greatly improved measurements on nuclei near stability and
making nuclei far from stability accessible to spectroscopy for the
first time.

The experiment described in Chapter~\ref{chap162er} was one of the
first experiments to be performed at the recently commissioned TRIUMF
Isotope Separator and Accelerator (ISAC)
facility~\cite{schmor1999:isac}, a new-generation ISOL facility.  The
experiment made use of an existing tape transport system at the
General Purpose Station (GPS)~\cite{hagberg1994:39ca,ball2001:74rb}.
After production in the ISAC surface ion
source~\cite{dombsky2000:isac}, spallation products are mass separated
by the ISAC separator and transported to the GPS at extraction energy,
29\,keV for the experiment in Chapter~\ref{chap162er}.

\begin{figure}
\begin{center}
\includegraphics*[width=0.55\textwidth]{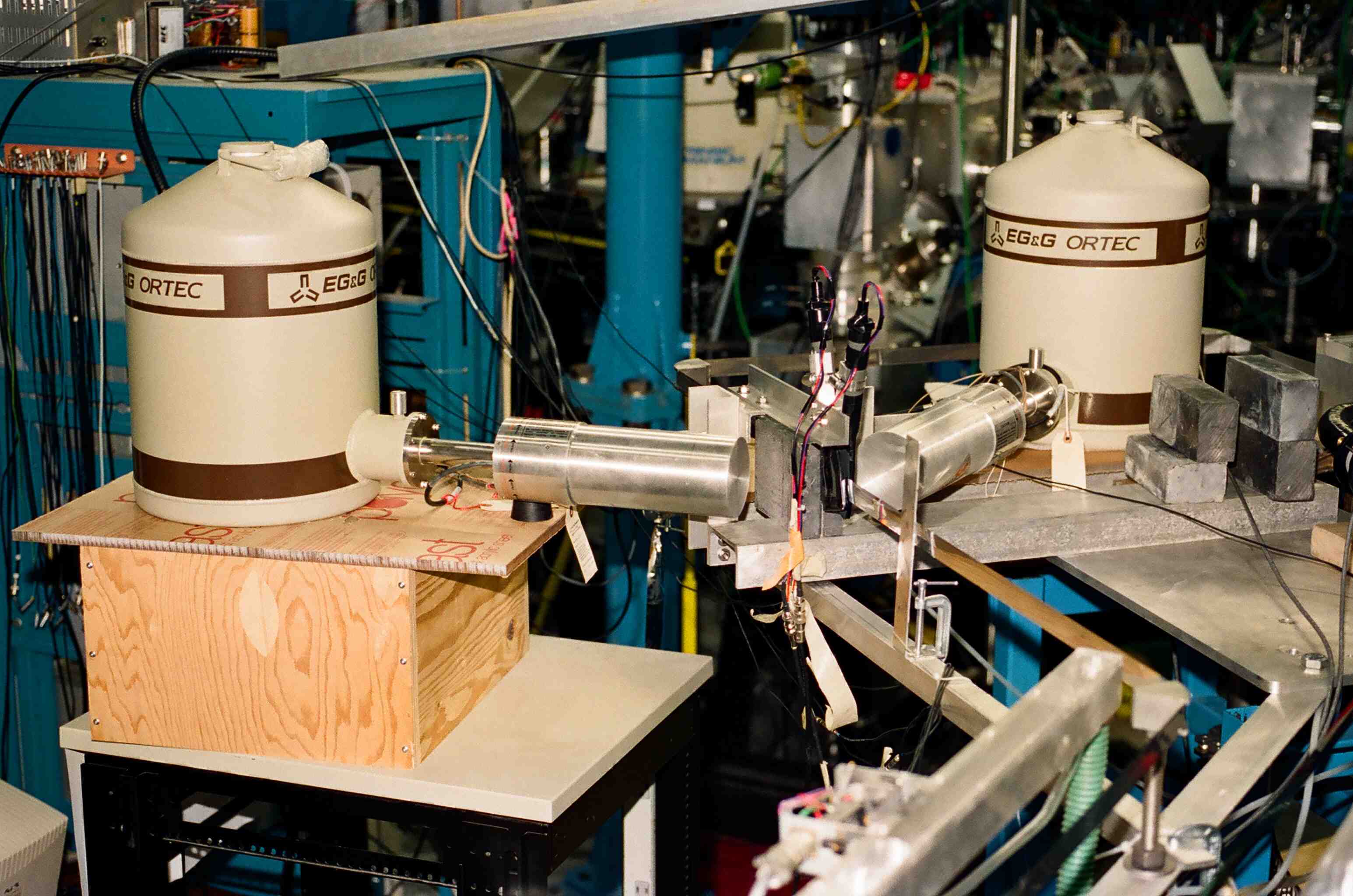}
\hfill
\includegraphics*[width=0.40\textwidth]{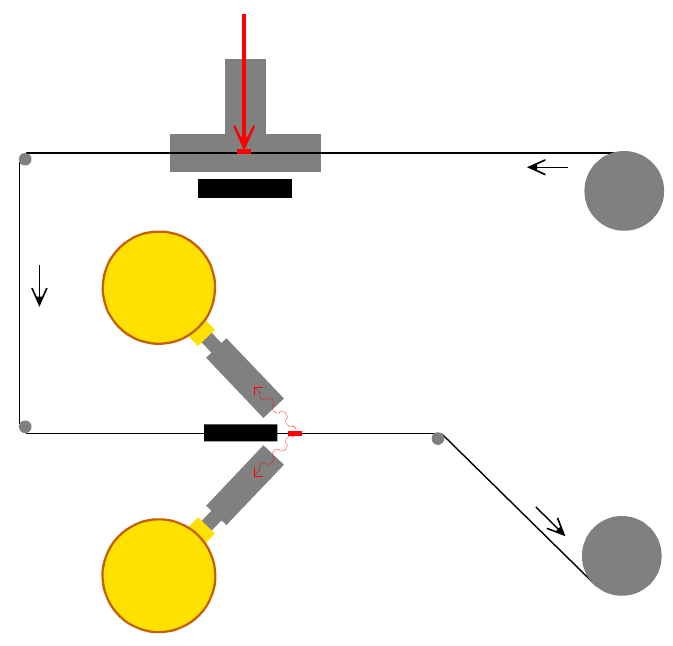}
\end{center}
\caption
[Photograph and schematic diagram of the ISAC GPS. ] {Photograph and
schematic diagram of the ISAC GPS in its $\gamma$-$\gamma$
spectroscopy configuration.
\label{figgps}
}
\end{figure}

A photograph and schematic diagram of the GPS are shown in
Fig.~\ref{figgps}.  The beam line terminates in a vacuum chamber, in
which the beam nuclei are embedded into a 25\,mm wide aluminized
Mylar tape.  Diagnostic scintillation detectors (plastic for $\beta$
detection and/or NaI for $\gamma$ detection) at the deposition point
are used to monitor the accumulated activity, providing feedback for
control of the beam intensity.  The tape exits the chamber through
100\,$\mu$m differentially-pumped slits and transports the activity in
air to a detector area.  The GPS transport system is based upon a
single-pass reel-to-reel tape design and is optimized for high-speed
operation with tape cycles of $\lesssim$0.25\,s.  The tape reels are
pneumatically actuated, controlled directly by the acquisition
computer, and the tape is fed over air-cushion bearings~--- curved
metal surfaces perforated with numerous holes through which
pressurized air is bled~\cite{hagberg1994:39ca}~--- instead of
conventional rollers.

The detector configuration used for the $\gamma$-$\gamma$ coincidence
spectroscopy experiment of Chapter~\ref{chap162er} consisted of two
large-volume coaxial Ge detectors of 80$\%$ relative efficiency
located 12\,cm from the source position.  The detectors were oriented
obliquely with respect to each other and separated by lead shielding,
in order to suppress unwanted coincidence events due to $\beta^+$
annihilation photon pairs and Compton cross-scattering.  The combined
photopeak efficiency was 0.8$\%$ at 1.3\,MeV.  A pair of thin plastic
scintillation detectors surrounding the tape at the source location
provided tagging of decays involving $\beta$ particle emission.

Detector energy and timing information at the GPS are recorded in
event mode using a CAMAC-based acquisition system read out by an
Intel/Linux front end computer running PSI/TRIUMF
MIDAS~\cite{ritt:midas}.  Deposition rates and other diagnostic data
are obtained using a CAMAC scaler module, also included in the data
stream.  The MIDAS system supports unlimited concurrent operations to
be performed on the data stream, interchangeably running on the front
end or remote networked machines, including data logging to primary
and backup storage media, graphical monitoring of experiment rates
(\eg, to assist ISAC operations staff in controlling deposition
rates), and online sorting.

Experimental $\beta$-decay work at an ISOL radioactive beam presents special
challenges.  A summary of some main considerations follows.

Beam steering and measurement diagnostics can be of limited use or
nonexistent, since beam currents of $\lesssim$10$^7$\,particles/s are
too low for conventional current-based devices (Faraday cups, slit
boxes, wire scanners, harp monitors) and optical phosphors to be
useful, and these devices may also be omitted to prevent the
accumulation of contamination.  For this reason, the deposition
monitoring detectors at the GPS were crucial for controlling beam
intensities.  However, much more sophisticated radioactive beam
diagnostic detectors are currently in use or under development (\eg,
Ref.~\cite{shapira2000:psmcp}).

ISOL beam purity can vary greatly with nuclide and with production
source details.  ISOL sources can provide beams of exceptional purity
if only one member of the nuclear mass chain selected by the separator
escapes easily from the ISOL target or, ideally, if mass separation of
the isobars {\it within} a mass chain is accomplished, as is possible
with a sufficiently high-resolution mass separator.  In general,
however, a beam containing multiple radionuclides (``cocktail'' beam)
must be expected.  Contaminant nuclei with half lives differing by at
least an order of magnitude from that of the parent nucleus of
interest do not usually present difficulties, since use of the proper
tape advance cycle (Appendix~\ref{appmtc}) strongly suppresses their
decay in the detector area relative to the decay of interest.  When
the contaminant nuclei decay primarily by electron capture and the
parent nucleus of interest decays by $\beta^+$ or $\beta^-$ decay, or
vice versa, $\beta$ particle tagging may be useful as a coincidence
requirement or veto, respectively.  The GPS setup incorporated and
tested this capability, but the extremely high beam purity encountered
in the actual experiment (Chapter~\ref{chap162er}) rendered it
unnecessary.

Making full use of the intense $\sim$10$^{10}$\,Bq $\beta$-decay
activities which can be obtained for
copiously-produced species of nuclei near stability presents a
challenge to the instrumentation, in the form of an ``embarassment of
riches''.  The experiment at the GPS was essentially detector-limited
in nature, due to the constraint of maintaining acceptable count rates
in individual detectors ($\lesssim$20\,kHz): the maximum beam
deposition rate which could be accomodated was $\sim$$10^5$/s,
although a beam intensity of $\sim$$10^9$/s was available from the ISAC
separator.  Arrays with high granularity are less subject to
limitation by single detector count rates and thus can benefit more
fully from the large beam intensities available.  Specifically,
coincidence spectroscopy experiments (Section~\ref{seccoin}) require
high-statistics two-fold $\gamma$-ray coincidence data.  For an array
of total efficiency $\varepsilon$ divided among $n$ similar array
elements, these data are collected with a rate $\propto
A\varepsilon^2(n-1)/n$, where $A$ is the source activity.  The $\gamma$
absorption rate in an individual array element is $\propto A
\varepsilon /n$.  If the maximum acceptable count rate for a single
element is $R_\text{max}$, then the corresponding maximum coincidence
data collection rate is $\propto R_\text{max} n \varepsilon$,
suppressing the $(n-1)/n$ factor relevant only at small $n$. Improved
data collection can be obtained by increasing any of the factors:
single-element rate capability, array granularity, or overall
efficiency.  For this reason, much improved coincidence sensitivity
should be obtained as large, Compton-suppressed, multi-detector
arrays, such as the Clarion array~\cite{gross2000:rms} at ORNL HRIBF
or the Miniball array~\cite{habs1997:miniball} at CERN REX-ISOLDE, are
implemented for decay studies at ISOL ion sources.

\chapter{Gamma-ray spectroscopy with modern high-efficiency arrays}
\label{chapspec}

\section{Coincidence spectroscopy for transition intensities}
\label{seccoin}

Gamma-ray spectroscopy, through the identification of $\gamma$-ray
transitions and measurement of their intensities, plays a central role
in obtaining experimental information necessary for understanding
nuclear structure, as described in Chapter~\ref{chapexpt}.  One
broadly-stated goal in the spectroscopic study of a nucleus is to
understand the placement of the observed $\gamma$ ray lines, in the
process establishing the level scheme of the nucleus.  Another is to
deduce the relative strengths of all transitions depopulating each
level of interest, since these relative intensities allow either
ratios of $B(\sigma\lambda)$ strengths or, if the level lifetime is
known, absolute $B(\sigma\lambda)$ strengths to be determined.

Although usually only the relative intensities of transitions sharing
the same parent level are needed for model analysis, it is often more
practical to measure the intensities of transitions from all levels
populated in the nucleus on a common scale~--- in $\beta$ decay, this
is the probability of the transition occuring per unit parent nucleus
decay.  The measurement techniques, with one exception described
below, yield separate intensities for each transition individually on
this common scale; the relative intensities for the transitions from a
given level are only subsequently deduced by comparison of these.
Intensities expressed on such a common scale can directly be compared
across all experiments which use the exact same mechanism to populate
the nucleus.  Thus, points of agreement or contradiction between
different sets of results can be identified, and results from separate
experiments can be assembled into a more comprehensive composite set
of information~--- for instance, a conversion electron intensity from
one experiment can be combined with a $\gamma$-ray intensity from
another to extract a conversion coefficient.

\begin{figure}
\begin{center}
\includegraphics*[width=0.8\textwidth]{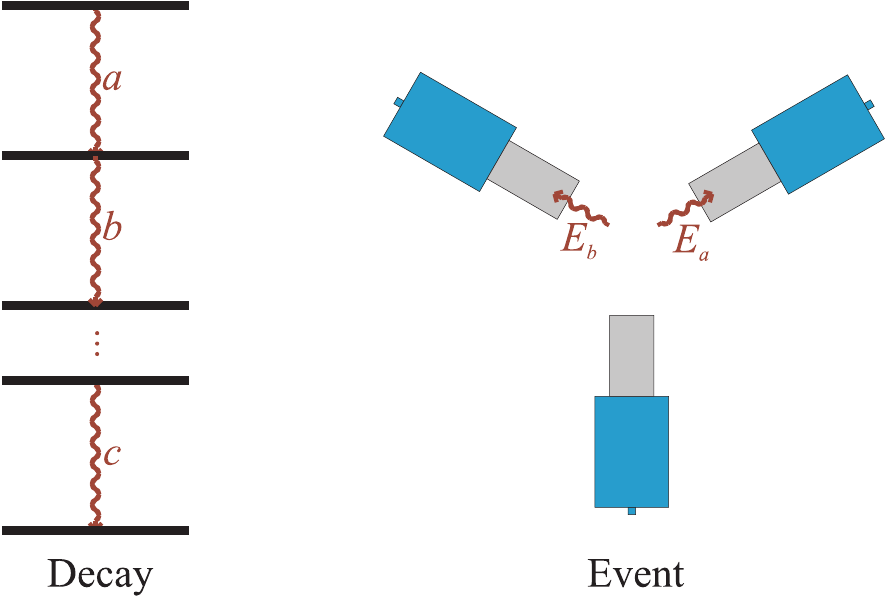}
\end{center}
\caption[Schematic illustration of a nuclear decay and the corresponding event.]  
{Schematic illustration of a nuclear decay (the physical
process) and the corresponding event (the observed data).
\label{figdecayevent}
}
\end{figure}

Spectroscopic measurements are indirect measurements of a physical
system via an instrument.  The physical system being analyzed is an
ensemble of nuclei which have been populated in various excited states
and which then deexcite by emission of a series of radiations.  The
decay process for each of these individual nuclei may, or may not,
result in an observed ``event'', that is, detection of one or more of
the emitted radiations in an array of detectors, as illustrated in
Fig.~\ref{figdecayevent}.  The experimental challenge is to understand
the instrument response, which maps decays to events, and then, given
a set of data, invert this process to extract the properties of the
underlying ensemble of decays.  Given the rich nature of the data
collected, involving simultaneous detection of multiple radiations
from the same decay (coincidences), there are often several possible
complementary ways of extracting the same piece of information, each
of which may be more or less effective according to the circumstances.
The following discussion summarizes some basic techniques as applied
in pure $\gamma$-ray spectroscopy following $\beta$ decay, but the
approaches presented are more generally applicable to a wide range of
spectroscopy experiments with multidetector arrays and reasonably low
detection multiplicity.  Some more technical issues are deferred to
Appendix~\ref{appcoin}.

Singles measurements, based upon analysis of a simple aggregate energy
spectrum of all $\gamma$ rays detected, in principle directly provide
both energies and intensities for all transitions.  However,
measurements of intensities from singles data are notoriously prone to
error due to the presence of unresolved contaminant transitions.
Beta-decay experiments with $Q_\beta$ values of a few MeV commonly
produce several hundred identifiable $\gamma$-ray transitions, many of
them yielding overlapping or unresolved peaks in the singles spectra.
For instance, in $^{162}$Er (Chapter~\ref{chap162er}), five of the ten
most strongly populated transitions in $\beta$-decay are
doublets~\cite{deboer1974:162er-beta,nds1999:162}.  Construction of a
level scheme from a singles spectrum in such a decay experiment is
extremely challenging, since it relies upon the recognition of groups
of $\gamma$-ray lines with energies which sum to yield the same
excited level energy, yet a virtual continuum energies of partially
resolved $\gamma$-ray lines is observed in the spectrum [see
Fig.~\ref{fig156dyspectra} on page~\pageref{fig156dyspectra} and
Fig.~\ref{fig162ersingles} on page~\pageref{fig162ersingles} for
examples].  Historically, decay experiments have relied heavily or
exclusively upon singles data.  Singles analysis is ``efficient'', in
that it makes use of all detected $\gamma$ rays, so for experiments
performed using only one or a few low-efficiency detectors, or when
only a small activity can be obtained, singles measurements may be the
only type possible.  A preponderance of the studies
contributing to the present corpus of decay data in medium-mass and
heavy nuclei were carried out several decades ago when only
small-volume Ge detectors were available.

\begin{figure}
\begin{center}
\raisebox{4ex}{
\includegraphics*[width=0.4\textwidth]{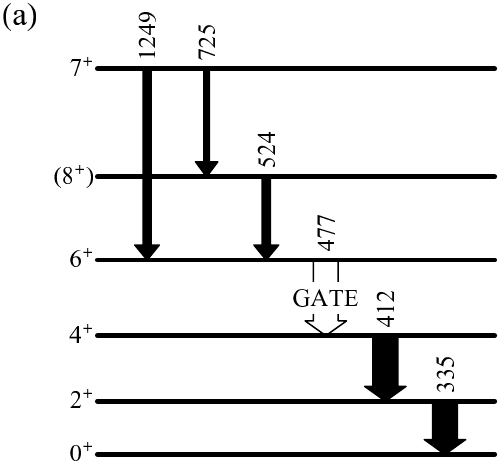}
}
\hfill
\includegraphics*[width=0.55\textwidth]{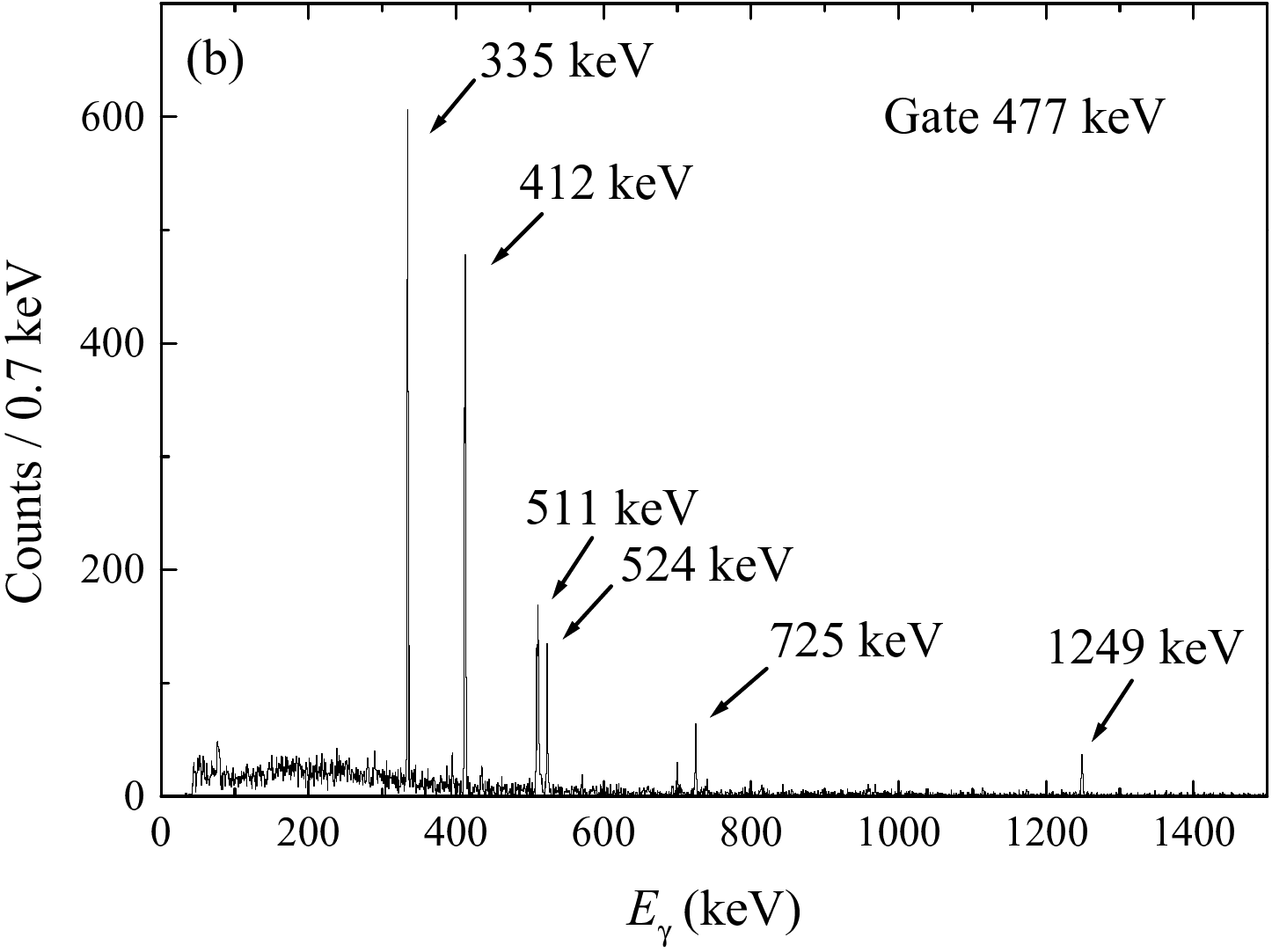}
\end{center}
\caption[Placement information
contained in a gated spectrum.]  {Placement information contained in a
gated spectrum: (a)~level scheme showing transitions feeding or
following the gate transition and (b)~a gated spectrum showing
transitions coincident with this transition.  Arrow thicknesses in
part~(a) are proportional to the expected intensity coincident with
the 477\,keV transition (Appendix~\ref{appcoin}).  The example gate is
on the 477\,keV transition in $^{154}$Dy, populated in $^{154}$Ho
$\beta^+/\varepsilon$ decay (Chapter~\ref{chap154dy}).
\label{figcoinplace}
}
\end{figure}

The advent of compact, high-efficiency arrays of large-volume Ge
detectors has permitted a new generation of $\gamma$-ray spectroscopy
$\beta$-decay experiments which produce high-statistics coincidence
data.  These data provide much more reliable information on placement
of transitions in the decay scheme and allow $\gamma$-ray transition
intensities to be determined from relatively clean gated spectra.  An
example gated coincidence spectrum in Fig.~\ref{figcoinplace}
illustrates how the transitions observed coincident with a $\gamma$
ray of interest are related to the placement of the transition in the
level scheme, showing how the gating transition is coincident both
with transitions feeding the initial level and depopulating the final
level.  

When the transition $x$ of interest directly feeds a level which
decays by a $\gamma$ radiation $b$ for which the intensity branching
fraction $B_b$ is known, then the intensity $I_x$ per parent
decay can be determined from
\begin{equation}
\label{eqngatebelow}
G_{b:x} = N I_x B_b \varepsilon(E_b,E_x)
\end{equation}
where $G_{b:x}$ is the number of detected coincidences between $b$ and
$x$ from a gated spectrum, $N$ is the number of parent decays, and
$\varepsilon(E_b,E_x)$ is the coincidence efficiency (Section~\ref{seccoineff}).
The branching fraction $B_b$ is calculated from known
intensities as $B_b = I_b / (\sum_i I_i+\sum_i I^\text{ce}_i)$, where
the sums are over all $\gamma$-ray and conversion electron transitions
depopulating the level.

The {\it relative} intensities of two branches $x$ and $y$ from a
level can also be obtained from a spectrum gated on a transition $a$
feeding the level, according to
\begin{equation}
\label{eqngateabove}
\frac{G_{a:x}}{G_{a:y}} = \frac{I_x \varepsilon(E_a,E_x)}{I_y \varepsilon(E_a,E_y)}.
\end{equation}
Intensities can only be measured in this way if there exist one or
more strong discrete feeding transitions to the level, and this method
therefore tends to be useful only for levels low in the excitation
spectrum.  This method in most cases provides lower statistics than
can be obtained by gating below the transition of interest --- the
intensity flow below the transition of interest is typically
concentrated in one or two strong branches, but the intensity flow
feeding the transition of interest is usually diluted among several
weak transitions or may come from direct $\beta$ feeding.  A gate on a
feeding transition can only provide intensities \textit{relative} to
other branches from the same level.

\begin{figure}
\begin{center}
\includegraphics*[width=0.65\hsize]{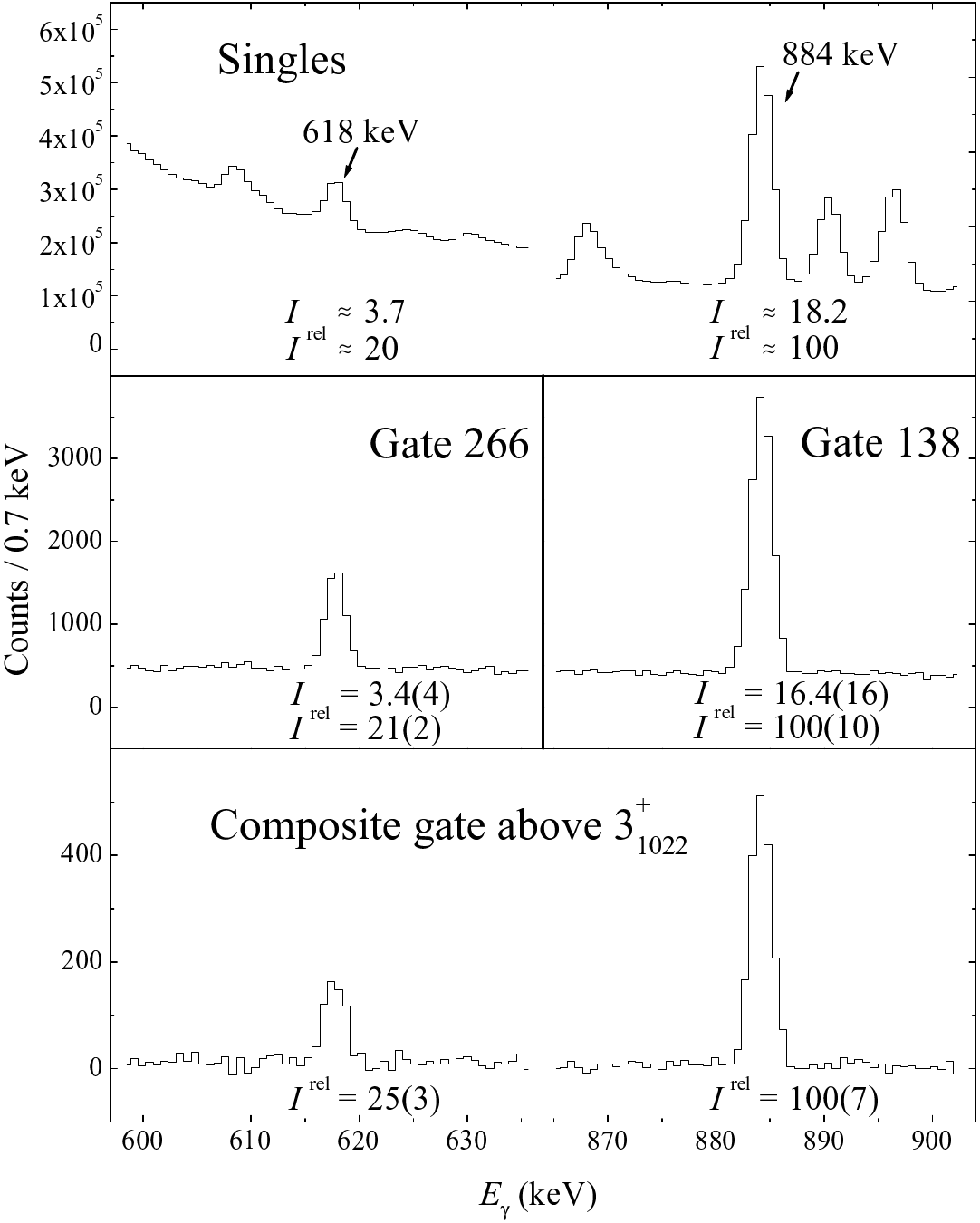}
\end{center}
\caption[Comparison of different sources of intensity information.]
{Comparison of the different sources of intensity information for the
618\,keV and 884\,keV branches from the $3^+$ level at 1022\,keV in
$^{156}$Dy.  The spectra are (top) from singles, (middle) gated below
the branch of interest, or (bottom) gated on transitions feeding the
$3^+$ level.  The composite spectrum on the bottom is gated on 655,
819, 1081, 1301, 1310, and 1386\,keV transitions.  
(Figure from Ref.~\cite{caprio2002:156dy-beta}.)
\label{figcoinlev1022}
}
\end{figure}

The relative quality of these different sources of intensity
information is illustrated in Fig.~\ref{figcoinlev1022}, which shows
the data used in the measurement of the intensities of the 617\,keV
and 884\,keV branches from the $3^+$ level at 1022\,keV in $^{156}$Dy
(Chapter~\ref{chap156dy}).

All intensity measurements rely upon an accurate knowledge of the
array efficiency.  Calibration of the singles and coincidence
efficiency of an array of detectors is discussed in
Section~\ref{seccoineff}.

\section{Angular correlation and polarization measurements}
\label{secangpol}

\subsection{Theoretical summary}

The angular correlation patterns and polarization properties of
radiation emitted by a nucleus are directly related to the spins of
the states in the emitting nucleus and to the multipolarities of the
emitted radiations (\eg,
Refs.~\cite{biedenharn1953:angcorr,fagg1959:polarization,frauenfelder1965:ang}).
Observation of these properties provides information on spin/parity
assignments of states as well as on the relative contributions of
different multipole matrix elements to the observed transition
intensities.

In the $\beta$-decay spectroscopy experiments considered here, the
sources are unoriented, \ie, posess no preferred direction in space.
More precisely, when the $\beta$-decay parent nuclei are produced in
fusion-evaporation or spallation (Sections~\ref{secmtc}
and~\ref{secgps}), they are actually created with a preferred
direction, namely the beam axis, and posess an anisotropic angular
momentum distribution~\cite{newton1974:higamma}, but this orientation
is lost well before $\beta$ decay occurs, on a time scale typically of
hundreds of ns~\cite{frauenfelder1965:ang}.  Consequently, all
radiations from the $\beta$ decay are emitted with isotropic angular
and polarization
\textit{distributions}.  However,
angular or polarization \textit{correlations} between pairs of emitted
radiations are in general anisotropic.  In $\beta$-decay experiments,
either $\gamma$-$\gamma$ correlations or $\beta$-$\gamma$ correlations
may be measured.

The basic angular correlation relations are well
known~\cite{biedenharn1953:angcorr,frauenfelder1965:ang}, so only a
brief summary is provided here.  For two radiations emitted in
cascade, the probability per unit solid angle of emission of radiation
1 in direction $\Omega_1$ and radiation 2 in direction $\Omega_2$ can
depend, by isotropy of the source, only upon the relative angle
$\theta$ between these two directions.  The probability distribution
can thus be entirely described by an angular correlation function
$W(\theta)$ on the interval [0$^\circ$,180$^\circ$].  It is convenient
to decompose this function as a Legendre series
\begin{equation}
W(\theta)=\sum_\nu A_\nu P_\nu(\cos \theta).
\end{equation}
The coefficents $A_\nu$ may be expressed in a simple fashion in terms
of Clebsch-Gordon and Racah coefficients involving the spins of the
nuclear states and the multipolarities of the radiations, and they can
be calculated for quite general pairs of radiations ($\beta$,
$\gamma$, conversion electron, ...).  For any particular set of spins
and multipolarities, the series terminates after only a few terms.
Only even-order terms have nonvanishing coefficients (this follows
from parity being a good quantum number for the states
involved~\cite{biedenharn1953:angcorr}), so the angular distribution
is symmetric about $\theta$=90$^\circ$.  Angular correlation patterns
for some commonly encountered spin and multipolarity combinations are
shown in Fig.~\ref{figangcorr}.  Calculation of the correlation
functions is discussed in Appendix~\ref{appangpol}.
\begin{figure}
\begin{center}
\includegraphics*[width=0.48\textwidth]{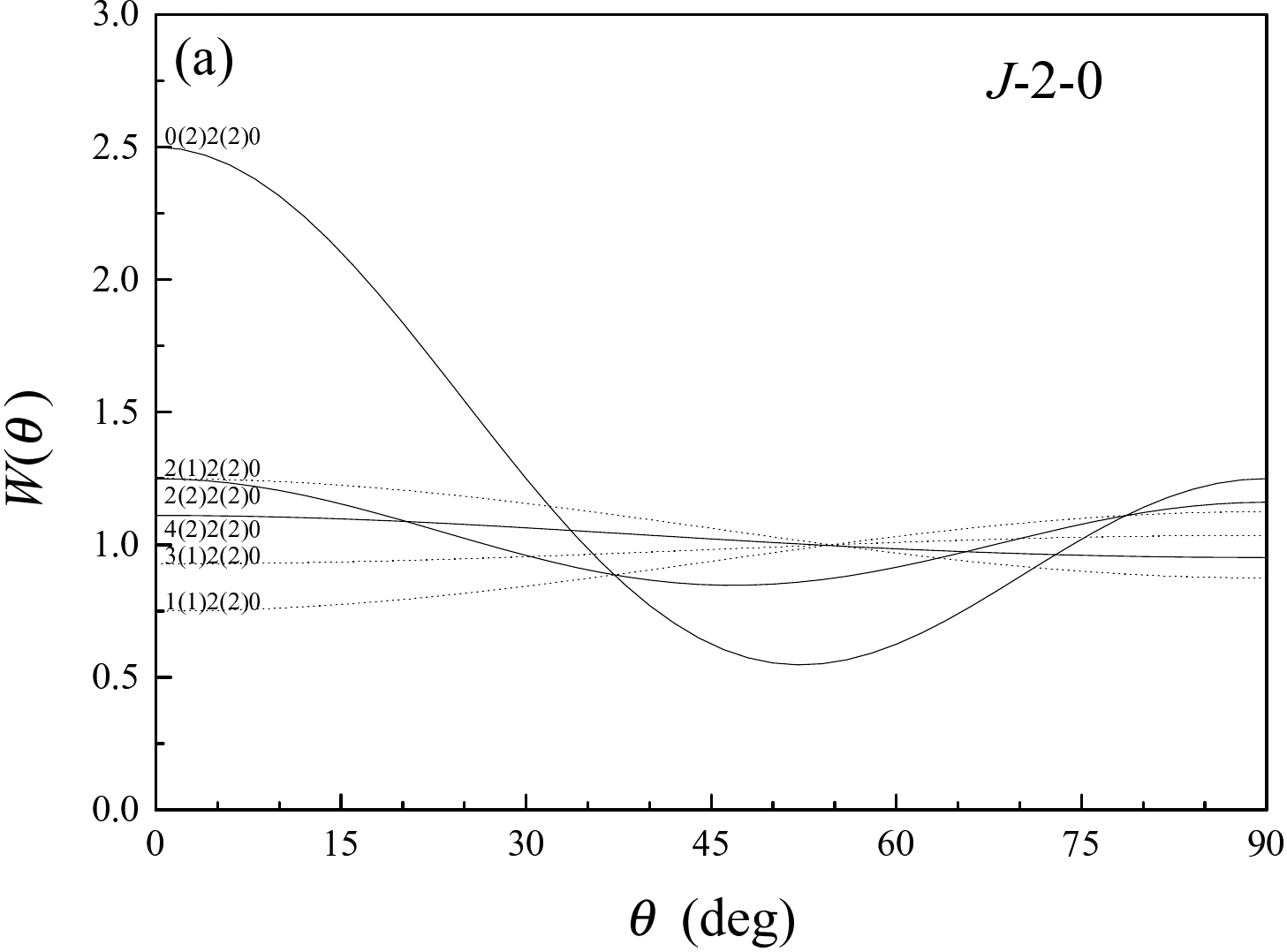}
\hfill
\includegraphics*[width=0.48\textwidth]{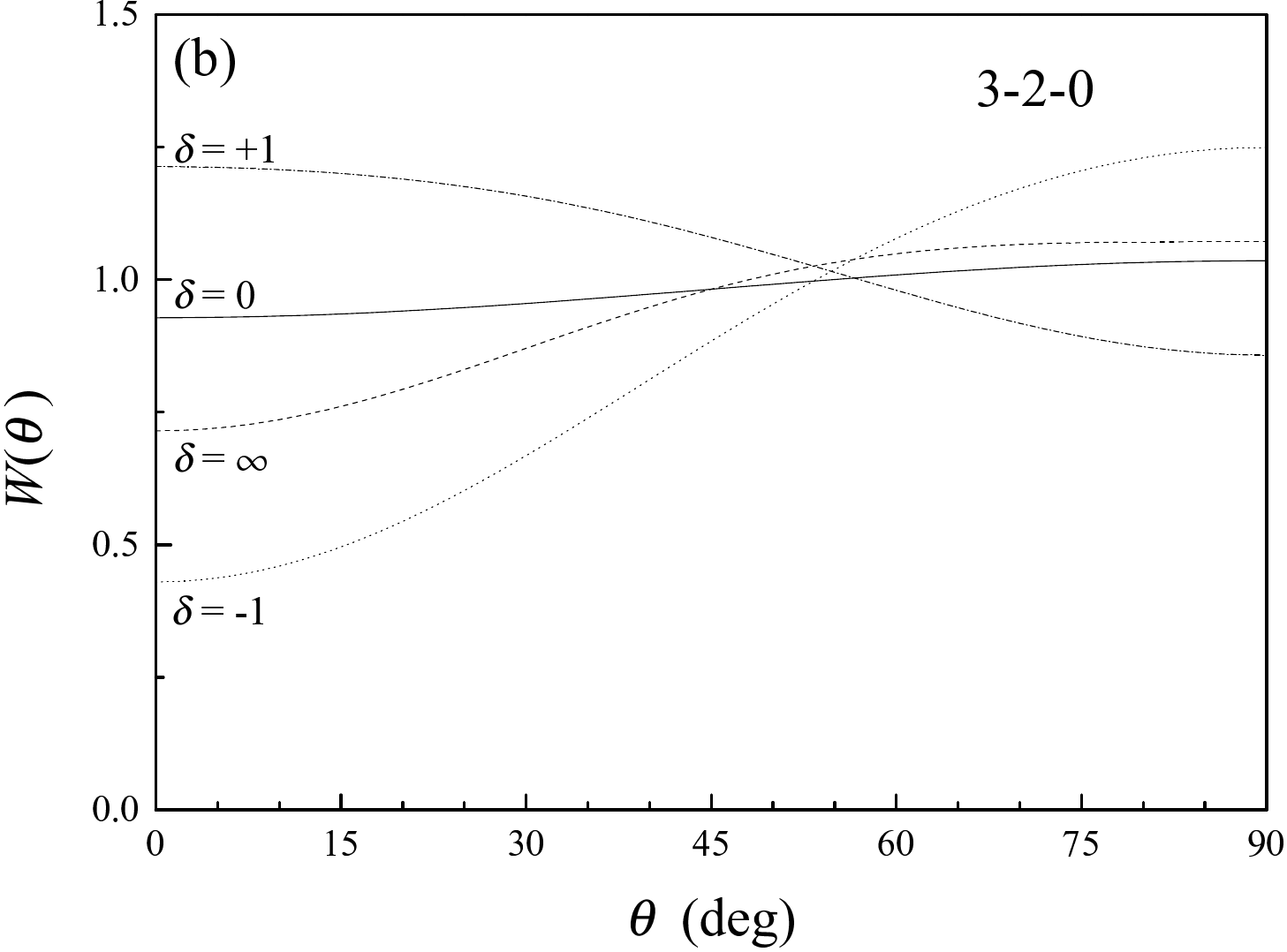}
\end{center}
\caption[Angular correlation patterns for some common $\gamma$-ray
cascades.]{Angular correlation patterns for some common $\gamma$-ray
cascades among low-spin states: (a) Cascades involving transitions of pure dipole or
quadrupole multipolarity among levels with spins $J$-2-0.  Curves are designated
$J_1(L_1)J(L_2)J_2$. (b) Cascades involving levels of spins 3-2-0,
for various $\delta_{E2/M1}$ mixing ratios in the $3\rightarrow2$ transition.
\label{figangcorr}
}
\end{figure}

The angular correlation formalism can account for arbitrary mixtures
of contributing radiation multipolarities.  As discussed in
Chapter~\ref{chapexpt}, $\gamma$ radiation often consists of an
$M\lambda$ and $E(\lambda+1)$ admixture, in which case interference
between the two multipoles occurs.  The correlation pattern then
depends strongly upon the mixing ratio $\delta$ [Fig.~\ref{figangcorr}(b)].

\begin{figure}
\begin{center}
\includegraphics*[width=0.75\textwidth]{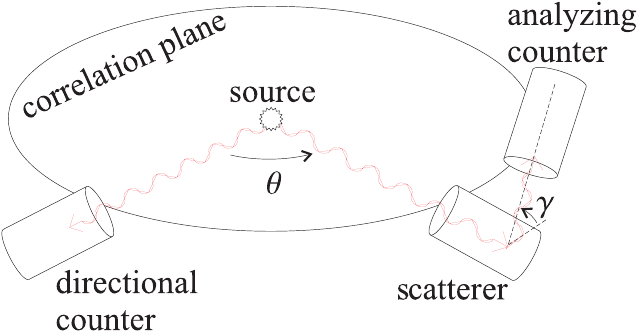}
\end{center}
\caption[Geometry for direction-polarization correlation measurements.]
{Geometry for direction-polarization correlation measurements.  The
basic experimental components are a directional counter and a Compton
polarization analyzer, consisting of a paired scatterer and counter.  
\label{figpolgeom}
}
\end{figure}

\begin{figure}
\begin{center}
\includegraphics*[width=0.48\textwidth]{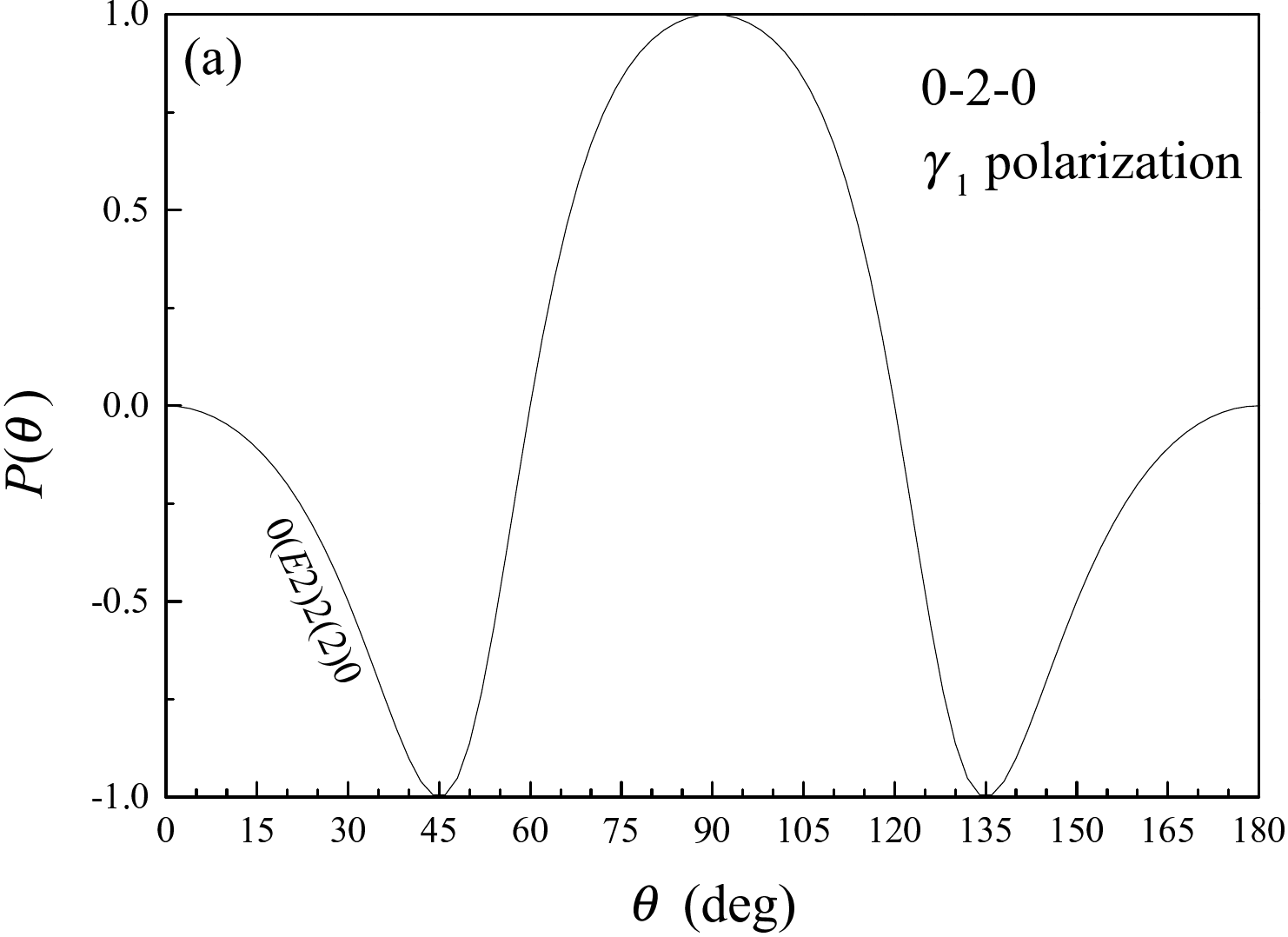}
\hfill
\includegraphics*[width=0.48\textwidth]{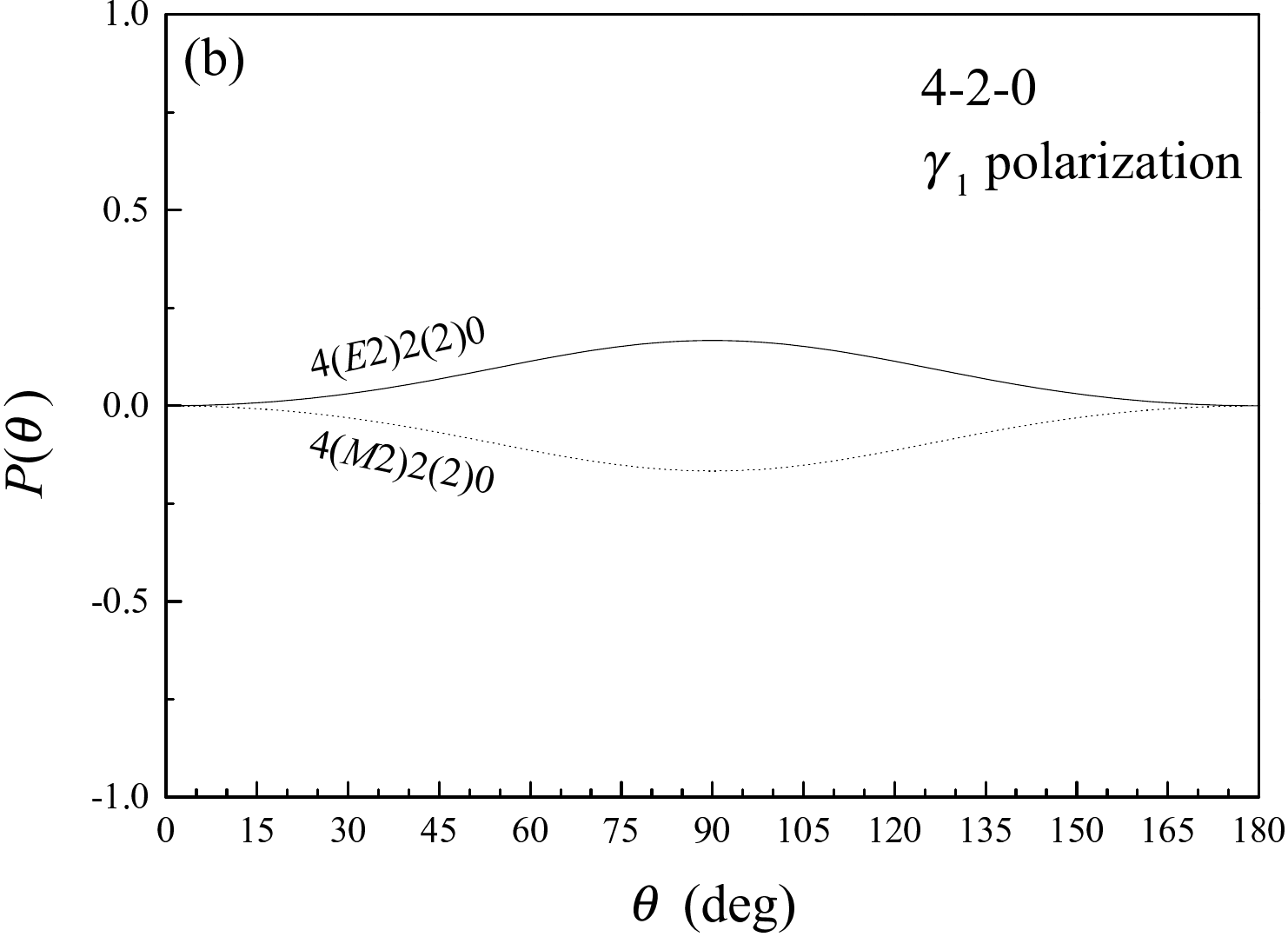}
\end{center}
\caption[Direction-polarization correlation patterns
for $0,4$$\rightarrow$$2$$\rightarrow$$0$
cascades.]{Direction-polarization correlation patterns for (a)
$0\xrightarrow[\text{POL}]{E2}2\xrightarrow{2}0$ and (b)
$4\xrightarrow[\text{POL}]{E2~\text{or}~M2}2\xrightarrow{2}0$ cascades,
illustrating the complete polarization found in 0--2--0 cascades (useful
for identification of $0^+$ states) and the overall sign reversal of
the polarization depending upon the magnetic or electric nature of the
measured radiation (useful for parity determinations when the angular
momenta are already known).
\label{figpolj20}
}
\end{figure}

\begin{figure}
\begin{center}
\includegraphics*[width=0.48\textwidth]{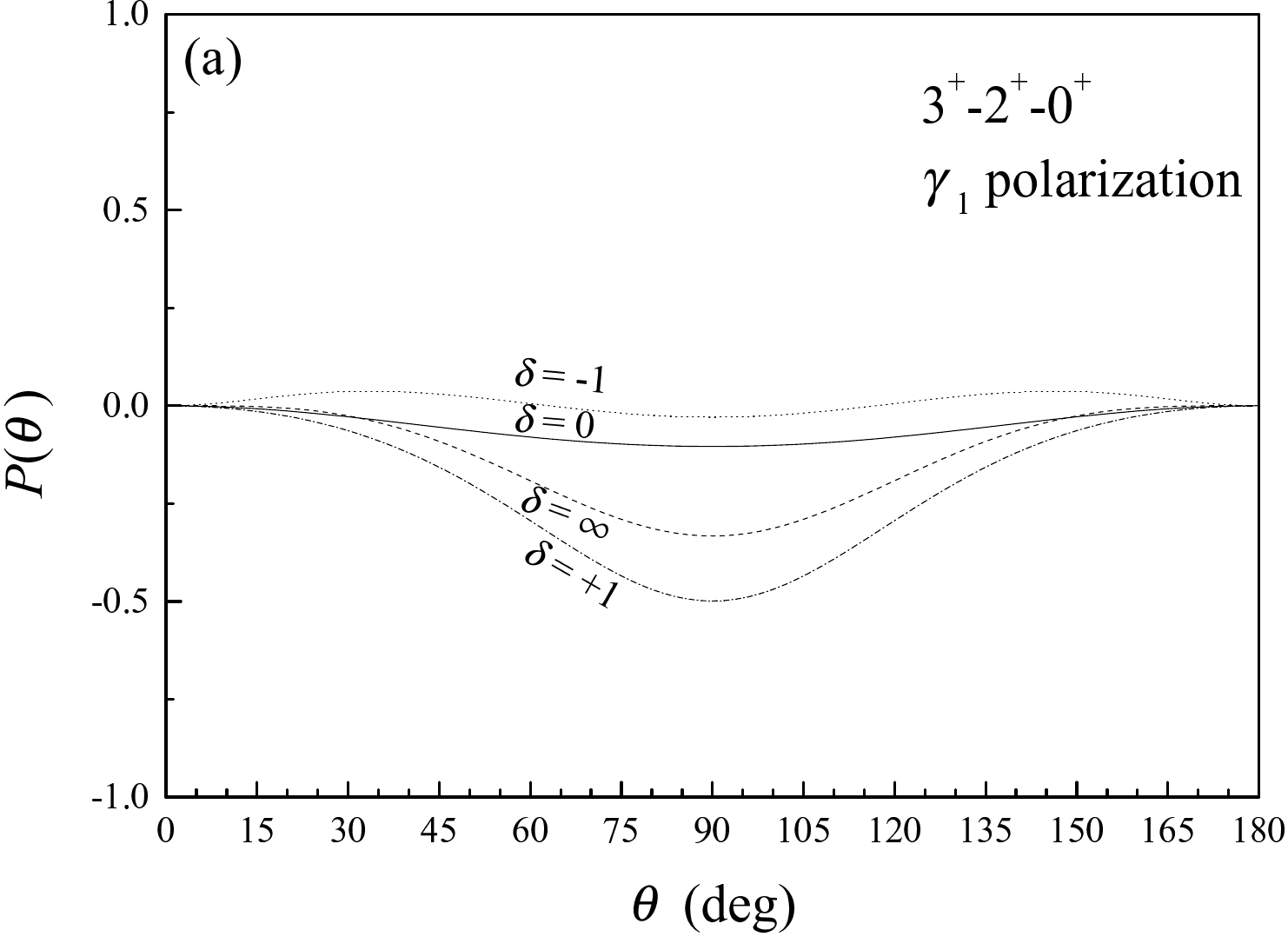}
\hfill
\includegraphics*[width=0.48\textwidth]{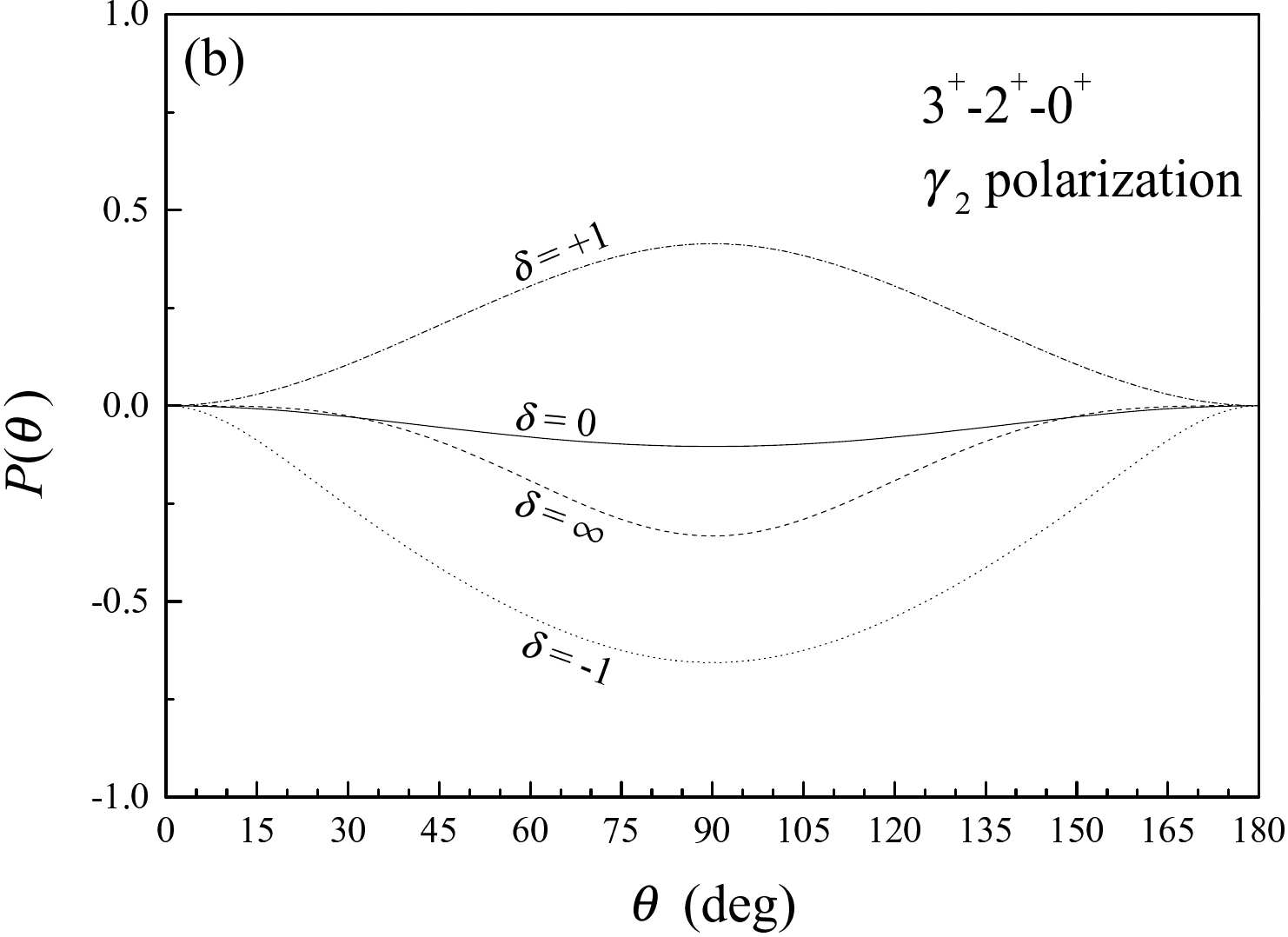}
\end{center}
\caption[Direction-polarization correlation patterns for 
mixed $3^+$$\rightarrow$$2^+$$\rightarrow$$0^+$ cascades.]{Direction-polarization correlation
patterns for mixed (a)
$3^+\xrightarrow[\text{POL}]{M1,E2}2^+\xrightarrow{E2}0^+$ and (b)
$3^+\xrightarrow{M1,E2}2^+\xrightarrow[\text{POL}]{E2}0^+$ cascades,
illustrating how measurement of the polarization for either transition
in the cascade can provide information on the mixing ratio $\delta$ of
the first transition. 
\label{figpol320}
}
\end{figure}

The distribution of a $\gamma$ radiation may be considered as a
function of $\gamma$-ray linear
polarization~\cite{fagg1959:polarization} in addition to the direction
of emission.  For two radiations emitted in a cascade, where one (at
least) is a $\gamma$ ray, the probability of emission of this $\gamma$
ray with polarization angle $\gamma$ relative to the plane containing
the emission directions of two radiations (Fig.~\ref{figpolgeom}) is
\begin{equation}
W(\theta,\gamma)=\left[ \sum_\nu A_\nu P_\nu(\cos
\theta)\right]+\left[ \sum_\nu \mathcal{A}_\nu  P_\nu^{(2)}(\cos \theta)\right]\cos 2 \gamma ,
\end{equation}
where the $\mathcal{A}_\nu$ are again simple functions of the spins
and multipolarities involved (Appendix~\ref{appangpol}).  By axial symmetry,
polarization effects vanish for back-to-back radiations
($\theta$=180$^\circ$).  Polarization effects dominated by the dipole
term [$P_2^{(2)}(\cos
\theta)=3\sin^2\theta$] are most pronounced at an angle of 90$^\circ$ between the emitted
radiations.  All polarization information is contained in a
decomposition of the polarization along the two directions
$\gamma$=0$^\circ$ and 90$^\circ$, conventionally summarized by the
polarization distribution function
\begin{equation}
\label{eqndefpol}
P(\theta)=\frac {W(\theta,0^\circ)-W(\theta,90^\circ)}
{W(\theta,0^\circ)+W(\theta,90^\circ)}.
\end{equation}
The polarization functions for some
representative spin and multipolarity combinations are shown in
Figs.~\ref{figpolj20} and~\ref{figpol320}.  

Polarization measurements provide several forms of useful information.
The degree of polarization in the spin 0--2--0 quadrupole cascade is
very high [Fig.~\ref{figpolj20}(a)], reaching complete polarization at
the angles of $45^\circ$, $90^\circ$, and $135^\circ$ so polarization
measurements can aid in the identification of $0^+$ states.  Unlike
the angular correlation function $W(\theta)$, which depends only upon
angular momenta, the polarization correlation function
$W(\theta,\gamma)$ is highly sensitive to the electric or magnetic
nature of the $\gamma$ radiation.  The polarization distribution
function for a $\gamma$ radiation flips sign depending upon whether
the radiation is electric or magnetic in character [see
(\ref{eqsubstpol})--(\ref{eqnpol3d22p0})], as shown in
Fig.~\ref{figpolj20}(b).  Consequently, if all spins and
multipolarities are known, polarization measurements can be used to
determine the parity of the radiation and thus to obtain level parity
assignments~\cite{stump1952:124sb-beta-gamma-pol,hamilton1953:beta-gamma-pol}.
For mixed multipolarity $\gamma$ radiation, the polarization depends
upon the mixing ratio $\delta$ (Fig.~\ref{figpol320}).  Polarization
measurements can be used to resolve ambiguities in measurements of
$\delta$ from angular correlations.

\subsection{Experimental considerations}

Classic angular correlation measurements~\cite{frauenfelder1965:ang}
commonly determined the function $W(\theta)$ by counting coincidences
between a fixed detector and a moveable detector for each of several
values of the relative angle $\theta$.  This approach has the benefit
of simplicity, since, with careful experiment design, detector
efficiencies remain constant between measurements at different
$\theta$ values.  Modern angular correlation measurements, in
contrast, make use of high-efficiency multidetector arrays.  With such
arrays, all detectors are held stationary, and coincidences at
multiple relative angles are measured simultaneously.  This approach
greatly improves the available statistics but introduces some
complexities to the data analysis.

For two detectors $i$ and $j$ at relative angle $\theta_{i:j}$, with radiation $a$
detected in one and radiation $b$ detected in the other, the number of
coincident counts detected is
\begin{equation}
G_{a:b}^{i:j}=N I_{a:b} W_{a:b}(\theta_{i:j})
\varepsilon^{i:j}(E_a,E_b),
\label{eqnaivegatew}
\end{equation}
where $N$ is the number of parent decays, $I_{a:b}$ is the number of
coincident radiations $a$ and $b$ emitted per parent decay, and
$\varepsilon^{i:j}$ is the detector pair efficiency.  In principle,
$W_{a:b}(\theta)$ can be extracted from this coincidence measurement
directly, since the activity and coincident intensity are known from
the spectroscopy experiment (Section~\ref{seccoin}) and the detector
efficiency is calibrated.  

In practice, however, uncertainties in the intensity and efficiency
can be large compared to the necessary accuracy of the
$W_{a:b}(\theta)$ measurement.  For example, small deviations of the
calibration source position from the experimental source position,
which have negligible effect on the overall array efficiency
calibration, can significantly disrupt the angular correlation pair
efficiency.  This method also contains no internal check
against drift in efficiency.  It is therefore useful to have a
measurement technique in which as many parameters as possible cancel
or are internally calibrated.  Several approaches exist for situations
in which only coincidence data are
available~\cite{wolf2002:128ba-beta,asai2001:angcorr}.  However, if singles
data are taken from all detectors simultaneously with the coincidence
data, an especially simple approach is possible, provided that the
coincidence efficiency for the two detectors factorizes as
$\varepsilon^{i:j}(E_a,E_b)=\varepsilon^{i}(E_a)\varepsilon^{j}(E_b)$
(Section~\ref{seccoineff}).  When \textit{ratios} of coincidence counts for radiations $a$
and $b$ in different detector pairs are used to obtain
\textit{ratios} of $W_{a:b}(\theta)$ values, the singles counts in the
individual detectors [$S_a^i=N\varepsilon^i(E_a) I_a$] provide an
internal calibration of the detector efficiencies, and all
efficiencies required in (\ref{eqnaivegatew}) cancel:
\begin{equation}
\frac{G_{a:b}^{i:j}}{G_{a:b}^{k:l}}
=
\frac
{W_{a:b}(\theta_{i:j})S_a^i S_b^j}
{W_{a:b}(\theta_{k:l})S_a^k S_b^l} .
\end{equation}
The internal consistency of this method can be checked by comparing
values obtained for different detector pairs at the same angles,
including comparison of the ``reverse'' gates in the same detector
pair [$W_{a:b}(\theta_{i:j})=W_{a:b}(\theta_{j:i})$].

With an array of clover detectors, as at the Yale Moving Tape
Collector, two approaches may be taken for angular correlation
measurements.  Each clover may be treated as four independent coaxial
detectors within a common cryostat (as in
Ref.~\cite{wolf2002:128ba-beta}), and angular correlation measurements
are in this case carried out simultaneously for all pairs of clover
elements in the array, excluding pairs within the same detector since
these are subject to excessive Compton cross scatter.  An array of
three clover detectors arranged in a plane at equal distances from the
source provides 48 correlation pairs at 18 distinct angles, once
symmetries of the element positions within the clovers are taken into
account.  Alternatively, each clover as a whole may be treated as a
single detector (as in the experiment of Chapter~\ref{chap162er}), in
which case an array of three clover detectors provides only three
correlation pairs.  This approach is sufficient if the goal of the
experiment is simply to distinguish spin 0--2--0 quadrupole cascades
from other cascades, due to the highly pronounced 0--2--0 correlation
pattern (Fig.~\ref{figangcorr}).  For this purpose, it is preferable
to have one clover pair positioned at a relative angle of
$\sim$180$^\circ$, equivalent to the first-quadrant angle of 0$^\circ$
in Fig.~\ref{figangcorr}, since this is the location of the maximum of
$W(\theta)$ for a 0--2--0 cascade.  The remaining clover can then be
inserted between these at an angle of approximately 52$^\circ$
relative to one of the first two clovers, the location of the minimum
of $W(\theta)$, and the remaining angle is then automatically the
supplementary angle 128$^\circ$, also equivalent to 52$^\circ$.
However, this configuration is not optimal for spectroscopy, since two
clovers diametrically opposite each other are susceptible to unwanted
coincidence events due to $\beta^+$ annihilation photon pairs and 
Compton cross-scattering.

The effects of finite detector opening angle and source size must be
considered in angular correlation experiments.  The measured angular
correlation $W^\text{expt}(\theta)$ is the result of the convolution
of the true angular correlation with the extended angular acceptances
of the detectors.  A single clover element spans $\sim$25$^\circ$ at
a distance of 10\,cm from the source; the additional spread due to the
$\lesssim$1\,cm source size contributes negligibly.  For a pair of
cylindrically symmetric detectors, the addition theorem for spherical
harmonics dictates that the convolution 
results simply in a term-by-term multiplicative
correction~\cite{frankel1951:angcorr-correct}
\begin{equation}
W^\text{expt}(\theta)=\sum_\nu A_\nu Q_\nu^i Q_\nu^j P_\nu(\cos \theta),
\end{equation}
where the solid angle correction factors $Q_\nu^i$ and $Q_\nu^j$ for
the two detectors are energy dependent and can be calculated by
integrations of the detector efficiency over the detector
volume~\cite{rose1953:angcorr,yates1965:solid-angle}.  These results are applicable to
correlations of individual clover elements, since these are
approximately cylindrical, but not to the entire clover used as a
detector for angular correlations.  The entire clover has a much
reduced symmetry~--- namely symmetry under horizontal and vertical
reflection~--- which is sufficient to guarantee that the convolution
process will not introduce odd-order Legendre polynomials or
higher-order terms than those already
present~\cite{feingold1955:angcorr-correct} but provides little
additional simplification.  Fortunately, for the 0--2--0 cascade and
the optimized detector configuration described above, averaging over
even tens of degrees of acceptance still leaves the 0--2--0 cascade
clearly distinguishable from the other likely cascades
(Fig.~\ref{figangcorr}), and sophisticated corrections are
unnecessary.

The linear polarization of a $\gamma$ ray is most easily measured by
means of ``Compton
polarimetry''~\cite{fagg1959:polarization,frauenfelder1965:ang}, a
method based upon the polarization dependence of the Compton
scattering cross section.  The Compton scattering probability, given
by the Klein-Nishina formula, is dependent upon the angle between the
scattering direction and the plane perpendicular to the incident
$\gamma$-ray polarization vector~--- it is maximal in this plane and
lowest directly out of this plane.  A minimal apparatus for Compton
polarimetry consists of a scatterer and a pair of analyzing counters
to measure the relative probability of scattering in two perpendicular
planes.  

\begin{figure}
\begin{center}
\includegraphics*[width=0.5\textwidth]{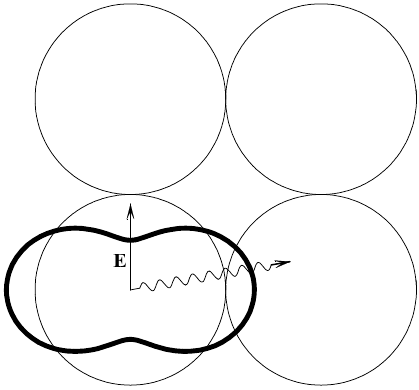}
\end{center}
\caption[The clover detector as a Compton polarimeter.]
{The clover detector as a Compton polarimeter, viewed from the source
position.  Each leaf serves as both a scatterer and an analyzing
counter.  A representative angular distribution pattern for the
scattered photon is indicated, illustrating the preference for
scattering perpendicular to the incident photon polarization vector.
\label{figclovpol}
}
\end{figure}

The clover detector may be used as a Compton
polarimeter~\cite{jones1995:clover-polarimeter,garci-raffi1997:clover-pol,duchene1999:clover}
in which each leaf acts both as a scatterer and as an analyzing
counter to measure scattering from the other two adjacent leaves (see
Fig.~\ref{figclovpol}).  In software analysis of the clover hit, the
total energy of the incident $\gamma$-ray is obtained as the sum of
the energies deposited in the two leaves, and the Compton scattering
direction is identified as being either in the correlation plane or
perpendicular, depending upon which pair of adjacent leaves fired
(hits involving scattering between diagonally opposite leaves are
discarded).  The incident $\gamma$ rays with out-of-plane polarization
preferentially scatter between in-plane pairs of leaves, and vice
versa, but both polarizations (parallel or perpendicular) contribute
to both types of scattering (parallel or perpendicular).
Consequently, the observed numbers of scattered counts depend upon a
total of four efficiencies:
\begin{align}
N_\parallel & 
= \varepsilon_\parallel(\parallel) I_\parallel 
+ \varepsilon_\parallel(\perp)I_\perp, \\
N_\perp & 
= \varepsilon_\perp(\parallel) I_\parallel 
+ \varepsilon_\perp(\perp)I_\perp, 
\end{align}
where $N_{\parallel(\perp)}$ is the number of scatterings between
in-plane (out-of-plane) pairs of leaves, $I_{\parallel(\perp)}$ is the
in-plane (out-of-plane) polarization component, and $\epsilon_x(y)$ is
the efficiency for a photon with polarization $y$ to result in an
observed scattering of type $x$.  The efficiencies are energy
dependent, due both to the intrinsic energy dependence in the
Klein-Nishina formula and to the detector properties.  Inverting these
relations provides an expression for the
polarization~(\ref{eqndefpol}) in terms of experimental quantities
\begin{equation}
P=\frac
{
[\varepsilon_\perp(\perp)+\varepsilon_\perp(\parallel)] N_\parallel 
-[\varepsilon_\parallel(\perp)+\varepsilon_\parallel(\parallel)]N_\perp
}
{
[\varepsilon_\perp(\perp)-\varepsilon_\perp(\parallel)] N_\parallel 
-[\varepsilon_\parallel(\perp)-\varepsilon_\parallel(\parallel)]N_\perp
}.
\end{equation}

For an ideal clover with four identical leaves,
$\varepsilon_\parallel(\perp)=\varepsilon_\perp(\parallel) \, (\equiv
\varepsilon_>)$ and
$\varepsilon_\parallel(\parallel)=\varepsilon_\perp(\perp) \, (\equiv
\varepsilon_<)$, so the polarization is simply extracted as
\begin{equation}
P = \frac{(1 + \varepsilon_< / \varepsilon_>)}
{(1 - \varepsilon_< /\varepsilon_> )}
 \frac{N_\perp-N_\parallel}{N_\perp+N_\parallel}.
\label{eqnpexpt}
\end{equation}
The reciprocal of the first factor, a measure of the polarimeter's
sensitivity, is commonly denoted by $Q$, and the second factor is
referred to as the experimental asymmetry ratio $A$.  $Q$ is 
energy dependent and  may be calibrated using a $\gamma$ ray of known
polarization near the energy of interest or may be calculated in
detector simulations~\cite{garci-raffi1997:clover-pol}.  The deviation
of a polarimeter from ideal symmetric response is conventionally
absorbed into an
\textit{ad hoc} instrumental asymmetry correction parameter~\cite{butler1973:pol-asymm}, by which $N_\perp$ must be
multiplied in~(\ref{eqnpexpt}), so
\begin{equation}
P = \frac{1}{Q}
 \frac{aN_\perp-N_\parallel}{aN_\perp+N_\parallel}.
\label{eqnpexptwitha}
\end{equation}
[This procedure is motivated by
assuming a specific form for the asymmetry,
$\mathbox{\varepsilon_\parallel(\perp)}=a\mathbox{\varepsilon_\perp(\parallel)}$ and
$\mathbox{\varepsilon_\parallel(\parallel)}=a\mathbox{\varepsilon_\perp(\perp)}$.]  The
intrumental asymmetry correction is calibrated by exposing the
detector to known unpolarized $\gamma$ rays, for which $A$ should be
vanishing: these may be singles $\gamma$ rays, $\gamma$ rays
coincident at $\theta$=180$^\circ$, or $\gamma$ rays coincident with
uncorrelated radiations, such as atomic x~rays or annihilation
radiation.  An experimental example is given in
Chapter~\ref{chap152sm}.

\section{Fast electronic scintillation timing measurements}
\label{secfest}

It is desirable to be able to measure the lifetimes of excited nuclear
states, since electromagnetic transition strengths are deduced from
these in combination with the $\gamma$-ray branching information.  For
a state populated in some form of decay process, the lifetime may be
determined by measuring the time difference between detection of a
radiation emitted in the process of population of the level and of a
radiation emitted as part of the subsequent decay of that level.

For measurement of short lifetimes, the timing response of the
detection system must be optimized in two respects: to minimize
statistical fluctuations in the timing response for the detected
radiations (timing jitter) and to minimize variation of this 
response with respect to the energy of the detected radiation (timing
walk).  To optimize either of these aspects of timing, it is
generally necessary that detectors be used which produce as
short an output pulse rise time as possible.

\begin{figure}[t]
\begin{center}
\includegraphics*[width=0.65\textwidth]{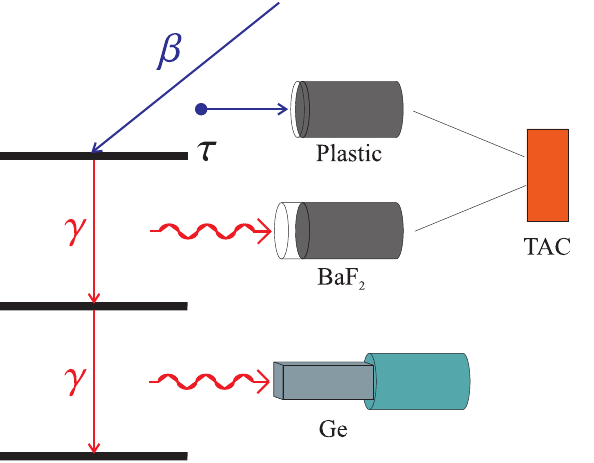}
\end{center}
\caption[Schematic representation of the FEST lifetime measurement method.]
{Schematic representation of FEST lifetime measurement method.  The
$\beta$ radiation is detected in a fast-timing plastic scintillation
detector and a $\gamma$ ray is detected in a fast-timing BaF$_2$
detector.  An additional $\gamma$ ray detected in a Ge detector may be
used when the energy resolution of the BaF$_2$ detector is insuffient
to isolate the $\gamma$ ray of interest in singles or for special
purposes involving cascade selection (\eg,
Fig.~\ref{fig154dycascades}).  The plastic-BaF$_2$ relative timing is
used for the determination of the level lifetime $\tau$.
\label{figfestcartoon}
}
\end{figure}

The fast electronic scintillation timing (FEST) techniques developed
by Mach, Moszy\'nski, Gill and
collaborators~\cite{mach1989:fest1,moszynski1989:fest2,mach1991:sr-fest}
make possible lifetime measurements for levels populated in $\beta$
decay with lifetimes as short as several ps.  These methods are based
upon electronic timing of the interval between $\beta$-particle
emission, detected with a fast plastic scintillation detector, and the
subseqent $\gamma$-ray decay, detected with a BaF$_2$ detector.  It is
also often useful to require a coincidence with an additional $\gamma$
ray in a Ge detector.  A schematic representation of the method is
shown in Fig.~\ref{figfestcartoon}.  To accomplish the measurement of
such short lifetimes, several special techniques are used.

The fast-timing $\beta$ detection in FEST makes use of the $\Delta E$
signal produced by the $\beta$ particle as it traverses a thin slice
of fast plastic scintillation material.  The energy dependence of a
timing signal results primarily from the pulse-height dependence of
the electronic timing discrimination process.  The advantage to using
the $\Delta E$ signal from a thin scintillator is that the $\Delta E$
signal is largely independent of incident $\beta$-particle
energy~\cite{knop1965:radiation-interaction}, and thus the timing has minimal
dependence upon the $\beta$-particle energy.  The energy spectrum of
emitted $\beta$ particles is continuous up to an end-point energy, and
so use of a $\beta$-particle full-energy absorption signal from a
thick scintillator would in contrast require timing of signals
covering a broad dynamic range.  Plastic scintillation materials with
rise times as short as hundreds of ps are
available~\cite{knoll1989:radiation-detection}.  The NE111
scintillators used for many FEST experiments have a measured rise time
of 200\,ps and decay time of
1.7\,ns~\cite{bengtson1974:organic-timing}.

For fast timing $\gamma$-ray detection, it is preferable to use a
high-$Z$ inorganic scintillator to provide appreciable efficiency for
full-energy $\gamma$ detection.  The timing performance of inorganic
scintillators typically lags that of organic scintillators by orders
of magnitude (\eg, Ref.~\cite{knoll1989:radiation-detection}).  To acheive the
necessary timing performance, FEST relies upon the fast ultraviolet
component of scintillation light from BaF$_2$~\cite{laval1983:baf2},
which has a decay constant of $\sim$600\,ps.  BaF$_2$ provides an
energy resolution marginally inferior to that obtained with NaI, due
to its lower photon yield.

A coincidence with an additional $\gamma$ ray detected in a Ge
detector can be required as a gating condition.  A Ge gating condition
is valuable when the energy resolution of the BaF$_2$ detector is
insuffient to isolate the $\gamma$ ray of interest in singles, as is
often the case.  Gating may also be used for cascade selection to
choose specific $\gamma$-ray feeding paths for the lifetime
measurement, as in Chapter~\ref{chap154dy}.

Additional technical innovations can contribute further moderate
performance improvements to FEST detector systems.  These include the
use of specially-shaped truncated conical BaF$_2$ crystals to control
the reflection of scintillation light into the photomultiplier
tube~\cite{mcgervey1977:conical} and extraction of the timing signal
from an early dynode stage of a modified photomultiplier voltage
divider chain~\cite{bengtson1982:dynode}.

\begin{figure}
\begin{center}
\includegraphics*[width=\textwidth]{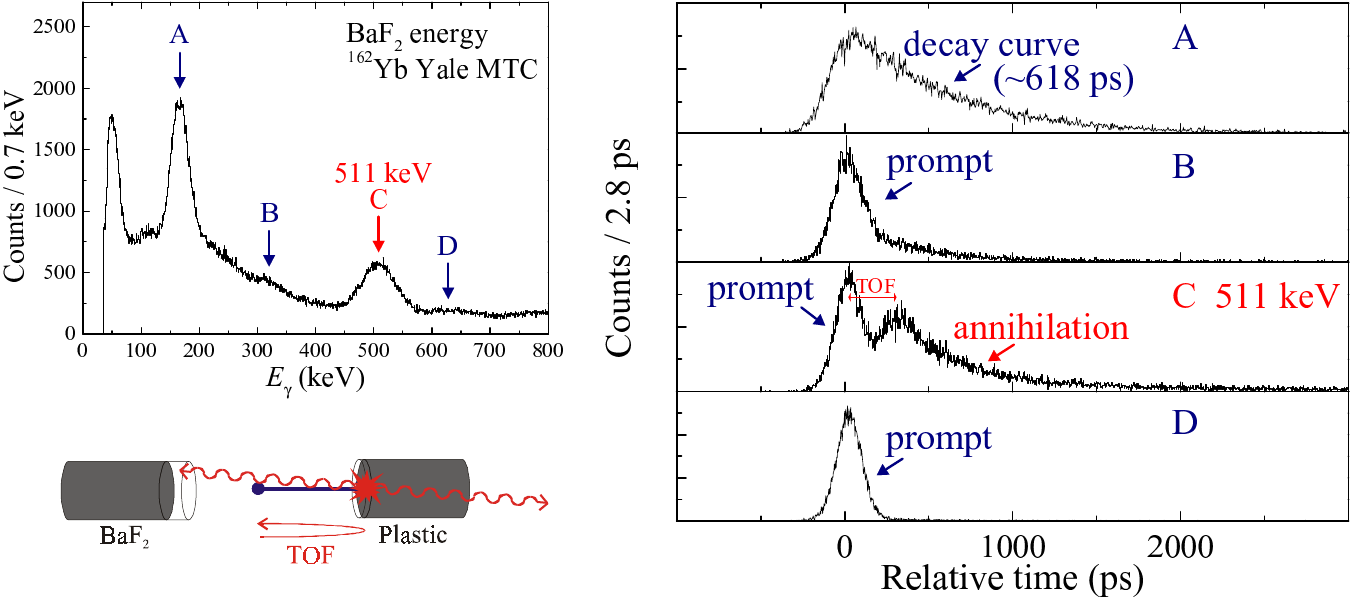}
\end{center}
\caption[Detection of delayed annihilation
radiation in the BaF$_2$ detector.]{Detection of delayed $\beta$-coincident annihilation
radiation in the BaF$_2$ detector for FEST experiments in
$\beta^+$ decay.  For gates
on each of the BaF$_2$ energies indicated by lettered arrows (top left), the
corresponding BaF$_2$-plastic TAC spectra are shown (right).  The
earliest-arriving annihilation radiation is delayed relative to the
prompt location by an additional $\beta$ and $\gamma$ time of flight
(TOF) (bottom left).
\label{figfestannihil}
}
\end{figure}

The FEST method was orginally designed for use with neutron-rich
nuclei provided by reactors or ISOL-type sources, but it has also been
successfully applied to the study of proton-rich
nuclei~\cite{morikawa1992:ba-fest,mantica1996:120xe,mach1997:fest}, as
in the present measurements at the Yale MTC (Chapters~\ref{chap162yb},
\ref{chap154dy}, and~\ref{chap162er}).  Challenges associated with
measurements of proton-rich nuclei include the presence of delayed
coincident background radiation from $\beta^+$ annihilation, with a
time profile (Fig.~\ref{figfestannihil}) which extends into the ns
range and is strongly dependent upon detector geometry and absorber
materials~\cite{eldrup1995:positron}, and competition from electron
capture, in which no emitted $\beta^+$ particle is available for
timing~\cite{joshi:fest-ec-e0}.

\begin{figure}[p]
\begin{center}
\includegraphics*[width=0.75\textwidth]{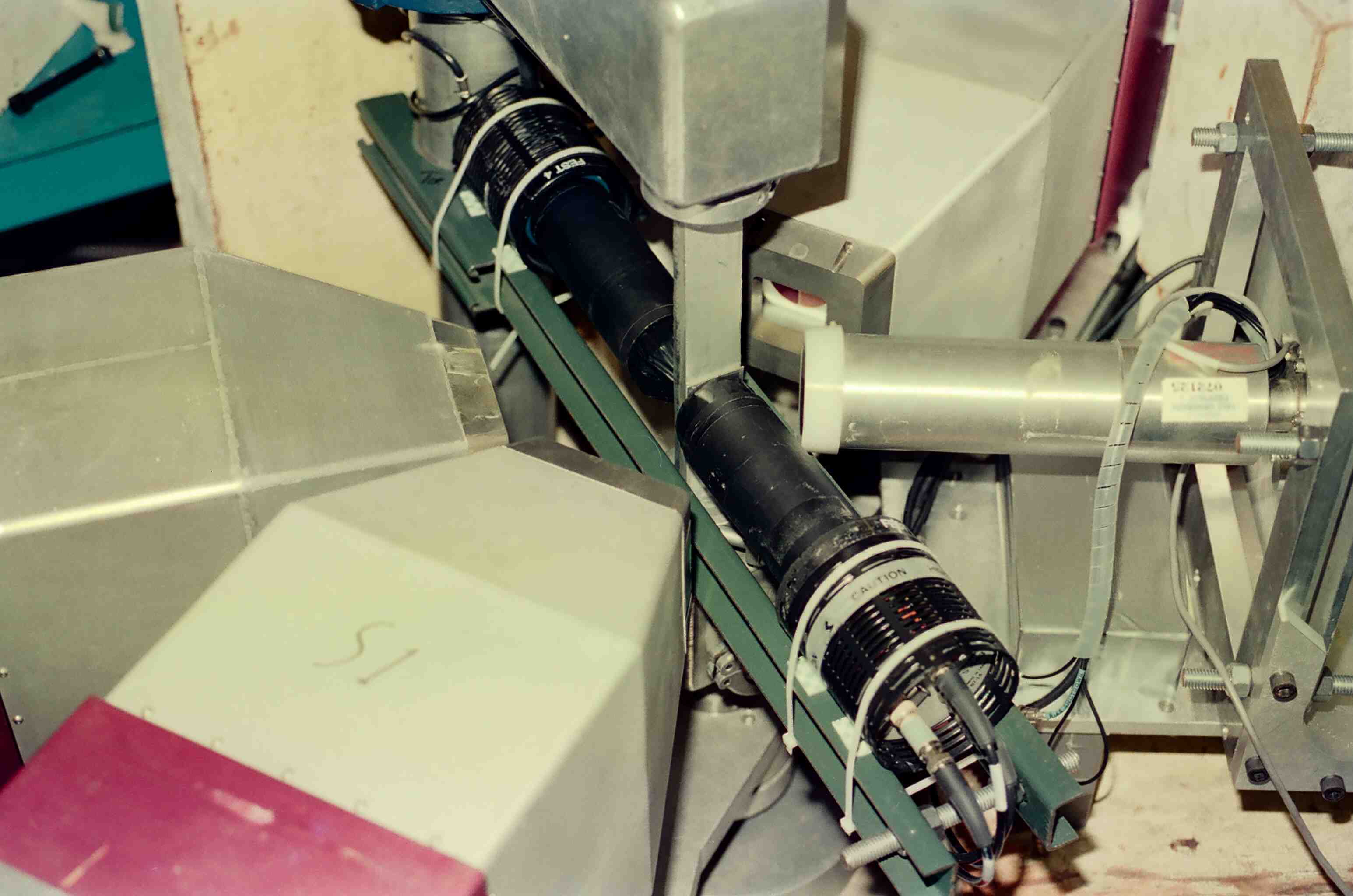}
\end{center}
\caption[Photograph of the Yale MTC FEST configuration.]
{Photograph of the Yale MTC FEST configuration.
\label{figfestphoto}
}
\end{figure}

\begin{figure}[p]
\begin{center}
\includegraphics*[width=0.8\textwidth]{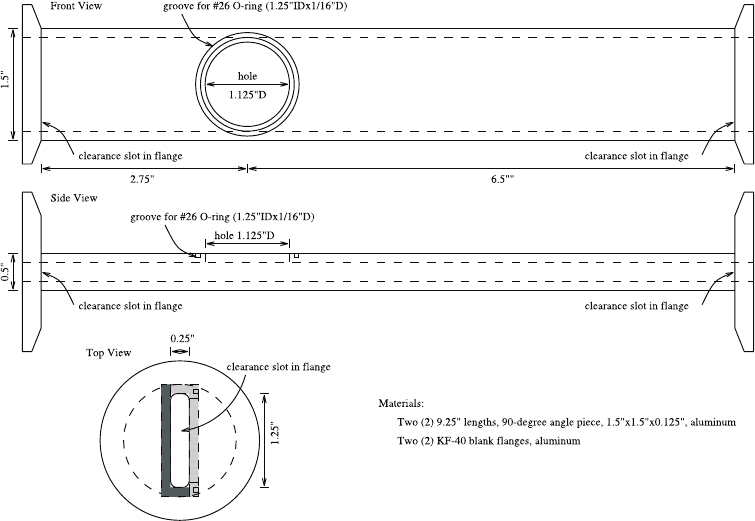}
\end{center}
\caption[Tape transport chamber for FEST experiments at the Yale MTC.]
{Tape transport chamber for FEST experiments at the Yale MTC.
\label{figfestchamber}
}
\end{figure}

The FEST system implemented at the Yale MTC uses an integrated
multidetector array consisting of three Compton-suppressed clover
detectors, a LEPS detector, and the plastic and BaF$_2$ fast timing
detectors (Fig.~\ref{figfestphoto}).  This configuration allows
$\gamma\gamma$ spectroscopy, angular correlation and polarization, and
FEST measurements to be carried out simultaneously.  Fast timing
$\beta$ particle detection is accomplished with a disk of NE111A
plastic scintillation material (1.3\,cm diameter and 3\,mm thick)
coupled to a Photonis XP2020 photomultiplier tube and covered only by
a 20\,$\mu$m aluminum foil, to minimize energy loss of $\beta$
particles entering the detector.  Fast timing $\gamma$-ray detection
is carried out with a BaF$_2$ crystal in the shape of a conical
frustum (1.91\,cm forward diameter, 2.54\,cm length, and 2.54\,cm rear
diameter) coupled using Viscasil silicone
fluid~\cite{klamra1987:uvgrease,ge:viscasil} for transmission of the
fast ultraviolet scintillation light to a quartz-windowed Photonis
XP2020Q photomultiplier tube.  The voltage divider chains for both
photomultiplier tubes are modified to provide timing signals from the
ninth dynode.  The tape carrying the activity passes through the
detector area inside a flat aluminum transport duct, designed to allow
the FEST detectors to be placed facing each other across the tape with
minimal separation (Fig.~\ref{figfestchamber}).  The $\beta$ particles
exit the chamber through a 51\,$\mu$m polypropylene vacuum
window~\cite{3m:scotch3750}.

The fast timing detectors are incorporated into the YRAST Ball
electronics suite~\cite{beausang2000:yrastball}.  Conventional energy and
logic/timing signals are acquired as for the Ge detectors.
Timing discrimination for the fast-timing detectors is carried out
using Tennelec TC454 constant fraction discriminators with minimal
(3\,cm) external wire delays.  The critical plastic-BaF$_2$ time
difference is measured using an Ortec 567 TAC acquired through an ADC,
calibrated using known delays.  

\begin{figure}[p]
\begin{center}
\includegraphics*[width=0.6\textwidth]{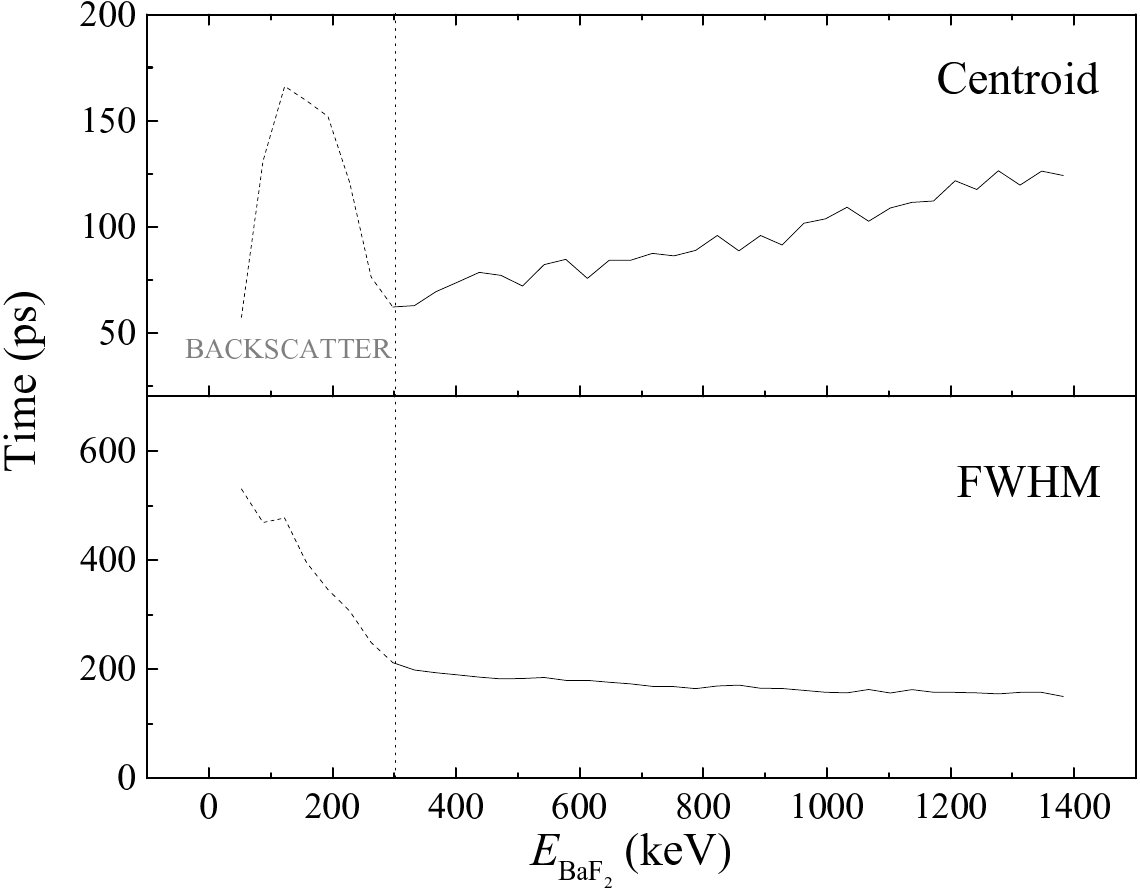}
\end{center}
\caption[Timing resolution and walk properties of the FEST detector system.]
{Timing resolution and walk properties of the Yale MTC FEST detector
system, as determined from $^{60}$Co $\gamma$-ray coincidences (see
text).  Some of the $^{60}$Co $\gamma$ rays which deposit energy in
the plastic scintillator backscatter and are detected in the BaF$_2$
detector, producing a coincidence involving a delay due to the extra
time of flight between the detectors.  These $\gamma$ rays have an
energy after scattering of $\sim$210\,keV, so the timing response
measured near this energy is artificially delayed and broadened
relative to the true response.
\label{figfestperformance}
}
\end{figure}

The timing properties of the Yale MTC FEST system are shown in
Fig.~\ref{figfestperformance}.  These are obtained using coincidences
between the 1173\,keV and 1332\,keV $\gamma$ rays emitted in $^{60}$Co
decay, making use of the plastic scintillator's albeit limited
efficiency for $\gamma$-ray detection through Compton interaction.
Compton and full energy deposition interactions provide a continuum of
energies in the BaF$_2$ detector, allowing the timing characteristics
to be measured as a function of BaF$_2$ energy up to
$\sim$1332\,keV.  The energy dependence of the BaF$_2$-plastic
relative time distribution centroid and width are shown in
Fig.~\ref{figfestperformance} as measures of the timing walk and
resolution, respectively.  The timing resolution under
experimental conditions, with $\Delta E$ signals from high-energy
$\beta$ particles in the plastic detector, is $\sim$140\,ps FWHM at
$\gamma$-ray energies $\gtrsim$1.5\,MeV.

\begin{figure}[p]
\begin{center}
\includegraphics*[width=0.8\textwidth]{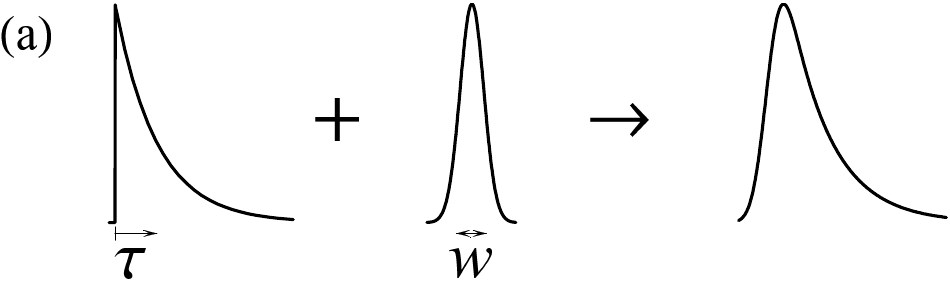}\\
\includegraphics*[width=0.8\textwidth]{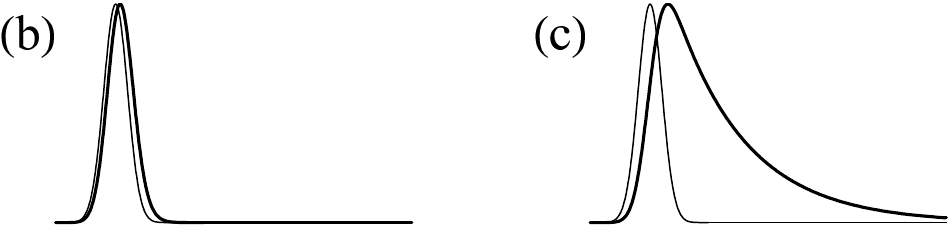}
\end{center}
\caption[Convolution of the decay curve and the instrument response.]
{Convolution of the decay curve and the instrument response: (a) schematic
diagram for lifetime $\tau$ and width $w$, (b) for a short lifetime
($\tau$$\ll$$w$), and (c) for a long lifetime ($\tau$$\gtrsim$$w$).
\label{figconvolution}
}
\end{figure}

The measured distribution for the relative time between $\beta$ and
$\gamma$ emission is the convolution of the true time distribution (a
decaying exponential) with the instrumental response
[Fig.~\ref{figconvolution}(a)].%
The level lifetime can be extracted by a
nonlinear least-squares fitting procedure, the ``deconvolution
method''~\cite{bostroem1966:delayconv1,olsen1966:delayconv2,mach1989:fest1},
provided the instrumental response is sufficiently well known.  In
practice, however, two specialized approaches, the ``centroid shift''
and ``slope'' methods, can be used for the extraction of relatively
short or relatively long lifetimes.

The $\beta\gamma$ delayed time distribution has a centroid shifted
relative to that of a prompt distribution by a time equal to the
lifetime of the intervening level.  If the
experimental prompt centroid time is known for a given $\gamma$-ray
energy, then the lifetime of a level depopulated by a $\gamma$-ray
transition of that energy can be measured simply by observing the
shift of its centroid~\cite{mach1989:fest1} relative to the prompt
centroid [Fig.~\ref{figconvolution}(b)].  Accurate calibration of the
prompt peak position can in general be challenging (see, \eg,
Ref.~\cite{mach1989:fest1} for a detailed discussion), but such
calibration is not always necessary for the analysis
(Chapter~\ref{chap154dy}).  This centroid shift method is effective
for the lifetimes ranging from $\sim$10\,ps to $\sim$100\,ps.  For
long lifetimes, the centroid shift method becomes impractical, since a
wide range of channels must be included in the centroid analysis, and
small fluctuations in the number of background counts in the tail
region have a large effect on the calculated centroid.  The centroid
shift method is also difficult to apply if a significant prompt
contamination is present in the time spectrum.

For decay times much longer than the width of the instrument response,
the tail of the measured decay curve approximates a true decaying
exponential.  The lifetime can then be extracted from the logarithmic
slope of this tail, most conveniently by linear least-squares fitting
of the logarithmic number of counts.  The
slope method is insensitive to prompt contamination, so long as the
prompt region is excluded from fitting.  Incorporation of time
channels containing a small number of counts into the analysis tends
to skew the extracted lifetime, since a simple square-root error
weighting can place far too much weight on such a channel.  A
practical solution is to exclude the extreme tail region from fitting
and to repeatedly compress the spectrum, so that each individual
channel has more counts, until the deduced lifetime is stable against
further compression.

\chapter{Gamma-ray induced Doppler broadening measurements}
\label{chapgrid}

Thermal neutron capture~\cite{motz1965:n-capture,casten2000:ns} on a
target nucleus can be an effective mechanism for populating low-spin
states of interest in the resulting product nucleus.  Since a thermal
neutron contributes a kinetic energy of $\lesssim$0.1\,eV to the
system, the product nucleus is produced at a tightly constrained
excitation energy very near the neutron separation energy
($\sim$5--10\,MeV).  Consequently, the capture reaction proceeds with
significant strength only to the one resonance (or at most a few
resonances) nearest this excitation energy in the product nucleus;
this resonance usually has a spin close ($\pm$1/2) to that of
the target ground state, since $s$-wave capture dominates.  The
capture resonance then decays by $\gamma$-ray emission to lower-lying
levels.  A few percent of this decay intensity proceeds for by strong
high-energy $\gamma$-ray transitions (``primary $\gamma$ rays'') from
the capture resonance to low-lying levels, while the rest proceeds by
cascades through the many densely-spaced levels present above
$\sim$2\,MeV excitation energy (``statistical cascades'').  At any
given excitation energy, the level population intensity is spin
dependent, and states with spins near that of the capture resonance
are most strongly populated (\eg,
Ref.~\cite{breitig1974:200hg-ngamma}).

Tremendous production rates can be achieved through thermal neutron
bombardment of bulk samples in a fission reactor.  Available neutron
fluxes are as high as $\sim$10$^{14}$\,cm$^{-2}$s$^{-1}$, and neutron capture
cross sections typically are in the range $\sim$1--10$^5$\,b.  The
correspondingly high $\gamma$-ray fluxes produced make feasible the
use of crystal spectrometers~\cite{knowles1965:crystal-spectrometer}
for $\gamma$-ray energy measurements.  These devices can acheive
extraordinarily fine energy resolutions, on the order of eV, but at
the cost of having detection efficiencies orders of magnitude below
those of conventional scintillation or semiconductor detectors.
Crystal spectrometers rely upon the dispersive nature of Bragg
diffraction of $\gamma$ rays by the internal lattice structure of
crystals, such as Si or Ge.  A beam of photons to be analyzed is
collimated and passed through a series of crystals and collimators
configured so that only $\gamma$ rays with a very specific Bragg
scattering angle, and hence specific energy, successfully transit the
system.  These $\gamma$ rays are then counted with a conventional
detector, for which the energy resolution is not critical.  To
construct a spectrum of the $\gamma$-ray intensity as a function of
energy, the crystal orientations are scanned to select a series of
different energies, and the $\gamma$ ray intensity is counted at each
energy.  

The GAMS4
spectrometer~\cite{greene1986:deuteron-gams4,dewey1989:gams4,boerner1993:grid},
located at the 58\,MW high-flux reactor of the Institut Laue-Langevin
(ILL) in Grenoble, France, achieves energy resolutions of $\sim$2
parts per million.  The target sample is located 50\,cm from the
reactor core, where the neutron flux is 5$\times$10$^{14}$\,cm$^{-2}$s$^{-1}$.
Capture $\gamma$ rays are collimated into a beam by a
2\,mm\,$\times$\,25\,mm slit collimator and travel 15\,m to the
spectrometer, where they are transmitted through a series of two flat
$\sim$5\,mm thick Si or Ge crystals separated by 50\,cm.  The analyzed
$\gamma$ rays exit through another collimator and are counted by a Ge
detector.  The crystal orientations are monitored using an optical
interferometer system, which allows the diffraction angles,
$\sim$$10^{-3}$\,rad, to be determined to a precision of
$\sim$$10^{-10}$\,rad.  The $\gamma$-ray detection efficiency of the
system is $\lesssim$$10^{-11}$.  The comparatively low efficiency of
crystal spectrometers such as GAMS4 arises from the extremely low
acceptance in the initial collimation stage, as well as the low
probability of Bragg scattering in the analyzing crystals.

\begin{figure}
\begin{center}
\includegraphics*[width=0.6\hsize]{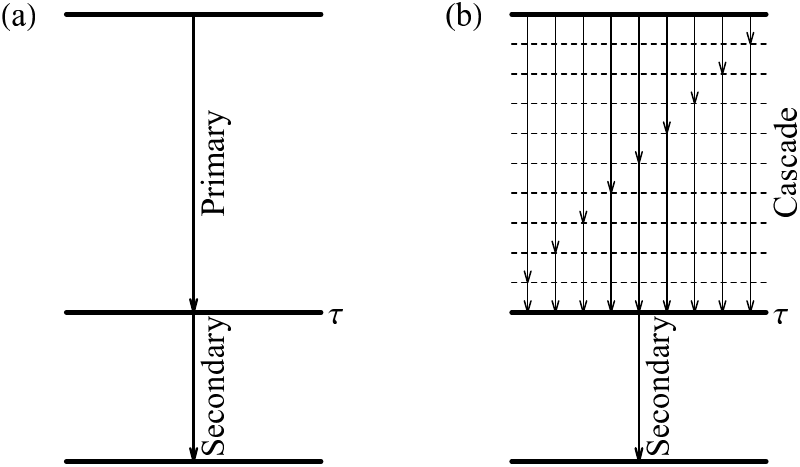}
\end{center}
\caption[Feeding of a level in thermal neutron capture.]{Feeding of a level in thermal
neutron capture by (a) direct primary feeding or (b) cascade feeding.
The feeding $\gamma$ rays impart a recoil momentum to the nucleus, so
that the secondary $\gamma$ ray is emitted with a Doppler shift.
\label{figgridfeeding}
}
\end{figure}

The parts-per-million $\gamma$-ray energy resolution available with
GAMS4, and related spectrometers at the ILL, allows lifetimes of
levels populated in thermal neutron capture to be measured using the
gamma-ray induced doppler broadening (GRID)
technique~\cite{boerner1993:grid}.  The capture product nucleus is
essentially at rest, except for thermal motion, when it is first
created in its capture resonance state.  Emission of a primary
$\gamma$ ray imparts to the nucleus a recoil velocity $v$ given by
\begin{equation}
\label{eqngammarecoil}
\frac{v}{c} = \frac{E_\gamma}{Mc^2},
\end{equation}
where $E_\gamma$ is the primary $\gamma$-ray energy and $M$ is the
nuclear mass (\eg, $E_\gamma$=5\,MeV and $M$=100\,u yield
$v/c$$\approx$5$\times$$10^{-5}$).  If the resonance decays through a
statistical cascade, a recoil velocity will similarly be imparted; in
this case the velocity is somewhat lower for the same total energy
emitted, since the recoil momenta imparted by the cascade $\gamma$
rays are not colinear but rather add as a random walk.  The nucleus
subsequently loses its recoil velocity by collisions with neighboring
nuclei in the target material.  The time scale at which slowing starts
is the time required to reach the nearest atomic neighbor, or
$\sim$10\,fs.  If the level populated by the primary or cascade
$\gamma$ rays emits a secondary $\gamma$ ray while the nucleus is
still recoiling, this $\gamma$ ray will be Doppler shifted
(Fig.~\ref{figgridfeeding}).  The recoil directions in the target are
isotropic relative to the emission direction of the secondary $\gamma$
ray detected in the spectrometer (excepting effects due to angular
correlations between the primary $\gamma$ rays and the $\gamma$ rays
observed by the spectrometer, which are negligible in most
cases~\cite{boerner1993:grid}), so the shift observed by the
spectrometer will be averaged over a full 4$\pi$ solid angle,
resulting in a symmetric Doppler broadening of the measured
$\gamma$-ray line.  If, however, the level lifetime is comparable to
slowing time of the recoiling nucleus, not all emitted $\gamma$ rays
will be fully Doppler shifted, and the level lifetime can be extracted
from the observed line shape (for examples, see
Section~\ref{sec152sm3plus}).  This method is practical for the
determination of lifetimes ranging from a few fs to a few ps.
Scanning a single $\gamma$-ray transition with a mesh of $\sim$40
diffraction angles, accumulating the few thousand
counts total needed for a GRID measurement, typically requires several
hours to one day of experiment time.

The dominant source of uncertainty in lifetime values determined with
the GRID method is usually the limited knowledge of the initial recoil
velocity distribution imparted to the level of interest at the time it
is populated.  This distribution depends both on the $\gamma$-ray
energies and intermediate level lifetimes for the feeding cascades.
The analysis can be carried out using extreme bounds on the velocity
distribution in order to place upper and lower limits on the deduced
lifetime: the scenario in which all the indirect feeding is via a
single two-step cascade through a level with zero lifetime (no
slowing) provides an upper bound~(\ref{eqngammarecoil}) on the recoil
velocity, and a conservative scenario involving a cascade through a
long-lived low-lying level (see Ref.~\cite{boerner1993:grid} for
details) gives a lower bound on the recoil velocity.  However, the
lifetimes deduced in these limiting cases typically differ by more
than a factor of two, leaving the actual lifetime poorly determined.
Statistical models of the cascade feeding can be used to extract a
more specific lifetime value.  A straightforward
approach~\cite{boerner1999:158gd-grid} for estimating the feeding
effects is based the consideration that all levels being analyzed in a
given nucleus usually lie withing a narrow window of spins and
excitation energies, compared to the energy of the neutron capture
resonance, and so should be subject to similar statistical feeding
patterns.  If the lifetimes of one or more levels being analyzed by
GRID are already known from other methods, these lifetimes can be used
to calibrate the GRID feeding assumptions.  Usually it has been found
that the upper lifetime limit, or lower velocity scenario, described
above provides a good estimate of the actual value.

\addtocontents{toc}{\protect\newpage}
\part{Experimental results}
\label{partexptresults}
\chapter{$^{152}$Sm}
\label{chap152sm}

\section{Experimental motivation}
\label{sec152smmotivation}

The nucleus $^{152}$Sm, with $N$=90 and $Z$=62, lies intermediate
between nuclei of known spherical shape and well-deformed
axially-symmetric rotor structure.  A sudden change in deformation
occurs at $N$$\approx$88--90 for the Sm and neighboring isotopic
chains [see Fig.~\ref{fig2vs0}(b) on page~\pageref{fig2vs0}].  The
nucleus $^{152}$Sm has played a central role in the recent
developments in the treatment of transitional structure outlined in
Section~\ref{sectrans}.  The energies of the first $2^+$ and $4^+$
states in $^{152}$Sm place it near the phenomenological ``critical
point'' (see Section~\ref{sectrans}) of the transition between
spherical and axially-symmetric deformed structure.  New data obtained
by Casten \etal~\cite{casten1998:152sm-beta} (as well as the results
of Section~\ref{sec152smspec}) constrain the description of this
nucleus within the IBM to parameter values near the critical point of
the transition from oscillator to rotor
structure~\cite{iachello1998:phasecoexistence}.  The nucleus
$^{152}$Sm was the first to be proposed for description by
the X(5) model~\cite{casten2001:152sm-x5}.

The nucleus $^{152}$Sm, especially among the $N$$\approx$88--90
transitional nuclei, has historically drawn attention due to the
availability of extensive and varied experimental data.  The common
$\gamma$-ray calibration source $^{152}$Eu populates excited states in
$^{152}$Sm through $\beta^+/\varepsilon$ decay, so spectroscopic data
from this decay has long been available.  The nucleus $^{152}$Sm is
stable, with a reasonably high abundance (27$\%$), and the neighboring
even-mass isotopes $^{150}$Sm and $^{154}$Sm are both stable as well.
These circumstances make $^{152}$Sm accessible to studies by Coulomb
excitation, inelastic scattering of various projectiles, ($t$,$p$) and
($p$,$t$) two-neutron transfer reactions, and neutron capture via
double capture on $^{150}$Sm, among other population mechanisms (see
Ref.~\cite{nds1996:152}).

The excited states of $^{152}$Sm exhibit a band structure roughly
similar to that of a rotational nucleus, but with significant
differences, suggesting it to be a transitional nucleus.  As early as
the 1960s, a full-fledged three-band mixing calculation performed
using detailed data from $\beta$
decay~\cite{riedinger1969:152sm154gd-beta} failed to determine
consistent mixing parameters within the conventional rotational
picture.  Experiments using $(p,t)$ and $(t,p)$ transfer reactions
suggested the ``coexistence'' of well-deformed states with undeformed
states in $^{152}$Sm as well as the presence of large shape
fluctuations in the ground
state~\cite{bjerregaard1966:sm-tp,mclatchie1970:sm-pt,debenham1972:sm-pt}.
The observation of unusually large electric monopole matrix elements
in $^{152}$Sm provides corroborating evidence indicating that both the
ground state and excited states may contain admixtures of
configurations of differing deformations~\cite{wood1999:e0}.

The recent developments provided motivation for several new
experiments involving $^{152}$Sm.  Although the $\gamma$-ray transitions most
strongly populated in $\beta$-decay had been well studied as
calibration standards, many of the weaker lines were subject to the
inaccuracies inherent to older experiments carried out without
high-statistics coincidence data (Section~\ref{seccoin}), suggesting
that model analyses of $^{152}$Sm might well be based in part upon
inaccurate data.  The $\gamma$-ray spectroscopy
experiment in $\beta$-decay by Casten
\etal~\cite{casten1998:152sm-beta} revealed that there was a genuine need
for careful remeasurement of observables involving the low-lying
states of $^{152}$Sm.  For full model treatments, it became apparent
that further lifetime and mixing ratio measurements were needed as
well.

The following sections present results from experiments involving the
population of $^{152}$Sm in $\beta$ decay and neutron capture.  These
results were reported in
Refs.~\cite{zamfir1999:152sm-beta,zamfir2002:152sm3plus-grid-pol}.
New lifetime values have also been determined by Klug
\etal~\cite{klug2000:152sm-rdm} from recoil decay method (RDM)
measurements on states populated by Coulomb excitation.
Interpretation of the experimental results is deferred to
Chapter~\ref{chapphenom}.

\section{Spectroscopic measurements}
\label{sec152smspec}

A high-statistics $\gamma$-ray coincidence spectroscopy experiment on
$^{152}$Sm was performed at the Yale YRAST Ball
array~\cite{beausang2000:yrastball}.  For this experiment, the array
consisted of three clover detectors with BGO Compton suppression
shields, 17 coaxial Ge detectors ($\sim$25$\%$ relative efficiency,
except for one with a relative efficiency of 70$\%$) with BGO or NaI
suppression shields, and one LEPS detector.  The array efficiency was
1.7$\%$ at 1.3\,MeV.  States in $^{152}$Sm were populated through
$\beta^+$/$\varepsilon$ decay of the $3^-$ ground state of $^{152}$Eu
($T_{1/2}$=13.5\,y), from a 7\,$\mu$Ci commercial calibration source.
Data were acquired with a $\gamma$-ray singles (or higher fold)
trigger, using the YRAST Ball FERA/VME acquisition system, for a
period of 25\,d.  The experiment yielded 2.0$\times$10$^9$~singles
counts and 1.3$\times$10$^8$~coincidence pairs.

\begin{figure}[t]
\begin{center}
\includegraphics*[width=0.75\hsize]{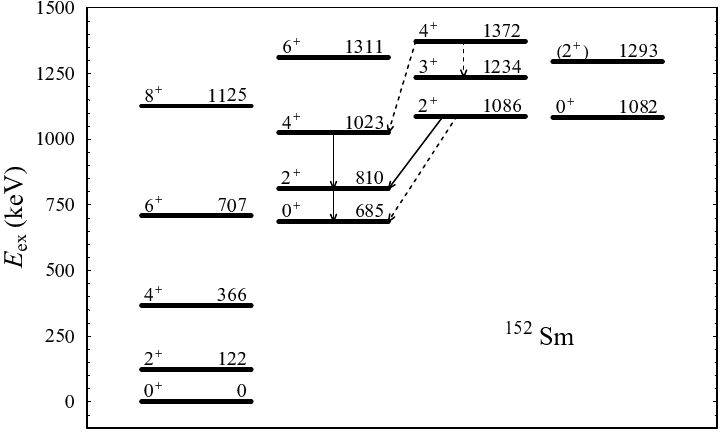}
\end{center}
\caption[Low-lying positive-parity states in $^{152}$Sm.]
{Low-lying positive-parity states in $^{152}$Sm.
Transitions for which intensity values or limits from the YRAST Ball
data are discussed in the text are indicated by solid and dashed arrows,
respectively.
\label{fig152smbands}
}
\end{figure}

Since the intensities for most of the stronger transitions from each
of the low-lying levels in $^{152}$Sm are reliably known, as are
precise lifetime values for several of these
levels~\cite{nds1996:152}, the following discussion will mainly serve
to highlight specific experimental issues regarding individual
$\gamma$-ray transitions of interest.  The low-lying levels of
$^{152}$Sm are summarized in Fig.~\ref{fig152smbands}.  In the
following discussion, the notation $J^\pi_{E_\text{ex}~\text{(keV)}}$
is used to denote the level of spin assignment $J^\pi$ at excitation
energy $E_\text{ex}$.  Relative intensities are normalized to
$I^\text{rel}$=100 for the strongest branch from each level.

\begin{figure}[t]
\begin{center}
\includegraphics*[width=0.75\hsize]{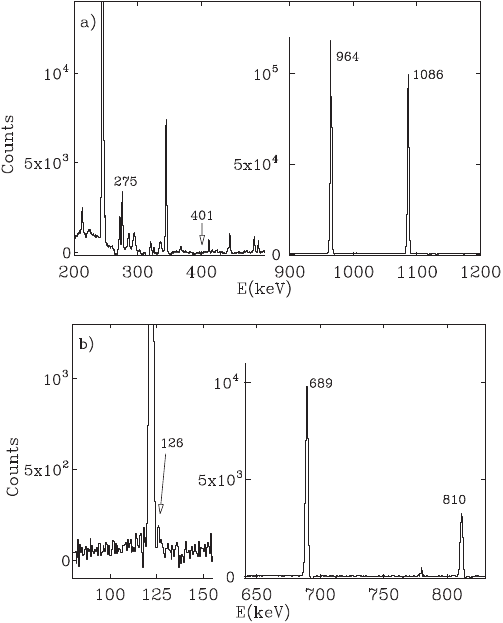}
\end{center}
\caption
[Spectra gated on transitions feeding the
$2^+_{810}$ and $2^+_{1086}$ levels in $^{152}$Sm.]  {Composite
coincidence spectra gated on (a) the 444, 494, 644, and 671\,keV
transitions feeding the $2^+_{810}$ level and (b) the 275, 482, 769,
and 839\,keV transitions feeding the $2^+_{1086}$ level in $^{152}$Sm,
used to deduce relative intensities for branches from these levels
(see text).  (Figure adapted from Ref.~\cite{zamfir1999:152sm-beta}.)
\label{fig152smspectra}
}
\end{figure}

$2^+_{810}$: The in-band 126\,keV $2^+_{810}\rightarrow0^+_{685}$
transition was previously attributed a relative intensity
$I^\text{rel}_{126}$=1.9(6)~\cite{nds1996:152} (normalized to the
$2^+_{810}\rightarrow2^+_{122}$ transition), based upon $\gamma$-ray
singles measurements in $\beta^+/\varepsilon$ decay from
Refs.~\cite{meyer1990:152sm-beta,stewart1990:152sm-beta}.  Combined
with the level lifetime value of 10.7(9)\,ps from Coulomb excitation
(see Ref.~\cite{nds1996:152}), this yielded a
$B(E2;2^+_{810}\rightarrow0^+_{685})$ strength of 520(170)\,\Wu.  Such
a strong $2^+\rightarrow0^+$ transition is unprecedented in the rare
earth nuclei and would be very difficult to explain.  For comparison,
the yrast $2^+\rightarrow0^+$ transition strength in this nucleus is
only 144(3)\,\Wu~\cite{nds1996:152}.  In the YRAST Ball experiment,
the 126\,keV transition is measured to have an intensity
$I^\text{rel}_{126}$=0.4(1), significantly lower than previously
reported, from a composite gate on transitions feeding the $2^+_{810}$
level [Fig.~\ref{fig152smspectra}(a)].  It is largely, but
not completely, resolved from the strong 122\,keV yrast
$2^+\rightarrow0^+$ transition in that gate.  Klug
\etal~\cite{klug2000:152sm-rdm} confirm the prior level
lifetime measurement, obtaining a value of 10.8(9)\,ps.  The resulting
$B(E2;2^+_{810}\rightarrow0^+_{685})$ value is a much more reasonable
111(28)\,\Wu.

$4^+_{1023}$: The prior intensity value for the 213\,keV
$4^+_{1023}\rightarrow2^+_{810}$ in-band transition was
$I^\text{rel}_{213}$=13(4) from $\beta$ decay, normalized to the
$4^+_{1023}\rightarrow4^+_{366}$ transition, but discrepant values had
been reported in various in-beam reactions (see
Ref.~\cite{nds1996:152}).  Only an estimate for the level lifetime, of
$\sim$7\,ps, had been obtained from multiple Coulomb
excitation~\cite{fraser1969:152sm-coulex}.  The YRAST Ball data yield
$I^\text{rel}_{213}$=11(1), consistent with the prior value.  Klug
\etal~\cite{klug2000:152sm-rdm} obtain an actual lifetime measurement
for the $4^+_{1023}$ level of 12.0(19)\,ps.  Together, these yield
$B(E2;4^+_{1023}\rightarrow0^+_{810})$=203(38)\,W.u., about half the
strength which would be obtained from the prior values.

$2^+_{1086}$: The present interest in $^{152}$Sm stemmed in part from
the attempt by Casten \etal~\cite{casten1998:152sm-beta} to deduce an
explicit limit for the intensity of the unobserved 401\,keV
$2^+_{1086}\rightarrow0^+_{685}$ transition.  In
Ref.~\cite{casten1998:152sm-beta}, a peak was observed in the singles
spectrum with an area corresponding to $I^\text{rel}_{401}$=0.023(2),
normalized to the $2^+_{1086}\rightarrow2^+_{122}$ transition, but
this peak contained a contribution from the
401.258(14)\,keV line in $^{154}$Eu decay~\cite{nds1998:154}.  The
YRAST Ball coincidence data [Fig.~\ref{fig152smspectra}(b)] provide a
more stringent limit $I^\text{rel}_{401}$\lt0.014.  The lifetime of
the $2^+_{1086}$ level is evaluated in Ref.~\cite{nds1996:152} to be
1.26(6)\,ps, from a weighted average of Coulomb excitation
measurements.  (The $2^+_{1086}$ lifetime is too short for RDM to be
an optimal measurement technique.  Klug
\etal~\cite{klug2000:152sm-rdm} obtain an approximately consistent lifetime value of
1.57(20)\,ps with inclusion of a correction for the nonzero stopping
time of the recoils.  See also Section~\ref{sec152sm3plus} for GRID
results.)  Together, these values yield
$B(E2;2^+_{1086}\rightarrow0^+_{685})$$<$0.11\,\Wu.

The 275\,keV $2^+_{1086}\rightarrow2^+_{810}$ transition was
previously attributed a predominantly $M1$
character~\cite{nds1996:152}, since the measured $\gamma$-ray
intensity in $\beta^+/\varepsilon$ decay from
Refs.~\cite{meyer1990:152sm-beta,stewart1990:152sm-beta} combined with
the measured $K$ conversion intensity from
Ref.~\cite{goswamy1991:152sm-beta} yielded a $K$ conversion
coefficient $\alpha^K_{275}$=0.106(20)~\cite{nds1996:152}, closely
matching the theoretical~\cite{hager1968:convcoeff} value 0.090 for
$M1$ multipolarity.  The YRAST Ball data
[Fig.~\ref{fig152smspectra}(b)], however, show the $\gamma$-ray
intensity to be substantially larger than previously reported
[$I^\text{rel}_{275}$=0.56(7), rather than
$I^\text{rel}_{275}$=0.229(15)~\cite{nds1996:152}], decreasing the
experimental conversion coefficient to $\alpha^K_{275}$=0.046(8),
which is more consistent with the theoretical value 0.059 for pure
$E2$ multipolarity.  The resulting strength is
$B(E2;2^+_{1086}\rightarrow2^+_{810})$=27(4)\,\Wu.

$4^+_{1372}$: Two spin-allowed transitions from this level~--- the
interband 349\,keV $4^+_{1372}\rightarrow4^+_{1023}$ transition and the
in-band 138\,keV $4^+_{1372}\rightarrow3^+_{1324}$ transition~--- had
not been reported in prior studies.  The YRAST Ball data were used to
provide explicit intensity limits $I^\text{rel}_{349}$\lt3 and
$I^\text{rel}_{138}$\lt0.2, normalized to the
$4^+_{1372}\rightarrow4^+_{366}$ transition.  The level lifetime is
known to be 2.0(6)\,ps from Coulomb
excitation~\cite{fraser1969:152sm-coulex}, as approximately confirmed
by the GRID measurement (Section~\ref{sec152sm3plus}).  Using this
lifetime value, the intensity limits correspond to
$B(E2;4^+_{1372}\rightarrow4^+_{1023})$\lt35\,\Wu and
$B(E2;4^+_{1372}\rightarrow3^+_{1324})$\lt250\,\Wu.

The spectroscopic results of this section substantially improve the
quality of the available data on transitions between low-lying states
in $^{152}$Sm.  The transition strengths obtained are necessary for
the interpretation and model analysis discussed in
Chapter~\ref{chapphenom}.

\section{Lifetime and polarization measurements}
\label{sec152sm3plus}

Gamma-ray direction-polarization correlation measurements were carried
out in a $\beta^+/\varepsilon$ decay experiment at the Yale SPEEDY
array~\cite{kruecken2001:rdm-speedy}, using seven YRAST Ball clover
detectors and one coaxial Ge detector (70$\%$
relative efficiency).  The clover detectors were used as Compton
polarimeters, as described in Section~\ref{secangpol}.  The total
array efficiency was $\sim$2$\%$ at 1.3\,MeV.  The detector positions
in the SPEEDY array, which are optimized for
Doppler-shift lifetime measurements, are also well-suited for
direction-polarization correlation measurements, providing four
detector pairs at a 98$^\circ$ relative angle.  A 7\,$\mu$Ci commercial
$^{152}$Eu calibration source, as used in the YRAST Ball experiment,
was placed at the center of the array, and data were acuired with a
$\gamma$-ray doubles (or higher fold) trigger, using the YRAST Ball
FERA/VME acquisition system, for a period of 10\,d.

The experiment provided useful data on the multipolarity of the
148\,keV $3^+_{1234}\rightarrow2^+_{1086}$ in-band transition, for
which no information had previously been available.  The polarization
of the 1086\,keV $2^+_{1086}\rightarrow0^+_{0}$ transition was measured
in coincidence with the 148\,keV transition in the detector pairs at
98$^\circ$ relative angle.  In the analysis, events involving Compton
cross-scatter between clover elements
\textit{within} the coincidence plane of each detector pair were
aggregated and compared to the sum of those involving Compton
cross-scatter \textit{perpendicular to} the coincidence plane.  The
instrumental asymmetry (Section~\ref{secangpol}) at a $\gamma$-ray
energy of 1.1\,MeV was calibrated using coincidences involving pairs of
detectors at a relative angle of 180$^\circ$, for which the
polarization must be vanishing by axial symmetry, and was found to be
$a$(1.1\,MeV)=1.00(3).  The instrumental polarization sensitivity
(Section~\ref{secangpol}) at this energy was calibrated using
coincidences between the 1112\,keV and 121\,keV $\gamma$-rays, for which
the multipolarities are well known~\cite{nds1996:152}, yielding
$Q$(1.1\,MeV)=0.115(12), in good agreement with Monte Carlo
simulations~\cite{garci-raffi1997:clover-pol}.  The experiment yielded
a polarization $P(98^\circ)=0.96(65)$.  

\begin{figure}[t]
\begin{center}
\includegraphics*[width=0.6\hsize]{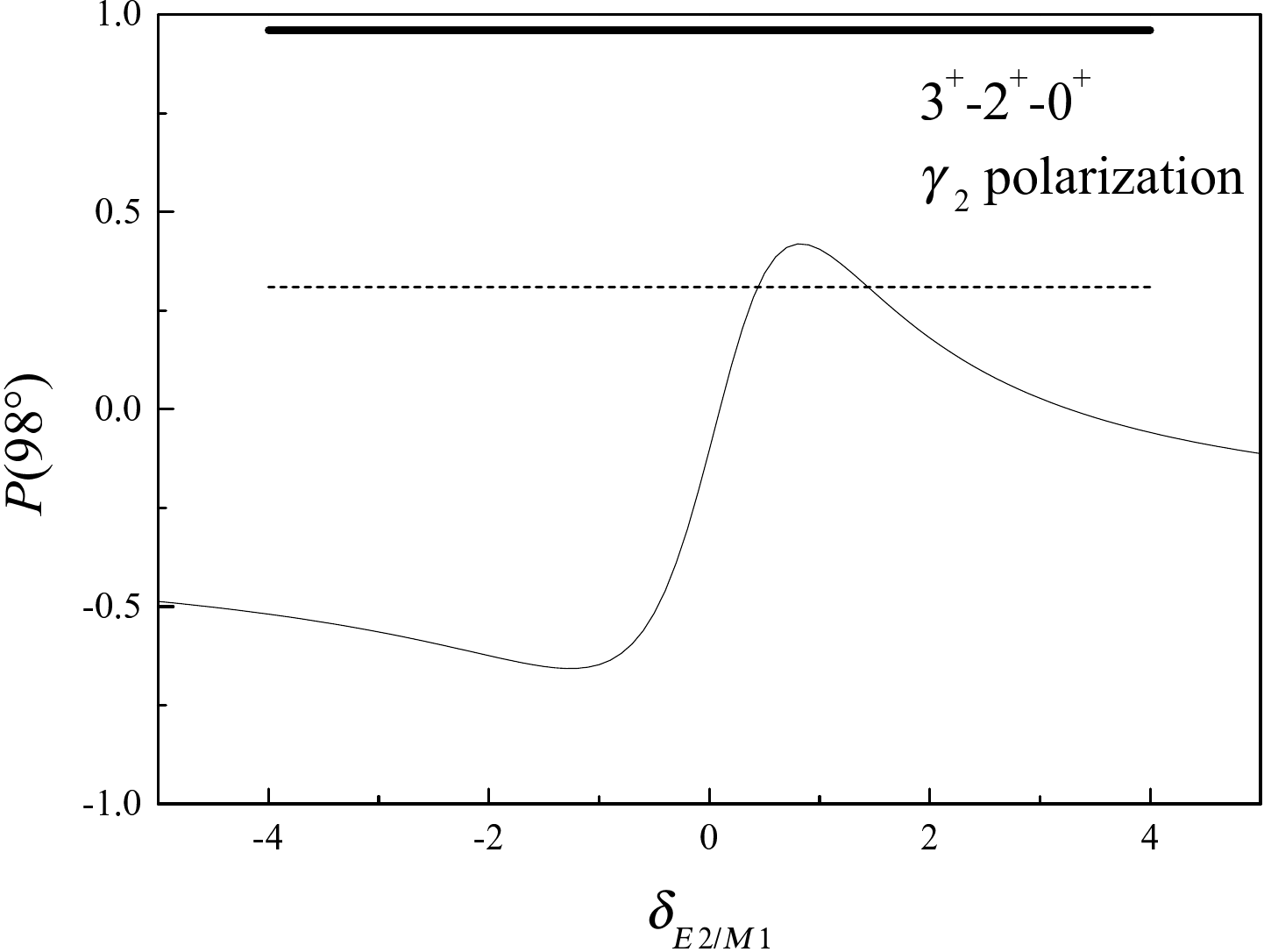}
\end{center}
\caption[Polarization for the
$3^+_{1234}\rightarrow2^+_{1086}\rightarrow0^+_{0}$ cascade in
$^{152}$Sm.]{Polarization at a relative angle of 98$^\circ$ for the
$3^+\xrightarrow{M1,E2}2^+\xrightarrow[\text{POL}]{E2}0^+$ cascade as
a function of $E2/M1$ mixing ratio.  The experimental value for the
$3^+_{1234}\rightarrow2^+_{1086}\rightarrow0^+_{0}$ cascade in
$^{152}$Sm is shown (solid line) together with the one-$\sigma$ error
limit (dotted line).
\label{fig152smpol}
}
\end{figure}

The predicted polarization for a
$3^+\xrightarrow{M1,E2}2^+\xrightarrow[\text{POL}]{E2}0^+$ cascade
[given by~(\ref{eqnpol3d22p0})] is shown as a function of mixing
ratio in Fig.~\ref{fig152smpol}, where it is overlayed, for
comparison, with the experimental value and one-$\sigma$ error band.
The experimental error band is consistent with an $E2/M1$ mixing ratio
for the 148\,keV transition of $\delta_{E2/M1}$=1.0(6).  However, it
should be noted that, given the properties of the polarization
function in Fig.~\ref{fig152smpol}, a relatively small increase in
uncertainty would admit a much larger range of $\delta$ values.

\begin{figure}[t]
\begin{center}
\includegraphics*[width=0.46\hsize]{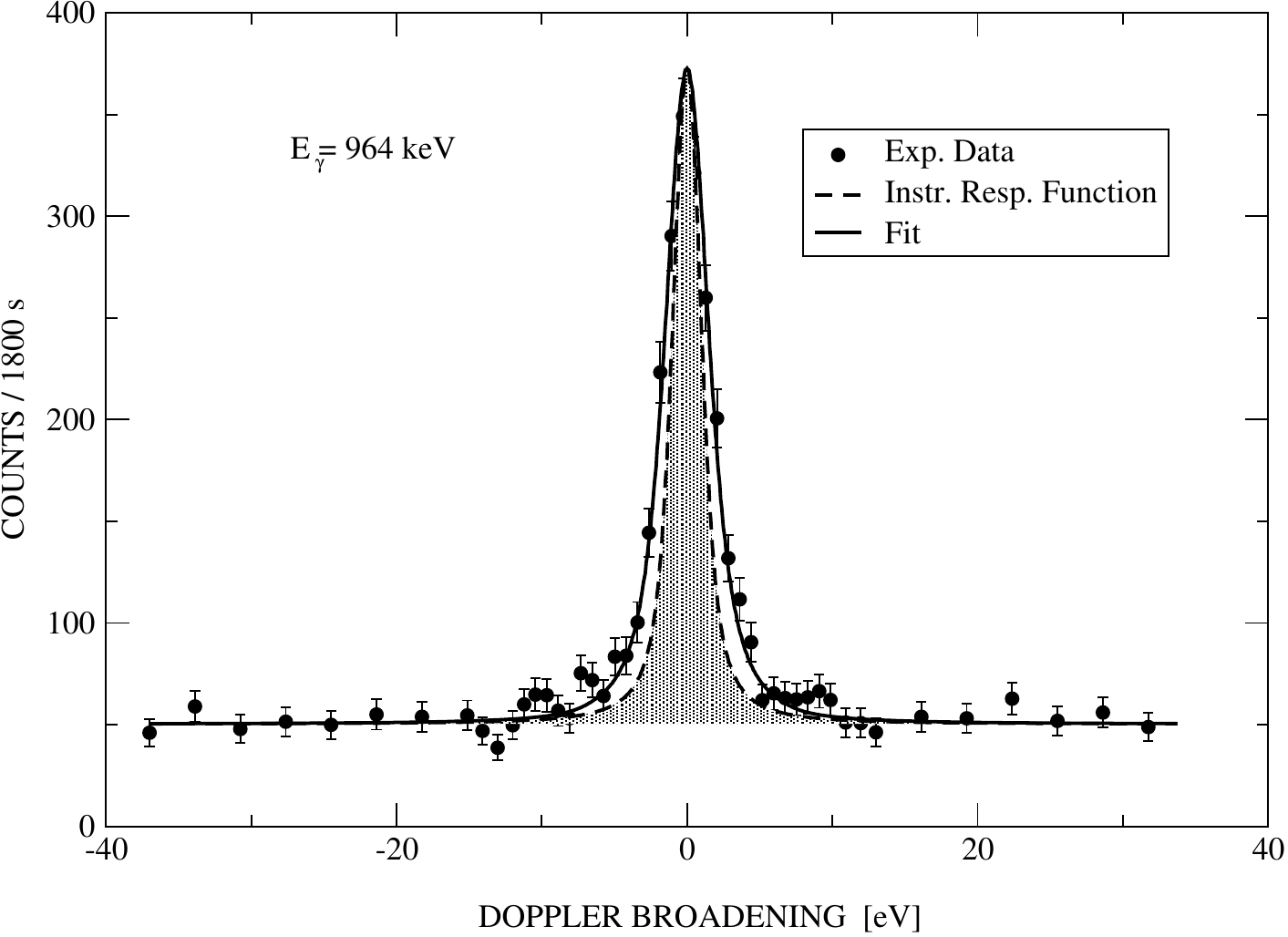}
\hfill
\includegraphics*[width=0.46\hsize]{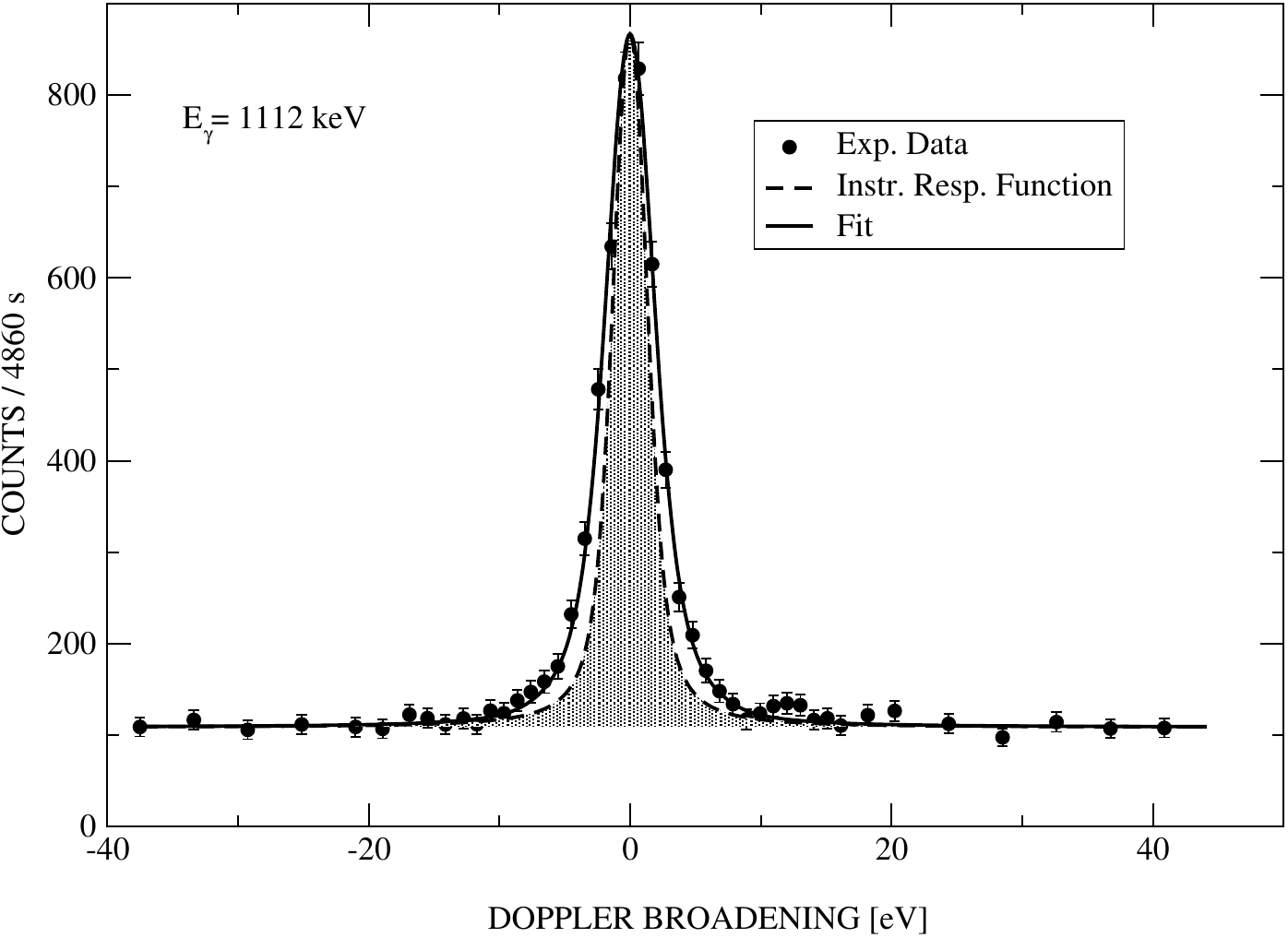}
\end{center}
\caption[Doppler-broadened line shapes for transitions in $^{152}$Sm.]{Doppler-broadened line shapes for the
(a) 964\,keV $2^+_{1086}\rightarrow2^+_{122}$ and (b) 1112\,keV
$3^+_{1234}\rightarrow2^+_{122}$ transitions in $^{152}$Sm.  The dotted line shows the instrumental response, and
the solid line represents the fit to the data incorporating Doppler
broadening.  (Left panel courtesy of H.~B\"orner.  Right panel from
Ref.~\cite{zamfir2002:152sm3plus-grid-pol}.)
\label{fig152smgrid}
}
\end{figure}

Gamma-ray induced doppler broadening (GRID) lifetime measurements on
levels in $^{152}$Sm were carried out using the GAMS4 spectrometer
(Chapter~\ref{chapgrid}).  An isotopically enriched $^{150}$Sm target
was placed near the reactor core, subject to a neutron flux of
\sci{5}{14}\,cm$^{-2}$s$^{-1}$.  The surface layer of this material was converted
through neutron capture ($\sigma$$\approx$1.0$\times$10$^2$\,b) to
long-lived $^{151}$Sm ($T_{1/2}$=90\,y), and $^{152}$Sm nuclei were
produced through subsequent neutron capture
($\sigma$$\approx$1.5$\times$10$^4$\,b).  The lifetime of the
$3^+_{1234}$ state was previously unknown; an upper limit of 9\,ps,
from BaF$_2$-BaF$_2$ electronic timing of $\gamma$-$\gamma$
coincidences~\cite{seo1993:152eu-bafl-decompose}, had been reported.
Lifetime measurements for the $2^+$, $3^+$, and $4^+$ members of the
$K^\pi=2^+$ band are summarized in Table~\ref{tab152smgrid}.
Doppler-broadened line shapes for the 964\,keV
$2^+_{1086}\rightarrow2^+_{122}$ and 1112\,keV
$3^+_{1234}\rightarrow2^+_{122}$ transitions used in this analysis are
shown in Fig.~\ref{fig152smgrid}.
\begin{table}
\begin{center}
\begin{tabular}{l=r=r=r@{\extracolsep{0pt}--}l=c}
\pseudoruledtabular
\multicolumn{1}{c}{$J^\pi$}&
\multicolumn{1}{c}{$E_\text{ex}$}&
\multicolumn{1}{c}{$E_\gamma$}&
\multicolumn{2}{c}{$\tau_\text{GRID}$}&
\multicolumn{1}{c}{$\tau_\text{\,lit}$}
\\
\multicolumn{1}{c}{}&
\multicolumn{1}{c}{(keV)}&
\multicolumn{1}{c}{(keV)}&
\multicolumn{2}{c}{(ps)}&
\multicolumn{1}{c}{(ps)}
\\ 
\colrule
$2^+$ & 1086 &  964 & 0.46 & 1.55 & 1.26(6)/1.57(20)\\
$3^+$ & 1234 & 1112 & 0.44 & 1.08 & \\
$4^+$ & 1372 & 1005 &0.54 & 1.78  & 2.0(6)\\
\pseudoruledtabular
\end{tabular}
\end{center}
\caption[Lifetime ranges deduced from GRID line shapes for levels in
$^{152}$Sm.]{\ssp Lifetime ranges deduced from GRID line shapes for levels
in $^{152}$Sm and literature values (see text) for comparison.  The
limits of the ranges correpond to the extreme feeding assumptions
(Chapter~\ref{chapgrid}).
\label{tab152smgrid}
}
\end{table}
As discussed in
Chapter~\ref{chapgrid}, GRID measurements produce a range of possible
lifetime values, depending upon the feeding scenario assumed, but
a value near the upper end of the range provides the best estimate
under typical feeding conditions.  Comparison of the limits
obtained for the $2^+_{1086}$ lifetime with the prior measurements
(Table~\ref{tab152smgrid}) supports this feeding scenario for
$^{152}$Sm.  Consequently, a more specific estimate
$\tau(3^+_{1234})$$\approx$1.1\,ps can be made.  Similar use of the
upper end of the lifetime range allows an estimate
$\tau(4^+_{1372})$$\approx$1.8\,ps, supporting a value near the center
of the uncertainty range for the prior value of 2.0(6)\,ps from Coulomb
excitation~\cite{fraser1969:152sm-coulex}.

The lifetime and multipolarity measurements discussed in this section
involve the $2^+_{1086}$, $3^+_{1243}$, and $4^+_{1372}$ states, all
of which are members of the first excited $K^\pi$=$2^+$ band, and so
will provide data for comparison with the X(5) predictions for the
$n_\gamma$=1 band~\cite{iachelloINPREP}.  Predictions of the GCM are
discussed in Chapter~\ref{chapphenom}.

\chapter{$^{156}$Dy}
\label{chap156dy}

\section{Experimental motivation}
\label{sec156dymotivation}

The nucleus $^{156}$Dy, with $N$=90 and $Z$=66, shows strong
similarities to the lower-$Z$ $N$=90 transitional nuclei, both in
level energies and in transition strengths.  The yrast band level
energies closely match those of $^{150}$Nd, $^{152}$Sm, and $^{154}$Gd
and are nearly identical to the X(5) model
predictions [Fig.~\ref{fign90yrast}(a) on page~\pageref{fign90yrast}].
These and other basic observables involving low-lying levels suggest
that $^{156}$Dy warrants further examination as a candidate for
description using the X(5) model (see Chapter~\ref{chapphenom}).

Further interpretation of the structure of $^{156}$Dy,
however, requires accurate information on the branching properties of
the low-lying non-yrast states.  The previous spectroscopic studies of
$^{156}$Dy, both from in-beam data and decay data, present seriously
contradictory results for some of the most basic observables
concerning the low-lying off-yrast states: values stated in the
literature for the intensities of some branching transitions from the
lowest excited $2^+$ and $4^+$ states~\cite{ndsboth:156} disagree by
factors of five or more, and values for others have uncertainties
which are too large for useful analysis.

The following sections describe the results of a spectroscopic study
of $^{156}$Dy performed to address some of the issues regarding
transition strengths in $^{156}$Dy and to provide a detailed data set
to serve as a basis for interpretation of $^{156}$Dy.  The
high-statistics $\gamma$-ray coincidence spectroscopy data allow
many of the ambiguities (\textit{e.g.}, contaminant transitions,
unresolved doublets) inherent to singles studies to be largely
avoided.  From this study, not only are substantially improved
measurements of the branching properties of low-lying levels obtained,
resolving the outstanding conflicts in the literature, but much of the
previous level scheme for excitation energies above $\sim$1200\,keV is found to be in error.
Results from this experiment were reported in
Refs.~\cite{caprio2001:156dy-beta-rjp,caprio2002:156dy-beta}.

\section{Spectroscopy experiment overview}
\label{sec156dyspec}

A $\gamma\gamma$ spectroscopy experiment to study $^{156}$Dy in
$\beta^+/\varepsilon$ decay was carried out at the Yale MTC.  Parent
$^{156}$Er nuclei were produced through the reaction
$^{148}\text{Sm}(^{12}\text{C},4n)^{156}\text{Er}$ at a beam energy of
73\,MeV, using an $\sim$10\,pnA beam provided by the Yale ESTU tandem
accelerator incident upon a 1.8\,mg/cm$^2$ $96\%$-isotopically-enriched
target.  The nucleus $^{156}$Er decays with a 19.5\,min half life to
$^{156}$Ho [$J^\pi$=(4$^+$)], which in turn decays with a 56\,min half
life to $^{156}$Dy~\cite{ndsboth:156}.  Two-step decay was chosen
since $\beta$ decay from the $0^+$ ground state of $^{156}$Er avoids
population of $\beta$-decaying high spin isomeric states in $^{156}$Ho
and thus enhances the population of low-spin off-yrast states in
$^{156}$Dy, as discussed in Section~\ref{secbetapop}.  The tape was
advanced, carrying the deposited activity to the detector area, at 1\,h
intervals.

\begin{figure}[t]
\begin{center}
\includegraphics*[width=1.0\hsize]{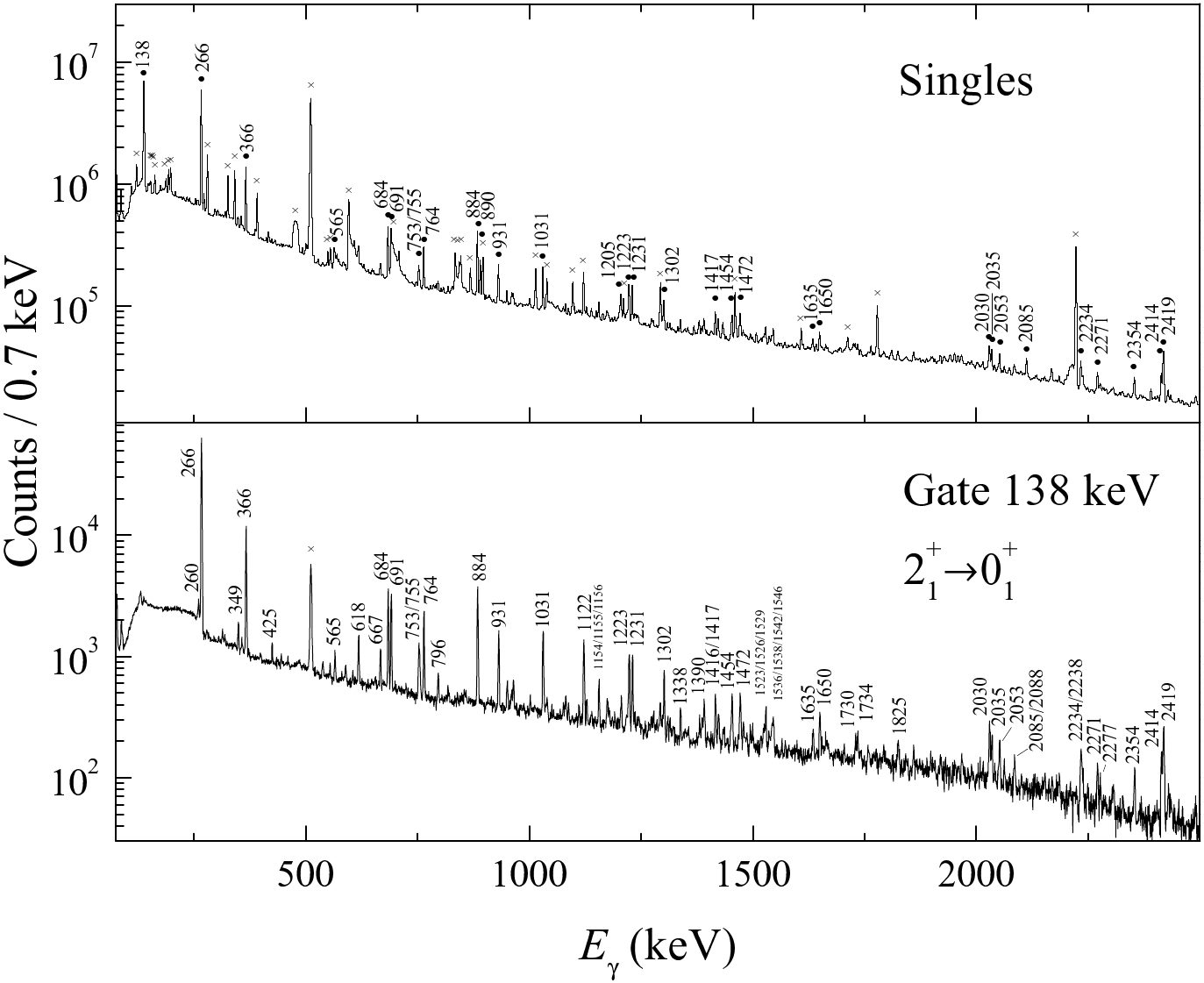}
\end{center}
\caption[Spectra from the $^{156}$Dy experiment.]
{Spectra from the $^{156}$Dy experiment.  (Top) Clover singles
spectrum.  Transitions from $^{156}$Dy are marked with a circle
(\captionsolidcircle), and contaminant lines from $^{157}$Ho, $^{157}$Dy, room
or neutron-induced background, and positron annihilation are indicated
with a cross ($\times$).  (Bottom) Clover-clover coincidence spectrum
gated on the 138\,keV $2^+_1\rightarrow0^+_1$ transition.  Annihilation
radiation is marked with a cross ($\times$).  (Figure from
Ref.~\cite{caprio2002:156dy-beta}.)
\label{fig156dyspectra}
}
\end{figure}

The Yale MTC was in its basic $\gamma\gamma$ spectroscopy
configuration (Section~\ref{secmtc}) for this experiment, with three
Compton-suppressed clover detectors and one LEPS detector.  The array
photopeak efficiency was $1.1\%$ at 1.3\,MeV and the dynamic range
extended from $\sim$35\,keV to 2650\,keV.  Data were acquired in event
mode with a singles (or higher fold) trigger, using the YRAST Ball
FERA/VME acquisition system~\cite{beausang2000:yrastball}.  The
experiment yielded $7.2\times10^8$ clover singles events and
$1.7\times10^7$ clover-clover coincidence pairs in 125\,h.  The
combined clover singles spectrum from this experiment is shown in
Fig.~\ref{fig156dyspectra}, along with an example gated spectrum.  The
main contaminant nuclei present in the experiment were in the
neighboring $A$=157 mass chain, with $\gamma$-ray energies largely
below $\sim$500\,keV.

The nucleus $^{156}$Dy is populated in $\beta^+/\varepsilon$ decay
with a $Q_\varepsilon$ value of $\sim$5\,MeV~\cite{ndsboth:156}.  As
described in Section~\ref{seccoin}, such population results in the
production of several hundred identifiable transitions, many of them
yielding overlapping or unresolved peaks in the singles spectra, and
measurements of intensities from such singles data are therefore not
generally reliable.  All intensity values reported in the present work
were deduced from the high-statistics coincidence data.  Information
was extracted from coincidences both with feeding transitions and with
transitions below the transition of interest whenever possible.  The singles
data were used primarily to provide corroboration of these intensities
and to deduce limits on the intensities of certain unobserved
transitions.  Details of these methods for measuring intensities are
discussed in Sections~\ref{seccoin} and~\ref{seccointechniques}.

An accurate knowledge of the array singles and coincidence
efficiencies is essential for intensity measurements.
The array singles and coincidence efficiencies were determined as
descibed in Section~\ref{seccoineff}.  The coincidence efficiency
calibration for this experiment was made using 35 known coincidences
in $^{152}$Sm and $^{152}$Gd from $^{152}$Eu calibration source
decay~\cite{nds1996:152}.  Use was also made of ``internal''
calibrator coincidences from the $^{156}$Dy data involving transitions
in the yrast cascade, since for these transitions the branching
fractions depend only upon $E2$ internal conversion coefficients,
which are reliably known from atomic physics.  The array exhibited a
coincidence efficiency attenuation at low $\gamma$-ray energies (see
Section~\ref{seccoineff}), with an attenuation factor of $w\approx0.65$
for $\gamma$-ray energies of $\sim$100\,keV, and reached ideal
efficiency ($w$$\approx$1) for energies above $\sim$300\,keV.

Intensities measured by coincidence methods can also be affected by
angular correlations between the emitted $\gamma$ rays.  For the
clover detector pair angles used in the present experiment
(approximately 110$^\circ$--125$^\circ$), the effect on the
$\gamma$-ray intensity measurement is $\sim$24$\%$ for a spin 0--2--0
cascade and substantially smaller ($\lesssim$5$\%$) for other common
cascades.  Since reliable multipolarity information is not
consistently available, intensities reported here are not corrected
for angular correlation effects.

\begin{figure}[p]
\begin{center}
\includegraphics*[width=1.0\hsize]{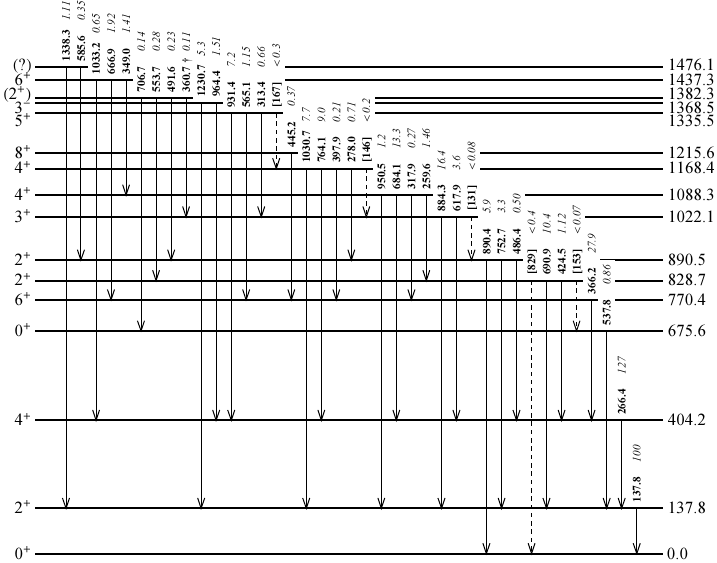}
\end{center}
\caption[Low-lying levels in $^{156}$Dy populated in $^{156g}$Ho
$\beta$ decay.]  {Low-lying levels in $^{156}$Dy populated in
$^{156g}$Ho $\beta$ decay and their depopulating $\gamma$-ray
transitions.  Unobserved transitions for which intensity limits are
obtained contradicting previously reported values are indicated by
dashed arrows, and a dagger ($\dagger$) indicates tentative placement.
Level energies (in keV), transition energies (in keV), and transition
intensities in $\beta$ decay (normalized to the
$2^+_{138}\rightarrow0^+_{0}$ transition) are from
Table~\ref{tabbranch}.  Level spin assignments are from
Ref.~\cite{ndsboth:156} except as noted in the text.  (Figure from
Ref.~\cite{caprio2002:156dy-beta}.)
\label{fig156dyscheme}
}
\end{figure}

The coincidence data provide placement and intensity information on
over 250 $\gamma$-ray transitions in $^{156}$Dy, and the level scheme
obtained constitutes a substantial revision to that found in the
literature~\cite{ndsboth:156}.  Over 50 new levels are identified,
numerous levels previously claimed from $\beta$-decay data are found
to be unsubstantiated, and the decay properties of many of the
remaining levels are substantially modified.  The $\gamma$-ray
transition energies, placements, and absolute and relative intensities
deduced in this experiment are summarized in Tables~\ref{tabline}
and~\ref{tabbranch}.  Intensity limits for unobserved transitions,
valuable for model analysis, are systematically included for low-lying
levels (Table~\ref{tabbranch}).  In the tables and in the following
discussion, all absolute intensities are normalized to
$I_{138}\equiv100$ for the $2^+_1\rightarrow0^+_1$ transition
intensity, and relative intensities are quoted normalized to
$I^\text{rel}$=100 for the strongest branch from each level.  The
level scheme for levels populated below 1500\,keV is shown in
Fig.~\ref{fig156dyscheme}.

\section{Transitions depopulating low-lying levels}
\label{sec156dytransitions}

This section summarizes the experimental results for levels which
are of current interest in the structural interpretation of
$^{156}$Dy, elaborating upon the basic information presented in
Table~\ref{tabbranch}.  The notation
$J^\pi_{E_\text{ex}~\text{(keV)}}$ is used to denote the level of spin
assignment $J^\pi$ at excitation energy $E_\text{ex}$.  Spin
assignments are taken from Ref.~\cite{ndsboth:156} unless 
information affecting the spin assignment has been obtained from the
present experiment.

\begin{figure}[t]
\begin{center}
\includegraphics*[width=0.48\hsize]{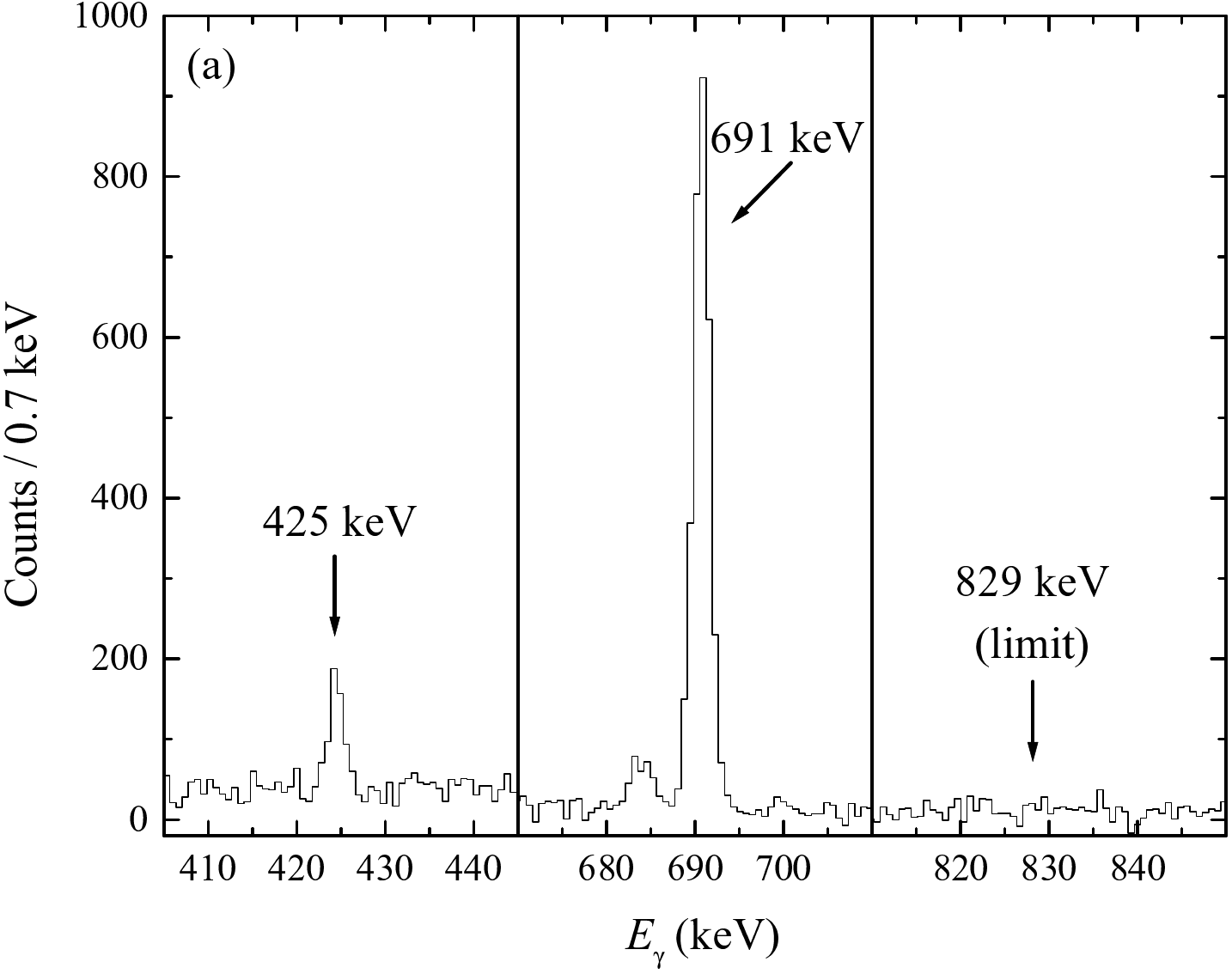}
\hfill
\includegraphics*[width=0.48\hsize]{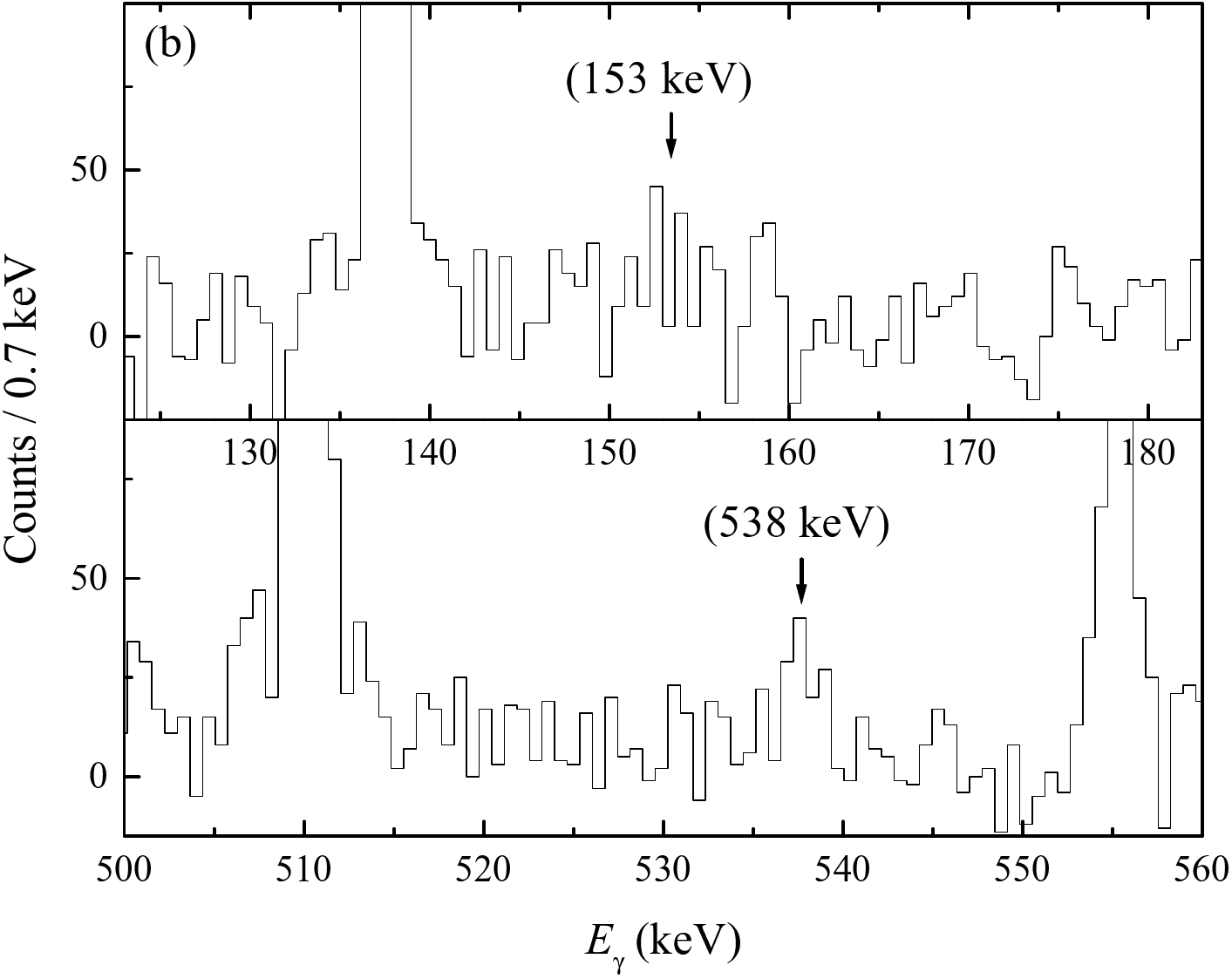}
\\
\includegraphics*[width=0.48\hsize]{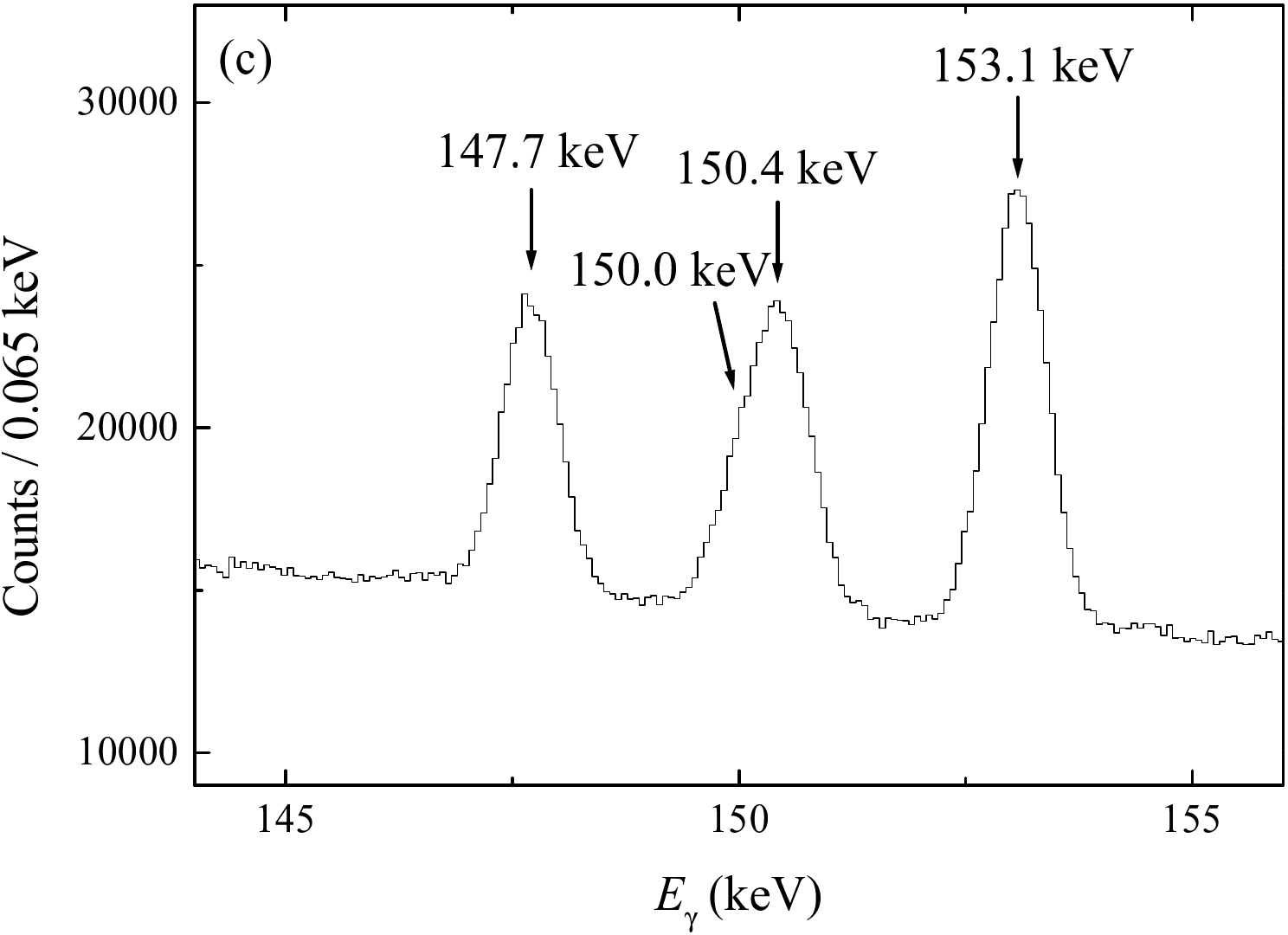}
\end{center}
\caption[Transitions from the $2^+$ level at 829\,keV in $^{156}$Dy.]
{Transitions from the $2^+$ level at 829\,keV in $^{156}$Dy.  (a)
Composite spectrum gated on 260, 796, 1416, and 1479\,keV transitions
feeding the $2^+_{829}$ level, showing the 424, 691, and unobserved
829 keV branches from this level.  (b) Spectra gated on the 538\,keV
$0^+_{676}\rightarrow2^+_{138}$ transition (top) and on the expected
energy of the 153\,keV $2^+_{829}\rightarrow0^+_{676}$ transition
(bottom), which allow a limit to be placed upon coincidences between
these transitions.  (c) LEPS detector singles spectrum showing the
strong contaminant peaks at 147.7\,keV ($^{157}$Dy), 150.0\,keV
($^{157}$Ho), 150.4\,keV ($^{157}$Dy), and 153.1\,keV ($^{157}$Dy).
(Figure from Ref.~\cite{caprio2002:156dy-beta}.)
\label{fig156dylev828}
}
\end{figure}

$2^+_{829}$: Measurement of the branching properties of this level
provided much of the initial motivation for the present experiment, as
this level is central to interpretation of the low-lying collective
structure of $^{156}$Dy and considerable ambiguities existed in the
literature (Table~\ref{tabbranch}).  The relative intensities of the
two strongest branches from this level, to the $2^+$ and $4^+$ members
of the yrast band, were confirmed.  However,
the 829\,keV transition to the ground state is highly suppressed, in
spite of its having a larger transition energy than the other
branches. Only a limit on its intensity could be obtained,
$I^\text{rel}_{829}<4$, from spectra gated on transitions feeding
$2^+_{829}$ [Fig.~\ref{fig156dylev828}(a)].  Previously an intensity
$I^\text{rel}_{829}=16(18)$ had been proposed from an $(\alpha,4n)$
study~\cite{deboer1977:156dy-a4np4n}.

Wildly discrepant intensities for the 153\,keV
$2^+_{829}\rightarrow0^+_{676}$ transition have been reported.  Much
of the confusion probably results from the presence of a strong
153.0\,keV transition in $^{157}$Dy --- relative intensities as large
as $I^\text{rel}_{153}\approx144$ were found in experiments subject to
$^{157}$Dy
contamination~\cite{deboer1977:156dy-a4np4n,elmasri1976:156dy-p4n}.
The prior $\beta$-decay work~\cite{gromov1976:156dy-beta} estimated
$I^\text{rel}_{153}\approx1.9$ from conversion electron singles data.
In singles, the $\gamma$-ray spectrum at this energy in the present
experiment is overwhelmingly dominated by the contaminant transition
in $^{157}$Dy [Fig.~\ref{fig156dylev828}(c)].  Coincidences with the
538\,keV $0^+_{676}\rightarrow2^+_{138}$ transition were not observed
[Fig.~\ref{fig156dylev828}(b)], allowing a limit of
$I^\text{rel}$$<$0.7 to be placed on the intensity of the
$2^+_{829}\rightarrow0^+_{676}$ transition, which eliminates the
various previously claimed intensities (Table~\ref{tabbranch}) for
this transition.  

$2^+_{890}$: The reported intensities for the transitions to the yrast
$0^+$, $2^+$, and $4^+$ states are essentially confirmed, with some
reduction in uncertainty for the intensity of the 486\,keV
$2^+_{890}\rightarrow4^+_{404}$ transition.  A weak ($\sim$5$\%$)
doublet contribution is found in the 890\,keV
$2^+_{890}\rightarrow0^+_{0}$ transition.  Limits are placed upon any
possible transitions to the $0^+_{676}$ and $2^+_{829}$ states.  A
$2^+_{890}\rightarrow2^+_{829}$ $\gamma$-ray transition deduced in
Ref.~\cite{gromov1976:156dy-beta} on the basis of conversion electron
data is not excluded by the present limit.

\begin{figure}[t]
\begin{center}
\includegraphics*[width=0.48\hsize]{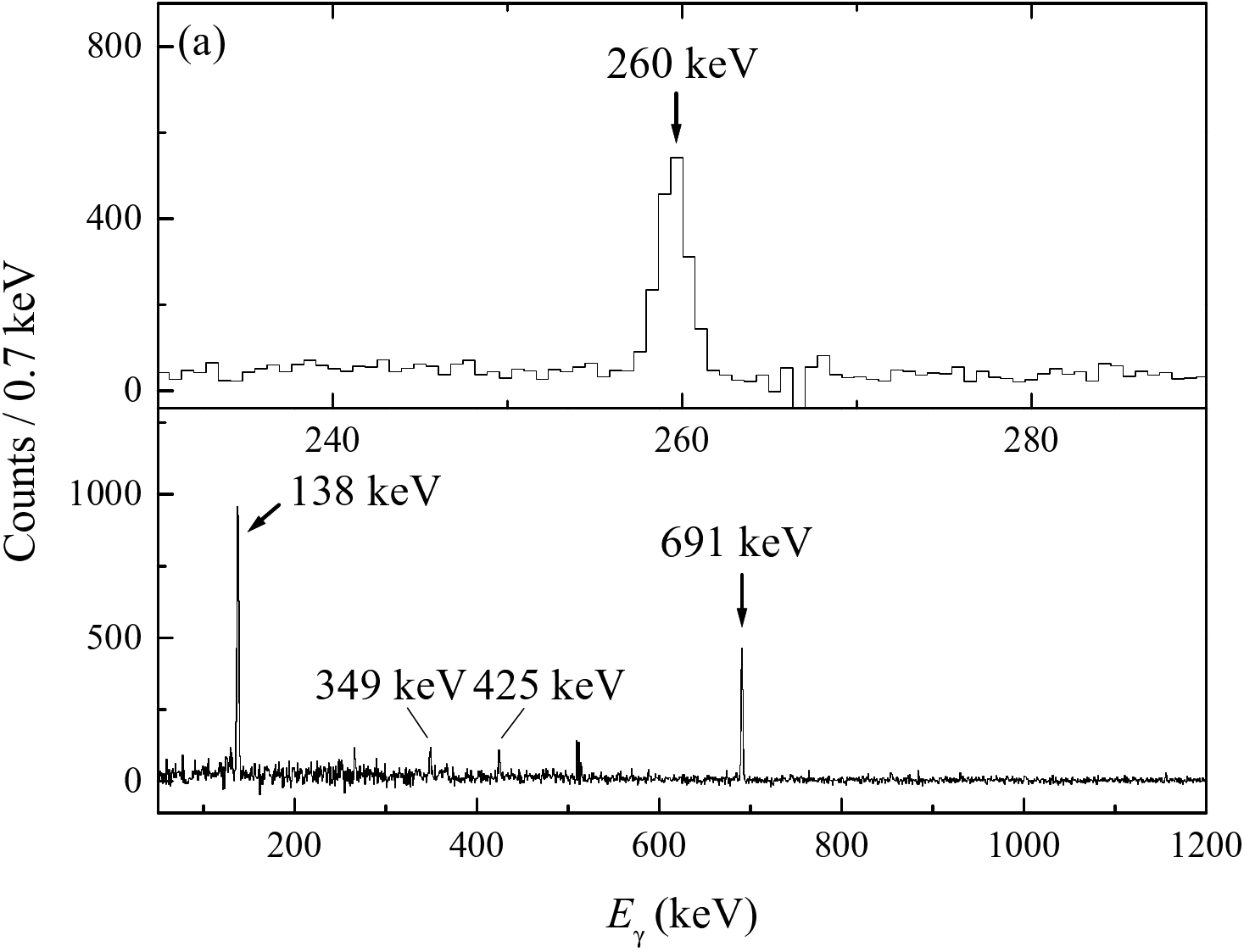}
\hfill
\includegraphics*[width=0.48\hsize]{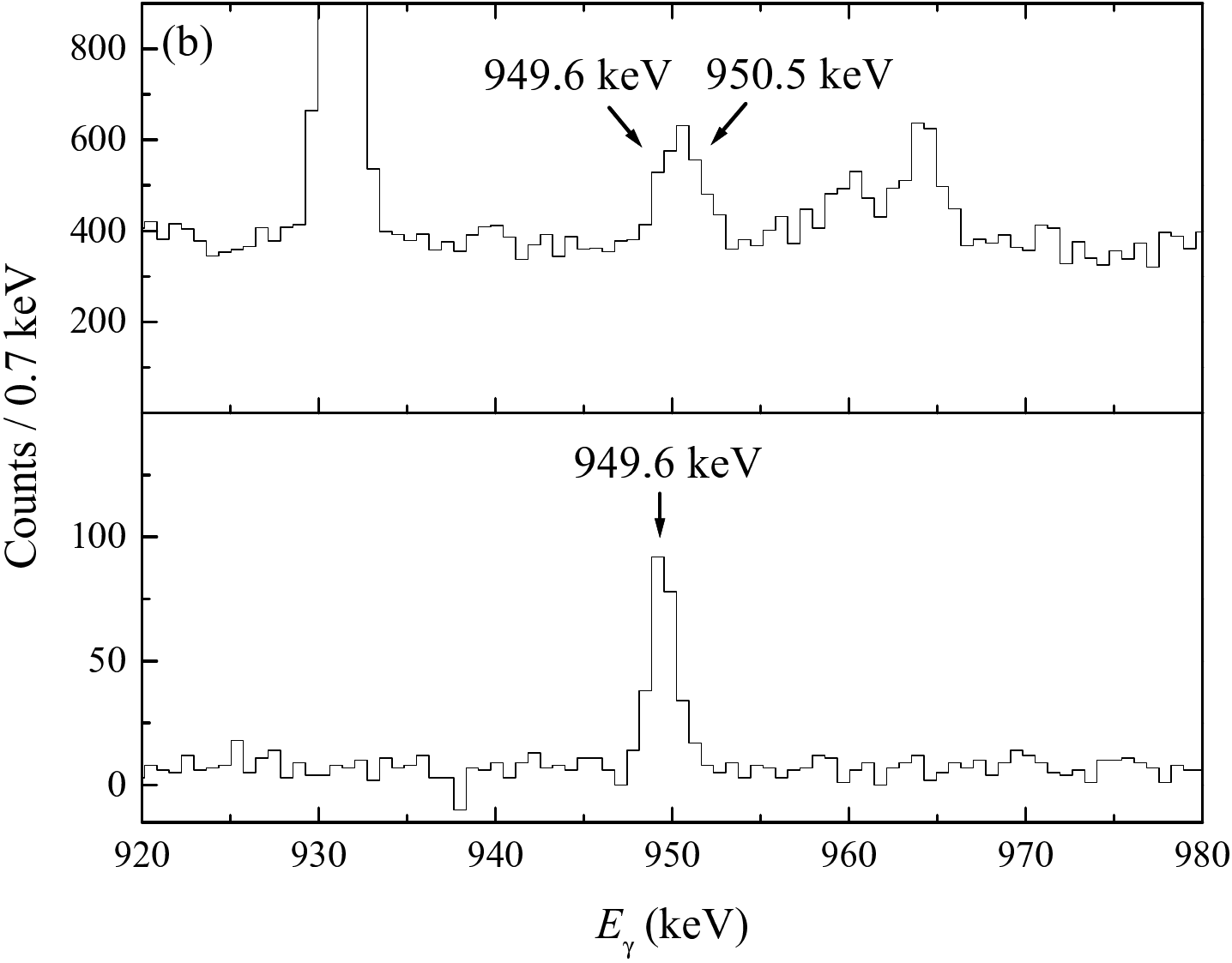}
\end{center}
\caption[Transitions from the $4^+$ level at 1088\,keV in $^{156}$Dy.]
{Transitions from the $4^+$ level at 1088\,keV in $^{156}$Dy.  (a)
Spectra gated on the 691\,keV $2^+_{829}\rightarrow2^+_{138}$ (top) and
260\,keV $4^+_{1088}\rightarrow2^+_{829}$ (bottom) transitions,
supporting the placement of the 260\,keV transition and allowing
measurement of its intensity.  (b) Spectra gated on the 138\,keV
$2^+_{138}\rightarrow0^+_{0}$ (top) and 890\,keV
$2^+_{890}\rightarrow0^+_{0}$ (bottom) transitions, illustrating the
950.5\,keV branch from the $4^+_{1088}$ level and the doublet 949.6\,keV
$(?)_{1840}\rightarrow 2^+_{890}$ transition.  (Figure from
Ref.~\cite{caprio2002:156dy-beta}.)
\label{fig156dylev1088}
}
\end{figure}

$4^+_{1088}$: The most important result obtained for this level is the
confirmation of the existence of the ``in-band'' 260\,keV
$4^+_{1088}\rightarrow2^+_{829}$ transition, together with a reliable
intensity measurement [$I^\text{rel}_{260}$=11.0(10)].  Coincidences
between this transition and transitions depopulating the $2^+_{829}$
level or feeding the $4^+_{1088}$ level are shown in
Fig.~\ref{fig156dylev1088}(a).  Prior values for the relative
intensity had ranged from about 11 to
45~\cite{deboer1977:156dy-a4np4n,elmasri1976:156dy-p4n}, with
nonobservation in $\beta$ decay~\cite{gromov1976:156dy-beta}.

The intensities of the transitions from the $4^+_{1088}$ state to the
yrast band members are also different from those previously reported.
The 950\,keV $4^+_{1088}\rightarrow2^+_{138}$ transition is found to
contain a $(?)_{1840}\rightarrow2^+_{890}$ doublet contributing
$\sim$37$\%$ of the total intensity [Table~\ref{tabline} and
Fig.~\ref{fig156dylev1088}(b)], and the uncertainty in the intensity
of the 318\,keV $4^+_{1088}\rightarrow6^+_{770}$ transition is
considerably reduced.

$4^+_{1168}$: The strong 1031\,keV and 764\,keV branches to the yrast
$2^+$ and $4^+$ states are essentially unchanged from
Ref.~\cite{gromov1976:156dy-beta}, although the
1031\,keV transition is found to have a weak ($\sim$1.4$\%$) doublet
contribution.  The 398\,keV
$4^+_{1168}\rightarrow6^+_{770}$ branch is now clearly identified
[$I^\text{rel}_{398}$=2.3(6)] from coincidences.  Previous claims for
the intensity of this transition had disagreed radically, with
nonobservation in $\beta$ decay~\cite{gromov1976:156dy-beta}.  The
intensity measured for the ``in-band'' 278\,keV
$4^+_{1168}\rightarrow2^+_{890}$ branch is modified ($\sim$37$\%$
decrease) relative to the value from Ref.~\cite{gromov1976:156dy-beta}.  A
limit placed on the possible 146\,keV branch to the $3^+_{1022}$ level
($I^\text{rel}_{146}<3$) excludes an extremely large relative
intensity ($\sim$530) reported in $(\alpha,4n)$
\cite{deboer1977:156dy-a4np4n}, which perhaps resulted from contamination by
the 147.7\,keV transition in $^{157}$Dy [Fig.~\ref{fig156dylev828}(c)].
A weaker $\gamma$-ray transition deduced from conversion electron
data~\cite{gromov1976:156dy-beta} is not excluded by the present
limit.

\begin{figure}[t]
\begin{center}
\includegraphics*[width=0.60\hsize]{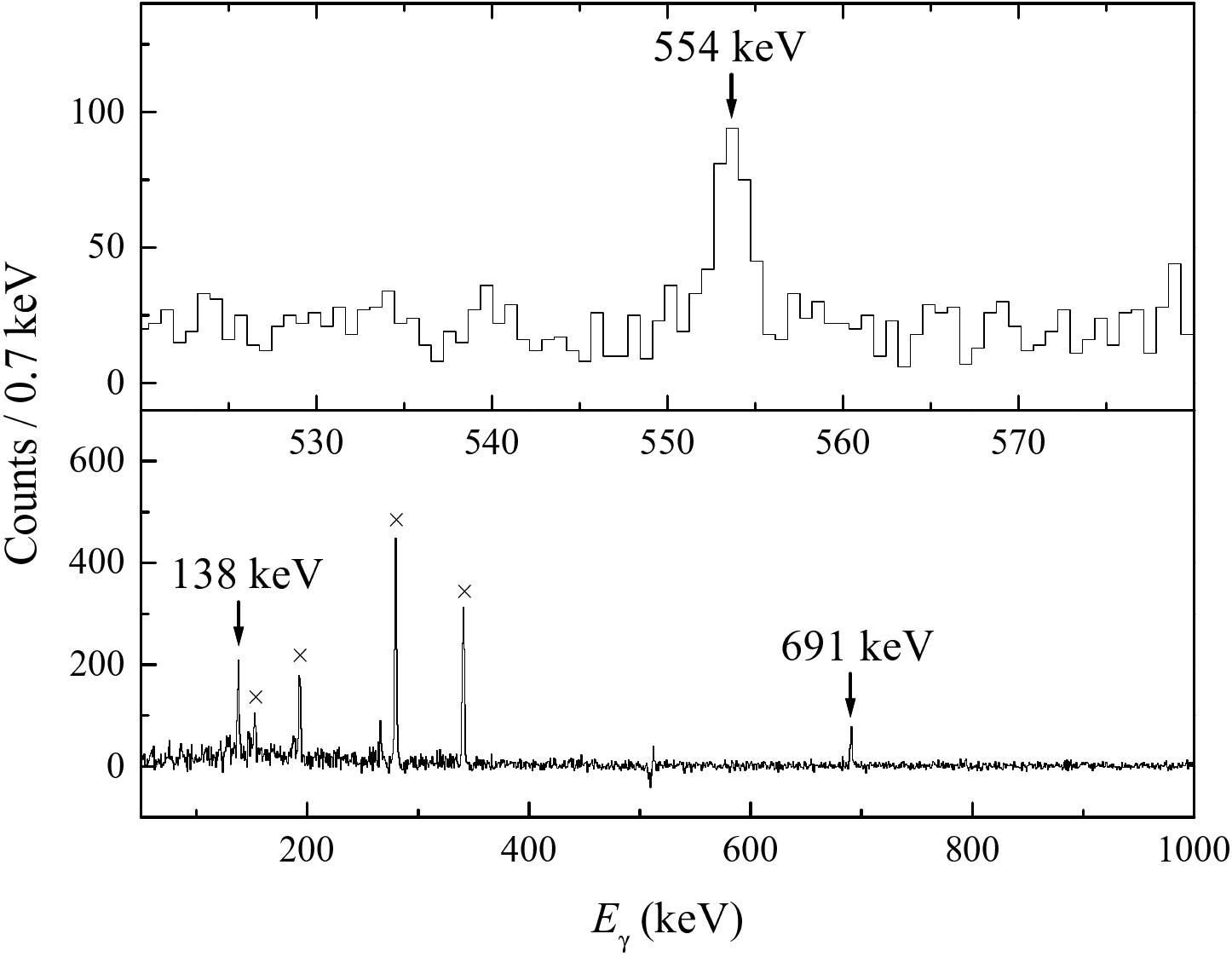}
\end{center}
\caption[Gated spectra for the
placement of the 554\,keV transition in $^{156}$Dy.]
{Gated coincidence spectra providing evidence for the
placement of the 554\,keV transition as a branch from the
$(2^+)_{1382}$ level to the $2^+_{829}$ level in $^{156}$Dy.  
Spectra gated on the 691\,keV $2^+_{829}\rightarrow2^+_{138}$ (top) and
554\,keV (bottom) transitions.  Contaminant transitions coincident with the
555\,keV transition in $^{157}$Dy are indicated with a cross
($\times$).  (Figure from Ref.~\cite{caprio2002:156dy-beta}.)
\label{fig156dylev1382}
}
\end{figure}

$(2^+)_{1382}$: A weakly-populated level is identified at
1382.3(2)\,keV on the basis of $\gamma$-ray branches to the
$0^+_{676}$, $2^+_{829}$, $2^+_{890}$, and possibly $3^+_{1022}$
levels deduced from coincidences with the
transitions depopulating these respective levels
(Fig.~\ref{fig156dylev1382} and Table~\ref{tabline}).  The present level may
be identified with the $(3^-)$ level at
1385(5)\,keV~\cite{ndsboth:156}, previously only reported in the
$(p,t)$ reaction~\cite{kolata1977:156dy-pt}, for which no prior $\gamma$-ray
information was known.  Observation of the $\gamma$-ray branch to an
excited $0^+$ state, together with the tentative transition to a $3^+$
state, suggests a $(2^+)$ spin assignment instead, although a spin of
$3^-$ cannot be excluded if the transition to the $0^+_{676}$ state is
taken to be of $E3$ multipolarity.

$6^+_{1437}$: The intensities found for the strong 667\,keV and 349\,keV
branches to the $6^+_{770}$ and $4^+_{1088}$ levels are in agreement
with the literature.  However, the intensity
of the 1033\,keV $6^+_{1437}\rightarrow4^+_{404}$ transition, which was
previously not observed in $\beta$-decay, is found to have about twice
the value obtained from $(\alpha,4n)$~\cite{deboer1977:156dy-a4np4n}.

\begin{figure}[t]
\begin{center}
\includegraphics*[width=0.48\hsize]{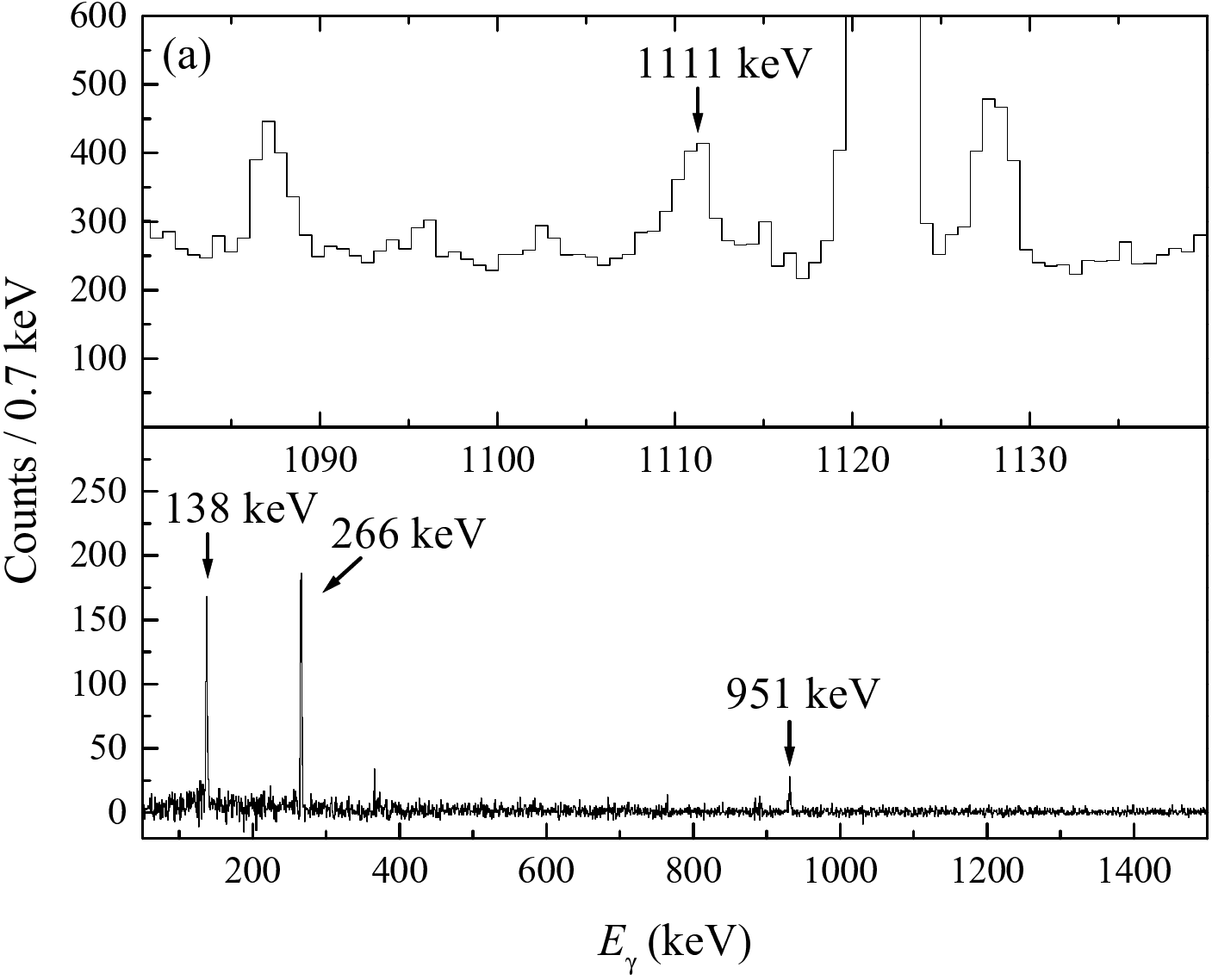}
\hfill
\includegraphics*[width=0.48\hsize]{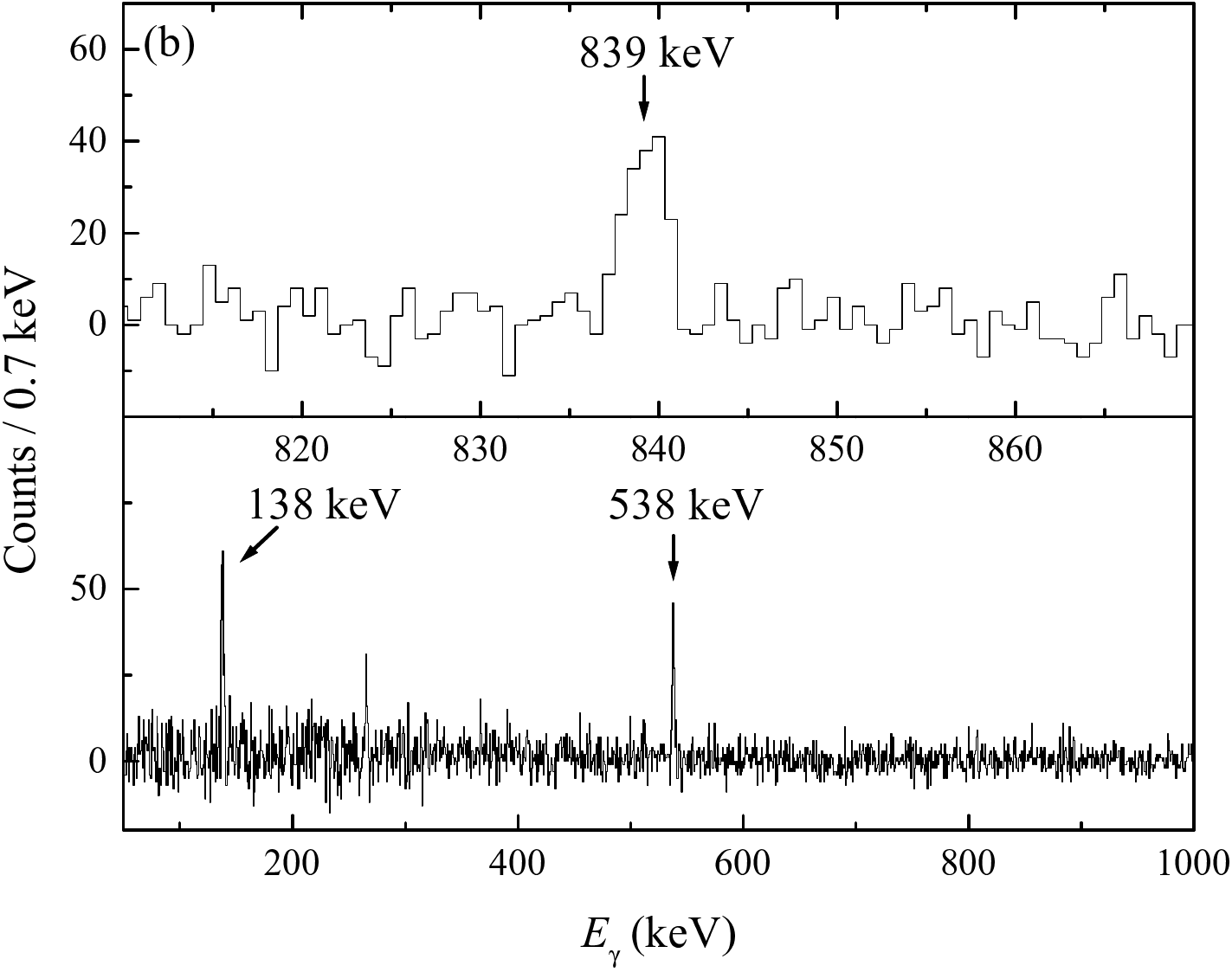}
\end{center}
\caption[Gated spectra for 
transitions from the level at 1515\,keV in $^{156}$Dy.]  {Gated
coincidence spectra providing evidence for branching transitions from
the level at 1515\,keV in $^{156}$Dy to $4^+$ and $0^+$ levels,
suggesting a $(2^+)$ spin assignment.  (a) Spectra gated on the
266\,keV $4^+_{404}\rightarrow2^+_{138}$ transition (top) and
1111.2\,keV $(2^+)_{1515}\rightarrow4^+_{404}$ transition (bottom),
supporting the placement of the latter transition.  The 1111\,keV peak
in the top spectrum is a doublet, containing a contribution from the
1110.7\,keV $(?)_{2445}\rightarrow5^+_{1336}$ transition, which also
produces the coincidence with the 951\,keV
$5^+_{1336}\rightarrow4^+_{404}$ transition observed in the bottom
spectrum.  (b) Spectra gated on the 538\,keV
$0^+_{676}\rightarrow2^+_{128}$ transition (top) and 839\,keV
$(2^+)_{1515}\rightarrow0^+_{676}$ transition (bottom), supporting the
placement of the latter transition.  (Figure from
Ref.~\cite{caprio2002:156dy-beta}.)
\label{fig156dylev1515}
}
\end{figure}

$(2^+)_{1515}$: A level at 1515.0(2)\,keV is identified on the
basis of transitions to the $4^+_{404}$, $0^+_{676}$, and $2^+_{890}$
states.  These transitions suggest a $2^+$ spin assignment (Fig.~\ref{fig156dylev1515}), although
a spin of $3^-$ cannot be excluded if $E3$ multipolarity is
considered.

$6^+_{1525}$ and $(5^-)_{1526}$: A closely spaced pair of levels lies
at 1525.3(2)\,keV and 1526.0(2)\,keV.  All transitions feeding or
depopulating these levels are potentially doublets at $\sim$0.7\,keV
separation, and division of the intensities has been a challenge to
all studies of these levels~\cite{deboer1977:156dy-a4np4n,elmasri1976:156dy-p4n}.
In fact, the prior published $\beta$-decay
work~\cite{gromov1976:156dy-beta} failed to identify the lower of the two
levels, but unpublished $\beta$-decay work~\cite{koldewijn:156dy-beta} cited
in Ref.~\cite{deboer1977:156dy-a4np4n} did observe both levels.  At least
one of these levels decays to each of the levels $4^+_{404}$,
$6^+_{770}$, $4^+_{1088}$, and $4^+_{1168}$.  The 190\,keV branch to
the $5^+_{1336}$ level reported in
$(\alpha,4n)$~\cite{deboer1977:156dy-a4np4n} is excluded by the present
data.  The present data are likewise inconclusive about the
assignment of intensities.  However, the centroid energies of the
observed peaks in gated spectra suggest that most of the intensity in
the 755\,keV transition to the $6^+_{770}$ level and in the 356\,keV
transition to the $4^+_{1168}$ level is depopulating the $6^+_{1525}$
level, while most of the observed intensity in the 1122\,keV transition
to the $4^+_{404}$ level and in the 438\,keV transition to the
$4^+_{1088}$ level is depopulating the $(5^-)_{1526}$ level.
(Estimated limits on the intensity in the other portion of each
potential doublet are given in Table~\ref{tabbranch}.)  Note that the
adopted spin assignments~\cite{ndsboth:156} of the two levels, which
are based upon measured conversion coefficients for the depopulating
transitions, are rather speculative given the uncertainties in
$\gamma$-ray intensities.

Among the higher-lying levels, several previously reported in $\beta$
decay~\cite{gromov1976:156dy-beta} are observed~--- $(3)^-_{1609}$,
$(4)^+_{1627}$, $2^+_{2090}$, $4^+_{2307}$, $(?)_{2409}$ [previously
$(2^-)$], $(?)_{2517}$ [previously $(1)^-$], $(?)_{2818}$, and
$(?)_{2823}$~--- albeit some with significantly modified decay
properties.  Three levels previously only identified in in-beam
studies~\cite{deboer1977:156dy-a4np4n,elmasri1976:156dy-p4n,riley1988:156dy-hixn}~---
$7^+_{1729}$, $7^-_{1810}$, and $(6)_{1898}$ [previously
$(6,7^-)$]~--- are also observed, yielding new information on their
branching properties.  The following comments address only a few of
the confirmed higher-lying levels, those for which the new data modify
the spin assignment, but detailed data for all may be found in
Table~\ref{tabbranch}.

$(6)_{1898}$: The adopted level at 1898.64(10)\,keV~\cite{ndsboth:156}
had been assigned a spin of $(6,7^-)$ based upon results from
$(\alpha,4n)$, $(p,4n)$, and
$(\text{HI},xn)$ studies~\cite{deboer1977:156dy-a4np4n,riley1988:156dy-hixn}.
The present data show a transition to a $5^+$ level, which eliminates
the possible $7^-$ assignment.

$(?)_{2409}$: The adopted level at 2409.64(20)\,keV~\cite{ndsboth:156}
had been assigned a spin of $(2^-)$, based upon a supposed 880\,keV
$M1$ $\gamma$-ray transition to a $(1^-)$ level at 1529\,keV and the
presence of $\gamma$-ray transitions to $4^+$ levels.  This spin
assignment would have required both transitions to $4^+$ levels to be
$E3$ in character, constituting a fairly unusual situation.  However,
the present data show that there is no evidence for the $(1^-)$ level
at 1529\,keV (see the following section) or for the 880\,keV branch
to this level.  Several new branches from the $(?)_{2409}$ level are
observed, and all branches to levels of known spin are to $2^+$,
$3^\pm$, or $4^+$ levels.

$(?)_{2517}$: The present level at 2516.6(7)\,keV may be identified
with the adopted level at 2517.55(16)\,keV~\cite{ndsboth:156}, which
had a spin assignment of $(1^-)$ based upon a supposed $E2$
$\gamma$-ray transition to a $(3)^-$ level and a $\gamma$-ray
transition to a $0^+$ level.  (There is an observed transition to a
$4^+$ level, which this assignment would have required to be $E3$ in
nature.)  However, the present data eliminate the claimed 907\,keV
transition to the $(3)^-_{1609}$ level and 1841\,keV transition to the
$0^+_{676}$ level.  This leaves only transitions to $2^+$, $3^+$, and
$4^+$ levels, and possibly a transition to a $3^-$ level
(Table~\ref{tabbranch}).  It also should be noted that the level
energy calculated from the transition energy of the 1493.8(10)\,keV
branch from this level (as measured in a spectrum gated on the 884\,keV
transition) disagrees with that calculated from the transition energy
of the 1348.9(5)\,keV branch (as measured in spectra gated on the 764
and 1031\,keV transitions) and from the other two tentatively-placed
branches.  Even though the discrepancy ($\sim$1.4\,keV) is within the
extreme range of the energy uncertainties, it calls into question the
identity of the level as a single level.

\begin{figure}[t]
\begin{center}
\includegraphics*[width=0.48\hsize]{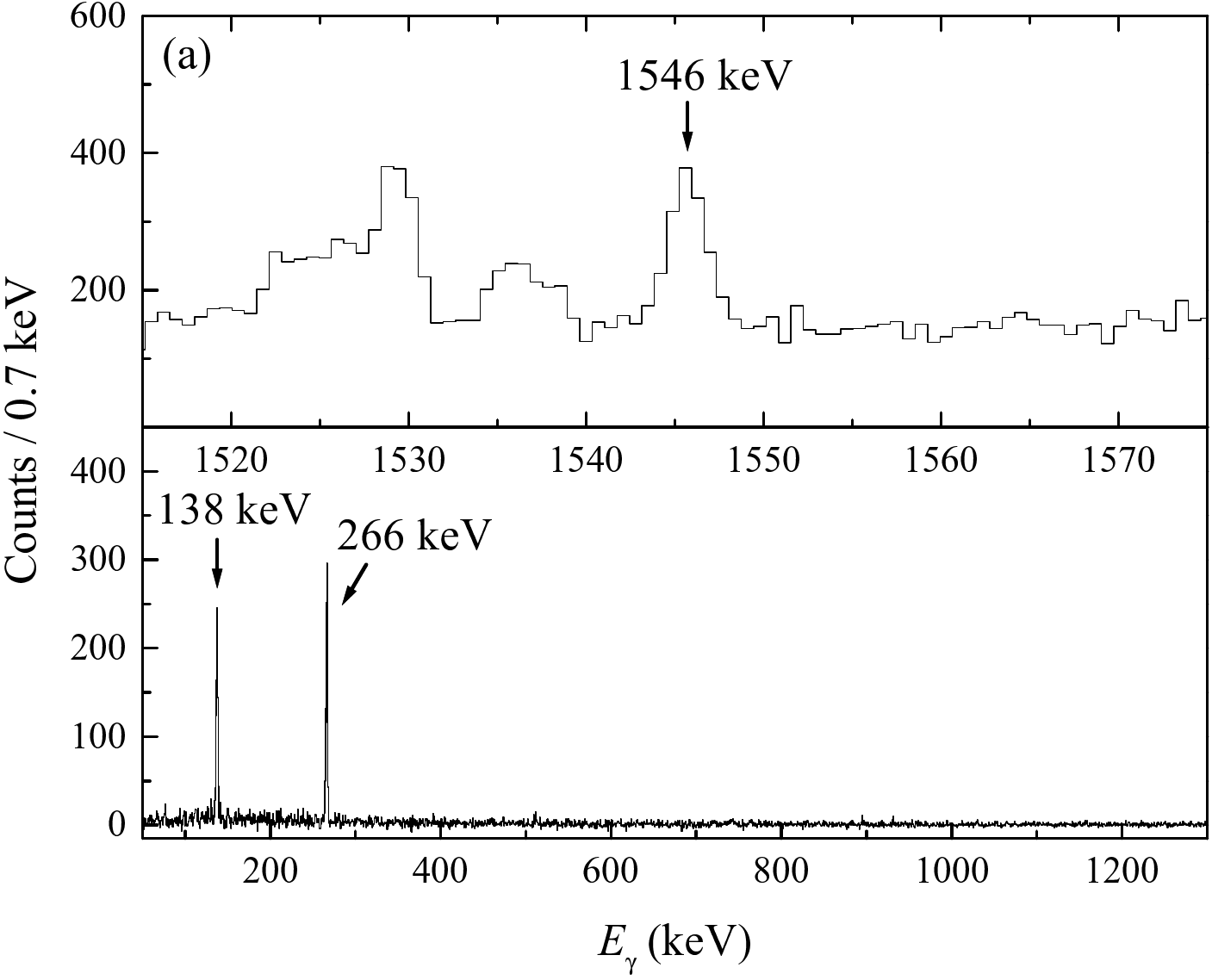}
\hfill
\includegraphics*[width=0.48\hsize]{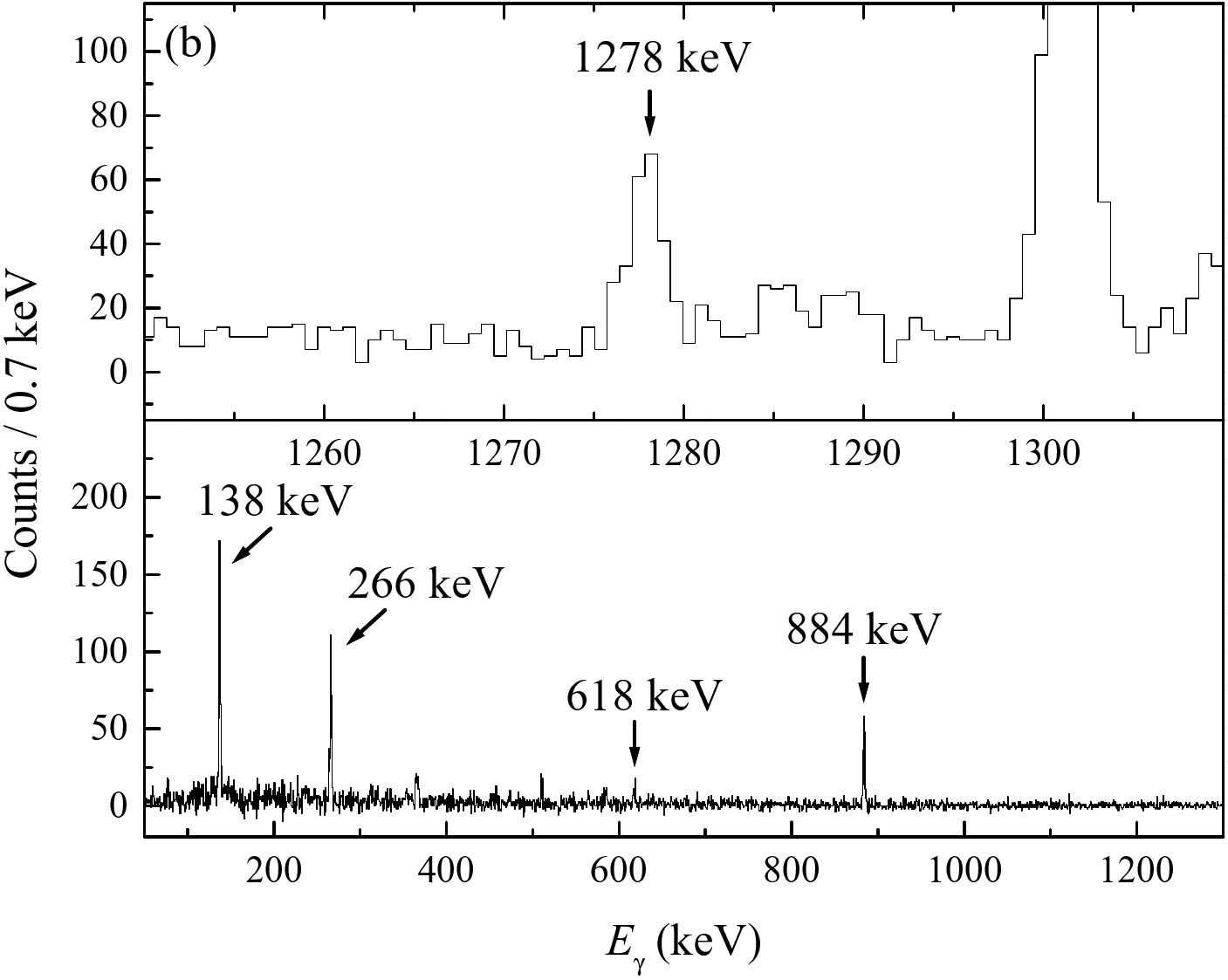}
\end{center}
\caption[Gated spectra for the 
levels at 1950\,keV and 2300\,keV in $^{156}$Dy.]  {Gated
coincidence spectra providing evidence for the existence of 
levels at 1950\,keV and 2300\,keV in $^{156}$Dy.  (a) Spectra gated on
the 266\,keV $4^+_{404}\rightarrow2^+_{128}$ transition (top) and
1546\,keV $(?)_{1950}\rightarrow4^+_{404}$ transition (bottom). (b)
Spectra gated on the 884\,keV $3^+_{1022}\rightarrow2^+_{138}$
transition (top) and 1278\,keV $(?)_{2300}\rightarrow3^+_{1022}$
transition (bottom), supporting its placement as directly feeding the
$3^+_{1022}$ level.  (Figure from Ref.~\cite{caprio2002:156dy-beta}.)
\label{fig156dylevnew}
}
\end{figure}

Several new levels are identified as well (Table~\ref{tabbranch}).
Many of the new levels are identified on the basis of several
corroborating branching or feeding transitions, each independently
placed from coincidence data.  Other levels are identified on the
basis of only one or two branches.  These have been retained in the
tabulation when there is fairly strong evidence for their placement
from coincidence relations (see Fig.~\ref{fig156dylevnew}). Some of the
``new'' levels below 2250\,keV likely correspond to levels previously
reported in $(d,d')$ or $(p,t)$
studies~\cite{grotdal1968:156dy-ddp,kolata1977:156dy-pt}, but the low energy
resolution of such studies precludes the establishment of an
unambiguous correspondence.

\section{Previously reported levels from $\beta$ decay for which no evidence is found}
\label{sec156dydestroy}

The previous $\beta$-decay study~\cite{gromov1976:156dy-beta} of
$^{156}$Dy identified levels on the basis primarily of singles
$\gamma$-ray data, together with singles conversion electron data and
some very limited coincidence data.  Each level was invoked to explain
several $\gamma$ rays observed in singles.  As outlined in
Section~\ref{seccoin}, this process relies upon the recognition of
groups of $\gamma$-ray lines with energies which sum to yield the same
excited level energy and is extremely challenging when the singles
spectrum is a near continuum of unresolved or overlapping lines.  In
contrast, the high-statistics coincidence data available in the
present study provide placement information directly from coincidence
relations.

A large number of levels previously proposed on the basis of
data from $\beta$ decay are found to be unsubstantiated by the present data, so
it is worth summarizing the evidence used in dismissing these levels.
The previously proposed placements of $\gamma$ rays implied
coincidences relations involving these $\gamma$ rays, and the present
data are used to verify whether or not these coincidences are actually
present with the necessary strength.  Also, if substantially all of
the singles intensity for a $\gamma$ ray line is now accounted for by
one or more new placements from coincidence information, the need for
the original proposed placement is obviated.
Detailed discussions of a few of the most important low-lying
dismissed levels follow.  

$(1,2^+)_{1219}$: This level was proposed in $\beta$
decay~\cite{gromov1976:156dy-beta} to explain six $\gamma$-ray
transitions, placed as two branches and four feeding transitions.  The
prior study identifies a 1218.9(4)\,keV line of intensity 0.74(11),
which it places as a branch to the ground state.  The present
coincidence data show this intensity to be accounted for, within
uncertainties, by two new transitions: a 1217.2(3)\,keV
$(?)_{2386}\rightarrow4^+_{1168}$ transition of intensity 0.25(7) and
a 1218.9(5)\,keV $4^+_{2307}\rightarrow4^+_{1088}$ transition of
intensity 0.39(10).  Ref.~\cite{gromov1976:156dy-beta} also reports a
1081.38(20)\,keV line of intensity 1.01(6), which it places as a
branch to the $2^+_{138}$ level, supported by an observed qualitative
coincidence with the $2^+_{138}\rightarrow0^+_{0}$ transition.  The
present coincidence data show there to be a 1081.18(9)\,keV
$(?)_{2103}\rightarrow3^+_{1022}$ transition of intensity 0.64(5).
This leaves a residual intensity of 0.4(2) feeding the $2^+_{138}$
level in this energy region, observed in a spectrum gated on the
138\,keV transition after subtraction of the known placement.  The two
stronger supposed feeding transitions, a 582.6(4)\,keV transition of
intensity 0.24(4) and a 2428.0(5)\,keV transition of intensity
0.35(4), are noncoincident with the depopulating transitions at a
level inconsistent with the prior decay scheme.  (Specifically, from
the present data, the coincident intensity of any possible 583\,keV
transition with a 1081\,keV transition is found to be $<$0.08, and
that with a 1219\,keV transition is $<$0.05.  The coincident intensity
of any 2428\,keV transition with a 1081\,keV transition is $<$0.06,
and that with a 1219\,keV transition is $<$0.07.)  The 2428\,keV
$\gamma$-ray line is now replaced by a 2429.5(7)\,keV
$(?)_{2833}\rightarrow4^+_{404}$ transition of intensity 0.63(9).

$1^-_{1293}$: This level was identified in $\beta$
decay~\cite{gromov1976:156dy-beta} on the basis of two depopulating
transitions and five feeding transitions.  (A discrepancy of
$\sim$0.67\,keV existed, however, between the level energies deduced
from the two different depopulating transitions.)
Ref.~\cite{gromov1976:156dy-beta} reported a 1292.85(22)\,keV transition of
intensity 1.28(6), which it gave a $1^-_{1293}\rightarrow0^+_{0}$
placement.  However, this singles intensity is now fully accounted for
by three transitions (Table~\ref{tabline})~--- a 1292.3(3)\,keV
$(?)_{2818}\rightarrow(5^-)_{1526}$ transition, a 1293.0(5)\,keV
$(?)_{2184}\rightarrow2^+_{890}$ transition (tentative), and a
1293.4(15)\,keV $(?)_{2818}\rightarrow6^+_{1525}$ transition~--- with a
combined intensity of 1.28(14).  Most ($\sim$85$\%$) of the singles
intensity observed at this energy in the present experiment comes from
a 1293.7(2)\,keV contaminant transition from $^{116}$Sn, most likely
arising from $^{115}$In neutron capture followed by $^{116m}$In
$\beta$ decay to $^{116}$Sn~\cite{nds1994:116}, identified by its
coincidences with the 818, 1097, 1507, and 1753\,keV transitions in
that nucleus.  Ref.~\cite{gromov1976:156dy-beta} also reported a
1155.72(14)\,keV transition of intensity 2.14(5), qualitatively
coincident with the $2^+_{138}\rightarrow0^+_{0}$ transition, which it
assigned a $1^-_{1293}\rightarrow2^+_{138}$ placement.  This intensity
is now mostly accounted for by three transitions~--- an 1154.4(8)\,keV
$(?)_{2490}\rightarrow5^+_{1336}$ transition (tentative), an 1155.3(2)
$(?)_{2324}\rightarrow4^+_{1168}$ transition, and an 1156.4\,keV
$(?)_{2245}\rightarrow4^+_{1088}$ transition~--- with a combined
intensity of 1.72(13).  The intensities of all five feeding
transitions are also accounted for (Table~\ref{tabline}), and all
are found to be noncoincident with 1293\,keV and 1156\,keV transitions
at intensity limits contradicting the earlier placements.

$(2^+)_{1447}$: This level was reported only in the $(a,4n)$
and $(p,4n)$ literature~\cite{deboer1977:156dy-a4np4n}, but the placement
given in that work was justified using information from an earlier
unpublished $\beta$-decay study~\cite{koldewijn:156dy-beta}.  No evidence is
found for the existence of this level in the present experiment.

$2^+_{1518}$: This level was identified in $\beta$
decay~\cite{gromov1976:156dy-beta} on the basis of two depopulating
transitions and two feeding transitions.  The reported singles
intensities of these transitions are now accounted for by other
placements (Table~\ref{tabline}), and both feeding transitions are
noncoincident with the depopulating transitions at intensity limits
contradicting the earlier placements.

$(1^-)_{1529}$: This level was proposed in $\beta$
decay~\cite{gromov1976:156dy-beta} to explain seven $\gamma$-ray
transitions, placed as two depopulating and five feeding transitions.
The reported intensities of the 1529 and 1392\,keV branching
transitions and the 1274, 1542, and 1900\,keV feeding transitions are now accounted for by
other placements (Table~\ref{tabline}), and
all feeding transitions are noncoincident with both depopulating
transitions at intensity limits contradicting the earlier placements.

By similar arguments, there is no evidence for the existence of the
levels claimed from $\beta$ decay~\cite{gromov1976:156dy-beta} at 1801,
1944, 2006, 2169, 2216, 2476, 2514, 2637, 2661, and 2803\,keV or above
2900\,keV excitation energy.

The level scheme and $\gamma$-ray decay data for $^{156}$Dy are
considerably modified by the results presented in this chapter.  The
results which have the most direct implications for the structural
interpretation of $^{156}$Dy are the intensity measurements for
transitions between low-lying levels and the clarification of
existence or nonexistence for several of the low-lying levels.  These
results are discussed within the context of the phenomenology of the
$N$$\approx$90 region and the X(5) model in Chapter~\ref{chapphenom}.

\chapter{$^{162}$Yb}
\label{chap162yb}

\begin{figure}
\begin{center}
\includegraphics[width=0.48\hsize]{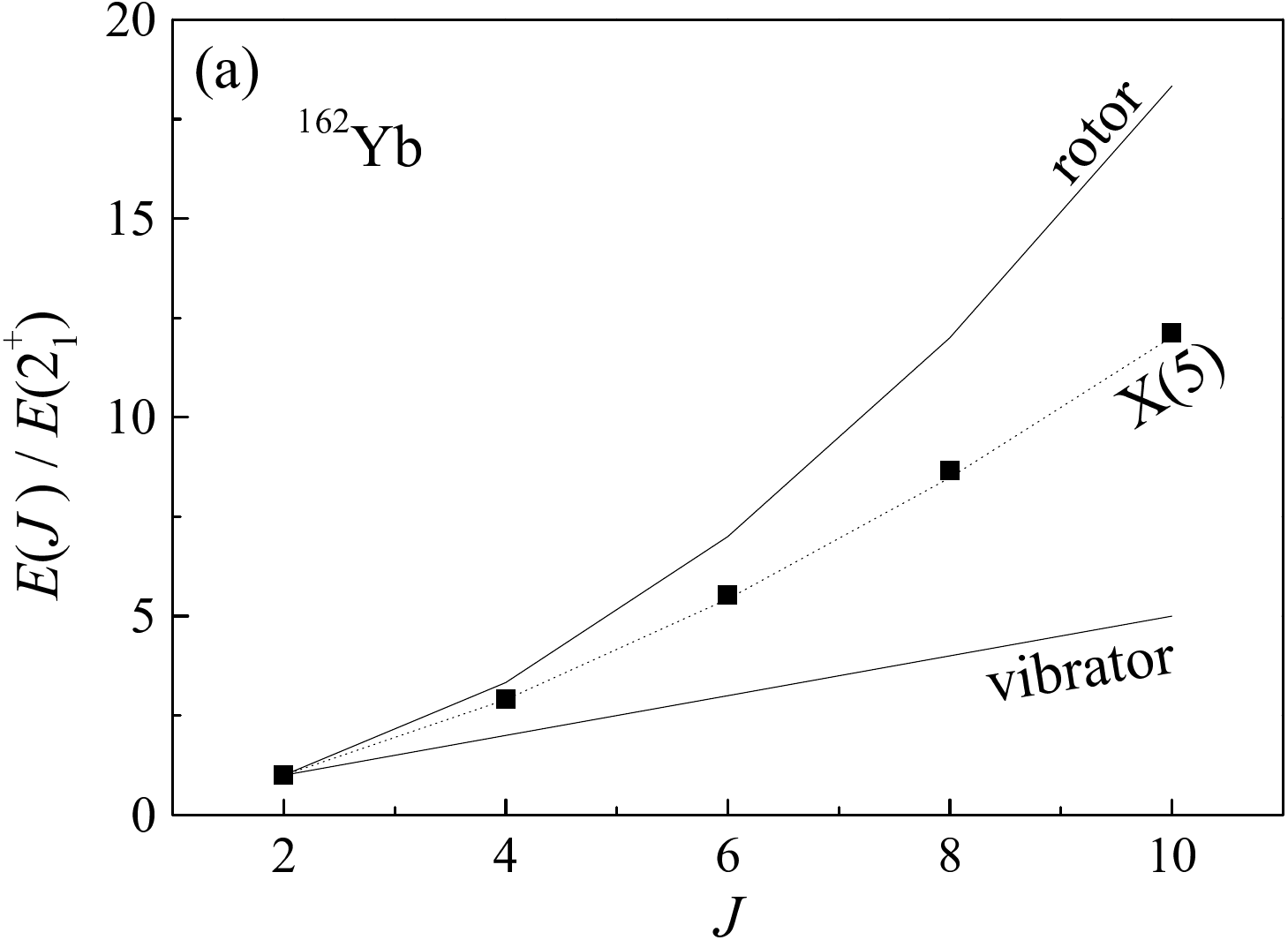}
\hfill
\includegraphics[width=0.48\hsize]{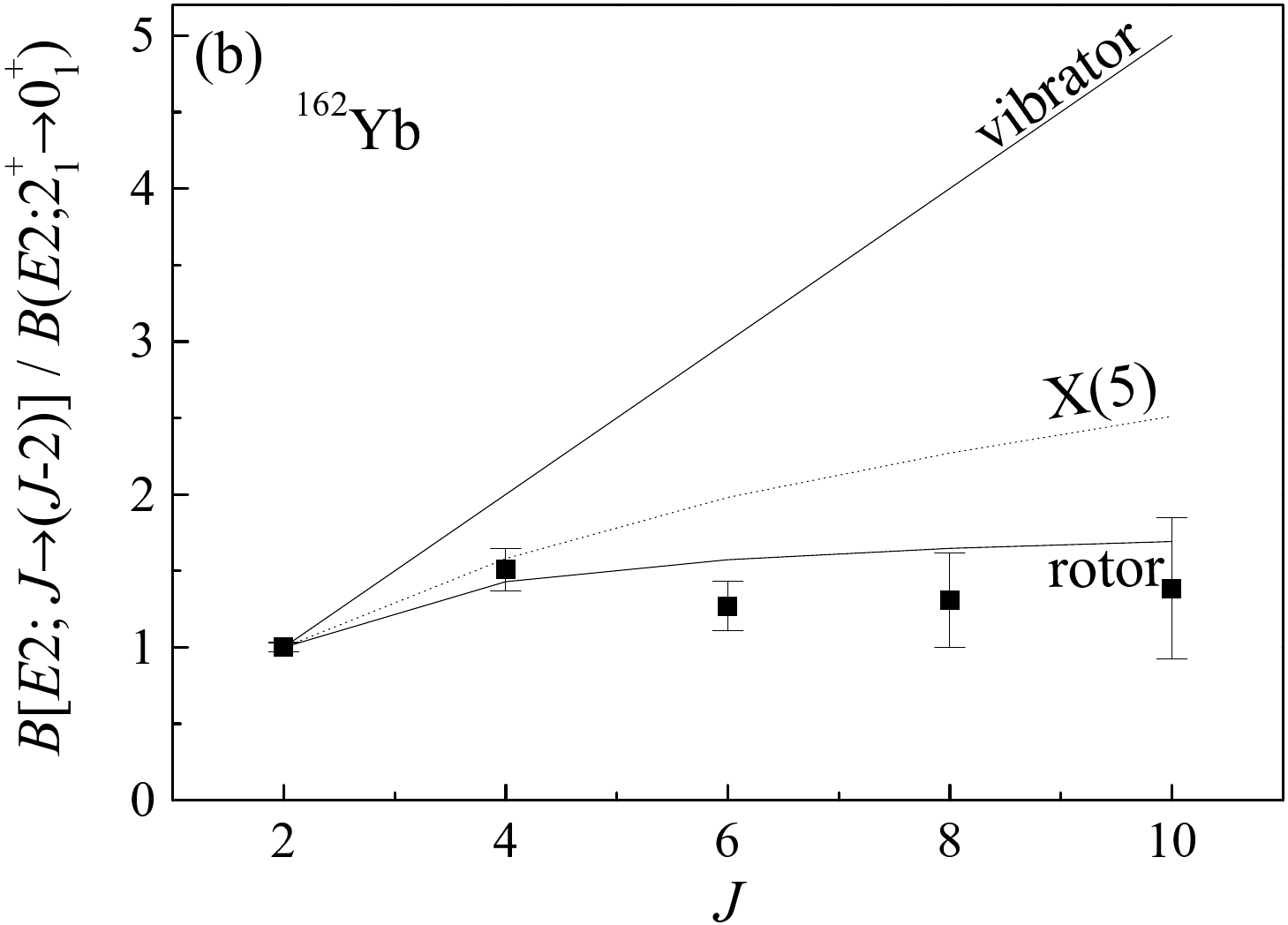}
\end{center}
\caption[Yrast band energies
and $B(E2)$ values in $^{162}$Yb.]{Yrast band (a) energies, normalized
to the $2^+$ band member, and (b) $B(E2)$ values, normalized to the
$2^+_1\rightarrow0^+_1$ transition, in $^{162}$Yb.  The rotor, X(5),
and vibrator predictions are shown for comparison.  The experimental
$B(E2)$ values are from Ref.~\cite{nds1999:162}.  [For the
$B(E2;2^+_1\rightarrow0^+_1)$ value, see text.]
\label{fig162ybyrast}
}
\end{figure}

The nucleus $^{162}$Yb shows a similar spin dependence in its yrast
band energies to the $N$=90 X(5) candidate nuclei
[Fig.~\ref{fig162ybyrast}(a)].  However, difficulties arise in
interpreting other aspects of the existing data, including $B(E2)$
values and off-yrast structure, in terms of an X(5) picture, as shown
in Fig.~\ref{fig162ybyrast}(b).  Since the overall scale of the X(5)
predictions is arbitrary, the experimental $B(E2)$ values are
normalized to the $B(E2;2^+_1\rightarrow0^+_1)$ value for comparison,
and an accurate $B(E2;2^+_1\rightarrow0^+_1)$ value is essential to
such a comparison.  A $\beta$-decay experiment on $^{162}$Yb was
performed to obtain a FEST measurement of the $2^+_1$ level lifetime.
(The experiment also yielded spectroscopic and angular correlation
information on low-lying non-yrast states, reported
elsewhere~\cite{mccutchanINPREP}.)  This lifetime measurement has been
reported in Ref.~\cite{caprio2002:162yb162er-fest}.

The experiment was carried out in $\beta^+$/$\varepsilon$ decay at the
Yale MTC in its FEST configuration (Section~\ref{secfest}), consisting
of three Compton-suppressed clover detectors, a LEPS detector, and the
plastic and BaF$_2$ fast-timing detectors. Parent $^{162}$Lu nuclei
were produced through the reaction $^{147}$Sm($^{19}$F,$4n$)$^{162}$Lu
at a beam energy of 95\,MeV, using an $\sim$7\,pnA beam incident upon a
1.8\,mg/cm$^2$ 98$\%$-isotopically-enriched target.  The $^{162}$Lu
parent nucleus decays to $^{162}$Yb with a half life of
$\sim$1.4\,min~\cite{nds1999:162}.  The tape was advanced at 125\,s
intervals.

This experiment was one of the first to make use of a new VME-based
acquisition system for WNSL nuclear structure experiments, with
readout by an Intel/Linux front end computer.  The data acquisition
and sorting software presently used for nuclear structure experiments
at WNSL (see Appendix~\ref{appacq}) was written in preparation for and
first tested in this experiment.  Data were acquired in event mode
with a Ge singles (or higher fold) or plastic scintillation detector
trigger.  In 79\,h, the experiment yielded 1.2$\times$10$^6$
plastic-BaF$_2$ coincidence events, including 1.5$\times$10$^5$
Ge-plastic-BaF$_2$ coincidence events.

\begin{figure}
\begin{center}
\includegraphics[width=0.7\hsize]{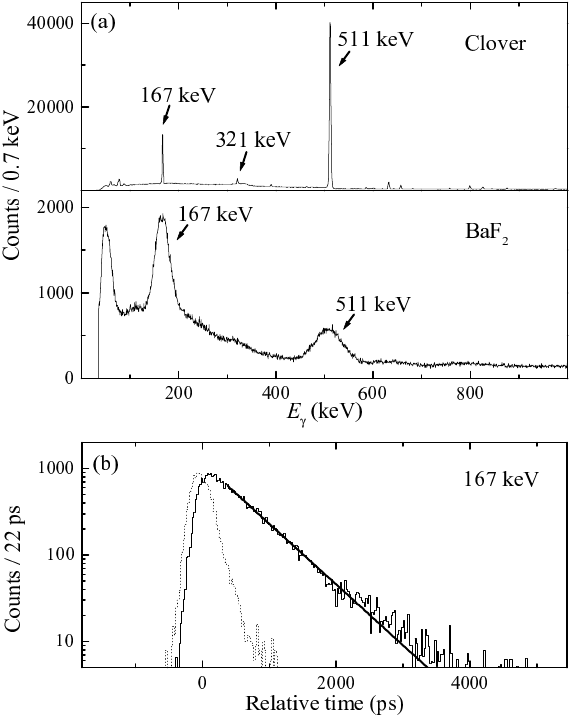}
\end{center}
\caption[Spectra from the $^{162}$Yb lifetime measurement.]
{Spectra from the $^{162}$Yb lifetime measurement. (a) Clover and BaF$_2$
detector energy spectra coincident with $\beta^+$ $\Delta E$ signals,
showing the 167\,keV $2^+_1\rightarrow0^+_1$ and 321\,keV
$4^+_1\rightarrow2^+_1$ $\gamma$-ray transitions as well as 511\,keV
annihilation radiation. (b) Measured time distribution for the 167\,keV
BaF$_2$ detector $\gamma$-ray time signal relative to the $\beta^+$
time signal, together with the fitted $\tau$=618\,ps decay curve (heavy 
line).  The prompt response curve (dotted) is shown for comparison
(see text).  (Figure from Ref.~\cite{caprio2002:162yb162er-fest}.)
\label{fig162ybfest}
}
\end{figure}

The 167\,keV $2^+_1\rightarrow0^+_1$ $\gamma$-ray transition in
$^{162}$Yb is by far the most intense observed transition in the
$\beta^+$ decay of $^{162}$Lu [Fig.~\ref{fig162ybfest}(a)].  
Consequently, a
high-statistics measurement of its decay time can be carried out from
the full plastic-BaF$_2$ coincidence data, without it being necessary
to require triple coincidences with a feeding $\gamma$-ray transition
detected in the Ge detector.  

The time distribution obtained for a 17\,keV wide energy gate on the
167\,keV transition in the BaF$_2$ detector, with a local background
subtraction on both the Ge and BaF$_2$ gates, is shown in
Fig.~\ref{fig162ybfest}(b).  The prompt timing response at this energy
is shown for comparison.  (This curve is obtained using $\gamma\gamma$
coincidences from $^{60}$Co decay, with Compton energy deposition in
both timing detectors, shifted~\cite{moszynski1989:fest2} so that the
time response centroids at higher energies coincide with those for
prompt Compton events in the $^{162}$Yb data.)  The plastic-BaF$_2$
time spectrum obtained for $\sim$167\,keV energy deposition in the
BaF$_2$ detector contains both a prompt background from partial-energy
deposition by higher-energy $\gamma$ rays from $^{162}$Yb and a small
delayed background from partial-energy deposition by 511\,keV
annihilation $\gamma$ rays.

The decay time of the 167\,keV transition deduced using the slope
method is 618(19)\,ps, where the uncertainty accounts for both
statistical and systematic contributions, including
walk~\cite{mach1991:sr-fest} and weighting uncertainties in the
background subtraction.  The slope was obtained by a
Gaussian-error-weighted least-squares fit of the time spectrum,
sufficiently compressed to provide a deduced lifetime stable against
counting fluctuations (Section~\ref{secfest}), over a range of time
channels excluding the prompt region and extreme tail region.
Corroborating results were obtained, but with larger statistical
uncertainties, using time spectra gated on specific feeding
$\gamma$-ray transitions detected in the Ge array as well as on
511\,keV annihilation radiation detected in the Ge array, which
selects events in which the partner annihilation $\gamma$ ray does
\textit{not} enter the BaF$_2$ detector, providing a reduced
background for the 167\,keV $\gamma$ ray.  

The decay time of the 167\,keV transition can be interpreted directly
as the lifetime of the $2^+_1$ level it depopulates, due to
comparatively short decay times of the higher-lying feeding
transitions.  Approximately 10$\%$ of the $\beta^+$-coincident feeding
of the $2^+_1$ level comes through the 321\,keV $4^+_1\rightarrow2^+_1$
transition [Fig.~\ref{fig162ybfest}(a)], and so the 15.3(14)\,ps lifetime of
the $4^+_1$ level~\cite{nds1999:162} introduces a small feeding delay
to the population of the $2^+_1$ level.  However, the contribution of
this delay to the effective lifetime, or reciprocal of the logarithmic
slope of the decay curve, for the 167\,keV
transition is much less than 1\,ps in the time region used for the
analysis.  The measured $2^+_1$ lifetime is comparable to prior
reported values (Table~\ref{tab162yblifetimes}) and has a 
low uncertainty due to the high statistics obtained and simplicity of
the background.  
\begin{table}
\begin{center}
\begin{tabular}{r@{}lll}
\pseudoruledtabular
\multicolumn{2}{c}{$\tau$ (ps)} &
\multicolumn{1}{l}{Reference} &
\multicolumn{1}{l}{Method}\\
\hline
633&(53)& Ref.~\cite{backe1978:162yb-rsm} & Recoil conversion electron shadow \\
577&(19)& Ref.~\cite{mcgowan1992:162yb-rdm} & Recoil Doppler \\
618&(19)& Present &\\
\pseudoruledtabular
\end{tabular}
\end{center}
\caption
[Values for the lifetime of the first excited $2^+$ state in $^{162}$Yb.]
{\ssp Values for the lifetime of the first excited $2^+$ state in $^{162}$Yb 
as determined by various methods in prior experiments and in the present work.
\label{tab162yblifetimes}
}
\end{table}

The spin dependence of the yrast band $B(E2)$ values in $^{162}$Yb
presents great difficulties for model interpretation.  The lifetime
measured in this experiment, combined with a total electron conversion
coefficient of 0.503~\cite{nds1999:162}, yields a
$B(E2;2^+_1\rightarrow0^+_1)$ value of 130(4)\,\Wu.
Figure~\ref{fig162ybyrast}(b) shows the $B(E2)$ values for transitions
in the yrast band normalized to this strength.  The data for $J$\gt2
are from Ref.~\cite{nds1999:162}, based upon the recoil Doppler method
data of Refs.~\cite{bochev1976:162yb-rdm,mcgowan1992:162yb-rdm}.  The
$B(E2)$ strengths depopulating the $6^+$ and higher band members fall
not only well below the X(5) predictions but even below the rotor
predictions.  With the $2^+_1$ lifetime well established, it would be
valuable to have verification of the lifetimes for higher band members
from an experiment making use of modern gated coincidence Doppler
shift techniques~\cite{boehm1993:ddcm}.

\chapter{$^{154}$Dy}
\label{chap154dy}

The nucleus $^{154}$Dy, with $N$=88 and $Z$=66, is structurally
intermediate between the oscillator nuclei and the $N$=90 transitional
nuclei.  This nucleus is therefore of interest as part of the program
to trace structural evolution in transitional regions, especially in
conjunction with the new data for the neighboring isotope $^{156}$Dy
described in Chapter~\ref{chap156dy}.  The first excited $0^+$ state
plays a special role in indicating the evolution of
$\beta$-deformation and $\beta$-softness, as discussed in
Section~\ref{sectrans} [see Fig.~\ref{fig2vs0} on
page~\pageref{fig2vs0}].  Lifetimes for excited $0^+$ states in
unstable, proton-rich nuclei (such as $^{154}$Dy) are not commonly
known, unless some form of isomerism makes the $0^+$ lifetime
accessible to electronic timing (as for $^{102}$Pd in
Chapter~\ref{chap102pd}), since Coulomb excitation, $(n,n')$
scattering, and neutron capture are not available as population
mechanisms for these nuclei.  The excited $0^+$ states in $^{154}$Dy
were not significantly populated in (HI,$xn$) Doppler shift lifetime
measurements (see Ref.~\cite{nds1998:154}).  However, the $0^+_2$
state in $^{154}$Dy can be populated in $\beta$ decay, and the
$0^+_2$-$2^+_1$ energy difference in $^{154}$Dy is unusually low
(326\,keV), so even with collective decay strength to the $2^+_1$
state (tens of \Wu), this state should have a lifetime near 100\,ps.
This is long enough to be accessible to a FEST measurement.

A $\gamma$$\gamma$ spectroscopy and FEST lifetime measurement
experiment on $^{154}$Dy was carried out in $\beta^+/\varepsilon$
decay at the Yale MTC in its FEST configuration
(Section~\ref{secfest}).  Parent $^{154}$Er nuclei were produced
through the reaction $^{144}$Sm($^{13}$C,3$n$)$^{154}$Er at a beam
energy of 62\,MeV, with a beam current of $\sim$20\,pnA.  A target stack
containing a total of $\sim$713\,$\mu$g/cm$^2$ thickness of $^{144}$Sm
(unknown enrichment) and 60\,$\mu$g/cm$^2$ thickness of C backing
material was used.  However, mechanical damage was evident in some of the layers,
leading this target to be replaced by a sandwich target consisting of
10.6\,mg/cm$^2$ $^{209}$Bi (placed upstream), 2.2\,mg/cm$^2$ $^{144}$Sm,
and 60\,$\mu$g/cm$^2$ Au for the final portion of the experiment.  The
beam energy was adjusted to 73\,MeV to allow for energy loss in the Bi
and current was limited to $\lesssim$4\,pnA by target heating
considerations.  The detector array consisted of three
Compton-suppressed clovers and a LEPS detector, as well as the
fast-timing detectors, but the array efficiency was much lower than
usual ($\sim$0.51$\%$ at 1.3\,MeV), since an interim mounting
used for the fast-timing detectors imposed constraints on detector
positioning.  Data were acquired in event mode with a Ge singles (or
higher fold) trigger using the YRAST Ball FERA/VME data acquisition
system~\cite{beausang2000:yrastball}.  In $\sim$50\,h, the experiment
yielded \sci{4.4}{8}~Ge singles counts and \sci{2.3}{5} Ge-plastic-BaF$_2$
triple-coincidence events.

The nucleus $^{154}$Er ($T_{1/2}$=3.7\,min) decays to $^{154}$Ho,
primarily populating the $2^-$ ($T_{1/2}$=11.8\,min) ground
state~\cite{nds1998:154}, and so a 25\,min tape advance cycle was used
for the MTC.  The nucleus $^{154}$Ho also posesses a low-lying $8^+$
($T_{1/2}$=3.1\,min) $\beta$-decaying isomer~\cite{nds1998:154}.  The
$\gamma$-ray and conversion electron spectroscopy of $^{154}$Dy
following the $\beta^+/\varepsilon$ decay of both ground state and
isomeric $^{154}$Ho has been studied in detail by Zolnowski
\etal~\cite{zolnowski1980:154dy156er-beta}, making heavy use of
$\gamma$$\gamma$ coincidence data.  The level scheme and branching
intensity results of Ref.~\cite{zolnowski1980:154dy156er-beta} are
generally supported by the present $\gamma$-ray data.  Levels
indicated in Ref.~\cite{zolnowski1980:154dy156er-beta} to be populated
in $^{154}$Ho ground state decay following $^{154}$Er decay were, as
expected, strongly populated.  However, higher-spin states reported in
Ref.~\cite{zolnowski1980:154dy156er-beta} as being present only in
$^{154}$Ho isomer decay were strongly populated as well (see
Fig.~\ref{figcoinplace} on page~\pageref{figcoinplace}).  This might
be attributable to direct production of isomeric $^{154}$Ho in the
present experiment through the $^{144}$Sm($^{13}$C,$p$2$n$)$^{154}$Ho
reaction channel, although this reaction channel was
calculated~\cite{gavron:pace} to constitute only $\sim$21\,mb out of
the total $\sim$600\,mb reaction cross section, compared to
$\sim$470\,mb for the main $^{144}$Sm($^{13}$C,3$n$)$^{154}$Er
channel.  The long tape-advance cycle used should also have strongly
emphasized the two-step decay over the one-step isomer decay.

\begin{figure}
\begin{center}
\includegraphics*[width=0.8\hsize]{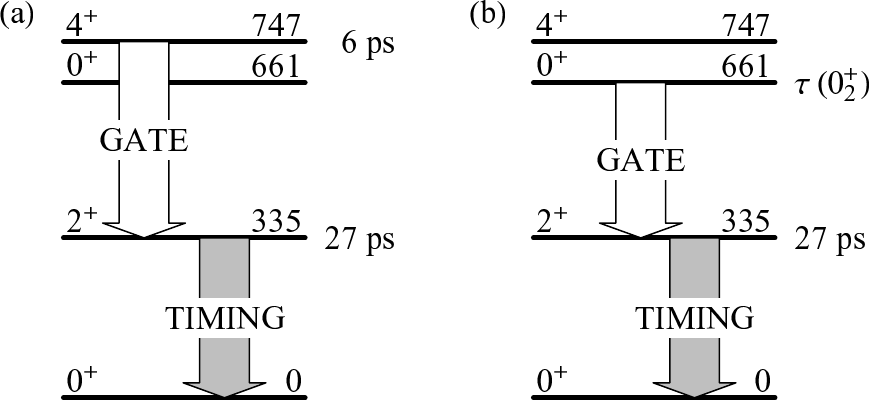}
\end{center}
\caption[Cascades involving the $4^+_1$ and $0^+_2$ states in $^{154}$Dy.]
{Cascades involving the (a) $4^+_1$ and (b) $0^+_2$ states, selected by a
Ge gating condition, for measurement of the $0^+_2$ lifetime in $^{154}$Dy.
\label{fig154dycascades}
}
\end{figure}

The $0^+_2$ level lifetime is short compared to the FEST system time
resolution, and so it must be determined by a centroid shift
measurement.  The decay scheme of $^{154}$Dy is such that the lifetime
can be extracted by using the Ge detectors to select specific feeding
paths and observing the difference between the timing curve centroids
for a
\textit{single} $\gamma$-ray transition, the 335\,keV
$2^+_1\rightarrow0^+_1$ transition, subject to two different Ge
detector gating conditions [Fig.~\ref{fig154dycascades}].  If a
coincidence with the 412\,keV $4^+_1\rightarrow2^+_1$ transition is
required [Fig.~\ref{fig154dycascades}(a)], the timing centroid for the 
335\,keV $\gamma$ ray detected in the BaF$_2$ detector is
\begin{equation}
x_{\text{gate}~412}=x_\text{prompt}(335\,\mathrm{keV}) + \tau_{4^+_1} + \tau_{2^+_1}.
\end{equation}
If, instead, a
coincidence with the 326\,keV $0^+_2\rightarrow2^+_1$ transition is
required [Fig.~\ref{fig154dycascades}(b)], the timing centroid for the 
335\,keV $\gamma$ ray detected in the BaF$_2$ detector is
\begin{equation}
x_{\text{gate}~326}=x_\text{prompt}(335\,\mathrm{keV}) + \tau_{0^+_2} + \tau_{2^+_1}.
\end{equation}
(These expressions assume that the feedings of the $4^+_1$ and $0^+_2$
levels contribute negligible additional delay.) The same
$\gamma$-ray energy is used for timing in both cases, so the prompt
position is the same for each, and the $2^+_1$ level lifetime
contributes the same delay in both cases, so the difference in
centroids is simply
\begin{equation}
x_{\text{gate}~326} - x_{\text{gate}~412}= \tau_{0^+_2} - \tau_{4^+_1}.
\end{equation}
The $4^+_1$ level lifetime is known to be 6.9(5)\,ps (see
Ref.~\cite{nds1998:154}), leaving the $0^+_2$ lifetime as the only
unknown quantity.  A measurement in which two different delayed
centroids are compared to each other without reference to the prompt
centroid position is termed a ``relative'' measurement.  Such a
measurement has the significant benefit that knowledge of the prompt
centroid position is irrelevant; the prompt centroid calibration can
otherwise be the dominant source of uncertainty.

Whereas the $4^+_1$ level receives considerable direct
$\beta^+/\varepsilon$ feeding, the feeding of the $0^+_2$ level is
essentially all via higher-lying levels ranging in excitation energy
from 905\,keV to 2272\,keV~\cite{zolnowski1980:154dy156er-beta}.
Since $^{154}$Ho ground state decay occurs with a $Q_\varepsilon$
value of 5751(11)\,keV, the $\beta^+$ fraction for feeding of these
levels falls off rapidly with excitation energy~\cite{nds1998:154},
and the average emitted $\beta^+$ particle energy is lower.
Consequently, much less of the $0^+_2$ level feeding is found to be
coincident with a $\beta^+$ particle over 500\,keV in energy than
is the case for the $4^+_1$ level, hindering timing with the plastic
scintillation detector $\Delta E$ signal.

\begin{figure}
\begin{center}
\resizebox{\hsize}{!}{
\includegraphics*[height=1in]{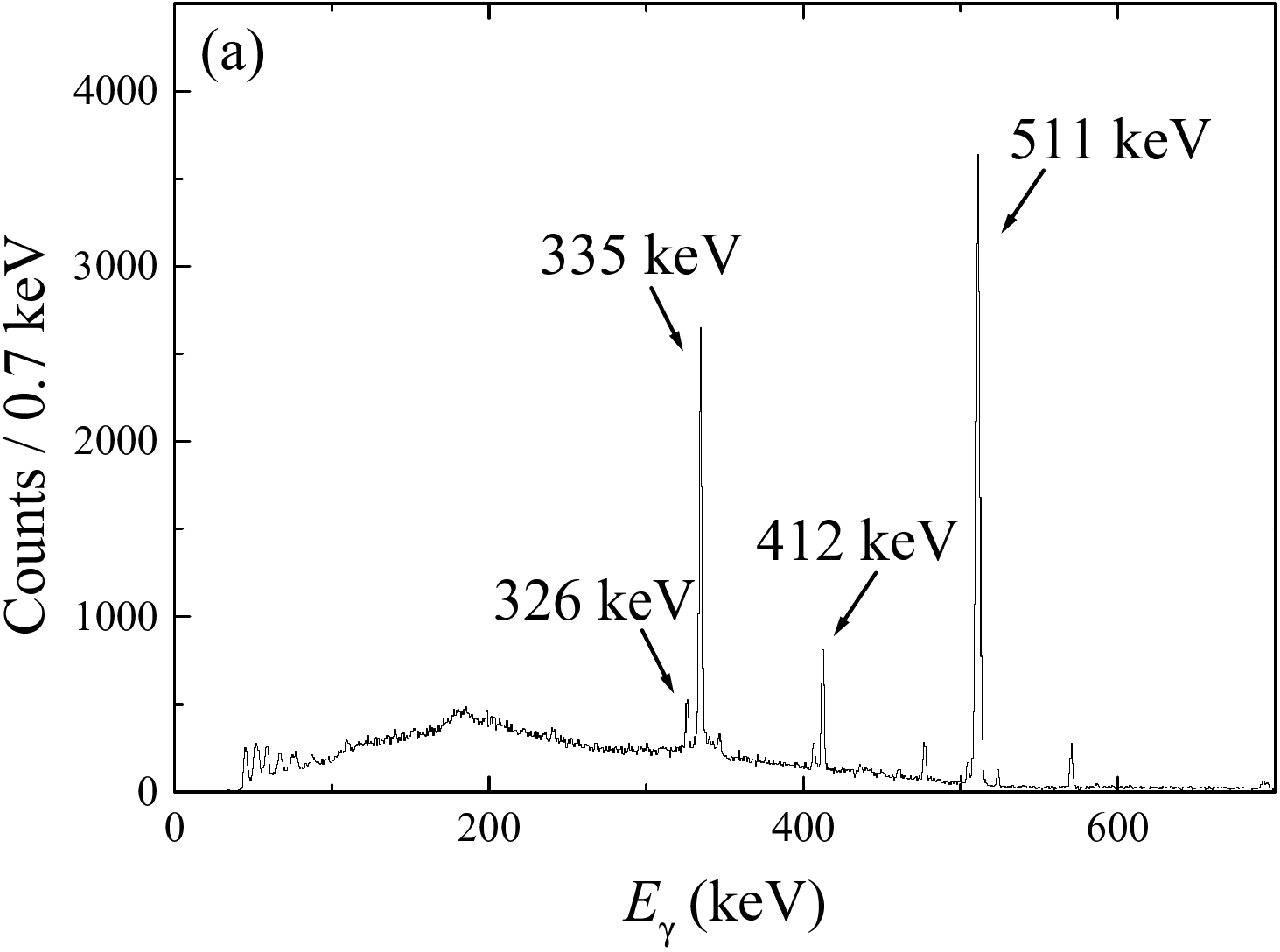}
\hspace{0.05in}
\raisebox{0.03in}{\includegraphics*[height=0.97in]{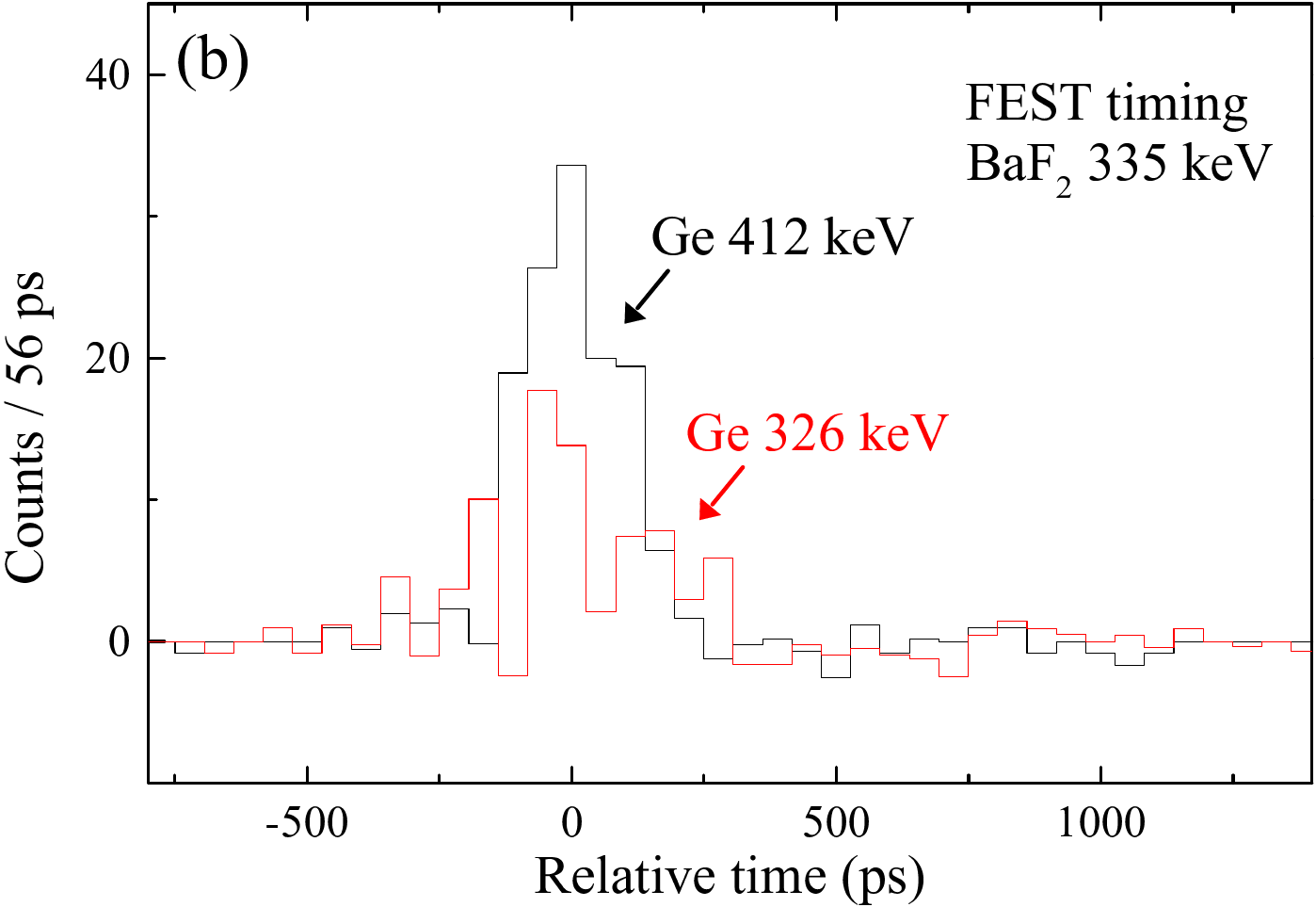}}
}
\end{center}
\caption[FEST data for the $0^+_2$ lifetime measurement in $^{154}$Dy.]
{FEST data for the $0^+_2$ lifetime measurement in $^{154}$Dy. (a) Clover detector energy projection for all clover-BaF$_2$-plastic
triple coincidence events, showing the 326\,keV $0^+_2\rightarrow2^+_1$, 335\,keV
$2^+_1\rightarrow0^+_1$, and 412\,keV $4^+_1\rightarrow2^+_1$
transitions.  (b) FEST timing distributions for the 335\,keV $\gamma$
ray subject to clover detector gates on the 326\,keV and 412\,keV transitions.
\label{fig154dyfest}
}
\end{figure}

The 326, 335, and 412 keV $\gamma$-ray lines are shown in the clover
spectrum of Fig.~\ref{fig154dyfest}(a).
The BaF$_2$ detector timing
distributions for the 335\,keV transition subject to 412\,keV and
326\,keV clover detector gating conditions, with local background
subtractions for the Ge and BaF$_2$ detectors, are shown in
Fig.~\ref{fig154dyfest}(b).  For increased stastics, these were
obtained with no energy cut placed on the plastic scintillation
detector signal.  Therefore, a significant background from $\gamma$ rays Compton
scattered in the plastic scintillator is present, and timing
walk effects over the broad energy range accepted degrade the timing
performance obtained from this detector.

The statistics obtained are insufficient for a meaningful measurement
of the $0^+_2$ level lifetime.  The absence of a visible difference
between the centroids of the time distributions in Fig.~\ref{fig154dyfest}(b)
suggests a $0^+_2$ level lifetime of under $\sim$100\,ps, but the use
of low-energy signals in the plastic scintillation detector likely
introduces systematic errors.  The decay of the $0^+_2$ level proceeds
not only by the 326\,keV $\gamma$-ray transition
[$I^\text{rel}$=100(16)~\cite{zolnowski1980:154dy156er-beta}] and its
associated conversion electrons, but by an $E0$ transition to the
ground state as well
[$I^\text{ce,rel}$=9.5(5)~\cite{zolnowski1980:154dy156er-beta,nds1998:154}].
With this branching information, an upper limit of 100\,ps on the
$0^+_2$ lifetime corresponds to a transition strength
$B(E2;0^+_2\rightarrow2^+_1)$\gt26\,\Wu.  By way of comparison, the
$B(E2;0^+_2\rightarrow2^+_1)$ strengths in the lower-$Z$ $N$=88
isotones $^{148}$Nd, $^{150}$Sm, and $^{152}$Gd have been measured to
be 31.4(22)\,\Wu, 54(5)\,\Wu, and 180(40)\,\Wu, respectively (see
Refs.~\cite{nds2000:148,nds1995:150,nds1996:152}).
It is likely that in a full-length $\sim$100\,h experiment, with more
optimal target and detector array conditions, about a factor of ten
more data could be obtained than in the present experiment, allowing
extraction of a value or useful limit for the $0^+_2$ level lifetime
in $^{154}$Dy.

\chapter{$^{102}$Pd}
\label{chap102pd}

A nucleus with an $R_{4/2}$ value near 2.20 in a known spherical to
$\gamma$-soft transition region is immediately of interest as a
prospective E(5) nucleus.  It must then be seen to what extent the
excited levels follow the energy and transition properties expected
for the E(5) multiplet structure (Section~\ref{sectrans}) and whether or
not an excited $0^+$ level with its associated family of levels can be
found.  

\begin{figure}[t]
\begin{center}
\includegraphics*[width=0.75\hsize]{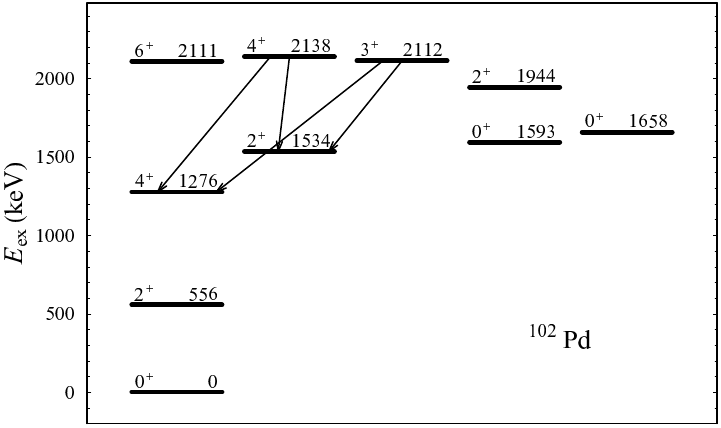}
\end{center}
\caption[Low-lying positive-parity states in $^{102}$Pd.]
{Low-lying positive-parity states in $^{102}$Pd.  Transitions for
which intensity values from the present data are discussed in the text
are indicated by arrows.
\label{fig102pdlevels}
}
\end{figure}

The Pd isotopes constitute an example of such a spherical to
$\gamma$-soft transition region and have been extensively modeled as
transitional nuclei using the
IBM~\cite{stachel1982:ibm,bucurescu1986:ibm,pan1998:so6u5}, the
IBM-2~\cite{vanisacker1980:ibm,kim1996:ibm,giannatiempo1998:ibm}, and
other models~\cite{vorov1985:quartic}.  The nucleus $^{102}$Pd lies in
the correct position along the isotopic chain to be a promising
candidate for E(5) structure.  This nucleus has an $R_{4/2}$ value of
2.29, and the levels form an approximate O(5) multiplet structure
(Section~\ref{secbenchgsoft}), with levels at the correct energies to
constitute the $4^+$-$2^+$ members of a $\tau$=2 multiplet and the
$6^+$-$4^+$-$3^+$ members of a $\tau$=3 multiplet
(Fig.~\ref{fig102pdlevels}).  Two $0^+$ levels are present at energies
close to that expected for the head of the first excited ($\xi$=2)
family in the E(5) description.  However, difficulties are encountered
with such an interpretation of the structure of $^{102}$Pd.  Several
of the transitions predicted by E(5) to be enhanced transitions were
missing from the known decay scheme~\cite{nds1998:102}, suggesting the
need for a spectroscopy experiment either to identify these
transitions or to place limits on their existence.

\begin{figure}[t]
\begin{center}
\includegraphics*[width=0.78\hsize]{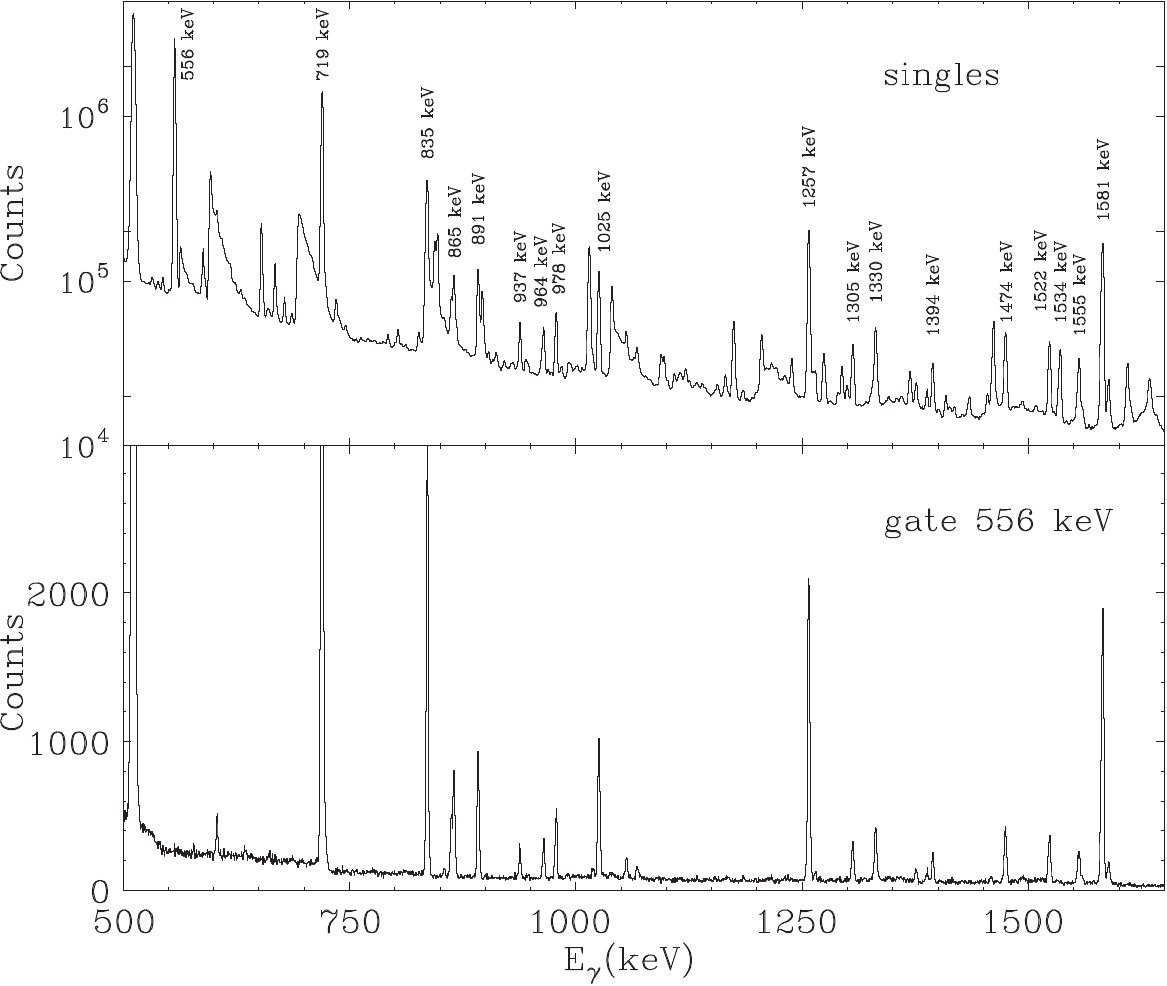}
\end{center}
\caption[Gamma-ray spectra from the Yale MTC $^{102}$Pd experiment.]
{Gamma-ray singles spectrum (top) and coincidence spectrum
gated on the 556\,keV $2^+_{556}\rightarrow0^+_{0}$ transition (bottom)
from the Yale MTC $^{102}$Pd experiment.  (Figure adapted from 
Ref.~\cite{zamfir2002:102pd-beta}.)
\label{fig102pdspectra}
}
\end{figure}

An experiment was carried out in $\beta^+$/$\varepsilon$ decay at the
Yale MTC in its standard spectroscopy configuration
(Section~\ref{secmtc}), consisting of three Compton-suppressed clover
detectors and a LEPS detector.  Parent $^{102}$Ag nuclei were produced
through the reaction $^{89}$Y($^{16}$O,$3n$)$^{102}$Ag at a beam
energy of 75\,MeV, using an $\sim$20\,pnA beam incident upon a
$\sim$5\,mg/cm$^2$ $^{89}$Y target (monoisotopic element) on a 2\,mg/cm$^2$ Au
backing.  Production in heavy-ion fusion-evaporation preferentially
populates the higher-spin $5^+$ ground state ($T_{1/2}$=12.9\,min) of
Ag rather than the $2^+$ $\beta$-decaying isomer
($T_{1/2}$=7.7\,min)~\cite{nds1998:102}.  The tape was advanced at
20\,min intervals.  Data were acquired in event mode with a Ge singles
(or higher fold) trigger using the YRAST Ball FERA/VME data
acquisition system~\cite{beausang2000:yrastball}.  The experiment
provided \sci{3.6}{8} singles counts and \sci{9.4}{6} coincidence
pairs in $\sim$100\,h (Fig.~\ref{fig102pdspectra}).  The results were reported in
Ref.~\cite{zamfir2002:102pd-beta}.

The following discussion summarizes the spectroscopic information
obtained most relevant to the interpretation of the low-lying
postitive-parity structure of $^{102}$Pd.  

$3^+_{2112}$: The $3^+_{2112}$ state was previously reported (see
Ref.~\cite{nds1998:102}) only to decay by a 1555\,keV transition to
the $2^+_{556}$ state.  In the context of the E(5) description, or in
any picture involving O(5) symmetry, this is the \textit{forbidden}
($\Delta\tau$=-2) transition.  In the present data, an
836\,keV $3^+_{2112}\rightarrow4^+_{1276}$ transition and a 577\,keV
$3^+_{2112}\rightarrow2^+_{1534}$ transition are observed.  These
transitions show a substantial $B(E2)$ enhancement relative to the
$3^+_{2112}\rightarrow2^+_{556}$ transition, as expected for
$\Delta\tau$=-1 transitions, provided $E2$ character is assumed for
these transitions (see Table~\ref{tab102pdbranch}).
\begin{table}
\begin{center}
\begin{tabular}{l.,c,.l_c_r.l_r.l_r.l_c_r.l}
\pseudoruledtabular
\multicolumn{3}{c}{}&
&
\multicolumn{6}{c}{Experiment}&
&
\multicolumn{2}{c}{E(5)}\\
\cline{5-10}\cline{12-13}
\multicolumn{3}{c}{Transition} &
&
\multicolumn{2}{c}{$E_\gamma$ (keV)} & 
\multicolumn{2}{c}{$I$} & 
\multicolumn{2}{c}{$B(E2)^\text{rel}$} & 
&
\multicolumn{2}{c}{$B(E2)^\text{rel}$} \\ 
\colrule
$2^+_{1534}$ & $\rightarrow$ & $0^+_{0}$    & & 1534&.3(1) & 1&.9(2)   &  12&.6(14)       & & \multi{}\\
             &               & $2^+_{556}$  & & 977&.75(5) & 1&.81(16) & 100&(9)$^a$      & & 100&\\ 
\colrule
$3^+_{2112}$ & $\rightarrow$ & $2^+_{556}$  & & 1555&.1(1) & 1&.44(15) & 6&.0(6)$^{b,c}$  & & \multi{}      \\
             &               & $4^+_{1276}$ & &  836&.0(5) & 0&.15(8)  & 14&(7)$^b$       & &  40&  \\
             &               & $2^+_{1534}$ & &  577&.1(1) & 0&.17(3)   & 100&(18)$^b$    & & 100& \\
\colrule
$4^+_{2138}$ & $\rightarrow$ & $2^+_{556}$  & & 1581&.1(1) & 12&.9(8) & 6&.5(4)           & & \multi{}      \\
             &               & $4^+_{1276}$ & &  861&.9(1) & 1&.59(14)  & 16&.6(15)$^b$   & &  90&  \\
             &               & $2^+_{1534}$ & &  603&.32(6)& 1&.61(14)   & 100&(9)       & & 100&\\
\pseudoruledtabular
\end{tabular}
\end{center}
\begin{tablenotes}
$^a$ Calculated using $\delta$=2.8(2) from low-temperature nuclear
orientation~\cite{wouters1987:102pd-beta-ltno}.\\ 
$^b$ Calculated assuming pure $E2$ character.\\ 
$^c$ The reported $\delta$ is 0.24(16) or \gt15 from the $\gamma$-ray singles angular distribution in
(HI,$xn$)~\cite{grau1976:102pd104pd106pd-hixn}, and $p\gamma$ angular
correlations in $(p,p')$ scattering support the
latter~\cite{lange1977:102pd-ppprime}.
\end{tablenotes}
\caption[Intensities of
transitions from the $2^+_{1534}$, $3^+_{2112}$ and $4^+_{2138}$ states in
$^{102}$Pd.]{\ssp Intensities and
relative $B(E2)$ strengths of transitions from the $3^+_{2112}$ and
$4^+_{2138}$ states in $^{102}$Pd, compared with the E(5) multiplet
member predictions.  All intensities are from the present
$\beta^+$/$\varepsilon$ decay data, normalized to 100 for the 556\,keV
$2^+_{556}\rightarrow0^+_{0}$ transition.
\label{tab102pdbranch}
}
\end{table}

$4^+_{2138}$: The 1581\,keV $4^+_{2138}\rightarrow2^+_{556}$ transition
was previously known, and the 603\,keV
$4^+_{2138}\rightarrow2^+_{1534}$ had been reported as well, with a
relatively large ($\sim$33$\%$) uncertainty on its intensity (see
Ref.~\cite{nds1998:102}).  Other transitions had been proposed from
the $4^+_{2138}$ level but with uncertain
placements~\cite{nds1998:102}.  The present coincidence data show the
existence of an 862\,keV $4^+_{2138}\rightarrow4^+_{1276}$ transition
and provide a much more precise determination of the 603\,keV
transition intensity (Table~\ref{tab102pdbranch}).  The intensity data
combined with the $B(E2;4^+_{2138}\rightarrow2^+_{556})$ value of
3.0(6)\,\Wu from Coulomb
excitation~\cite{luontama1986:102pd104pd-p2n-ppprime-coulex} yield
$B(E2;4^+_{2138}\rightarrow2^+_{1534})$=46(11)\,\Wu, in good agreement
with the E(5) prediction of 51(4)\,\Wu [based upon a
$B(E2;2^+_{556}\rightarrow0^+_{0})$ strength of
32.6(23)\,\Wu~\cite{nds1998:102}].  The
$4^+_{2138}\rightarrow4^+_{1276}$ transition is somewhat weaker than
expected (Table~\ref{tab102pdbranch}).  Again, the transitions which
are characterized as $\Delta\tau$=-1 transitions in the E(5) picture
show enhanced $B(E2)$ strength over the forbidden $\Delta\tau$=-2
$4^+_{2138}\rightarrow2^+_{556}$ transition. 

\begin{figure}
\begin{center}
\includegraphics*[width=0.75\hsize]{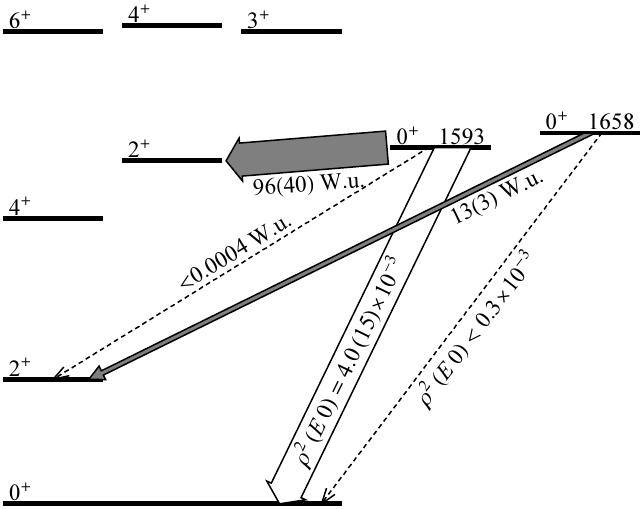}
\end{center}
\caption[Status of $E2$ and $E0$ strengths from low-lying $0^+$ excitations in
$^{102}$Pd.]
{Status of $E2$ and $E0$ strengths from low-lying $0^+$ excitations in
$^{102}$Pd, based upon data from
Ref.~\cite{luontama1986:102pd104pd-p2n-ppprime-coulex} (see text).
\label{fig102pd0plus}
}
\end{figure}

\begin{figure}[t]
\begin{center}
\includegraphics*[width=0.48\hsize]{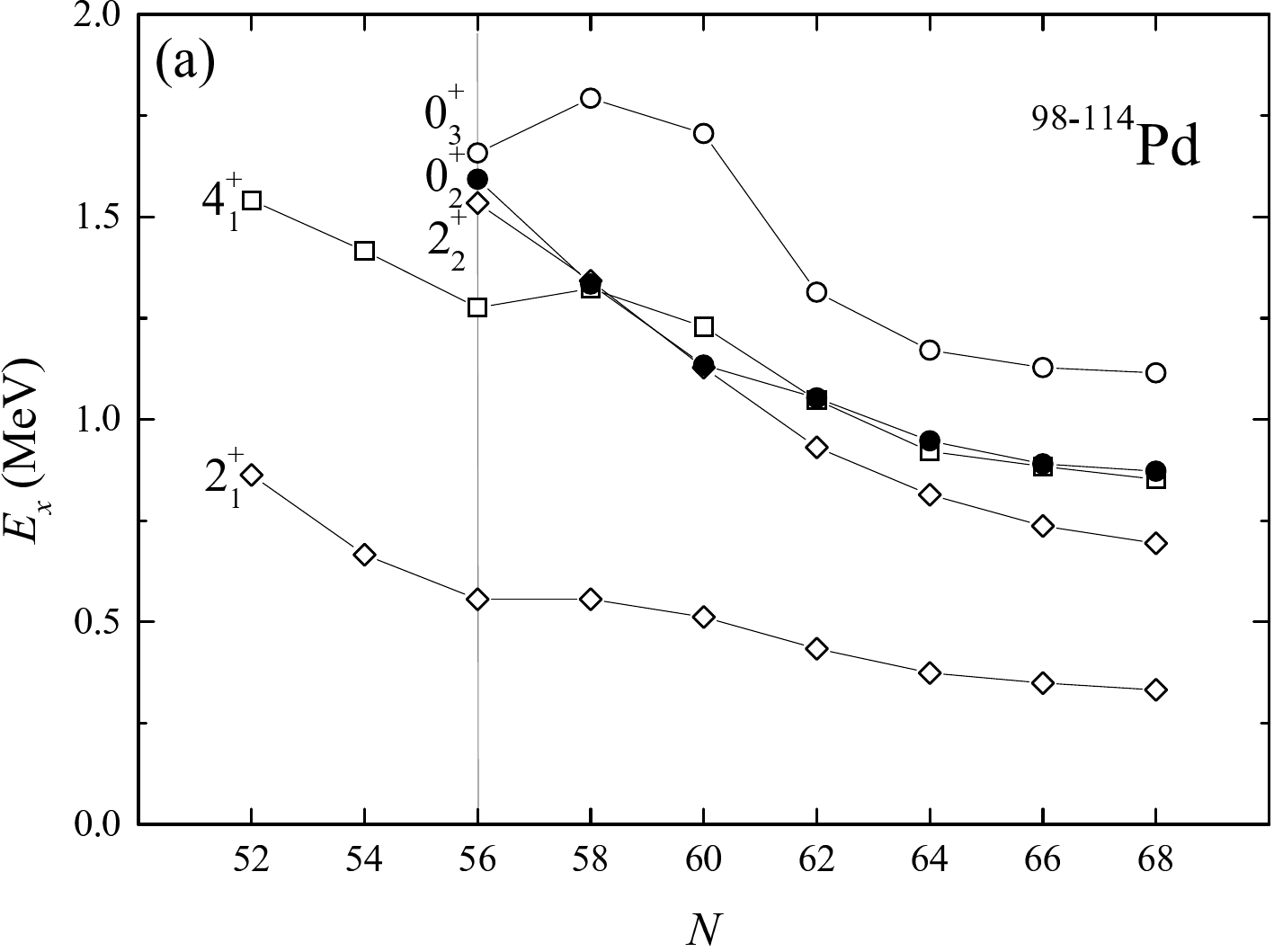}
\hfill
\includegraphics*[width=0.48\hsize]{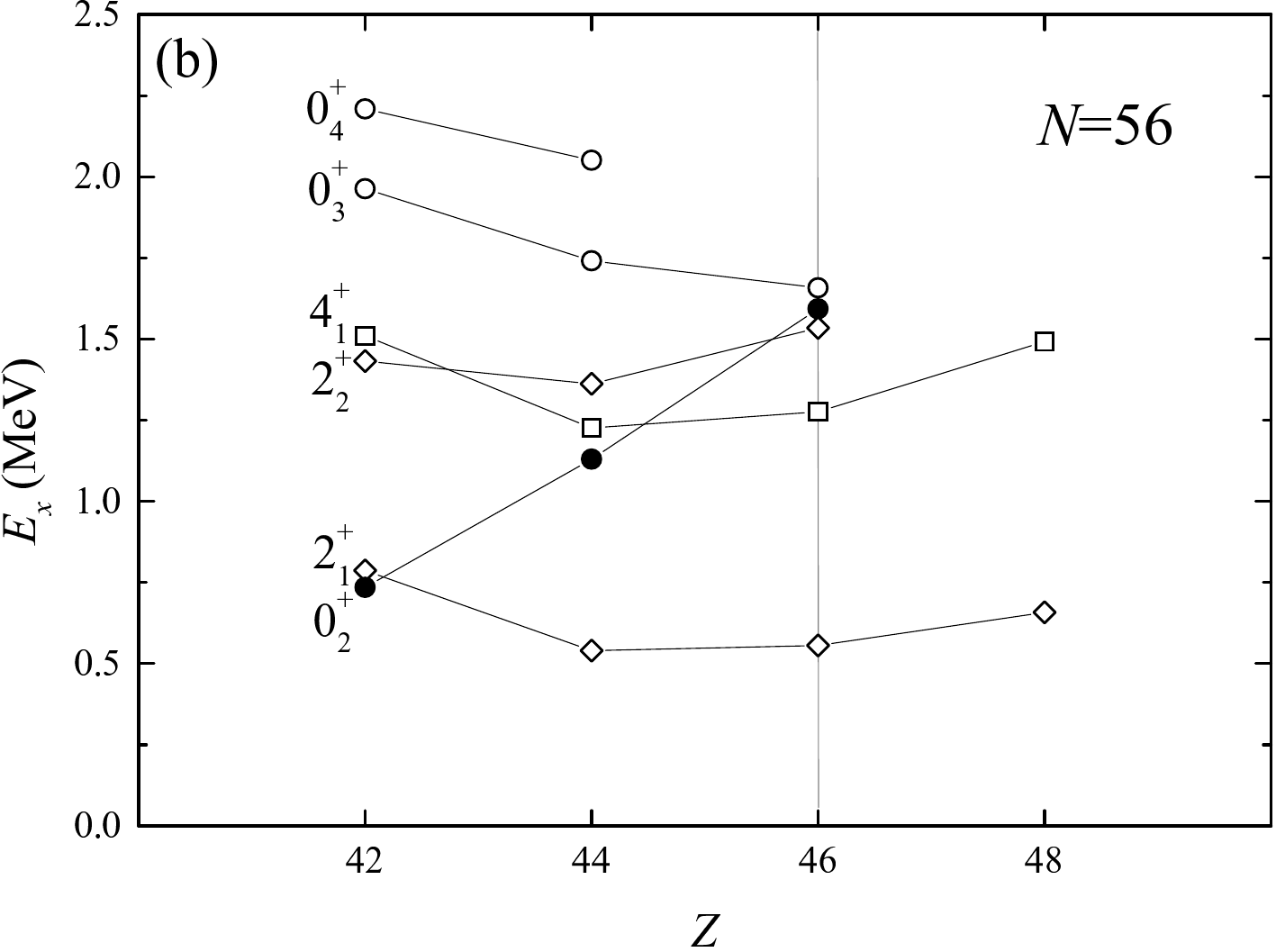}
\end{center}
\caption[Evolution of level energies along the Pd
and $N$=56 chains.]  {Evolution of level energies for low-lying states
along (a) the Pd ($Z$=46) isotopic chain and (b) $N$=56 isotonic
chain.  The contrary behavior of the $0^+_2$ state (filled symbols)
in part (b), with a decreasing energy towards the $Z$=40 shell
closure, indicates a cross-shell intruder nature.
\label{fig102pdsyst}
}
\end{figure}

The nucleus $^{102}$Pd provides an example of an important
consideration relating to the interpretation of excited states,
especially low-lying $0^+$ states, namely the possibility of intruder
character.  The low-lying $0^+$ states of $^{102}$Pd are not
significantly populated in $\beta$ decay from the $^{102}$Ag $5^+$ ground
state.  However, these states have been well-studied in several
experiments involving $^{102}$Ag $2^+$
decay~\cite{cornelis1979:102pd-beta,cornelis1981:102pd108cd-beta}, the
$(p,2n)$ transfer
reaction~\cite{luontama1986:102pd104pd-p2n-ppprime-coulex}, $(p,p')$
scattering~\cite{farzin1980:102pd-ppprime,farzin1987:102pd-ppprime,luontama1986:102pd104pd-p2n-ppprime-coulex},
and Coulomb
excitation~\cite{luontama1986:102pd104pd-p2n-ppprime-coulex}.  The
reported $E2$ and $E0$ decay properties of these states are summarized
in Fig.~\ref{fig102pd0plus}.
The lowest excited $0^+$ state ($0^+_{1593}$) is an isomer, with a
lifetime of 21(6)\,ns~\cite{nds1998:102}.  It decays primarily by $E0$
decay to the ground state.  An exceedingly stringent limit has been
deduced on any $E2$ strength to the $2^+_{556}$ level, and a
low-energy transition to the $2^+_{1534}$ level has been
reported~\cite{luontama1986:102pd104pd-p2n-ppprime-coulex}.  A $0^+$
level at this low energy, with these decay properties, is completely
inexplicable within the E(5) picture.  However, an inspection of the
evolution of level energies for the low-lying levels
(Fig.~\ref{fig102pdsyst}) along the isotopic and isotonic chains
containing $^{102}$Pd indicates an anomalous behavior for the first
excited $0^+$ state along the $N$=56 isotonic chain.  The energy
\textit{decreases} towards the $Z$=40 shell closure, indicating that
this is not an ordinary collective state constructed from the
$Z$=40--50 valence space.  The evolution is consistent with that of a
collective excitation involving the entire $Z$=28--50 shell, breaking
the subshell closure at $Z$=40.  Collective intruder $0^+$ states have
previously been identified in the higher-$N$ Pd
isotopes~\cite{wood1992:coexistence}.  
The next excited $0^+$ level
($0^+_{1658}$) lies at an energy [$E(0^+_{1658})$=2.98$E(2^+_{556})$]
in near perfect agreement with the E(5) prediction
[$E(0^+_2)$=3.03$E(2^+_1)$].  This level decays with collective $E2$
strength to the $2^+_{556}$ level
$B(E2;0^+_{1658}\rightarrow2^+_{556})$=13.3\,\Wu, as expected for the
head of the $\xi$=2 family, though this strength is lower than
predicted [28(2)\,\Wu] in E(5).

The E(5) model reproduces many of the characteristics of the low-lying
levels of $^{102}$Pd.  The main deviations from the E(5) picture
include the splitting between the $4^+$ and $2^+$ members of the
$\tau$=2 multiplet and several quantitative differences between
predicted and observed $B(E2)$ strengths, which are enumerated more
fully in Section~\ref{secphenomgsoft}.  The divergences from the E(5)
model predictions are generally consistent with a slight preference
for axial symmetry, signaling the incipient formation of a $\gamma$
band.  This possibility is addressed quantitatively in
Section~\ref{secphenomgsoft}.

\chapter{$^{162}$Er}
\label{chap162er}

\section{Experimental motivation}

The nucleus $^{162}$Er, with $N$=94, is the lowest neutron number
member of the Er isotopic chain for which the ground state band
observables indicate a well-deformed rotational structure, with a
$2^+_g$ level energy of 102\,keV, an $E(4^+_g)/E(2^+_g)$ energy ratio
of 3.23, and a $B(E2;2^+_g\rightarrow0^+_g)$ strength of
191(1)\,\Wu~\cite{nds1999:162}.  This nucleus is therefore of interest
as a bridge between the transitional nuclei and true rotor nuclei
(Section~\ref{sectrans}).  The $\beta$-vibrational mode (see
Section~\ref{secbenchrotor}) has been widely accepted as a benchmark
for the interpretation of the lowest $K^\pi=0^+$ excitation in
rotational nuclei.  However, difficulties arise in the categorical
interpretation of the lowest $K^\pi=0^+$ excitation as a $\beta$
vibrational or as any form of collective excitation at
all~\cite{garrett2001:0plus,casten1994:robust,casten1994:newinterpretation}.
The conclusive identification of examples of $\beta$-vibrational
excitations in rotational nuclei~--- especially as the lowest-lying
$K=0^+$ excitation~--- has been elusive.  Substantial evidence has
been obtained in only a few
cases~\cite{garrett2001:0plus,aprahamian2002:178hf-grid}, including
for a high-lying $\beta$-vibrational excitation in the higher-$N$ Er
isotope $^{166}$Er~\cite{garrett1997:166erbeta-nnprime}.  
        
The nucleus $^{162}$Er is of special interest since the $2^+$ state at
1171\,keV, assigned~\cite{deboer1974:162er-beta} to the $K^\pi=0^+_2$ band,
is reported~\cite{ronningen1982:coulex} to decay to the ground state
band with a collective $B(E2)$ strength corresponding to a squared
intrinsic matrix element
\mbox{$|\langle K^\pi=0^+_2|\mathfrak{M}'(E2)|K^\pi=0^+_g \rangle|^2$}=8.0(13)\,\Wu.  This 
is comparable to the largest such values for any $K^\pi=0^+$ excitation in the
rare earth region~\cite{garrett2001:0plus} and fully half as large
as for the $\gamma$-vibrational
excitation in this nucleus \mbox{$|\langle K^\pi=2^+_\gamma|E2|K^\pi=0^+_g
\rangle|^2$}=15.5(13)\,\Wu~\cite{ronningen1982:coulex}.  However, the reported relative $B(E2)$ 
strengths~\cite{deboer1974:162er-beta} of the transitions depopulating this
level deviate from the Alaga rules by nearly an order of magnitude and
can be reconciled with the Alaga rules through mixing of the ground
and $K^\pi=0^+_2$ bands only by invoking interaction strengths
($\sim$50\,keV) for the $2^+$ state an order of magnitude stronger than
are typical for rotation-vibration interactions in the rare earth
region.  The decay properties of this level thus precluded any simple
interpretaton of the $2^+$ level and consequently of the $K=0^+_2$
excitation.  

A $\gamma$-ray spectroscopy experiment was carried out under the
TRIUMF E801 proposal~\cite{castenUNP:e801-proposal}, with the original
intent of looking for significant strengths for transitions from the
$K^\pi=0^+_2$ band members to the $\gamma$ band members, which would
indicate possible two-phonon $\gamma$ vibrational character
(Section~\ref{secbenchrotor}).  Such strong interband strengths
have been identified for excitations in the higher-mass isotopes
$^{166,168}$Er~\cite{garrett1997:166erdblgamma-nnprime,davidson1981:168er-ngamma,boerner1991:168er-ngamma,lehmann1998:168er164dy0plus}.
Although limits on these strengths
were obtained, the most interesting data turned out to be of an
entirely different nature (Section~\ref{sec162ergps}).  Subseqently,
FEST and angular correlation measurements were carried out at the Yale
MTC (Section~\ref{sec162ermtc}).  Interpretation of the $K^\pi=0^+_2$
excitation based upon the present data is discussed in
Section~\ref{sec162erinterp}.

\section{Spectroscopic measurements}
\label{sec162ergps}

The nucleus $^{162}$Er was populated in $\beta^+/\epsilon$ decay and
studied through $\gamma$-ray coincidence spectroscopy in one of the
first experiments to be carried out at the ISAC radioactive ion beam
facility.  The results of this
experiment were reported in Ref.~\cite{caprio2002:162er-beta}.

A beam of $^{162}$Yb parent nuclei was produced in the ISAC ion
source, through spallation of a Ta production target by a 500\,MeV
proton beam from the TRIUMF Cyclotron, and transported to the GPS tape
collector end station.  Details of the experimental configuration are
given in Section~\ref{secgps}.  The beam was of very high purity, and
no contaminants were present at observable levels.  Beam intensities
for $^{162}$Yb of $\sim$$10^9$/s were available from the ion source,
but, as discussed in Section~\ref{secgps}, the maximum beam deposition
rate which could be accommodated was $\sim$$10^5$/s due to
detector limitations.
  
\begin{figure}
\begin{center}
\includegraphics*[width=0.75\hsize]{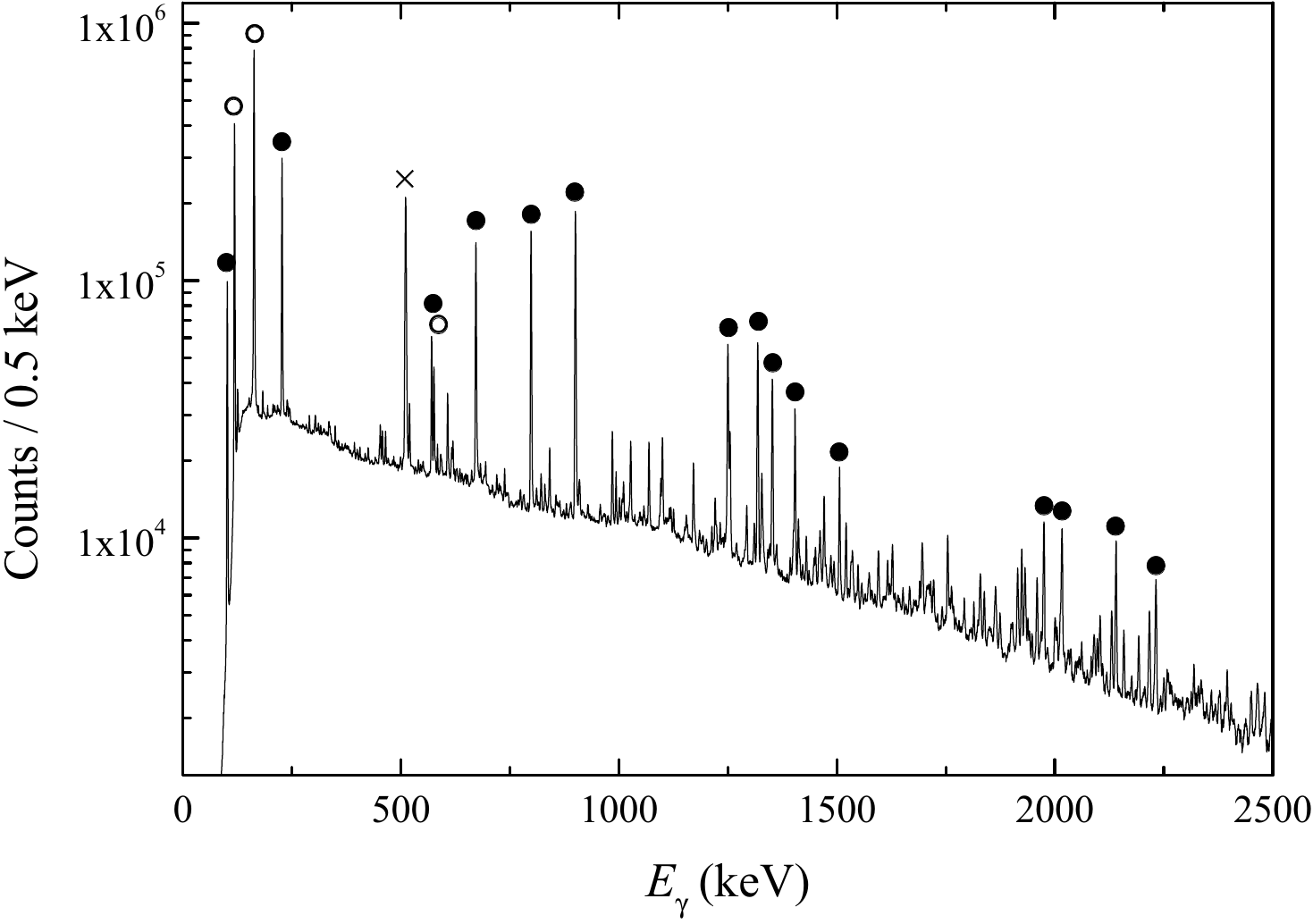}
\end{center}
\caption[Ge detector singles spectrum for $^{162}$Er.]
{ Ge detector singles spectrum for $^{162}$Er.  Intense transitions
from $^{162}$Tm (open circles) and $^{162}$Er (filled circles) and
511\,keV annihilation radiation (cross) are marked.  (Figure
from Ref.~\cite{caprio2002:162er-beta}.)
\label{fig162ersingles}
}
\end{figure}

The nucleus $^{162}$Yb decays with an 18.9\,m half life through
$\beta^+/\epsilon$ decay to $^{162}$Tm, the ground state of which in
turn decays with a 21.7\,m half life to $^{162}$Er~\cite{nds1999:162}.
The GPS tape was therefore advanced, carrying the deposited activity
to the detector area, at approximately 1\,h intervals.  Data were
taken with a Ge singles trigger for 16\,h, yielding $1.5\times10^8$
events, and with a doubles trigger for 120\,h, yielding $6\times10^7$
coincidence events.  A Ge detector singles spectrum from the
experiment is shown in Fig.~\ref{fig162ersingles}.  Transitions from
both members of the decay chain~--- $^{162}$Tm and $^{162}$Er~--- are
clearly visible, with essentially no other background.

\begin{figure}[t]
\begin{center}
\includegraphics*[width=0.6\hsize]{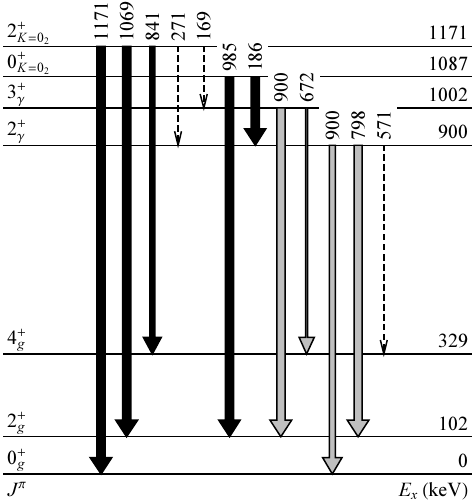}
\end{center}
\caption[Level scheme for the low-spin band members in $^{162}$Er.]{ Level scheme for the low-spin
members of the ground, $\gamma$, and $K^\pi=0^+_2$ bands in $^{162}$Er
with transition intensities obtained from the present work.
Transitions from the $K^\pi=0^+_2$ band members (black arrows) and
from the $\gamma$ band members (gray arrows) are shown.  Arrow widths
are proportional to $\gamma$-ray intensity, normalized to the
strongest transition from each level. Unobserved transitions for which
intensity limits were obtained are indicated by dashed arrows.  (Figure
adapted from Ref.~\cite{caprio2002:162er-beta}.)
\label{fig162erdecay}
}
\end{figure}

Two members of the $K^\pi=0^+_2$ band are populated in $\beta$ decay:
the $0^+$ level at 1087\,keV and the $2^+$ level at 1171\,keV.  The
level scheme for the low-spin members of the $\gamma$ and
$K^\pi=0^+_2$ bands, including these levels, is shown in
Fig.~\ref{fig162erdecay}.  The coincidence data yield
intensities for the transitions depopulating these levels, as well as
limits for the intensities of transitions which are not observed,
summarized in Table~\ref{tab162erbranchbeta}.  
\begin{table}[t]
\begin{center}
\begin{tabular}{l.,c.,l_c_r.l_r.l_r.l_c_r.l}
\pseudoruledtabular
\multicolumn{3}{c}{}&
&
\multicolumn{6}{c}{Experiment}&
&
\multicolumn{2}{c}{Alaga}\\
\cline{5-10}\cline{12-13}
\multicolumn{3}{c}{Transition} &
&
\multicolumn{2}{c}{$E_\gamma$} & 
\multicolumn{2}{c}{$I^\text{rel}$} & 
\multicolumn{2}{c}{$B(E2)$} & 
&
\multicolumn{2}{c}{$B(E2)$} \\ 
\multicolumn{3}{c}{} &
&
\multicolumn{2}{c}{(keV)} & 
\multicolumn{2}{c}{} & 
\multicolumn{2}{c}{(W.u.)} & 
&
\multicolumn{2}{c}{(W.u.)} \\ 
\colrule
$0^+_{K=0_2}$ & $\rightarrow$ & $2^+_g$ &      & 985&.2(2) & 100&(5)      \\
             &                & $2^+_\gamma$ & & [186&]      & \lt3&.4    \\
\colrule                                         
$2^+_{K=0_2}$ & $\rightarrow$ & $0^+_g$   &    & 1171&.05(15)  & 100&(5)  & $\equiv$1&.6(3)$^a$    & & $\equiv$1&.6$^a$\\
             &                & $2^+_g$ &      & 1069&.05(15)   & 100&(5)   & 2&.5(5)$^{b,c}$        & & 2&.3\\
             &                & $4^+_g$ &      & 841&.37(18)    & 59&(3)    & 4&.9(10)$^b$           & & 4&.1\\
             &                & $2^+_\gamma$ & & [271&]    & \lt 1&.5  & \lt 37&    \\       
             &                & $3^+_\gamma$ & & [169&]    & \lt2&.6   & \multi{\lt6.6$\times10^2$}\\
\pseudoruledtabular
\end{tabular}
\end{center}
\begin{tablenotes}
$^a$Normalized to the literature $B(E2;0^+_g\rightarrow2^+_{K=0_2})$
value of 8.0(13)\,\Wu~\cite{ronningen1982:coulex} (see text).\\
$^b$The dominant contribution to the uncertainties in deduced $B(E2)$
values is from normalization to the
$B(E2;0^+_g\rightarrow2^+_{K=0_2})$ value of
Ref.~\cite{ronningen1982:coulex}.  Ratios of these $B(E2)$ values,
\eg, for comparison with the Alaga rules or use in the mixing analysis
(see text), may be obtained with much smaller uncertainties directly
from the intensity values in column 3.\\ 
$^c$$B(E2)$ value calculated assuming pure $E2$ multipolarity.
\end{tablenotes}
\caption[Intensities of transitions depopulating the $K^\pi=0^+_2$ band in $^{162}$Er.]
{\ssp Relative intensities of
transitions depopulating members of the $K^\pi=0^+_2$ band in
$^{162}$Er and intensity limits for unobserved transitions.  The
$B(E2)$ strengths obtained by combining these with the
$B(E2;0^+_g\rightarrow2^+_{K=0_2})$ value from Coulomb excitatation
are shown, along with the Alaga rule predictions.
\label{tab162erbranchbeta}
}
\end{table}

\begin{figure}
\begin{center}
\includegraphics*[width=0.5\hsize]{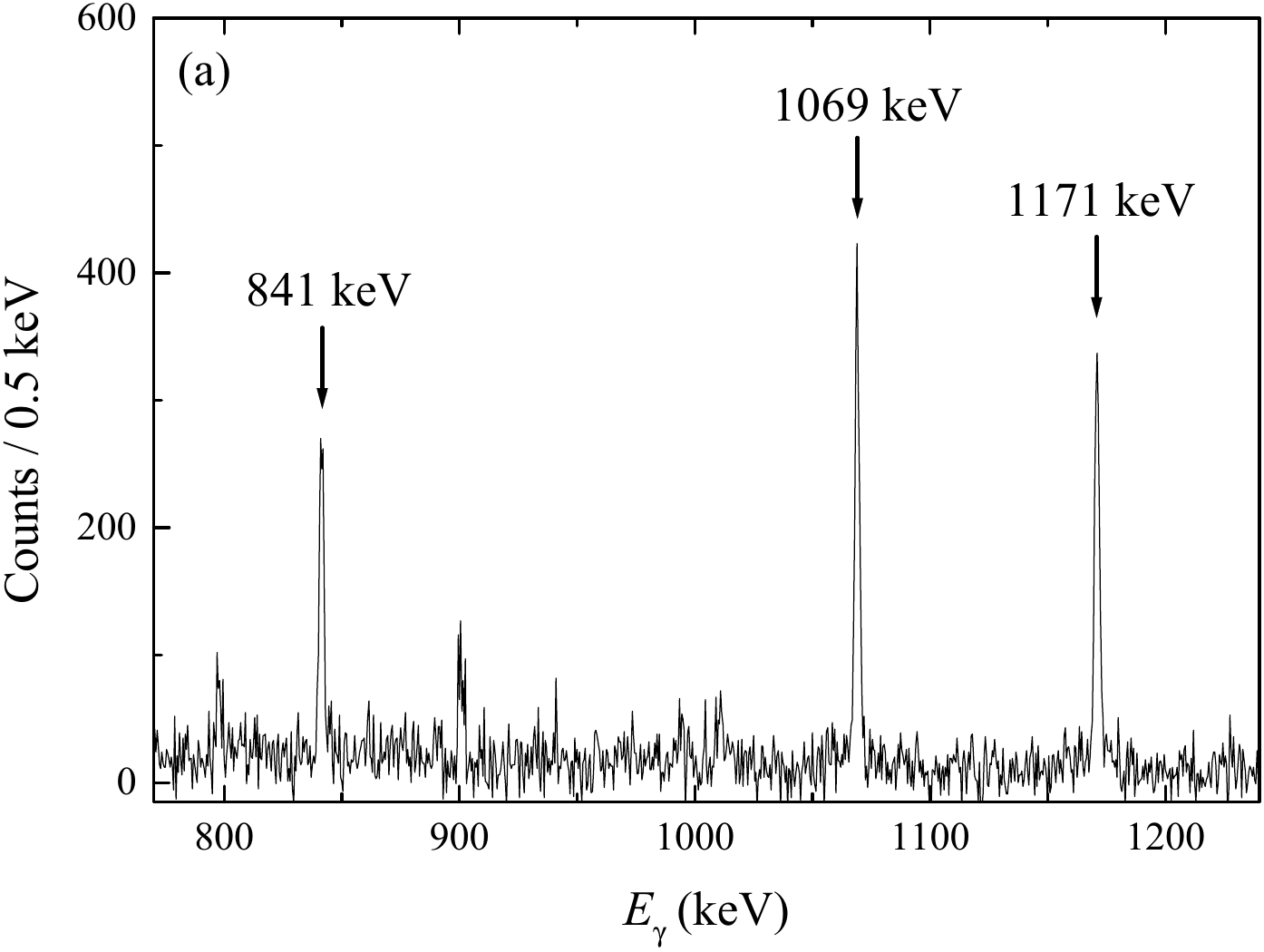}
\hfill
\includegraphics*[width=0.47\hsize]{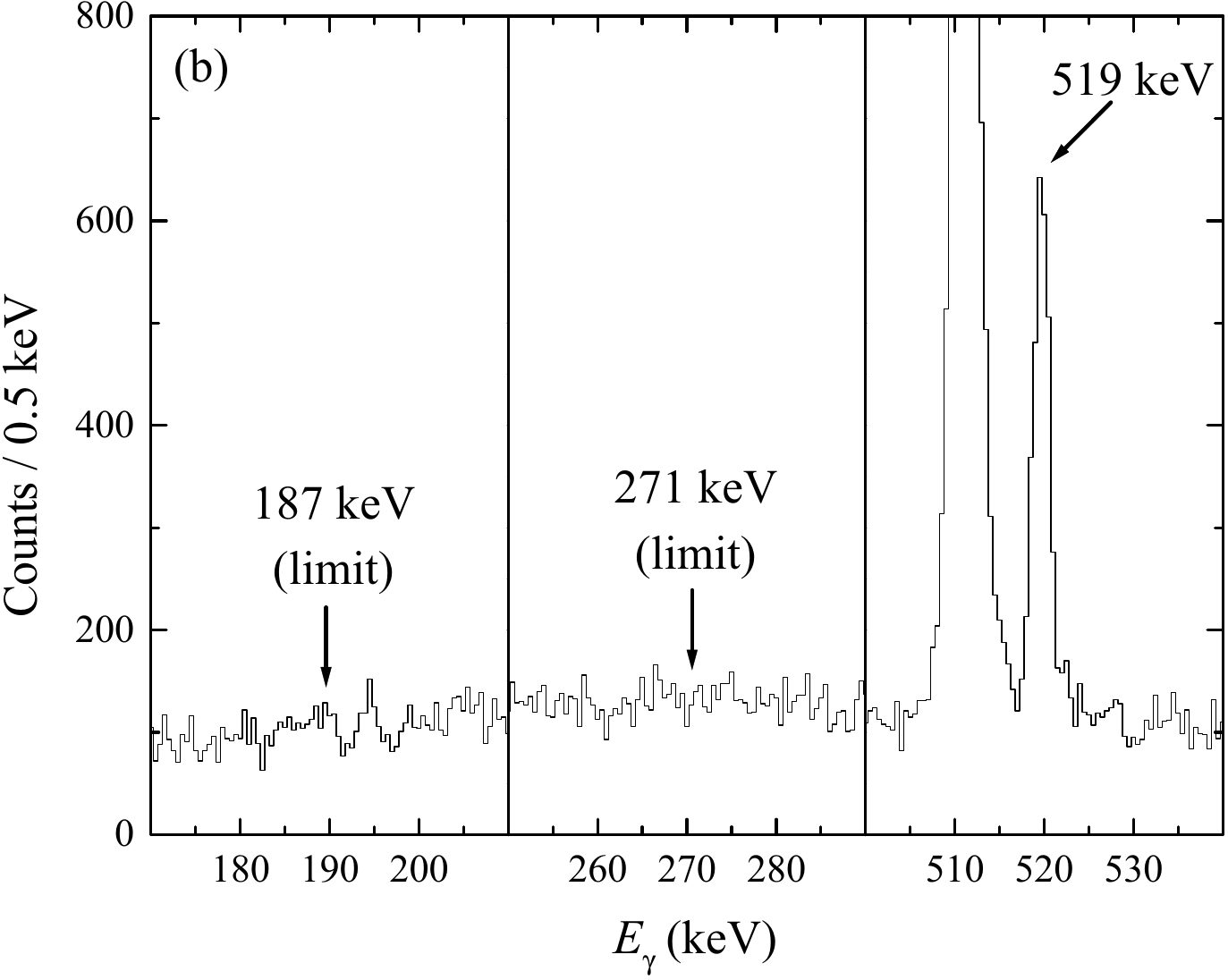}
\end{center}
\caption[Gated spectra showing transitions from $K^\pi=0^+_2$ band
members in $^{162}$Er.]
{Gated spectra showing transitions from $K^\pi=0^+_2$ band members in $^{162}$Er.  (a) Composite spectrum gated on 551,
639, 1224, 1236, and 2096\,keV transitions feeding the $2^+_{K=0_2}$ level at
1171\,keV, showing the 841, 1069, and 1171\,keV branches from this
level.  (b) Spectrum gated on the 798\,keV
$2^+_\gamma\rightarrow2^+_\mathrm{g}$ transition.  Nonobservation of
186 or 271\,keV transitions from $0^+_{K=0_2}$ or $2^+_{K=0_2}$ to
$2^+_\gamma$ in this spectrum allows limits to be placed upon their
intensities.  The observed 519\,keV transition
(see Table~\ref{tab162erbranch1420}) from the misassigned
$(0^+)$ level at 1420\,keV is shown for comparison. (Figure
from Ref.~\cite{caprio2002:162er-beta}.)
\label{fig162ergatesbeta}
}
\end{figure}

Substantially revised $\gamma$-ray branching properties are measured
for the $2^+_{K=0_2}$ level.  The intensity of the 1171\,keV
$2^+_{K=0_2}\rightarrow0^+_g$ transition, relative to the other
branches depopulating the $2^+_{K=0_2}$ level, can be obtained in a
straightforward fashion from spectra gated on transitions feeding this
level~[Fig.~\ref{fig162ergatesbeta}(a)].  The data provide a relative intensity about
7 times greater than previously reported.  (Most of the singles
intensity of the 1171\,keV line was previously
assigned~\cite{deboer1974:162er-beta} to a placement elsewhere in the level
scheme, which the present coincidence data eliminate.)
With this newly measured transition intensity, the relative $B(E2)$
values for the branches from the $2^+_{K=0_2}$ level are now found to
be in reasonably good agreement with the Alaga rules
(Table~\ref{tab162erbranchbeta}), as is discussed further in
Section~\ref{sec162erinterp}.

The other low-lying $K^\pi=0^+$ excitation in $^{162}$Er identified in the
literature is based upon a spin assignment of $(0^+)$ for the level at
1420\,keV.  This excitation, from its reported decay properties, would
have been of interest as a possible two-$\gamma$-phonon excitation
candidate.  

There are several reasons given in Ref.~\cite{deboer1974:162er-beta}
for a $0^+$ assignment for the level at 1420\,keV.  The only observed
$\gamma$ rays were to $2^+$ states~--- $2^+_g$ and $2^+_\gamma$.
Conversion electron data suggested a possible $E2$ character for the
$(0^+)_{1420}\rightarrow2^+_g$ transition
[$\alpha_K$=1.6(7)$\times10^{-3}$], although the stated uncertainty
does not exclude $E1$ character at two standard deviations.  There was
also a possible $E0$ transition from this level to the ground state,
though this transition was at the limit of observation.  Because of
the uncertainties in these arguments,
Ref.~\cite{deboer1974:162er-beta} also suggests $2^-$ as a possible
spin for this level.

\begin{figure}
\begin{center}
\includegraphics*[width=0.65\hsize]{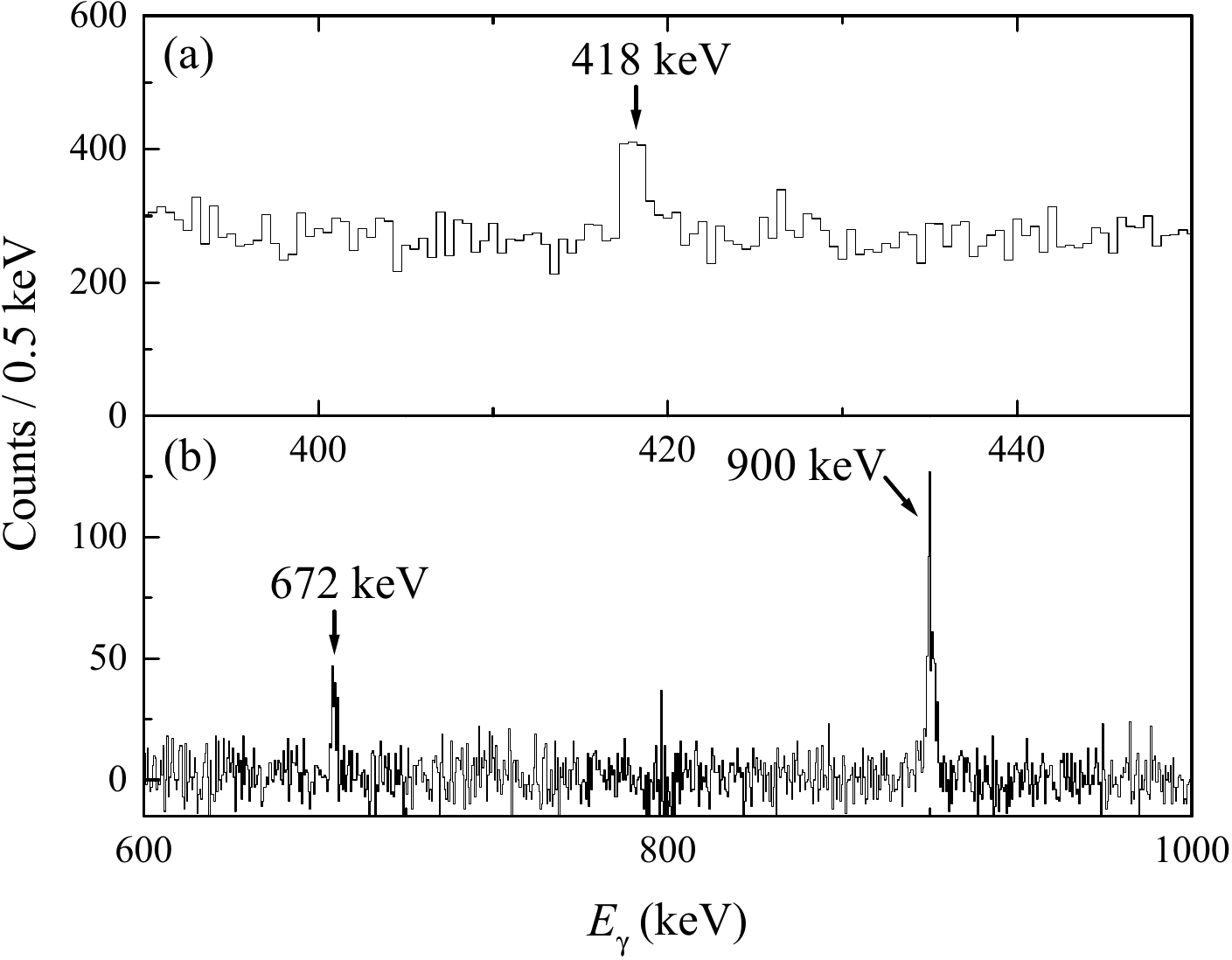}
\end{center}
\caption[Spectra gated on the 900\,keV and 418\,keV transitions in
$^{162}$Er.]
{
Specta gated on the (a)~900\,keV and (b)~418\,keV transitions in $^{162}$Er,
supporting the placement of the 418\,keV transition as directly feeding 
the $3^+_\gamma$ level (see Fig.~\ref{fig162erdecay}).  (Figure
from Ref.~\cite{caprio2002:162er-beta}.)
\label{fig162ergatebelow418}
}
\end{figure}

The present coincidence data show the existence of a weak 418.1(2)\,keV
transition from the level at 1420\,keV to the $3^+$ member of the
$\gamma$ band (Fig.~\ref{fig162ergatebelow418}).  The existence of
such a transition is inconsistent with a $0^+$ assignment for the
level at 1420\,keV.  (Corroborating evidence that this level does not
have spin $0^+$ is obtained from angular correlation results described
in Section~\ref{sec162ermtc}.)  A summary of intensities for
transitions depopulating the level at 1420\,keV is given in
Table~\ref{tab162erbranch1420}.
\begin{table}
\begin{center}
\begin{tabular}{l.,c,.l_r.l_r.l}
\pseudoruledtabular
\multicolumn{3}{c}{Transition} &
\multicolumn{2}{c}{$E_\gamma$ (keV)} & 
\multicolumn{2}{c}{$I_\text{rel}$} 
 \\ 
\colrule
$(2^-)$ & $\rightarrow$ & $0^+_g$ &    [1420&]       &   \lt1&.1\\
             &             & $2^+_g$ &        1318&.42(11)  &    100&(4)\\
             &             & $4^+_g$ &        [1090&]       &   \lt0&.6\\
             &             & $6^+_g$ &        [753&]        &   \lt0&.15\\
             &             & $2^+_\gamma$ &   519&.54(13)   &    11&.9(4)\\
             &             & $3^+_\gamma$ &   418&.1(2)     &    0&.95(13)\\
             &             & $0^+_{K=0_2}$ &    [333&]        &    \lt0&.18\\
             &             & $4^+_\gamma$ &   [292&]        &    \lt0&.8\\
             &             & $2^+_{K=0_2}$ &    [249&]        &    \lt0&.9\\
\pseudoruledtabular
\end{tabular}
\end{center}
\caption[Intensities of transitions depopulating the level
at 1420\,keV in $^{162}$Er.]
{\ssp Relative intensities of transitions depopulating the level
at 1420\,keV in $^{162}$Er, including intensity limits on unobserved transitions. 
\label{tab162erbranch1420}
}
\end{table}

A spin assignment of $2^-$ for the level at 1420\,keV is reasonable in
the context of the systematics of octupole excitations in the deformed
rare earth region, where a $K^\pi=2^-$ band at comparable excitation
energy is known in several of the neighboring
nuclei~\cite{cottle1996:octupole}.  The
$B(E1;2^-_{K=2}\rightarrow3^+_\gamma)/B(E1;2^-_{K=2}\rightarrow2^+_\gamma)$
ratios for the neighboring nuclei cluster in the range 0.4--0.6,
close to the Alaga value of 0.5.  In $^{162}$Er, this ratio has a
somewhat lower value of 0.15(2), as extracted from the intensities in
Table~\ref{tab162erbranch1420}.

\section{Lifetime and angular correlation measurements}
\label{sec162ermtc}

A complementary experiment, also in $\beta^+$/$\varepsilon$ decay, was
carried out at the Yale MTC in its FEST configuration
(Section~\ref{secfest}), consisting of three Compton-suppressed clover
detectors, a LEPS detector, and the plastic and BaF$_2$ fast-timing
detectors.  This experiment allowed measurement of the $2^+_1$ level
lifetime by FEST and also provided angular correlation data sensitive
to $0^+$--$2^+$--$0^+$ cascades.  The lifetime measurement has
been reported in Ref.~\cite{caprio2002:162yb162er-fest}.

Parent $^{162}$Yb nuclei were produced through the reaction
$^{155}$Gd($^{12}$C,5$n$)$^{162}$Yb at a beam energy of 86\,MeV, with
an $\sim$20\,pnA beam incident upon a 5\,mg/cm$^2$
99.8$\%$-isotopically-enriched target.  The tape was advanced at 1\,h
intervals, as in the ISAC GPS experiment. Data were acquired in event
mode with a Ge singles (or higher fold) trigger using the YRAST Ball
FERA/VME data acquisition system~\cite{beausang2000:yrastball},
yielding \sci{1.0}{9} clover singles events and \sci{1.3}{7}
clover-clover coincidences in 110\,h.  The experiment produced
5.3$\times$$10^5$ Ge-plastic-BaF$_2$ triple coincidence events for the
FEST analysis.

\begin{figure}
\begin{center}
\includegraphics[width=0.7\hsize]{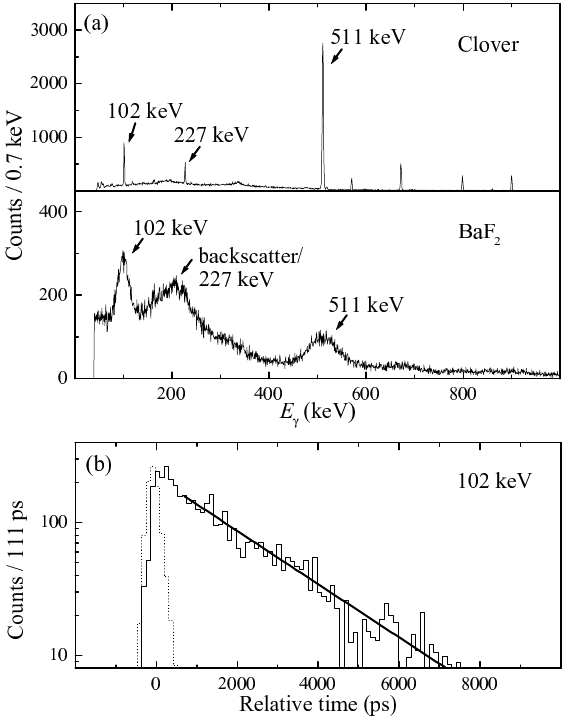}
\end{center}
\caption[Spectra from the $^{162}$Er lifetime measurement.]
{Spectra from the $^{162}$Er lifetime measurement. (a) Clover and
BaF$_2$ detector energy spectra from clover-plastic-BaF$_2$ triples
events, 102\,keV $2^+_g\rightarrow0^+_g$ and 227\,keV
$4^+_g\rightarrow2^+_g$ $\gamma$-ray transitions as well as 511\,keV
annihilation radiation and backscatter (see text).  (b) Measured time
distribution for 102\,keV $\gamma$-ray detection in the BaF$_2$
detector relative to the $\beta^+$ time signal, shown with the fitted
$\tau$=2.2\,ns decay curve (heavy line).  The prompt response curve
(dotted) is shown for comparison (see Chapter~\ref{chap162yb}).
(Figure from Ref.~\cite{caprio2002:162yb162er-fest}.)
\label{fig162erfest}
}
\end{figure}

The lifetime of the $2^+_g$ level in $^{162}$Er is deduced from
$\beta\gamma$ coincidences involving the 102\,keV
$2^+_g\rightarrow0^+_g$ transition by an analysis similar to that
described in Chapter~\ref{chap162yb}.  The decay of $^{162}$Tm
to $^{162}$Er proceeds predominantly by electron capture, with a
$\beta^+$ decay fraction of only $\sim$6$\%$
\cite{strusny1975:160er162er-beta}.  Since in the vast majority of decays only
$\gamma$-rays are present, the detection of true $\beta^+$ $\Delta E$
signals in the plastic scintillation detector competes with a
substantial background of $\gamma$-ray Compton scattering interations
in the plastic scintillator depositing energies in the same energy
range.  Detection of the corresponding Compton-scattered $\gamma$ rays
in the BaF$_2$ detector gives rise to a strong coincident backscatter
peak in the BaF$_2$ energy spectrum [Fig.~\ref{fig162erfest}(a)].
However, the 102\,keV $2^+_1\rightarrow0^+_1$ transition is largely
resolved from the backscatter peak in the BaF$_2$ spectrum
[Fig.~\ref{fig162erfest}(a)], and the time spectrum from the
backscatter events is essentially prompt, with an excess flight time
of only $\sim$100\,ps relative to prompt $\gamma$ rays, so this
background has little influence on the lifetime measurement.  The time
distribution obtained for a 30\,keV wide energy gate on the 102\,keV
transition in the BaF$_2$ detector with a local background subtraction
is shown in Fig.~\ref{fig162erfest}(b).  

The lifetime deduced for the $2^+_g$ level in $^{162}$Er, using
the slope method, is 2.2(4)\,ns.  The larger relative uncertainty
obtained for this lifetime compared to the lifetime in $^{162}$Yb
(Chapter~\ref{chap162yb}) is a result both of lower statistics and
uncertainties in the local background subtraction in the BaF$_2$
detector due to the adjacent backscatter peak.  The present
measurement is compared with prior values in
Table~\ref{tab162erlifetimes}.  Due to its large uncertainty, this
measurement does not resolve the discrepancy between the prior values.
\begin{table}
\begin{center}
\begin{tabular}{r@{}lll}
\pseudoruledtabular
\multicolumn{2}{c}{$\tau$ (ns)} &
\multicolumn{1}{l}{Reference} &
\multicolumn{1}{l}{Method}\\
\hline
 1&.69(14)&      Ref.~\cite{morozov1970:162er-beta} & Electronic timing\\
 1&.96(6)\,$^a$&      Ref.~\cite{ronningen1979:coulex} & Coulomb excitation\\
 2&.2(4)& Present & \\
\pseudoruledtabular
\end{tabular}\\
\end{center}
\begin{tablenotes}
$^a$Deduced from the reported
$B(E2;0^+_g\rightarrow2^+_g)$ value using a total electron
conversion coefficient of 2.76(8)~\cite{nds1999:162}.
\end{tablenotes}
\caption[Values for the lifetime of the first excited $2^+$ state in $^{162}$Er.]
{\ssp Values for the lifetime of the first excited $2^+$ state in
$^{162}$Er as determined by various methods in prior experiments and
in the present work.
\label{tab162erlifetimes}
}
\end{table}

The clover detectors for this experiment were positioned at relative
angles optimized for the identification of spin 0--2--0 cascades, as
described in Section~\ref{secangpol}, separated by angles of 65$^\circ$,
112$^\circ$, and 177$^\circ$.  These are equivalent for angular correlation
analysis to first quadrant angles of 3$^\circ$, 65$^\circ$, and
68$^\circ$.  The clover detectors were located with their faces 13~cm
from the activity, giving an array efficiency of 0.6$\%$ at 1.3\,MeV.

\begin{figure}
\begin{center}
\includegraphics*[width=0.75\hsize]{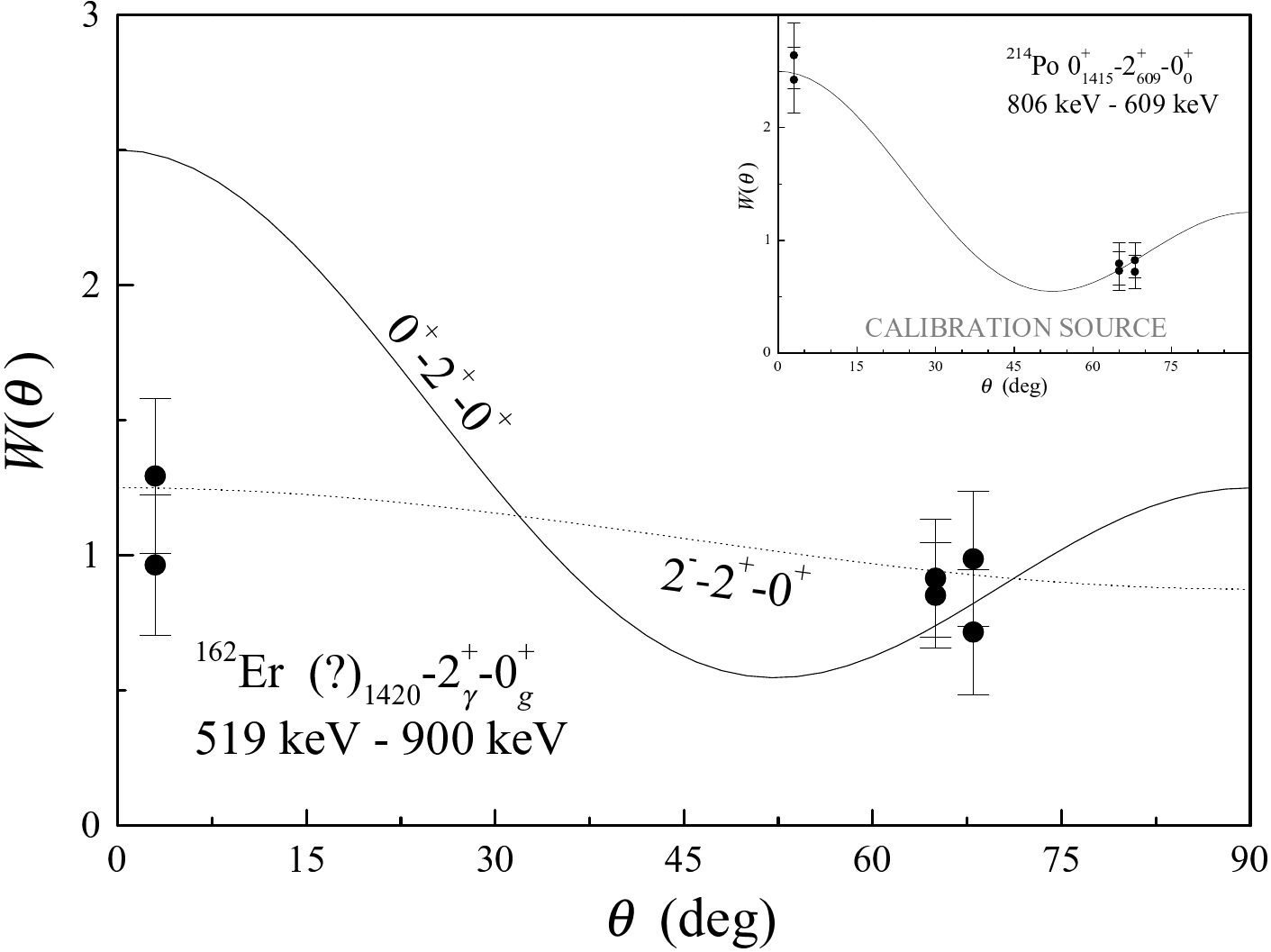}
\end{center}
\caption[Angular correlation data for the
$(?)_{1420}\rightarrow2^+_\gamma\rightarrow0^+_g$ cascade in $^{162}$Er.]
{Angular correlation data for the
$(?)_{1420}\rightarrow2^+_\gamma\rightarrow0^+_g$ cascade in $^{162}$Er
(see text), together with the theoretical curves for the previously
proposed $0^+$ or $2^-$ spin assignments.  The correlation pattern
measured with the same array geometry for the known
$0^+_{1415}\rightarrow2^+_{609}\rightarrow0^+_{0}$ cascade in
$^{214}$Po, involving similar $\gamma$-ray energies, is shown for
comparison (inset).
\label{fig162erangcorr}
}
\end{figure}

Angular correlation data for the 519\,keV and 900\,keV $\gamma$ rays
of the $(?)_{1420}\rightarrow2^+_\gamma\rightarrow0^+_g$ cascade are
shown in Fig.~\ref{fig162erangcorr}.  Ratios of $W(\theta)$ values
were obtained according to the singles internal calibration procedure
of Section~\ref{secangpol}.  [The overall normalization for the
experimental $W(\theta)$ values shown in Fig.~\ref{fig162erangcorr} is
therefore arbitrary and has been adjusted to approximately match the
$0^+$--$2^+$--$0^+$ and $2^-$--$2^+$--$0^+$ theoretical functions at
$\sim$65$^\circ$ for comparison purposes.]  For each clover detector
pair, coincidences involving detection of the 519\,keV $\gamma$ ray in
one detector and the 900\,keV $\gamma$ ray in the other have been
distinguished from the reverse ordering, yielding a total of six data
points each involving $\sim$50 counts.  The angular correlation
pattern is inconsistent with a spin assignment of $0^+$ for the level
at 1420\,keV but consistent with the suggested assignment of $2^-$, as
well as with other assignments (Fig.~\ref{figangcorr}).  Insufficient
statistics are available in the
$(?)_{1420}\rightarrow2^+_g\rightarrow0^+_g$ cascade for useful
analysis in this experiment, but results from perturbed angular
correlations in $\beta$ decay~\cite{berantPC} indicate that the
$(?)_{1420}\rightarrow2^+_g\rightarrow0^+_g$ $\gamma$-ray cascade
similarly does not exhibit the characteristic $0^+$--$2^+$--$0^+$
angular correlation pattern.

\section{Interpretation}
\label{sec162erinterp}

The revised branching data for the $2^+_{K=0_2}$ level qualititively
affect several different aspects of the interpretation of this level.
The intensity values from Table~\ref{tab162erbranchbeta} yield a
\mbox{$B(E2;2^+_{K=0_2}\rightarrow4^+_g)$}/\mbox{$B(E2;2^+_{K=0_2}\rightarrow0^+_g)$}
ratio of 3.1(2).  The deviation of this from the Alaga value of 2.6
can be attributed to relatively minor mixing effects.  In contrast,
the prior data~\cite{deboer1974:162er-beta} corresponded to a $B(E2)$
ratio of 20(15), which would have suggested an order of magnitude
larger mixing and thus would have rendered implausible any
straightforward interpretation of the $2^+$ level as a simple
$K^\pi=0^+$ level.  The $\gamma$-ray branching data also have major
effects on the interpretation of the existing Coulomb excitation and
conversion electron data.

The $B(E2;0^+_g\rightarrow2^+_{K=0_2})$ transition strength is
reported from a Coulomb excitation cross section
measurement~\cite{ronningen1982:coulex} to be 8.0(13)\,\Wu.  The
procedure used in Ref.~\cite{ronningen1982:coulex} for extracting the
$B(E2;0^+_g\rightarrow2^+_{K=0_2})$ value from the Coulomb excitation
cross section was noted, in Ref.~\cite{wood1999:e0}, to be suspect
since it ignores multi-step processes.  However, since the present
branching ratio data show the relative $2^+_{K=0_2}\rightarrow0^+_g$
$\gamma$-ray branch to be much stronger than previously thought,
direct excitation is greatly enhanced over any possible two-step
excitation.  Coulomb excitation yield calculations using the present
intensities and the semiclassical formalism of Alder and
Winther~\cite{alder1956:coulex,alder1975:book} show that excitation via the $2^+_g$
state and reorientation effects can produce at most a 10$\%$ change in
the $B(E2)$ value obtained.

A new $\rho^2(E0;2^+_{K=0_2}\rightarrow 2^+_g)$ value can be deduced
by combining existing conversion electron data with the present
$\gamma$-ray data.    The $K$-conversion intensity for the
$2^+_{K=0_2}\rightarrow2^+_g$ transition in $^{162g}$Tm $\beta$ decay
was obtained from singles conversion electron data in
Ref.~\cite{deboer1974:162er-beta}.  The value in that reference is
given indirectly, as a $\gamma$-ray intensity together with a
$K$-conversion coefficient, but it is straightforward to extract the
result $I^K(2^+_{K=0_2}\rightarrow2^+_g)$=0.43(5), normalized to
$I^\gamma(4^+_g\rightarrow2^+_g)$$\equiv$100.  In these same units,
the present $\gamma$-ray intensity for the
$2^+_{K=0_2}\rightarrow0^+_g$ transition is 17.3(15).  Since the ratio
of the intensities of two transitions depopulating a level is equal to
the ratio of their transition probabilities per unit time,
\begin{equation}
\frac{I^K(E0;2^+_{K=0_2}\rightarrow2^+_g)}{I^\gamma(E2;2^+_{K=0_2}\rightarrow0^+_g)}
=
\frac{T^K(E0;2^+_{K=0_2}\rightarrow2^+_g)}{T^\gamma(E2;2^+_{K=0_2}\rightarrow0^+_g)}.
\end{equation}
Of these quantities, electron and $\gamma$-ray intensities have just been
discussed and the transition probability per unit time for the
$E2$ $\gamma$-ray is directly related to the
known $B(E2;2^+_{K=0_2}\rightarrow0^+_g)$ value by~(\ref{eqntfromb}).  The transition probability per unit time
for $K$-shell $E0$ electron emission is related to $\rho^2(E0)$ by
\begin{equation}
T^K(E0) = \rho^2(E0)\Omega_K,
\end{equation}
where $\Omega_K$ is the $K$-shell electronic factor for the $E0$
transition~\cite{bell1970:e0}.  Thus, the
$\rho^2(E0;2^+_{K=0_2}\rightarrow 2^+_g)$ value may be expressed in
terms of measured quantities as
\begin{equation}
\rho^2(E0;2^+_{K=0_2}\rightarrow 2^+_g) = 
\frac{T^\gamma(E2;2^+_{K=0_2}\rightarrow0^+_g)}{\Omega_K(2^+_{K=0_2}\rightarrow
2^+_g)}
\frac{I^K(E0;2^+_{K=0_2}\rightarrow2^+_g)}{I^\gamma(E2;2^+_{K=0_2}\rightarrow0^+_g)}.
\end{equation}
[It is best to express $\rho^2(E0;2^+_{K=0_2}\rightarrow 2^+_g)$
directly in terms of the measured quantities in this fashion, rather
than first converting the data to a lifetime for the $2^+_{K=0_2}$
level and using this lifetime to deduce
$\rho^2(E0;2^+_{K=0_2}\rightarrow 2^+_g)$, since use of the lifetime
as an intermediate result leads to a greatly exaggererated uncertainty
estimate unless correlated uncertainties are properly taken into
account.]  The experimental $2^+_{K=0_2}\rightarrow2^+_g$ $K$-shell
electron intensity includes $E0$, $M1$, and $E2$ contributions, and so
the $M1$ and $E2$ contributions~\cite{hager1968:convcoeff} must be
subtracted on the basis of the known $2^+_{K=0_2}\rightarrow2^+_g$
$\gamma$-ray intensity in order to recover the $E0$ contribution.  The
$M1/E2$ mixing ratio for this transition is not known, but the value
of the mixing ratio affects the calculated $\rho^2(E0)$ by at most
10$\%$.  Assuming no $M1$ contribution, the extracted $\rho^2(E0)$ is
0.061(14).  In comparison, the previously existing branching data
would have led to a $\rho^2(E0)$ value of 0.5(4), far larger than any
other reported $\rho^2(E0)$ value in the deformed rare earth
nuclei~\cite{wood1999:e0}.

If the $K^\pi=0^+_2$ excitation is assumed to be a $\beta$ vibration,
the rotation vibration model~(RVM)~\cite{faessler1965:rvm} relates the
$E0$ strength to $E2$ strengths as~\cite{wood1999:e0}
\begin{equation}
\rho^2(E0;\beta\rightarrow g) = \frac{4
B(E2;0^+_g\rightarrow2^+_\beta)\beta_0^2}{e^2r_0^4A^{4/3}},
\end{equation}
where the equilibrium quadrupole deformation $\beta_0$ can be
extracted from the ground state intraband
$B(E2;2^+_g\rightarrow0^+_g)$ strength~\cite{bohr1998:v2}.  The RVM
value of 0.077(12) for $\rho^2(E0;\beta\rightarrow g)$ is in excellent
agreement with the present experimental value of 0.061(14).

It is useful to assess the extent to which \mbox{$\Delta K=0$} mixing
between the $K^\pi=0^+_2$ and ground state bands affects the value
deduced for the intrinsic interband matrix element.  The necessary
relations are summarized or derived in Appendix~\ref{appmixing}.  The
mixing parameter $a_0$ deduced from the observed
$B(E2;2^+_{K=0_2}\rightarrow4^+_g)/B(E2;2^+_{K=0_2}\rightarrow0^+_g)$
ratio, using (\ref{eqnmixinga0x}),
is $a_0$=-0.0044(19).  (This would be induced by an intrinsic
interaction strength $| \langle
K^\pi=0^+_2|h_0|K^\pi=0^+_g\rangle|$=0.44(19)\,keV and corresponds to a
mixing matrix element of $\sim$2.6\,keV for the $2^+$ state.)  With
this mixing, the squared intrinsic matrix element
\mbox{$|\langle K^\pi=0^+_2|\mathfrak{M}'(E2)|K^\pi=0^+_g\rangle|^2$} is 8.4(16)\,\Wu.

While it would be interesting to have the results of a full three-band
mixing
calculation~\cite{lipas1962:threeband,riedinger1969:152sm154gd-beta}
involving the ground, $\gamma$, and $K^\pi=0^+_2$ excitations, there
is currently insufficient information on higher-lying band members and
on $M1/E2$ mixing ratios~\cite{nds1999:162} for such an analysis to be
feasible.  Mixing between the $\gamma$ and $K^\pi=0^+_2$ bands is only
expected to have a significant effect upon the $K^\pi=0^+_2$ to ground
state band transition strengths in nuclei for which the $K^\pi=0^+_2$
and $\gamma$ bands are nearly degenerate.  For mixing between the
$\gamma$ and $K^\pi=0^+_2$ bands to account for the observed
$B(E2;2^+_{K=0_2}\rightarrow0^+_g)$ strength in $^{162}$Er would
require nearly complete mixing of the $2^+_{K=0_2}$ and $2^+_\gamma$
states, corresponding to an interaction matrix element of
$\sim$100\,keV.  Such mixing would also result in a strong
$2^+_{K=0_2}\rightarrow2^+_\gamma$ transition with
$B(E2;2^+_{K=0_2}\rightarrow2^+_\gamma)\approx$176\,\Wu.  This would
correspond to a $\gamma$-ray intensity for the 271\,keV
$2^+_{K=0_2}\rightarrow2^+_\gamma$ transition about 7$\%$ as strong as for the
1171\,keV $2^+_{K=0_2}\rightarrow0^+_g$ transition.  Experimentally, no
$\gamma$-ray transition was observed between the $2^+_{K=0_2}$ and
$2^+_\gamma$ states~ [Fig.~\ref{fig162ergatesbeta}(b)], and the
intensity limit obtained~(Table~\ref{tab162erbranchbeta}) is
inconsistent with such a mixing picture.

Transitions between the $K^\pi=0^+_2$ and $\gamma$ bands are also
important for interpretation within the framework of the interacting
boson model (IBM), since the IBM can predict substantial transition
strengths between these two
bands~\cite{warner1981:168er-ibm,casten1994:newinterpretation}.  For IBM
parameter values relevant to $^{162}$Er~\cite{chou1997:ibafits}, the
predicted interband to in-band transition strength ratio is
$B(E2;2^+_{K=0_2}\rightarrow2^+_\gamma)/B(E2;2^+_{K=0_2}\rightarrow2^+_g)\approx6$,
which is not inconsistent with the present experimental limit of $<$15
(Table~\ref{tab162erbranchbeta}).

In summary, revised $\gamma$-ray branching data show that the decays
from the $2^+$ state at 1171\,keV in $^{162}$Er are in good agreement
with the Alaga rules for a $K^\pi=0^+$ excitation, and the analysis
used in obtaining the previously measured
$B(E2;0^+_g\rightarrow2^+_{K=0_2})$ value [8.0(15)\,\Wu] from Coulomb
excitation~\cite{ronningen1982:coulex} is validated by the present
branching results.  These results provide evidence for a collective
intrinsic matrix element between the ground and $K^\pi=0^+_2$
excitations, as expected for a $\beta$ vibration.  This case is
therefore one of only a few in which strong evidence for such
$\beta$-vibrational structure exists, especially for the lowest $0^+$
excitation.  Also, the resulting $\rho^2(E0)$ value is in agreement
with the RVM prediction for a $\beta$ vibration, and moderately
restrictive limits are placed upon transitions to the $\gamma$ band.
Since the structural interpretation of the $K^\pi=0^+_2$ excitation
depends heavily upon a single $B(E2)$ determination, it would,
however, be valuable to obtain confirmation of the $2^+_{K=0_2}$
lifetime, either from Coulomb excitation or by direct measurement.

The present results highlight the need to obtain a more complete set
of information on the low-lying excitations of $^{162}$Er.  The $4^+$
member of the $K^\pi=0^+_2$ band has only tentatively been
reported~\cite{tjom1968:er-ddprime} and none of the higher band
members are known.  Since the spin assignment of the previously
reported $K^\pi=(0^+_3)$ excitation was altered by the present
results, the $K^\pi=0^+_2$ excitation is left as the only identified
low-lying $K^\pi=0^+$ excitation in $^{162}$Er.  However, the
neighboring higher-mass Er isotopes ($^{164,166,168}$Er) are all
known~\cite{deboer1971:164er-beta,garrett1997:166erbeta-nnprime,garrett1997:166erdblgamma-nnprime,davidson1981:168er-ngamma}
to have several excited $0^+$ states below 2\,MeV, and so it is likely
that $^{162}$Er does as well.  Recent results elsewhere in the rare
earth region of nuclei~\cite{aprahamian2002:178hf-grid} indicate that
a low-lying $\beta$-vibrational excitation can be accompanied by a
well-defined higher-lying two-phonon $\beta$-vibrational excitation.
Further experiments using other low-spin population mechanisms, such
as $(p,t)$ transfer reactions or $(n,n'\gamma)$ scattering, will be
required to obtain this information.

Identification of the $K^\pi=0^+_2$ excitation in $^{162}$Er as having
probable $\beta$-vibrational character is, in and of itself, relevant
to the study of the existence and properties of vibrational modes in
deformed nuclei.  From a broader perspective, though, confirmation of
the collective character of this excitation allows the evolution of
the lowest collective $0^+$ excitation to be traced across the
$N$$\approx$90 transition region to the well-deformed rotor nuclei.

\part{Model analysis}
\label{partmodelanalysis}
\chapter{The geometric collective model parameter space
}
\label{chapgcm}
\pseudofootnotetext{The
results of this chapter were subsequently reported in M.~A.~Caprio,
Phys.~Rev.~C \textbf{68}, 054303 (2003).}

\section{Introduction to the GCM}
\label{secgcmintro}

\subsection{The geometric Hamiltonian}
\label{subsecgcmhamiltonian}

For a collective nucleus undergoing quadrupole deformation
(Section~\ref{seccollective}), it is natural to describe the dynamics
of the system by using the deformation coordinates $\alpha_{2\mu}$ as
the relevant dynamical variables.  In order for predictions of the
nuclear excitation properties to be made, an expression for the
nuclear Hamiltonian in terms of these deformation variables must
somehow be deduced or postulated.  This Hamiltonian can then be solved
for its eigenvalues and eigenfunctions.  

The Hamiltonian will in general be a function of both the coordinates
$\alpha_{2\mu}$ and their conjugate momenta $\pi_{2\mu}$.  The
geometric collective model (GCM), developed by Gneuss, Mosel, and
Greiner~\cite{gneuss1969:gcm,gneuss1970:gcm,gneuss1971:gcm},
is based upon an approach in which the Hamiltonian is expanded as a
power series in the coordinates and momenta.  The requirements of
rotational, reflection, and time reversal invariance constrain the
possible terms which may be present in this Hamiltonian, yielding a
series of the form
\begin{equation}
\begin{split}
\label{eqngcmhamiltonian}
H=&\frac{1}{B_2}[\pi\times\pi]^{(0)}+B_3[[\pi\times\alpha]^{(2)}\times\pi]^{(0)}+\cdots\\
&+C_2[\alpha\times\alpha]^{(0)}+C_3[[\alpha\times\alpha]^{(2)}\times\alpha]^{(0)}+\cdots.
\end{split}
\end{equation}
The terms may be classified as either ``kinetic energy'' terms,
involving the momenta $\pi$, or ``potential energy'' terms, involving
only the coordinates $\alpha$.

The leading-order kinetic energy term, known as the ``harmonic'' term,
arises directly when the classical expression for the kinetic energy
of a liquid drop undergoing small-amplitude vibrations is
quantized~\cite{eisenberg1987:v1}.  If only this term is used as the
kinetic energy operator, then the canonical momenta are
$\pi_{2\mu}$$=$$-i\hbar\partial/\partial\alpha_{2\mu}$, and the kinetic
energy operator is thus
\begin{equation}
\label{eqnharmonicterm}
-\frac{\hbar^2}{\sqrt{5}B_2}\sum_\mu
\frac{\partial^2}{\partial\alpha_{2\mu}\partial\alpha^*_{2\mu}}.
\end{equation}
The eigenproblem for a Hamiltonian with this kinetic energy operator
is equivalent to the Schr\"odinger equation in Cartesian coordinates
in five-dimensional space (Appendix~\ref{appscaling}).  The harmonic
kinetic energy operator may alternatively be written (see
Ref.~\cite{eisenberg1987:v1}) in terms of the Euler angles $\theta_i$
and the shape coordinates $\beta$ and $\gamma$
(Section~\ref{secbenchintro}) as
\begin{equation}
\label{eqnbohrkinetic}
- \frac{\hbar^2}{2B} \left[ \frac{1}{\beta^4} \frac{\partial}{\partial \beta}
\beta^4  \frac{\partial}{\partial \beta} + \frac{1}{\beta^2 \sin 3\gamma} 
\frac{\partial}{\partial \gamma} \sin 3\gamma \frac{\partial}{\partial \gamma} 
 - \frac{1}{4 \beta^2} \sum_\kappa \frac{M_\kappa^2}{\sin^2(\gamma - \frac{2}{3} \pi \kappa)} \right],
\end{equation}
where $B\equiv\sqrt{5}B_2/2$ and the $M_\kappa$ are angular momentum
operators which may be expressed in terms of the $\theta_i$ and
$\partial/\partial\theta_i$.  This expression for the kinetic energy
was first used in the Bohr Hamiltonian~\cite{bohr1952:vibcoupling},
for the description of rotational nuclei.

The potential energy depends only upon the shape of the nucleus, not
its orientation in space, and so the potential can be expressed purely
in terms of the shape coordinates $\beta$ and $\gamma$.  The series
expression for the potential energy in terms of these variables is
\begin{multline}
\label{eqnpotlseries}
V(\beta,\gamma)= \frac{1}{\sqrt{5}} C_2\beta^2 
- \sqrt{\frac{2}{35}} C_3 \beta^3 \cos{3\gamma}
+\frac{1}{5}C_4\beta^4\\
-\sqrt\frac{2}{175}C_5\beta^5\cos3\gamma+\frac{2}{35}C_6\beta^6\cos^2
3\gamma
+\frac{1}{5\sqrt{5}}D_6\beta^6+\cdots.
\end{multline}

Once wave functions for the nuclear eigenstates are calculated
(Subsection~\ref{subsecgcmnumerical}), electromagnetic matrix elements
between these states can be calculated.  This requires that the
electromagnetic multipole operators, which depend upon the charge and
current distributions in the nucleus, be expressed in terms of the
quadrupole collective coordinates, something for which there is no
unambiguous recipe without a full knowledge of the underlying
single-particle dynamics.  The most commonly used expression for the
electric quadrupole operator is deduced using the mathematically
simple, but somewhat arbitrary, assumption that the nuclear charge is
uniformly distributed within a radius
$R=R_0(1+\sum_\mu\alpha_{2\mu}Y^*_{2\mu})$~\cite{eisenberg1987:v1,hess1981:gcm-pt-os-w}.
This charge distribution leads to a series expression in powers of
$\alpha$
\begin{equation}
\label{eqnqe2coll}
Q_{2\mu}=\frac{3ZR_0^2}{4\pi}\left[\alpha^*_{2\mu}
-\frac{10}{\sqrt{70\pi}}[\alpha\times\alpha]^{(2)\,*}_\mu+\cdots \right],
\end{equation}
where $R_0\equiv r_0A^{1/3}$ (with $r_0=1.1$\,fm in Ref.~\cite{hess1981:gcm-pt-os-w}).

\subsection{Numerical solution of the GCM eigenproblem}
\label{subsecgcmnumerical}

The geometric Hamiltonian~(\ref{eqngcmhamiltonian}) is defined as a
differential operator in coordinate space.  The Schr\"odinger-like
partial differential equation for this Hamiltonian may be solved
directly by finite difference methods.  Such a technique has been
applied by Kumar and
Barranger~\cite{kumar1967:bohr-numerical}.
However, an alternative approach, which can be considerably more
computationally efficient, is to first calculate the matrix elements of
the Hamiltonian with respect to a complete set of functions on the
coordinate space and to then diagonalize the Hamiltonian in this
basis.  The eigenfunctions of the five-dimensional harmonic oscillator
constitute an especially convenient basis for diagonalization,
since these are known in closed
form~(Section~\ref{secbenchosc}),
and group-theoretic results can be applied in calculation of matrix
elements of various operators with respect to this basis.  Essentially
all calculations using the GCM have been performed using the
oscillator basis~\cite{gneuss1969:gcm,gneuss1970:gcm,gneuss1971:gcm}.

The GCM calculations disscussed in the present work were performed using a
computer code written by Troltenier, Maruhn, and
Hess~\cite{troltenier1991:gcm} which calculates the GCM Hamiltonian
eigenvalues and eigenstates by diagonalization in a truncated basis of
oscillator eigenfunctions.  The code constructs matrix elements of the
Hamiltonian operator (\ref{eqngcmhamiltonian}) from precomputed matrix
elements for each of the potential and kinetic energy terms, which
were originally obtained either analytically or by numerical
integration~\cite{hess1980:gcm-details-238u,troltenier1991:gcm}.  The
code includes the necessary matrix elements for calculations
using basis functions with phonon numbers up to $N$=30 and angular
momenta up to $L$=10.  The code accomodates calculations with kinetic
energy terms through second order in $\pi$ and potential energy terms
through sixth order in $\beta$.  It calculates electric quadrupole
matrix elements between eigenstates using the first two terms of the
operator given in (\ref{eqnqe2coll}) and deduces $E2$ transition
strengths and quadrupole moments from these.  For the present work,
the code has been adapted to also perform calculations using only the
linear term, for reasons discussed in Section~\ref{secgcmscaling}.

If diagonalization of the GCM Hamiltonian in a truncated basis is to
yield eigenvalues and eigenfunctions which reasonably reproduce the
true ones, the set of basis functions chosen should be adequately matched to
the eigenfunctions they are being used to approximate.  The
eigenfunctions of the harmonic oscillator Hamiltonian
\begin{equation}
H_\text{osc}=\frac{1}{B_2}\left( [\pi\times\pi]^{(0)}+B_2 C_2[\alpha\times\alpha]^{(0)}\right),
\end{equation}
depend upon the ``stiffness'' $s\equiv(B_2C_2/\hbar^2)^{1/4}$ of the
oscillator from which they are constructed.  The oscillator
eigenfunctions for different stiffnesses are related to each other by
simple dilation,
$\Phi^s(\beta,\gamma,\underline{\theta})=s^{5/2}\Phi(s\beta,\gamma,\underline{\theta})$
(Section~\ref{secbenchosc}), so the stiffness parameter controls the
radial ($\beta$) extent of these
functions.  For diagonalization
in a truncated basis to produce accurate results, the $s$ for the
basis must chosen so that the radial extent of the basis functions
matches the radial extent of the GCM eigenfunctions they are being
used to approximate.  The code of Ref.~\cite{troltenier1991:gcm} uses
variational procedures~\cite{margetan1982:basis-optimization} to automatically optimize
$s$ for each calculation.

Even if the stiffness of the basis functions is chosen as well as
possible, the diagonalization process will fail to produce accurate
results if the basis is truncated at too low a phonon number.
Solution wave functions which are sharply peaked at some nonzero
deformation must be obtained from basis functions which vary rapidly
at that deformation, and superposition of a large
number of basis functions is required in order for cancellation to be obtained
throughout the region of $\beta$ less than that deformation.  Thus,
stiffly-deformed wave functions are likely to require basis functions
of high phonon number for their description.  Basis functions of high
phonon number are also needed to produce $\gamma$-rigid angular wave
functions.  For successfully convergent diagonalizations, most
of the probability density is usually accounted for by the lowest $\sim$10$\%$
of the basis functions in order of increasing phonon
number~\cite{gneuss1971:gcm,margetan1982:basis-optimization}.
Significant probability in the highest basis functions is an indicator
that the solution may have ``overflowed'' the truncated basis,
requiring additional basis functions for proper description, and that
the results of the diagonalization are not reliable.  Detailed
discussions, and illustrative plots, of the dependence of convergence
properties upon the basis $s$ and $N_\text{max}$ may be found in
Refs.~\cite{gneuss1971:gcm,margetan1982:basis-optimization}.

\subsection{Application of the GCM}

Attempts have been made to derive the parameters in the collective
Hamiltonian operator~(\ref{eqngcmhamiltonian}) from models of the
underlying single particle dynamics (\eg,
Refs.~\cite{mosel1968:gcm-microscopic,kumar1974:150sm152sm-ppq,eisenberg1976:v3}).
However, the necessary theory is not sufficiently well developed to
provide a full description of the nuclear phenomenology.  An
alternative, more pragmatic, approach is to choose the collective
Hamiltonian so as to best reproduce observed nuclear properties.

The GCM Hamiltonian with eight parameters ($B_2$, $B_3$, $C_2$, $C_3$,
$C_4$, $C_5$, $C_6$, and $D_6$) accomodated by the existing codes is
capable of describing a rich variety of nuclear structures and can
flexibly reproduce many details of potential energy surface
shapes~\cite{vonbernus1975:gcm,eisenberg1987:v1}.  Manual selection of
parameter values in this full eight-parameter model is impractical, so
parameter values for the description of a particular nucleus must be
found through automated
fitting~\cite{hess1980:gcm-details-238u,troltenier1991:gcm} of the
nuclear observables.  This process introduces technical difficulties
associated with reliable minimization in eight-dimensional space, and
often the parameter values appropriate to a nucleus are
underdetermined by the available observables~\cite{eisenberg1987:v1}.
This is not necessarily an impediment to physical interpretation of
the fitted nuclei, since, as discussed in
Ref.~\cite{hess1980:gcm-details-238u}, the available data are usually
sufficient to establish the qualitative nature of the GCM potential
and determine the coefficients of lower order terms, while the
uncertainties in the values of higher-order parameters represent only
fine adjustments in the predicted structure.  However, a more
tractable form of the model is desirable.

A GCM Hamiltonian truncated to the harmonic term in the kinetic energy
and to the three lowest-order terms in the
potential,
\begin{equation}
\label{eqnpotltrunc}
V(\beta,\gamma)= \frac{1}{\sqrt{5}} C_2\beta^2 
- \sqrt{\frac{2}{35}} C_3 \beta^3 \cos{3\gamma}
+\frac{1}{5}C_4\beta^4,
\end{equation}
is sufficient (see Refs.~\cite{eisenberg1987:v1,zhang1997:gcm-trunc})
to produce the rotor, oscillator, and $\gamma$-soft structures
discussed in Section~\ref{seccollective} as well as various more
exotic possibilities involving shape coexistence.  With fewer
parameters in the Hamiltonian, it is more feasible to survey the full
range of phenomena accessible in the parameter space, and the
parameter values applicable to a given nucleus are much more fully
determined by the available observables.  These benefits must be
weighed against the limitations inherent in using the truncated model:
the full generality of the GCM is forsaken, precluding, for instance,
the description of rigid triaxiality (\eg,
Refs.~\cite{troltenier1996:gcm-ru,starostaINPREP}), and, even within
its qualitative domain of applicability, the truncated model can be
expected to have reduced flexibility in reproducing subtleties of the
potential energy surface.

\section{Scaling properties for the GCM}
\label{secgcmscaling}

The truncated form of the GCM, with potential~(\ref{eqnpotltrunc}),
still contains four parameters ($B_2$, $C_2$, $C_3$, and $C_4$).  The
relationship between a set of values for these parameters and the
structure of the resulting predictions is not evident without detailed
calculations.  It would be useful to have a model which covers the
full range of features needed for description of the physical system
but which simultaneously has a dependence upon its parameters which is
simple, qualitatively predictable by inspection, and directly
understandable.  It is therefore desirable to further simplify the GCM
parameter space, but
\textit{without} additional truncation of the model.

The GCM predictions for different sets of parameter values are not
entirely independent.  Proper use of analytic relations can
considerably simplify the use of the model.

First, observe that an essentially trivial manipulation saves
considerable computational effort.  Overall multiplication of any
Hamiltonian by a constant factor $b$ results in multiplication of all
eigenvalues by $b$ and leaves the eigenstates unchanged.  This
transformation thus leaves unchanged all energy \textit{ratios}, as
well as all observables which depend only upon the wave functions.  A
practical consequence is that the truncated GCM Hamiltonian is reduced
from having four active parameters to effectively having three
parameters.  The transformation
\begin{equation}
\label{eqnscalingbbccc}
B_2'= \frac{1}{b}B_2 \qquad C_2'=bC_2 \qquad C_3'=bC_3 \qquad C_4'=bC_4
\end{equation}
has the effect only of rescaling all energies by $b$.  Therefore, all
calculations can be performed for some reference value of $B_2$,
varying only $C_2$, $C_3$, and $C_4$, and any calculation with another
value of $B_2$ would be equivalent to one of these to within a
rescaling of energies.

\begin{figure}
\begin{center}
\resizebox{1.0\hsize}{!}
{
\includegraphics*[height=1.4in]{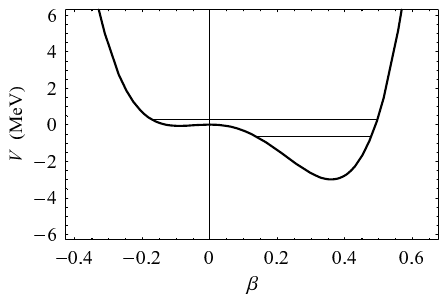}
\raisebox{0.7in}{\scalebox{1}{
$\xrightarrow{\substack{\text{deepen}\\\text{by 2}}}$
}} 
\includegraphics*[height=1.4in]{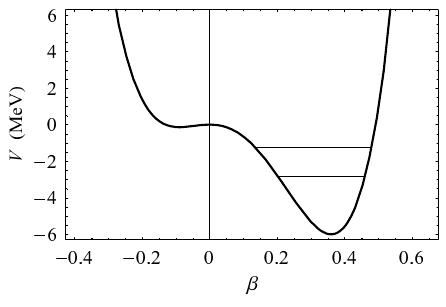}
\raisebox{0.7in}{\scalebox{1}{
$\xrightarrow{\substack{\text{squeeze}\\\text{by $\sqrt{2}$}}}$
}} 
\includegraphics*[height=1.4in]{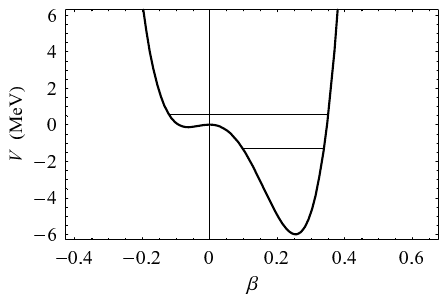}
}
\end{center}
\vspace{-12pt}
\caption[Successively deepening and squeezing a potential.]
{Successively deepening and squeezing a potential in the correct
proportion (see text) have effects which exactly cancel each other, in
that the level energies are simply renormalized and all wave functions
(not shown) are dilated by the same amount.  Under the scaling
transformation, the potential energy function is multiplied by the
same factor as the level energies.  The rightmost panel can
thus be obtained simply by magnifying the vertical axis and compressing the
horizontal axis of the leftmost panel.
(This is \textit{not} not the case for the intermediate
stage shown in the middle panel.)
\label{figscaling}
}
\end{figure}

There is a second simplification which can be made.  It is a
well-known heuristic that ``deepening'' a potential lowers the
energies of levels confined within the potential, while ``narrowing''
a potential raises the level energies.  It would thus be expected that
successively deepening and then squeezing a given potential, if
performed in the correct proportion, could have effects which roughly
offset each other, as illustrated in Fig.~\ref{figscaling}.

For a large class of problems, of which the GCM Hamiltonian with
harmonic kinetic energy is a special case (Appendix~\ref{appscaling}),
there is an exact analytic result to this effect.  Consider the
transformation in which the potential is multiplied by a factor $a^2$
while also dilated by a factor $1/a$, \ie, multiplying $\beta$ by $a$
in the argument to $V$,
\begin{equation}
\label{eqnscalingpotential}
V'(\beta,\gamma)=a^2V(a\beta,\gamma).
\end{equation}
The effect of this transformation is
simply to multiply all eigenvalues by a factor $a^2$ and radially
dilate the wave functions by $1/a$, as shown in
Appendix~\ref{appscaling}.  Thus, for the truncated GCM Hamiltonian,
the transformation
\begin{equation}
\label{eqnscalingabccc}
B_2'= B_2 \qquad C_2'=a^4C_2 \qquad C_3'=a^5C_3 \qquad C_4'=a^6C_4
\end{equation}
results in a rescaling of all level energies by $a^2$ and dilation of all
the solution wave functions by $1/a$.  This is illustrated in
Fig.~\ref{figscaling}.  Observe that, since level energies are
multiplied by the same scaling factor $a^2$ as the potential itself,
they retain their positions relative to the recognizable ``features''
of the potential, such as barriers or inflection points.

If the electric quadrupole transition operator~(\ref{eqnqe2coll}) is
truncated to its linear term, then all matrix elements of this
operator change by the same factor under wave function dilation
(Appendix~\ref{appscaling}), so $B(E2)$ ratios are left unchanged.
Many GCM studies have retained the second-order
term~\cite{troltenier1991:gcm}, but, since the different terms
in~(\ref{eqnqe2coll}) scale by different powers of $a$ under dilation
(see Appendix~\ref{appscaling}), inclusion of the second-order or
higher-order terms destroys the simple invariance of $B(E2)$ ratios.
The second-order term usually provides only a small correction to the
linear term, and the correct coefficient by which it should be
normalized is highly
uncertain~\cite{eisenberg1987:v1,hess1981:gcm-pt-os-w}.  Comparative
studies by Petkov, Dewald, and Andrejtscheff~\cite{petkov1995:gcm-ba}
have shown no clear benefit to its inclusion.  In light of the simple
scaling properties obtained by its omission, calculations of $B(E2)$
strengths are carried out using a linear electric
quadrupole operator throughout the present work.

A scaling result equivalent to that just discussed has been used in
earlier
work~\cite{vonbernus1971:gcm-scaling,habs1974:gcm-n50to82,troltenier1991:gcm}
to simplify the automated fitting of experimental data with the full
eight-parameter GCM Hamiltonian.  In that work, the canonical
transformation $\pi\rightarrow\frac{1}{a}\pi$ and $x\rightarrow a
x$~\cite{troltenier1991:gcm} was used to produce wave function
dilation with no change in energy scale, equivalent to the
transformation~(\ref{eqnprimedefs}) followed by an overall
multiplication of the Hamiltonian by $a^{-2}$.  Since in
Refs.~\cite{vonbernus1971:gcm-scaling,habs1974:gcm-n50to82,troltenier1991:gcm}
the second-order form of the quadrupole operator was used, $B(E2)$
ratios \textit{were} changed under scaling.  Therefore, the fitting
procedure was carried out using only energy ratios, and, after
fitting, the wave functions were dilated to reproduce a single $B(E2)$
strength or quadrupole moment.

Systematic use of the scaling relations just discussed is facilitated
by the adoption of a simple reparametrization of the GCM potential.
Specifically, let us rewrite (\ref{eqnpotltrunc}) as
\begin{equation}
\label{eqnpotldef}
V(\beta,\gamma)=f \left[
\frac{9}{112} d \left(\frac{\beta}{e}\right)^2
- \sqrt{\frac{2}{35}} \left(\frac{\beta}{e}\right)^3 \cos{3\gamma}
+ \frac{1}{5} \left(\frac{\beta}{e}\right) ^4
\right].
\end{equation}
This expression has been constructed so that varying each of the parameters
$d$, $e$, and $f$ controls one specific aspect of the potential:
\begin{dissenumeratelist}
\item[$d$~--] \underline{d}escribes the shape of the potential, \ie, $d$
uniquely defines the shape to within a horizontal and vertical scaling
\item[$e$~--] \underline{e}xpands the potential horizontally, \ie,
varying $e$ scales $V$ in the radial coordinate $\beta$
\item[$f$~--] is a \underline{f}actor multiplying the entire potential, \ie,
varying $f$ scales the magnitude of $V$.
\end{dissenumeratelist}
Comparison of the coefficients in (\ref{eqnpotldef}) with those in the
original parametrization (\ref{eqnpotltrunc})
yields the conversion formulae
\begin{equation}
\label{eqndefccc}
\begin{alignedat}{3}
d &= \frac{112}{9\sqrt{5}}\frac{C_2C_4}{C_3^2}&
\qquad e&=\frac{C_3}{C_4}&
\qquad f&=\frac{C_3^4}{C_4^3}\\
C_2&=\frac{9\sqrt{5}}{112}\frac{fd}{e^2}&
\qquad C_3&=\frac{f}{e^3}&
\qquad C_4&=\frac{f}{e^4}.
\end{alignedat}
\end{equation}

\begin{figure}
\begin{center}
\includegraphics*[width=0.8\hsize]{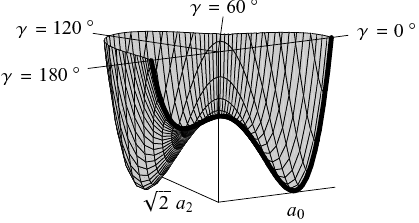}
\end{center}
\caption[The GCM potential function for $d$\lt0.]
{The GCM potential function for $d$\lt0 posesses a global minimum, a
saddle point, and a local maximum at $\beta=0$.  The case $d=-5$ is
shown here as an example, plotted as a function of the Cartesian
coordinates $(a_0,\sqrt{2}a_2)$ or polar coordinates $(\beta,\gamma)$,
for $\gamma$=0$^\circ$ to 180$^\circ$.  The cut along the $a_0$-axis
(thick line) is the same as is shown in the second panel of
Fig.~\ref{figgcmpotld}.  The global minimum is visible in this figure
at $\gamma$=0$^\circ$ and 120$^\circ$, and the saddle point is visible
at $\gamma$=60$^\circ$ and 180$^\circ$.  The saddle point is a local
minimum with respect to $\beta$ and a local maximum with respect to
$\gamma$.
\label{figgcmsaddle}
}
\end{figure}

The extremum structure of the truncated GCM potential, investigated in
Refs.~\cite{greiner1963:gcm-dneg,acker1965:gcm-dpos}, can be expressed
very concisely in terms of the present parametrization
(\ref{eqnpotldef}).  Extrema occur where $V(\beta,\gamma)$ is locally
extremal with respect to both $\beta$ and $\gamma$ individually.
Thus, they are possible where $\cos 3\gamma$ attains its maximum value
of +1 ($\gamma$=0$^\circ$, 120$^\circ$, and 240$^\circ$) or its
minimum value of -1 ($\gamma$=60$^\circ$, 180$^\circ$, and
300$^\circ$).  Since the potential (\ref{eqnpotldef}) repeats every
120$^\circ$ in $\gamma$, it suffices to locate the extrema on one
particular ray in the $\beta\gamma$-plane with $\cos 3\gamma$=+1 (\eg,
$\gamma$=0$^\circ$) and on one ray with $\cos 3\gamma$=-1 (\eg,
$\gamma$=180$^\circ$).  Thus, extrema need only be sought on a
cut through the potential along the $a_0$-axis, and these extrema
will then be duplicated along the other rays of $\cos 3\gamma$=$\pm1$, as
illustrated in Fig.~\ref{figgcmsaddle}.

Extrema along the two rays $\gamma$=0$^\circ$ and $\gamma$=180$^\circ$
are found by identification of the zeroes of $\partial
V/\partial\beta$ for $\cos 3\gamma$=+1 and for $\cos 3\gamma$=-1.  The
variable $\beta$ is a radial coordinate (see
Section~\ref{secbenchintro}) and so takes on only positive values.
However, the form of the following results is considerably simplified
by noting that the only occurrence of $\cos 3\gamma$ in
(\ref{eqnpotldef}) is in a product also containing the only occurrence
of an odd power of $\beta$, so substituting $\cos 3\gamma$=-1 is
algebraically equivalent to setting $\cos 3\gamma$=+1 and negating
$\beta$.  Any extremum occuring for $\cos 3\gamma$=-1 will thus be found
when the extrema for $\cos 3\gamma$=+1 are sought, but at a fictitious
``negative'' $\beta$ value.  A simple expression for the $\beta$
values yielding extrema along $\gamma$=0$^\circ$ follows, but it must
be interpreted with the proviso that when a negative $\beta$ value is
encountered it actually represents a positive $\beta$ value along
$\gamma$=180$^\circ$.  The only extrema of $V$ along the $a_0$-axis cut are
located at
\begin{displaymath}
\begin{alignedat}{2}
&\beta=0&     \qquad&\text{any $d$}\\
&\beta=\beta_-&     \qquad&d\leq1\\
&\beta=\beta_+&     \qquad&d\leq1,
\end{alignedat}
\end{displaymath}
where
\begin{equation}
\label{eqnbetapm}
\beta_\pm=\frac{3}{4}\sqrt{\frac{5}{14}}\left(1\pm r\right) e,
\end{equation}
in terms of $r\equiv\sqrt{1-d}$.  The 
extremal values of the potential are  
\begin{equation}
V(\beta_\pm)=-\frac{135}{50176}(r\pm1)^3(3r\mp1)f.
\end{equation}

The nature of the extrema~--- whether they are minima, maxima, saddle
points, or inflection points~--- can be ascertained from the signs of
the partial derivatives.  The extremum structure of the potential
depends only upon the value of $d$, as summarized in
Table~\ref{tabdefextrema} and Fig.~\ref{figgcmpotld}.  For $d$\lt0,
the potential has both a global minimum and a saddle point at nonzero
$\beta$ (Fig.~\ref{figgcmsaddle}).  For 0\lt$d$\lt1, minima are
present at both at nonzero $\beta$ and at $\beta$=0, with
the deformed minimum lower for 0\lt$d$\lt8/9 and the undeformed
minimum lower for 8/9\lt$d$\lt1.  For $d$\gt1, there is only one minimum, located at
$\beta$=0.
%
\begin{table}[p]
{\ssp
\begin{center}
\begin{tabular}{p{0.55in}.p{0.65in}.p{1.55in}.p{0.2in}.p{0.55in}.p{0.65in}.p{1.55in}}
\pseudoruledtabular
$d$ & $\beta$ & Extremum type & & $d$ & $\beta$ & Extremum type\\
\hline
(-$\infty$,0)   & $\beta_-$\,(\lt0)$^a$ & Saddle point$^{b}$    & & (8/9,1)         & 0                     & Minimum (global)      \\ 
                & 0                     & Maximum               & &                 & $\beta_-$             & Saddle$^c$            \\  
                & $\beta_+$             & Minimum (global)      & &                 & $\beta_+$             & Minimum (not global)  \\ 
0               & 0\,(=$\beta_-$)       & Inflection point      & & 1               & 0                     & Minimum (global)      \\ 
                & $\beta_+$             & Minimum (global)      & &                 & $\beta_-$(=$\beta_+$) & Inflection point    \\ 
(0,8/9)         & 0                     & Minimum (not global)  & & (1,$\infty$)    & 0                     & Minimum (global)      \\
                & $\beta_-$             & Saddle point$^c$      & \\
                & $\beta_+$             & Minimum (global)      & \\
8/9             & 0                     & Minimum (global)      & \\ 
                & $\beta_-$             & Saddle point$^c$      & \\ 
                & $\beta_+$             & Minimum (global)      & \\
\pseudoruledtabular
\end{tabular}
\end{center}
}
\begin{tablenotes}
$^a$A ``negative'' solution for $\beta$ indicates an extremum at
$\gamma$=180$^\circ$, \ie, on the negative $a_0$-axis (see
text).\\ 
$^b$Local minimum with respect to $\beta$ and local maximum
with respect to $\gamma$.\\ 
$^c$Local maximum with respect to $\beta$
and local minimum with respect to $\gamma$.
\end{tablenotes}
\caption
[Locations and types of extrema in $V(\beta,\gamma)$.]  {\ssp The
locations and types of each of the extrema in $V(\beta,\gamma)$ along
the $a_0$-axis cut, for the different ranges of $d$ values.  Extrema
are given in order along the $a_0$-axis cut.
\label{tabdefextrema}
}
\end{table}

%
%
\newcommand{\gcmpanel}[1]{\includegraphics*[height=1in]{figures/theory/diss_gcmpotl_#1}}
\begin{figure}[p]
\begin{center}
\resizebox{1.0\hsize}{!}
{
\tiny
\begin{tabular}{cccc}
\gcmpanel{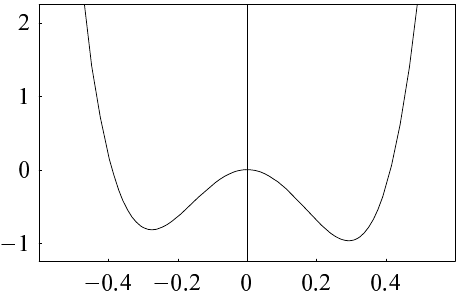} & \gcmpanel{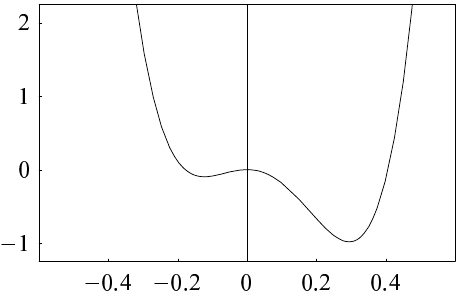} & \gcmpanel{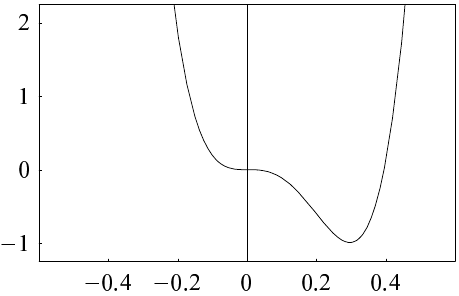} & \gcmpanel{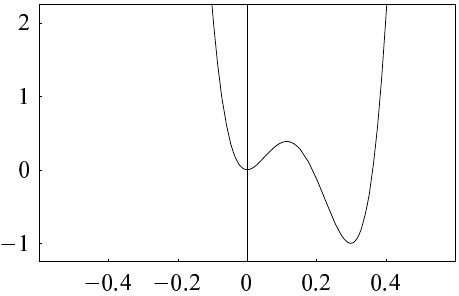} \\
$\gamma$-soft limit & & $\leftarrow$rotor$\rightarrow$ & \\
$d\rightarrow - \infty$ & $d<0$ & $d=0$ & $0<d<8/9$ \\
$(d=-1000,~e=0.02,~f=1.1\times10^{-4})$ & $(d=-5,~e=0.19,~f=1.4)$& 
$(d=0,~e=0.33,~f=23)$ & $(d=0.8,~e=0.46,~f=359)$
\vspace{0.3in}\\
\gcmpanel{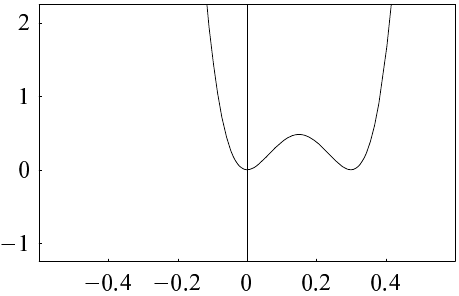} & \gcmpanel{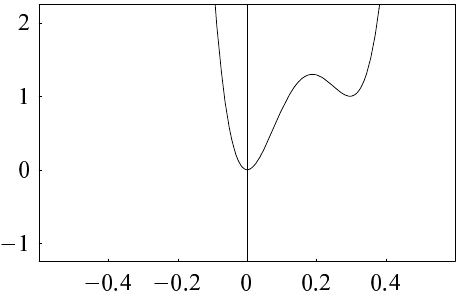} & \gcmpanel{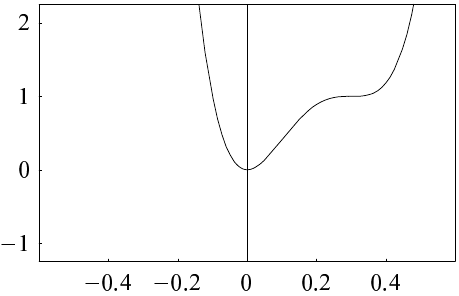} & \gcmpanel{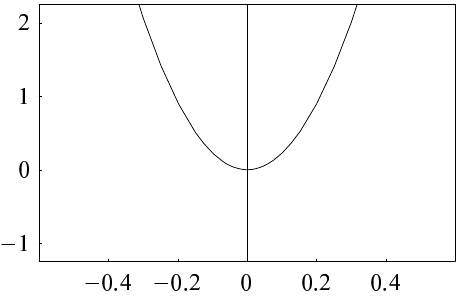} \\
&&&harmonic oscillator\\
$d=8/9$& $8/9<d<1$ & $d=1$ & $d\rightarrow+\infty$\\
$(d=8/9,~e=0.50,~f=300)$ & $(d=0.95,~e=0.54,~f=616)$ &
$(d=1,~e=0.67,~f=371)$ & $(d=15,~e=4,~f=300)$
\end{tabular}
}
\end{center}
\caption[The shapes of the GCM potential.]
{Illustration of the qualitatively different shapes of the GCM
potential function (\ref{eqnpotldef}) obtained for different ranges of
values for the parameter $d$.  Potentials are shown as a function of
$\beta$ along the $a_0$-axis cut (see text).  For $f$ in MeV, the energy scale is also in MeV.
\label{figgcmpotld}
}
\end{figure}
\newcommand{\gcmpanelef}[1]{\includegraphics*[height=1.5in]{figures/theory/diss_gcmpotl_#1}}
\begin{figure}
\begin{center}
{\tiny
\begin{tabular}{c@{\hspace{0.5in}}c}
\gcmpanelef{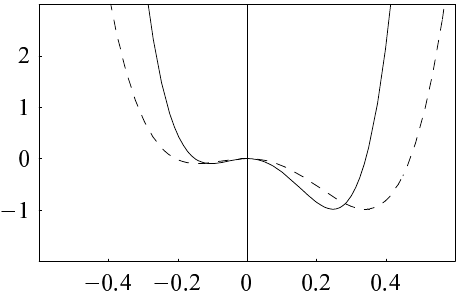} & \gcmpanelef{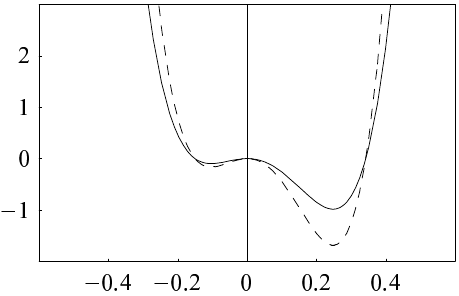}\\
$e$ scales $\beta$ radially & $f$ scales $V$\\
(\underline{e}xpands horizontally) & (\underline{f}actor multiplying
potential)\\
$d=-5,~e=\left\lbrace\begin{aligned}0&.16~\text{solid}\\0&.22~\text{dashed}\end{aligned}\right\rbrace,~f=1.4$
& 
$d=-5,~e=0.16,~f=\left\lbrace\begin{aligned}1&.4~\text{solid}\\2&.4~\text{dashed}\end{aligned}\right\rbrace$  \\ 
\end{tabular}
}
\end{center}
\caption[Horizontal and vertical scaling of the GCM potential function.]
{Horizontal and vertical scaling of the GCM potential function
(\ref{eqnpotldef}), obtained as the effects of changing the values of
the parameters $e$ and $f$, respectively.  Potentials are shown as a function of
$\beta$ along the $a_0$-axis cut (see text).  For $f$ in MeV, the energy scale is also in MeV.
\label{figgcmpotlef}
}
\end{figure}

Let us briefly address the ranges of definition for the parameters $e$
and $f$.  Negating $e$ reflects the potential about $\beta$=0.  A
positive value of $e$ places the deformed minimum on the prolate side
of the cut ($\beta_+$\gt0), while negative $e$ places the deformed
minimum on the oblate side of the cut ($\beta_+$\lt0).  All model
predictions for energies and transition strengths are unchanged under
interchange of prolate and oblate deformations, and only the signs of
quadrupole matrix elements and thus quadrupole moments are affected.
Throughout the discussions and examples in the present work, $e$ will
be taken positive without loss of generality.  Only positive values of
$f$ are meaningful, since for $f$ negative the coefficient on the
$\beta^4$ term in the potential is negative.  This makes
$V\rightarrow-\infty$ as $\beta\rightarrow\infty$, leaving the system
globally unbound.  The effects on the potential of varying the
parameters $e$ and $f$ are illustrated in Fig.~\ref{figgcmpotlef}.

In terms of the new parameters, overall multiplication of the
Hamiltonian by $b$ is obtained by the transformation
\begin{equation}
\label{eqntransfbdef}
B_2'=\frac{1}{b}B_2 \qquad d'=d \qquad e'=e \qquad f'=bf,
\end{equation}
and deepening the potential by $a^2$ while dilating by $1/a$ is
accomplished by the transformation
\begin{equation}
\label{eqntransfadef}
B_2'=B_2 \qquad d'=d \qquad e'=\frac{1}{a}e \qquad f'=a^2f.
\end{equation}
If two sets of parameter values, call them $(B_2,d,e,f)$ and
$(B_2',d',e',f')$, can be transformed into each other by any
combination of these scaling relations, the solutions for these
parameter sets will be identical, to within energy normalization and
wave function dilation.  If, however, $(B_2,d,e,f)$ and
$(B_2',d',e',f')$ cannot be transformed into each other, the solutions
for these parameters will be distinct.  Parameter sets are thus
naturally grouped into ``families'', where $(B_2,d,e,f)$ and
$(B_2',d',e',f')$ are members of the same family if and only if they
are related by the scaling transformations~(\ref{eqntransfbdef})
and~(\ref{eqntransfadef}).

We are now equipped to construct a ``structure parameter'' $S$ which
is invariant under the transformations~(\ref{eqntransfbdef})
and~(\ref{eqntransfadef}).  Let
\begin{equation}
\label{eqngcms}
S\equiv\frac{1}{B_2e^2f}.
\end{equation}
Under overall multiplication of the Hamiltonian
(\ref{eqntransfbdef}),
\begin{equation}
\begin{aligned}
S'
&=\frac{1}{\left(\frac{1}{b}B_2\right)\left(e\right)^2\left(bf\right)}\\
&=\frac{1}{B_2e^2f}
=S,
\end{aligned}
\end{equation}
and, under deepening and narrowing of the potential (\ref{eqntransfadef}),
\begin{equation}
\begin{aligned}
S'
&=\frac{1}{\left(B_2\right)\left(\frac{1}{a}e\right)^2\left(a^2f\right)}\\
&=\frac{1}{B_2e^2f}
=S,
\end{aligned}
\end{equation}
so $S$ is unchanged under both transformations.
If two points in parameter space are characterized by the same
values of $d$ and $S$, they yield identical energy spectra, to
within an overall normalization factor, and identical wave functions, to
within dilation, and consequently identical $B(E2)$ ratios.  Two points
characterized by different values of $d$ or of $S$ will in
general give different energy spectra, wave functions, and $B(E2)$
ratios.  Increasing $S$ corresponds to decreasing the mass parameter
$B_2$, making the potential narrower, or making the potential shallower, all
of which have the effect of raising the energy levels within the
potential.  Thus, for a given potential shape, given by $d$, the
parameter $S$ determines how ``high'' the levels lie relative to the
features of the potential.

\section{Mapping the GCM parameter space}
\label{secgcmmapping}

The truncated GCM described in Section~\ref{secgcmscaling} is
effectively a {\it two-parameter} model, with parameters $d$ and $S$.
The two other degrees of freedom remaining from the four original
parameters provide only an overall normalization factor on the energy
scale and a radial scaling for the wave functions in the coordinate
$\beta$.  Because of the simplicity of this model, the behavior of an
observable over the {\it entire model space} can be summarized on a
single contour plot.  However, the parameter values needed to cover
the structural features of interest span many orders of magnitude, so
it is necessary to make some preliminary analytic estimates to guide
the numerical calculations if an effective and comprehensive survey of
the parameter space is to be made.

Let us first consider what qualitatively different types of behavior
are possible within the model space.  Each of the different potential
``shapes'' depicted in Fig.~\ref{figgcmpotld} can give rise to a
several different types of structure, depending upon the excitation energy
of the ground state and other low-lying levels relative to the minimum of
the potential.  These structural regimes are illustrated in
Table~\ref{tabdeltav}.  For potentials with $d$\lt0:
\begin{dissenumeratelist}
\item If level energies lie well below the saddle point, the states will be
energetically confined to the deformed minimum, yielding rotational behavior.
\item If level energies lie between the saddle point and the local
maximum at $\beta$=0, all $\gamma$ values will be energetically
accessible, but $\beta$=0 will still not be accessible.  In this case,
deformed $\gamma$-soft structure is possible.
\item If level energies lie well above the local
maximum at $\beta$=0, the potential controlling the behavior of these
states is dominantly a $\beta^4$ quartic oscillator well.
\end{dissenumeratelist}
For potentials with 0\lt$d$\lt1, two minima are present, one at zero
deformation and one at nonzero deformation:
\begin{dissenumeratelist}
\item If level energies lie well below the higher minimum,
the states will be energetically confined to the global minimum, yielding
rotational behavior for 0\lt$d$\lt8/9 or approximately harmonic
oscillator behavior for 8/9\lt$d$\lt1.
\item If level energies lie above both minima but below the saddle point barrier, the
energetically-accessible regions around the two
minima will be separated from each other by this barrier.  States
involving mixing through the barrier may be possible.
\item If level energies lie well above the 
barrier, the behavior again approaches that of a quartic
oscillator.
\end{dissenumeratelist}
Finally, for potentials with $d$\gt1, deformed structure is not possible.
If level energies are low in the well, so the states are confined to a
region of small $\beta$, where the $\beta^2$ term in the potential
dominates, harmonic oscillator behavior arises, but for level energies
much higher in the well the $\beta^4$ term dominates, yielding quartic
oscillator behavior.
\newcommand{\gcmpaneldeltav}[1]{\includegraphics*[height=0.9in]{figures/theory/diss_gcmpotl_#1}}
\begin{table}
\begin{center}
\begin{tabular}{lcccc}
\pseudoruledtabular
& $d\leq0$ & $0\leq d\leq8/9$ & $8/9\leq d\leq1$ & $d\geq1$\\
\hline
~\\
& \gcmpaneldeltav{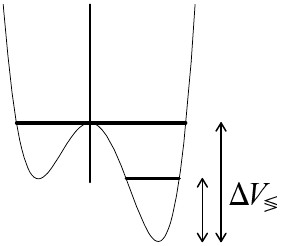}& \gcmpaneldeltav{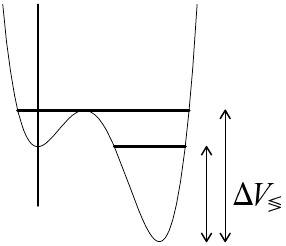} & \gcmpaneldeltav{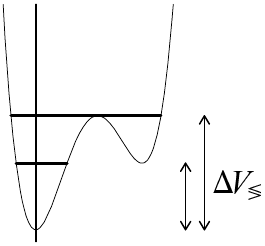} & \gcmpaneldeltav{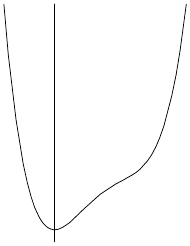}
\\
$\Delta V_<$ &   $V(\beta_-) - V(\beta_+)$     & $- V(\beta_+)$ & $V(\beta_+)$ & ---\\
$\Delta V_>$ &   $-V(\beta_+)$     & $V(\beta_-) - V(\beta_+)$ & $V(\beta_-)$ & ---\\
\pseudoruledtabular
\end{tabular}
\end{center}
\caption
[Structures possible for the GCM potential.] {\ssp Structures possible
for each of the qualitatively different shapes of the truncated GCM
potential, shown on the cut along the $a_0$-axis.  The quantities
$\Delta V_\lessgtr$ represent the approximate ground state energies, relative
to the global minimum of the potential, at which the boundaries
between the structural regimes occur.  For energies substantially less
than $\Delta V_<$, the low-lying levels are ``trapped'' in the global
minimum.  For energies substantially greater than $\Delta V_>$,
quartic anharmonic vibrational structure dominates.
\label{tabdeltav}
}
\end{table}

\begin{figure}
\begin{center}
\resizebox{1.0\textwidth}{!}
{
\includegraphics*[height=1in]{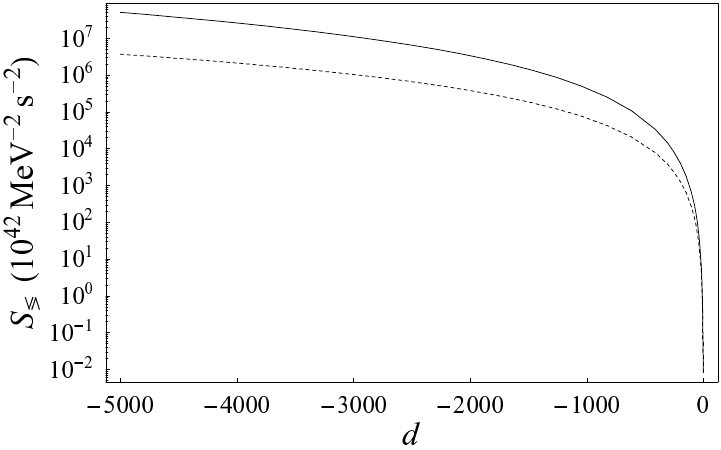}
\hspace{0.1in}
\includegraphics*[height=1in]{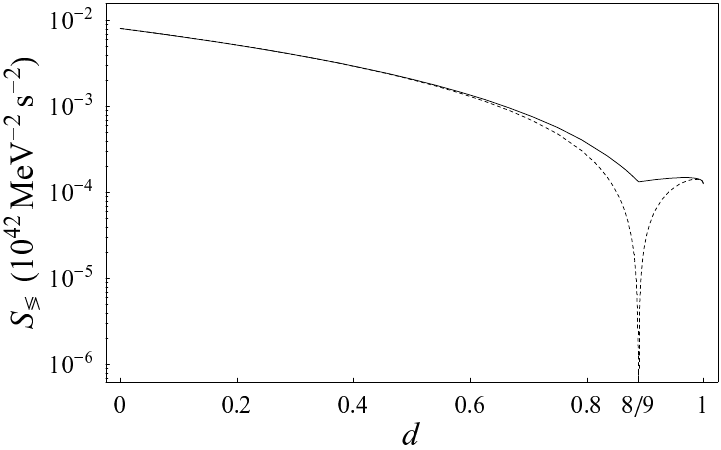}
}
\end{center}
\caption[The quantities $S_<$ and $S_>$ as functions of $d$.]
{The quantities $S_<$ (dotted line) and $S_>$ (solid line) as
functions of $d$.  The intervals $d\leq0$ and $0\leq d\leq1$ are
plotted separately for clarity.
\label{figslessgtr}
}
\end{figure}

Thus, for the potentials with $d$\lt1, the qualitative nature of the
low-lying levels is expected to depend upon the excitation energy of
these levels relative to specific extremum features of the potential
well, and three possible different structural regimes are possible at
each given value of $d$.  We can estimate the values for $S$, at a
given $d$, at which we expect the transition to occur from the
lowest-energy structural regime to the intermediate regime [call this
$S_<(d)$] and at which we expect the transition to occur from this
intermediate regime to the anharmonic oscillator
regime [call this $S_>(d)$].  The energy which the ground state must
have, relative to the lowest point of the potential, in order to be at
the approximate boundary between each type of structure is given
Table~\ref{tabdeltav}, where it is denoted $\Delta V_<$ or $\Delta
V_>$, respectively.  The Wentzel-Kramers-Brillouin approximation (\eg,
Ref.~\cite{messiah1999:qm}) yields a quantization condition
\begin{equation}
\label{eqnfivedimwkb}
\int_0^{\beta_\text{max}}d\beta
\sqrt{B[E-V(\beta)]}\approx\left(n+\frac{3}{4}\right)\hbar\pi
\qquad n=0,1,2,\ldots
\end{equation}
on the radial coordinate in the five-dimensional Schr\"odinger
equation, ignoring here the five-dimensional equivalent of the
centrifugal potential, where the mass $B$ appearing in the usual form
of the Schr\"odinger equation is $\sqrt{5}B_2/2$
(Appendix~\ref{appscaling}).  Since we seek only an order-of-magnitude
estimate, let us consider only the ground state and replace
$E-V(\beta)$ in this equation by a constant value $E-\overline{V}$
representing the excitation energy relative to the
``average'' floor of the well.  Then~(\ref{eqnfivedimwkb}) reduces to the 
relation,
\begin{equation}
\beta_\text{max}\approx\frac{1}{2}\frac{2\pi\hbar}{\sqrt{2B(E-\overline{V})}},
\end{equation} 
which simply states that, for the ground state, approximately one-half
of a de~Broglie wavelength must fit within the width
$\beta_\text{max}$ of the well.  For
the GCM potential, let us approximate the width of the well by
$\beta_+$ and require that the ground state lie at the energy $\Delta
V_<$ or $\Delta V_>$ relative to the floor of the well.  This provides the
desired estimates for the parameter values at which the transition
between structural regimes occurs,
\begin{equation}
\label{eqnsestimate}
S_\lessgtr\approx\frac{1}{\pi^2\hbar^2}\left(\frac{\Delta V_\lessgtr}{f}\right)\left(\frac{\beta_+}{e}\right)^2,
\end{equation}
where factors of order unity have been suppressed in this derivation.
The values of $S_\lessgtr(d)$ obtained by substituting the expressions
for $\Delta V_\lessgtr$ from Table~\ref{tabdeltav} are plotted in
Fig.~\ref{figslessgtr}

For $d$\lt8/9, rotational behavior occurs when the levels are at
a sufficiently low energy with respect to the
potential.  For these rotational nuclei, several useful analytic
estimates can be made of the quantitative dependence of observables on
$d$ and $S$.  For states sufficiently low-lying (well-confined) in the
deformed minimum, the structure approaches that of small oscillations
about a deformed equilibrium, which is described analytically in the
rotation-vibration model (RVM)~\cite{faessler1965:rvm}.  The potential
in the region of the minimum is treated as a paraboloid, and, to first
approximation, pure harmonic oscillations occur in the $\beta$ and
$\gamma$ degrees of freedom.  Using the leading-order approximations
from this model, the $2^+$ state energy for the yrast band is
determined by the moment of inertia, giving
\begin{equation}
E_{2g}\approx \frac{\hbar^2}{B \beta_0^2},
\end{equation}
where $\beta_0$ is the equilibrium deformation.  

The $\beta$-vibrational and $\gamma$-vibrational excitation energies
are determined by the curvature of the potential in each of these
degrees of freedom, yielding
\begin{equation}
\label{eqnrvmbetagamma}
E_\beta\approx\hbar\sqrt{\frac{V_{\beta\beta}}{B}}
\qquad
E_\gamma\approx\hbar\sqrt{\frac{V_{\gamma\gamma}}{B\beta_0^2}},
\end{equation}
where $V_{\beta\beta}$$\equiv$$\partial^2V/\partial\beta^2$ and
$V_{\gamma\gamma}$$\equiv$$\partial^2V/\partial\gamma^2$.  The
bandhead state energies are related to $E_\beta$ and $E_\gamma$ by
$E(0^+_\beta)\approx E_\beta$ and $E(2^+_\gamma)\approx
E_\gamma+\frac{1}{3}E_{2g}$~\cite{faessler1965:rvm}.
For the GCM potential (\ref{eqnpotldef}), the curvatures in the $\beta$
and $\gamma$ directions, evaluated at the global minimum
$\beta=\beta+$, are
\begin{align}
\label{eqnvbbpvggp}
V_{\beta\beta}(\beta_+,0)&=\frac{9}{28}r(1+r)\frac{f}{e^2}\\
\beta_+^{-2}V_{\gamma\gamma}(\beta_+,0)&=\frac{27}{28}(1+r)\frac{f}{e^2},
\end{align}
in terms of $r=\sqrt{1-d}$.  Note that these $\beta$ and
$\gamma$ curvatures have
\textit{different} dependences upon $d$, through $r$, so their
\textit{ratio} varies with $d$.
Substituting these expressions for $V_{\beta\beta}$ and
$\beta^{-2}V_{\gamma\gamma}$, along
with the value $\beta_0\approx\beta_+$, into
(\ref{eqnrvmbetagamma}) yields the estimates
\begin{align}
E_{2g}&\approx \frac{224}{45}\frac{1}{(1+r)^2}\frac{\hbar^2}{Be^2}\\
E_\beta&\approx \sqrt{\frac{9}{28}}\sqrt{r(1+r)\frac{\hbar^2f}{Be^2}}\\
E_\gamma&\approx \sqrt{\frac{27}{28}}\sqrt{(1+r)\frac{\hbar^2f}{Be^2}}.
\end{align}
From the ratios of these expressions, the vibrational energies
normalized to the yrast $2^+$ energy are
\begin{align}
\label{eqnbetagammas}
\frac{E_\beta}{E_{2g}}&\approx
\frac{135}{448\sqrt{7}}\sqrt{\frac{\sqrt{5}}{2}\frac{r(1+r)^5}{\hbar^2 S}}\\
\frac{E_\gamma}{E_{2g}}&\approx
\frac{135\sqrt{3}}{448\sqrt{7}}\sqrt{\frac{\sqrt{5}}{2}\frac{(1+r)^5}{\hbar^2S}},
\end{align}
and the ratio of the $\beta$ and $\gamma$ vibration excitation
energies is
simply
\begin{equation}
\label{eqnbetagammaratio}
\frac{E_\beta}{E_\gamma}\approx\sqrt{\frac{r}{3}}.
\end{equation}
These ratios of energies
(\ref{eqnbetagammas})--(\ref{eqnbetagammaratio}) depend only upon $d$
and $S$, as expected from the scaling properties of
Section~\ref{secgcmscaling}.
These estimates provide guidance (needed below) as to both the range of $d$ values of
physical interest and the appropriate axis scale or calculational mesh
spacing for the parameter $d$.  Observe that,
by~(\ref{eqnbetagammaratio}), it is expected that ``$\beta$-stiff''
rotors, with $E_\beta$$>$$E_\gamma$, occur for $d$\lt-8, while
``$\gamma$-stiff'' rotors, with $E_\gamma$$>$$E_\beta$, occur for $d$\gt-8.

\begin{figure}[p]
\begin{center}
\includegraphics*[width=1.0\hsize]{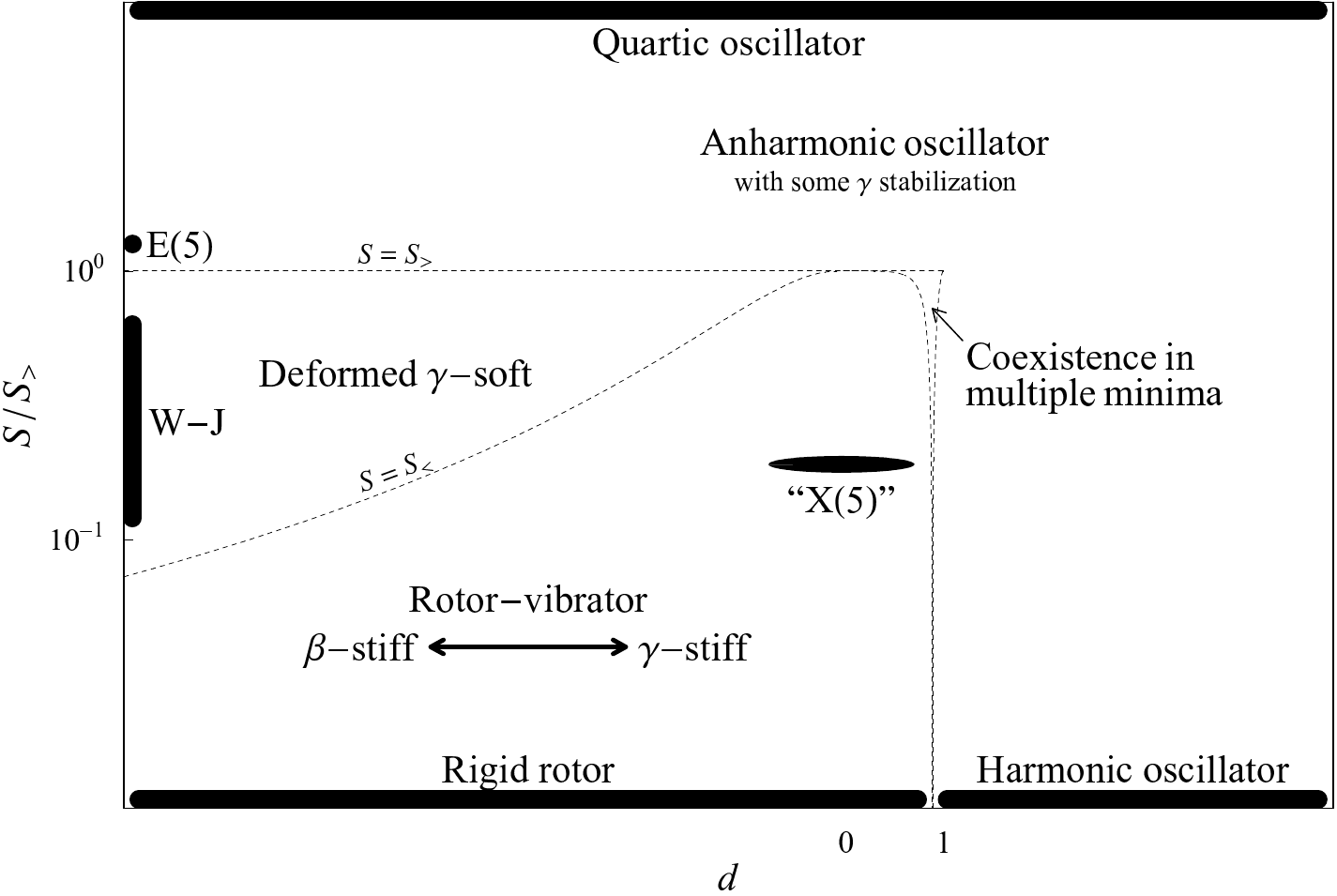}
\end{center}
\caption[Map of the GCM $(d,S)$ parameter space.]
{Map of the GCM $(d,S)$ parameter space.  The regions in which
qualitatively different structures occur are indicated.  The curves
$S=S_<$ and $S=S_>$ (dotted lines) provide estimates for the
approximate boundaries between these regions.  Within the
rotor-vibrator region, the stiffness for $\beta$ and $\gamma$
vibrations varies with $d$ (double arrow) with a dependence given
approximately by (\ref{eqnbetagammaratio}).  Bars along the edges of
the plot represent structures which occur in their ideal form at
$d\rightarrow\pm\infty$ or $S\rightarrow0$~or~$\infty$, as discussed
in more detail in Section~\ref{secgcmgsoft}.  (``W-J'' denotes
Wilets-Jean rigidly-deformed $\gamma$-soft structure.)  Parameter
values which reproduce E(5)-like and X(5)-like structures are also
investigated in Chapter~\ref{chapgcmspecial}.  The $d$- and $S$-axis
scales match those of
Figs.~\ref{figgcmcontour_full_e}--\ref{figgcmcontour_full_be2} (see
text) to facilitate direct comparison.
\label{figgcmmap}
}
\end{figure}

The combined results of this section give a detailed picture of the
qualitative characteristics expected for predictions of the truncated
GCM and provide quantitative estimates as to where in the
$(d,S)$ parameter space these properties are to be found.  The results
are summarized graphically as a ``map'' of the parameter space in
Fig.~\ref{figgcmmap}.

\begin{figure}[p]
\begin{center}
\includegraphics*[width=1.0\hsize]{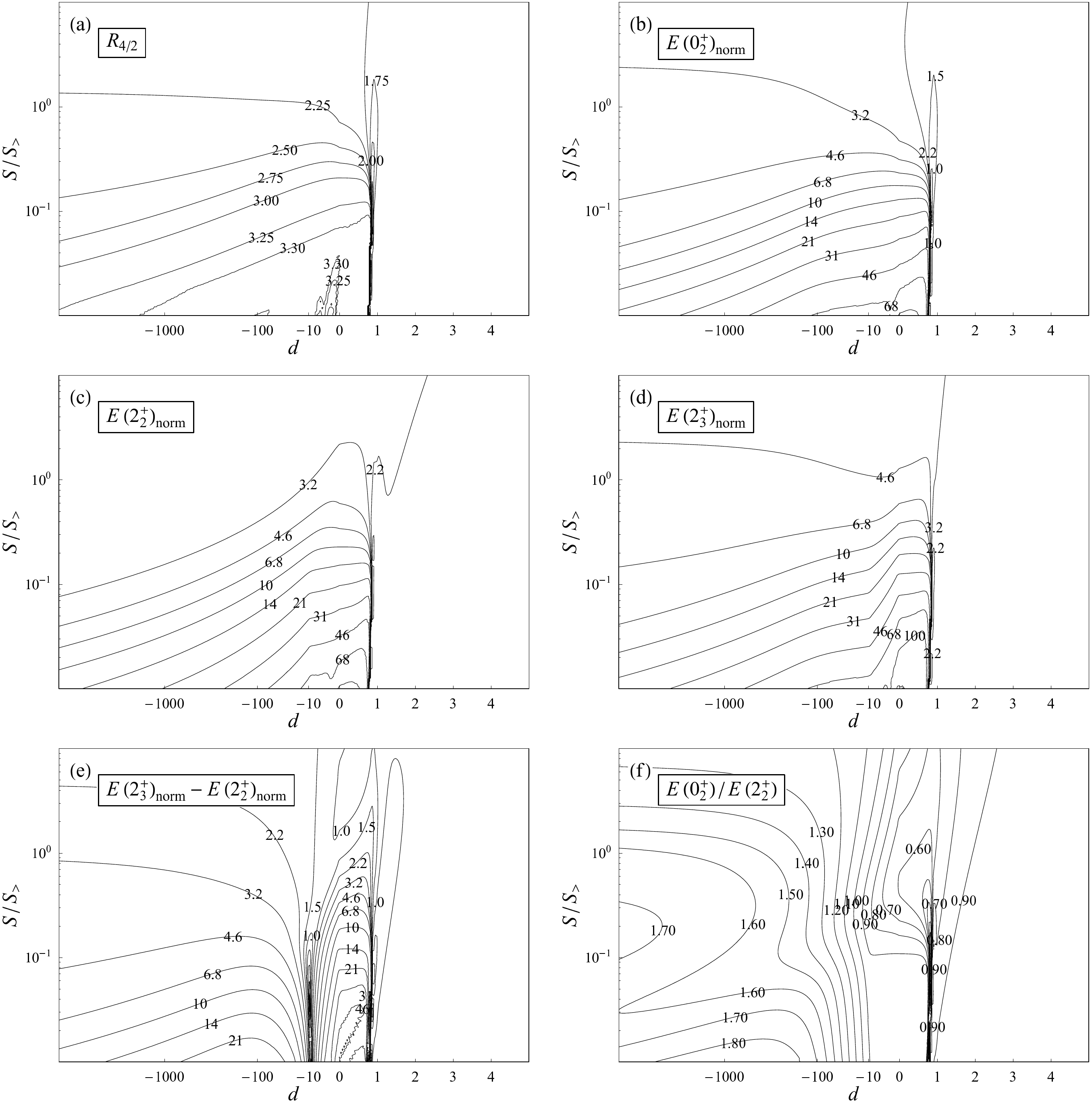}
\end{center}
\caption[Energy observable predictions for $-5000\leq d \leq 5$.]
{Energy observable predictions of the GCM for $-5000\leq d \leq 5$ and
$10^{-2}S_>\leq S \leq10^{+1}S_>$: (a)~$R_{4/2}\equiv
E(4^+_1)/E(2^+_1)$, (b)~$E(0^+_2)/E(2^+_1)$, (c)~$E(2^+_2)/E(2^+_1)$,
and (d)~$E(2^+_3)/E(2^+_1)$.  The observables
(e)~$[E(2^+_3)-E(2^+_2)]/E(2^+_1)$ and (f)~$E(0^+_2)/E(2^+_2)$ can be
deduced from these.
\label{figgcmcontour_full_e}
}
\end{figure}

\begin{figure}[p]
\begin{center}
\includegraphics*[width=1.0\hsize]{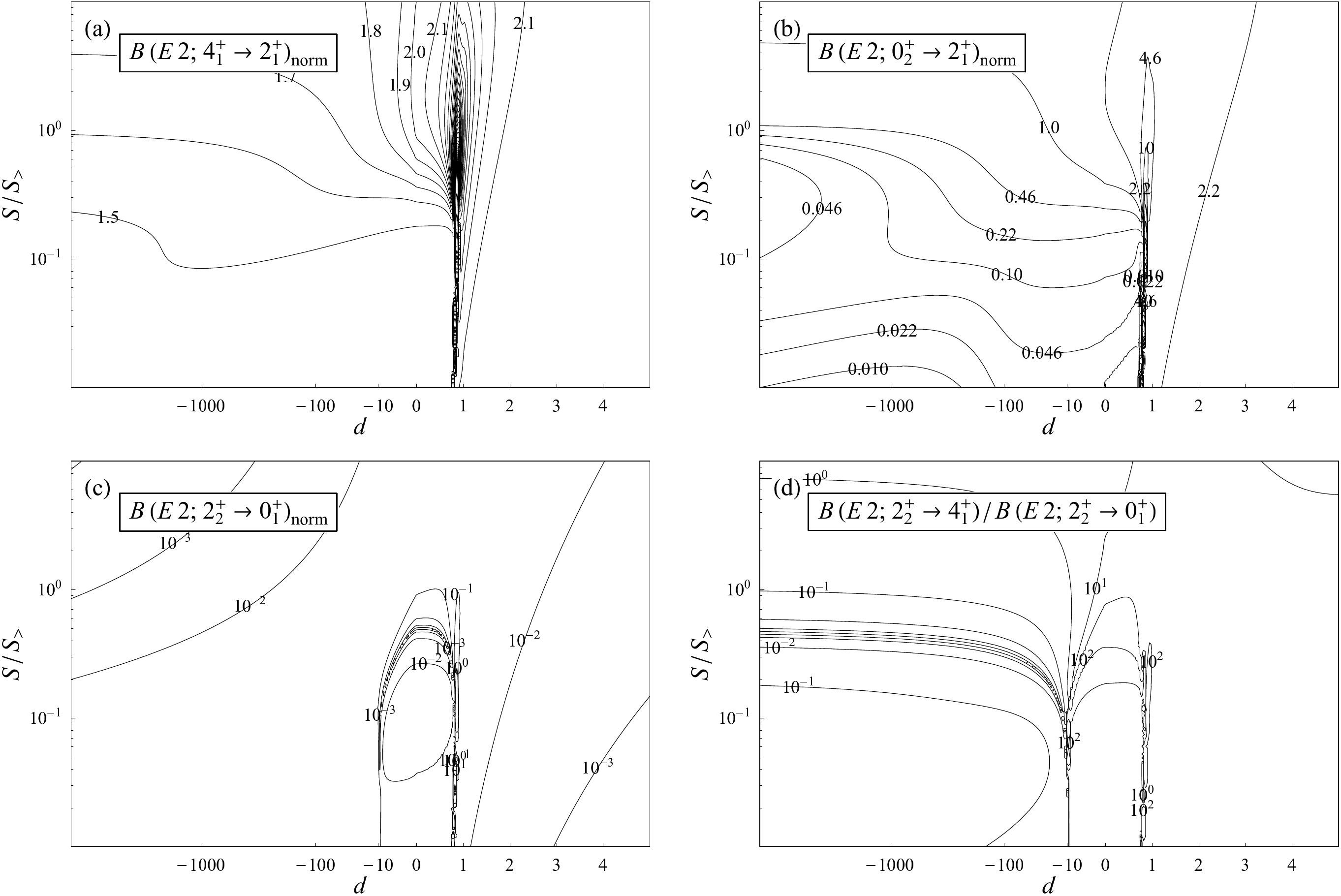}
\end{center}
\caption[$B(E2)$ observable predictions for $-5000\leq d \leq 5$.]
{$B(E2)$ observable predictions of the GCM for $-5000\leq d \leq 5$
and $10^{-2}S_>\leq S \leq10^{+1}S_>$:
(a)~$B(E2;4^+_1\rightarrow2^+_1)/B(E2;2^+_1\rightarrow0^+_1)$,
(b)~$B(E2;0^+_2\rightarrow2^+_1)/B(E2;2^+_1\rightarrow0^+_1)$,
(c)~$B(E2;2^+_2\rightarrow0^+_1)/B(E2;2^+_1\rightarrow0^+_1)$, and
(d)~$B(E2;2^+_2\rightarrow4^+_1)/B(E2;2^+_2\rightarrow0^+_1)$.
\label{figgcmcontour_full_be2}
}
\end{figure}

Contour plots of several observables over the $(d,S)$ parameter space
are shown in
Figs.~\ref{figgcmcontour_full_e}--\ref{figgcmcontour_full_be2}.  All observables plotted are \textit{ratios} of energies or of $B(E2)$
values.  As discussed in the previous section, at a
given $(d,S)$
any desired overall normalization for the energy and $B(E2)$
scales can then be obtained by proper rescaling of the well width, well depth,
and mass parameter.
\begin{figure}[p]
\begin{center}
~~~\vspace{12pt}\\
\includegraphics*[width=1.0\hsize]{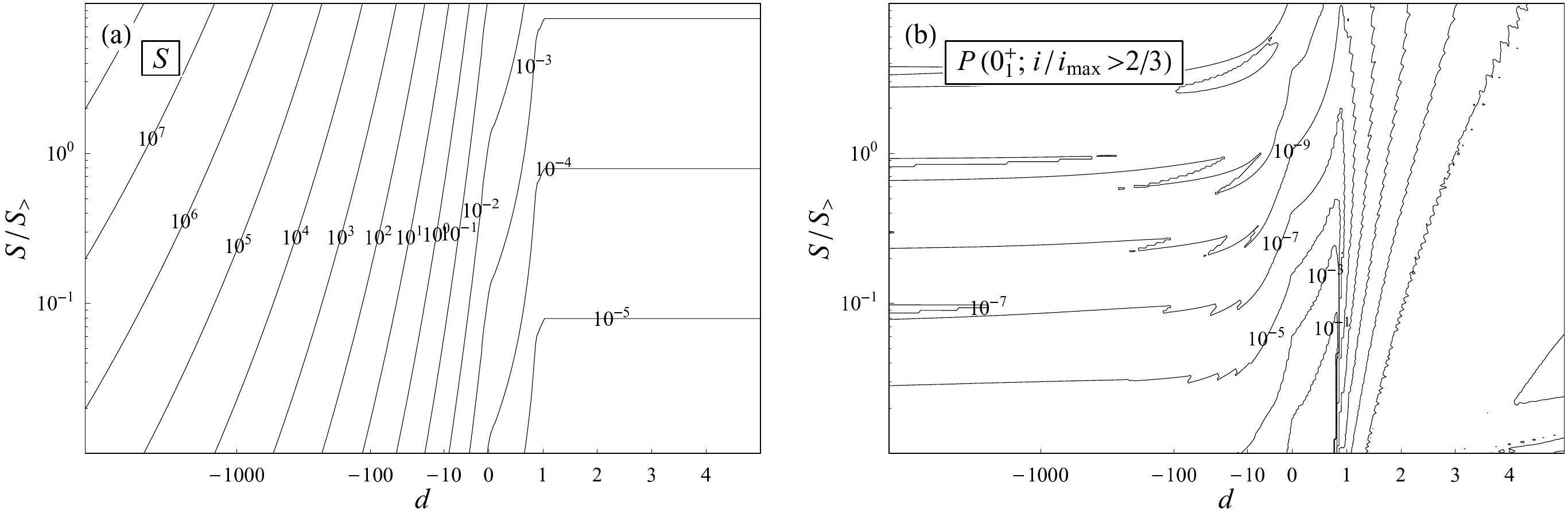}
\end{center}
\caption[Auxiliary plots for $-5000\leq d \leq 5$.]
{Auxiliary plots for $-5000\leq d \leq 5$ and $10^{-2}S_>\leq S
\leq10^{+1}S_>$: (a)~$S$ values for the calculations of
Figs.~\ref{figgcmcontour_full_e}--\ref{figgcmcontour_full_be2}, to allow $S$
parameter values to be read directly, rather than as $S/S_>$, for
points on these plots, and (b)~the total probability content of the
highest-indexed 1/3 of basis states for the calculated $0^+_1$ state,
as an indicator of convergence (see text).
\label{figgcmcontour_full_aux}
}
\end{figure}

The $d$-axis in
Figs.~\ref{figgcmcontour_full_e}--\ref{figgcmcontour_full_be2} extends from
$d$=-5000 to 5.  Inclusion of the low end of this range is necessary
to allow description of rotational nuclei with high-lying $\beta$
vibrations~(\ref{eqnbetagammaratio}).  In order to encompass this full
range while maintaining a reasonably detailed view of the region
around $d$=0 in the plots, it is helpful to use a nonlinear $d$-axis
scale for $d\lt0$.  The estimate~(\ref{eqnbetagammaratio}) indicates
that for rotor-vibrator nuclei the observable $E_\beta/E_\gamma$
varies as $(1-d)^{1/4}$, so for $d<0$ the $d$-axis of
Figs.~\ref{figgcmcontour_full_e}--\ref{figgcmcontour_full_be2} is chosen to
be linear in $(1-d)^{1/4}$.

In the range of $d$ values being considered, the $S$ values resulting
in pheonomena of interest span approximately \textit{fourteen} orders
of magnitude, as is seen from Fig.~\ref{figslessgtr}.  This occurs
since $S$ is defined in terms of $e$ and $f$, and the values of these
parameters needed to construct a reasonably-sized potential vary
greatly with $d$ (see Fig.~\ref{figgcmpotld} for examples).  However, at
any particular value of $d$, only about three decades in $S$, those
immediately surrounding $S$=$S_>(d)$, contain predictions of
interest.  To make effective use of plotting
space, the $S$-axis in
Figs.~\ref{figgcmcontour_full_e}--\ref{figgcmcontour_full_be2} is
expanded to show only $10^{-2}S_>(d)\leq S
\leq10^{+1}S_>(d)$ at each point along the $d$-axis.
Fig.~\ref{figgcmcontour_full_aux}(a) facilitates the reading of $S$ directly
off the contour plots.

Details of the calculational mesh used to generate
Figs.~\ref{figgcmcontour_full_e}--\ref{figgcmcontour_full_be2} are given
in Table~\ref{tabgcmmesh}.  A single point in $(d,S)$ space
corresponds to an entire two-parameter family of points related by
scaling transformations in the $(B_2,C_2,C_3,C_4)$ parameter space,
but for numerical diagonalization to be carried out by the code of
Ref.~\cite{troltenier1991:gcm} a specific set of values for the
Hamiltonian coefficients $B_2$, $C_2$, $C_3$, and $C_4$ must be
chosen.  The choice is largely irrelevant, but the diagonalization
code~\cite{troltenier1991:gcm} can fail to produce convergent results
if the eigenstates have $\langle\beta\rangle$ many orders of magnitude
away from 1.  The actual calculations for
Figs.~\ref{figgcmcontour_full_e}--\ref{figgcmcontour_full_be2} were
performed by setting $B_2$=50\,MeV.  For $d\leq1$, $e$ was chosen such that
$\beta_+$=0.3, and, for $d>1$, $e$ was set to 0.669.
\begin{table}
\begin{center}
\begin{tabular}{l=c=r,c,l=r=l}
\pseudoruledtabular
\multicolumn{1}{c}{Figure}&
\multicolumn{1}{c}{Axis}&
\multicolumn{3}{c}{Range}&
\multicolumn{1}{r}{Steps}&
\multicolumn{1}{l}{Distribution function}
\\
\hline
Figs.~\ref{figgcmcontour_full_e}--\ref{figgcmcontour_full_aux} &
$d$&
-5000&--&5&
300&
\ensuremath{\begin{array}[t]{l}\left\lbrace\begin{array}{@{}ll}1-(1-d)^{1/4}&d<0\\
d&d\geq0\end{array}\right.\end{array}}
\\
&
$S$&
$10^{-2}S_>$&--&$10^{+1}S_>$&
200&
$\log S$\\
\hline
Fig.~\ref{figgcmcontour_zoom} &
$d$&
0&--&1&
100&
$d$\\
&
$S$&
$10^{-2}S_>$&--&$10^{+1}S_>$&
50&
$\log S$\\
\pseudoruledtabular
\end{tabular}
\end{center}
\caption
[Calculational meshes used to generate contour plots.] {\ssp 
Calculational meshes used to generate the contour plots of
Figs.~\ref{figgcmcontour_full_e}--\ref{figgcmcontour_zoom}.
Calculation mesh points are equally spaced with respect to the
distribution function in the rightmost column.
\label{tabgcmmesh}
}
\end{table}

Convergence failure of the code occurs for some of the most sharply
deformed structures covered by
Figs.~\ref{figgcmcontour_full_e}--\ref{figgcmcontour_full_be2}.  At very low
$S$ values in the rotor region, where the most clearly well-deformed
structure occurs and $R_{4/2}$ should approach 3.33, the numerical
calculations instead exhibit sporadic patches of sharp fall-off in the
$R_{4/2}$ value [Fig.~\ref{figgcmcontour_full_e}(a), at $d\lesssim0$].
As outlined in Subsection~\ref{subsecgcmnumerical}, an indication of
the convergence properties of a calculation is given by the 
total fraction of the probability contained in the high-phonon-number
basis states.  Fig.~\ref{figgcmcontour_full_aux}(b) provides a contour
plot of the probability content of the highest 1/3 of basis states, in
order of increasing phonon number~\cite{troltenier1991:gcm}, for the
calculated $0^+_1$ state.  The calculations in the region of
convergence failure exhibit relatively high values ($\gtrsim10^{-3}$)
for this quantity.

\begin{figure}[p]
\begin{center}
\includegraphics*[width=1.0\hsize]{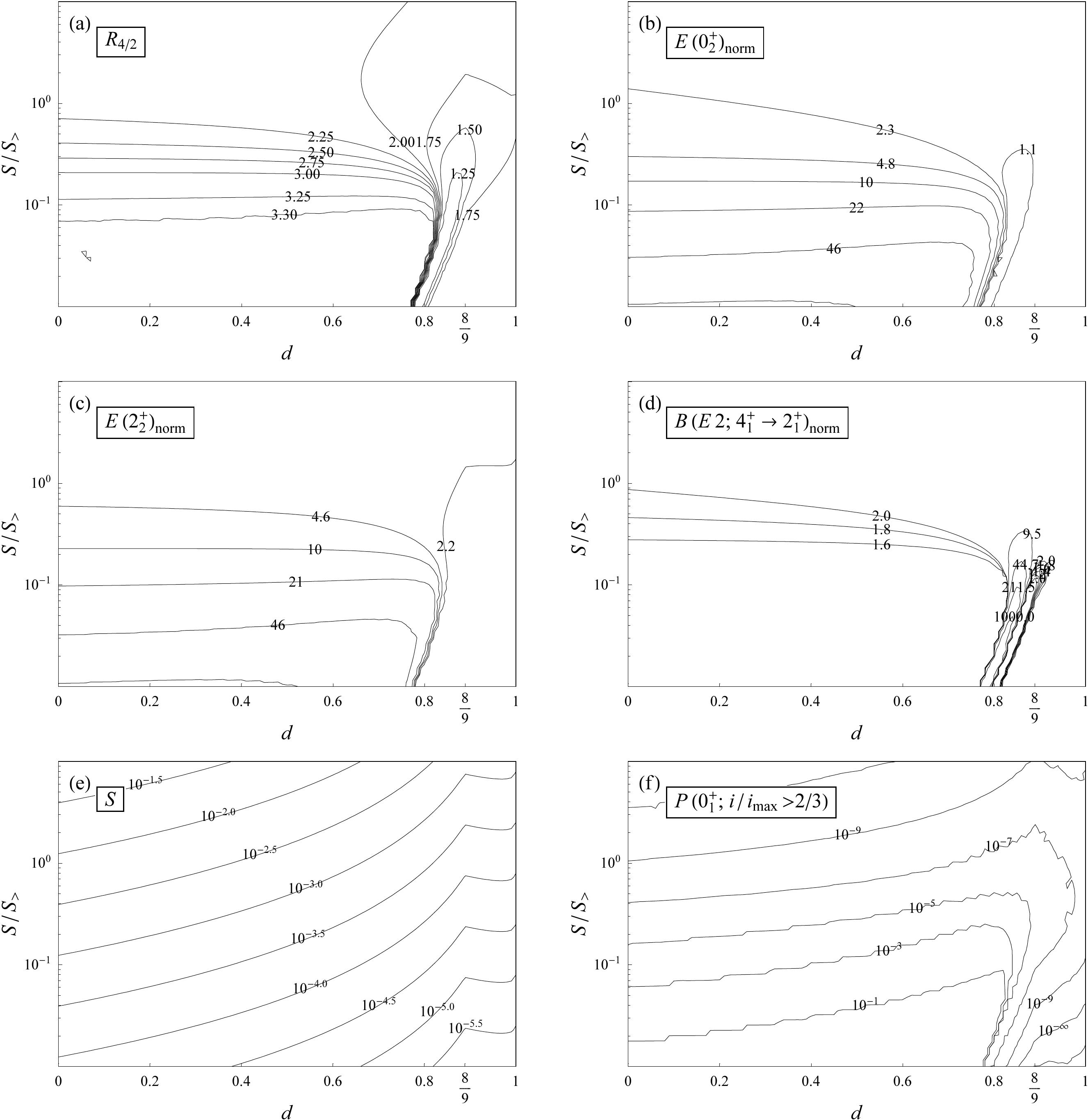}
\end{center}
\caption[Selected predictions of the GCM for $0\leq d \leq 1$.]
{Selected prediction of the GCM for $0\leq d \leq 1$ and
$10^{-2}S_>\leq S \leq10^{+1}S_>$, with additional auxiliary plots:
(a)~$R_{4/2}\equiv E(4^+_1)/E(2^+_1)$, (b)~$E(0^+_2)/E(2^+_1)$,
(c)~$E(2^+_2)/E(2^+_1)$,
(d)~$B(E2;4^+_1\rightarrow2^+_1)/B(E2;2^+_1\rightarrow0^+_1)$,
(e)~$S$, (f)~probability content of the highest-indexed 1/3 of basis
states for the $0^+_1$ state calculation, as an indicator of convergence.
\label{figgcmcontour_zoom}
}
\end{figure}

Since many observables vary rapidly in the interval 0\lt$d$\lt1, a few
selected contour plots showing an expanded view of this region are
given in Fig.~\ref{figgcmcontour_zoom}.  Caution must be exercised in
using the results produced by the code of
Ref.~\cite{troltenier1991:gcm} for parameter sets near $d=0.8$, since
Fig.~\ref{figgcmcontour_zoom}(f) indicates possible convergence failure
in this region.  Detailed convergence studies must be carried out to
determine whether, and to what extent, this is attributable to a
failure in the automatic optimization of the basis stiffness or to a
genuine need for basis functions with $N>30$ in the diagonalization
(Subsection~\ref{subsecgcmnumerical}).

Energy observables are shown in Fig.~\ref{figgcmcontour_full_e}.  The
values of the ratio $R_{4/2}\equiv
E(4^+_1)/E(2^+_1)$~[Fig.~\ref{figgcmcontour_full_e}(a)], an observable
which serves as a basic indicator of structure
(Appendix~\ref{appbench}), closely match the values expected from the
$S_<$ and $S_>$ estimates.  The region with 2.2\lt$R_{4/2}$\lt2.6
corresponds approximately to that between the dotted lines
representing $S_<$ and $S_>$ in Fig.~\ref{figgcmmap}, the region in
which deformed $\gamma$-soft structure is expected.  For the
rotational-vibrational nuclei, found in the lower left-hand region of
the plots, the observables $E(0^+_2)/E(2^+_1)$, $E(2^+_2)/E(2^+_1)$,
and $E(2^+_3)/E(2^+_1)$ [Fig.~\ref{figgcmcontour_full_e}(b-d)] reflect
the basic dependences of the $\beta$ and $\gamma$ excitation energies
estimated in~(\ref{eqnbetagammas}) and~(\ref{eqnbetagammaratio}).  At a
given $d$, the excited band energies increase relative to $E(2^+_1)$
as $S$ decreases.  Degeneracy of
the $\beta$ and $\gamma$ excitations is expected at $d$$\approx$$-8$,
and this behavior is clearly visible from the sharp minimum in
$[E(2^+_3)-E(2^+_2)]/E(2^+_1)$ [Fig.~\ref{figgcmcontour_full_e}(e)],
where an avoided level crossing occurs.  To the left of this division,
the $2^+_2$ state is the $\gamma$-vibrational bandhead; whereas, to
the right of this division, the $2^+_2$ state is the
$\beta$-vibrational $2^+$ band member.  Proceeding to the left of the
band crossing, the $0^+_2$ level energy rises relative to the $2^+_2$
energy [Fig.~\ref{figgcmcontour_full_e}(f)], as expected, due to
increasing $\beta$ stiffness.  The $0^+_2$ energy saturates at
$\lesssim 2E(2^+_2)$, however, since as the $\beta$ vibrational
excitation continues to rise it leaves the two-phonon
$\gamma$-vibrational state as the lowest $K=0$ excitation (see
Ref.~\cite{faessler1965:rvm}).

$B(E2)$ observable predictions are shown in
Fig.~\ref{figgcmcontour_full_be2}.  The ratio
$B(E2;4^+_1\rightarrow2^+_1)/B(E2;2^+_1\rightarrow0^+_1)$
[Fig.~\ref{figgcmcontour_full_be2}(a)] varies essentially smoothly from
rotational values to harmonic oscillator
values~(Appendix~\ref{appbench}), except that extreme large and small
values are encountered in a narrow region between $d\approx0.8$ and
$d\approx0.9$.  In this region, structures involving coexistence in
multiple minima are expected, so the $2^+_1$ and $4^+_1$ levels do not
necessarily correspond to the same structure.  Numerical convergence
may also not be occurring for certain calculations in this region, as
discussed above.  The observable
$B(E2;0^+_2\rightarrow2^+_1)/B(E2;2^+_1\rightarrow0^+_1)$
[Fig.~\ref{figgcmcontour_full_be2}(b)] is of interest as the
$\beta$-vibrational decay strength for rotational nuclei.  In the
region $d\lesssim-8$,
$B(E2;2^+_2\rightarrow0^+_1)/B(E2;2^+_1\rightarrow0^+_1)$
[Fig.~\ref{figgcmcontour_full_be2}(c)] is the $\gamma$-vibrational decay
strength.  The predictions for the observable
$B(E2;2^+_2\rightarrow4^+_1)/B(E2;2^+_2\rightarrow0^+_1)$
[Fig.~\ref{figgcmcontour_full_be2}(d)] may be compared with the ``Alaga
ratio'' predictions for true rotors (Section~\ref{secbenchrotor}),
those for the $2^+_\gamma$ state if $d\lesssim-8$ or for the
$2^+_\beta$ state if $d\gtrsim-8$.

The general approach and specific techniques discussed in this chapter
recast the GCM, with truncated Hamiltonian, as a very tractable model
for theoretical studies and practical application.
The parameter values most
appropriate for the description of a given nucleus can be deduced by
inspection of contour plots such as those in
Figs.~\ref{figgcmcontour_full_e}--\ref{figgcmcontour_zoom}.  Some
specialized results pertaining to specific classes of potential are
addressed in the following chapter.  Examples of the phenomenological
application of the model are discussed in Chapter~\ref{chapphenom}.

\chapter{Special potentials in the geometric collective model}
\label{chapgcmspecial}

\section{The quadratic-quartic potential}
\label{secgcmgsoft}

Gamma-independent potentials have special symmetry properties.  The
Hamiltonian consisting of a $\gamma$-independent potential with the
harmonic kinetic energy operator is invariant under the group O(5) of
rotations in the five-dimensional space of the coordinates
$\alpha_{2\mu}$~(Section~\ref{secbenchgsoft}).  This has several
consequences, including that the level energies follow a multiplet
structure and the wave functions are separable into radial and angular
factors.  The harmonic oscillator, E(5), and well-deformed
$\gamma$-soft structures introduced in Sections~\ref{seccollective}
and~\ref{sectrans} are all generated by $\gamma$-independent
potentials.

\begin{figure}
\begin{center}
\includegraphics*[width=0.8\hsize]{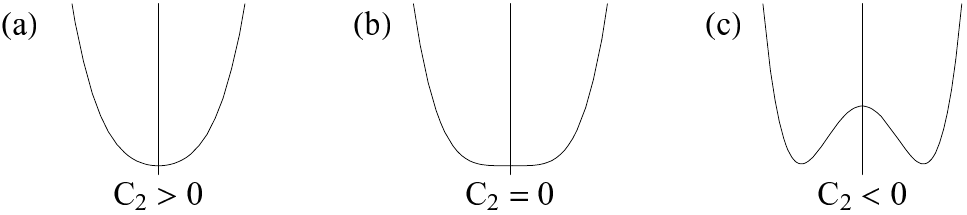}
\end{center}
\caption[Possible shapes of the quadratic-quartic potential.]
{Possible shapes of the quadratic-quartic potential, shown along the $a_0$-axis cut: (a)~$C_2>0$,
(b)~$C_2=0$, and (c)~$C_2<0$.  $C_4$ is positive in all cases.
\label{figpotlgsoft}
}
\end{figure}

If the coefficient $C_3$ in the truncated GCM potential
(\ref{eqnpotltrunc}) is chosen equal to
zero, so
\begin{equation}
\label{eqnpotlgsoft}
V(\beta,\gamma)= \frac{1}{\sqrt{5}} C_2\beta^2 +\frac{1}{5}C_4\beta^4,
\end{equation}
then the potential is $\gamma$-independent.  When $C_2$ is positive
and $C_4$ is positive [Fig.~\ref{figpotlgsoft}(a)], the potential has
only a minimum at $\beta$=0.
The quadratic term dominates at small $\beta$, and the quartic term
dominates at large $\beta$.  For $C_2$=0 [Fig.~\ref{figpotlgsoft}(b)],
the special case of a quartic oscillator occurs.  (This potential has
recently found application in the description of anharmonic oscillator
nuclei~\cite{caprio2002:110cd-anomaly}.)  When $C_2$ is negative and $C_4$ is
positive [Fig.~\ref{figpotlgsoft}(c)], the $\beta^2$ term produces an inverted paraboloid ``hump''
at small $\beta$, while the positive quartic term ensures that the
system is globally bound.  The resulting potential has a continuous
``ring'' of minima, located at the same
$\beta$ value,
\begin{equation}
\beta_0=\sqrt{-\frac{\sqrt{5}}{2}\frac{C_2}{C_4}},
\end{equation}
and at all possible $\gamma$ values.  The depth for these minima is
\begin{equation}
V(\beta_0)=-\frac{1}{4}\frac{C_2^2}{C_4}.
\end{equation}

In principle, the quadratic-quartic GCM potential~(\ref{eqnpotlgsoft})
can be treated using the parameters $d$, $e$, and $f$, as well as the
structure parameter $S=1/(B_2e^2f)$, of Section~\ref{secgcmscaling}.
Potentials with $C_3$=0 can be approached to any desired degree of
numerical accuracy by taking $d\rightarrow\pm\infty$, $e\rightarrow0$,
and $f\rightarrow+\infty$, as is seen from the conversion
formulae~(\ref{eqndefccc}).

However, if strictly potentials with $C_3=0$ are of interest, it is
more straightforward to apply the scaling properties of
Section~\ref{secgcmscaling} directly to the parameters $C_2$ and
$C_4$.  Overall multiplication of the Hamiltonian by a factor $b$ is
obtained by the parameter transformation
\begin{equation}
\label{eqntransfbgsoft}
B_2'= \frac{1}{b}B_2 \qquad C_2'=bC_2 \qquad C_4'=bC_4,
\end{equation}
while deepening the potential by $a^2$ and narrowing it by $a$ is
accomplished by the parameter transformation
\begin{equation}
\label{eqntransfagsoft}
B_2'= B_2 \qquad C_2'=a^4C_2 \qquad C_4'=a^6C_4.
\end{equation}
Since these two scaling tranformations leave energy spectra unchanged,
to within overall normalization, and wave functions unchanged, to
within dilation, there remains only one degree of freedom in the
parameter space to control the structure of the solutions.  Let us
define a structure parameter
\begin{equation}
S_\gamma \equiv - \frac{B_2C_2^3}{C_4^2},
\end{equation}
analogous to the parameter $S$ of Section~\ref{secgcmscaling}.  It is
easily verified that $S_\gamma$ is invariant under both tranformations
(\ref{eqntransfbgsoft}) and (\ref{eqntransfagsoft}).

The model consisting of a quadratic-quartic potential together with
the harmonic kinetic energy term is therefore a \textit{one-parameter}
model, producing results which are dependent only upon $S_\gamma$ to
within overall normalization for the energies and wave
function dilation.  Let us consider the qualitative dependence of the
predicted structure on $S_\gamma$.  For $S_\gamma$ large negative,
$C_2$ is positive, giving the potential of Fig.~\ref{figpotlgsoft}(a),
and the levels are low-lying in the potential, where the $\beta^2$
term dominates.  Thus, harmonic oscillator structure arises.  As
$S_\gamma$ increases towards zero, the levels rise in the potential,
and so the quartic term induces increasingly anharmonic behavior.  At
$S_\gamma$=0, the potential is a pure quartic
potential~[Fig.~\ref{figpotlgsoft}(b)].  For $S_\gamma$ positive,
$C_2$ is negative, producing the potential of
Fig.~\ref{figpotlgsoft}(c).  As $S_\gamma$ increases from zero, the
levels become progressively lower in this potential, and for
sufficiently large $S_\gamma$ the states are confined in the
$\beta$-deformed minimum.  In the limit $S_\gamma\rightarrow\infty$,
the rigidly-deformed $\gamma$-soft structure described analytically by
Wilets and Jean~(Section~\ref{secbenchgsoft}) is obtained.  Observe
that the structures obtained by varying $S_\gamma$ from $-\infty$ to
$+\infty$ occur, as limiting cases of the $(d,S)$ model, successively
along the right, top, and upper left edges of the parameter space
``map'' diagram (Fig.~\ref{figgcmmap}).

\begin{figure}[p]
\begin{center}
\includegraphics*[width=0.9\hsize]{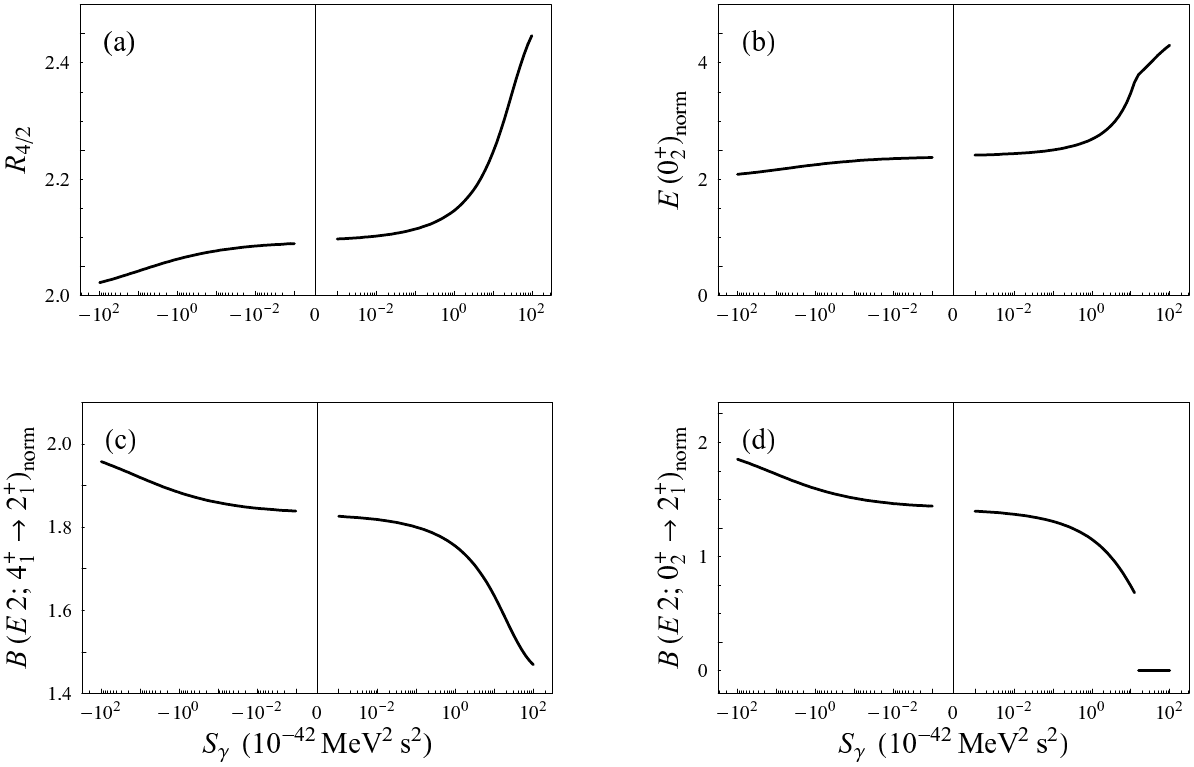}
\end{center}
\caption[Energy and $B(E2)$ observables for the
quadratic-quartic potential.]  {Some basic energy and $B(E2)$
observables for the quadratic-quartic potential, as a function of
$S_\gamma$: (a)~$R_{4/2}\equiv E(4^+_1)/E(2^+_1)$,
(b)~$E(0^+_2)/E(2^+_1)$,
(c)~$B(E2;4^+_1\rightarrow2^+_1)/B(E2;2^+_1\rightarrow0^+_1)$, and
(d)~$B(E2;0^+_2\rightarrow2^+_1)/B(E2;2^+_1\rightarrow0^+_1)$.
\label{figgsoftobs}
}
\end{figure}

The behavior of an observable over the entire model space can be
summarized on a single graph as a function of $S_\gamma$.  Plots of
some basic energy and $B(E2)$ observables are shown in
Fig.~\ref{figgsoftobs}.
\begin{figure}[p]
\begin{center}
\includegraphics*[width=0.65\hsize]{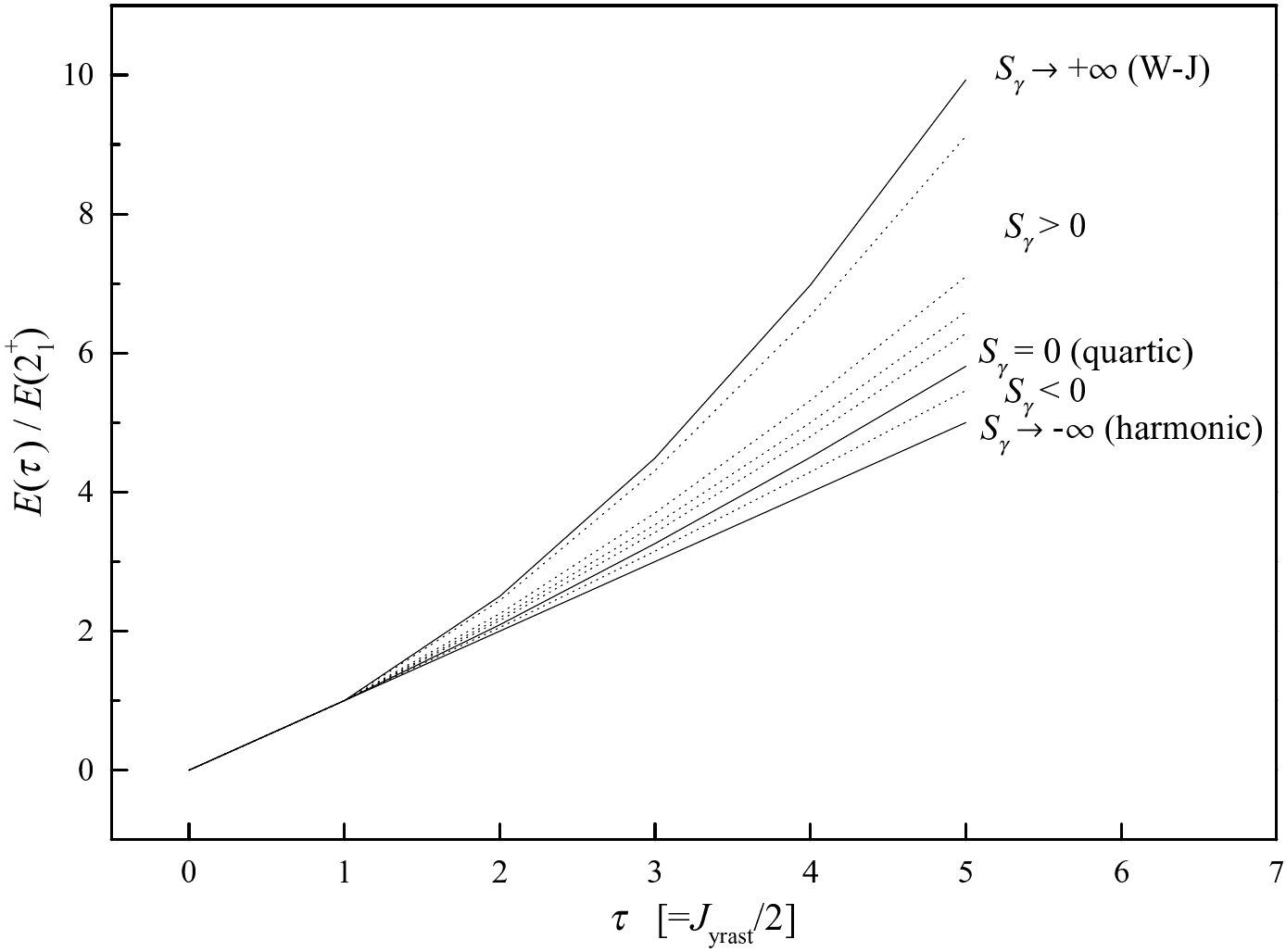}
\end{center}
\caption[Yrast level energies for the quadratic-quartic
potential.]  {Yrast level energies predicted by for the
quadratic-quartic potential as a function of multiplet number $\tau$,
or, equivalently, $J_\text{max}/2$, for various values of $S_\gamma$.
This curve is a straight line at $S_\gamma=-\infty$ (harmonic
oscillator) but becomes continuously more upward-curving as $S_\gamma$
is increased towards $S_\gamma=+\infty$ (Wilets-Jean limit).  (Figure
adapted from Ref.~\cite{caprio2002:110cd-anomaly}.)
\label{figgsoftyraste}
}
\end{figure}%
The observables $R_{4/2}$ and
$B(E2;4^+_1\rightarrow2^+_1)/B(E2;2^+_1\rightarrow0^+_1)$
[Fig.~\ref{figgsoftobs}(a,c)] vary monotonically from their values for
the harmonic oscillator to their values for the Wilets-Jean
rigidly-deformed structure.  Observables involving the $0^+_2$ state
exhibit a kink~[Fig.~\ref{figgsoftobs}(b)] or
discontinuity~[Fig.~\ref{figgsoftobs}(d)] at
$S_\gamma\approx$14\timesSgammaunits due to the occurence of a
crossing of the $0^+_2$ and $0^+_3$ levels.  For $S_\gamma$ less than
this value, the $0^+_2$ state is the $\tau=0$ head of the second
family of levels, but, for $S_\gamma$ greater than this value, the
$0^+_2$ state is the $\tau=3$ multiplet member of the first family of
level.  The curve of yrast energies as a function of spin or,
equivalently, of multiplet number $\tau$, is given in
Fig.~\ref{figgsoftyraste}.

\begin{figure}[p]
\begin{center}
\includegraphics*[width=1.0\hsize]{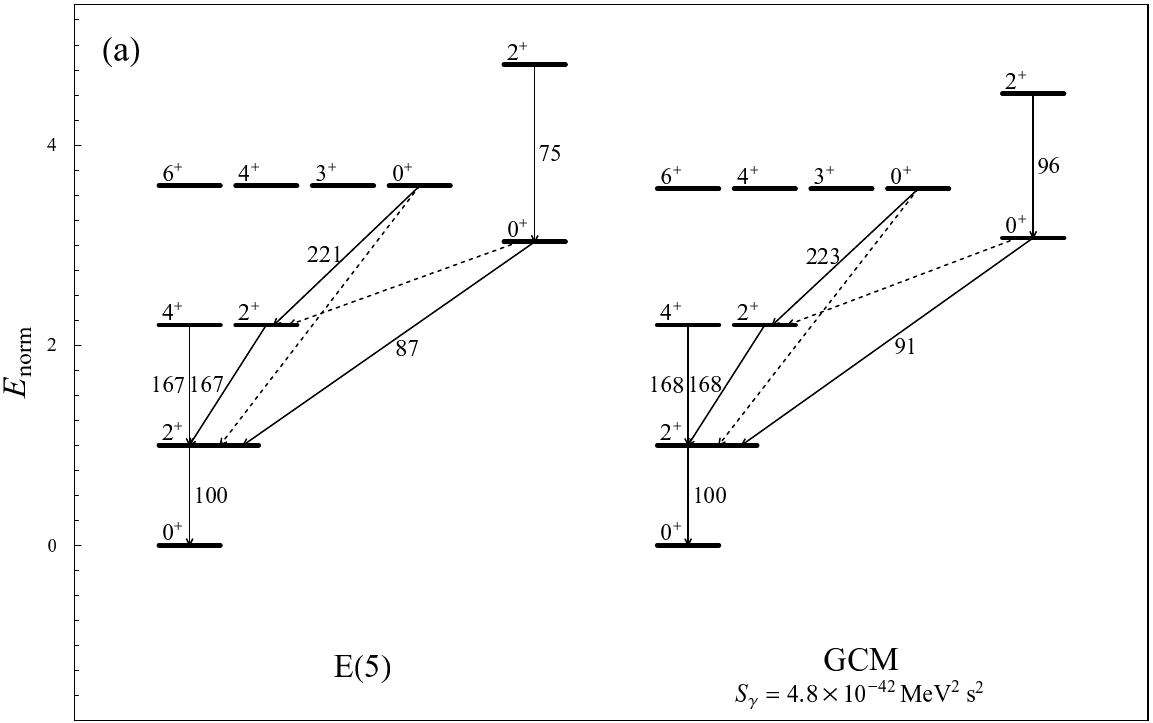}
\vspace{20pt}\\
\includegraphics*[height=2.5in]{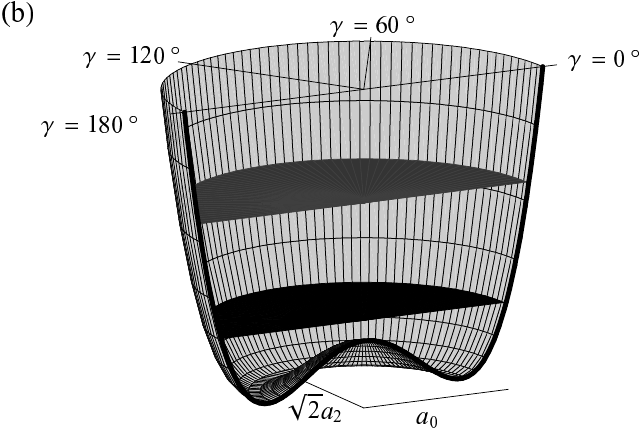}
\vspace{11pt}\\
\end{center}
\caption[E(5)-like structure in the GCM.]
{E(5)-like structure in the GCM, for $S_\gamma$=4.8\timesSgammaunits:
(a)~Level energies and $B(E2)$ strengths predicted in E(5) and the
GCM, normalized to $E(2^+_1)$ and $B(E2;2^+_1\rightarrow0^+_1)$.
Dotted arrows represent forbidden transitions.  (b)~A plot of the
potential for the GCM calculation as a function of the shape
coordinates, showing $0^+_1$ and $0^+_2$ level energies (black and
gray laminae).
\label{figgcme5}
}
\end{figure}

It is of interest to see to what extent, and for what parameter
values, $E(5)$-like behavior can be reproduced with the
quadratic-quartic potential in the GCM.  These results were reported
in part in Ref.~\cite{zhang2001:gcm-random}.  $R_{4/2}$ values of
2.18--2.22, approximately the value for E(5), occur for
$S_\gamma$$\approx$3.2--6.4\timesSgammaunits
[Fig.~\ref{figgsoftobs}(a)].  The level scheme for
$S_\gamma$=4.8\timesSgammaunits is shown alongside the E(5)
predictions (Section~\ref{secbenche5x5}) in Fig.~\ref{figgcme5}(a).
The predictions, especially those for level energies and transition
strengths within the ground state family of levels, match those for
the E(5) potential quite well.  The properties of successively
higher-lying families of levels reproduce those of E(5) less closely.
Note that the transitions which are strictly forbidden under the
$\Delta\tau=\pm1$ selection rule for the E(5) model with linear
transition operator (Section~\ref{secbenchgsoft}) are strictly
forbidden in the present calculations as well, but use of a
second-order transition operator in either
model~\cite{arias2001:134ba-e5,zhang2001:gcm-random} leads to
nonvanishing strengths in these transitions.

The $S_\gamma$ value used to produce this E(5)-like behavior results
in a ground state energy lying somewhat above the local maximum in the
potential at $\beta$=0 [Fig.~\ref{figgcme5}(b)].  The effect of the
maximum and minimum structure is essentially to provide a relatively
flat region in the potential before the walls of the well rise sharply
at large $\beta$.  The potential, for excitation energies in this
region, is thus conceptually similar to the square well potential (see
also Fig.~\ref{figflatbottom}).

\section{X(5)-like structure}
\label{secgcmx5}

The X(5) model is based upon the assumptions of a square well
potential with respect to $\beta$, extending from 0 to some
$\beta_\text{max}$, stability about $\gamma$=0$^\circ$, and complete
decoupling of the $\beta$ and $\gamma$ degrees of freedom.  The
resulting level structure consists of a series of $K^\pi=0^+$ bands,
but with energy spacings and $B(E2)$ strengths different from those of
a conventional rotor (Section~\ref{secbenche5x5}).  Similar structural
properties would seem most likely to occur in the GCM for potentials
which provide strong $\gamma$ confinement, and hence energetic
separation between the $\beta$ and $\gamma$ degrees of freedom, while
being relatively flat with respect to $\beta$.  These criteria are met
by the potentials with $d\approx8/9$ [Fig.~\ref{figgcmpotld} on
page~\pageref{figgcmpotld}], provided that the low-lying levels occur
at energies moderately above the extrema of the potential but well
below the onset of quartic oscillator structure,
\ie, for $S$ of roughly the same order of magnitude as $S_>$.

\begin{figure}[t]
\begin{center}
\includegraphics*[width=0.9\hsize]{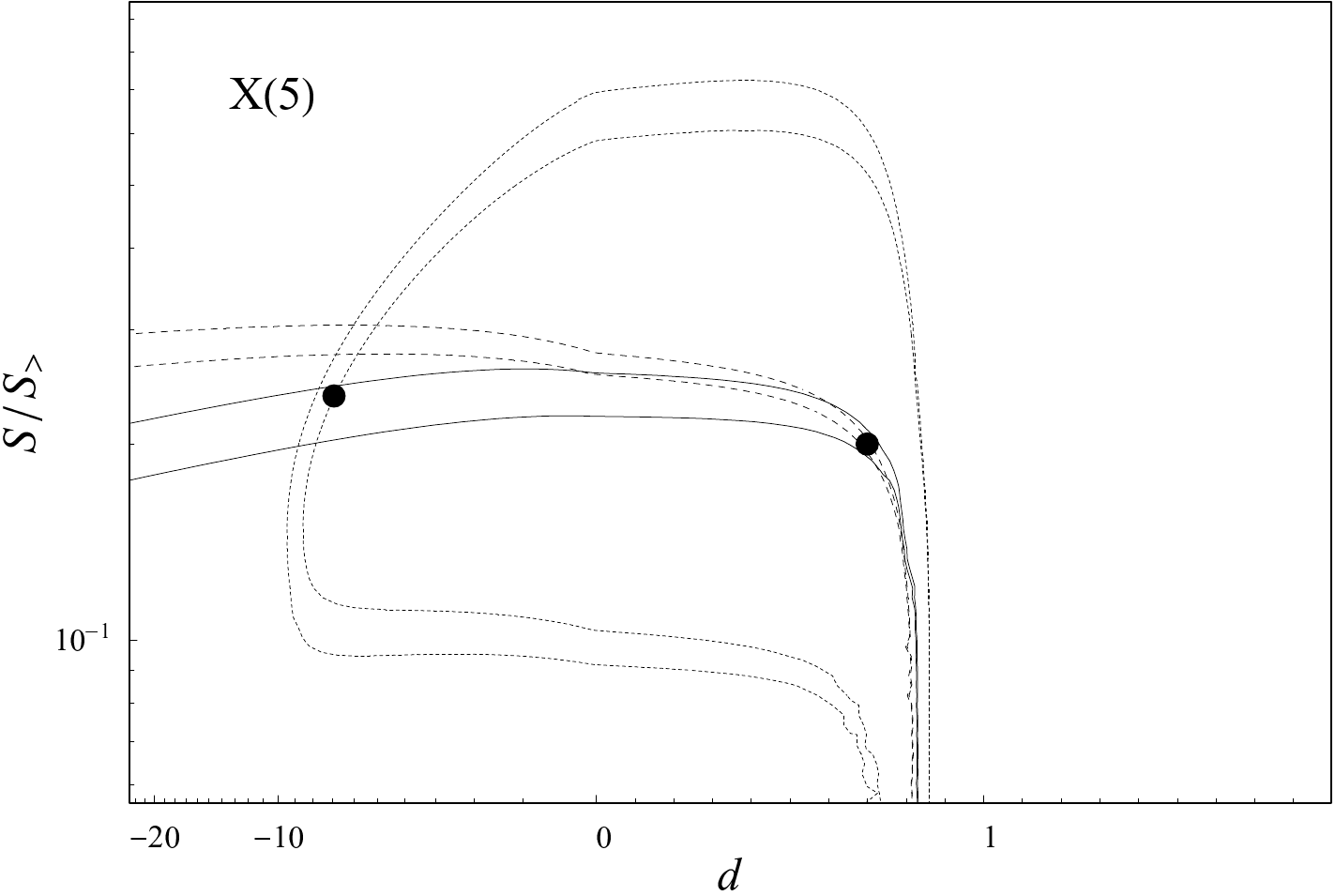}
\end{center}
\caption[Regions for which GCM predictions match
those of X(5).]  {Regions in the ($d$,$S$) parameter space for which
the GCM predictions of selected observables match those of X(5):
$R_{4/2}$=2.91 to within 2$\%$ (solid line), $E(0^+_2)/E(2^+_1)$=5.67
to within 5$\%$ (dashed line), and $[E(2^+_2)-E(0^+_2)]/E(2^+_1)$=1.80 to within
5$\%$ (dotted line).  The solid circles indicate the points in
parameter space corresponding to the calculations discussed in the
text.
\label{figgcmcontour_x5}
}
\end{figure}

Two basic predictions of the X(5) model are that $R_{4/2}$=2.91
and $E(0^+_2)/E(2^+_1)$=5.67.  There is only one point in the
truncated GCM parameter space, $d$=0.70 and
$S$=0.00016\timesSunits ($S/S_>$=0.20), for which the GCM
predictions for both these quantities match the X(5) predictions
simultanously (see Fig.~\ref{figgcmcontour_x5}), although reasonably
close agreement can be found along a trajectory in parameter space
extending to $d\approx-10$.

\begin{figure}[p]
\begin{center}
\includegraphics*[width=1.0\hsize]{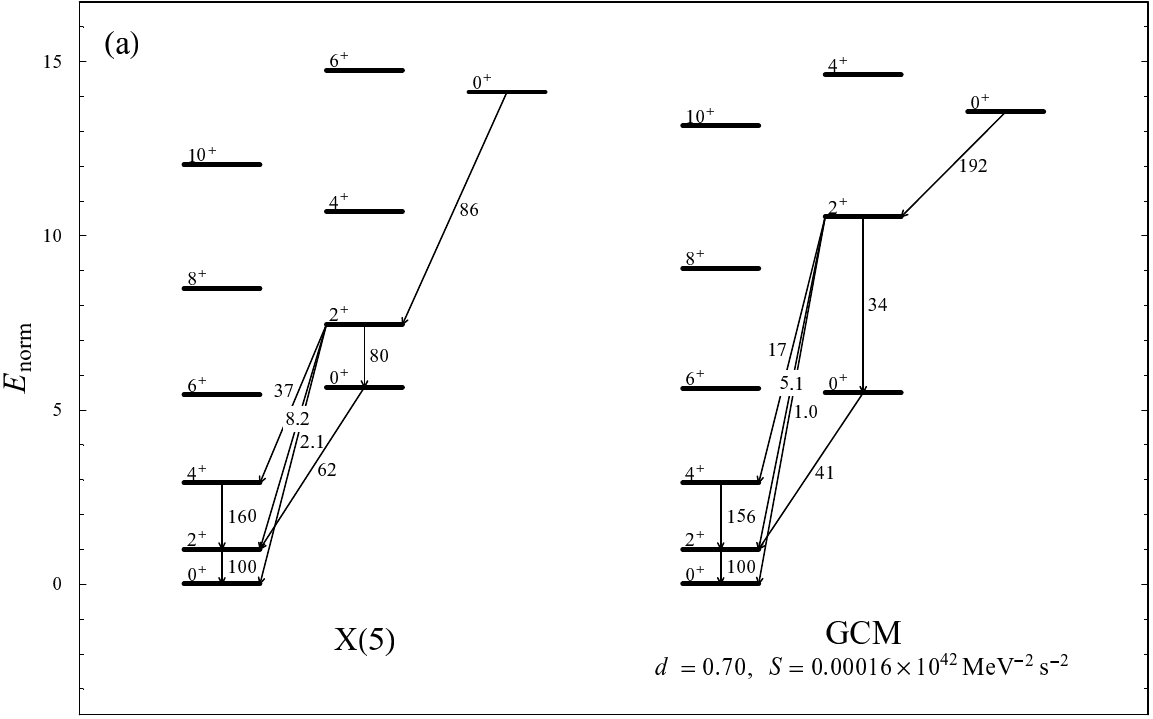}
\vspace{20pt}
\\
\includegraphics*[height=2.5in]{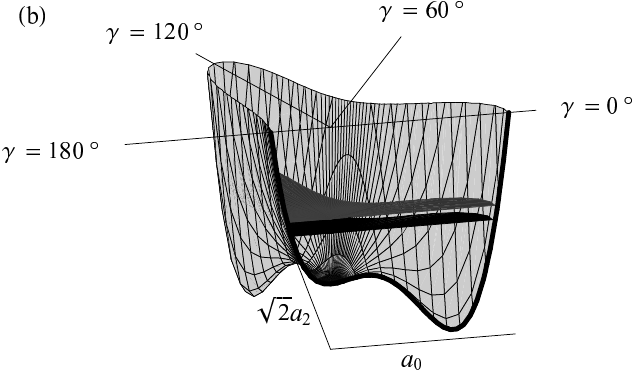}
\vspace{11pt}\\
\end{center}
\caption[X(5)-like structure in the GCM.]
{X(5)-like structure in the GCM, for $d=0.70$ and
$S=$0.00016\timesSunits: (a)~level energies and $B(E2)$ strengths
predicted in X(5) and the GCM, normalized to $E(2^+_1)$ and
$B(E2;2^+_1\rightarrow0^+_1)$, and (b)~a plot of the potential for the
GCM calculation as a function of the shape coordinates, showing
$0^+_1$ and $0^+_2$ level energies.
\label{figgcmx5}
}
\end{figure}

Let us compare the predictions of the GCM for $d$=0.70 and
$S$=0.00016\timesSunits with those of X(5).  Excitation energies and
transition strengths for the low-lying levels are shown in
Fig.~\ref{figgcmx5}(a).
The energies of the ground state and $0^+_2$ state are shown within a
plot of the potential in Fig.~\ref{figgcmx5}(b).  It is seen that both
states have energies which make both zero and nonzero $\beta$
accessible but with $\gamma$ confinement at the nonzero $\beta$,
consistent with the qualitative discussion above.  For the GCM
calculation, energies and $B(E2)$ strengths within the yrast band are
similar to those of X(5), though with noticeable differences at higher
spins.  The bandhead of the third $K$=0 band has an energy
in this calculation [$E(0^+_3)/E(2^+_1)$=13.6] which closely matches that in
X(5).  The qualitative features of reduced $R_{4/2}$ ratio, increased
$2^+-0^+$ level spacing, and reduced $B(E2;2^+\rightarrow0^+)$
strength within the excited bands relative to the yrast band also
appear, but in a greatly exaggerated fashion: the $R_{4/2}$ for the
second band is only 1.81, the $2^+-0^+$ level spacing is 5.05 times
that of the yrast band, and the $B(E2;2^+\rightarrow0^+)$ strength is
only $\sim1/3$ that in the yrast band.  These are far more extreme
changes relative to the yrast band than are encountered in X(5)
(Fig.~\ref{figgcmx5}).  Absolute interband and intraband transition
$B(E2)$ strengths involving the low-lying excited bands agree only to
within a factor of three or so between the two models.  However, the
\textit{ratios} of interband transition strengths are extremely
similar in both calculations, showing a characteristic pattern of
suppression of the spin-descending transitions.  A quantitative
comparison is given in Table~\ref{tabbranchgcmx5}.
\begin{table}
\begin{center}
\begin{tabular}{l,c,l_c_r.l_c_r.l}
\pseudoruledtabular
\multicolumn{3}{c}{}&
&
\multicolumn{5}{c}{$B(E2)^\text{rel}$} \\
\cline{5-9}
\multicolumn{3}{c}{Transition} &
&
\multicolumn{2}{c}{X(5)}&
&
\multicolumn{2}{c}{GCM} \\ 
\colrule
$2^+_2$ & $\rightarrow$ & $0^+_1$  & & 100&   & &  100&\\
        &               & $2^+_1$  & &  22&   & &   30&\\
        &               & $4^+_1$  & &   5&.8 & &    6&.0\\
\colrule
$4^+_2$ & $\rightarrow$ & $2^+_1$  & & 100&   & &  100&\\
        &               & $4^+_1$  & &  22&   & &   28&\\
        &               & $6^+_1$  & &   3&.4 & &    9&.2\\
\pseudoruledtabular
\end{tabular}
\end{center}
\caption
[Calculated branching ratios in X(5) and in the GCM.] {\ssp 
Calculated $E2$ branching ratios, from $K^\pi=0^+_2$ band
members to $K^\pi=0^+_1$ band
members, in X(5) and in the GCM for $d=0.70$ and
$S=$0.00016\timesSunits. 
\label{tabbranchgcmx5}
}
\end{table}

The GCM calculation just described closely reproduces several X(5)
predictions but exhibits extreme discrepancies from X(5) in the energy
and transition strength scales for the excited bands.  It is possible
to seek a point in parameter space for which the calculated $R_{4/2}$
and $E(0^+_2)/E(2^+_1)$ values match those of X(5) less closely but
for which the scale $[E(2^+_2)-E(0^+_2)]/E(2^+_1)$ is similar to that
in X(5).  An example of such a point, $d\approx$-6.9 and
$S\approx$0.18\timesSunits, is indicated in
Fig.~\ref{figgcmcontour_x5}.  However, the structural situation in
this region of parameter space differs substantially from that
encountered in the X(5) picture.  As discussed in
Section~\ref{secgcmmapping}, degeneracy of the $\beta$ and $\gamma$
bands for rotational nuclei occurs at $d\approx$-8
[see~(\ref{eqnbetagammaratio}) and
Fig.~\ref{figgcmcontour_full_e}(e)].  Although the $2^+_2$-$0^+_2$
energy difference does match that of X(5), the $2^+_2$ state
properties indicate that this state involves a substantial mixing of
$K^\pi=0$ and $K^\pi=2$ contributions.  For instance,
$B(E2;2^+_2\rightarrow0^+_2)$ and $B(E2;2^+_3\rightarrow0^+_2)$ are
both of ``in-band'' strength and differ from each other by less than a
factor of two.

Thus, while the E(5) predictions could be matched very well by a GCM
calculation, it appears that the X(5) predictions can only be
identified with a general region in the truncated GCM parameter space.
Other examples of GCM calculations for parameter values in this
general region are considered in the context of the $N$=90
transitional nuclei in Chapter~\ref{chapphenom}.

\chapter{Finite well solution for the E(5) Hamiltonian}
\label{chapfwell}

The E(5) model (Sections~\ref{sectrans} and~\ref{secbenche5x5}),
recently proposed by Iachello~\cite{iachello2000:e5}, provides an
analytic description of structure near the critical point of the
transition between spherical oscillator structure and rigidly-deformed
$\gamma$-soft structure.  This description is obtained by considering
the geometric Hamiltonian (Section~\ref{secgcmintro}) with a potential
function which is an infinite square well in quadrupole deformation
space.  The solution wave functions are analytic in form, consisting
of the spherical Bessel functions, and the eigenvalues are simply
given in terms of the zeros of the Bessel functions
(Section~\ref{secbenche5x5}).  The use of an infinite well potential
is a convenient calculational approximation, allowing extremely simple
solutions to be obtained.  However, actual potentials describing
nuclei are expected to be finite, not infinite in depth, and the
interacting boson model at finite boson number along the
U(5)--SO(6) transition~\cite{pan1998:so6u5} leads to potentials
which may be approximated by a finite-depth square well.  It is
therefore of interest, as suggested in Ref.~\cite{iachello2000:e5}, to
assess the extent to which the results from the E(5) description are
sensitive to finite well depth.  In the present work, the evolution of
nuclear observables as a function of well depth is investigated, and
observables sensitive to finite well depth are identified.  These
results were reported in Ref.~\cite{caprio2002:fwell}.

\begin{figure}
\begin{center}
\includegraphics*[width=0.65\hsize]{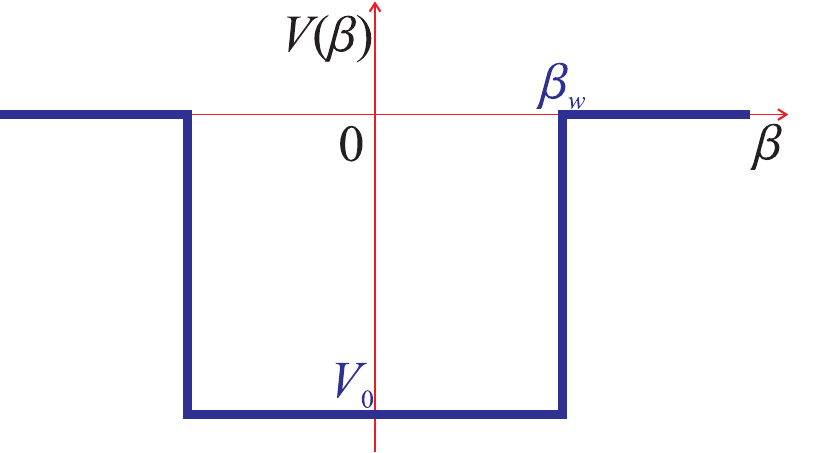}
\end{center}
\caption[Finite square well potential.]
{Finite square well potential in deformation space, shown along the
$a_0$-axis cut.
\label{figfwell}
}
\end{figure}

The five-dimensional finite square well potential (Figure~\ref{figfwell}), 
\begin{equation}
V(\beta) = \begin{cases} V_0 & \beta \leq \beta_w \\ 0 & \beta >
\beta_w, \end{cases}
\end{equation}
is, like the E(5) potential, independent of $\gamma$.  
Consequently,
the Hamiltonian,
\begin{equation}
-\,\frac{\hbar^2}{2B}\sum_\mu
\frac{\partial^2}{\partial\alpha_{2\mu}\partial\alpha^*_{2\mu}} +
V(\beta),
\end{equation}
is separable (Section~\ref{secbenchgsoft}), and the
eigenfunctions are of the form $f(\beta)\Phi(\gamma,\underline{\theta})$.  The
solutions for the ``angular'' ($\gamma$,$\underline{\theta}$) wave
functions~\cite{bes1959:gamma} are common to all $\gamma$-soft
problems, while the dependence upon the potential $V(\beta)$ is
isolated in the ``radial'' ($\beta$) wave function.  
The equation for $f(\beta)$ is
\begin{equation}
\label{eqnradial}
\left[ \frac{\hbar^2}{2B}\left( -\frac{1}{\beta^4}
\frac{\partial}{\partial\beta}\beta^4\frac{\partial}{\partial\beta}
+\frac{\tau(\tau+3)}{\beta^2} \right) + V(\beta) \right] f(\beta) = E
f(\beta),
\end{equation}
where the separation constant $\tau$ assumes the values
$\tau=0,1,\ldots$.  The notation is simplified if an overall factor of
$\hbar^2/(2B)$ is extracted from the Hamiltonian, leaving the equation
in terms of the reduced eigenvalue $\varepsilon \equiv
\frac{2B}{\hbar^2} E$ and the reduced potential
$v(\beta)\equiv\frac{2B}{\hbar^2} V(\beta)$, of depth
$v_0\equiv\frac{2B}{\hbar^2} V_0$.  Bound state solutions can only
occur with eigenvalues in the range $v_0<\varepsilon<0$.  Each
solution of the $\beta$ equation results in a multiplet of solutions
to the full problem, degenerate with respect to angular momentum
according to the $\gamma$-soft $\tau$ multiplet structure
(Section~\ref{secbenchgsoft}).  The notation $J_{\xi,\tau}^+$ is used
to designate the states corresponding to the $\xi$th radial solution
for separation constant $\tau$.

The finite well potential is piecewise constant as a function of
$\beta$.  Within a region of constant potential, the radial
equation~(\ref{eqnradial}) reduces to the Bessel equation of
half-integer order, as outlined in Ref.~\cite{iachello2000:e5}.  In
the interior of the well $(\beta<\beta_w)$, where the difference
$\varepsilon-v_0$ is positive, the solutions involve the ordinary
Bessel functions, while in the classically forbidden region exterior
to the well $(\beta>\beta_w)$, where the difference $\varepsilon-v_0$
is negative, the solutions involve the modified Bessel functions.
Although the equations are satisfied for an arbitrary linear
combination of the Bessel functions of the first and second kind, the
solution chosen for the interior region must have the correct
convergence properties at the origin~\cite{fluegge1971:qm}, and the
solution for the exterior region must have the proper asymptotic
behavior (convergence at $\beta\rightarrow\infty$).  The Bessel
functions of half-integer order may be expressed in terms of the
spherical Bessel functions of integer order~(\eg,
Refs.~\cite{arfken1995:mathmethods,abramowitz1965}), and the correct
combinations for this problem involve
\begin{equation}
\label{eqnbesseldef}
j_n(x)\equiv\sqrt\frac{\pi}{2} x^{-1/2}J_{n+1/2}(x) \qquad
k_n(x)\equiv\sqrt\frac{2}{\pi} x^{-1/2}K_{n+1/2}(x) .
\end{equation}
The wave function in $\beta$ for the $\xi$th solution with separation
constant $\tau$ is
\begin{equation}
f_{\xi,\tau}(\beta)=\begin{cases} A_{\xi,\tau} \beta^{-1}
j_{\tau+1}[(\varepsilon_{\xi,\tau}-v_0)^{1/2}\beta] & \beta \leq
\beta_w\\ B_{\xi,\tau} \beta^{-1}
k_{\tau+1}[(-\varepsilon_{\xi,\tau})^{1/2}\beta] & \beta >
\beta_w. \end{cases}
\end{equation}

The eigenvalues for the finite well are determined by the requirement
that $f(\beta)$ be continuous and smooth at the matching point
$\beta=\beta_w$.  The eigenvalue condition for $\varepsilon$ can be
obtained in a manner analogous to that for the three-dimensional
square well ({\it e.g.}, Ref.~\cite{fluegge1971:qm}).  Since the
spherical Bessel functions may be expressed in terms of trigonometric
and exponential functions, the matching condition assumes the form of
a transcendental equation.  This condition may be derived by direct
algebraic manipulation, for each value of $\tau$, as in
Ref.~\cite{fluegge1971:qm}.  However, the resulting expressions
rapidly become cumbersome beyond the lowest few values for $\tau$.  It
is more convenient to instead derive a general form for the condition,
expressed in terms of the coefficients in polynomial expansions of the
spherical Bessel functions.  Let us define a dimensionless energy
variable,
\begin{equation}
\eta(\varepsilon)\equiv\left[ 1 - \frac{\varepsilon}{v_0}\right]^{1/2},
\end{equation}
and a ``well size'' parameter,
\begin{equation}
\label{eqnx0def}
x_0\equiv (-v_0)^{1/2}\beta_w.
\end{equation} 
The matching condition, in terms of these quantities, is the
transcendental equation
\begin{equation}
\label{eqntranscendental}
-\frac{
\sum_{i,j=0}^{\tau+2} \left[c_{(\tau+1)i}e'_{(\tau+1)(j+1)} - c'_{(\tau+1)(i+1)}e_{(\tau+1)j} \right] x_0^{-i-j}\eta^{-i}[(1-\eta^2)^{1/2}]^{-j}
}{
\sum_{i,j=0}^{\tau+2} \left[s_{(\tau+1)i}e'_{(\tau+1)(j+1)} - s'_{(\tau+1)(i+1)}e_{(\tau+1)j} \right] x_0^{-i-j}\eta^{-i}[(1-\eta^2)^{1/2}]^{-j}
} =
\tan(x_0\eta),
\end{equation}
where $c_{ni}$, $s_{ni}$, and $e_{ni}$ are the coefficients in the
expansions
\begin{equation}
\begin{gathered}
j_n(x)=\left(\sum_{i=1}^{n+1}c_{ni}x^{-i}\right)\cos x +
\left(\sum_{i=1}^{n+1}s_{ni}x^{-i}\right)\sin x \\
k_n(x)=\left(\sum_{i=1}^{n+1}e_{ni}x^{-i}\right)e^{-x},
\end{gathered}
\end{equation}
and $c'_{ni}$, $s'_{ni}$, and $e'_{ni}$ are defined similarly for the
derivative functions $j'_n(x)$ and $k'_n(x)$.  Recurrence relations
for the calculation of these coefficients may be derived in a
straightforward fashion from the Bessel function recurrence relations.

\begin{figure}[t]
\begin{center}
\includegraphics*[width=0.8\hsize]{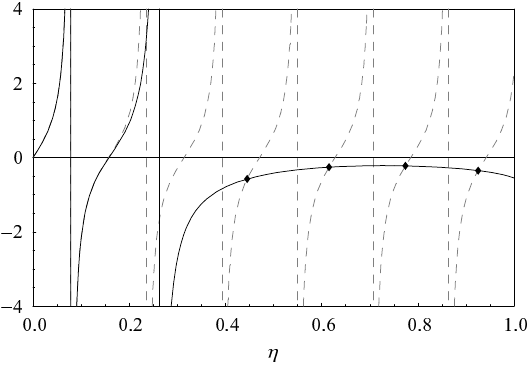}
\end{center}
\caption[Graphical representation of the finite well eigenvalue problem.]
{Graphical representation of the finite well eigenvalue problem, for
$x_0$=20 and $\tau$=4.  The left hand side (solid line) and right hand
side (dashed line) of the eigenvalue
equation~(\ref{eqntranscendental}) are plotted as functions of $\eta$.
The intersection points (diamonds) indicate eigenvalues.
\label{figrootfind}
}
\end{figure}

The equation~(\ref{eqntranscendental}) must be solved numerically for
the eigenvalues of $\varepsilon$.  An example case is shown
graphically in Fig.~\ref{figrootfind}.
Once an eigenvalue $\varepsilon_{\xi,\tau}$ is found, the coefficients
$A_{\xi,\tau}$ and $B_{\xi,\tau}$ follow from the continuity condition
at $\beta=\beta_w$ and the normalization condition $\int_0^\infty
\beta^4 d\beta |f(\beta)|^2=1$.  The normalization integrals can be
evaluated analytically, giving
\newcommand{\jfunction}[2]{j_{#1}[(\varepsilon-v_0)^{1/2}{#2}]}
\newcommand{\kfunction}[2]{k_{#1}[(-\varepsilon)^{1/2}{#2}]}
\begin{align}
c_1
&\begin{aligned}[t]
&\equiv \int_0^{\beta_w}\beta^4 d\beta
\left[\beta^{-1}\jfunction{\tau+1}{\beta}\right]^2\\ 
&=
\frac{\beta_w^3}{2}\left[\jfunction{\tau+1}{\beta_w}^2-\jfunction{\tau}{\beta_w}\jfunction{\tau+2}{\beta_w}\right]
\end{aligned}\\
\intertext{and}
c_2
&
\begin{aligned}[t]
&\equiv \int_{\beta_w}^\infty\beta^4 d\beta \left[\beta^{-1}\kfunction{\tau+1}{\beta}\right]^2\\
&=- \frac{\beta_w^3}{2}\left[\kfunction{\tau+1}{\beta_w}^2-\kfunction{\tau}{\beta_w}\kfunction{\tau+2}{\beta_w}\right].
\end{aligned}
\end{align}
In terms of $c_1$, $c_2$, $y_1\equiv\beta_w^{-1}\jfunction{\tau+1}{\beta_w}$, and
$y_2\equiv\beta_w^{-1}\kfunction{\tau+1}{\beta_w}$, the
continuity and normalization conditions are $Ay_1=By_2$ and $A^2c_1+B^2c_2=1$.
The coefficients satifying these are 
\begin{equation}
A=\pm \frac{y_2}{[y_1^2c_2+y_2^2c_1]^{1/2}} \qquad
B=\pm \frac{y_1}{[y_1^2c_2+y_2^2c_1]^{1/2}}.
\end{equation}
Note that the numerical values for these coefficients depend upon the
normalization convention~(\ref{eqnbesseldef}) chosen for the spherical
Bessel functions, which varies in the literature.

The eigenvalue spectrum of the solution depends upon the parameters
$\beta_w$ and $v_0$ exclusively in the combination $x_0$, as can be
seen from the eigenvalue equation (\ref{eqntranscendental}).  That is,
if two wells $(\beta_w,v_0)$ and $(\beta_w',v_0')$ have the same value
for $x_0$, they will have identical energy spectra, to within an
overall normalization factor.  Two wells with different $x_0$ values
will have different energy spectra.  The parameter $x_0$ is invariant
under multiplication of the potential by a factor $a^2$ followed by
dilation by $1/a$, so this result is actually a special case of the
scaling relation discussed in Appendix~\ref{appscaling}, and $x_0$
plays a role analogous to the structure parameters $S$ and $S_\gamma$
of the preceding chapters.  

Therefore, for a given value of $x_0$, the numerical solution
procedure need only be carried out once, at some ``reference'' choice
of the well width and depth ({\it e.g.}, $\beta_w=1$), and the
solution for any other well of the same $x_0$ can be deduced by a
simple rescaling of all energies and dilation of all wave functions.
To state the relations explicitly,
consider a reference calculation performed at $\beta_w=1$ (and thus
$v_0=-x_0^2$), and suppose this calculation produces an eigenvalue
$\varepsilon$ and normalized wave function $f(\beta)$.  Then for a
well of the same $x_0$ but a different width $\beta_w'$ (and thus
$v_0'=-x_0^2/\beta_w'^2$), the corresponding eigenvalue $\varepsilon'$
and normalized wave function $f'(\beta)$ are given by the simple
rescalings
\begin{equation}
\begin{gathered}
\label{eqnfwellscale}
\varepsilon'=\beta_w'^{-2}\varepsilon\\
f'(\beta)=\beta_w'^{-5/2} f(\beta/\beta_w').
\end{gathered}
\end{equation}
The scaling properties of matrix elements of the operator $\beta^m$,
which are encountered in the calculation of electromagnetic transition
strengths, are given in Appendix~\ref{appscaling}.  

\begin{figure}[t]
\begin{center}
\resizebox{\hsize}{!}
{
\includegraphics*[height=1in]{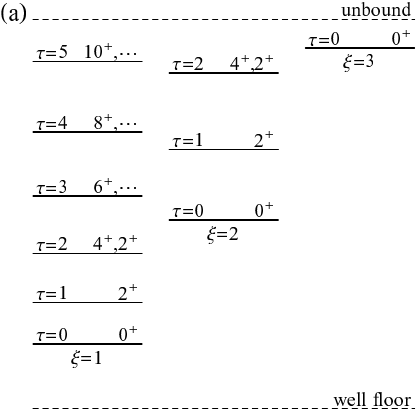}
\hspace{0.02in}
\includegraphics*[height=1in]{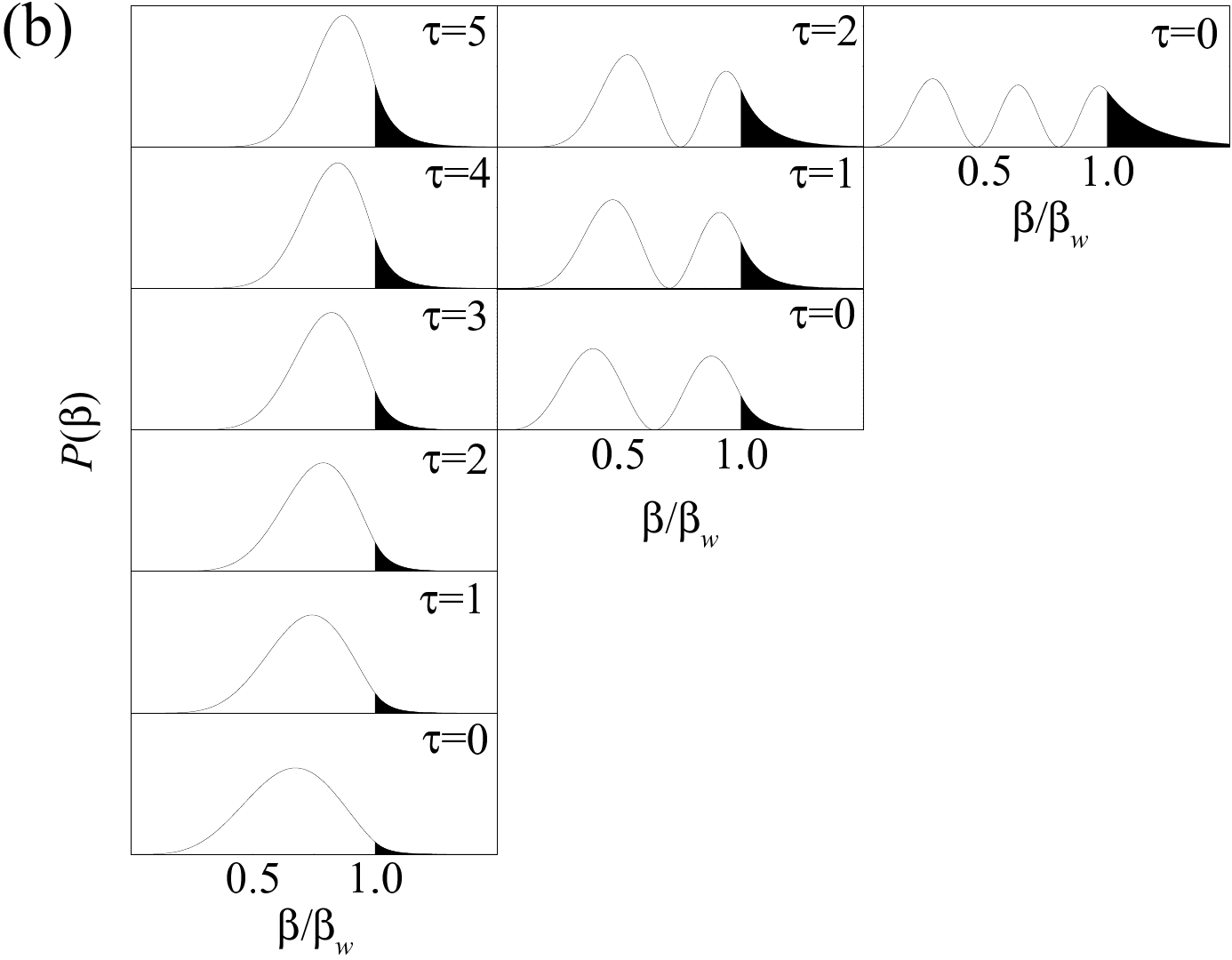}
}
\end{center}
\caption[Bound states of the $x_0=10$ well.]
{Bound states of the $x_0=10$ well: (a) excitation energies and (b) probability
density functions $P(\beta) \equiv \beta^4|f(\beta)|^2$.  The shaded areas under the probability
density functions indicate penetration into classically forbidden
$\beta$ values ($\beta>\beta_w$).  (Figure from Ref.~\cite{caprio2002:fwell}.)
\label{figfwellxten}
}
\end{figure}

\begin{figure}[p]
\begin{center}
\includegraphics*[width=0.5\hsize]{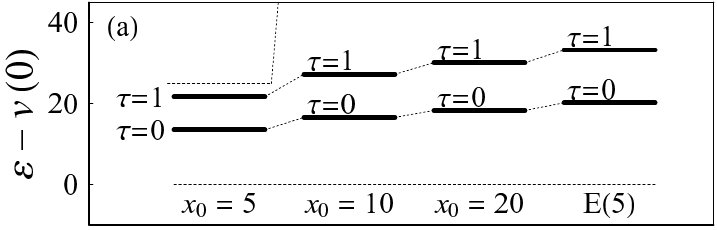}\\
\hspace{0.002\hsize}\includegraphics*[width=0.5\hsize]{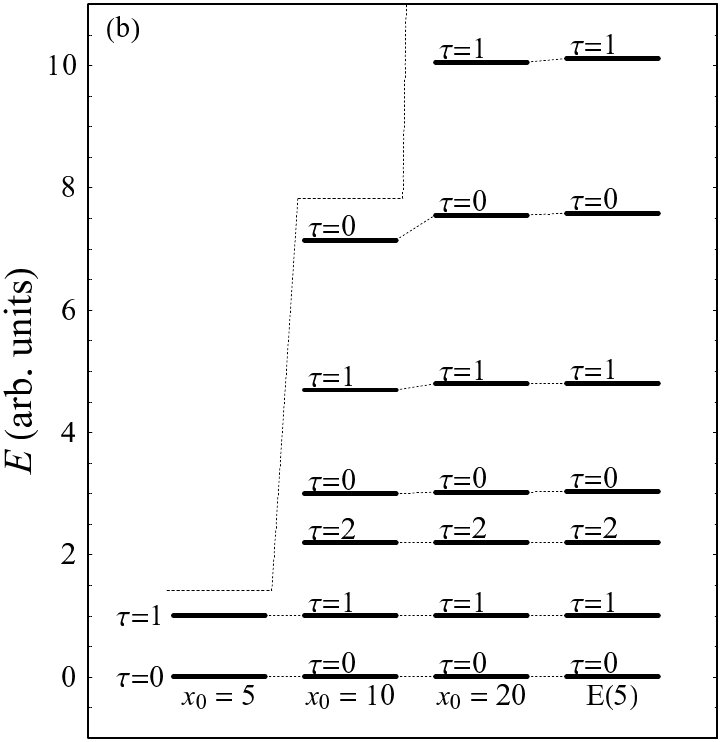}
\end{center}
\caption[Evolution of level excitation energies as a function of well size.]
{Evolution of level excitation energies as a function of well
size parameter $x_0$ for selected low-lying levels.  (a) Absolute
eigenvalue relative to floor of well [$\varepsilon-v(0)$].  (b)
Excitation energy normalized to the first excited state.  The upper
dashed line indicates the energy at which the system becomes unbound
(top of well).  (Figure adapted from Ref.~\cite{caprio2002:fwell}.)
\label{figfwellevoln}
}
\end{figure}

The solution for $x_0=10$ is illustrative of the main effects of
finite well depth.  The level energies and wave functions for this
solution are shown in Fig.~\ref{figfwellxten}.  The main consequences of
the finite well depth are not unexpected:
\begin{dissenumeratelist}
\item There are only a finite number of bound states.  For this example,
only members of the lowest few $\xi$ families are bound
[Fig.~\ref{figfwellxten}(a)].  A summary of the number of bound states for
other well sizes is given in Table~\ref{tabfwellbound}.
\item The wave functions penetrate the classically forbidden region
$\beta>\beta_w$.  For the highest-lying states, a substantial portion of the
probability distribution in $\beta$ lies outside $\beta_w$.  This is
indicated by the shaded areas of Fig.~\ref{figfwellxten}(b).
\item  The eigenvalues are lowered relative to those for the infinite
E(5) well of the same $\beta_w$ [Fig.~\ref{figfwellevoln}(a)].  This is
a natural consequence of the finite well depth:  the wave functions
are given the freedom to spread into the region $\beta>\beta_w$, and
this is analogous in effect to a widening of the well, causing the energies 
to ``settle'' lower.
\end{dissenumeratelist}
\begin{table}[t]
\begin{center}
\begin{tabular}{r=r=r=r=r=r} 
\pseudoruledtabular
$x_0$& $\tau=0$& $\tau=1$& $\tau=2$& $\tau=3$ & $\cdots$\\
\colrule
5 & 1 & 1 &  &     \\
10 & 3 & 2 & 2 & 1       & $\cdots$\\
20 & 6 & 5 & 5 & 4       & $\cdots$\\
50 & 15 & 14 & 14 & 14   & $\cdots$\\
\pseudoruledtabular
\end{tabular}
\end{center}
\caption
[Number of bound $\beta$ solutions for selected $x_0$.] 
{\ssp 
Number of bound $\beta$ solutions, by $\tau$ quantum number,
for selected $x_0$.
\label{tabfwellbound}
}
\end{table}

Some interesting properties, however, are revealed by an examination
of the systematic evolution of the solution with changing well size.
A series of calculations ($x_0=5,10,20$) spanning the physical range
of interest in the study of nuclei are presented alongside the E(5)
solution in Fig.~\ref{figfwellevoln}.  At a fixed width, $x_0$ is a
measure of the depth of the well, and the infinite E(5) well is
obtained in the limit $x_0\rightarrow\infty$.  Although the energy
eigenvalues do experience a lowering as the well depth decreases
[Fig.~\ref{figfwellevoln}(a)], it turns out that the level energies are
nearly {\it uniformly} lowered by the same {\it factor} for all levels
in the well, leaving energy {\it ratios} virtually unchanged.  A plot
of excitation energies normalized to the first excited state
[Fig.~\ref{figfwellevoln}(b)] reveals these energies to be essentially
insensitive to the well depth.  Some relevant energy ratios are
summarized in Table~\ref{tabfwellenergies}.
\begin{table}[p]
\begin{center}
\begin{tabular}{r_c_r.l_c_r.l_r.l_r.l_c_r.l_r.l_r.l}
\pseudoruledtabular
&
&
\multicolumn{2}{c}{$\xi=1$}&
&
\multicolumn{6}{c}{$\xi=2$}&
&
\multicolumn{6}{c}{$\xi=3$}\\
\cline{3-4}\cline{6-11}\cline{13-18}
$x_0$&
&
\multicolumn{2}{c}{$R_{4/2}$}&
&
\multicolumn{2}{c}{$E_{\tau=0}$}& 
\multicolumn{2}{c}{$E_{\tau=1}$} &
\multicolumn{2}{c}{$R_{4/2}$}&
&
\multicolumn{2}{c}{$E_{\tau=0}$}& 
\multicolumn{2}{c}{$E_{\tau=1}$} &
\multicolumn{2}{c}{$R_{4/2}$}\\
\colrule
10&     &  2&.19&   &  2&.99&   4&.69&   2&.09&   &  7&.14&   \multi{---}& \multi{---}\\    
20&     &  2&.20&   &  3&.02&   4&.79&   2&.12&   &  7&.55&   10&.05&  2&.08\\
E(5)& &  ~~2&.20&   &  ~~3&.03&   ~~4&.80&   ~~2&.12&   &  ~~7&.58&   ~~10&.11&  ~~2&.09\\
\pseudoruledtabular
\end{tabular}
\end{center}
\caption
[Excitation energy observables for selected $x_0$.] {\ssp 
Excitation energy observables for selected $x_0$.
Excitation energies are normalized to $E_{\xi=1,\tau=1}=1$.  The quantity
$R_{4/2}$ is defined for each $\xi$ family as
$(E_{\tau=2}-E_{\tau=0})/(E_{\tau=1}-E_{\tau=0})$.
\label{tabfwellenergies}
}
\end{table}

\begin{figure}[t]
\begin{center}
\includegraphics*[width=0.7\hsize]{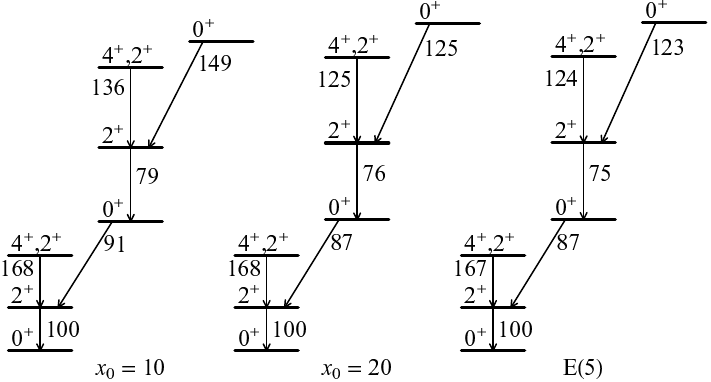}
\end{center}
\caption[Evolution of $B(E2)$ strengths.]
{Evolution of $B(E2)$ strengths as a function of well parameter $x_0$.
Values are normalized to $B(E2;2^+_{1,1}\rightarrow0^+_{1,0})=100$.  (Figure adapted from Ref.~\cite{caprio2002:fwell}.)
\label{figfwellevolne2}
}
\end{figure}

\begin{figure}[t]
\begin{center}
\includegraphics*[width=0.7\hsize]{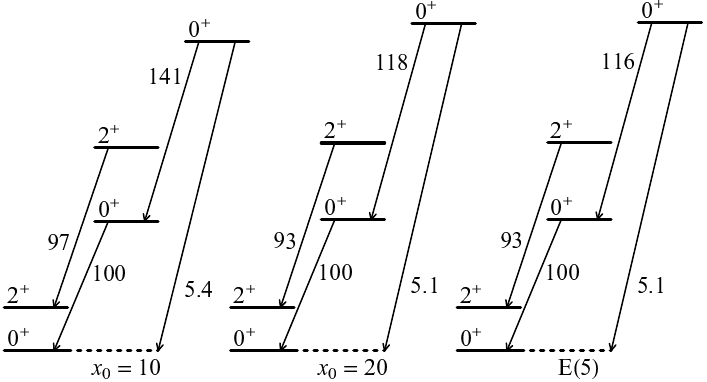}
\end{center}
\caption[Evolution of $B(E0)$ strengths.]
{Evolution of $B(E0)$ strengths as a function of well parameter $x_0$.
Values are normalized to $B(E0;0^+_{2,0}\rightarrow0^+_{1,0})=100$.
(Figure adapted from Ref.~\cite{caprio2002:fwell}.)
\label{figfwellevolne0}
}
\end{figure}

Electromagnetic transition strengths can be calculated from the matrix
elements of the collective multipole operators.  Expressed in terms of
the intrinsic coordinates $\beta$ and $\gamma$, the $E2$ and $E0$
transition operators are~\cite{bohr1998:v2}
\begin{equation}
\begin{gathered}
\mathfrak{M}(E2;\mu)\propto\beta\bigl[D^{2\,*}_{\mu,0}\cos\gamma+\frac{1}{\sqrt{2}}(D^{2\,*}_{\mu,2}+D^{2\,*}_{\mu,-2})\sin\gamma\bigr]\\
\mathfrak{M}(E0;0)\propto\beta^2,
\end{gathered}
\end{equation}
and the transition strengths are $B(E\lambda;J_i\rightarrow
J_f)=\frac{1}{2J_i+1} |\left<J_f||
\mathfrak{M}(E\lambda)||J_i\right>|^2$ (Appendix~\ref{appgamma}).
Only integrals of the $\beta$-dependent factors with respect to the
radial wave functions need be carried out, since the angular integrals
of $\mathfrak{M}(E2)$~\cite{arima1979:o6} are
common to all $\gamma$-soft problems.  The evolution of key $E2$ and
$E0$ transition strengths is shown in Fig.~\ref{figfwellevolne2} and
Fig.~\ref{figfwellevolne0}.  The absolute transition strengths are
larger at finite well depth than for the infinite well, at the same
width $\beta_w$, but the increase is, again, largely a uniform
overall increase, leaving $B(E2)$ or $B(E0)$ {\it ratios} nearly
unchanged from the E(5) limit.

The uniform reduction of all energies and enhancement of all
transition matrix elements does not serve as a useful identifying
feature of finite well depth, since arbitrary energy and
transition strength normalizations can be obtained for the infinite
E(5) well simply by varying the parameters $\beta_w$ and $B$.

Only the very highest energy levels, just short of being
unbound, show appreciable deviations from the E(5) normalized
energies and transition strengths.  The third $0^+$ state at $x_0=10$
demonstrates these effects nicely: lowered energy
(Fig.~\ref{figfwellevoln}), enhanced $E2$ transitions (Fig.~\ref{figfwellevolne2}),
and enhanced $E0$ transitions (Fig.~\ref{figfwellevolne0}).  

The results found for the finite well present a challenge from an
experimental viewpoint.  There are few clear signatures of finite well
depth.  Those signature which are present consist of moderate
modifications to energies or transition strengths for high-lying
levels, but such levels are typically the least accessible
experimentally and most subject to contamination from degrees of
freedom outside the collective model framework.

The results are, however, reassuring from a theoretical perspective.
They suggest that the E(5) description is ``robust'' in nature.  The
main features of the E(5) solutions remain virtually unchanged under
radical modification of the depth of the potential, from the ideal
infinite well to the realistic finite well likely to be applicable to
actual nuclei.

\chapter{Phenomenological interpretation}
\label{chapphenom}

\section{The $N$$\approx$90 transitional nuclei}
\label{secphenomaxial}

A rapid change in deformation occurs at $N$$\approx$90 along the Nd,
Sm, Gd, and Dy isotopic chains, as a transition occurs from spherical
to axially symmetric deformed structure.  Many of the transitional
phenomena described in Section~\ref{sectrans} find some of their best
candidates among the $N$=90 isotopes $^{150}$Nd,
$^{152}$Sm, $^{154}$Gd, and $^{156}$Dy.  The description of these
nuclei in the
IBM~\cite{scholten1978:ibm-fits,iachello1998:phasecoexistence,zamfir1999:152sm-beta}
requires parameter values near those of the phase transition in the
IBM, and the low-lying level energies of these nuclei place them at or
near the phenomenological ``critical point'' in the evolution of
observables (see Section~\ref{sectrans}).  These nuclei are thus now of
great interest for comparison to the X(5) model predictions, since the first
proposed candidates for description by the X(5) model were
$^{152}$Sm~\cite{casten2001:152sm-x5} and
$^{150}$Nd~\cite{kruecken2002:150nd-rdm}, and all four $N$=90 isotopes
have very similar characteristics with respect to the relevant
observables.  Unusual experimental results have repeatedly drawn
attention to the $N$=90 nuclei.  These have included observed $(p,t)$
and $(t,p)$ cross sections and $E0$ strengths suggestive of shape
coexistence, as well as spectroscopic results indicating the breakdown
of conventional rotational band mixing models, as outlined in
Section~\ref{sec152smmotivation}.

With both new model descriptions and extensive new sets of
spectroscopic and lifetime data available, let us now survey the
phenomenology of the $N$=90 nuclei.  It is expecially of interest to consider
how the basic observables of low-lying levels in these nuclei compare
to the X(5) predictions.

\begin{figure}[p]
\begin{center}
\includegraphics*[width=0.48\hsize]{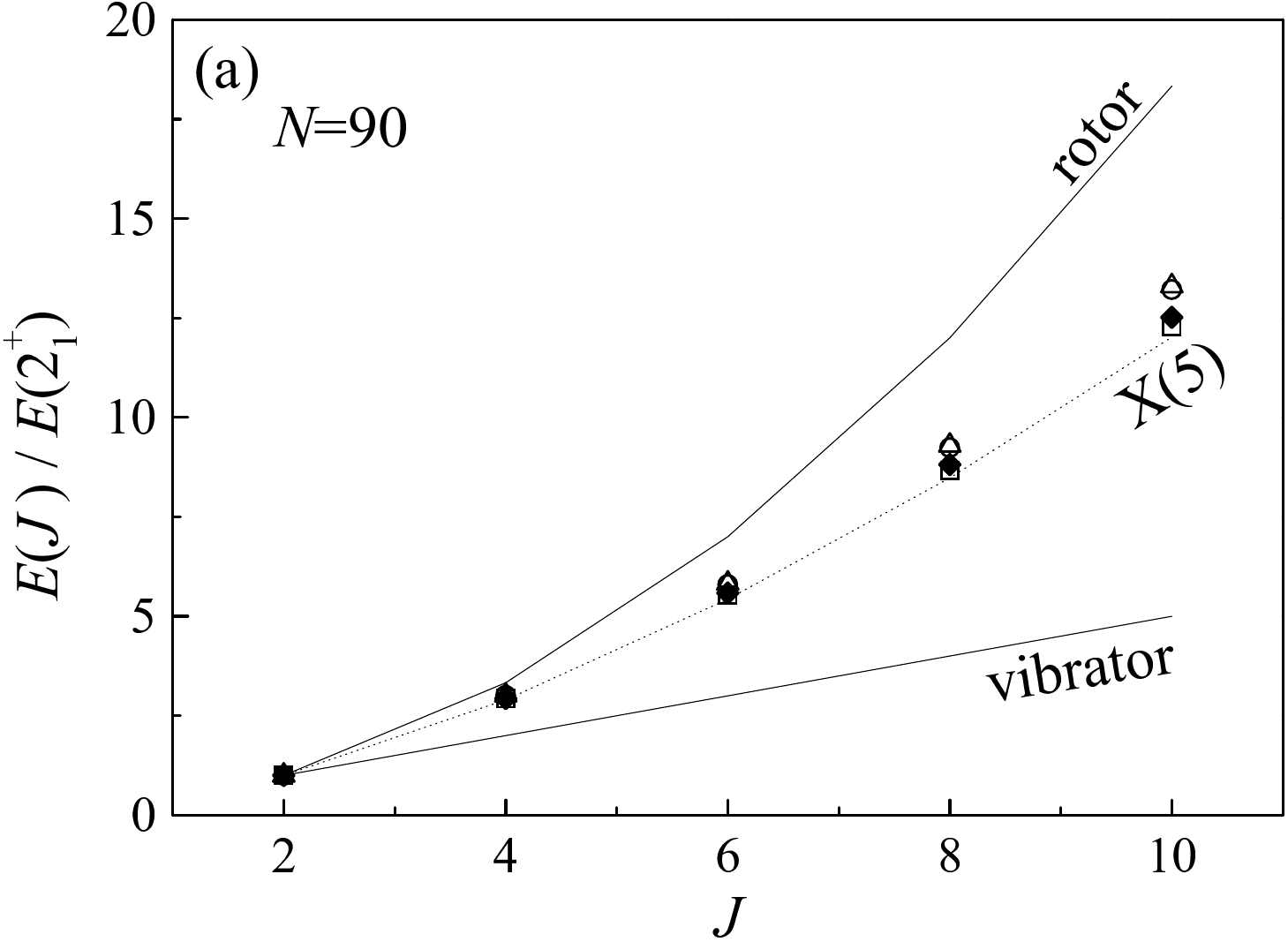}
\hfill
\includegraphics*[width=0.48\hsize]{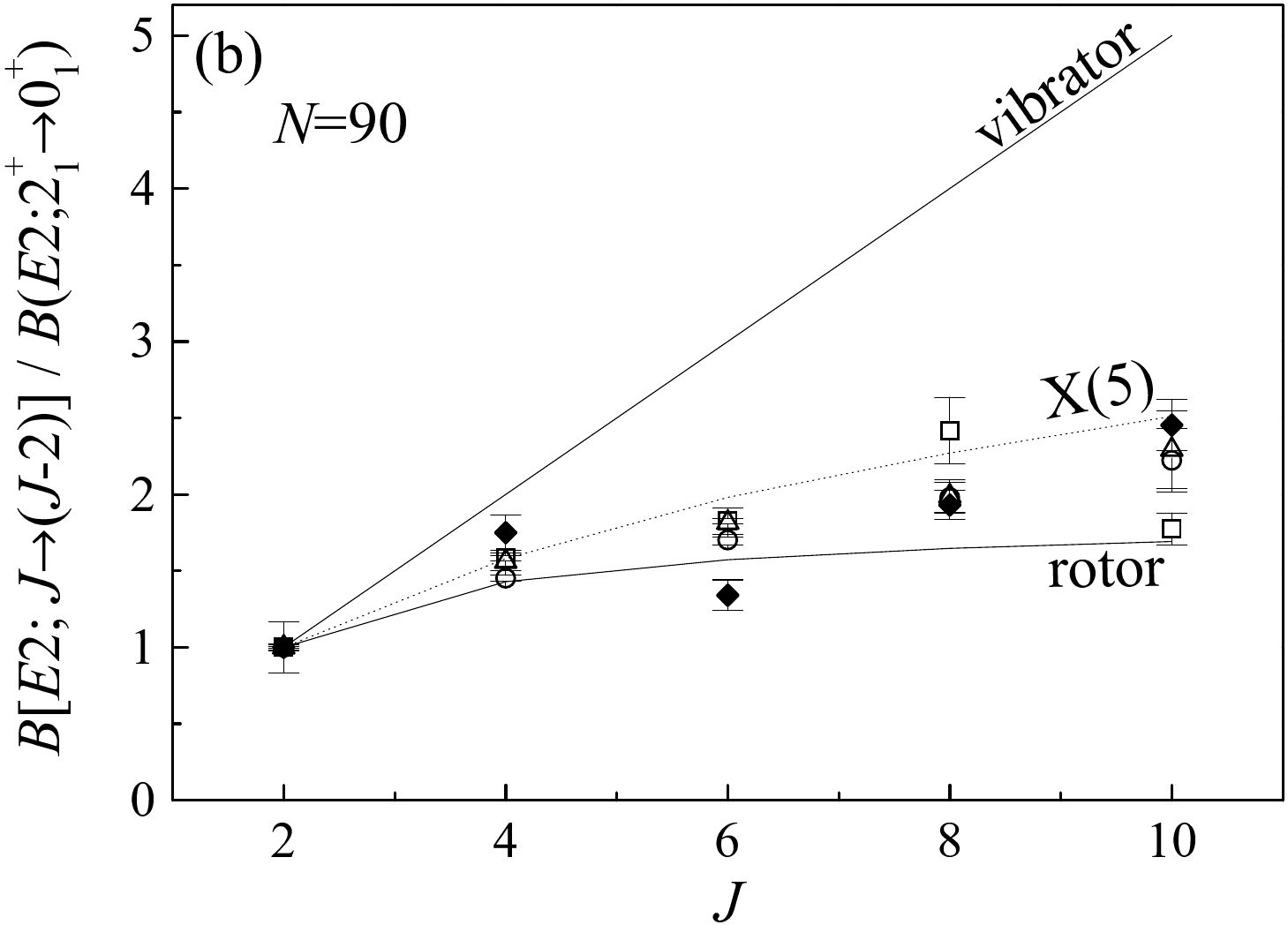}
\end{center}
\caption[Yrast band level energies and $B(E2)$ values for the $N$=90 isotones.]
{Yrast band (a) level energies and (b) $B(E2)$ values, normalized to
the those for $2^+$ band member, for the $N$=90 isotones
$^{150}$Nd~(\captionopensquare), $^{152}$Sm~(\captionopencircle),
$^{154}$Gd~(\captionopenuptriangle), and
$^{156}$Dy~(\captionsoliddiamond).  The rotor, X(5), and harmonic
vibrator predictions are shown for comparison.  Data are from
Refs.~\cite{nds1995:150,nds1996:152,nds1998:154,ndsboth:156,kruecken2002:150nd-rdm}.
(Figure adapted from Ref.~\cite{caprio2002:critex}.)
\label{fign90yrast}
}
\end{figure}

\begin{figure}[p]
\begin{center}
\includegraphics*[width=0.48\hsize]{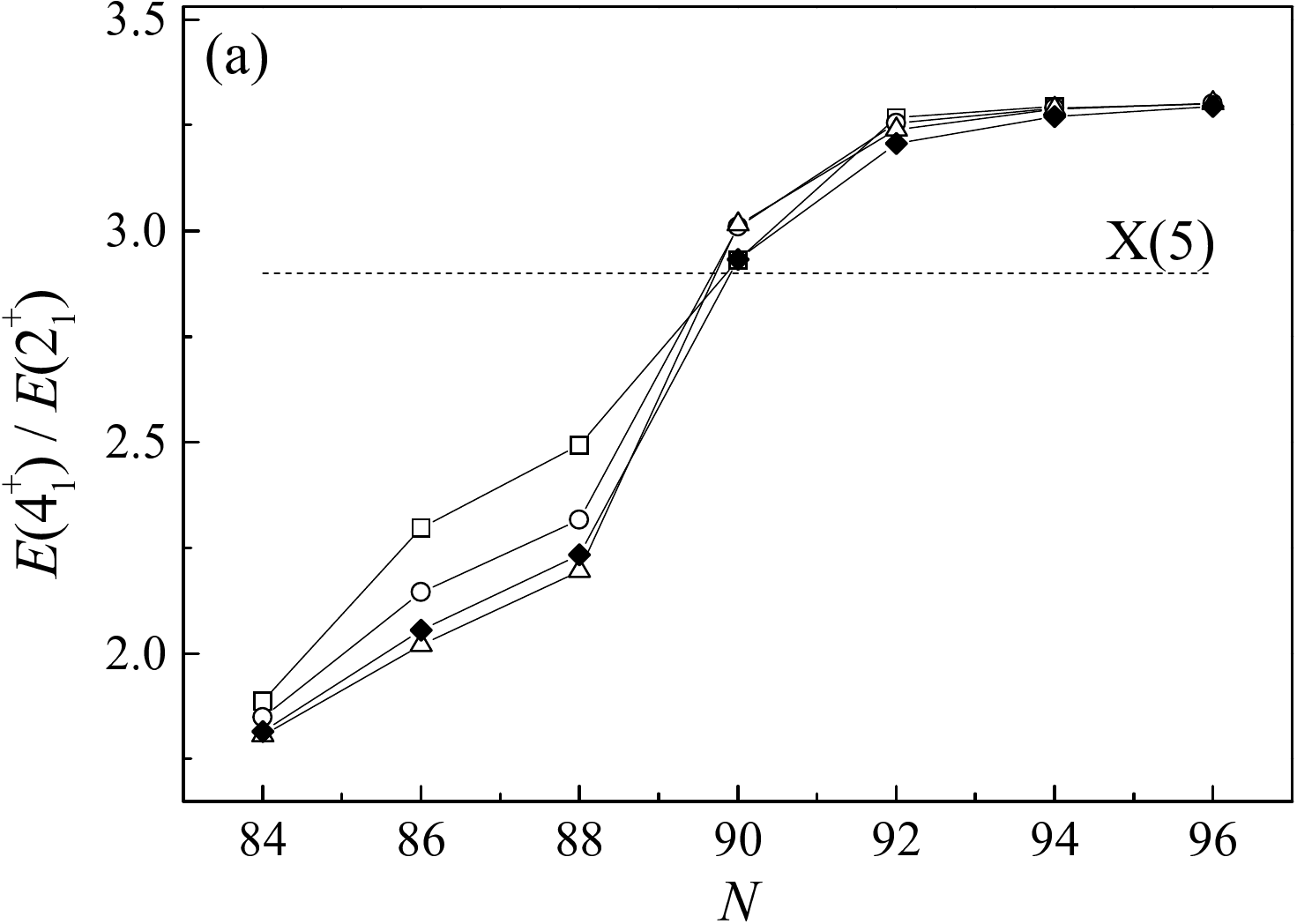}
\hfill
\includegraphics*[width=0.48\hsize]{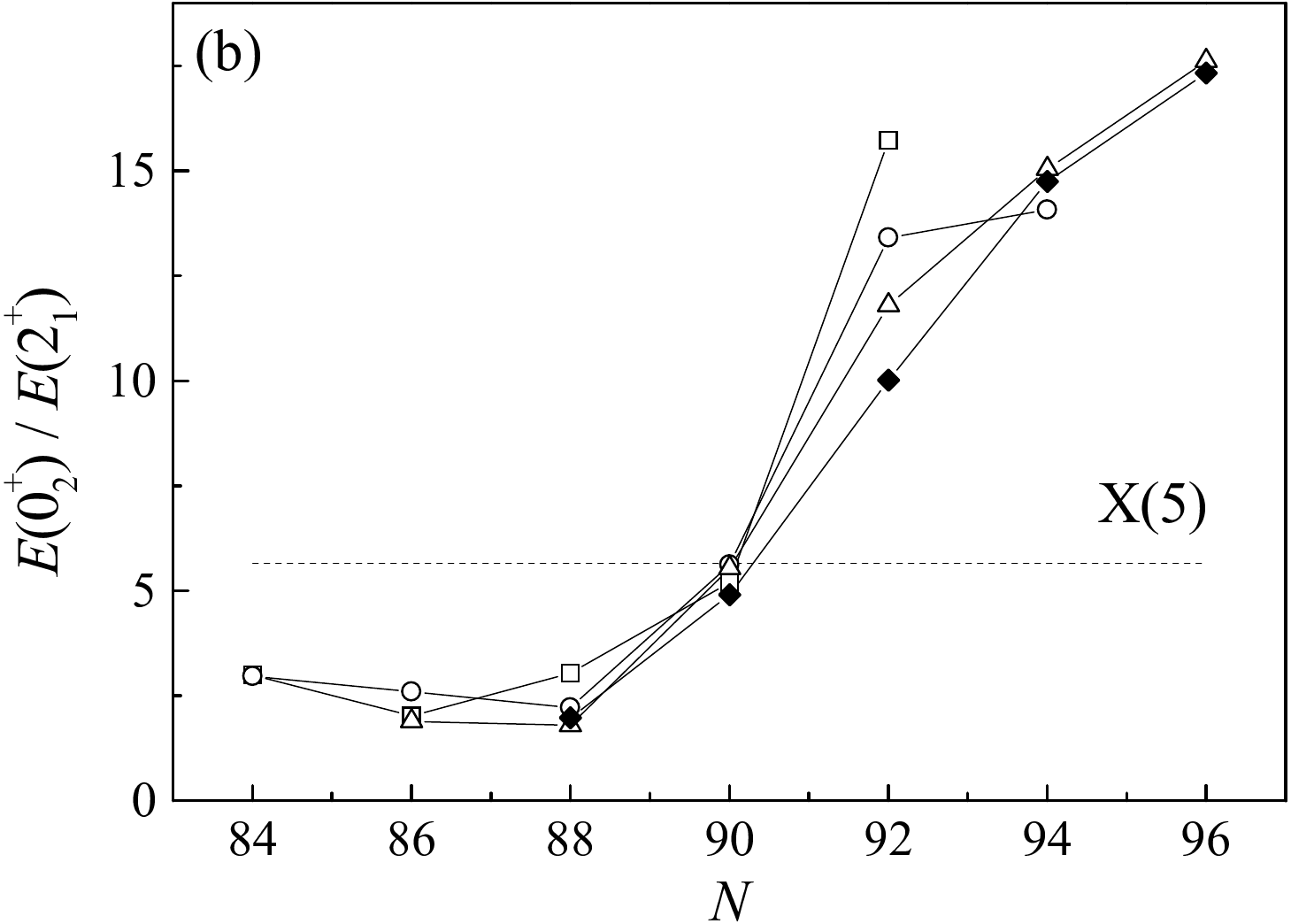}
\end{center}
\caption[Evolution of the $4^+_1$ and $0^+_2$ energies across the
$N$=90 region.]
{Evolution of the (a) $4^+_1$ and (b)
$0^+_2$ energies (normalized to the $2^+_1$ energy) across the
$N$=90 transition region, for the
Nd~(\captionopensquare),
Sm~(\captionopencircle),
Gd~(\captionopenuptriangle), and
Dy~(\captionsoliddiamond) isotopic chains,
compared with the X(5) prediction.  (Figure from Ref.~\cite{caprio2002:156dy-beta}.)
\label{fign90syst}
}
\end{figure}

The $N$=90 Nd--Dy nuclei have yrast band level energies closely
matching those of the X(5) model [Fig.~\ref{fign90yrast}(a)].  The
evolution of $R_{4/2}$$\equiv$$E(4^+_1)/E(2^+_1)$ along the Nd--Dy
isotopic chains is shown in Fig.~\ref{fign90syst}(a).  It can be seen
that the agreement of yrast energies with the X(5) predictions is
localized to $N$=90, as substantial changes occur by $N$=88 or
$N$=92.

The situation for the $B(E2)$ strengths within the yrast bands of the
$N$=90 Nd--Dy isotones [Fig.~\ref{fign90yrast}(b)] is less clear. The
spin-dependence of the yrast band $B(E2)$ values is generally
intermediate between the ideal vibrator and rotor limits, and the
qualitative trend is consistent with the X(5) predictions.  However,
many of the data points lie somewhat below the X(5) predictions.  Some
of the experimental $B(E2)$ values shown were deduced using highly
reliable techniques, \eg, the differential decay curve
method~\cite{boehm1993:ddcm} for recoil distance method lifetime
measurements.  However, for others, reported values differ or the
measurements involved experimental ambiguities.  The very low value
for
\mathbox{B(E2;6^+_1\rightarrow4^+_1)} in $^{156}$Dy~\cite{ward1979:156dy-rdm,emling1984:156dy-rdm},
which lies below even the rotor model predictions
[Fig.~\ref{fign90yrast}(b)], provides an example of such a case (see
Ref.~\cite{caprio2002:156dy-beta} for details).  Remeasurement could provide
valuable clarification.

The excitation energy of the first excited $0^+$ sequence is very
similar for all four $N$=90 Nd--Dy isotones and closely matches the
prediction for the first excited $K^\pi=0^+$ band in the X(5) model
[Fig.~\ref{fign90syst}(b)].  The observable $E(0^+_2)/E(2^+_1)$ is
rapidly changing along the isotopic chains [see also
Fig.~\ref{fig2vs0}(b) on page~\pageref{fig2vs0}], so the agreement with
X(5) is again highly localized to $N$=90.

\begin{figure}
\begin{center}
\includegraphics*[width=0.48\hsize]{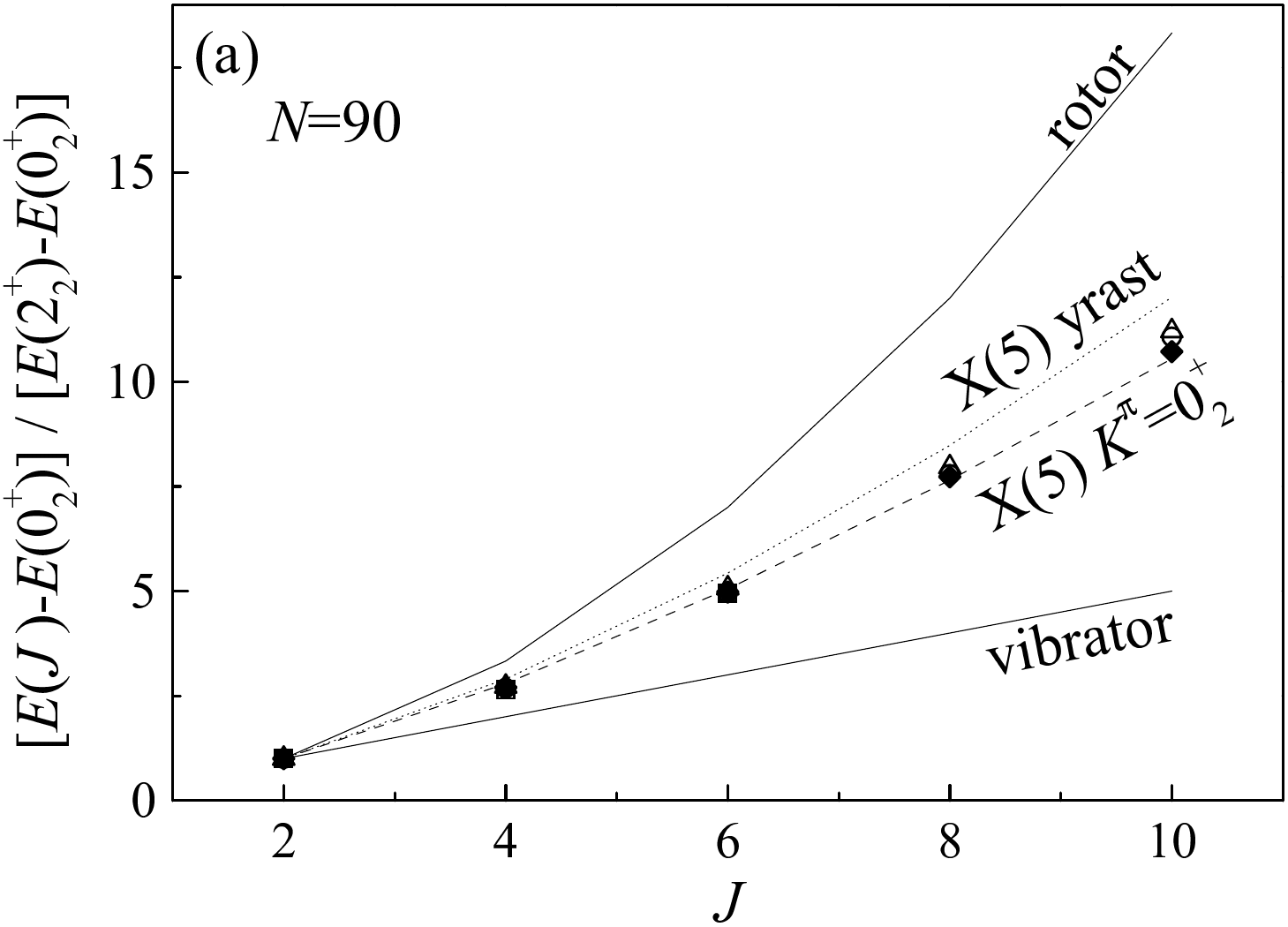}
\hfill
\includegraphics*[width=0.48\hsize]{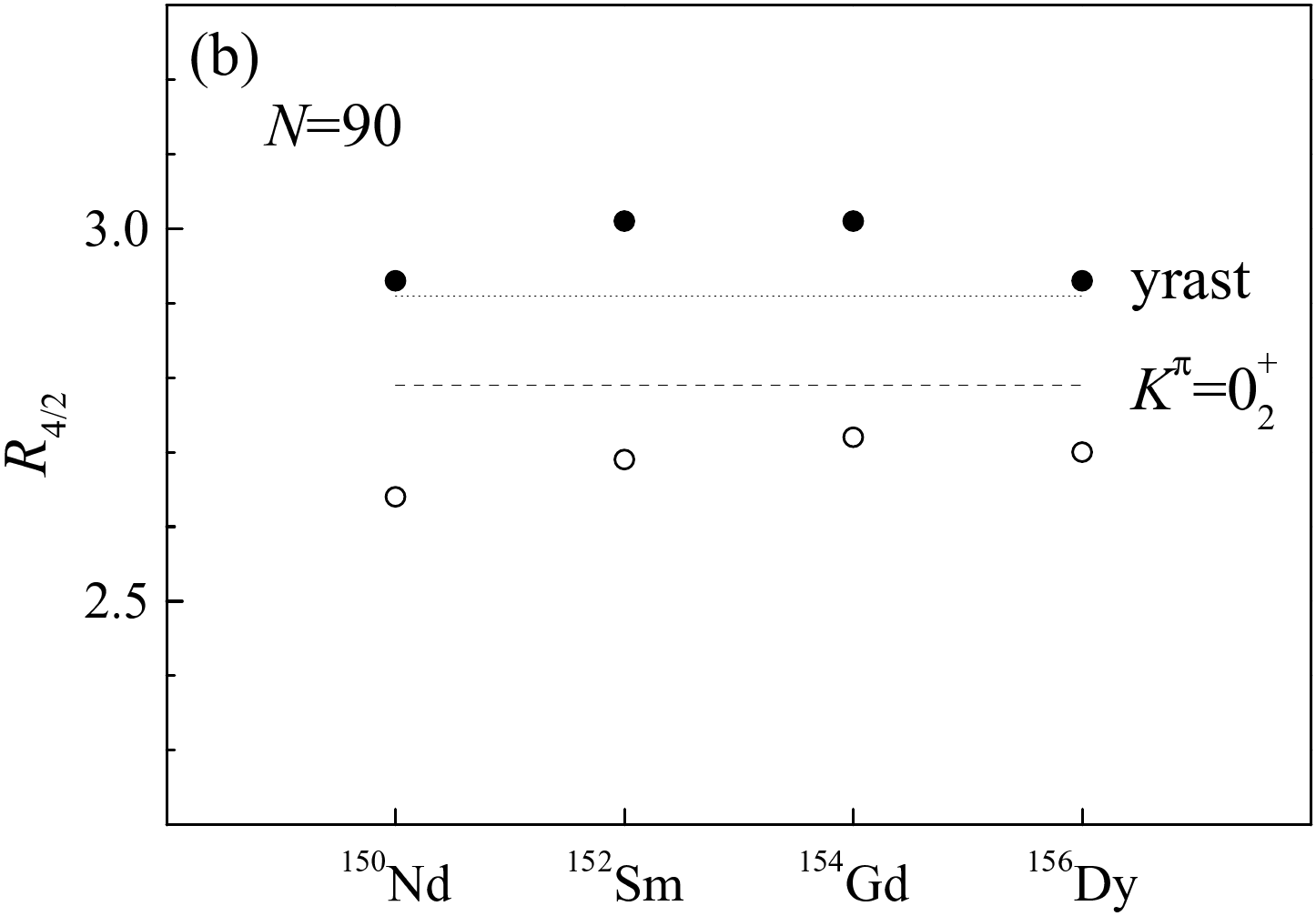}
\end{center}
\caption[$K^\pi=0^+_2$ band level
energies for the $N$=90 isotones.]  {$K^\pi=0^+_2$ band level
energies: (a)~Energies taken relative to the band head and normalized
to the $2^+$ band member, for the $N$=90 isotones
$^{150}$Nd~(\captionopensquare), $^{152}$Sm~(\captionopencircle),
$^{154}$Gd~(\captionopenuptriangle), and
$^{156}$Dy~(\captionsoliddiamond).  The X(5) predictions both for this
band (dashed line) and for the yrast band [$E(J)/E(2^+_1)$] (dotted
line) are shown, illustrating the differing spin dependences discussed
in the text. The rotor and harmonic vibrator predictions are also
indicated. (b)~$R_{4/2}$ values for the yrast~(\captionsolidcircle) and
$K^\pi=0^+_2$~(\captionopencircle) bands in these nuclei, along with the
X(5) predictions.  [Part~(a) from Ref.~\cite{caprio2002:156dy-beta}.]
\label{fign90betae}
}
\end{figure}

In the X(5) predictions, the successive $K^\pi=0^+$ bands differ from
each other in both the energy ratios and energy spacing scale of
levels within the band.  The dependence of the level energies upon
spin becomes successively more linear, or oscillator-like, for more
excited $K^\pi=0^+$ bands.  This effect is apparent in the X(5)
predictions shown in Fig.~\ref{fign90betae}(a) for the first two
$K^\pi=0^+$ bands, where the curve for the $K^\pi=0^+_2$ band is less
sharply upward-curving than is the curve for the yrast band (compare
the dashed and dotted lines).  The $N$=90 Nd--Dy nuclei exhibit this
behavior; while the yrast band level energies follow the X(5)
predictions for the yrast band [Fig.~\ref{fign90yrast}(a)], the
$K^\pi=0^+_2$ band level energies have a spin dependence which closely
matches the X(5) predictions for the $K^\pi=0^+_2$ band
[Fig.~\ref{fign90betae}(a)].  For the $2^+$ and $4^+$ band members
specifically, this effect corresponds to a reduction in the ratio of
$4^+$ state energy to $2^+$ state energy (taken relative to the band
head) for the higher bands.  The values for the $N$=90 nuclei are
summarized in Fig.~\ref{fign90betae}(b).

\begin{figure}
\begin{center}
\includegraphics*[width=0.48\hsize]{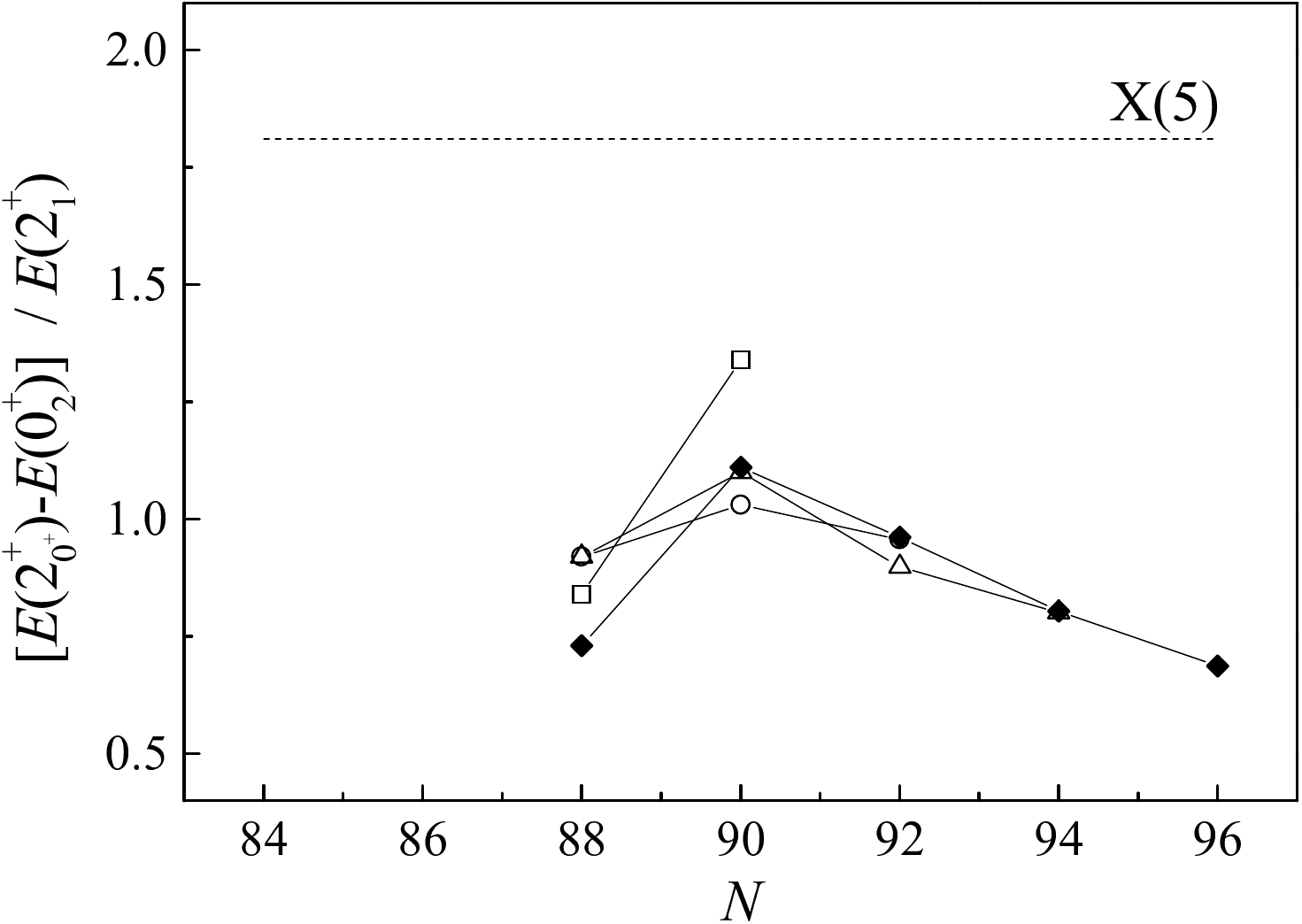}
\end{center}
\caption[Energy spacing scale of the excited $0^+$ sequence across the
$N$=90 region.]
{Energy spacing scale of the excited $0^+$ sequence relative to that of the 
yrast sequence, for the
Nd~(\captionopensquare),
Sm~(\captionopencircle),
Gd~(\captionopenuptriangle), and
Dy~(\captionsoliddiamond) isotopic chains,
compared with the X(5) prediction.  (Figure from Ref.~\cite{caprio2002:156dy-beta}.)
\label{fign90systrmom}
}
\end{figure}

The energy spacing \textit{scale} for the $K^\pi=0^+_2$ band is
predicted in X(5) to be much larger than for the yrast band, with
$[E(2^+_2)-E(0^+_2)]/E(2^+_1)\approx1.81$.  Only in $^{150}$Nd is the
$K^\pi=0^+_2$ band $2^+$-$0^+$ energy spacing significantly larger
than the yrast band spacing (Fig.~\ref{fign90systrmom}).  For the Sm,
Gd, and Dy isotopic chains, $N$=90 is the location of a local maximum
$[E(2^+_2)-E(0^+_2)]/E(2^+_1)$ in this ratio
(Fig.~\ref{fign90systrmom}), but only relative to neighboring isotopes
for which the $K^\pi=0^+_2$ band spacing is \textit{smaller} than the
yrast band spacing.  The actual spacing is still much less than
predicted by the X(5) model.  This substantial discrepancy is also
encountered in descriptions of the $N=90$ nuclei using the
IBM~\cite{scholten1978:ibm-fits,zamfir1999:152sm-beta}, truncated GCM
(see below), and other collective
models~\cite{kishimoto1976:boson-expansion}, which likewise lead to
predictions of a much larger spacing scale for the $0^+_2$ sequence.

The new intensity and lifetime results for the $N$=90 nuclei provide
greatly improved information on $B(E2)$ strengths for transitions from
the $2^+$, $4^+$, and $6^+$ members of the first excited $K^\pi=0^+$
band, including both in-band and interband transitions.  Absolute
$B(E2)$ strengths are known for band members in $^{150}$Nd,
$^{152}$Sm, and $^{154}$Gd, while, due to a lack of lifetime data,
only relative $B(E2)$ strengths are available for $^{156}$Dy (see
Refs.~\cite{kruecken2002:150nd-rdm,nds1996:152,nds1998:154,ndsboth:156}).

\begin{figure}
\begin{center}
\includegraphics*[width=0.75\hsize]{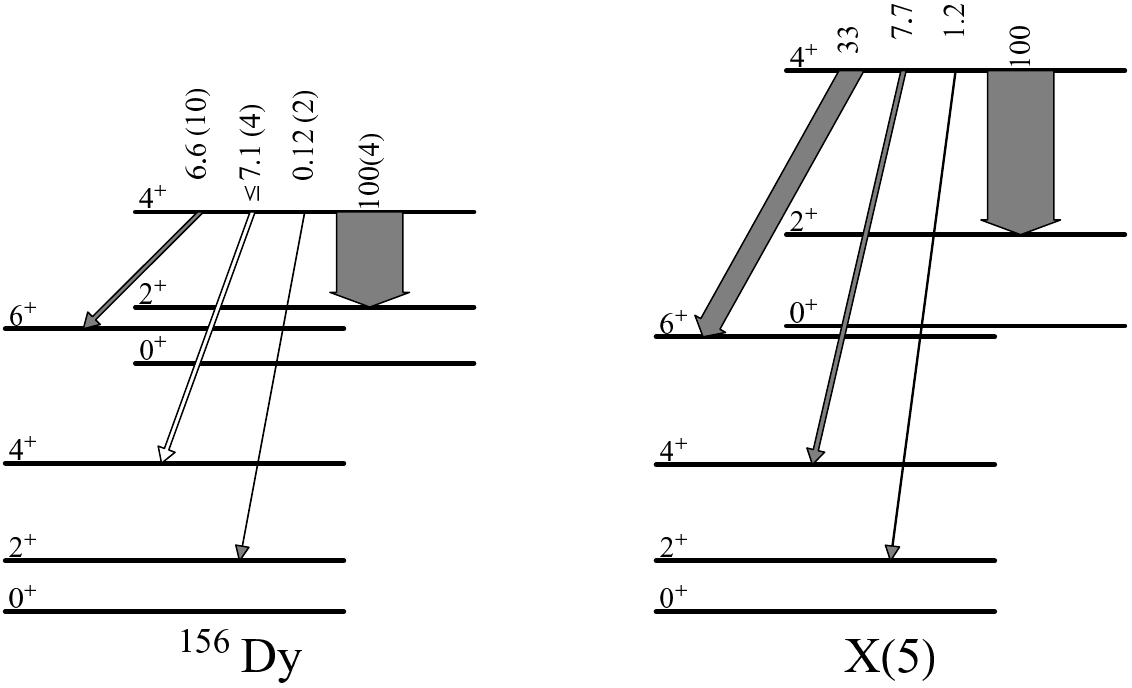} 
\end{center}
\caption[Relative $B(E2)$ strengths for transitions from the $4^+_{K^\pi=0^+_2}$ 
state in $^{156}$Dy.]  {Relative $B(E2)$ strengths for transitions
from the $4^+$ member of the $K^\pi=0^+_2$ band in $^{156}$Dy, based
upon data from Chapter~\ref{chap156dy}.  The X(5) predictions are
shown for comparison.  The open arrow for the spin-unchanging
transition in $^{156}$Dy indicates the possibility of $M1$
contamination.  (Figure adapted from
Ref.~\cite{caprio2002:156dy-beta}.)
\label{fig156dybranch4}
}
\end{figure}

Let us first consider the overall strength scales of the in-band and
interband transitions.  In $^{150}$Nd, $^{152}$Sm, and $^{154}$Gd, the
interband transition strengths from the $0^+_2$ and $2^+_2$ states to
the yrast band are consistently weaker than expected for X(5) by about
a factor of two.  The pattern of interband transition strengths from
the $4^+_2$ level is less regular: in $^{150}$Nd, the $4^+_2\rightarrow6^+_1$
strength is about a factor of two
\textit{above} the X(5) prediction, but in $^{152}$Sm it is a factor of 10
\textit{below} the X(5) prediction.   In $^{156}$Dy,
the $2^+_2$ and $4^+_2$ level lifetimes are unknown, so the absolute
strengths of the transitions cannot be determined.  For the $2^+$ band
member, not even the relative strengths of the in-band and interband
transitions can be compared, since only a limit can be placed on the intensity of
the in-band $2^+_2\rightarrow0^+_2$ transition
(Chapter~\ref{chap156dy}).  For the $4^+_2$ state in
$^{156}$Dy, such a comparison of relative strengths can be
made (Fig.~\ref{fig156dybranch4}), and it is found that the
interband transitions are much weaker relative to the
in-band transition than would be expected from the X(5) predictions.
Thus, the general trend for these nuclei seems to be that the
$K^\pi=0^+_2$$\rightarrow$$K^\pi=0^+_1$ interband transition strength
scale is weaker than predicted in X(5), albeit with some exceptions
and some missing data.

\begin{figure}[p]
\begin{center}
\includegraphics*[width=1.0\hsize]{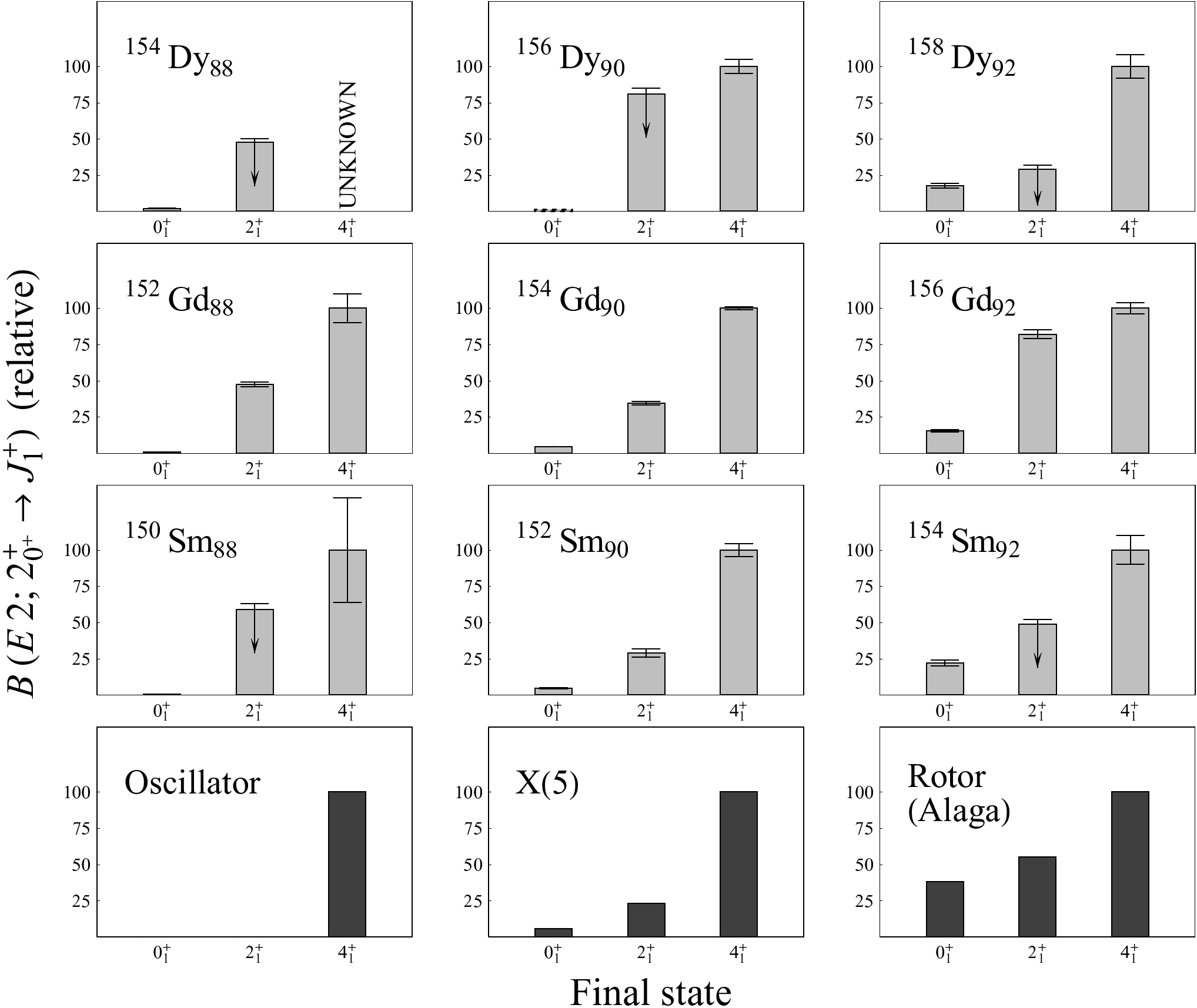}
\end{center}
\caption[Relative $B(E2)$ branching strengths for transitions from 
the $2^+$ state.]
{Relative $B(E2)$ branching strengths for the ``interband'' transitions from 
the $2^+$ state built upon the first excited $0^+$ state to the
yrast $0^+$, $2^+$, and $4^+$ states, shown for the $N$=88, 90, and 92
isotopes of Dy, Gd, and Sm.  The extreme harmonic
oscillator limit, X(5), and pure rotor limit predictions are
provided for comparison at the bottom.  Data are adopted values from Nuclear Data
Sheets~\cite{nds1995:150,nds1996:152,nds1998:154,ndsboth:156,nds1996:158}
except those for $^{152}$Sm (Chapter~\ref{chap152sm}) and
$^{156}$Dy (Chapter~\ref{chap156dy}).  Error bars on the relative $B(E2)$ strengths
include contributions from the experimental uncertainties in intensity
and $E2/M1$ mixing ratios.  In the case of spin-unchanging transitions
for which the $E2/M1$ mixing ratio is unknown, the $B(E2)$ strength is
deduced assuming pure $E2$ multipolarity, and the possibility of
arbitrarily large $M1$ contamination is indicated by a downward
arrow.  Since the strength of the $2^+_2\rightarrow4^+_1$ transition
in $^{154}$Dy is unknown, the strengths of the $2^+_2\rightarrow0^+_1$
and $2^+_2\rightarrow2^+_1$ transitions are normalized for comparison
purposes by setting the relative $2^+_2\rightarrow2^+_1$ strength
equal to that in the neighboring isotone $^{152}$Gd.  (Figure from Ref.~\cite{caprio2002:156dy-beta}.)
\label{fign90branch2}
}
\end{figure}

The X(5) model predicts a characteristic branching pattern for the
decay of members of the first excited band to the yrast band, in which
the spin-descending branches are highly suppressed.
Fig.~\ref{fign90branch2} summarizes the branching properties for the
$2^+$ state in the first excited $0^+$ sequence (denoted by
$2^+_{0^+}$) for the $N$=88, 90, and 92 isotopes of Dy, Gd, and Sm.
In going from $N$=88 to $N$=92, the $2^+_{0^+}\rightarrow0^+_1$
transition evolves from being highly suppressed relative to the
$2^+_{0^+}\rightarrow4^+_1$ transition
[$B(E2;2^+_{0^+}\rightarrow0^+_1)/B(E2;2^+_{0^+}\rightarrow4^+_1)\approx0.01$
at $N$=88] to having substantial strength
[$B(E2;2^+_{0^+}\rightarrow0^+_1)/B(E2;2^+_{0^+}\rightarrow4^+_1)\approx0.20$
at $N$=92].  The suppression for $N$=88 is reminiscent of the
situation in the pure oscillator limit, although clearly these nuclei
are far from achieving pure oscillator structure.  In this limit, the
$2^+$ state directly above the $0^+_2$ state is the $2^+_3$ state, and
the only one of the three transitions which is phonon-allowed is the
$2^+_{0^+}\rightarrow4^+_1$ transition (Fig.~\ref{fign90branch2},
bottom left panel).  For the $N$=92 nuclei, the relative $B(E2)$
strengths begin to resemble the Alaga-rule strengths expected for a
pure rotor (Fig.~\ref{fign90branch2}, bottom right panel), though they
do not fully match these values.  For the $N$=90 nuclei $^{152}$Sm and
$^{154}$Gd (as well as $^{150}$Nd~\cite{kruecken2002:150nd-rdm}, not
shown) the agreement with the X(5) predictions
(Fig.~\ref{fign90branch2}, bottom center panel) is quite good.  In
$^{156}$Dy, the $2^+_2\rightarrow0^+_1$ transition is extremely weak,
with a relative $B(E2)$ strength at least a factor of 3--5 weaker than
in the neighboring $N$=90 isotones or the X(5) predictions.

Of the $N$=90 isotones, $^{152}$Sm is the most useful for comparison
to model predictions, due to the availability of a full set of data
from complementary types of experiments.  Since the other $N$=90
isotones have similar properties, many aspects of the analysis of
$^{152}$Sm carry over directly to these nuclei as well.  Comparison
with the predictions of the X(5) model is straightforward, since the
predictions for the yrast and $K^\pi=0^+_2$ bands have essentially no
free parameters (Section~\ref{secbenche5x5}).  However, comparison
with the predictions of the truncated form of the GCM discussed in
Chapter~\ref{chapgcm} first requires a suitable set of parameter
values to be identified.  

\begin{figure}[t]
\begin{center}
\includegraphics*[width=0.9\hsize]{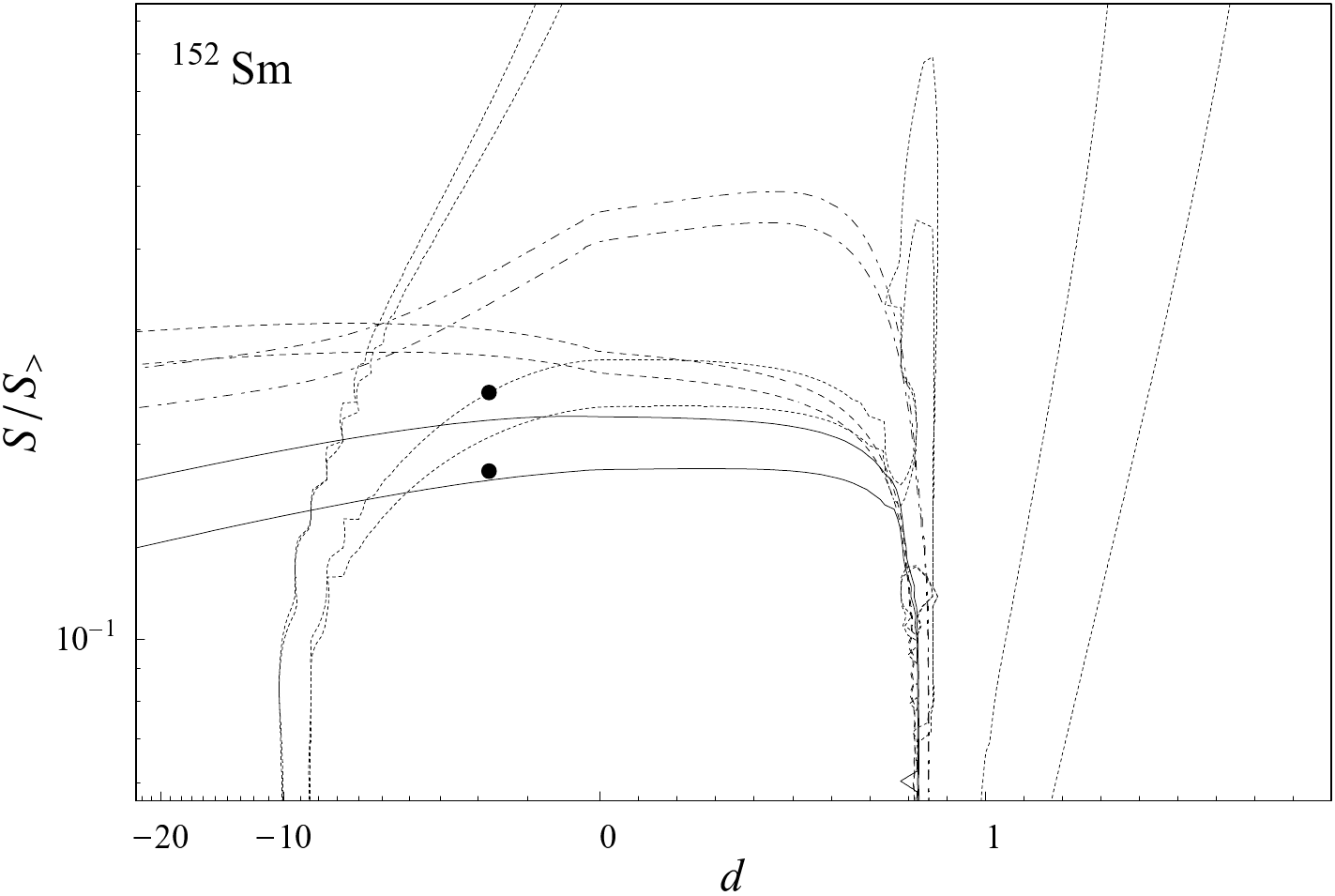}
\end{center}
\caption[Regions for which GCM predictions match
$^{152}$Sm.]  {Regions in the ($d$,$S$) parameter space for which the
GCM predictions of selected observables match the values found in
$^{152}$Sm: $R_{4/2}$=3.02 to within 2$\%$ (solid line),
$E(0^+_2)/E(2^+_1)$=5.61 to within 5$\%$ (dashed line),
$E(2^+_3)/E(2^+_1)$=8.90 to within 5$\%$ (dashed-dotted line), and
$B(E2;2^+_2\rightarrow4^+_1)/B(E2;2^+_2\rightarrow0^+_1)$=21 to
within 25$\%$ (dotted line).  The solid circles indicate 
($d$=-1.75, $S$=0.0121\timesSunits) and
($d$=-1.75, $S$=0.018\timesSunits), the parameter values discussed in
the text.
\label{figgcmcontour_152sm}
}
\end{figure}

Some of the most basic observables used to characterize the structure
of the $N$=90 nuclei in the preceding discussion were the energy
ratio $R_{4/2}$, the $0^+_2$ level energy, and the branching ratio
$B(E2;2^+_2\rightarrow4^+_1)/B(E2;2^+_2\rightarrow0^+_1)$.  The
$2^+_3$ level energy is strongly influenced by the $\gamma$-dependence
of the potential, which is irrelevant in the X(5) description due to a separation of variables
but which must be addressed in the GCM
description.  Fig.~\ref{figgcmcontour_152sm} shows the regions in the
$(d,S)$ parameter space for which the GCM predictions of these
observables match the values found in $^{152}$Sm.  
No point of simultaneous agreement for all four observables exists,
but reasonable compromises may be found.  The GCM predictions for
$d$=-1.75 and $S$=0.0121\timesSunits were reported
in Ref.~\cite{zhang1999:152sm-gcm}.  The predictions for $d$=-1.75 and
$S$=0.018\timesSunits, discussed here, are similar, differing somewhat
in the energy scale of the excited $0^+$ sequence and in the overall
normalization of the interband $B(E2)$ strengths.

\begin{figure}[p]
\begin{center}
\includegraphics*[width=1.0\hsize]{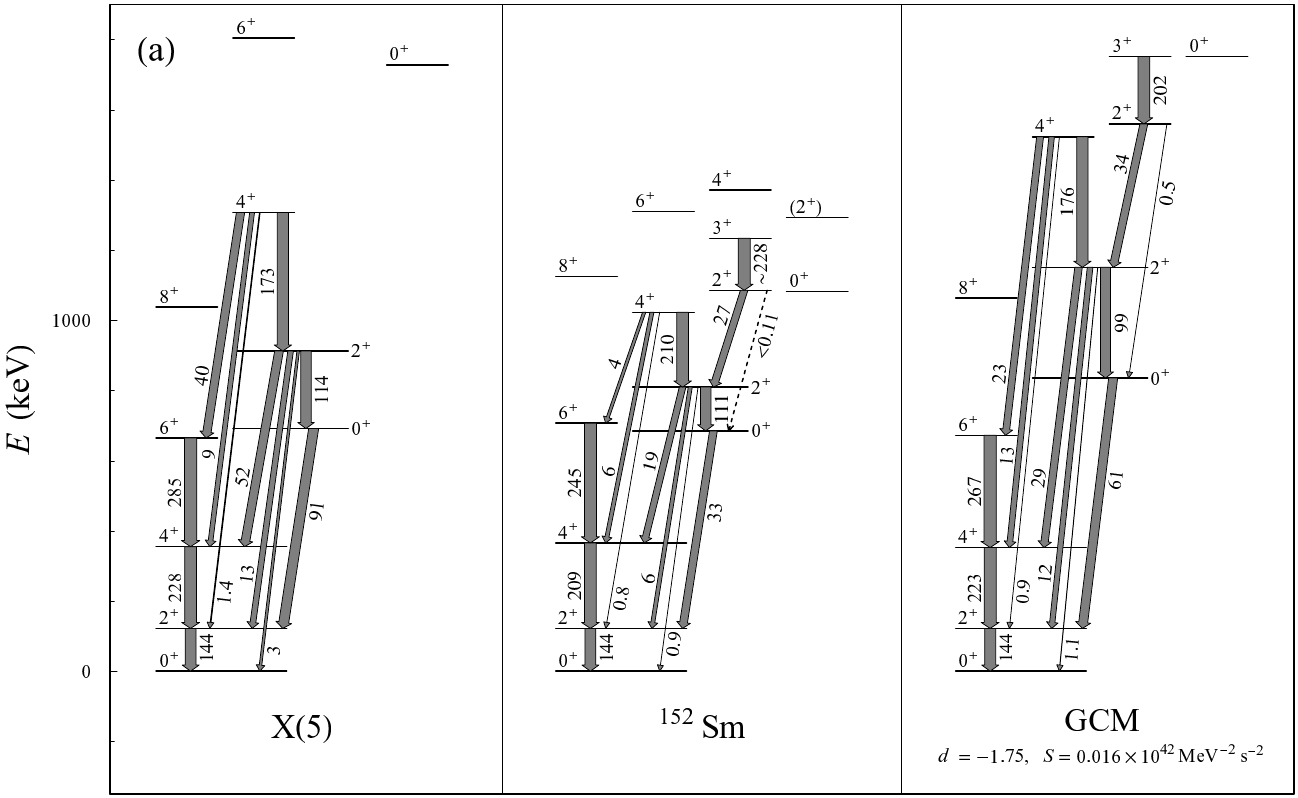}
\vspace{20pt}\\
\includegraphics*[height=2.5in]{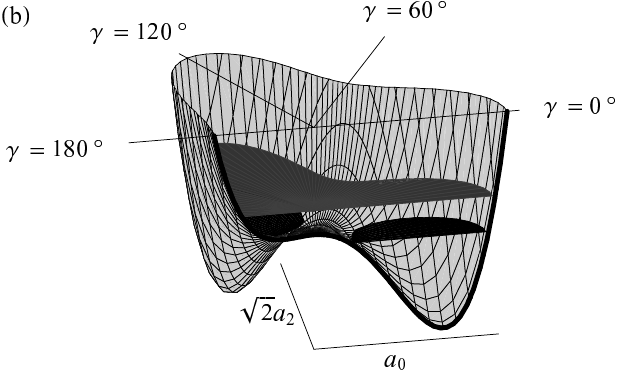}
\vspace{11pt}\\
\end{center}
\caption[Level scheme for $^{152}$Sm, with X(5) and GCM predictions; GCM potential.]  {Model
description of $^{152}$Sm: (a)~Experimental level scheme and selected
$B(E2)$ strengths for $^{152}$Sm.  The X(5) predictions and the GCM
predictions for $d$=-1.75 and $S$=0.016\timesSunits, normalized to the
experimental $E(2^+_1)$ and $B(E2;2^+_1\rightarrow0^+_1)$ values, are
shown for comparison.  All $B(E2)$ strengths are indicated in
\Wu, and transition arrow widths are proportional to the logarithm of
the $B(E2)$ strength. Experimental values are from data in
Refs.~\cite{nds1996:152,klug2000:152sm-rdm} and
Chapter~\ref{chap152sm}.  (See these references for uncertainties.)
(b)~A plot of the potential for the GCM calculation as a function of the shape
coordinates, showing $0^+_1$ and $0^+_2$ level energies.
\label{figgcm152sm}
}
\end{figure}

The predictions of the GCM for $d$=-1.75 and $S$=0.018\timesSunits are
shown in Fig.~\ref{figgcm152sm}(a), together with the observed values
in $^{152}$Sm and the predictions of the X(5) model.  
The GCM calculation yields an energy spacing scale for the excited
$0^+$ sequence which is greatly expanded compared to that of the yrast
sequence.  This effect is much more exaggerated than in the X(5)
model, as was also found in Section~\ref{secgcmx5}.  The $R_{4/2}$
value for the excited $0^+$ band, at 2.18, is much lower than the
experimental or X(5) values [Fig.~\ref{fign90betae}(b)].  The $B(E2)$
predictions from the GCM calculation are similar to those of X(5),
providing comparable or, for several transitions, better agreement
with the experimental values.  The GCM calculation reproduces the
decay properties of the $2^+_3$ level to the $0^+_2$ sequence,
providing a good quantitative prediction for the relatively strong
$2^+_3\rightarrow2^+_2$ transition and yielding a suppressed
$2^+_3\rightarrow0^+_2$ transition.  (The prediction for the
$2^+_3\rightarrow0^+_2$ transition's strength varies rapidly in this
region of parameter space and is much lower for $d$=-1.75 and
$S$=0.0121\timesSunits.  See Ref.~\cite{zhang1999:152sm-gcm}.)  The
GCM calculation, like the X(5) model, yields a relatively high-lying
$0^+_3$ state.  It is not clear to what extent this state can be
structurally identified with the observed $0^+_3$ level in $^{152}$Sm.

The potential for the GCM calculation at $d$=-1.75 and
$S$=0.018\timesSunits is shown in Fig.~\ref{figgcm152sm}(b), together
with the $0^+_1$ and $0^+_2$ level energies relative to this
potential.  At these energies, the potential is markedly soft in
$\beta$, and, for the higher-energy of these two levels, $\beta=0$ is
even classically energetically accessible.

It apprears from the phenomenological comparisons that the gross
qualitative features of the $N$=90 nuclei can be explained using a
collective model with a $\beta$-soft, $\gamma$-stabilized deformation
potential.  Such a potential is a fundamental feature common to the
descriptions of the $N$=90 nuclei provided by the X(5) model and the
GCM calculations, as well as by IBM
calculations~\cite{scholten1978:ibm-fits}.  Description of the $N$=90
nuclei with a $\beta$-soft potential is also consistent with
predictions based upon the underlying single-particle structure, in
the models of Kumar and Barranger and of Kishimoto and Tamura and in
the Nilsson-Strutinsky-BCS model (see
Refs.~\cite{kumar1974:150sm152sm-ppq,kishimoto1976:boson-expansion,zhang1999:152sm-gcm}).
However, major aspects of the structure and of structural evolution in
the region remain to be understood, as evidenced by, for instance, the
substantial misprediction by the collective models of the energy scale
for the excited sequence and by the erratic variations among the
nuclei in the strengths for transitions from the $4^+_2$ state.

\section{$^{102}$Pd}
\label{secphenomgsoft}

\begin{figure}[tb]
\begin{center}
\includegraphics*[width=1.0\hsize]{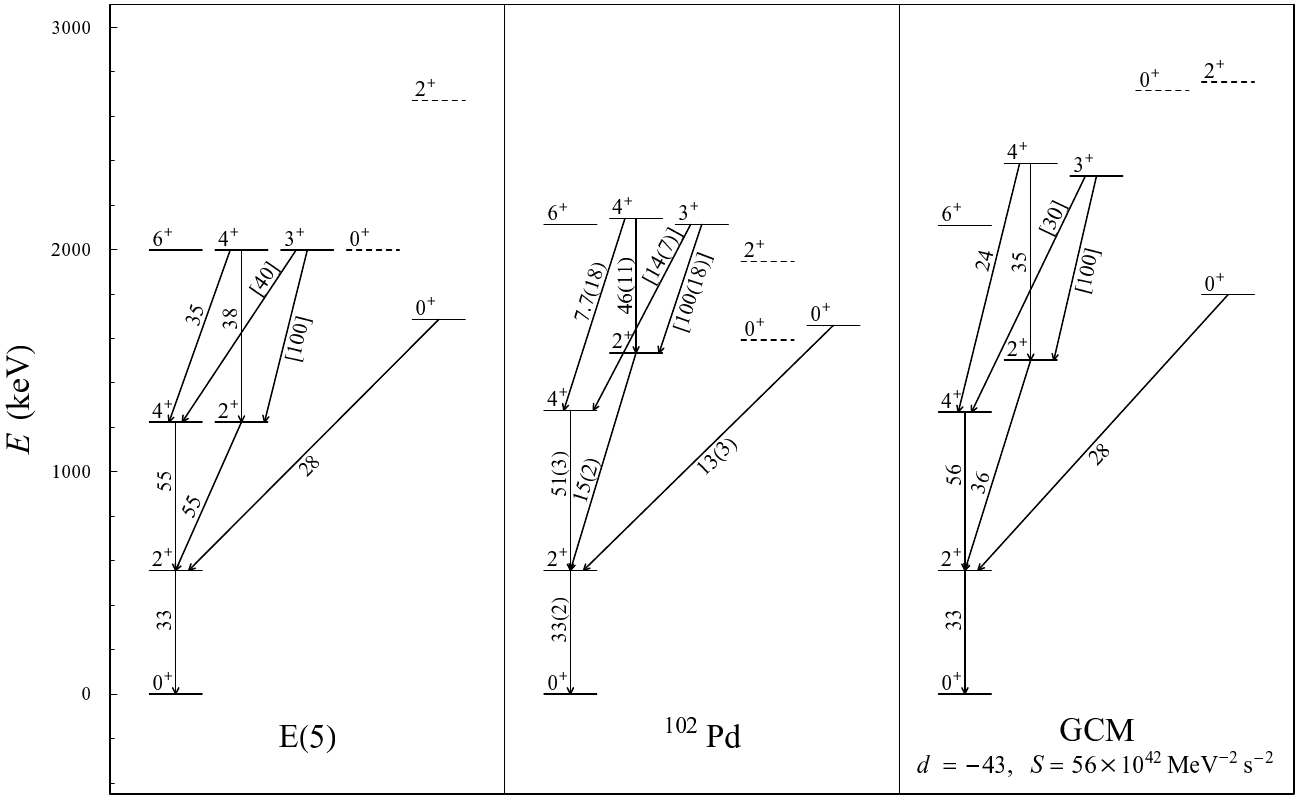}
\end{center}
\caption[Level scheme for $^{102}$Pd,
with E(5) and GCM predictions.]  {Experimental level scheme and
$B(E2)$ strengths for $^{102}$Pd, alongside the E(5) predictions and
the GCM predictions for $d$=-43 and $S$=56\timesSunits, normalized to
the experimental $E(2^+_1)$ and $B(E2;2^+_1\rightarrow0^+_1)$ values.
Observed levels with no clear theoretical counterpart or calculated
levels with no clear experimental counterpart (see text) are indicated
with dashed lines.  Experimental $B(E2)$ strengths are from
Refs.~\cite{nds1998:102,luontama1986:102pd104pd-p2n-ppprime-coulex}
and Chapter~\ref{chap102pd}, with the assumption of pure $E2$
multipolarity if the multipolarity is not otherwise known (see
Table~\ref{tab102pdbranch}).  $B(E2)$ values in brackets are relative
values, while all others are absolute values in \Wu. 
\label{figgcm102pd}
}
\end{figure}

The level energies and $B(E2)$ strengths for the nucleus $^{102}$Pd
qualitatively resemble those predicted by the E(5) model, as discussed
in Chapter~\ref{chap102pd}.  Let us summarize the comparison between
the $^{102}$Pd observed properties and the E(5) predictions
[Fig.~\ref{figgcm102pd}~(left, middle)] permitted by the new
spectroscopic data.  In the proposed $\xi$=1 family, spin multiplets
following the O(5) structure are present for $\tau$=0, 1, 2, and 3.
These occur at energies ($R_{4/2}$, $E_{\tau=3}$/$E_{\tau=1}$) close
to those expected in E(5) (Fig.~\ref{figgcm102pd}), although the
splitting of the $\tau$=2 multiplet in $^{102}$Pd leaves only the
$4^+$ member closely matching the predicted $E_{\tau=2}$/$E_{\tau=1}$
of E(5).  A $0^+$ member of the $\tau$=3 multiplet must be found for
this picture to be complete.  The observed strengths for $E2$
transitions which are allowed in an O(5) scheme ($\Delta\tau=\pm1$)
are consistently enhanced over the forbidden transitions
($\Delta\tau=\pm2$), pending experimental $\delta$ values.  The
lifetime of the $4^+$ member of the $\tau$=3 multiplet is
known~\cite{luontama1986:102pd104pd-p2n-ppprime-coulex}, yielding a
$B(E2)$ strength to the $2^+_{\tau=2}$ level in agreement with E(5)
(Fig.~\ref{figgcm102pd}); lifetimes must be measured for the other
multiplet members.  With the lowest $0^+$ state dismissed as a
probable intruder on the basis of its energy evolution along the
isotonic chain, the next higher $0^+$ state is a good candidate for
the $\xi$=2 family head, showing the $E2$ collectivity required in an
O(5) picture and lying close to the energy predicted by E(5).%

However, several differences in quantitative detail are present, as
can be seen from Fig.~\ref{figgcm102pd}.  Some of the main deviations
from the E(5) picture are:
\begin{dissenumeratelist}
\item The $4^+$ and $2^+$
members of the proposed $\tau$=2 multiplet are split in energy, with the $4^+$ level lower.
\item The absolute
\mathbox{B(E2;2^+_{\tau=2}\rightarrow2^+_{\tau=1})} strength is lower than the
\mathbox{B(E2;4^+_{\tau=2}\rightarrow2^+_{\tau=1})} strength.
\item Both the
$3^+_{\tau=3}$ and $4^+_{\tau=3}$ states show a strong preference to
decay to the $2^+_{\tau=2}$ state rather than to the $4^+_{\tau=2}$
state.  Only a slight such preference is expected in E(5).  (See also
Table~\ref{tab102pdbranch} on page~\pageref{tab102pdbranch}.)
\end{dissenumeratelist}
These deviations from the E(5) model predictions all are qualitatively
consistent with the incipient formation of a $\gamma$ band.  Such
structure would be expected if the potential were
perturbed from E(5) to impose a slight preference for axial symmetry.

\begin{figure}[p]
\begin{center}
\includegraphics*[width=0.9\hsize]{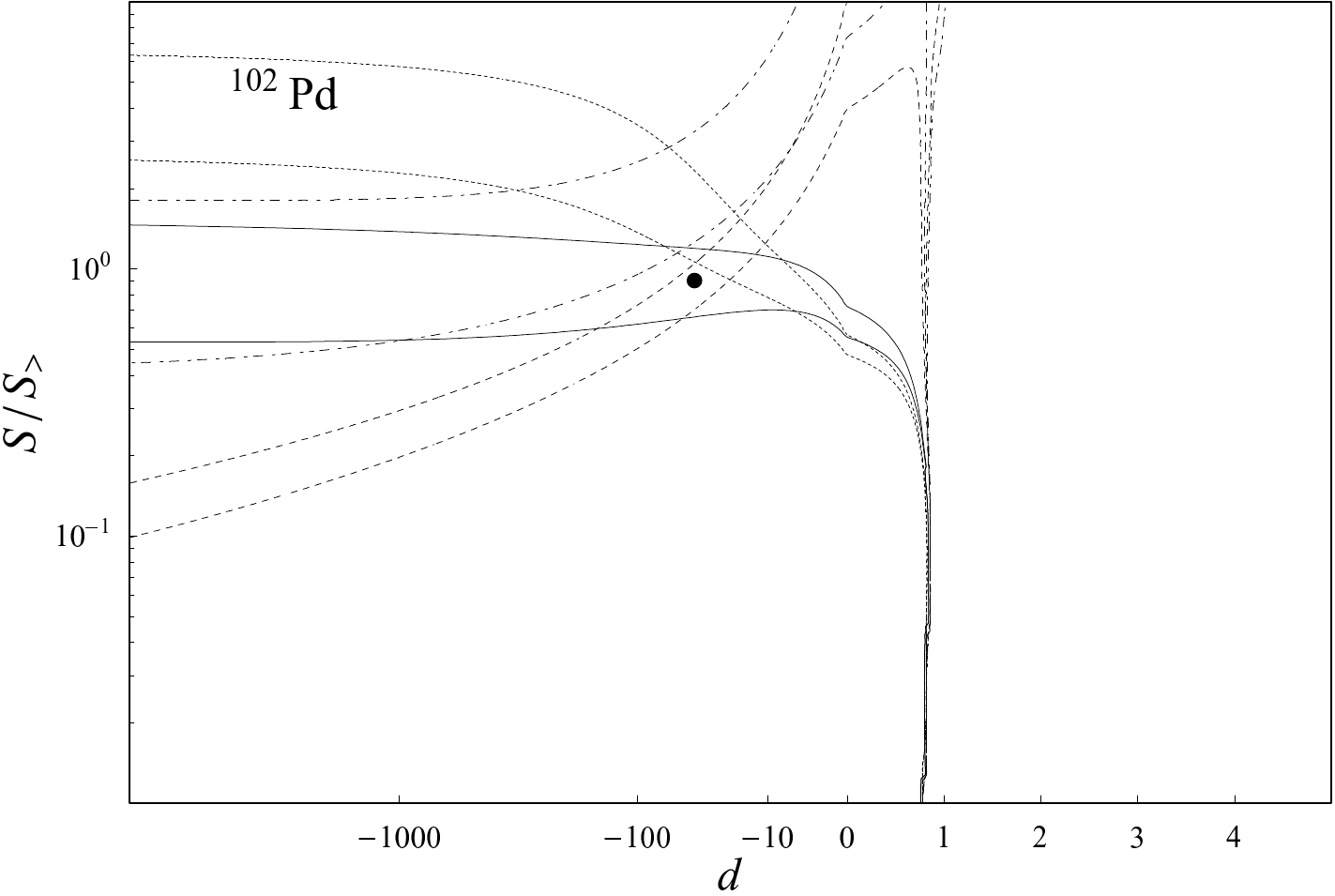}
\end{center}
\caption[Regions for which GCM predictions match
$^{102}$Pd.]  {Regions in the ($d$,$S$) parameter space for which the
GCM predictions of selected observables match the values found in
$^{102}$Pd: $R_{4/2}$=2.29 to within 2$\%$ (solid line),
$E(2^+_2)/E(2^+_1)$=2.75 to within 5$\%$ (dashed line),
$E(4^+_2)/E(2^+_1)$=3.84 to within 5$\%$ (dashed-dotted line), and
$E(0^+_2)/E(2^+_1)$=2.98 to within 5$\%$ (dotted line).  The solid circle
indicates $d$=-43 and $S$=56\timesSunits, the parameter values
discussed in the text.
\label{figgcmcontour_102pd}
}
\end{figure}

To consider this perturbation quantitatively, let us make use of the
truncated GCM.  The choice of values for the parameters $d$ and $S$
can be constrained simply by requiring that the low-lying energy
observables be reproduced.  It is desirable to retain the good
agreement with the experimental $4^+_{\tau=2}$ level energy, $\tau$=3
multiplet energy, and first excited $0^+$ level energy obtained with
the E(5) description while also reproducing the splitting of the
$\tau=2$ multiplet.  Contours indicating the regions in GCM ($d$,$S$)
parameter space for which the predictions match the experimental
$R_{4/2}$, $E(2^+_2)/E(2^+_1)$, $E(4^+_2)/E(2^+_1)$, and
$E(0^+_2)/E(2^+_1)$ values are shown in
Fig.~\ref{figgcmcontour_102pd}, leading to parameter values
$d$$\approx$-43 and $S$$\approx$56\timesSunits.

The GCM predictions for $d$=-43 and $S$=56\timesSunits are shown in
Fig.~\ref{figgcm102pd}~(right).  The $4^+_1$-$2^+_2$ splitting and
low-lying level energies are well reproduced, although some splitting
is introduced between the $6^+_1$, $4^+_2$, and $3^+_1$ ``multiplet''
states.  The $B(E2)$ strengths for the $2^+_2\rightarrow2^+_1$,
$4^+_2\rightarrow4^+_1$, and $3^+_1\rightarrow4^+_1$ transitions are
all reduced relative to the E(5) predictions, while the other
transition strengths in Fig.~\ref{figgcm102pd} are relatively
unaffected.  The change in the predicted $2^+_2\rightarrow2^+_1$,
$4^+_2\rightarrow4^+_1$, and $3^+_1\rightarrow4^+_1$ $B(E2)$ strengths
is in the correct sense to bring them closer to the experimental
values but leaves these strengths still much greater than observed.
As noted above, one of the other outstanding issues regarding the
interpretation of $^{102}$Pd in the context of the E(5) picture is the
nonobservation of any $0^+$ level degenerate with the proposed
$\tau$=3 multiplet.  In the GCM calculation, the $0^+_3$ level is
predicted to be substantially higher in energy than the $6^+_1$,
$4^+_2$, and $3^+_1$ states.

\begin{figure}[p]
\begin{center}
\includegraphics*[height=2.5in]{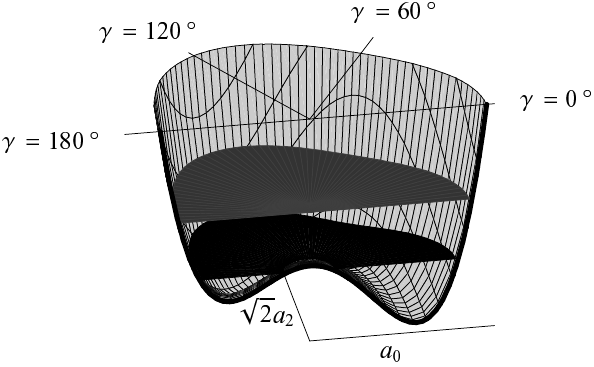}
\end{center}
\caption[Potential
for the GCM calculation for $^{102}$Pd.]  {Plot of the potential
for the GCM calculation for $^{102}$Pd ($d$=-43, $S$=56\timesSunits)
as a function of the shape coordinates, showing the $0^+_1$ and
$0^+_2$ level energies.
\label{figgcm102pdpotl}
}
\end{figure}

The potential for the GCM calculation for $^{102}$Pd is
plotted in Fig.~\ref{figgcm102pdpotl}, 
showing the extent of the
gentle minimum with respect to $\gamma$ at $\gamma$=0$^\circ$.  This potential may be compared with the E(5)-like potential of
Fig.~\ref{figgcme5}(b) on page~\pageref{figgcme5}.  In the present
calculation, the ground state energy is just below the local maximum
at $\beta$=0, making zero deformation energetically inaccessible.
This does not, however, necessarily imply any major structural change
to the states: even for a pure square well, with no maximum in the
potential at $\beta$=0, the wave functions for low-lying states posess
a low probability density at $\beta$=0 (see Fig.~\ref{figfwellxten}
on~\pageref{figfwellxten} for an illustration in the case of a finite
well).

\addtocontents{toc}{\protect\newpage}
\part{Conclusion}
\chapter{Conclusion}
\label{chapconcl}

The collective structure and excitation modes of nuclei intermediate
between spherical and well-deformed structure present special
challenges to categorization and to theoretical understanding.  The
recently proposed E(5) and X(5) models provide a new perspective on
structure near the critical points of shape transitions, suggesting
the relevance of an extreme ``$\beta$-soft'' form of structure.
Examples of nuclei to which these models are relevant have been
identified, including several in the $A$$\approx$100 and
$N$$\approx$90 ($A$$\approx$150) regions considered in the present
work.

For collective modes in transitional nuclei to be understood, solid
experimental information must be obtained, especially for the
low-spin, non-yrast states.  Detailed and reliable knowledge of level
properties is necessary both for an initial characterization of the
structure of the nuclear states and for in-depth comparison with model
predictions.  Acquisition of this information requires varied and
complementary experimental techniques, making use of the diverse
available population mechanisms and detection resources.  Several
experimental approaches have been utilized in the present work, all of
them involving $\gamma$-ray detection, based upon coincidence
spectroscopy, electronic timing, and Doppler broadening methods in
heavy-ion accelerator, reactor, and radioactive beam environments.

However, the necessary theoretical tools must also be present if an
understanding of transitional structure is to be extracted from the
experimental data.  The E(5) and X(5) models provide benchmarks for
structural comparison of transitional nuclei.  The present work
addressed the robustness of the predictions of the E(5) and X(5)
models and identified the extent to which these predictions change
under perturbations of the model Hamiltonian.  The present work also
included the development of an approach simplifying the use of an
existing model, the geometric collective model, making this model more
straightforward to apply to transitional and deformed nuclei.  This
approach facilitates the study of nuclei with structures near, but not
necessarily identical to, the E(5) and X(5) descriptions.

The present investigations demonstrate the relevance of $\beta$-soft
potentials to transitional nuclei.  The $N$=90 isotopes of Nd, Sm, Gd,
and Dy all have characteristics reproduced by the X(5) model, with its
$\beta$-soft, $\gamma$-localized potential, and by similar model
descriptions.  The nucleus $^{162}$Er, bordering the $N$$\approx$90
transition region, appears to support an extremely low-lying
$\beta$-vibrational excitation, implying a comparatively $\beta$-soft
rotational structure.  And the properties of $^{102}$Pd match those of
the E(5) model and similar descriptions involving a $\beta$-soft,
nearly $\gamma$-independent potential.  Future work must place these
$\beta$-soft forms of nuclear structure more fully in the context of
structural evolution.  It must be seen in what form and to what extent
these structures occur in other experimentally accessible transition
regions.

\phantomsection  
\addcontentsline{toc}{partstar}{Appendices}
\part*{Appendices}
\appendix

\chapter{Collective nuclear structure benchmarks}
\label{appbench}

\section{Preliminary definitions}
\label{secbenchintro}

Several simple analytical models of collective structure are used as
benchmarks for comparison throughout the preceding discussions.  In
this appendix, the basic results of these models are presented for
reference.  Further details may be found in, \eg,
Refs.~\cite{eisenberg1987:v1,bohr1998:v2,casten2000:ns} or in the
original literature.  Detailed level schemes and sets of $B(E2)$
values for the E(5) and X(5) models are also included.

The expansion of the nuclear surface shape in spherical harmonics was
introduced in Section~\ref{seccollective}.  For quadrupole
deformation, the expansion (\ref{eqnmultipoleshape}) reduces to
\begin{equation}
R(\theta,\varphi) = R_0 \left [1
+\sum_{\mu=-2}^{2} \alpha_{2\mu} Y_{2\mu}^*(\theta,\varphi).
\right]
\end{equation}
The coefficients $\alpha_{2\mu}$ and the functions
$Y_{2\mu}^*(\theta,\varphi)$ are in general both complex valued.  For
this ``radius'' $R(\theta,\varphi)$ to be a real number, the
constraint $\alpha_{2\mu}^*=(-)^\mu\alpha_{2-\mu}$ must be imposed upon the
coefficients.  

The ``instantaneous principal axis system'' is the coordinate system
defined by the conditions
\begin{equation}
\label{eqnaxiscond}
a_{2-1}=a_{21}=0 \qquad a_{2-2}=a_{22},
\end{equation}
so that the coordinate axes are aligned along the quadrupole shape's
principal axes.  The Euler angles $\underline{\theta}$ describing the
rotation which relates this coordinate system ($a_{\lambda\mu}$) to the
laboratory coordinate system ($\alpha_{\lambda\mu}$) therefore specify
the ``orientation'' of the nucleus.  Two degrees of
freedom in the coordinates remain, describing the ``shape'' of the nucleus.
These are the real variables $a_0\equiv a_{20}$ and $a_2\equiv
a_{22}$, which may be reexpressed in terms of the ``Bohr
shape variables'' $\beta$ and $\gamma$ as
\begin{equation}
a_0=\beta\cos\gamma \qquad a_2=\frac{1}{\sqrt{2}}\beta\sin\gamma.
\end{equation}
The coordinate $\beta$ plays the role of a ``radial'' variable in the
five-dimensional space of the coordinates $\alpha_{2\mu}$, since in
general
\begin{equation}
\beta^2=\sum_{\mu=-2}^2 \alpha_{2\mu}^*\alpha_{2\mu}.
\end{equation}
The coordinate system defined by~(\ref{eqnaxiscond}) is ambiguous, in
that any of the three coordinate axis may be chosen to lie along any
of the three nuclear principal axes.  The shapes described by
$(\beta,\gamma)$=$(\beta_0,\gamma_0+2n\pi/3 )$ or
$(\beta,\gamma)$=$(\beta_0,-\gamma_0)$ are equivalent to that of
$(\beta,\gamma)$=$(\beta_0,\gamma_0)$, but aligned along a different
permutation of the axes.  Thus, any function which is to depend upon
the nuclear shape only must be invariant under these transformations.

The benchmark models described in the following sections may be
defined in terms of a geometric Hamiltonian acting on the deformation
coordinates, consisting of a kinetic energy operator and a potential
energy operator which is a function of the nuclear shape only.  For these models, the
``harmonic'' kinetic energy operator
\begin{equation}
-\frac{\hbar^2}{2B}\sum_\mu
\frac{\partial^2}{\partial\alpha_{2\mu}\partial\alpha^*_{2\mu}}
\end{equation}
is used.  Several of these models also occur as dynamical symmetries
of the IBM (Section~\ref{seccollective}) in the limit of infinite
boson number.

\section{Spherical oscillator}
\label{secbenchosc}

The five-dimensional harmonic oscillator Hamiltonian,
\begin{equation}
H=-\frac{\hbar^2}{2B}\sum_\mu
\frac{\partial^2}{\partial\alpha_{2\mu}\partial\alpha^*_{2\mu}}
+\frac{1}{2}C\beta^2,
\end{equation}
has equally-spaced eigenvalues
\begin{equation}
E=\hbar \omega (N+\frac{5}{2}),
\end{equation}
where $\omega\equiv(C/B)^{1/2}$ and $N$=0,1,$\ldots$, analogous to
those of the one-dimensional harmonic oscillator.\footnote{The parameters $B$
and $C$ are related to the GCM parameters $B_2$ and $C_2$
(Chapter~\ref{chapgcm}) by $B=(\sqrt5/2)B_2$ and $C=(2/\sqrt5)C_2$.}
The eigenfunctions are known in closed
form~\cite{eisenberg1987:v1,chacon1976:oscillator,chacon1977:oscillator,corrigan1976:oscillator,gheorghe1978:ho},
consisting of a radial factor,
\begin{equation}
f(\beta)=A_{n,\tau}\beta^\tau e^{-\frac{s^2\beta^2}{2}}L_n^{\tau+\frac{3}{2}}(s^2\beta^2),
\end{equation}
where $A_{n,\tau}$ is a normalization constant,
$s\equiv(B_2C_2/\hbar^2)^{1/4}$ is the oscillator ``stiffness'',
$L_n^{\tau+\frac{3}{2}}$ is a Laguerre polynomial, $\tau$ is a
separation constant, and $n$ is the index for the radial solution,
multiplied by a more complicated angular wave function
$\Phi(\gamma,\underline{\theta})$.  The oscillator eigenstates
constitute a convenient basis for the expansion of other states in
numerical work, as discussed in Subsection~\ref{subsecgcmnumerical}.
The harmonic oscillator arises in the IBM as well, as the infinite
boson number limit of the U(5) dynamical symmetry~\cite{arima1976:u5}.

The eigenstates of the harmonic oscillator fall into degenerate
multiplets, equally spaced in excitation energy, containing levels of
spins $0^+$, $2^+$, $4^+$-$2^+$-$0^+$, $6^+$-$4^+$-$3^+$-$2^+$-$0^+$,
\etc, as illustrated in Fig.~\ref{figbasicschemes}(a) on
page~\pageref{figbasicschemes}.  For comparison to other models,
observe that $R_{4/2}$[$\equiv$$E(4^+_1)/E(2^+_1)$]=2.00.  The $E2$
transitions between these levels follow a ``phonon'' creation or
destruction selection rule, in which transitions are only possible
between states in adjacent multiplets.  Specific $B(E2)$ strength
predictions exist (\eg, Ref.~\cite{arima1976:u5}).  Here let us just
note that
\mathbox{B(E2;4^+_1\rightarrow2^+_1)}/\mathbox{B(E2;2^+_1\rightarrow0^+_1)}=2.

\section{Deformed $\gamma$-soft structure}
\label{secbenchgsoft}

If the potential $V(\beta,\gamma)$ is a function of $\beta$ only,
independent of $\gamma$, the five-dimensional equivalent of a
``central force problem'' arises, and the Hamiltonian is invariant
under the group O(5) of rotations in five-dimensional space.  A
separation of variables
occurs~\cite{wilets1956:oscillations,rakavy1957:gsoft}, giving
eigenfunctions of the form $f(\beta)\Phi(\gamma,\underline{\theta})$.
The solutions for the ``angular'' ($\gamma$,$\underline{\theta}$) wave
functions~\cite{bes1959:gamma} are common to all $\gamma$-independent
(``$\gamma$-soft'') problems.  The equation for $f(\beta)$ is
\begin{equation}
\label{eqnradialprime}
\left[ \frac{\hbar^2}{2B}\left( -\frac{1}{\beta^4}
\frac{\partial}{\partial\beta}\beta^4\frac{\partial}{\partial\beta}
+\frac{\tau(\tau+3)}{\beta^2} \right) + V(\beta) \right] f(\beta) = E
f(\beta),
\end{equation}
where the separation constant $\tau$ assumes the values
$\tau=0,1,\ldots$.  Each radial solution gives rise to a degenerate
multiplet of states, with each of the angular wave functions obtained
for that value of $\tau$.  The first few $\tau$ multiplets contain
members of spins $0^+$, $2^+$, $4^+$-$2^+$, $6^+$-$4^+$-$3^+$-$0^+$,
$\ldots$.  For a linear electric quadrupole operator, the $E2$
transitions follow a $\delta \tau$=$\pm$1 selection rule.  Also, since
all $E2$ matrix elements between members of two specific multiplets
share the same radial matrix element, their relative strengths depend
only upon the angular matrix elements, which are universal to all
$\gamma$-soft problems.  Several examples of $\gamma$-soft
Hamiltonians are encountered in the preceding chapters, including the
E(5) finite well Hamiltonian and the GCM quadratic-quartic
Hamiltonian.  The harmonic oscillator is a special example, in which
extra degeneracies occur.

Wilets and Jean~\cite{wilets1956:oscillations} consider the limiting
case of near-rigid $\beta$ deformation, in which harmonic oscillations
in $\beta$ occur, independent of the separation constant $\tau$.  This
yields an analytic solution for the level excitation energies
\begin{equation}
E=\hbar\left(\frac{B}{C}\right)^{1/2}\left[\frac{\tau(\tau+3)}{2x_0^2}+n_\beta\right]
\end{equation}
where $n_\beta$=0,1,$\ldots$, $\tau$=0,1,$\ldots$, and the parameters
$B$, $C$, and $x_0$$\gg$1 arise in the particular
parametrization of the potential 
used~\cite{wilets1956:oscillations} to obtain this limiting case.
Thus, for each value of $n_\beta$, a ``family'' of $\tau$ multiplets
arises, as illustrated in Fig.~\ref{figbasicschemes}(c) on
page~\pageref{figbasicschemes}, sharing the same radial wave function,
and the $B(E2)$ strengths within a family are completely
determined~\cite{arima1979:o6}.  For reference, let us note that
$R_{4/2}$=2.50 and
\mathbox{B(E2;4^+_1\rightarrow2^+_1)}/\mathbox{B(E2;2^+_1\rightarrow0^+_1)}=10/7.
The Wilets-Jean deformed $\gamma$-soft structure arises in the IBM as
the infinite boson number limit of the O(6) dynamical
symmetry~\cite{arima1979:o6}.

\section{Deformed axially symmetric rotor}
\label{secbenchrotor}

If $\beta$ is rigidly confined to some nonzero value $\beta_0$, and
$\gamma$ is rigidly confined to 0$^\circ$, the nucleus posesses a fixed
axially-symmetric shape, and it can undergo rotations in space.  The
wave functions for the rotational motion in terms of the Euler angles,
$\phi(\underline{\theta})$, have closed-form solutions in terms of the
Wigner $D$ functions.  

A more general case occurs if the ``confinement'' in $\beta$ and
$\gamma$ is not strictly rigid but rather allows small oscillations to
occur, decoupled from the overall rotational motion.  The relevant
modes are illustrated in Fig.~\ref{figquadrupolemodes} on
page~\pageref{figquadrupolemodes}.  The wave function then may be
decomposed as the product of an intrinsic ($\beta$, $\gamma$) wave
function with the rotational wave function (see
Ref.~\cite{bohr1998:v2} for details of symmetrization).  The intrinsic
wave function describes the state of the system as viewed in the
body-fixed principal axis system (Subsection~\ref{secbenchintro})
rotating along with the nucleus.  Since the Hamiltonian is invariant
under rotations about the symmetry axis, the projection $K$ of angular
momentum along this axis must be a good quantum number.  The
one-phonon excitation for vibrations in the $Y_{20}$, or $\beta$, mode
has $K$=0, while that for the $Y_{2\pm2}$, or $\gamma$, mode has
$K$=2.  Each intrinsic state gives rise to a sequence, or ``band'', of
rotational states, with spins $J$=$K,(K+1),\ldots$, except that $K$=0
yields only even spins $J$=0,2,$\ldots$.  The excitation energies
within a band depend upon $J$ as $\hbar^2J(J+1)/(2\mathcal{J})$, where
$\mathcal{J}$ is a moment of inertia.  A schematic rotational level
scheme is shown in Fig.~\ref{figbasicschemes}(b) on
page~\pageref{figbasicschemes}.  The basic picture of intrinsic
vibrational excitations provided here is somewhat schematic, but the
topic can be treated quantitatively, in terms of a harmonic oscillator
potential at $\beta$=$\beta_0$, within the rotation vibration model
(RVM)~\cite{faessler1965:rvm}.  Rotational structure arises in the IBM
as the infinite boson number limit of the SU(3) dynamical
symmetry~\cite{arima1978:su3}.  For the rotor, $R_{4/2}$=3.33 and
\mathbox{B(E2;4^+_1\rightarrow2^+_1)}/\mathbox{B(E2;2^+_1\rightarrow0^+_1)}=10/7.

The matrix element of an electromagnetic transition operator between
two states is a product of a factor depending upon the intrinsic
structure, namely, the matrix element of an ``intrinsic frame''
multipole operator $\mathfrak{M}'(\sigma\lambda)$ between the
respective intrinsic states, and a factor depending upon the
rotational wave functions, which depends only upon the $J$ and $K$
quantum numbers involved.  For two states, $\alpha_1K_1J_1$ and
$\alpha_2K_2J_2$, $K_1\leq K_2$, the matrix element of the multipole
operator $\mathfrak{M}(\sigma\lambda)$ is
\begin{multline}
\label{eqnalaga}
\langle \alpha_2K_2J_2 || \mathfrak{M}(\sigma\lambda) || \alpha_1K_1J_1\rangle =
(2J_1+1)^{1/2} \langle
J_1K_1\lambda(K_2-K_1)|J_2K_2\rangle \\
\times \langle\alpha_2K_2|\mathfrak{M}'(\sigma\lambda;K_2-K_1)|\alpha_1K_1\rangle
\begin{disscases}\sqrt2 &K_1=0~\text{and}~K_2\neq0\\ 1 & \text{otherwise}\end{disscases},
\end{multline}
provided that, if $K_1$ and $K_2$ are both nonzero, $K_1+K_2>\lambda$
(otherwise, an extra ``signature'' cross term is present).  Since all
states within a band share the same intrinsic state, the intrinsic
matrix element is common to all pairs of states chosen from two given
bands.  This leads to the Alaga strength relations for interband
transitions, summarized for selected $E2$ transitions in
Table~\ref{tabalaga}.  The Alaga rules are reasonably well reproduced
in many rotational nuclei.  However, the rotational model is most
successful not in its pure form but rather when it incorporates band
mixing, according to a mechanism described by Bohr and Mottelson,
which can allow large sets of transition strengths to be described
with just a single mixing parameter~\cite{bohr1998:v2}.  The specific
case of mixing between bands with $\Delta K$=0 is considered in
Appendix~\ref{appmixing}.
\begin{table}[p]
{
\ssp
\small
\begin{center}
\begin{tabular}{r,c,r==r,c,r==r.l,l====r,c,r==r,c,r==r.l,l}
\pseudoruledtabular
$K_i$ &
$\rightarrow$&
$K_f$ &
$J_i$ &
$\rightarrow$&
$J_f$ &
\multicolumn{3}{r}{$\frac{B(E2;i\rightarrow f)}{\langle i | \mathfrak{M}'(E2)
|f\rangle^2}$}&

$K_i$ &
$\rightarrow$&
$K_f$ &
$J_i$ &
$\rightarrow$&
$J_f$ &
\multicolumn{3}{r}{$\frac{B(E2;i\rightarrow f)}{\langle i | \mathfrak{M}'(E2)
|f\rangle^2}$}\\

\hline
 0 & $\rightarrow$ &  0 &   0 & $\rightarrow$ &   2 &  1&.0000 &            &  2 & $\rightarrow$ &  0 &   2 & $\rightarrow$ &   0 &  0&.2000 &  $\times$ 2 \\ 
   &             &    &   2 & $\rightarrow$ &   0 &  0&.2000 &            &    &             &    &     &             &   2 &  0&.2857 &  $\times$ 2 \\ 
   &             &    &     &             &   2 &  0&.2857 &            &    &             &    &     &             &   4 &  0&.0143 &  $\times$ 2 \\ 
   &             &    &     &             &   4 &  0&.5143 &            &    &             &    &   3 & $\rightarrow$ &   2 &  0&.3571 &  $\times$ 2 \\ 
   &             &    &   4 & $\rightarrow$ &   2 &  0&.2857 &            &    &             &    &     &             &   4 &  0&.1429 &  $\times$ 2 \\ 
   &             &    &     &             &   4 &  0&.2597 &            &    &             &    &   4 & $\rightarrow$ &   2 &  0&.1190 &  $\times$ 2 \\ 
   &             &    &     &             &   6 &  0&.4545 &            &    &             &    &     &             &   4 &  0&.3506 &  $\times$ 2 \\ 
   &             &    &   6 & $\rightarrow$ &   4 &  0&.3147 &            &    &             &    &     &             &   6 &  0&.0303 &  $\times$ 2 \\ 
   &             &    &     &             &   6 &  0&.2545 &            &    &             &    &   5 & $\rightarrow$ &   4 &  0&.3182 &  $\times$ 2 \\ 
   &             &    &     &             &   8 &  0&.4308 &            &    &             &    &     &             &   6 &  0&.1818 &  $\times$ 2 \\ 
   &             &    &   8 & $\rightarrow$ &   6 &  0&.3294 &            &    &             &    &   6 & $\rightarrow$ &   4 &  0&.0979 &  $\times$ 2 \\ 
   &             &    &     &             &   8 &  0&.2526 &            &    &             &    &     &             &   6 &  0&.3636 &  $\times$ 2 \\ 
   &             &    &     &             &  10 &  0&.4180 &            &    &             &    &     &             &   8 &  0&.0385 &  $\times$ 2 \\ 
 0 & $\rightarrow$ &  2 &   0 & $\rightarrow$ &   2 &  1&.0000 &  $\times$ 2  &    &             &    &   7 & $\rightarrow$ &   6 &  0&.3000 &  $\times$ 2 \\ 
   &             &    &   2 & $\rightarrow$ &   2 &  0&.2857 &  $\times$ 2  &    &             &    &     &             &   8 &  0&.2000 &  $\times$ 2 \\ 
   &             &    &     &             &   3 &  0&.5000 &  $\times$ 2  &    &             &    &   8 & $\rightarrow$ &   6 &  0&.0882 &  $\times$ 2 \\ 
   &             &    &     &             &   4 &  0&.2143 &  $\times$ 2  &    &             &    &     &             &   8 &  0&.3684 &  $\times$ 2 \\ 
   &             &    &   4 & $\rightarrow$ &   2 &  0&.0079 &  $\times$ 2  &    &             &    &     &             &  10 &  0&.0433 &  $\times$ 2 \\ 
   &             &    &     &             &   3 &  0&.1111 &  $\times$ 2  &  2 & $\rightarrow$ &  2 &   2 & $\rightarrow$ &   2 &  0&.2857 &           \\ 
   &             &    &     &             &   4 &  0&.3506 &  $\times$ 2  &    &             &    &     &             &   3 &  0&.5000 &           \\ 
   &             &    &     &             &   5 &  0&.3889 &  $\times$ 2  &    &             &    &     &             &   4 &  0&.2143 &           \\ 
   &             &    &     &             &   6 &  0&.1414 &  $\times$ 2  &    &             &    &   3 & $\rightarrow$ &   2 &  0&.3571 &           \\ 
   &             &    &   6 & $\rightarrow$ &   4 &  0&.0210 &  $\times$ 2  &    &             &    &     &             &   3 &  0&.0000 &           \\ 
   &             &    &     &             &   5 &  0&.1538 &  $\times$ 2  &    &             &    &     &             &   4 &  0&.3429 &           \\ 
   &             &    &     &             &   6 &  0&.3636 &  $\times$ 2  &    &             &    &     &             &   5 &  0&.3000 &           \\ 
   &             &    &     &             &   7 &  0&.3462 &  $\times$ 2  &    &             &    &   4 & $\rightarrow$ &   2 &  0&.1190 &           \\ 
   &             &    &     &             &   8 &  0&.1154 &  $\times$ 2  &    &             &    &     &             &   3 &  0&.2667 &           \\ 
   &             &    &   8 & $\rightarrow$ &   6 &  0&.0294 &  $\times$ 2  &    &             &    &     &             &   4 &  0&.0416 &           \\ 
   &             &    &     &             &   7 &  0&.1765 &  $\times$ 2  &    &             &    &     &             &   5 &  0&.2333 &           \\ 
   &             &    &     &             &   8 &  0&.3684 &  $\times$ 2  &    &             &    &     &             &   6 &  0&.3394 &           \\ 
   &             &    &     &             &   9 &  0&.3235 &  $\times$ 2  \\   
   &             &    &     &             &  10 &  0&.1022 &  $\times$ 2  \\

\pseudoruledtabular
\end{tabular}
\end{center}
}
\caption
[Ratio of $B(E2)$ strength to squared intrinsic matrix element (Alaga
rules).]  {\ssp Ratio of the $B(E2)$ strength to the squared intrinsic
matrix element for an ideal rotor,
$B(E2;\alpha_iK_iJ_i\rightarrow\alpha_fK_fJ_f)/\langle\alpha_fK_f|\mathfrak{M}'(E2;K_f-K_i)|\alpha_iK_i\rangle^2$,
shown for selected $K$ and $J$ values.  The values for this ratio
yield the Alaga interband strength rules, as well as in-band strength
relations.  Values are calculated from (\ref{eqnalaga}) and the
definition (\ref{eqnbe2defn}) of $B(E2)$.
\label{tabalaga}
}
\end{table}

\section{Square well potentials: E(5) and X(5)}
\label{secbenche5x5}

The E(5) and X(5) models of
Iachello~\cite{iachello2000:e5,iachello2001:x5}, introduced in
Section~\ref{sectrans}, are both based upon a square well potential in
$\beta$.
For the E(5) model, the potential is $\gamma$-independent, 
\begin{equation}
V(\beta) = \begin{disscases} 0 & \beta \leq \beta_w \\ \infty & \beta >
\beta_w, \end{disscases}
\end{equation}
and so the
separation of variables described in Section~\ref{secbenchgsoft}
occurs.  The radial equation for $\beta$$<$$\beta_w$ is
equivalent~\cite{iachello2000:e5} to the Bessel equation of
half-integer order $\nu$=$\tau$+3/2.  The solution wave function may
be be written in terms of the spherical Bessel function of the first
kind, $j_n(x)\equiv\sqrt\frac{\pi}{2} x^{-1/2}J_{n+1/2}(x)$, of
integer order $n$=$\tau$+1.  In terms of the reduced energy eigenvalue
$\varepsilon_{\xi,\tau}\equiv (2B/\hbar^2)E_{\xi,\tau}$, the wave
function in $\beta$ for the $\xi$th solution with separation constant
$\tau$ is
\begin{equation}
\label{eqne5wavefcn}
f_{\xi,\tau}(\beta)=\begin{disscases} A_{\xi,\tau} \beta^{-1}
j_{\tau+1}(\varepsilon_{\xi,\tau}^{1/2}\beta) & \beta \leq \beta_w\\
0 & \beta > \beta_w, \end{disscases}
\end{equation}
where $A_{\xi,\tau}$ is a normalization constant.  The wave function
must vanish at $\beta=\beta_w$, yielding the eigenvalue condition
\begin{equation}
\varepsilon_{\xi,\tau}=\beta_w^{-2}x_{\tau+3/2,\xi}^2,
\end{equation}
$\xi=1,2,\ldots$, where $x_{\nu,i}$ is the $i$th zero of the ordinary
Bessel function $J_\nu(x)$.  The value of $A$ is determined
by the normalization condition $\int_0^\infty
\beta^4 d\beta |f(\beta)|^2=1$.  The normalization integral can be evaluated
analytically~\cite[(5.54.2)]{gradshteyn1994:table}, yielding
\begin{equation}
\label{eqne5norm}
A=\left[ \frac{\beta_w^3}{2}\left[j_n(\varepsilon^{1/2}\beta_w)^2-j_{n-1}(\varepsilon^{1/2}\beta_w)j_{n+1}(\varepsilon^{1/2}\beta_w)\right] \right]^{-1/2},
\end{equation}
where $n$=$\tau$+1.  

\begin{figure}[p]
\begin{center}
\includegraphics*[angle=90,totalheight=8in]{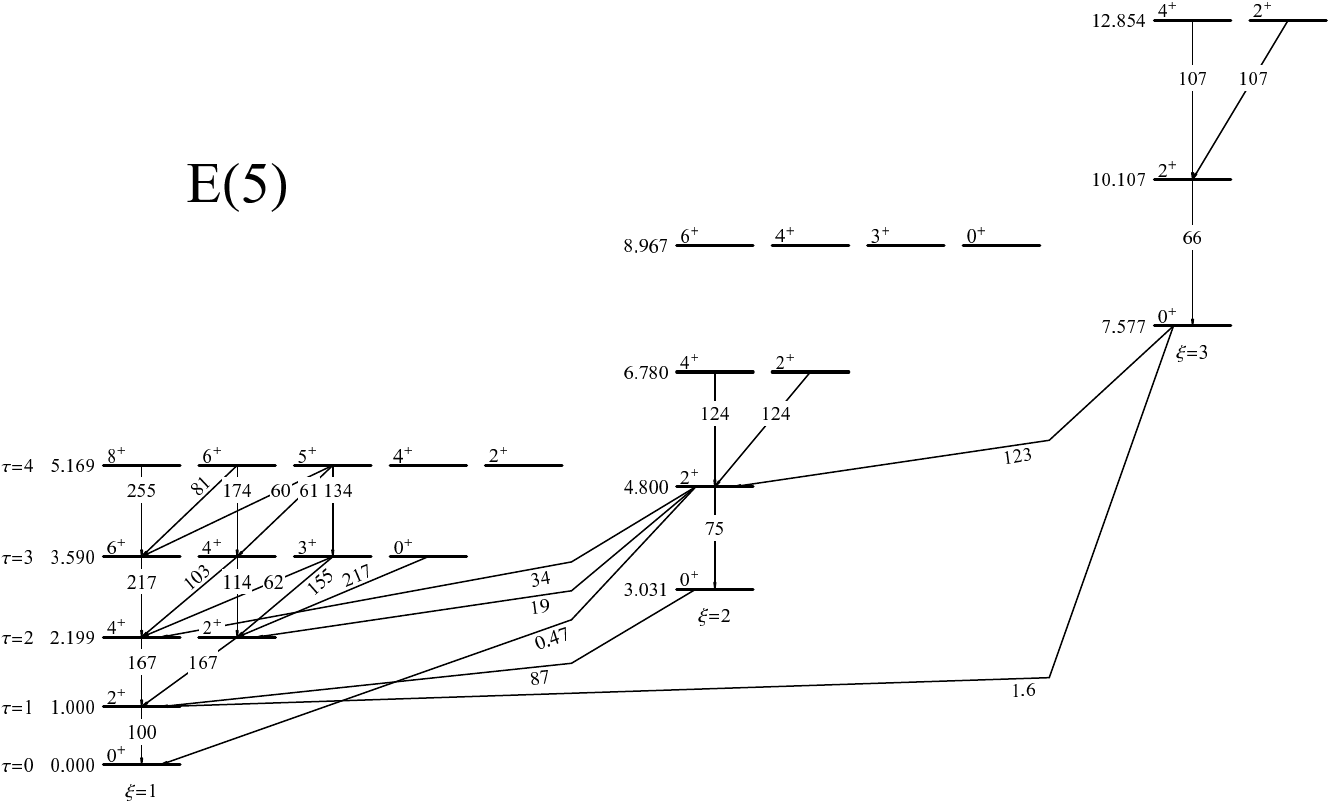}
\end{center}
\caption[Level scheme for the E(5) model.]
{Level scheme for the E(5) model, showing the lowest $\tau$ multiplets
of the first three $\xi$ families. Level energies and selected $B(E2)$
strengths are indicated.
\label{fige5reference}
}
\end{figure}

Electromagnetic transition strengths can be calculated from the matrix
elements of the collective multipole operators.  In terms of the
intrinsic frame coordinates $\beta$ and $\gamma$, the $E2$ transition
operator is\footnote{The Wigner function
$D^j_{mm'}(\underline{\theta})$ used here is that of
Rose~\cite{rose1957:am}, consistent with Ref.~\cite{eisenberg1987:v1},
which is related to the ``script''
$\mathscr{D}^{(j)}_{mm'}(\underline{\theta})$ of both
Rose~\cite{rose1957:am} and Edmonds~\cite{edmonds1960:am} by
$D^j_{mm'}(\underline{\theta})=(-)^{m'-m}\mathcal{D}^{(j)\,*}_{mm'}(\underline{\theta})$
and is related to the $\mathscr{D}^{j}_{mm'}(\underline{\theta})$ of
Bohr and Mottelson~\cite{bohr1998:v1,bohr1998:v2} by complex
conjugation.  }~\cite{eisenberg1987:v1,bohr1998:v2}
\begin{equation}
\mathfrak{M}(E2;\mu)\propto\beta\biggl[D^{2\,*}_{\mu,0}(\underline{\theta})\cos\gamma
+\frac{1}{\sqrt{2}}\bigl[D^{2\,*}_{\mu,2}(\underline{\theta})+D^{2\,*}_{\mu,-2}(\underline{\theta})\bigr]\sin\gamma\biggr],
\end{equation}
and the transition strengths are $B(E2;J_i\rightarrow
J_f)=\frac{1}{2J_i+1} |\left<J_f||
\mathfrak{M}(E2)||J_i\right>|^2$ (Appendix~\ref{appgamma}).  Matrix
elements factorize into a radial integral
\begin{equation}
\label{eqne5meintegral}
\int_0^\infty \beta^4 d\beta \,f^\mathrm{I}(\beta) \, \beta \, f^\mathrm{II}(\beta),
\end{equation}
which may be evaluated using the wave functions (\ref{eqne5wavefcn}),
and an angular integral common to all $\gamma$-soft problems,
tabulated in, \eg, Ref.~\cite{arima1979:o6}.  Multiplet energies and
$B(E2)$ strengths for the E(5) model are summarized in
Fig.~\ref{fige5reference}.

\begin{figure}[t]
\begin{center}
\includegraphics*[width=\hsize]{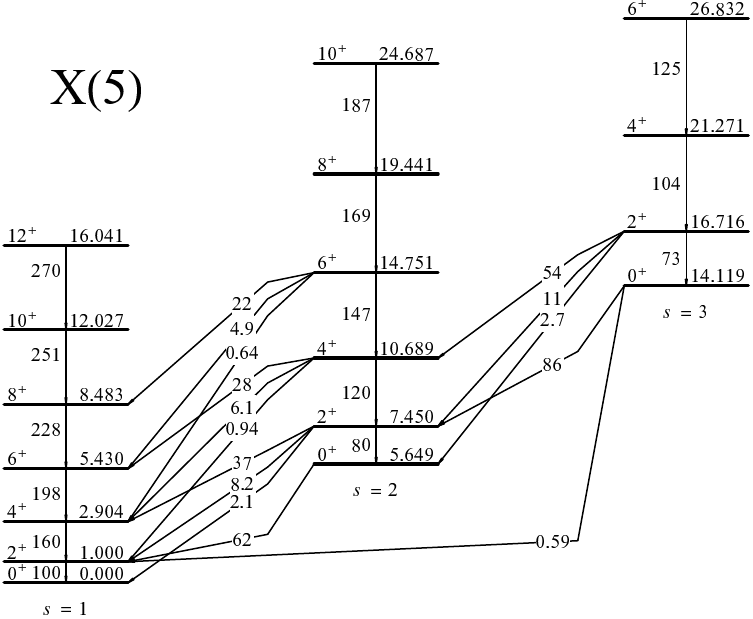}
\end{center}
\caption[Level scheme for the X(5) model.]
{Level scheme for the X(5) model, showing the lowest members of the
first three $s$ bands.  Level energies and selected $B(E2)$ strengths
are indicated.
\label{figx5reference}
}
\end{figure}

In the X(5) model, a square well potential in $\beta$ is combined with
a term $V_\gamma(\gamma)$ providing stabilization about $\gamma$=0, so
\begin{equation}
V(\beta,\gamma) = \begin{disscasesclosed} 0 & \beta \leq \beta_w \\ \infty & \beta >
\beta_w \end{disscasesclosed}+V_\gamma(\gamma).
\end{equation}
An approximate separation of the wave function into radial, $\gamma$,
and rotational factors, $f(\beta)\eta(\gamma)\phi(\underline{\theta})$
occurs~\cite{iachello2001:x5}.  The radial equation for $f(\beta)$ is
equivalent to the Bessel equation of order
\begin{equation}
\label{eqnx5nu}
\nu=\left( \frac{L(L+1)}{3}+\frac{9}{4}\right)^{1/2},
\end{equation}
where the separation constant $L$ is the angular momentum
quantum number.  The radial wave function for the $s$th radial
excitation for separation constant $L$ is
\begin{equation}
\label{eqnx5wavefcn}
f_{s,L}(\beta)=\begin{disscases} A_{s,L} \beta^{-1}
j_{\nu-1/2}(\varepsilon_{s,L}^{1/2}\beta) & \beta \leq \beta_w\\
0 & \beta > \beta_w, \end{disscases}
\end{equation}
where $A_{s,L}$ is a normalization constant.  The condition that the wave function
vanish at $\beta=\beta_w$ yields the eigenvalue condition
\begin{equation}
\label{eqnx5energy}
\varepsilon_{s,L}=\beta_w^{-2}x_{\nu,s}^2,
\end{equation}
$s=1,2,\ldots$, and the value of $A$ is again determined
from~(\ref{eqne5norm}), but now using $n$=$\nu-1/2$ with $\nu$ given
by~(\ref{eqnx5nu}).  To the extent that the approximate separation of
variables holds, the details of the potential $V_\gamma$ are
irrelevant to the calculation of the properties of states involving
only rotational and $\beta$ excitations.  A band structure analogous
to that of the rigid rotor arises, with $K$=0 for all bands involving
only radial excitations.  However, the energies~(\ref{eqnx5energy})
differ substantially from those of a rigidly-deformed rotor, and the
intrinsic state~(\ref{eqnx5wavefcn}) is different for each member of a
band.  As in the E(5) calculations, the $E2$ matrix elements factor into a radial
part~(\ref{eqne5meintegral}), involving $f(\beta)$, and an angular
part.  The angular part in the X(5) case is the same as for the rigid
rotor~(\ref{eqnalaga}).
Level energies and
$B(E2)$ strengths for the X(5) model are summarized in
Fig.~\ref{figx5reference}.

\chapter{Gamma-ray transition rates and matrix elements}
\label{appgamma}

Several pieces of spectroscopic and lifetime data must usually be
combined to extract an experimental value for single transition matrix
element, as discussed in Chapter~\ref{chapexpt}.  This appendix
provides precise statements of the relations between the multipole
matrix elements, transition rates, lifetimes, branching fractions, and
mixing ratios used throughout this work, as well as a definition for
the ``Weisskopf unit'' used for reporting $B(\sigma\lambda)$
strengths.  These relations are also restated in forms more directly
useful in practical spectroscopic data analysis.

For a $\gamma$-ray transition of multipolarity $\sigma\lambda$ between
two levels, the transition probability per unit time is, in first
order perturbation theory~(\eg,
Refs.~\cite{moskowski1965:multipole,eisenberg1988:v2,bohr1998:v2}),
\begin{equation}
\label{eqntfromb}
T(\sigma\lambda;i\rightarrow f)= 
\frac{8\pi(\lambda+1)}{\lambda[(2\lambda+1)!!]^2}
\frac{1}{\hbar}
\left(\frac{E_\gamma}{\hbar c}\right)^{2\lambda+1}B(\sigma\lambda;i\rightarrow f),
\end{equation}
where $E_\gamma$ is the transition energy and $B(\sigma\lambda;i\rightarrow f)$
is the squared matrix element of the multipole operator
$\mathfrak{M}(\sigma\lambda)$, averaged over initial and summed over final spin
projection substates,\footnote{The reduced matrix element normalization used
  here is that of Edmonds~\cite{edmonds1960:am}, which, for
  spherical tensor operators of integer rank, is consistent with that of Bohr and
  Mottelson~\cite{bohr1998:v1,bohr1998:v2} and related to that of
  Rose~\cite{rose1957:am}, as used in Ref.~\cite{eisenberg1987:v1}, by a factor
  of $(2J_f+1)^{1/2}$.  }
\begin{equation}
\label{eqnbe2defn}
\begin{split}
B(\sigma\lambda;i\rightarrow f) 
&\equiv
\frac{1}{2J_i+1}\sum_{m_i,\mu,m_f} \left| \left \langle fJ_fm_f |
\mathfrak{M}(\sigma\lambda;\mu) | iJ_im_i\right\rangle \right|^2 \\ 
&=
\frac{1}{2J_i+1} \left| \left \langle f || \mathfrak{M}(\sigma\lambda)
|| i \right\rangle \right|^2.
\end{split}
\end{equation}
Specifically, for $E2$ and $M1$ transitions, the decay rates obtained are
\begin{equation}
\begin{aligned}
T(E2)&\approx(1.225\times10^{-2}/\mathrm{s})
\left( \frac{E_\gamma}{\mathrm{keV}}\right)^5 
\,\left[\frac{B(E2)}{e^2\mathrm{b}^2}\right]\\
T(M1)&\approx(1.758\times10^{4}/\mathrm{s}) 
\left( \frac{E_\gamma}{\mathrm{keV}}\right)^3 
\,\left[\frac{B(M1)}{\mu_N^2}\right]  ,
\end{aligned}
\end{equation}
making use of the relations $e^2$=$\alpha\hbar c$ and
$\mu_N=e\hbar/(2m_pc)$ in Gaussian units.

Weisskopf~\cite{weisskopf1951:estimate,bohr1998:v2} introduced  
estimates $B_W(\sigma\lambda)$ for the transition strength induced by a
single proton moving in a schematic orbit,
\begin{equation}
\begin{aligned}
B_W(E\lambda)&\equiv\frac{(0.12)^{2\lambda}}{4\pi}
\left(\frac{3}{\lambda+3}\right)^2A^{2\lambda/3}\,e^2\mathrm{b}^\lambda
\\
B_W(M\lambda)&\equiv\frac{10}{\pi}(0.12)^{2(\lambda-1)}
\left(\frac{3}{\lambda+3}\right)^2A^{2(\lambda-1)/3}\,\mu_N^2\mathrm{b}^{(\lambda-1)}
,
\end{aligned}
\end{equation}
and it is common to use these estimates as units (``W.u.'') for
reporting $B(\sigma\lambda)$ strengths.  To provide a concrete example, a
1\,MeV $E2$ $\gamma$-ray of 1\,\Wu strength in a mass 100 nucleus
produces a transition rate of \sci{3.4}{10}/s; a level decaying only
by this transition would have a lifetime of 30\,ps.  If the same
transition were an $M1$ transition of 1\,\Wu strength, the lifetime
would be 32\,fs.

The partial transition probabilities for the branches combine to
produce an aggregate decay rate
\begin{equation}
T=\sum_i T^\gamma_i + \sum_i T^\text{c.e.}_i,
\end{equation}
and hence lifetime $\tau=T^{-1}$, for the level.  The total time-integrated probability
$B_a$ that
the level will decay by any one particular branch $a$ is simply
the ratio of the partial transition rate for that branch to the
total rate, so
\begin{equation}
B_a=\frac{T_a}{T}.
\end{equation}

Each individual $\gamma$-ray transition can in general proceed by
multiple radiation multipolarities simultaneously, subject to angular
momentum and parity selection rules.  However, as a consequence of the
finite size of the nucleus, each successively higher multipole of
$\gamma$ radiation is suppressed by a factor of
$\sim$$k_\gamma$$R_\text{nucl}$ (typically $\sim$10--100), where $k_\gamma$
is the photon wave number and $R_\text{nucl}$ is the approximate
nuclear radius.  Therefore, in practice, $\gamma$ radiation is usually
encountered either with pure multipolarity or in an admixture of two
successive multipolarities~--- specifically, if the lowest
multipolarity capable of connecting the initial and final states is
magnetic in nature ($M\lambda$), the next higher electric multipole
[$E(\lambda+1)$] often contributes substantially.  The
``amplitude mixing ratio'' is $\delta\equiv{\left
\langle f ||E(\lambda+1) || i\right \rangle}$ / ${\left \langle f
||M\lambda || i \right \rangle}$, where the ``multipole operators''
used here are the $\mathfrak{M}(\sigma\lambda)$ renormalized to give
the transition rates~(\ref{eqntfromb}) directly.  Phase conventions
for $\delta$ are discussed further in Appendix~\ref{appangpol}.  The
relative contribution of the two multipolarities to the transition
rate, or the ``multipole radiation intensity ratio'', is then $\delta^2 =
{T[E(\lambda+1);i\rightarrow f]}/ {T(M\lambda;i\rightarrow f)}$.  Some
experimental methods for determining $\delta$ are described in
Section~\ref{secangpol}.

Let us now combine the information presented so far into expressions
of direct use to the spectroscopist: for a mixed $E2$/$M1$ transition
$a$ from level $i$ to level $f$, the $B(M1)$ and $B(E2)$ transition strengths in
W.u.\ are related to experimentally accessible quantities as
\begin{equation}
\begin{aligned}
B(E2;{i\rightarrow f})&\approx\frac{(\sci{1.374}{7}~\mathrm{W.u.})}{A^{4/3}}
\left(\frac{E_a}{\mathrm{keV}}\right)^{-5}
\left(\frac{\tau_i}{\mathrm{s}}\right)^{-1}
B_a \frac{\delta_a^2}{1+\delta_a^2} 
\\
B(M1;{i\rightarrow f})&\approx(\sci{3.176}{-5}~\mathrm{W.u.})
\left(\frac{E_a}{\mathrm{keV}}\right)^{-3}
\left(\frac{\tau_i}{\mathrm{s}}\right)^{-1}
B_a \frac{1}{1+\delta_a^2} ,
\end{aligned}
\end{equation}
where $A$ is the atomic mass, $E_a$ is the transition energy, $\tau_i$
is the initial level lifetime, $B_a$ is the $\gamma$-ray branching
fraction, and $\delta_a$ is the mixing ratio.

\chapter{Rotational $\Delta K=0$ mixing relations}
\label{appmixing}

Relations describing the mixing of two
rotational bands of the same $K$ quantum number ($\Delta K$=0 mixing) are needed
for the interpretation of the $^{162}$Er data in
Section~\ref{sec162erinterp}.  The purpose of this appendix is to
collect and summarize the relevant relations from
Ref.~\cite{bohr1998:v2} and to present additional specialized results needed for
Section~\ref{sec162erinterp}.

Bohr and Mottelson~\cite{bohr1998:v2} deduce a mixing operator
\begin{equation}
H_c=h_0(\mathbf{J}^2-J_3^2)
\end{equation}
between the same-spin levels of two bands related by $\Delta K$=0.
Here $h_0$ is an ``intrinsic'' operator (\ie, the matrix element of
$h_0$ between two levels depends only upon their intrinsic states),
and $J_3$ is the projection of angular momentum on the nuclear
symmetry axis.  If the bands have the same moment of inertia, the energy
difference $E(\alpha_2KJ)-E(\alpha_1KJ)$ between same-spin levels is
independent of $J$, and the mixing ``energy denominator'' for the
perturbation calculation can be absorbed into a redefined operator
\begin{equation}
\varepsilon_0\equiv\frac{h_0}{E(\alpha_2K)-E(\alpha_1K)},
\end{equation}
where band 2 is the \textit{higher} energy band. 
The states in the mixed bands still have good quantum number $K$, but
there is no longer a single intrinsic state common to all members of
the band.  The mixed states, as a function of spin, are
\begin{equation}
\begin{aligned}
\left| \hat{\alpha}_1KJ \right\rangle 
&= | \alpha_1KJ \rangle  \,-\, \langle \alpha_2K | \varepsilon_0 |
\alpha_1K \rangle \,J (J+1)\, | \alpha_2KJ\rangle \\
\left| \hat{\alpha}_2KJ \right\rangle 
&= | \alpha_1KJ \rangle  \,+\, \langle \alpha_2K | \varepsilon_0 |
\alpha_1K \rangle \,J (J+1)\, | \alpha_2KJ\rangle 
\end{aligned}
\end{equation}
in the limit of small mixing.  

The interband $E2$ matrix elements between members of mixed band 2 and
members of mixed band 1 are
\begin{multline}
\label{eqnmixedampl}
\langle \alpha_2KJ_2 || \mathfrak{M}(E2) || \alpha_1KJ_1\rangle \\
= (2J_1+1)^{1/2} \langle J_1K20|J_2K\rangle 
\left[M_1 + M_2 \left[J_2(J_2+1)-J_1(J_1+1)\right]\right], 
\end{multline}
where
\begin{equation}
\begin{aligned}
M_1 &\equiv \langle\alpha_2K|\mathfrak{M}'(E2;0)|\alpha_1K\rangle\\
M_2 &\equiv \langle \alpha K|\mathfrak{M}'(E2;0)|\alpha K\rangle \langle\alpha_2K|\varepsilon_0|\alpha_1K\rangle,
\end{aligned}
\end{equation}
provided both of the original bands have the same intraband
matrix element $\langle\alpha K|\mathfrak{M}'(E2;0)|\alpha K\rangle$.
Experimental papers often report the ratio
\begin{equation}
a_0 \equiv \frac{M_2}{M_1}.
\end{equation}
In practice, (\ref{eqnmixedampl}) is rewritten as
\begin{equation}
\frac{\langle \alpha_2KJ_2 || \mathfrak{M}(E2) || \alpha_1KJ_1\rangle }{(2J_1+1)^{1/2} \langle J_1K20|J_2K\rangle}
=
M_1 ~+~ M_2~ \left[J_2(J_2+1)-J_1(J_1+1)\right],
\end{equation}
casting $M_1$ and $M_2$ as the slope and intercept parameters for a
linear relation between a spin-corrected matrix element (LHS) and a
quadratic expression in terms of spins (on the RHS).  These quantities
are used as variables for plotting the entire set of data on interband
transitions for a band (``Mikhailov plot''),
and $M_1$ and $M_2$ are obtained by a linear fit (see
Ref.~\cite{bohr1998:v2} for examples).  If \textit{relative}
transition strengths are known experimentally (\eg, from a branching
ratio measurement) but absolute strengths are not, only the ratio
$a_0=M_2/M_1$ can be obtained.  Even if absolute strengths are known,
but with less precision than the relative strengths, it can still be
beneficial to perform such an analysis to extract $M_2/M_1$ with
higher accuracy.

The value of the unmixed intraband matrix element $\langle\alpha
K|\mathfrak{M}'(E2;0)|\alpha K\rangle$ is only needed for the analysis if
the mixing strength $\langle\alpha_2K|\varepsilon_0|\alpha_1K\rangle$
is to be extracted explicitly.  This matrix element is usually
expressed in terms of the ``intrinsic quadrupole moment'' $Q_0$, where
\begin{equation}
eQ_0\equiv \left(\frac{5}{16\pi}\right)^{1/2}\langle\alpha
K|\mathfrak{M}'(E2;0)|\alpha K\rangle.
\end{equation}
For the ground state band ($K^\pi=0^+$), the magnitude of the
intrinsic quadrupole moment is deduced most simply from the
experimental $2^+_g\rightarrow0^+_g$ transition strength. Since
$\langle\alpha 0 0 ||\mathfrak{M}(E2)||\alpha 0 2\rangle =
\langle\alpha 0|\mathfrak{M}'(E2;0)|\alpha 0\rangle$ by
(\ref{eqnalaga}) and
$B(E2;\alpha02^+\rightarrow\alpha00^+)=(1/5)\langle\alpha 0 0
||\mathfrak{M}(E2)||\alpha 0 2\rangle^2$ by (\ref{eqnbe2defn}),
\begin{equation}
B(E2;\alpha02^+\rightarrow\alpha00^+)=\frac{1}{16\pi}|eQ_0|^2.
\end{equation}

In Section~\ref{sec162erinterp}, an experimental value for $a_0$ is
deduced from the
\mbox{$B(E2;2^+_{K=0_2}\rightarrow4^+_g)$}/\mbox{$B(E2;2^+_{K=0_2}\rightarrow0^+_g)$}
branching ratio.  Specializing (\ref{eqnmixedampl}) to the cases
$\langle2^+_{K=0_2}||\mathfrak{M}(E2)||4^+_g\rangle$ and 
$\langle2^+_{K=0_2}||\mathfrak{M}(E2)||4^+_g\rangle$, and solving the resulting
system of two equations for $a_0$, yields the value and uncertainty
\begin{equation}
\label{eqnmixinga0x}
a_0=\frac{1-x}{14+6x} \qquad \sigma_{a_0}=\frac{20}{(14+6x)^2}\sigma_x
\end{equation}
in terms of the matrix element ratio
$x$$\equiv$$(7/18)^{1/2}\langle2^+_{K=0_2}||\mathfrak{M}(E2)||4^+_g\rangle$\linebreak[0]$/\langle2^+_{K=0_2}||\mathfrak{M}(E2)||0^+_g\rangle$
and its uncertainty $\sigma_x$.

\chapter{Calculation of angular correlation and polarization functions}
\label{appangpol}

The purpose of this appendix is to condense the results directly
needed for the interpretation of $\gamma$-$\gamma$ angular correlation
and polarimetry measurements following $\beta$ decay, as well as to
summarize the necessary steps to be taken for the derivation of
related results.  

An extensive formalism~\cite{biedenharn1953:angcorr}
exists for the prediction of the angular correlations between
sucessive radiations emitted in nuclear decay.  Quite general
results can be deduced for correlations between different types of
emitted particle ($\alpha$, $\beta$, $\gamma$, $e^-$), in various
cascades, with specific properties (\eg, propagation direction, spin
polarization) detected for each particle.  The main constraint to this
formalism is that initial, final, and intervening states all be states
of good spin and parity and that the initial and final states be
unoriented, so this formalism applies to essentially all decay
experiments but not to in-beam experiments.
The angular correlation functions $W(\theta)$ or
direction-polarization correlation functions $W(\theta,\gamma)$
(Section~\ref{secangpol}) for the various special cases
are very similar in form.  

The results may be summarized most
compactly if the correlation function for the ``simplest'' case, that of
two consecutive $\gamma$-ray transitions of pure multipolarity with
only their propagation directions measured,
is stated explicitly.  Then all the correlation functions for more
complicated cases are obtained from this simple function by
straightforward ``prescriptions''.  
For the $J_1\xrightarrow{L_1}J\xrightarrow{L_2}J_2$ cascade,
\begin{equation}
\label{eqw}
W(\theta)=\sum_\nu F_\nu(L_1L_1J_1J)F_\nu(L_2L_2J_2J)P_\nu(\cos
\theta),
\end{equation}
where the sum is over $\nu$=0,2,...  The $F_\nu$ coefficients are
given by simple expressions in terms of Clebsch-Gordan and Racah
coefficients; the zeroth order coefficients are
$F_0(LL'J_xJ)=\delta_{LL'}$, and the higher orders coefficients are
widely available in tabulated
form~\cite{wapstra1959:tables,ferentz1965:angcoeff}.  The highest
nonvanishing term is of order $\nu_\text{max}$=$\min (2L_1,2J,2L_2)$.
It is apparent from this expression that the angular correlation
functions are the same for the cascade
$J_1\xrightarrow{L_1}J\xrightarrow{L_2}J_2$ and the reversed cascade
$J_2\xrightarrow{L_2}J\xrightarrow{L_1}J_1$.  It can also be
shown~\cite{weneser1953:angcorr-stretched} that the correlation
patterns for all ``stretched'' cascades
$(J_0+L_1+L_2)\xrightarrow{L_2}(J_0+L_1)\xrightarrow{L_1}J_0$ are also
identical, independent of the base spin $J_0$.

Let us now summarize the prescriptions for modifying (\ref{eqw}) to
cover more general cases: 
\begin{itemize}
\item If one or more of the radiations involved is
not a $\gamma$ ray, corresponding ``particle factors'' $b_\nu$ must be
included (see Ref.~\cite{biedenharn1953:angcorr}). 
\item If unobserved
intervening radiations are present in the cascade, this is
accomodated by insertion of a factor $U_\nu$ for each such
radiation~(see
Refs.~\cite{satchler1954:angcorr-triple-unobs,fagg1959:polarization}).
\item If one of the observed transitions $i$ is of mixed multipolarity, the corresponding factor
$F_\nu(L_iL_iJ_iJ)$ is replaced by
\begin{equation}
\label{eqsubstmixingratio}
\frac{1}{1+\delta_i^2} 
\left[
F_\nu(L_iL_iJ_iJ) \pm 2 \delta_i F_\nu(L_i L_i' J_i J) + \delta_i^2
F_\nu(L_i' L_i' J_i J)
\right]
,
\end{equation}
where $\delta_i$ is the admixture of multipolarity $L_i'$, as defined
in Appendix~\ref{appgamma}.  Since $\delta_i$ is a ratio of multipole
operator matrix elements, the sign to be used on the interference term
in~(\ref{eqsubstmixingratio}) depends upon the sign conventions
involved in the definition of these operators.  The convention of
Krane and Steffen~\cite{krane1970:angcorr-110cd} dictates that, for
mixing of two consecutive multipoles, the $(-)$ sign be chosen when
the mixed radiation is the first radiation in the cascade and that the
$(+)$ sign be chosen when the mixed radiation is the final radiation.
A variety of alternate conventions are also in use, and these are not
necessarily even self-consistent between when the same radiation is
considered as the first or second radiation in a cascade (\eg,
Ref.~\cite{biedenharn1953:angcorr}); care must therefore be taken to
explicitly specify the sign convention adopted when quoting a value of
$\delta_i$ extracted from angular correlation data.
\item If the polarization of one of the $\gamma$-ray transitions, $i$, possibly a
mixed-multipolarity transition, is observed, then, in each term
involving the multipolarities $L_iL_i'$, $F_\nu(L_iL_i'J_iJ)P_\nu(\cos
\theta)$ must be replaced with
\begin{equation}
\label{eqsubstpol}
F_\nu(L_iL_i'J_iJ)P_\nu(\cos \theta) +
F_\nu(L_iL_i'J_iJ)P_\nu(\cos \theta) \kappa(L_iL_i')
(\pm)_{\sigma_i'}P_\nu^{(2)}(\cos \theta) \cos 2\gamma,
\end{equation}
where $\kappa(L_iL_i')$ is a simple ratio of Clebsch-Gordon
coefficients (tabulated in Ref.~\cite{fagg1959:polarization}),
$(\pm)_{\sigma_i'}$ is $(+)$ for electric and $(-)$ for magnetic
radiation, and the angle $\gamma$ is the polarization angle relative
to the coincidence plane (Section~\ref{secangpol}).  $P_\nu^{(2)}$ is
the associated Legendre polynomial of order 2, which is undefined for
$\nu$=0.  Note that $L_i$ and $L_i'$ are used as dummy variables in
(\ref{eqsubstpol}): when mixed radiation is involved, either one
may correpond to either $L_i$ or $L_i'$ of
(\ref{eqsubstmixingratio}) on a term-by-term basis.  (Note also that
the sign on the polarization term is indicated as the opposite by
authors who define $\gamma$ as the complementary angle.)
\end{itemize}

Two especially common cases of the direction-polarization correlation
function, needed for Section~\ref{secangpol} and
Section~\ref{sec152sm3plus}, are presented here for reference.  For a
$\gamma$-ray cascade $J_1\xrightarrow[\text{POL}]{L_1,L_1'}J\xrightarrow{L_2}J_2$
\mbox{($L_1'=L_1+1$)}, \ie, where the first radiation is a mixture of
consecutive multipoles and the polarization of the first transition is
measured,
\begin{equation}
\begin{split}
W(\theta,\gamma)&=\frac{1}{1+\delta_1^2} 
\bigg[
\sum_\nu 
\Big(
 F_\nu(L_1L_1J_1J)-2\delta_1 F_\nu(L_1L_1'J_1J)
\\
&\quad\quad\quad
+\delta_1^2 F_\nu(L_1'L_1'J_1J)
\Big)
F_\nu(L_2L_2J_2J)P_\nu(\cos
\theta)
\bigg] 
\\
&\quad 
+\frac{1}{1+\delta_1^2} 
(\pm)_{\sigma_1}
\bigg[
\sum_\nu 
\Big(
\kappa(L_1L_1)F_\nu(L_1L_1J_1J)+2\delta_1 \kappa(L_1L_1')F_\nu(L_1L_1'J_1J)
\\
&\quad\quad\quad
-\delta_1^2 \kappa(L_1'L_1')F_\nu(L_1'L_1'J_1J)
\Big)
F_\nu(L_2L_2J_2J)
P_\nu^{(2)}(\cos\theta)
\bigg]
\cos 2\gamma.
\end{split}
\end{equation}
For a
$\gamma$-ray cascade $J_1\xrightarrow{L_1,L_1'}J\xrightarrow[\text{POL}]{L_2}J_2$
\mbox{($L_1'=L_1+1$)}, \ie, where the first radiation is a mixture of consecutive
multipoles and the polarization of the second transition is measured,
\begin{equation}
\label{eqnpol3d22p0}
\begin{split}
W(\theta,\gamma)&=\frac{1}{1+\delta_1^2} 
\bigg[
\sum_\nu 
\Big(
 F_\nu(L_1L_1J_1J)-2\delta_1 F_\nu(L_1L_1'J_1J)
\\
&\quad\quad\quad
+\delta_1^2 F_\nu(L_1'L_1'J_1J)
\Big)
F_\nu(L_2L_2J_2J)P_\nu(\cos
\theta)
\bigg]
 \\
&\quad+
\frac{1}{1+\delta_1^2} 
(\pm)_{\sigma_2}
\bigg[
\sum_\nu 
\Big(
F_\nu(L_1L_1J_1J)-2\delta_1 F_\nu(L_1L_1'J_1J)
\\
&\quad\quad\quad
+\delta_1^2 F_\nu(L_1'L_1'J_1J)
\Big)
\kappa(L_2L_2)F_\nu(L_2L_2J_2J)P_\nu^{(2)}(\cos
\theta)
\bigg]
\cos 2\gamma.
\end{split}
\end{equation}

\chapter{Scaling relation for the $n$-dimensional Schr\"odinger equation}
\label{appscaling}
\pseudofootnotetext{The results of this appendix were subsequently reported in M.~A.~Caprio, 
Phys.~Rev.~C \textbf{68}, 054303 (2003).}

For the Schr\"odinger equation in $n$ dimensions, ``deepening'' and
``narrowing'' a potential in the correct proportion simply multiplies
all eigenvalues by an overall factor and dilates all eigenfunctions.
In this appendix, a simple proof of the scaling relation is given.
The behavior of the matrix elements of certain operators under this
transformation and the adaptations necessary for application to the
special case of the GCM Hamiltonian are also discussed.  These results
serve as the foundation for the approach to using the GCM developed in
Chapters~\ref{chapgcm}--\ref{chapgcmspecial}.  They are also applied
in the treatment of the finite square well potential in
Chapter~\ref{chapfwell}.

Consider the Schr\"odinger equation in the $n$ Cartesian coordinates
$x_1,\ldots,x_n$, with a kinetic energy operator having the standard
Cartesian form and a potential energy operator which depends only upon
the coordinates,
\begin{equation}
\label{eqnse}
\left[\sum_{i=1}^n \left( - \frac{\hbar^2}{2 m_i} \frac{\partial^2}{\partial
x_i^2}\right) + V(x_1,\ldots,x_n) - E\right]\Psi(x_1,\ldots,x_n)=0,
\end{equation}
where the $m_i$ are constants, $V$ is the potential energy operator,
and $E$ is the energy eigenvalue for wave function $\Psi$.  Suppose
that this equation is satisfied by a particular function $\Psi$, with
eigenvalue of $E$, for a specific potential $V$.  Now consider the
related expression obtained by substituting the quantities
\begin{equation}
\begin{aligned}
\label{eqnprimedefs}
V'(x_1,\ldots,x_n)&=a^2V(ax_1,\ldots,ax_n)\\
\Psi'(x_1,\ldots,x_n)&=\sqrt{a^n}\Psi(ax_1,\ldots,ax_n)\\
E'&=a^2E
\end{aligned}
\end{equation}
for the corresponding quantities on the left hand side of
(\ref{eqnse}).  This yields
\begin{align}
\Bigg[\sum_{i=1}^n \Bigg( - &\frac{\hbar^2}{2 m_i} \frac{\partial^2}{\partial
x_i^2}\Bigg) + a^2V(ax_1,\ldots,ax_n) -
a^2E\Bigg]\sqrt{a^n}\Psi(ax_1,\ldots,ax_n)\\
\intertext{which, with the substitution $u_i=ax_i$, becomes}
&=
\Bigg[\sum_{i=1}^n \Bigg( - \frac{\hbar^2}{2 m_i} a^2\frac{\partial^2}{\partial
u_i^2}\Bigg) + a^2V(u_1,\ldots,u_n) -
a^2E\Bigg]\sqrt{a^n}\Psi(u_1,\ldots,u_n)\notag\\
&=
a^2\sqrt{a^n}\Bigg[\sum_{i=1}^n \Bigg( - \frac{\hbar^2}{2 m_i} \frac{\partial^2}{\partial
u_i^2}\Bigg) + V(u_1,\ldots,u_n) -
E\Bigg]\Psi(u_1,\ldots,u_n)\notag\\
&=0 
\qquad\qquad\text{for all values of $(u_1,\ldots,u_n)$, by
(\ref{eqnse})}. \notag
\end{align}
Thus, the dilated wave function $\Psi'$ satisfies the Schr\"odinger
equation with the same kinetic energy operator as in~(\ref{eqnse}) but
with the modified potential $V'$, and $\Psi'$ has eigenvalue $E'$.
The factor $\sqrt{a^n}$ included in the definition of $\Psi'$
in~(\ref{eqnprimedefs}) serves to preserve normalization with respect
to the volume element $dx_1\cdots dx_n$.

It is often convenient to use $n$-dimensional spherical coordinates $(r,\Omega)$, where
$r=(x_1^2+\cdots+x_n^2)^{1/2}$ and $\Omega$ represents the angular
coordinates.  (This corresponds to polar or spherical coordinates in
the specific cases $n$=2 or $n$=3, respectively.)  In these coordinates, the
transformation is $V'(r,\Omega)=a^2V(ar,\Omega)$ and
$\Psi'(r,\Omega)=\sqrt{a^n}\Psi(ar,\Omega)$, which preserves
normalization with respect to the volume element $r^{n-1}drd\Omega$.

The matrix elements of the operator $r^q$ between two eigenfunctions
are frequently encountered, \eg, in the determination of transition
strengths, so it is useful to derive their properties under dilation.
Consider the radial integral
\begin{equation}
I^{AB}=\int_0^\infty r^{n-1} dr \, [\Psi^{A}(r)]^*\,r^q\,[\Psi^B(r)],
\end{equation}
where the angular variables have been suppressed for simplicity.
Then the radial integral for the corresponding transformed
eigenfunctions,
\begin{equation}
{I^{AB}}'=\int_0^\infty r^{n-1} dr \, [\sqrt{a^n}\Psi^{A}(ar)]^*\,r^q\,[\sqrt{a^n}\Psi^B(ar)],
\end{equation}
is related to the original integral by
\begin{equation}
\label{eqniscaling}
{I^{AB}}'=a^{-q} I^{AB},
\end{equation}
as can be shown by means of a simple change of variable $u=ar$.

The harmonic kinetic energy term~(\ref{eqnharmonicterm}) for the GCM,
\begin{equation}
-\frac{\hbar^2}{\sqrt{5}B_2}\sum_\mu
\frac{\partial^2}{\partial\alpha_{2\mu}\partial\alpha^*_{2\mu}},
\end{equation}
is not manifestly Cartesian in form, due to the presence of mixed
partial derivatives.  However, the preceding proof can readily be adapted to
apply in these coordinates.  Alternatively, a
change of variables~\cite{eisenberg1987:v1}
\begin{equation}
\begin{aligned}
x_0&=\alpha_{20}            && = \Real \alpha_{20} \\
x_1&=\frac{1}{\sqrt{2}}(\alpha_{21}-\alpha_{2-1})  && = \sqrt{2} \Real \alpha_{21} \\
x_{-1}&=\frac{1}{i\sqrt{2}}(\alpha_{21}+\alpha_{2-1})  && = \sqrt{2} \Imag \alpha_{21} \\
x_2&=\frac{1}{\sqrt{2}}(\alpha_{22}+\alpha_{2-2}) && = \sqrt{2} \Real \alpha_{22}\\
x_{-2}&=\frac{1}{i\sqrt{2}}(\alpha_{22}-\alpha_{2-2}) && = \sqrt{2} \Imag \alpha_{22},
\end{aligned}
\end{equation}
replaces the five complex coordinates $\alpha_{2\mu}$, subject to the
five constraints $\alpha_{2\mu}^*=(-)^\mu\alpha_{2-\mu}$, with five real
coordinates $x_i$ and
explicitly recasts the kinetic energy operator in the Cartesian form
\begin{equation}
-\frac{\hbar^2}{\sqrt{5}B_2}\sum_{i=-2}^2
\frac{\partial^2}{\partial x_i^2}.
\end{equation}
The ``radial'' coordinate to which the scaling property applies is $\beta$.

\chapter{Moving tape collector cycle optimization}
\label{appmtc}

Optimization of the tape advance cycle for moving tape collector
experiments of the type discussed in Chapter~\ref{chapbeta} is
summarized in this appendix.  Experiments in which spectroscopy is
performed on the daughter or granddaughter of the deposited parent
nucleus are considered.

\begin{figure}[t]
\begin{center}
\includegraphics*[width=0.7\hsize]{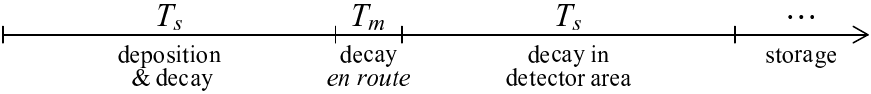}
\end{center}
\caption[Time line for a moving tape collector activity spot.]
{Time line for a moving tape collector activity spot.
\label{figmtcspottimeline}
}
\end{figure}

The ``life'' of a spot of activity deposited on the tape consists of
four stages (Fig.~\ref{figmtcspottimeline}):
\begin{dissenumeratelist}
\item Activity is deposited on the tape in the target box or other
deposition location (Chapter~\ref{chapbeta}), usually at a
constant rate $R$, for the time period $T_s$ that the tape is
stationary.  The activity immediately begins to suffer losses due to
decay in the deposition location.
\item The tape is advanced, requiring a time $T_m$ in which the tape
is in motion.  Activity is lost due to decay \textit{en route}.  For
the experiments discussed in the present work, $T_m$ is $\lesssim$1\,s
while $T_s$ is several minutes, so this
loss is negligible, but the loss can be significant in experiments involving
short half lives.
\item The decay radiations from the activity are observed in the
detector area, for a time $T_s$.  \item The activity is transported to
a storage area.
\end{dissenumeratelist} 
The number of radioactive nuclei present at a given time, $N(t)$, is
governed by 
\begin{equation}
\label{eqnactivityde}
\frac{d}{dt} N(t) = Q(t)-\lambda N(t),
\end{equation}
where $Q(t)$ is the rate at which new nuclei are being added and
$\lambda$ is the decay probability per unit time, related to the mean
and half lives by $\lambda$=$1/\tau$=$\ln 2/T_{1/2}$.
This is an inhomogeneous first-order differential equation, and can
thus be solved in closed form (\eg, Ref.~\cite{spiegel1968:handbook}),
yielding
\newcommand{\stackcenter}[1]{{\renewcommand{\arraystretch}{0}\begin{array}{c}#1\end{array}}}
\begin{equation}
N(t)=\underbrace{N(0)}_{\stackcenter{\text{original}\\\text{activity}}}
\underbrace{e^{-\lambda t}}_{\stackcenter{\text{decay
of}\\\text{original}\\\text{activity}}}
+
\int_0^t
\underbrace{dt'\,Q(t')}_{\stackcenter{\text{activity}\\\text{deposited}\\\text{at
time $t'$}}}
\underbrace{e^{-\lambda (t-t')}}_{\stackcenter{\text{decay
of}
\\\text{activity}\\\text{deposited}\\\text{at time $t'$}}}.
\end{equation}

\begin{figure}
\begin{center}
\includegraphics*[width=1.0\hsize]{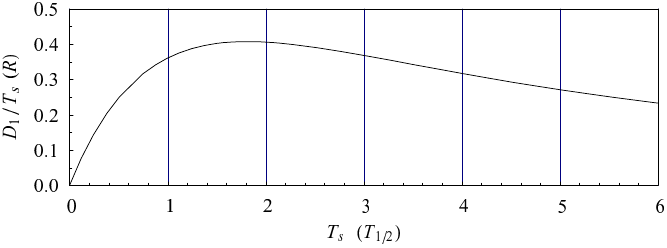}
\end{center}
\caption[Number of decays in the detector area per experiment time.]
{Number of decays in the detector area per experiment time, $D_1/T_s$,
as a fraction of the initial deposition rate $R$, plotted as a
function of the tape cycle $T_s$, for $T_m$=0.  This graph applies to
decay of the original deposited parent (nucleus 1), populating its
daughter (nucleus 2).
\label{figmtconestepcurve}
}
\end{figure}

For the nuclide deposited on the tape, denoted here as the
first member of the decay chain by a subscript 1, the rate $Q_1(t)$ at which activity
is added is $R$ for $0\leq t \lt T_s$ and zero after.  Consequently,
\begin{equation}
\label{eqnactivityn1}
N_1(t) = \begin{disscases}
R\frac{1}{\lambda_1} (1-e^{-\lambda_1 t}) & 0\leq t\lt T_s\\
R\frac{1}{\lambda_1} (1-e^{-\lambda_1 T_s})e^{-\lambda_1 (t-T_s)}
&t>T_s.
\end{disscases}
\end{equation}
The number of decays occurring in the detector area during the
observation time is
\begin{equation}
\begin{split}
D_1&=\int_{T_s+T_m}^{T_s+T_m+T_s} dt \, \lambda_1 N_1(t)\\
&=R \frac{1}{\lambda_1}\left[1-e^{-\lambda_1 T_s}\right]^2 e^{-\lambda_1T_m}.
\end{split}
\end{equation}
As longer tape cycle times are chosen, more decays will occur
in each individual cycle, but fewer cycles will fit into an experiment
of fixed length.  We are interested in maximizing the number of decays in the detector
area during the \textit{experiment}, not during one particular tape
cycle.  The maximum yield per experiment time is obtained by
maximizing $D_1/(T_s+T_m)$.  This quantity is plotted in
Fig.~\ref{figmtconestepcurve} as a function of $T_s$, for the case in which $T_m$ is
negligible.  The maximum occurs for a tape cycle duration of
$T_s$$\approx$$1.81T_{1/2}$, for which about 41$\%$ of deposited
nuclei decay in the detector area.

For the daughter nuclide, denoted here by a subscript 2, the activity
as a function of time is again dictated by~(\ref{eqnactivityde}).  In
this case the source of new nuclei is the decay of nuclide 1, so
$Q_2(t)$=$\lambda_1 N_1(t)$, with $N_1(t)$ given
by~(\ref{eqnactivityn1}).  The resulting function $N_2(t)$ is
\begin{equation}
N_2(t)=\begin{disscases}
R \Big[ \frac{1}{\lambda_2}-\frac{1}{\lambda_2-\lambda_1}
e^{-\lambda_1 t}
+\frac{\lambda_1}{\lambda_2(\lambda_2-\lambda_1)}e^{-\lambda_2
t}\Big] & 0\leq t \lt T_s\\
\\
R  \Bigg[
\left(
 \frac{1}{\lambda_2}-\frac{1}{\lambda_2-\lambda_1}
e^{-\lambda_1 T_s}
+\frac{\lambda_1}{\lambda_2(\lambda_2-\lambda_1)}e^{-\lambda_2
T_s}\right) e^{-\lambda_2(t-T_s)} & \\
\qquad+\frac{1}{\lambda_2-\lambda_1} \left(e^{\lambda_1T_s}-1\right)
\left[e^{-\lambda_1t}-e^{-\lambda_2t}e^{(\lambda_2-\lambda_1)T_s}
\right]
\Bigg]
&t>T_s.
       \end{disscases}
\end{equation}
The number of decays of the daughter in
the detector area during one tape cycle is
\begin{equation}
D_2=R \frac{1}{\lambda_1-\lambda_2}\left[
-\frac{\lambda_1}{\lambda_2}\left(1-e^{-\lambda_2T_s}\right)^2e^{-\lambda_2T_m}
+
\frac{\lambda_2}{\lambda_1}\left(1-e^{-\lambda_1T_s}\right)^2e^{-\lambda_2T_m}
\right].
\end{equation}
In order to optimize the observation of daughter decays populating the
granddaughter, for spectroscopy experiments making use of two-step
$\beta$ decay, the quantity $D_2/(T_s+T_m)$ must be maximized.

\chapter{Data acquisition and the \texttt{cscan} sorting package}
\label{appacq}

Within the timespan covered by the work described here, the facilities
available for nuclear structure experiments at WNSL expanded
considerably, in both the detection resources available and variety of
experiment types possible.  The demands upon the acquisition
electronics and flexibility required of the data analysis software
have increased correspondingly.  At the time of YRAST Ball's early
runs~\cite{beausang2000:yrastball}, the detection array consisted of a
relatively homogeneous set of Ge elements for $\gamma$-ray coincidence
spectroscopy, and analysis required only the construction of basic 1-,
2-, and 3-dimensional $\gamma$-ray energy histograms.  Recent combined
FEST and spectroscopy experiments at the Yale MTC entail the
simultaneous analysis of clover, LEPS, BaF$_2$, plastic scintillator,
and fast-timing TAC signals, with wall-time clock time stamps for
$\gamma$-ray multiscaling.  Similarly, a typical current experiment using the SASSYER
recoil separator~\cite{ressler2002:sassyer} involves approximately 40
channels of clover and LEPS data from a target-position array, several
additional Ge channels from an isomer array, and implantation and
$\alpha$-decay tagging signals from highly-pixelated particle
detectors at the focal plane.  The addition of time-of-flight
detectors is planned.  A correspondingly diverse range of diagnostic
and analysis histograms must be constructed by the sorting software.

This appendix summarizes the acquisition and sorting software
suite~\cite{caprioUNP:cscan,caprioUNP:vmeacquire} developed, as part
of the present work, for WNSL nuclear structure experiments.  This
suite consists of three main elements: the
\texttt{cscan} sorting package, an acquisition readout code with
event builder, and specialized data sorting routines.  Portions of
this software have also been used in $\beta$-decay experiments at the
TRIUMF ISAC and ORNL HRIBF radioactive beam facilities.

The \texttt{cscan} sorting package is a generic framework for online
and offline sorting.  This portion of the software suite is not
specialized to the specifics of the present WNSL acquisition
environment but rather is a control program, which calls external
routines for the site-specific tasks.  This same program, used with
different site-specific routines, provided the online and offline
sorting capability for TRIUMF E801 (Chapter~\ref{chap162er}) and
offline sorting for HRIBF RIB-095~\cite{zamfirINPREP}.  The spirit of
this package is similar to that of earlier packages (\textit{e.g.},
Ref.~\cite{jin:tscan}), used for tape-based sorting in nuclear
spectroscopy, in which the control and site-specific functions are
separated.  The main differences lie in the provisions for
communication with the acquisition code in an online environment and
for the flexible input and output of file-based data formats.  For
online use, the \texttt{cscan} control code calls a site-specific
routine at startup to establish a data connection to the acquisition
code.  This connection can be, \textit{e.g.}, via a first-in-first-out (FIFO)
pipe on the local host (as at WNSL) or via a remote procedure call
(RPC) server to receive data from an acquisition computer located
elsewhere on the network (as at TRIUMF).  For offline use, the
\texttt{cscan} control program calls site-specific routines to handle
all data file input tasks, and thus does not impose any specific
limitations on the data format, such as the requirements of fixed-size
block structure or event separator codes typical of tape-based
systems.  The site-specific input routines are free to support
variable-length events, whether or not they are stored in fixed-sized
blocks, and can also provide on-the-fly decompression of data files
during the sorting process.  The \texttt{cscan} package outputs
histograms in the
RadWare~\cite{radford1995:radware,radfordUNP:radware} spectrum,
matrix, and cube formats.

The expanded acquisition hardware needs for WNSL spectroscopy
experiments were addressed by the installation of a current-generation
acquisition system, based on 32-channel VME analog to digital
converter (ADC), time to digital converter (TDC), and scaler modules
with internal multievent storage buffers~\cite{caen:modules}, read out
via a fiber-optic link~\cite{sbs:module} by an Intel/Linux workstation.
The task of the acquisition code, which runs on the Linux workstation,
is to read data from the VME acquisition modules~\cite{caen:modules},
construct events from this data, and save these events to disk and/or
transmit them to the \texttt{cscan} online sorting code.  The VME modules
each store data from several successive events in an internal
multievent buffer.  The acquisition code then periodically reads out
all data from all modules, and an event builder routine
cross-correlates data items from the different modules to reassemble
the contents of individual events.  Although the modules tag each data item
with an internally-generated event serial number, the event builder
must carry out validation to robustly handle the drifts between
counters which occur at the level of $\sim$1 per $10^3$--$10^4$
events.  The ADC, TDC, and scaler data are stripped of
hardware-specific encoding information, and the events are written in
a header-plus-data format, along with run marker and diagnostic error
message pseudo-events sharing the same header format.  These events
may be sorted by
\texttt{cscan}, in conjunction with WNSL site-specific routines, or by the custom
sorting codes of outside research groups~\cite{kumbartzkiUNP}.

The sorting routine used with \texttt{cscan} for analyzing data from
WNSL spectroscopy experiments is based upon an approach which greatly
simplifies the processing of data from an inhomogeneous assortment of
detectors.  Commonly, sorting routines intended for use with only a few types of
detectors contain separate segments of code to perform the
basic processing for each detector type and use separate sets of
variables to store the data from each detector type.  As more detector
types are defined, this approach leads to a cumbersome and highly
redundant code structure.  Maintenance or improvement of the code is
impractical and error prone, since even minor changes must be manually
duplicated for each detector type, and the implementation of new
sorting tasks which encompass whole classes of detector types, \textit{e.g.},
all Ge detectors, requires separate code to be written for each
detector type in the class.  

\begin{figure}
\begin{center}
\includegraphics*[width=0.7\hsize]{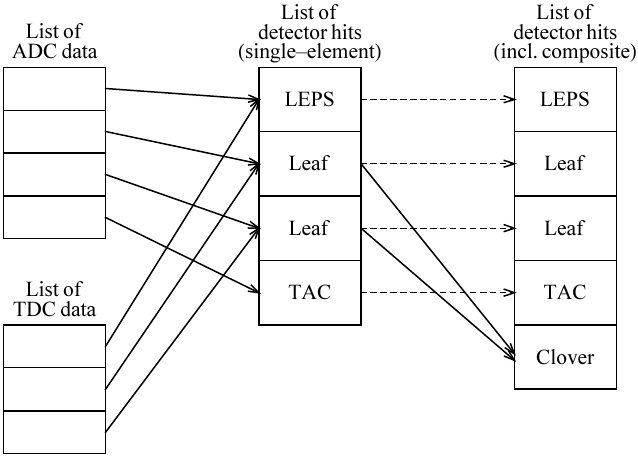}
\end{center}
\caption[Event processing scheme for \texttt{cscan} WNSL sorting
routine.]  {Event processing scheme for the \texttt{cscan} WNSL
site-specific sorting routine, in which ``hits'' for all detector
types are processed using the same code and stored in one common list.
ADC and TDC data are first scanned to construct hits of single-element
detectors, and these are in turn used to construct hits of composite
detectors.  
\label{figcscanhits}
}
\end{figure}

The \texttt{cscan} WNSL sorting routine instead uses one compact segment
of code to process all detector types.  The code can accommodate an
arbitrary number of detectors of an arbitrary number of types with no
modification.  At the beginning of a sorting run, a list of detector
definitions is read from an input file provided by the user,
specifying the type, ADC channel (if applicable), TDC channel (if
applicable), and such auxiliary information as array angle or pixel
grid position for each detector.  Separately, a list of corrections,
such as gain matching or Doppler correction, to be applied to all
detectors of a given type is constructed from information provided by
the user.  Then, the code processes the list of ADC data and TDC data
for each event in three steps, as summarized in
Fig.~\ref{figcscanhits}.
The ADC and TDC data are scanned to identify ``hits'' in any of the
detectors in the detector definition list.  If a hit occurs, the
corrections list is consulted, and the corrections appropriate to this
particular detector type are applied to the ADC and TDC data.  A
record of the hit, containing the energy and timing information
together with the detector ID and all auxiliary information included
in the detector definition, is added to a common list of hits shared
by all detector types.  After all hits of single-element detectors
are identified from the ADC and TDC data, a second pass is made
through these hits to construct hits in composite detectors built from
these individual array elements (\textit{e.g.}, clover detectors built from clover
leaves), and the resulting hit records are appended to the common list
of detector hits.  All sorting tasks are then carried out using the
common list of hits, typically looping over the hits and selecting
data from those for which the detector type and data values match
certain criteria.

\chapter{Gamma-ray coincidence intensity spectroscopy}
\label{appcoin}

\section{Coincidence techniques}
\label{seccointechniques}

The advent of compact, high-efficiency arrays of large-volume Ge
detectors makes possible a new generation of $\gamma$-ray spectroscopy
$\beta$-decay experiments, in which high-statistics $\gamma$-ray
coincidence data are available.  In order to obtain $\gamma$-ray
intensity measurements comparatively free of the contamination endemic
to singles data, the present work relies upon coincidence data for the
determination of intensities (Section~\ref{seccoin}).  In this
appendix, a ``formalism'' for the extraction of intensities from
coincidence data is presented, and basic examples of its application
are given.  The properties and calibration of the coincidence
efficiency for a multi-detector array are also discussed.

The methods used in the present work to accurately extract the
intensities of specific, often weak, transitions differ greatly from
the global least-squares intensity fitting techniques commonly applied
in high-spin studies with large multi-detector
arrays~\cite{radford1995:radware}.  The methods for extracting
intensities from coincidence data are somewhat more involved than
those for the simple conversion of a singles spectrum peak area to an
intensity.  However, the application of a few basic principles makes
reliable analysis of intensities from coincidence data a
straightforward process.

As discussed in Section~\ref{seccoin}, in a spectroscopy experiment, a set of
\textit{decays} in an ensemble of nuclei results in a set of
\textit{events} observed in the detector array, in
a way which depends upon the instrument response.  Let us establish
the relevant quantities describing the ensemble of decays and set of
events.  Let $N$ be the total number of decays in the ensemble.  For a
$\gamma$-ray transition $x$, we are interested in the fraction $I_x$
of decays which involve the emission of radiation $x$, that is,
the ``intensity'' of $x$.  For two $\gamma$-ray transitions, $x$ and
$y$, we can consider the fraction $I_{x:y}$ of decays which involve
emission of both transitions $x$ and $y$ in coincidence, and so on for
higher numbers of coincident radiations.  In the observed data, there
are analogous quantities: the number $S_x$ of observed singles counts
of $x$ and the number $G_{x:y}$ of detected coincidences between $x$ and
$y$, usually extracted by observing the peak area for $y$ in a
spectrum gated on $x$.

The observed quantities are related to the physical quantities by the
array efficiencies, technical aspects of which are discussed in the
following section.  In general, the efficiency for detecting a
specific radiation or coincident set of radiations depends upon many
parameters (relative directions of emission, isomeric delays between
emission, other correlated radiations which may block their detection,
\etc), but, under most routine circumstances, the efficiencies
depend, to a good approximation, upon just the energies of the
radiations involved.  Thus,
\begin{equation}
S_x = N I_x\varepsilon(E_x)
\end{equation}
and
\begin{equation}
\label{eqngateany}
G_{x:y} = N I_{x:y} \varepsilon(E_x,E_y).
\end{equation}
Note that $I_{x:y}=I_{y:x}$ by definition.  If symmetric gating is
performed, \ie, the set of detectors used for gating is the same
as that included in the gated spectra, then also $G_{x:y}=G_{y:x}$ and
$\varepsilon(E_x,E_y)=\varepsilon(E_y,E_x)$.  Coincidences between
distinct detector sets, \eg, LEPS detectors and clover detectors, may
also be considered, in which case such symmetry is not present in
$G$ and $\varepsilon$.

Two simple situations in which intensities may be extracted from
coincidence data were introduced in Section~\ref{seccoin}.
\begin{figure}
\begin{center}
\includegraphics*[width=1.0\hsize]{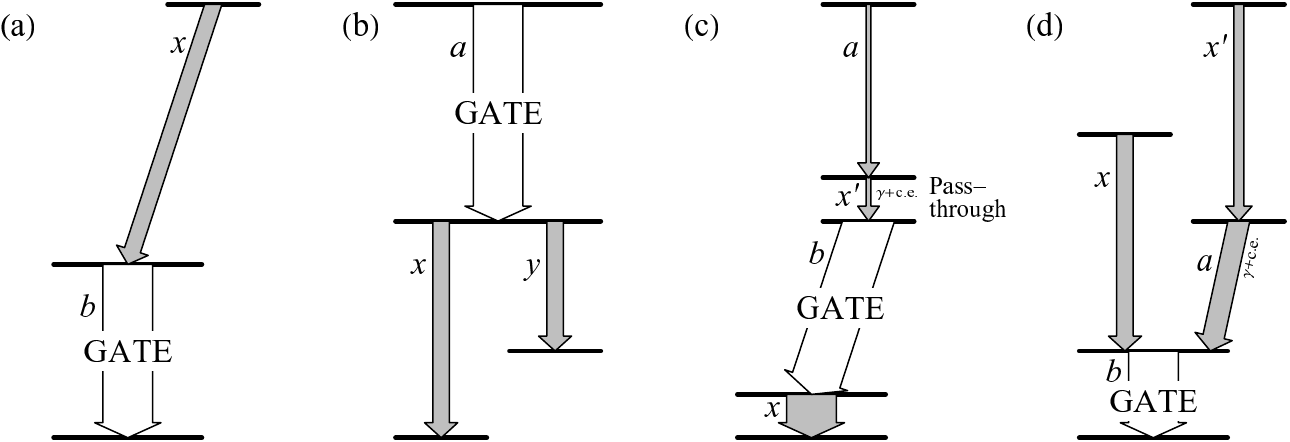}
\end{center}
\caption[Examples of the extraction of intensities
from coincidence data.]  {Examples of the use of the formalism for
extraction of intensities from coincidence data: (a)~a simple gate
below, (b)~a simple gate above, (c)~a pass-through
measurement to find the intensity of an otherwise
``hidden'' transition $x'$, and (d)~a gate below in which the
contribution of the contaminant $x'$ must be subtracted to obtain the
intensity of $x$.  The observed
transitions (gray) are drawn with widths indicating their intensity
\textit{coincident} with the gating transition (white).
\label{figcoinmeth}
}
\end{figure}
If a transition $x$ directly feeds a level which decays by a
transition $b$ [Fig.~\ref{figcoinmeth}(a)], then the probability of
emission of $x$ followed by emission of $b$ is $I_{x:b}$$=$$I_xB_b$,
where $B_b$ is the branching probability for the level to decay by
$b$.  Thus,~(\ref{eqngatebelow}) follows as a special case
of~(\ref{eqngateany}).  As noted in Section~\ref{seccoin}, $B_b$ is in
practice calculated from $B_b = I_b / (\sum_i I_i+\sum_i
I^\text{ce}_i)$, where the sums are over all $\gamma$-ray and
conversion electron transitions $i$ depopulating the level.
Similarly, when a transition $a$ directly feeds a level which decays
by two transitions $x$ and $y$ [Fig.~\ref{figcoinmeth}(b)], we have
\begin{equation}
\frac{I_{a:x}}{I_{a:y}}=\frac{I_aB_x}{I_aB_y}=\frac{I_x}{I_y},
\end{equation}
so~(\ref{eqngateabove}) also follows as a special case
of~(\ref{eqngateany}).  

However, much more complicated situations may be reliably addressed by
the methodical use of the present formalism.  Let us conclude this
section with three representative examples.

Consider the decay scheme of Fig.~\ref{figcoinmeth}(c), in which the
transition of interest, $x'$, is degenerate in energy with a low-lying
and much more strongly populated transition $x$.  Such doublet pairs
are inevitable in ``good'' rotor nuclei with the same moment of
inertia for different bands.  The transition $x'$ is completely obscured by
$x$ in spectra gated on $a$ or $b$, and so its intensity cannot be
measured directly.  However, the transitions $a$ and $b$ can only be
coincident with each other if ``pass-though'' via the transition $x'$, 
or its associated conversion electrons, occurs.  Quantitatively,
$I_{a:b}=I_a(B_{x'}+B_{x'}^\text{ce})B_b$, so, if the intensities of
$I_a$ and $B_b$ are known, $B_{x'}+B_{x'}^\text{ce}$ can be deduced
from $G_{a:b}$.

The decay scheme of Fig.~\ref{figcoinmeth}(d) illustrates a situation
in which the intensity of a transition $x$ cannot be deduced trivially
from its coincidences with a gating transition $b$ below, because of
contamination from a coincident doublet transition $x'$.  The measured
combined area is
$G_{b:x}+G_{b:x'}$$=$$N(I_{b:x}+I_{b:x'})\varepsilon(E_b,E_x)$.
However, since $I_{b:x'}=I_{x'}(B_{a}+B_{a}^\text{ce})B_b$, if
$I_{x'}$ can be determined independently, and the combined $\gamma$
and conversion electron branching strength of $a$ is known, then the
contribution of $x'$ to the gate can be subtracted.  Thus, $I_x$ can
be recovered.  Such a situation occurs for the 950.5\,keV
$4^+_{1088}\rightarrow2^+_{138}$ transition in $^{156}$Dy (see
Fig.~\ref{fig156dylev1088} on page~\pageref{fig156dylev1088}), which
is unresolved, in the spectrum gated on the 138\,keV
$2^+_{138}\rightarrow0^+_{0}$ transition, from the higher-lying
949.6\,keV $(?)_{1840}\rightarrow 2^+_{890}$ transition.

As a final example, and cautionary note, consider the case in which
the transitions $b$ and $x$ of Fig.~\ref{figcoinmeth}(a) are
degenerate to within the detector resolution.  Such is the case for
the 79.6\,keV and 81.0\,keV transitions in $^{133}$Ba source decay, or
for the 1030.7\,keV $4^+_{1088}\rightarrow2^+_{138}$ and 1031.8\,keV
$(?)_{2200}\rightarrow4^+_{1088}$ transitions in $^{156}$Dy
(Table~\ref{tabline}).  A naive application of~(\ref{eqngatebelow})
yields an intensity $I_x$ which is a factor of two too high.  Any gate
on $b$ is also a gate on $x$, so the observed counts are actually
$G_{b:x}+G_{x:b}$$=$$I_{b:x}\varepsilon(E_b,E_x)+I_{x:b}\varepsilon(E_x,E_b)$,
but this is just $2I_{b:x}\varepsilon(E_b,E_b)$.  Since both lines
contribute an exactly equal number of counts to the gated peak, the
peak centroid energy is the arithmetic mean of the individual energies
($E_\text{cent}$$=$$(E_x+E_b)/2$).  Thus, if $E_b$ is measured
independently, $E_x$ can be recovered from the observed peak centroid.

\section{Array coincidence efficiency}
\label{seccoineff}

The probability for two $\gamma$-rays emitted simultaneously to both
be detected in a detector array, and registered as coincident, is
\textit{approximately} just the probability that each will be detected
individually in distinct array elements.  Thus, the coincidence
efficiency of an array of detectors is approximately the sum of the
pairwise products of efficiencies of the individual array elements:
\begin{equation}
\label{eqncoineff}
\varepsilon(E_x,E_y) = \sum _ {i,j\,(i\neq j) } \varepsilon_i(E_x) \varepsilon_j(E_y),
\end{equation}
where the sum is over individual detectors $i$ and $j$, $i$$\neq$$j$.
For an array of $n$ identical detectors of efficiency
$\varepsilon_0(E)$, this reduces to $\varepsilon(E_x,E_y) =
n(n-1)\varepsilon_0(E_x)
\varepsilon_0(E_y)$.  For asymmetric gating, in which a gate condition
is placed on detectors in class $A$, and the resulting counts in
detectors in class $B$ are measured, the sum runs over $i$$\in$$A$ and
$j$$\in$$B$.

However, deviations from this ideal product efficiency occur.  At low
$\gamma$-ray energies, electronic timing jitter and walk can cause the
signals from two coincident $\gamma$ rays to fall outside the
acquisition system's timing acceptance.  In this case, the
expression~(\ref{eqncoineff}) serves as a useful baseline for the
calculation of $\varepsilon(E_x,E_y)$, valid at high $\gamma$-ray
energies, but the attenuation of efficiency due to such time
``windowing'' effects at low energies must be calibrated against known
coincidences, yielding a correction factor
$w(E_x,E_y)$$\equiv$$\varepsilon(E_x,E_y)/\sum\varepsilon_i(E_x)
\varepsilon_j(E_y)$.  Time windowing attenuation
introduces not only a loss of statistics but also analysis
uncertainties resulting from the correction factor $w(E_x,E_y)$, and
thus it is undesirable in the $\gamma$-ray energy range of interest.
The attenuation can be avoided if it is possible to set constant
fraction discriminator thresholds sufficiently below the energies of
interest to prevent leading-edge walk effects~\cite{ortec1982:timing} from
occuring and if generous timing acceptances
are allowed in the hardware trigger, time to digital converters, and
offline analysis cuts.

Deviations from the product form~(\ref{eqncoineff}) also occur if
additional correlations are present between the emitted $\gamma$ rays.
Angular correlation effects depend upon the array geometry and are
usually most important for $0^+$--$2^+$--$0^+$ cascades (see
Section~\ref{sec156dyspec}).  Additional $\gamma$-rays emitted in the
same decay suppress detection of coincidences of the $\gamma$-rays of
interest, since they may directly interact with the same detector as
one of the $\gamma$-rays of interest (conventional ``summing'') or
scatter into the Compton suppression shield of that detector.  This
effect is most important for high-multiplicity decays and for arrays
in which individual detector elements have efficiencies of
$\gtrsim$1$\%$.

\begin{figure}
\begin{center}
\includegraphics*[width=0.6\hsize]{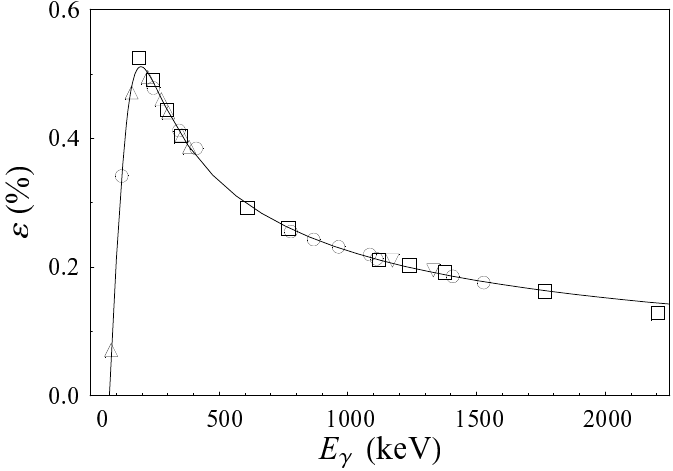}
\end{center}
\caption[Efficiency curve for a clover
detector at the Yale MTC.]  {Efficiency curve for a clover detector at
the Yale MTC, at a face distance of 13\,cm from the source
position, without add-back, fitted using the
parametrization~(\ref{eqnwseff}).  Calibration points from
$^{60}$Co~(\captionopendowntriangle),
$^{133}$Ba~(\captionopenuptriangle), $^{152}$Eu~(\captionopencircle),
and $^{226}$Ra~(\captionopensquare) are matched in normalization as
described in the text, and the overall normalization is deduced by the
coincidence method.
\label{figcloveff}
}
\end{figure}

For~(\ref{eqncoineff}) to be applied, the single-detector efficiency
functions $\varepsilon_i(E)$ must first be determined.  A parametrized
form which reproduces the efficiency characteristics of
both the LEPS and clover detectors used at the Yale MTC (Fig.~\ref{figcloveff})
is
\begin{equation}
\label{eqnwseff}
\varepsilon(E)= A E^{p_1}\left[1-u_{a,d}(E)\right]
+B E^{-p_2}u_{a,d}(E),
\end{equation}
where
\begin{equation}
u_{a,d}(E)\equiv 1 - \frac{1+e^{-a/d}}{1+e^{(E-a)/d}}.
\end{equation}
This function has of a power law dependence at low energies and a
power law dependence at high energies, joined smoothly by a
``softened'' step function $u_{a,d}(E)$ constructed from the Woods-Saxon
potential function.

No single $\gamma$-ray calibration standard (\eg,
Ref.~\cite{knoll1989:radiation-detection}) covers the full energy
range of interest ($\sim$100\,keV--2500\,keV) in the experiments in
the present work, so multiple sources ($^{60}$Co, $^{133}$Ba,
$^{152}$Eu, $^{226}$Ra), covering different portions of the energy
range, are used to calibrate $\varepsilon_i(E)$ in a piecewise
fashion.  The absolute activities of calibration sources are usually
only known to within $\sim$5--10$\%$, with possible further
uncertainties arising from the acquisition dead time determination in
calibrations using strong sources, so these calibrations generally do
not match in normalization at the points of overlap and provide only a
moderately accurate overall efficiency normalization.  To address
these considerations, a special approach has been used in the present
work.  The singles data from the sources are used only to determine
the \textit{shape} of the efficiency curve.  The relative strengths of
the sources are matched by allowing the normalizations of the
calibration points from different data sets to vary as a power law
function is locally fit through them in their region of overlap.  The
renormalized calibration points then yield a
\textit{relative} efficiency function $f_i(E)$, related to the true
efficiency by $\varepsilon_i(E)=Kf_i(E)$, where the factor $K$ is
common to all detectors in the array if the same source strength
renormalization factors are consistently used for all.

The normalization factor $K$ is then determined by a natural extension
of the standard ``coincidence method'' (\eg,
Ref.~\cite{knoll1989:radiation-detection}) for calibrating
two-detector systems.  The coincidence efficiency~(\ref{eqncoineff})
contains products involving two factors of $K$, while singles
efficiency, $\varepsilon(E)$$=$$\sum_i Kf_i(E)$, contains only one.
Thus, $K$ can be determined by considering the ratio of any gated
coincident area and any singles area, eliminating any need for
reference to the poorly-known integrated source activity $N$:
\begin{equation}
\label{eqncoinkratio}
\frac{G_{x:y}}{S_z}=\frac{I_{x:y}}{I_z} \frac{K\sum_{i,j\,(i\neq
j)}f_i(E_x) f_j(E_y)}{\sum_i f_i(E_z)}.
\end{equation}
Known coincidences in several common calibration sources, deduced from
$\gamma$-ray intensity and $\gamma$-ray and conversion electron
branching information in
Refs.~\cite{nds1993:60,nds1995:133,nds1996:152,nds1995:214}, are
summarized in Table~\ref{tabcoincal}.  If intense transitions from
strong sources are used, the calibration gate will contain a
substantial time randoms contribution, which must be subtracted either
by the standard relations for time
randoms~\cite{knoll1989:radiation-detection} or through graphical
subtraction of a singles spectrum from the gated spectrum, weighted so
that false coincidences exactly disappear.

\begin{figure}[t]
\begin{center}
\includegraphics*[width=0.7\hsize]{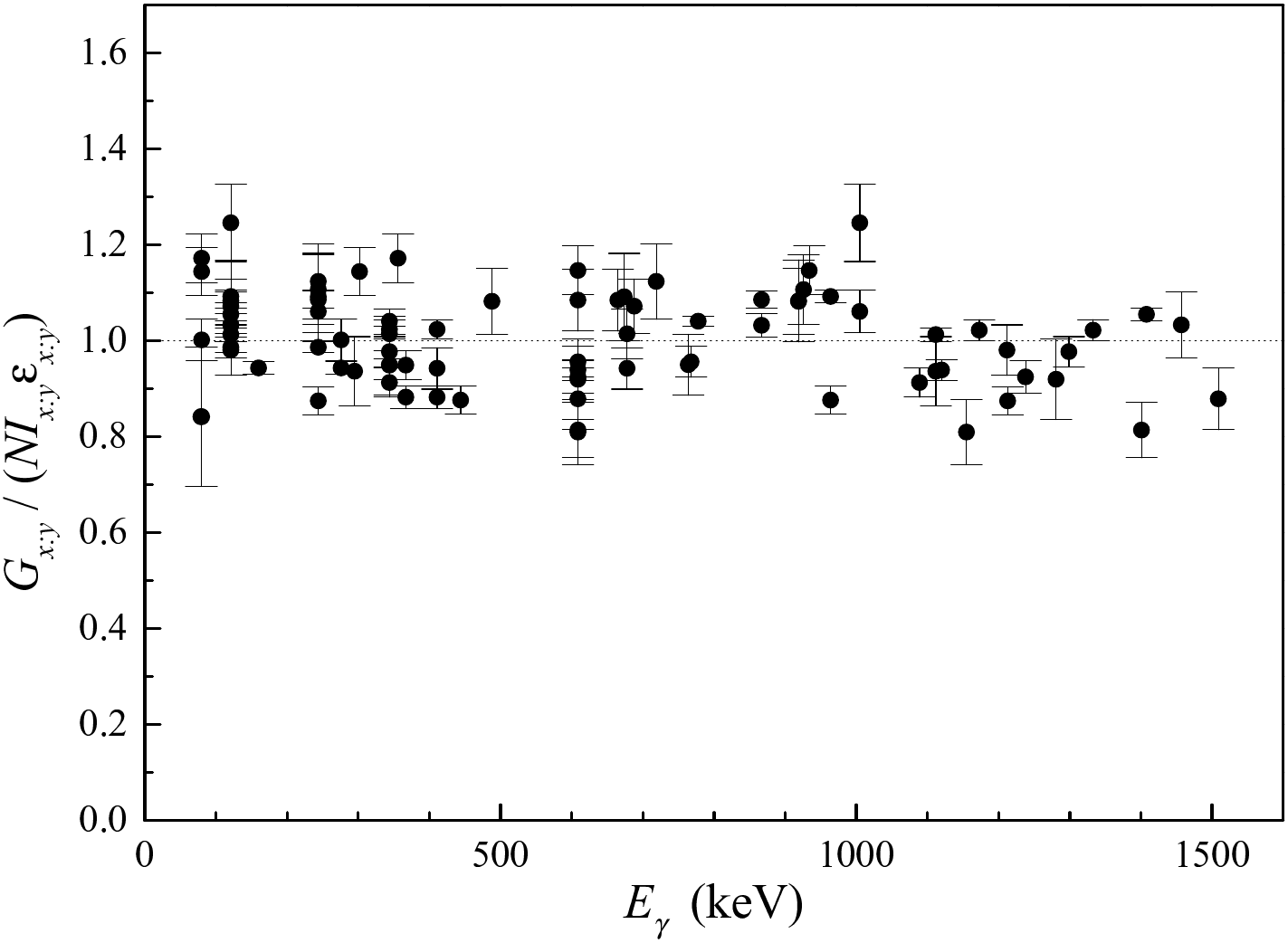}
\end{center}
\caption[Ratio of the observed coincident counts
to the number expected.]  {Ratio of the observed coincident counts to
the number expected from the known decay scheme and the assumed
product form~(\ref{eqncoineff}) for the array coincidence efficiency.
Data are shown for 43 pairs of coincident $\gamma$-ray transitions in
$^{60}$Co, $^{133}$Ba, $^{152}$Eu, and $^{226}$Ra source decay
(Table~\ref{tabcoincal}), detected in the array of three clover
detectors used for the experiment of Section~\ref{sec162ermtc}.  The
ratio for each coincidence is plotted both for the lower
$\gamma$-ray energy and higher $\gamma$-ray
energy.
\label{figcoineffcheck}
}
\end{figure}

Actual calibration data, showing the ratio of the measured $G_{x:y}$
to that expected from the known $I_{x:y}$ and assumed array
efficiency~(\ref{eqncoineff}), are plotted in
Fig.~\ref{figcoineffcheck}.  The values are for 43 calibrator
coincidences in $^{60}$Co, $^{133}$Ba, $^{152}$Eu, and $^{226}$Ra
source decay (Table~\ref{tabcoincal}), detected in an array of three
clover detectors at the Yale MTC.  
Such a calibration serves as a test for energy-dependent attenuation
effects.  These are not significantly present for the setup of
Fig.~\ref{figcoineffcheck}, \ie, $w$$\approx$1 throughout the energy
range calibrated.  The narrow spread of values indicates the
reliability of intensities extracted using the form~(\ref{eqncoineff})
for the coincidence efficiency.

\clearpage  
%
\begin{table}[H]
\caption[Coincident intensities in $^{60}$Co, $^{133}$Ba, $^{152}$Eu, and
$^{226}$Ra source decay.]  {\ssp Coincident intensities of pairs of
transitions in $^{60}$Co, $^{133}$Ba, $^{152}$Eu, and $^{226}$Ra
source decay deduced from the decay schemes of
Refs.~\cite{nds1993:60,nds1995:133,nds1996:152,nds1995:214}, for use
in calibration of coincidence efficiencies.  All intensities are per
parent decay.
\label{tabcoincal}
}
\end{table}
\addtocounter{table}{-1}
{
\vspace{-30pt}
\ssp\small
\begin{center}
\begin{tabular}{r.l_r.l_r.l_r.l_r.l==r.l_r.l_r.l_r.l_r.l}
\pseudoruledtabular
\multicolumn{2}{c}{$E_x$}&
\multicolumn{2}{c}{$I_x$}&
\multicolumn{2}{c}{$E_y$}&
\multicolumn{2}{c}{$I_y$}&
\multicolumn{2}{c}{$I_{x:y}$}&
\multicolumn{2}{c}{$E_x$}&
\multicolumn{2}{c}{$I_x$}&
\multicolumn{2}{c}{$E_y$}&
\multicolumn{2}{c}{$I_y$}&
\multicolumn{2}{c}{$I_{x:y}$}\\
\multicolumn{2}{c}{(keV)}&
\multicolumn{2}{c}{($\%$)}&
\multicolumn{2}{c}{(keV)}&
\multicolumn{2}{c}{($\%$)}&
\multicolumn{2}{c}{($\%$)}&
\multicolumn{2}{c}{(keV)}&
\multicolumn{2}{c}{($\%$)}&
\multicolumn{2}{c}{(keV)}&
\multicolumn{2}{c}{($\%$)}&
\multicolumn{2}{c}{($\%$)}
\\\hline
&                                                                                                \\
\multicolumn{10}{c}{$^{152}$Eu source decay\ensuremath{^{a}}}&                                           411&&           2&.232(4)&      368&&           0&.860(5)&              0&.840\\
\multicolumn{10}{c}{}&                                                                                                &&              &&              679&&           0&.471(4)&              0&.460\\
122&&           28&.58(9)&      245&&           7&.58(3)&               3&.49&                   586&&           0&.459(5)\ensuremath{^{b}}& 503&&  0&.148(8)&              0&.155\\
&&              &&              689&&           0&.857(8)&              0&.395&                  \\\cline{11-20}                                                                                  
&&              &&              867&&           4&.25(2)&               1&.96&                   \\                                                                                               
&&              &&              919&&           0&.427(6)&              0&.196&                  \multicolumn{10}{c}{$^{60}$Co source decay\ensuremath{^{c}}}\\                                            
&&              &&              964&&           14&.60(4)&              6&.73&                   \\                                                                                               
&&              &&              1005&&          0&.646(5)&              0&.298&                  1332&&       99&.9826(6)&   1173&&   99&.85(3)&                     99&.8\\                      
&&              &&              1112&&          13&.64(4)&              6&.29&                   \\\cline{11-20}                                                                                  
&&              &&              1213&&          1&.422(6)&              0&.656&                  \\                                                                                               
&&              &&              1408&&          21&.00(6)&              9&.68&                   \multicolumn{10}{c}{$^{133}$Ba source decay\ensuremath{^{d}}}\\                                          
&&              &&              1458&&          0&.502(5)&              0&.231&                  \\                                                                                               
&&              &&              1528&&          0&.281(5)&              0&.130&                  81&.0&          34&.1(3)\ensuremath{^{e}}& 79&.6&         2&.62(6)\ensuremath{^{f}}&        0&.964\ensuremath{^{g}}\\       
245&&           7&.58(3)&       656&&           0&.1448(19)&            0&.131\ensuremath{^{h}}&     &&              &&              276&&           7&.16(2)&               2&.63\ensuremath{^{i}}\\       
&&              &&              675&&           0&.172(5)&              0&.155&                  &&              &&              303&&           18&.33(6)&              6&.75\\                  
&&              &&              867&&           4&.25(2)&               3&.83&                   &&              &&              356&&           62&.05(19)&             22&.8\\                  
&&              &&              719&.4&         0&.278(8)\ensuremath{^{j}}&      0&.251\ensuremath{^{k}}&      161&&           0&.645(8)&      276&&           7&.16(2)&               0&.544\\                 
&&              &&              926&&           0&.278(5)&              0&.251&                  303&&           18&.33(6)&      53&&            2&.20(2)&               1&.40\\                  
&&              &&              1005&&          0&.646(5)&              0&.583&                  384&&           8&.94(3)&       53&&            2&.20(2)&               0&.684\\                 
&&              &&              1213&&          1&.422(6)&              1&.28&                   \\\cline{11-20}                                                                                  
919&&           0&.427(6)&      489&&           0&.419(3)&              0&.298&                  \\                                                                                               
675&&           0&.172(4)&      489&&           0&.419(3)&              0&.121&                  \multicolumn{10}{c}{$^{226}$Ra source decay\ensuremath{^{l}}}\\                                          
964&.1&         14&.58(2)\ensuremath{^{m}}&  444&&       2&.82(2)\ensuremath{^{n}}&       1&.64&                   \\                                                                                               
&&              &&              564&&           0&.489(6)\ensuremath{^{o}}&      0&.284\ensuremath{^{p}}&      609&&           46&.1(5)&       665&&           1&.46(3)&       1&.43\\                          
1112&&          13&.64(2)&      296&&           0&.447(5)&              0&.340&                  &&              &&              768&&           4&.94(6)&       4&.84\\                          
&&              &&              416&&           0&.110(2)&              0&.083&                  &&              &&              806&&           1&.22(2)&       1&.20\ensuremath{^{q}}\\                   
867&&           4&.246(19)&     296&&           0&.447(5)&              0&.106&                  &&              &&              934&&           3&.03(4)&       2&.97\\                          
344&&           26&.5(3)&       271&&           0&.073(3)&              0&.070\ensuremath{^{q}}&           &&              &&              1120&&          15&.1(2)&       14&.8\\                          
&&              &&              368&&           0&.860(5)&              0&.827&                  &&              &&              1155&&          1&.63(2)&       1&.60\\                          
&&              &&              411&&           2&.232(4)&              2&.15&                   &&              &&              1238&&          5&.79(8)&       5&.67\\                          
&&              &&              586&&           0&.459(5)&              0&.442\ensuremath{^{b}}&    &&              &&              1281&&          1&.43(2)&       1&.30\\                          
&&              &&              679&&           0&.471(4)&              0&.453&                  &&              &&              1402&&          1&.27(2)&       1&.24\\                          
&&              &&              765&&           0&.21(2)&               0&.202&                  &&              &&              1408&&          2&.15(5)&       2&.11\ensuremath{^{q}}\\                   
&&              &&              779&&           12&.93(2)&              12&.4&                   &&              &&              1509&&          2&.11(4)&       2&.07\\                          
&&              &&              1090&&          1&.726(6)&              1&.66&                   &&              &&              1583&&          0&.690(15)&     0&.68\\                          
&&              &&              1299&&          1&.622(8)&              1&.56\\

\pseudoruledtabular
\end{tabular}
\end{center}
}
\begin{widetablenotes}
\ensuremath{^{a}}Radiations are from
$^{152}$Eu($\beta^+$/$\varepsilon$)$^{152}$Sm and
$^{152}$Eu($\beta^-$)$^{152}$Gd.
\\ 
\ensuremath{^{b}}Area measurements for $S_{586}$ and $G_{344:586}$ are
impeded by the Compton edge of the 778\,keV line.
\\ 
\ensuremath{^{c}}Radiations are from $^{60}$Co($\beta^-$)$^{60}$Ni.
\\ 
\ensuremath{^{d}}Radiations are from $^{133}$Ba($\beta^+$/$\varepsilon$)$^{133}$Cs.
\\ 
\ensuremath{^{e}}If 79.6\,keV/81.0\,keV transitions are not resolved, the
measured $S_{81.0}$ must be multiplied by 0.929.
\\ 
\ensuremath{^{f}}If 79.6\,keV/81.0\,keV transitions are not resolved, the
measured $S_{79.6}$ must be multiplied by 0.0714.
\\ 
\ensuremath{^{g}}If 79.6\,keV/81.0\,keV transitions are not resolved, the
measured $G_{81.0:79.6}$ must by multiplied by 1/2.
\\ 
\ensuremath{^{h}}Insufficiently well resolved in a typical gated
spectrum to be of use as a calibrator.
\\ 
\ensuremath{^{i}}If 79.6\,keV/81.0\,keV transitions are not resolved, the
measured $G_{81.0:276}$ must by multiplied by 0.507.
\\ 
\ensuremath{^{j}}The measured $S_{719.4}$ must be multiplied by 0.825 to
eliminate the 719.3\,keV contribution.
\\ 
\ensuremath{^{k}}The measured $G_{245:719.4}$ must be multiplied by 0.96(4) to
eliminate the 719.3\,keV contribution.
\\ 
\ensuremath{^{l}}Radiations are from $^{214}$Bi($\beta^-$)$^{214}$Po.
\\ 
\ensuremath{^{m}}The measured $S_{964.1}$ must be multiplied by 0.991 to
eliminate the 963.4\,keV contribution.
\\ 
\ensuremath{^{n}}The measured $S_{444}$ must be multiplied by 0.896 to
eliminate the doublet contribution.
\\ 
\ensuremath{^{o}}The 564.0\,keV line is only partially resolved from the
566.4\,keV line in singles, rendering intensity measurement difficult.
\\ 
\ensuremath{^{p}}The measured $G_{964.1:564}$ should not be used for
calibration, since there is partially-resolved contamination from the
963.4\,keV and 566\,keV coincident pair of transitions.
\\ 
\ensuremath{^{q}}The coincidence is a $0^+$--$2^+$--$0^+$ cascade, so strong
angular correlation effects are present.
\\ 

\end{widetablenotes}

\chapter{Spectroscopic data for $^{156}$Dy}
\label{app156dy}

\newcommand{\seeepaps}{(See also Table~\ref{tabline}.)}

The $\gamma$-ray intensity data obtained in the experiment of
Chapter~\ref{chap156dy} are summarized in Tables~\ref{tabline}
and~\ref{tabbranch}.

{
\ssp
\footnotesize
\begin{table}[H]
\caption[Observed $\gamma$-ray transitions in $^{156}$Dy.]
{Observed $\gamma$-ray transitions in $^{156}$Dy,
arranged in order of increasing transition energy. Intensities,
normalized to $I_{137}\equiv100$, and 
coincidence relations\ensuremath{^{a}} are given.
\label{tabline}
}
\end{table}
\addtocounter{table}{-1}
\begin{longtable}{r.l_r.l_r.l_r.l_l}
\colrule \vspace{-9pt} \\ \colrule
\multicolumn{2}{c}{$E_\gamma$} &
\multicolumn{2}{c}{$E_i$} & 
\multicolumn{2}{c}{$E_f$} & 
\multicolumn{2}{c}{$I$} &
Coincidences\ensuremath{^{a}}
\\ 
\multicolumn{2}{c}{(keV)} & 
\multicolumn{2}{c}{(keV)} & 
\multicolumn{2}{c}{(keV)} &
\\ 
\colrule
\endhead
\colrule
\endfoot
\colrule \vspace{-9pt} \\ \colrule
\endlastfoot

137&.80(10)& 137&.80& 0&.00& 100&(7)& 266, 366, 684, 691, 764, 884\\
259&.59(15)\ensuremath{^{b}}& 1088&.28& 828&.66& 1&.46(13)& 138, 349, 424, 589, 691, 855, 1156, 1002, 1730\\
266&.38(10)& 404&.18& 137&.80& 127&(6)& 138, 366, 684, 764, 931, 1122, 1223\\
277&.96(18)& 1168&.4& 890&.50& 0&.71(7)& 138, 753, 890, 1174, 1650\\
304&.6(7)\ensuremath{^{c}}\ensuremath{^,}\ensuremath{^{d}}& 2408&.7& 2103&.3& 0&.10(3)& 1081\\
313&.4(2)& 1335&.51& 1022&.10& 0&.66(5)& 138, 618, 884, 1483\\
317&.9(2)& 1088&.28& 770&.40& 0&.27(4)& 266, 366\\
348&.96(14)& 1437&.28& 1088&.28& 1&.41(7)& 138, 266, 260, 684, 951, 1381\\
356&.5(3)\ensuremath{^{b}}& 1525&.3& 1168&.4& 0&.53(5)\ensuremath{^{e}}& 764, 1031\\
360&.7(12)\ensuremath{^{c}}\ensuremath{^,}\ensuremath{^{d}}& 1382&.3& 1022&.10& 0&.11(4)& 884\\
366&.22(12)& 770&.40& 404&.18& 27&.9(16)& 138, 266, 565, 667, 755, 1087, 1128, 2053\\
393&.2(6)\ensuremath{^{b}}& 1728&.7& 1335&.51& 0&.09(4)& 931\\
397&.9(2)\ensuremath{^{b}}& 1168&.4& 770&.40& 0&.21(5)& 266, 366\\
424&.5(2)& 828&.66& 404&.18& 1&.12(6)& 138, 260, 266, 796, 1174, 1416\\
437&.6(6)\ensuremath{^{c}}\ensuremath{^,}\ensuremath{^{d}}& 1526&.0& 1088&.28& 0&.08(6)\ensuremath{^{e}}& 684\\
445&.23(17)\ensuremath{^{b}}& 1215&.6& 770&.40& 0&.37(3)& 138, 266, 366, 595, 1572\\
456&.2(8)\ensuremath{^{d}}& 1624&.6& 1168&.4& 0&.09(3)& 138, 266, 764, 1031\\
458&.9(4)\ensuremath{^{d}}& 1627&.5& 1168&.4& 0&.20(6)& 138, 266, 764, 1031\\
486&.4(3)& 890&.50& 404&.18& 0&.50(8)& 138, 266, 1416\\
491&.6(3)\ensuremath{^{d}}& 1382&.3& 890&.50& 0&.23(6)& 138, 753, 890\\
537&.8(2)& 675&.6& 137&.80& 0&.86(12)& 138, 707, 839\\
553&.7(2)\ensuremath{^{d}}& 1382&.3& 828&.66& 0&.28(3)& 138, 691\\
562&.6(5)\ensuremath{^{d}}& 1898&.5& 1335&.51& 0&.12(5)& 931\\
565&.07(17)& 1335&.51& 770&.40& 1&.15(6)& 138, 266, 366, 768, 858, 1483\\
585&.6(2)\ensuremath{^{d}}& 1476&.10& 890&.50& 0&.35(7)& 138, 753, 890\\
588&.88(14)\ensuremath{^{d}}& 1677&.2& 1088&.28& 0&.52(4)& 138, 260, 266, 684, 691\\
594&.9(6)\ensuremath{^{b}}& 1809&.8& 1215&.6& 0&.047(13)& 445\\
605&.5(3)& 1627&.5& 1022&.10& 0&.36(7)& 138, 618, 884\\
617&.88(12)& 1022&.10& 404&.18& 3&.6(4)& 138, 266, 655, 818, 1081, 1277, 1301, 1386\\
620&.1(8)\ensuremath{^{d}}& 2244&.7& 1624&.6& 0&.10(3)& 796\\
624&.4(3)\ensuremath{^{d}}& 1515&.0& 890&.50& 0&.11(5)& 753, 890\\
654&.9(4)\ensuremath{^{d}}& 1677&.2& 1022&.10& 0&.33(9)& 138, 266, 618, 884\\
666&.88(15)& 1437&.28& 770&.40& 1&.92(10)& 138, 266, 366, 1381\\
671&.2(2)\ensuremath{^{d}}& 1840&.1& 1168&.4& 0&.18(4)& 138, 266, 764, 1031\\
684&.10(10)& 1088&.28& 404&.18& 13&.3(9)& 138, 266, 349, 589, 855, 1002, 1156, 1219, 1235, 1730\\
688&.9(5)\ensuremath{^{c}}\ensuremath{^,}\ensuremath{^{d}}& 1857&.82& 1168&.4& 0&.15(9)& 764, 1031\\
690&.86(13)& 828&.66& 137&.80& 10&.4(5)& 138, 260, 349, 554, 796, 1174, 1416, 1479, 1494, 1825\\
706&.74(16)\ensuremath{^{d}}& 1382&.3& 675&.6& 0&.14(2)& 138, 538\\
722&.3(7)\ensuremath{^{d}}& 2058&.6& 1335&.51& 0&.13(4)& 931\\
723&.5(4)\ensuremath{^{d}}& 2199&.7& 1476&.10& 0&.14(4)& 138, 1338\\
752&.67(15)& 890&.50& 137&.80& 3&.3(3)& 138, 278, 492, 586, 950, 1417, 1433, 1519\\
754&.9(2)\ensuremath{^{b}}& 1525&.3& 770&.40& 1&.75(11)\ensuremath{^{e}}& 138, 266, 366\\
764&.12(13)& 1168&.4& 404&.18& 9&.0(5)& 138, 266, 356, 1076, 1139, 1155, 1174, 1650\\
767&.8(4)\ensuremath{^{d}}& 2103&.3& 1335&.51& 0&.16(4)& 565, 931\\
786&.1(5)\ensuremath{^{c}}\ensuremath{^,}\ensuremath{^{d}}& 1677&.2& 890&.50& 0&.10(3)& 890\\
796&.03(15)\ensuremath{^{d}}& 1624&.6& 828&.66& 0&.98(6)& 138, 424, 620, 691\\
818&.1(2)\ensuremath{^{d}}& 1840&.1& 1022&.10& 0&.26(6)& 138, 884\\
818&.7(4)\ensuremath{^{c}}\ensuremath{^,}\ensuremath{^{d}}& 2445&.2& 1627&.5& 0&.19(5)& 1223\\
820&.9(6)\ensuremath{^{c}}\ensuremath{^,}\ensuremath{^{d}}& 2445&.2& 1624&.6& 0&.08(2)& 796\\
829&.4(7)& && && 0&.13(4)\ensuremath{^{f}}& 1122\\
839&.3(2)\ensuremath{^{d}}& 1515&.0& 675&.6& 0&.20(2)& 138, 538\\
845&.3(3)\ensuremath{^{d}}& 1933&.6& 1088&.28& 0&.11(2)& 684\\
848&.2(5)\ensuremath{^{d}}& 1677&.2& 828&.66& 0&.12(5)& 138, 691\\
851&.0(12)\ensuremath{^{c}}\ensuremath{^,}\ensuremath{^{d}}& 1679&.9& 828&.66& 0&.07(4)& 138, 691\\
854&.6(3)\ensuremath{^{d}}& 1942&.9& 1088&.28& 0&.34(5)& 138, 260, 266, 684\\
858&.0(3)\ensuremath{^{d}}& 2193&.5& 1335&.51& 0&.34(5)& 138, 266, 931\\
863&.3(10)\ensuremath{^{d}}& 2199&.7& 1335&.51& 0&.10(4)& 931\\
871&.6(5)\ensuremath{^{d}}& 2207&.4& 1335&.51& 0&.18(5)& 931\\
884&.30(10)& 1022&.10& 137&.80& 16&.4(16)& 138, 313, 655, 819, 1081, 1278, 1301, 1386\\
884&.3(8)\ensuremath{^{d}}& 2818&.4& 1933&.6& 0&.11(5)& 1529\\
890&.2(4)\ensuremath{^{d}}& 2058&.6& 1168&.4& 0&.27(10)& 138, 266, 764, 1031\\
890&.44(12)& 890&.50& 0&.00& 5&.9(9)& 277, 492, 586, 950, 1417, 1433, 1519\\
907&.2(4)\ensuremath{^{d}}& 1677&.2& 770&.40& 0&.14(5)& 138, 266, 366\\
908&.0(10)\ensuremath{^{c}}\ensuremath{^,}\ensuremath{^{d}}& 2433&.8& 1526&.0& 0&.19(6)& 1122\\
911&.5(6)\ensuremath{^{d}}& 1933&.6& 1022&.10& 0&.15(4)& 138, 884\\
914&.6(3)\ensuremath{^{d}}& 2003&.0& 1088&.28& 0&.14(5)& 138, 266, 684\\
919&.7(15)\ensuremath{^{d}}& 2818&.4& 1898&.5& 0&.13(5)& 366, 1128\\
921&.2(3)& 2089&.9& 1168&.4& 0&.26(6)& 138, 266, 764, 1031\\
931&.35(16)& 1335&.51& 404&.18& 7&.2(4)& 138, 266, 768, 858, 872, 1111, 1483\\
935&.0(4)\ensuremath{^{d}}& 2103&.3& 1168&.4& 0&.19(6)& 764, 1031\\
939&.2(11)\ensuremath{^{d}}& 2307&.4& 1368&.53& 0&.17(6)& 964, 1231\\
944&.3(4)\ensuremath{^{d}}& 2572&.0& 1627&.5& 0&.15(3)& 266, 1223\\
949&.60(16)\ensuremath{^{d}}& 1840&.1& 890&.50& 0&.71(5)& 138, 753, 890\\
950&.5(2)& 1088&.28& 137&.80& 1&.2(2)& 138, 349, 1730\\
955&.4(4)\ensuremath{^{d}}& 2323&.6& 1368&.53& 0&.19(4)& 964, 1231\\
958&.3(8)\ensuremath{^{b}}& 1728&.7& 770&.40& 0&.22(7)\ensuremath{^{g}}& 366\\
960&.6(3)\ensuremath{^{d}}& 2818&.4& 1857&.82& 0&.69(7)& 266, 366, 1087, 1454\\
964&.36(18)& 1368&.53& 404&.18& 1&.51(12)& 138, 266, 939\\
965&.3(8)\ensuremath{^{d}}& 2823&.3& 1857&.82& 0&.10(5)& 1454\\
970&.4(18)\ensuremath{^{c}}\ensuremath{^,}\ensuremath{^{d}}& 2058&.6& 1088&.28& 0&.06(4)& 684\\
988&.7(5)\ensuremath{^{c}}\ensuremath{^,}\ensuremath{^{d}}& 1878&.8& 890&.50& 0&.14(3)& 890\\
996&.1(4)\ensuremath{^{d}}& 2331&.7& 1335&.51& 0&.14(5)& 266, 931\\
1001&.7(3)\ensuremath{^{d}}& 2089&.9& 1088&.28& 0&.43(6)& 138, 260, 266, 684\\
1011&.7(2)\ensuremath{^{d}}& 1840&.1& 828&.66& 0&.10(3)& 138, 691\\
1024&.6(6)\ensuremath{^{d}}& 1794&.6& 770&.40& 0&.12(5)& 138, 266, 366\\
1030&.7(2)& 1168&.4& 137&.80& 7&.7(4)& 138, 356, 1076, 1139, 1155, 1174, 1650\\
1031&.8(8)\ensuremath{^{d}}& 2199&.7& 1168&.4& 0&.11(3)\ensuremath{^{g}}& 1031\\
1033&.2(3)\ensuremath{^{b}}& 1437&.28& 404&.18& 0&.65(13)& 266, 1381\\
1036&.4(2)\ensuremath{^{d}}& 2058&.6& 1022&.10& 0&.32(6)& 138, 618, 884\\
1038&.6(8)& && && 0&.12(4)\ensuremath{^{f}}& 931\\
1039&.3(2)\ensuremath{^{b}}& 1809&.8& 770&.40& 0&.33(5)& 138, 266, 366\\
1040&.0(7)\ensuremath{^{d}}& 2408&.7& 1368&.53& 0&.11(4)& 1231\\
1049&.6(15)\ensuremath{^{c}}\ensuremath{^,}\ensuremath{^{d}}& 1878&.8& 828&.66& 0&.12(5)& 691\\
1050&.0(5)\ensuremath{^{d}}& 2385&.6& 1335&.51& 0&.11(3)& 931\\
1050&.3(7)& && && 0&.07(3)\ensuremath{^{f}}& 1223\\
1076&.2(5)\ensuremath{^{d}}& 2244&.7& 1168&.4& 0&.42(8)& 764, 1031\\
1081&.2(4)\ensuremath{^{d}}& 2103&.3& 1022&.10& 0&.64(5)& 138, 266, 618, 884\\
1087&.40(16)\ensuremath{^{d}}& 1857&.82& 770&.40& 0&.62(4)& 138, 266, 366, 961\\
1094&.8(10)\ensuremath{^{c}}\ensuremath{^,}\ensuremath{^{d}}& 2264&.1& 1168&.4& 0&.15(5)& 764, 1031\\
1095&.9(5)\ensuremath{^{c}}\ensuremath{^,}\ensuremath{^{d}}& 2183&.8& 1088&.28& 0&.10(6)& 684\\
1110&.7(7)\ensuremath{^{d}}\ensuremath{^,}\ensuremath{^{h}}& 2445&.2& 1335&.51& 0&.29(6)& 931\\
1111&.2(6)\ensuremath{^{d}}& 1515&.0& 404&.18& 0&.53(13)\ensuremath{^{g}}& 138, 266\\
1121&.8(2)& 1526&.0& 404&.18& 8&.2(8)\ensuremath{^{e}}& 138, 266\\
1128&.07(15)\ensuremath{^{b}}& 1898&.5& 770&.40& 0&.89(5)& 138, 266, 366, 920\\
1137&.0(5)& && && 0&.21(7)\ensuremath{^{f}}& 890\\
1139&.0(6)\ensuremath{^{d}}& 2307&.4& 1168&.4& 0&.32(9)& 266, 764, 1031\\
1154&.4(8)\ensuremath{^{c}}\ensuremath{^,}\ensuremath{^{d}}& 2489&.5& 1335&.51& 0&.14(6)& 931\\
1155&.3(2)\ensuremath{^{d}}& 2323&.6& 1168&.4& 1&.26(9)& 764, 1031\\
1156&.4(3)\ensuremath{^{d}}& 2244&.7& 1088&.28& 0&.32(7)& 260, 684\\
1163&.1(6)\ensuremath{^{c}}\ensuremath{^,}\ensuremath{^{d}}& 2331&.7& 1168&.4& 0&.10(5)& 764, 1031\\
1172&.5(16)\ensuremath{^{c}}\ensuremath{^,}\ensuremath{^{d}}& 1942&.9& 770&.40& 0&.20(6)& 138, 266, 366\\
1174&.2(2)\ensuremath{^{d}}& 2342&.6& 1168&.4& 0&.42(7)& 764, 1031\\
1174&.5(8)\ensuremath{^{d}}& 2003&.0& 828&.66& 0&.22(8)& 138, 691\\
1176&.8(8)& && && 0&.10(4)\ensuremath{^{f}}& 1231\\
1177&.6(2)\ensuremath{^{d}}& 2199&.7& 1022&.10& 0&.29(5)& 138, 618, 884\\
1179&.4(8)& && && 0&.27(8)\ensuremath{^{f}}& 931\\
1185&.6(5)\ensuremath{^{d}}& 2207&.4& 1022&.10& 0&.22(4)& 884\\
1191&.1(5)& 2818&.4& 1627&.5& 0&.43(6)& 138, 266, 1223\\
1199&.0(3)& && && 0&.20(5)\ensuremath{^{f}}\ensuremath{^,}\ensuremath{^{i}}& 931, (1231)\\
1205&.2(2)& 1609&.4& 404&.18& 1&.27(11)& 138, 266\\
1217&.2(3)\ensuremath{^{d}}& 2385&.6& 1168&.4& 0&.25(7)& 138, 266, 764, 1031\\
1218&.9(5)\ensuremath{^{d}}& 2307&.4& 1088&.28& 0&.39(10)& 266, 684\\
1222&.8(3)\ensuremath{^{d}}& 2244&.7& 1022&.10& 0&.37(8)& 618, 884\\
1223&.36(18)& 1627&.5& 404&.18& 5&.6(4)& 138, 266, 944, 1191\\
1230&.72(14)& 1368&.53& 137&.80& 5&.3(5)& 138, 939, 955, 1450\\
1235&.3(2)\ensuremath{^{d}}& 2323&.6& 1088&.28& 0&.43(9)& 138, 266, 260, 684, 951\\
1241&.2(6)\ensuremath{^{c}}\ensuremath{^,}\ensuremath{^{d}}& 2264&.1& 1022&.10& 0&.15(6)& 884\\
1241&.3(12)\ensuremath{^{c}}\ensuremath{^,}\ensuremath{^{d}}& 2408&.7& 1168&.4& 0&.14(6)& 764, 1031\\
1245&.7(8)& && && 0&.13(5)\ensuremath{^{f}}& 1122\\
1259&.1(7)\ensuremath{^{d}}& 2594&.3& 1335&.51& 0&.19(8)& 931\\
1272&.8(3)\ensuremath{^{d}}& 1677&.2& 404&.18& 0&.32(8)& 138, 266\\
1276&.8(5)& && && 0&.12(4)\ensuremath{^{f}}& 266, 366\\
1278&.0(3)\ensuremath{^{d}}& 2300&.1& 1022&.10& 0&.52(14)& 138, 618, 884\\
1285&.4(4)\ensuremath{^{d}}& 2307&.4& 1022&.10& 0&.18(7)& 884\\
1289&.1(8)& && && 0&.16(6)\ensuremath{^{f}}& 884\\
1292&.3(3)\ensuremath{^{d}}& 2818&.4& 1526&.0& 0&.87(11)& 1122\\
1293&.0(5)\ensuremath{^{c}}\ensuremath{^,}\ensuremath{^{d}}& 2183&.8& 890&.50& 0&.14(8)& 890\\
1293&.4(15)\ensuremath{^{d}}& 2818&.4& 1525&.3& 0&.27(4)& 356, 366, 755\\
1297&.3(2)\ensuremath{^{d}}& 2823&.3& 1526&.0& 0&.33(8)& 1122\\
1301&.5(4)\ensuremath{^{d}}& 2323&.6& 1022&.10& 2&.58(14)& 138, 266, 618, 884\\
1309&.7(4)\ensuremath{^{d}}& 2331&.7& 1022&.10& 0&.37(8)& 618, 884\\
1314&.7(2)\ensuremath{^{d}}& 2085&.1& 770&.40& 0&.52(5)& 138, 266, 366\\
1320&.3(15)& 2408&.7& 1088&.28& 0&.12(5)& 138, 266, 684\\
1323&.2(4)\ensuremath{^{d}}& 2492&.0& 1168&.4& 0&.17(5)& 138, 266, 764, 1031\\
1338&.31(17)\ensuremath{^{d}}& 1476&.10& 137&.80& 1&.11(11)& 138, 723\\
1345&.6(3)\ensuremath{^{d}}& 2433&.8& 1088&.28& 0&.19(5)& 138, 266, 684\\
1348&.9(5)& 2516&.6& 1168&.4& 0&.19(5)& 764, 1031\\
1351&.3(6)\ensuremath{^{c}}\ensuremath{^,}\ensuremath{^{d}}& 2439&.2& 1088&.28& 0&.10(4)& 684\\
1354&.1(2)\ensuremath{^{d}}& 2244&.7& 890&.50& 0&.41(5)& 138, 753, 890\\
1355&.1(4)\ensuremath{^{d}}& 2183&.8& 828&.66& 0&.21(5)& 691\\
1363&.4(7)\ensuremath{^{c}}\ensuremath{^,}\ensuremath{^{d}}& 2385&.6& 1022&.10& 0&.08(3)& 884\\
1380&.9(2)\ensuremath{^{d}}& 2818&.4& 1437&.28& 0&.65(6)& 138, 266, 349, 366, 667, 684, 1033\\
1386&.3(2)& 2408&.7& 1022&.10& 0&.67(6)& 138, 266, 618, 884\\
1390&.33(17)\ensuremath{^{d}}& 1794&.6& 404&.18& 2&.07(12)& 138, 266\\
1393&(2)& && && 0&.13(5)\ensuremath{^{f}}\ensuremath{^,}\ensuremath{^{i}}& 684, 753, 890\\
1393&.9(7)\ensuremath{^{c}}\ensuremath{^,}\ensuremath{^{d}}& 2164&.3& 770&.40& 0&.09(4)& 366\\
1415&.9(2)\ensuremath{^{d}}& 2244&.7& 828&.66& 1&.50(9)& 138, 266, 424, 691\\
1416&.8(2)& 2307&.4& 890&.50& 0&.92(10)& 138, 486, 753, 890\\
1421&.2(6)& && && 0&.12(4)\ensuremath{^{f}}& 366\\
1423&.0(2)\ensuremath{^{d}}& 2445&.2& 1022&.10& 0&.68(9)& 138, 618, 884\\
1423&.3(6)\ensuremath{^{d}}& 2193&.5& 770&.40& 0&.16(6)& 366\\
1425&.9(4)\ensuremath{^{d}}& 2594&.3& 1168&.4& 0&.20(5)& 764, 1031\\
1432&.8(2)\ensuremath{^{d}}& 2323&.6& 890&.50& 1&.00(10)& 138, 486, 753, 890\\
1435&.7(5)\ensuremath{^{d}}& 1840&.1& 404&.18& 0&.47(9)& 138, 266\\
1450&.0(3)\ensuremath{^{d}}& 2220&.4& 770&.40& 0&.22(6)\ensuremath{^{g}}& 138, 266, 366\\
1450&.0(8)\ensuremath{^{c}}\ensuremath{^,}\ensuremath{^{d}}& 2818&.4& 1368&.53& 0&.15(6)& 1231\\
1453&.65(15)\ensuremath{^{d}}& 1857&.82& 404&.18& 2&.5(3)& 138, 266, 961\\
1460&.5(3)\ensuremath{^{d}}& 2230&.9& 770&.40& 0&.22(4)& 366\\
1467&.1(8)\ensuremath{^{d}}& 2489&.5& 1022&.10& 0&.10(5)& 138, 884\\
1469&.9(5)\ensuremath{^{d}}& 2492&.0& 1022&.10& 0&.19(6)& 138, 884\\
1471&.5(2)& 1609&.4& 137&.80& 2&.5(3)\ensuremath{^{g}}& 138\\
1471&.9(6)& && && 0&.22(13)\ensuremath{^{f}}& 266\\
1474&.2(4)\ensuremath{^{d}}& 1878&.8& 404&.18& 0&.56(14)& 138, 266\\
1478&.7(2)& 2307&.4& 828&.66& 0&.28(3)& 138, 424, 691\\
1480&.6(7)& && && 0&.11(3)\ensuremath{^{f}}& 1231\\
1482&.7(2)\ensuremath{^{d}}& 2818&.4& 1335&.51& 0&.30(5)& 138, 266, 931\\
1486&.4(7)\ensuremath{^{d}}& 1624&.6& 137&.80& 0&.54(16)& 138\\
1493&.8(10)& 2516&.6& 1022&.10& 0&.20(5)& 884\\
1494&.5(5)\ensuremath{^{d}}& 2323&.6& 828&.66& 0&.29(7)& 138, 691\\
1499&.6(3)\ensuremath{^{d}}& 2270&.0& 770&.40& 0&.62(9)& 138, 266, 366\\
1518&.3(7)& && && 0&.38(12)\ensuremath{^{f}}\ensuremath{^,}\ensuremath{^{g}}& 138\\
1518&.7(3)\ensuremath{^{d}}& 2408&.7& 890&.50& 0&.25(7)& 753, 890\\
1523&.0(3)\ensuremath{^{d}}& 2293&.4& 770&.40& 0&.38(6)& 138, 266, 366\\
1526&.1(6)\ensuremath{^{d}}& 1930&.0& 404&.18& 0&.64(16)& 138, 266\\
1529&.4(2)\ensuremath{^{d}}& 1933&.6& 404&.18& 1&.52(13)& 138, 266, 884\\
1536&.0(4)& 2307&.4& 770&.40& 0&.49(8)& 138, 266, 366\\
1538&.0(12)\ensuremath{^{c}}\ensuremath{^,}\ensuremath{^{d}}& 1942&.9& 404&.18& 0&.41(13)& 138, 266\\
1542&.1(8)\ensuremath{^{d}}& 1679&.9& 137&.80& 0&.80(16)\ensuremath{^{h}}& 138\\
1545&.8(2)\ensuremath{^{d}}& 1950&.0& 404&.18& 1&.44(8)& 138, 266\\
1572&.0(5)\ensuremath{^{d}}& 2594&.3& 1022&.10& 0&.13(5)& 884\\
1572&.5(8)\ensuremath{^{d}}& 2788&.1& 1215&.6& 0&.08(2)& 366, 445\\
1580&.3(4)\ensuremath{^{d}}& 2408&.7& 828&.66& 0&.11(3)& 138, 691\\
1598&.7(5)\ensuremath{^{d}}& 2003&.0& 404&.18& 0&.25(7)& 138, 266\\
1626&.8(6)\ensuremath{^{c}}& 2516&.6& 890&.50& 0&.16(6)& 890\\
1633&.5(10)& && && 0&.18(5)\ensuremath{^{f}}& 366\\
1634&.6(10)\ensuremath{^{d}}& 1772&.4& 137&.80& 1&.1(3)\ensuremath{^{g}}& 138\\
1642&.0(10)& && && 0&.36(17)\ensuremath{^{f}}& 138, 266\\
1648&.1(7)\ensuremath{^{c}}\ensuremath{^,}\ensuremath{^{d}}& 2418&.9& 770&.40& 0&.19(6)& 366\\
1649&.7(2)& 2818&.4& 1168&.4& 1&.37(11)& 138, 266, 278, 764, 1031\\
1653&.4(13)& && && 0&.13(4)\ensuremath{^{f}}& 691\\
1654&.0(11)\ensuremath{^{c}}\ensuremath{^,}\ensuremath{^{d}}& 2823&.3& 1168&.4& 0&.14(6)& 764, 1031\\
1654&(2)& && && 0&.13(5)\ensuremath{^{f}}& 366\\
1658&(2)& && && 0&.13(5)\ensuremath{^{f}}& 366\\
1663&.3(2)\ensuremath{^{d}}& 2433&.8& 770&.40& 0&.52(10)& 138, 266, 366\\
1668&.7(2)\ensuremath{^{d}}& 2439&.2& 770&.40& 0&.32(7)& 138, 266, 366\\
1688&.2(15)\ensuremath{^{c}}& 2516&.6& 828&.66& 0&.07(5)& 691\\
1704&(2)& && && 0&.09(3)\ensuremath{^{f}}& 366\\
1730&.1(2)& 2818&.4& 1088&.28& 0&.57(6)& 138, 260, 266, 684, 951\\
1733&.4(7)& && && 0&.22(5)\ensuremath{^{f}}& 366\\
1734&.3(5)& && && 0&.7(2)\ensuremath{^{f}}\ensuremath{^,}\ensuremath{^{g}}& 138, 266\\
1735&.7(5)\ensuremath{^{d}}& 2757&.8& 1022&.10& 0&.18(5)& 884\\
1741&.5(7)\ensuremath{^{d}}& 1878&.8& 137&.80& 0&.36(9)& 138\\
1757&.8(4)& && && 0&.37(9)\ensuremath{^{f}}\ensuremath{^,}\ensuremath{^{i}}& 138, 266, (366)\\
1760&.1(4)\ensuremath{^{d}}& 2164&.3& 404&.18& 0&.31(9)& 138, 266\\
1791&.9(9)\ensuremath{^{d}}& 1930&.0& 137&.80& 0&.50(18)& 138\\
1795&.6(5)\ensuremath{^{d}}& 2199&.7& 404&.18& 0&.42(15)& 266\\
1824&.7(5)\ensuremath{^{d}}& 2228&.9& 404&.18& 0&.63(9)& 138, 266\\
1824&.7(6)\ensuremath{^{d}}& 2653&.4& 828&.66& 0&.20(5)& 424, 691\\
1840&.5(8)\ensuremath{^{c}}\ensuremath{^,}\ensuremath{^{d}}& 2244&.7& 404&.18& 0&.22(9)& 138, 266\\
1843&.6(7)& && && 0&.07(3)\ensuremath{^{f}}& 884\\
1860&.1(5)\ensuremath{^{d}}& 2264&.1& 404&.18& 0&.81(13)& 138, 266\\
1867&.6(5)& && && 0&.30(8)\ensuremath{^{f}}& 138, 266\\
1872&.9(4)\ensuremath{^{d}}& 2894&.9& 1022&.10& 0&.21(5)& 138, 618, 884\\
1888&.8(15)\ensuremath{^{c}}\ensuremath{^,}\ensuremath{^{d}}& 2293&.4& 404&.18& 0&.27(10)& 266\\
1898&.8(8)& && && 0&.32(8)\ensuremath{^{f}}& 138, 266\\
1902&.5(5)& 2307&.4& 404&.18& 0&.42(10)& 138, 266\\
1919&.8(4)\ensuremath{^{d}}& 2323&.6& 404&.18& 0&.61(13)& 138, 266\\
1952&.3(9)\ensuremath{^{c}}& 2089&.9& 137&.80& 0&.24(10)& 138\\
1959&.1(9)& && && 0&.09(4)\ensuremath{^{f}}& 366\\
1959&.7(6)& && && 0&.25(8)\ensuremath{^{f}}\ensuremath{^,}\ensuremath{^{g}}& 266\\
1961&.8(15)& && && 0&.05(3)\ensuremath{^{f}}& 366\\
1967&.9(3)\ensuremath{^{d}}& 2372&.1& 404&.18& 0&.59(16)& 138, 266\\
1989&.2(12)& && && 0&.39(16)\ensuremath{^{f}}\ensuremath{^,}\ensuremath{^{i}}& 266, (366)\\
2003&.7(7)\ensuremath{^{c}}& 2408&.7& 404&.18& 0&.34(10)& 138, 266\\
2004&.2(9)\ensuremath{^{c}}\ensuremath{^,}\ensuremath{^{d}}& 2894&.9& 890&.50& 0&.10(4)& 753, 890\\
2008&.9(5)& && && 0&.26(8)\ensuremath{^{f}}& 138, 266\\
2014&.9(6)\ensuremath{^{d}}& 2418&.9& 404&.18& 0&.42(10)& 138, 266\\
2029&.70(18)\ensuremath{^{d}}& 2433&.8& 404&.18& 2&.17(16)& 138, 266\\
2035&.0(2)\ensuremath{^{d}}& 2439&.2& 404&.18& 1&.7(2)& 138, 266\\
2039&.9(10)\ensuremath{^{c}}\ensuremath{^,}\ensuremath{^{d}}& 2810&.4& 770&.40& 0&.11(4)& 266, 366\\
2041&.1(10)& && && 0&.32(10)\ensuremath{^{f}}& 138\\
2048&.0(2)\ensuremath{^{d}}& 2818&.4& 770&.40& 0&.19(6)& 266, 366\\
2052&.8(2)\ensuremath{^{d}}& 2823&.3& 770&.40& 0&.69(11)& 266, 366\\
2063&.2(4)\ensuremath{^{d}}& 2833&.6& 770&.40& 0&.23(4)& 266, 366\\
2064&.6(8)& && && 0&.13(8)\ensuremath{^{f}}& 138\\
2085&.4(5)\ensuremath{^{d}}& 2489&.5& 404&.18& 0&.49(10)& 138, 266\\
2088&.2(6)\ensuremath{^{d}}& 2492&.0& 404&.18& 0&.37(15)& 138, 266\\
2134&.1(5)& && && 0&.09(3)\ensuremath{^{f}}& 691\\
2135&.6(15)\ensuremath{^{h}}& && && 0&.45(12)\ensuremath{^{f}}& 266\\
2168&.9(7)\ensuremath{^{c}}\ensuremath{^,}\ensuremath{^{d}}& 2572&.0& 404&.18& 0&.23(8)& 266\\
2185&.6(6)\ensuremath{^{d}}& 2323&.6& 137&.80& 0&.31(10)& 138\\
2234&.2(4)\ensuremath{^{d}}& 2372&.1& 137&.80& 1&.7(4)& 138\\
2238&.3(2)\ensuremath{^{d}}& 2642&.5& 404&.18& 0&.77(13)& 138, 266\\
2249&(2)\ensuremath{^{c}}\ensuremath{^,}\ensuremath{^{d}}& 2653&.4& 404&.18& 0&.32(15)& 138, 266\\
2271&.0(2)\ensuremath{^{d}}& 2408&.7& 137&.80& 0&.88(13)& 138\\
2277&.4(4)& && && 0&.42(13)\ensuremath{^{f}}& 138\\
2286&.3(10)& && && 0&.26(9)\ensuremath{^{f}}& 266\\
2303&.9(6)& && && 0&.22(8)\ensuremath{^{f}}& 138\\
2306&.7(15)& && && 0&.33(11)\ensuremath{^{f}}& 266\\
2307&.4(8)\ensuremath{^{d}}& 2445&.2& 137&.80& 0&.27(11)& 138\\
2339&.2(9)& && && 0&.18(7)\ensuremath{^{f}}& 266\\
2354&.1(2)\ensuremath{^{d}}& 2492&.0& 137&.80& 0&.90(8)& 138\\
2406&.2(7)\ensuremath{^{d}}& 2810&.4& 404&.18& 0&.38(11)& 266\\
2414&.2(2)\ensuremath{^{d}}& 2818&.4& 404&.18& 1&.60(18)& 138, 266\\
2419&.2(2)& 2823&.3& 404&.18& 3&.3(3)& 138, 266\\
2429&.5(7)\ensuremath{^{d}}& 2833&.6& 404&.18& 0&.63(9)& 138, 266\\
2481&.4(6)& && && 0&.18(5)\ensuremath{^{f}}& 266\\
2490&.7(6)\ensuremath{^{d}}& 2894&.9& 404&.18& 0&.21(7)& 138, 266\\
2577&.3(13)\ensuremath{^{d}}& 2981&.5& 404&.18& 0&.33(7)& 138, 266\\
2579&.8(10)& && && 0&.08(3)\ensuremath{^{f}}& 366\\
2590&.8(7)& && && 0&.06(2)\ensuremath{^{f}}& 366\\
2613&.4(9)& && && 0&.09(2)\ensuremath{^{f}}& 366\\

\end{longtable}
\begin{flushleft}
\footnotesize
\ensuremath{^{a}}In the entries for many of the less intense higher-lying
transitions, coincidences with the low-lying 138\,keV and 266\,keV
transitions are not explicitly listed.  This indicates that the
corresponding region of the spectrum gated on the 138\,keV or 226\,keV
transition is sufficiently complicated~--- either dominated by a more
intense close doublet transition or obscured by multiple smaller
background peaks~--- that coincidence with 138\,keV or 226\,keV
transition cannot be unambiguously confirmed and therefore does not
provide useful additional placement or intensity information.  This
should \textit{not} be construed as indicating
\textit{noncoincidence}.  Also, to make efficient use of space,
coincidences with weaker feeding transitions are omitted in the
entries for low-lying transitions coincident with a large number of
feeding transitions and are included only in the entries for these
weaker feeding transitions.
\\ 
\ensuremath{^{b}}Transition was previously reported, but not in $\beta$
decay.
\\ 
\ensuremath{^{c}}Identification of transition is tentative.
\\ 
\ensuremath{^{d}}Gamma-ray line was not previously reported or was not
reported in this placement.
\\ 
\ensuremath{^{e}}Transitions from the closely-spaced pair of levels at
1525.3(2) and 1526.0(2)\,keV are all potentially doublets.  Each
transition is assigned a primary placement as depopulating one of
these levels on the basis of transition energy as measured in gated
spectra but may contain a significant unresolved contribution
depopulating the other member of the pair.  (See text.)
\\ 
\ensuremath{^{f}}Transition was observed in a gated spectrum, but its
placement is unknown, and its absolute intensity cannot therefore be
uniquely determined from the data.  The intensity value reported here
is an estimate, calculated from the intensity of coincidences between
this transition and the other coincident transition, corrected for the
branching fraction of the coincident transition.
\\ 
\ensuremath{^{g}}Energy and intensity are deduced from gated spectrum after
subtraction of contribution(s) from other placement(s).  \seeepaps
\\ 
\ensuremath{^{h}}Peak in gated spectrum has abnormally large width.
\\ 
\ensuremath{^{i}}Multiple gamma-ray transitions may be present on the
basis of the coincidences observed or their relative intensities, but
the transitions have been consolidated into one entry.  Any
coincidence significantly weaker than the others is enclosed in
parentheses.
\\ 

\end{flushleft}

\clearpage
\fontsize{8}{10}\selectfont
\begin{table}[H]
\caption[Branching properties of levels populated in
$^{156}$Dy.]
{
Branching properties of levels populated in
$^{156}$Dy.  Both absolute (in $\beta$ decay) and relative intensities
are given for $\gamma$-ray transitions depopulating the levels, and
these intensities are compared with literature
values~\cite{ndsboth:156} where available.  Intensity limits are given
for many unobserved transitions, in which case the approximate
transition energy expected from the level energy difference is shown
in brackets.  (Transitions on which limits were placed include
spin-allowed but unobserved transitions between low-lying levels
relevant to the structural interpretation of the nucleus and
presently-unobserved transitions which were reported to have been
observed in the prior literature.) 
\label{tabbranch}
} 
\end{table}
\addtocounter{table}{-1}
\begin{longtable}{l,r.l__l,r.l_c_r.l_r.l_r.l_c_r.l_r.l}
\colrule \vspace{-9pt} \\ \colrule
\multicolumn{6}{c}{Transition}&&\multicolumn{6}{c}{Experiment}&&\multicolumn{4}{c}{Literature}\\ 
\cline{1-6}\cline{8-13}\cline{15-18} 
\multicolumn{1}{c}{$J^\pi_i$~\ensuremath{^{a}}\hspace{-1.5em}} &
\multicolumn{2}{c}{$E_i$} & 
\multicolumn{1}{c}{$J^\pi_f$} & 
\multicolumn{2}{c}{$E_f$} & 
\multicolumn{1}{c}{} & 
\multicolumn{2}{c}{$E_\gamma$} &
\multicolumn{2}{c}{$I$} &
\multicolumn{2}{c}{$I^\text{rel}$} & 
\multicolumn{1}{c}{} & 
\multicolumn{2}{c}{$I$~\ensuremath{^{b}}} &
\multicolumn{2}{c}{$I^\text{rel}$~\ensuremath{^{c}}} 
\\ 
\multicolumn{1}{c}{} &
\multicolumn{2}{c}{(keV)} & 
\multicolumn{1}{c}{} & 
\multicolumn{2}{c}{(keV)} & 
&
\multicolumn{2}{c}{(keV)} &
\multicolumn{2}{c}{} &
\multicolumn{2}{c}{} & 
\\ 
\colrule
\endhead
\colrule
\endfoot
\colrule \vspace{-9pt} \\ \colrule
\endlastfoot

$2^+$& 137&.80(10)& $0^+$& 0&.00& & 137&.80(10)& 100&(7)& 100&(7)& & 100&& ~~~~~~100&\\

 \colrule 
$4^+$& 404&.18(14)& $2^+$& 137&.80& & 266&.38(10)& 127&(6)& 100&(5)& & 107&.1(11)& 100&\\

 \colrule 
$0^+$& 675&.6(2)& $2^+$& 137&.80& & 537&.8(2)& 0&.86(12)& 100&(15)& & 0&.53(7)& 100&\\

 \colrule 
$6^+$& 770&.40(17)& $4^+$& 404&.18& & 366&.22(12)& 27&.9(16)& 100&(6)& & 21&.0(2)& 100&\\

 \colrule 
$2^+$& 828&.66(15)& $0^+$& 0&.00& & [829&]& \lt0&.4& \lt4&& & && \multisplit{16(18),}{not~obs.}\\
 & && $2^+$& 137&.80& & 690&.86(13)& 10&.4(5)& 100&(5)& & 8&.46(13)& 100&(2)\\
 & && $4^+$& 404&.18& & 424&.5(2)& 1&.12(6)& 10&.8(5)& & 0&.77(13)& 9&(2)\\
 & && $0^+$& 675&.6& & [153&]& \lt0&.07& \lt0&.7& & 0&.19\ensuremath{^{d}}& \multisplit{36(7),}{1.9\ensuremath{^{d}}\ensuremath{^,}\ensuremath{^{e}}}\\

 \colrule 
$2^+$& 890&.50(11)& $0^+$& 0&.00& & 890&.44(12)& 5&.9(9)& 100&(6)& & 5&.21(8)& 100&(2)\\
 & && $2^+$& 137&.80& & 752&.67(15)& 3&.3(3)& 56&(5)& & 3&.05(18)& 59&(3)\\
 & && $4^+$& 404&.18& & 486&.4(3)& 0&.50(8)& 8&.5(10)& & 0&.31(12)& 6&.0(23)\\
 & && $0^+$& 675&.6& & [214&]& \lt0&.05& \lt0&.8& & && &\\
 & && $2^+$& 828&.66& & [62&]& \lt0&.06& \lt1&.0& & 0&.05\ensuremath{^{d}}& 0&.9\ensuremath{^{d}}\\

 \colrule 
$3^+$& 1022&.10(14)& $2^+$& 137&.80& & 884&.30(10)& 16&.4(16)& 100&(7)& & 13&.86(11)& 100&.0(8)\\
 & && $4^+$& 404&.18& & 617&.88(12)& 3&.6(4)& 22&(2)& & 2&.68(5)& 19&.3(4)\\
 & && $2^+$& 828&.66& & [193&]& \lt0&.07& \lt0&.4& & && &\\
 & && $2^+$& 890&.50& & [131&]& \lt0&.08& \lt0&.5& & 0&.48\ensuremath{^{d}}& 3&.5\ensuremath{^{d}}\ensuremath{^,}\ensuremath{^{f}}\\

 \colrule 
$4^+$& 1088&.28(14)& $2^+$& 137&.80& & 950&.5(2)& 1&.2(2)& 9&.0(15)& & 1&.39(6)& 13&.2(6)\\
 & && $4^+$& 404&.18& & 684&.10(10)& 13&.3(9)& 100&(7)& & 10&.54(14)& 100&.0(13)\\
 & && $6^+$& 770&.40& & 317&.9(2)& 0&.27(4)& 2&.0(3)& & 0&.33(13)& 3&.1(12)\\
 & && $2^+$& 828&.66& & 259&.59(15)\ensuremath{^{g}}& 1&.46(13)& 11&.0(10)& & && \multisplitthree{45(14),\ensuremath{^{h}}}{11(2),\ensuremath{^{h}}}{not~obs.}\\
 & && $2^+$& 890&.50& & [197&]& \lt0&.19& \lt1&.4& & && &\\
 & && $3^+$& 1022&.10& & [66&]& \lt0&.3& \lt2&& & && &\\

 \colrule 
$4^+$& 1168&.4(2)& $2^+$& 137&.80& & 1030&.7(2)& 7&.7(4)& 86&(4)& & 6&.19(9)& 89&.1(13)\\
 & && $4^+$& 404&.18& & 764&.12(13)& 9&.0(5)& 100&(6)& & 6&.91(7)& 100&.0(10)\\
 & && $6^+$& 770&.40& & 397&.9(2)\ensuremath{^{g}}& 0&.21(5)& 2&.3(6)& & && \multisplit{557(90),\ensuremath{^{i}}}{not~obs.}\\
 & && $2^+$& 828&.66& & [340&]& \lt0&.17& \lt1&.9& & && &\\
 & && $2^+$& 890&.50& & 277&.96(18)& 0&.71(7)& 7&.9(8)& & 0&.87(9)& 12&.5(13)\\
 & && $3^+$& 1022&.10& & [146&]& \lt0&.2& \lt3&& & 0&.11\ensuremath{^{d}}& \multisplit{530(60),\ensuremath{^{j}}}{1.7\ensuremath{^{d}}}\\
 & && $4^+$& 1088&.28& & [80&]& \lt0&.3& \lt3&& & [0&.6]\ensuremath{^{k}}& [9&]\ensuremath{^{k}}\\

 \colrule 
$8^+$& 1215&.6(2)\ensuremath{^{l}}& $6^+$& 770&.40& & 445&.23(17)\ensuremath{^{g}}& 0&.37(3)& 100&(8)& & && 100&\\

 \colrule 
$5^+$& 1335&.51(18)& $4^+$& 404&.18& & 931&.35(16)& 7&.2(4)& 100&(6)& & 5&.83(6)& 100&.0(10)\\
 & && $6^+$& 770&.40& & 565&.07(17)& 1&.15(6)& 16&.0(8)& & 1&.25(5)& 21&.4(9)\\
 & && $3^+$& 1022&.10& & 313&.4(2)& 0&.66(5)& 9&.2(7)& & 0&.68(13)& 11&.7(22)\\
 & && $4^+$& 1088&.28& & [247&]& \lt0&.2& \lt3&& & && &\\
 & && $4^+$& 1168&.4& & [167&]& \lt0&.3& \lt4&& & 0&.66\ensuremath{^{d}}& 11&\ensuremath{^{d}}\\

 \colrule 
$3^-$& 1368&.53(18)& $2^+$& 137&.80& & 1230&.72(14)& 5&.3(5)& 100&(10)& & 4&.20(13)& 100&(3)\\
 & && $4^+$& 404&.18& & 964&.36(18)& 1&.51(12)& 29&(2)& & 1&.22(8)& 29&(2)\\
 & && $2^+$& 828&.66& & [540&]& \lt0&.12& \lt2&& & && &\\
 & && $2^+$& 890&.50& & [478&]& \lt0&.13& \lt3&& & && &\\
 & && $3^+$& 1022&.10& & [346&]& \lt0&.10& \lt1&.9& & && &\\
 & && $4^+$& 1088&.28& & [280&]& \lt0&.19& \lt4&& & && &\\
 & && $4^+$& 1168&.4& & [200&]& \lt0&.2& \lt4&& & && &\\

 \colrule 
$(2^+)\ensuremath{^{m}}$& 1382&.3(2)\ensuremath{^{l}}\ensuremath{^,}\ensuremath{^{n}}& $0^+$& 0&.00& & [1382&]& \lt0&.6& \lt207&& & && &\\
 & && $2^+$& 137&.80& & [1245&]& \lt0&.9& \lt314&& & && &\\
 & && $4^+$& 404&.18& & [978&]& \lt0&.4& \lt150&& & && &\\
 & && $0^+$& 675&.6& & 706&.74(16)\ensuremath{^{o}}& 0&.14(2)& 50&(7)& & && &\\
 & && $2^+$& 828&.66& & 553&.7(2)\ensuremath{^{o}}& 0&.28(3)& 100&(11)& & && &\\
 & && $2^+$& 890&.50& & 491&.6(3)\ensuremath{^{o}}& 0&.23(6)& 82&(21)& & && &\\
 & && $3^+$& 1022&.10& & 360&.7(12)\ensuremath{^{o}}\ensuremath{^,}\ensuremath{^{p}}& 0&.11(4)& 39&(14)& & && &\\
 & && $4^+$& 1088&.28& & [294&]& \lt0&.11& \lt39&& & && &\\
 & && $4^+$& 1168&.4& & [214&]& \lt0&.2& \lt71&& & && &\\

 \colrule 
$6^+$& 1437&.28(17)& $4^+$& 404&.18& & 1033&.2(3)\ensuremath{^{g}}& 0&.65(13)& 34&(7)& & && 16&(8)\\
 & && $6^+$& 770&.40& & 666&.88(15)& 1&.92(10)& 100&(5)& & 2&.07(9)& 100&(4)\\
 & && $4^+$& 1088&.28& & 348&.96(14)& 1&.41(7)& 73&(4)& & 1&.46(8)& 71&(5)\\
 & && $4^+$& 1168&.4& & [268&]& \lt0&.4& \lt20&& & && &\\
 & && $8^+$& 1215&.6& & [222&]& \lt0&.11& \lt6&& & && &\\
 & && $5^+$& 1335&.51& & [101&]& \lt0&.7& \lt37&& & && &\\

 \colrule 
$(?)$& 1476&.10(16)\ensuremath{^{q}}& $2^+$& 137&.80& & 1338&.31(17)\ensuremath{^{o}}& 1&.11(11)& 100&(10)& & && &\\
 & && $2^+$& 890&.50& & 585&.6(2)\ensuremath{^{o}}& 0&.35(7)& 32&(6)& & && &\\

 \colrule 
$(2^+)\ensuremath{^{m}}$& 1515&.0(2)\ensuremath{^{q}}& $4^+$& 404&.18& & 1111&.2(6)\ensuremath{^{o}}& 0&.53(13)\ensuremath{^{r}}& 100&(25)& & && &\\
 & && $0^+$& 675&.6& & 839&.3(2)\ensuremath{^{o}}& 0&.20(2)& 37&(4)& & && &\\
 & && $2^+$& 890&.50& & 624&.4(3)\ensuremath{^{o}}& 0&.11(5)& 21&(9)& & && &\\

 \colrule 
$6^+$& 1525&.3(2)\ensuremath{^{l}}\ensuremath{^,}\ensuremath{^{s}}& $4^+$& 404&.18& & [1121&]& \lt3&\ensuremath{^{t}}& \lt149&\ensuremath{^{t}}& & && $\leq$52&\\
 & && $6^+$& 770&.40& & 754&.9(2)\ensuremath{^{g}}& 1&.75(11)\ensuremath{^{t}}& 100&(6)\ensuremath{^{t}}& & && 27&(14)\\
 & && $4^+$& 1088&.28& & [437&]& \lt0&.13\ensuremath{^{t}}& \lt7&\ensuremath{^{t}}& & && &\\
 & && $4^+$& 1168&.4& & 356&.5(3)\ensuremath{^{g}}& 0&.53(5)\ensuremath{^{t}}& 30&(3)\ensuremath{^{t}}& & && 43&(9)\\
 & && $5^+$& 1335&.51& & [190&]& \lt0&.2& \lt13&& & && 100&(5)\ensuremath{^{u}}\\

 \colrule 
$(5^-)$& 1526&.0(2)& $4^+$& 404&.18& & 1121&.8(2)& 8&.2(8)\ensuremath{^{t}}& 100&(10)\ensuremath{^{t}}& & 6&.54(11)& 100&.0(17)\\
 & && $6^+$& 770&.40& & [755&]& \lt0&.6\ensuremath{^{t}}& \lt7\ensuremath{^{t}}&& & 1&.39(11)& 21&.2(17)\\
 & && $4^+$& 1088&.28& & 437&.6(6)\ensuremath{^{o}}\ensuremath{^,}\ensuremath{^{p}}& 0&.08(6)\ensuremath{^{t}}& 1&.0(7)\ensuremath{^{t}}& & && &\\
 & && $4^+$& 1168&.4& & [357&]& \lt0&.2\ensuremath{^{t}}& \lt3\ensuremath{^{t}}&& & 0&.42(8)& 6&.4(12)\\

 \colrule 
$(3)^-$& 1609&.4(2)& $0^+$& 0&.00& & [1609&]& \lt1&.3& \lt52&& & 0&.14(3)& 6&.5(14)\\
 & && $2^+$& 137&.80& & 1471&.5(2)& 2&.5(3)\ensuremath{^{r}}& 100&(12)& & 2&.17(8)& 100&(4)\\
 & && $4^+$& 404&.18& & 1205&.2(2)& 1&.27(11)& 51&(4)& & 0&.97(9)& 45&(4)\\

 \colrule 
$(?)$& 1624&.6(2)\ensuremath{^{q}}& $2^+$& 137&.80& & 1486&.4(7)\ensuremath{^{o}}& 0&.54(16)& 55&(16)& & && &\\
 & && $2^+$& 828&.66& & 796&.03(15)\ensuremath{^{o}}& 0&.98(6)& 100&(6)& & && &\\
 & && $4^+$& 1168&.4& & 456&.2(8)\ensuremath{^{o}}& 0&.09(3)& 9&(3)& & && &\\

 \colrule 
$(4)^+$& 1627&.5(2)& $4^+$& 404&.18& & 1223&.36(18)& 5&.6(4)& 100&(7)& & 4&.75(13)& 100&(3)\\
 & && $3^+$& 1022&.10& & 605&.5(3)& 0&.36(7)& 6&.4(13)& & 0&.47(9)& 10&(2)\\
 & && $4^+$& 1168&.4& & 458&.9(4)\ensuremath{^{o}}& 0&.20(6)& 3&.6(11)& & && &\\
 & && $$& && & [178&]\ensuremath{^{v}}& \lt0&.07\ensuremath{^{w}}& \lt1&.2& & && $\leq$133&(43)\ensuremath{^{x}}\\

 \colrule 
$(4^+)\ensuremath{^{m}}$& 1677&.2(2)\ensuremath{^{q}}& $4^+$& 404&.18& & 1272&.8(3)\ensuremath{^{o}}& 0&.32(8)& 62&(15)& & && &\\
 & && $6^+$& 770&.40& & 907&.2(4)\ensuremath{^{o}}& 0&.14(5)& 29&(4)& & && &\\
 & && $2^+$& 828&.66& & 848&.2(5)\ensuremath{^{o}}& 0&.12(5)& 23&(10)& & && &\\
 & && $2^+$& 890&.50& & 786&.1(5)\ensuremath{^{o}}\ensuremath{^,}\ensuremath{^{p}}& 0&.10(3)& 19&(6)& & && &\\
 & && $3^+$& 1022&.10& & 654&.9(4)\ensuremath{^{o}}& 0&.33(9)& 63&(17)& & && &\\
 & && $4^+$& 1088&.28& & 588&.88(14)\ensuremath{^{o}}& 0&.52(4)& 100&(8)& & && &\\

 \colrule 
$(?)$& 1679&.9(8)\ensuremath{^{q}}& $2^+$& 137&.80& & 1542&.1(8)\ensuremath{^{o}}& 0&.80(16)\ensuremath{^{y}}& 100&(20)& & && &\\
 & && $2^+$& 828&.66& & 851&.0(12)\ensuremath{^{o}}\ensuremath{^,}\ensuremath{^{p}}& 0&.07(4)& 9&(5)& & && &\\

 \colrule 
$7^+$& 1728&.7(5)\ensuremath{^{l}}& $6^+$& 770&.40& & 958&.3(8)\ensuremath{^{g}}& 0&.22(7)\ensuremath{^{r}}& 100&(32)& & && 100&(10)\\
 & && $5^+$& 1335&.51& & 393&.2(6)\ensuremath{^{g}}& 0&.09(4)& 41&(18)& & && 52&(8)\\

 \colrule 
$(?)$& 1772&.4(10)\ensuremath{^{q}}& $2^+$& 137&.80& & 1634&.6(10)\ensuremath{^{o}}& 1&.1(3)\ensuremath{^{r}}& 100&(27)& & && &\\

 \colrule 
$(?)$& 1794&.6(2)\ensuremath{^{q}}& $4^+$& 404&.18& & 1390&.33(17)\ensuremath{^{o}}& 2&.07(12)& 100&(6)& & && &\\
 & && $6^+$& 770&.40& & 1024&.6(6)\ensuremath{^{o}}& 0&.12(5)& 6&(2)& & && &\\

 \colrule 
$7^-$& 1809&.8(3)\ensuremath{^{l}}& $6^+$& 770&.40& & 1039&.3(2)\ensuremath{^{g}}& 0&.33(5)& 100&(15)& & && obse&rved\ensuremath{^{z}}\\
 & && $8^+$& 1215&.6& & 594&.9(6)\ensuremath{^{g}}& 0&.047(13)& 14&(4)& & && obse&rved\ensuremath{^{z}}\\

 \colrule 
$(?)$& 1840&.1(2)\ensuremath{^{q}}& $4^+$& 404&.18& & 1435&.7(5)\ensuremath{^{o}}& 0&.47(9)& 66&(23)& & && &\\
 & && $2^+$& 828&.66& & 1011&.7(2)\ensuremath{^{o}}& 0&.10(3)& 14&(4)& & && &\\
 & && $2^+$& 890&.50& & 949&.60(16)\ensuremath{^{o}}& 0&.71(5)& 100&(7)& & && &\\
 & && $3^+$& 1022&.10& & 818&.1(2)\ensuremath{^{o}}& 0&.26(6)& 37&(8)& & && &\\
 & && $4^+$& 1168&.4& & 671&.2(2)\ensuremath{^{o}}& 0&.18(4)& 25&(6)& & && &\\

 \colrule 
$(?)$& 1857&.82(17)\ensuremath{^{q}}& $4^+$& 404&.18& & 1453&.65(15)\ensuremath{^{o}}& 2&.5(3)& 100&(12)& & && &\\
 & && $6^+$& 770&.40& & 1087&.40(16)\ensuremath{^{o}}& 0&.62(4)& 24&.8(16)& & && &\\
 & && $4^+$& 1168&.4& & 688&.9(5)\ensuremath{^{o}}\ensuremath{^,}\ensuremath{^{p}}& 0&.15(9)& 6&(4)& & && &\\

 \colrule 
$(?)$& 1878&.8(6)\ensuremath{^{q}}& $2^+$& 137&.80& & 1741&.5(7)\ensuremath{^{o}}& 0&.36(9)& 64&(16)& & && &\\
 & && $4^+$& 404&.18& & 1474&.2(4)\ensuremath{^{o}}& 0&.56(14)& 100&(25)& & && &\\
 & && $2^+$& 828&.66& & 1049&.6(15)\ensuremath{^{o}}\ensuremath{^,}\ensuremath{^{p}}& 0&.12(5)& 21&(9)& & && &\\
 & && $2^+$& 890&.50& & 988&.7(5)\ensuremath{^{o}}\ensuremath{^,}\ensuremath{^{p}}& 0&.14(3)& 25&(5)& & && &\\

 \colrule 
$(6)$& 1898&.5(2)\ensuremath{^{l}}\ensuremath{^,}\ensuremath{^{aa}}& $6^+$& 770&.40& & 1128&.07(15)\ensuremath{^{g}}& 0&.89(5)& 100&(6)& & && 100&(5)\\
 & && $5^+$& 1335&.51& & 562&.6(5)\ensuremath{^{o}}& 0&.12(5)& 13&(6)& & && &\\
 & && $(4)^+$& 1627&.5& & [271&]& \lt0&.16& \lt18&& & && $\leq$88&\\

 \colrule 
$(?)$& 1930&.0(5)\ensuremath{^{q}}& $2^+$& 137&.80& & 1791&.9(9)\ensuremath{^{o}}& 0&.50(18)& 78&(28)& & && &\\
 & && $4^+$& 404&.18& & 1526&.1(6)\ensuremath{^{o}}& 0&.64(16)& 100&(25)& & && &\\

 \colrule 
$(?)$& 1933&.6(2)\ensuremath{^{q}}& $4^+$& 404&.18& & 1529&.4(2)\ensuremath{^{o}}& 1&.52(13)& 100&(9)& & && &\\
 & && $3^+$& 1022&.10& & 911&.5(6)\ensuremath{^{o}}& 0&.15(4)& 10&(3)& & && &\\
 & && $4^+$& 1088&.28& & 845&.3(3)\ensuremath{^{o}}& 0&.11(2)& 7&.2(13)& & && &\\

 \colrule 
$(?)$& 1942&.9(3)\ensuremath{^{q}}& $4^+$& 404&.18& & 1538&.0(12)\ensuremath{^{o}}\ensuremath{^,}\ensuremath{^{p}}& 0&.41(13)& 121&(38)& & && &\\
 & && $6^+$& 770&.40& & 1172&.5(16)\ensuremath{^{o}}\ensuremath{^,}\ensuremath{^{p}}& 0&.20(6)& 59&(19)& & && &\\
 & && $4^+$& 1088&.28& & 854&.6(3)\ensuremath{^{o}}& 0&.34(5)& 100&(15)& & && &\\

 \colrule 
$(?)$& 1950&.0(2)\ensuremath{^{q}}& $4^+$& 404&.18& & 1545&.8(2)\ensuremath{^{o}}& 1&.44(8)& 100&(6)& & && &\\

 \colrule 
$(?)$& 2003&.0(3)\ensuremath{^{q}}& $4^+$& 404&.18& & 1598&.7(5)\ensuremath{^{o}}& 0&.25(7)& 100&(28)& & && &\\
 & && $2^+$& 828&.66& & 1174&.5(8)\ensuremath{^{o}}& 0&.22(8)& 88&(32)& & && &\\
 & && $4^+$& 1088&.28& & 914&.6(3)\ensuremath{^{o}}& 0&.14(5)& 56&(20)& & && &\\

 \colrule 
$(?)$& 2058&.6(2)\ensuremath{^{q}}& $3^+$& 1022&.10& & 1036&.4(2)\ensuremath{^{o}}& 0&.32(6)& 100&(19)& & && &\\
 & && $4^+$& 1088&.28& & 970&.4(18)\ensuremath{^{o}}\ensuremath{^,}\ensuremath{^{p}}& 0&.06(4)& 19&(13)& & && &\\
 & && $4^+$& 1168&.4& & 890&.2(4)\ensuremath{^{o}}& 0&.27(10)& 84&(31)& & && &\\
 & && $5^+$& 1335&.51& & 722&.3(7)\ensuremath{^{o}}& 0&.13(4)& 41&(13)& & && &\\

 \colrule 
$(?)$& 2085&.1(3)\ensuremath{^{q}}& $6^+$& 770&.40& & 1314&.7(2)\ensuremath{^{o}}& 0&.52(5)& 100&(10)& & && &\\

 \colrule 
$2^+$& 2089&.9(3)& $0^+$& 0&.00& & [2089&]& \lt0&.6& \lt144&& & 0&.31(13)& 27&(12)\\
 & && $2^+$& 137&.80& & 1952&.3(9)\ensuremath{^{p}}& 0&.24(10)& 56&(23)& & 0&.29(4)& 26&(4)\\
 & && $4^+$& 1088&.28& & 1001&.7(3)\ensuremath{^{o}}& 0&.43(6)& 100&(14)& & && &\\
 & && $4^+$& 1168&.4& & 921&.2(3)& 0&.26(6)& 60&(14)& & 0&.14(6)& 12&(5)\\
 & && $$& && & [796&]\ensuremath{^{v}}& && && & 1&.13(4)& 100&(4)\\

 \colrule 
$(?)$& 2103&.3(3)\ensuremath{^{q}}& $3^+$& 1022&.10& & 1081&.2(4)\ensuremath{^{o}}& 0&.64(5)& 100&(8)& & && &\\
 & && $4^+$& 1168&.4& & 935&.0(4)\ensuremath{^{o}}& 0&.19(6)& 30&(9)& & && &\\
 & && $5^+$& 1335&.51& & 767&.8(4)\ensuremath{^{o}}& 0&.16(4)& 25&(6)& & && &\\

 \colrule 
$(?)$& 2164&.3(4)\ensuremath{^{q}}& $4^+$& 404&.18& & 1760&.1(4)\ensuremath{^{o}}& 0&.31(9)& 100&(29)& & && &\\
 & && $6^+$& 770&.40& & 1393&.9(7)\ensuremath{^{o}}\ensuremath{^,}\ensuremath{^{p}}& 0&.09(4)& 29&(12)& & && &\\

 \colrule 
$(?)$& 2183&.8(4)\ensuremath{^{q}}& $2^+$& 828&.66& & 1355&.1(4)\ensuremath{^{o}}& 0&.21(5)& 100&(23)& & && &\\
 & && $2^+$& 890&.50& & 1293&.0(5)\ensuremath{^{o}}\ensuremath{^,}\ensuremath{^{p}}& 0&.14(8)& 67&(38)& & && &\\
 & && $4^+$& 1088&.28& & 1095&.9(5)\ensuremath{^{o}}\ensuremath{^,}\ensuremath{^{p}}& 0&.10(6)& 48&(29)& & && &\\

 \colrule 
$(?)$& 2193&.5(3)\ensuremath{^{q}}& $6^+$& 770&.40& & 1423&.3(6)\ensuremath{^{o}}& 0&.16(6)& 47&(18)& & && &\\
 & && $5^+$& 1335&.51& & 858&.0(3)\ensuremath{^{o}}& 0&.34(5)& 100&(15)& & && &\\

 \colrule 
$(?)$& 2199&.7(3)\ensuremath{^{q}}& $4^+$& 404&.18& & 1795&.6(5)\ensuremath{^{o}}& 0&.42(15)& 100&(36)& & && &\\
 & && $3^+$& 1022&.10& & 1177&.6(2)\ensuremath{^{o}}& 0&.29(5)& 69&(12)& & && &\\
 & && $4^+$& 1168&.4& & 1031&.8(8)\ensuremath{^{o}}& 0&.11(3)\ensuremath{^{r}}& 26&(7)& & && &\\
 & && $5^+$& 1335&.51& & 863&.3(10)\ensuremath{^{o}}& 0&.10(4)& 24&(10)& & && &\\
 & && $(?)$& 1476&.10& & 723&.5(4)\ensuremath{^{o}}& 0&.14(4)& 33&(10)& & && &\\

 \colrule 
$(?)$& 2207&.4(5)\ensuremath{^{q}}& $3^+$& 1022&.10& & 1185&.6(5)\ensuremath{^{o}}& 0&.22(4)& 100&(18)& & && &\\
 & && $5^+$& 1335&.51& & 871&.6(5)\ensuremath{^{o}}& 0&.18(5)& 82&(23)& & && &\\

 \colrule 
$(?)$& 2220&.4(4)\ensuremath{^{q}}& $6^+$& 770&.40& & 1450&.0(3)\ensuremath{^{o}}& 0&.22(6)\ensuremath{^{r}}& 100&(27)& & && &\\

 \colrule 
$(?)$& 2228&.9(5)\ensuremath{^{q}}& $4^+$& 404&.18& & 1824&.7(5)\ensuremath{^{o}}& 0&.63(9)& 100&(14)& & && &\\

 \colrule 
$(?)$& 2230&.9(4)\ensuremath{^{q}}& $6^+$& 770&.40& & 1460&.5(3)\ensuremath{^{o}}& 0&.22(4)& 100&(18)& & && &\\

 \colrule 
$(?)$& 2244&.7(3)\ensuremath{^{q}}& $4^+$& 404&.18& & 1840&.5(8)\ensuremath{^{o}}\ensuremath{^,}\ensuremath{^{p}}& 0&.22(9)& 15&(6)& & && &\\
 & && $2^+$& 828&.66& & 1415&.9(2)\ensuremath{^{o}}& 1&.50(9)& 100&(6)& & && &\\
 & && $2^+$& 890&.50& & 1354&.1(2)\ensuremath{^{o}}& 0&.41(5)& 27&(3)& & && &\\
 & && $3^+$& 1022&.10& & 1222&.8(3)\ensuremath{^{o}}& 0&.37(8)& 25&(5)& & && &\\
 & && $4^+$& 1088&.28& & 1156&.4(3)\ensuremath{^{o}}& 0&.32(7)& 21&(5)& & && &\\
 & && $4^+$& 1168&.4& & 1076&.2(5)\ensuremath{^{o}}& 0&.42(8)& 28&(5)& & && &\\
 & && $(?)$& 1624&.6& & 620&.1(8)\ensuremath{^{o}}& 0&.10(3)& 7&(2)& & && &\\

 \colrule 
$(?)$& 2264&.1(6)\ensuremath{^{q}}& $4^+$& 404&.18& & 1860&.1(5)\ensuremath{^{o}}& 0&.81(13)& 100&(16)& & && &\\
 & && $3^+$& 1022&.10& & 1241&.2(6)\ensuremath{^{o}}\ensuremath{^,}\ensuremath{^{p}}& 0&.15(6)& 19&(7)& & && &\\
 & && $4^+$& 1168&.4& & 1094&.8(10)\ensuremath{^{o}}\ensuremath{^,}\ensuremath{^{p}}& 0&.15(5)& 19&(6)& & && &\\

 \colrule 
$(?)$& 2270&.0(4)\ensuremath{^{q}}& $6^+$& 770&.40& & 1499&.6(3)\ensuremath{^{o}}& 0&.62(9)& 100&(15)& & && &\\

 \colrule 
$(?)$& 2293&.4(4)\ensuremath{^{q}}& $4^+$& 404&.18& & 1888&.8(15)\ensuremath{^{o}}\ensuremath{^,}\ensuremath{^{p}}& 0&.27(10)& 71&(26)& & && &\\
 & && $6^+$& 770&.40& & 1523&.0(3)\ensuremath{^{o}}& 0&.38(6)& 100&(16)& & && &\\

 \colrule 
$(?)$& 2300&.1(4)\ensuremath{^{q}}& $3^+$& 1022&.10& & 1278&.0(3)\ensuremath{^{o}}& 0&.52(14)& 100&(27)& & && &\\

 \colrule 
$4^+$& 2307&.4(3)& $2^+$& 137&.80& & [2169&]& \lt0&.4& \lt39&& & 0&.31(4)& 14&(2)\\
 & && $4^+$& 404&.18& & 1902&.5(5)& 0&.42(10)& 46&(11)& & 0&.48(6)& 22&(3)\\
 & && $6^+$& 770&.40& & 1536&.0(4)& 0&.49(8)& 53&(9)& & 0&.57(8)& 26&(4)\\
 & && $2^+$& 828&.66& & 1478&.7(2)& 0&.28(3)& 30&(3)& & 0&.66(8)& 40&(4)\\
 & && $2^+$& 890&.50& & 1416&.8(2)& 0&.92(10)& 100&(11)& & 2&.21(4)& 100&(2)\\
 & && $3^+$& 1022&.10& & 1285&.4(4)\ensuremath{^{o}}& 0&.18(7)& 20&(8)& & && &\\
 & && $4^+$& 1088&.28& & 1218&.9(5)\ensuremath{^{o}}& 0&.39(10)& 42&(10)& & && &\\
 & && $4^+$& 1168&.4& & 1139&.0(6)\ensuremath{^{o}}& 0&.32(9)& 35&(10)& & && &\\
 & && $3^-$& 1368&.53& & 939&.2(11)\ensuremath{^{o}}& 0&.17(6)& 18&(7)& & && &\\
 & && $(4)^+$& 1627&.5& & [680&]& \lt0&.10& \lt11&& & 0&.48(10)& 22&(5)\\

 \colrule 
$(?)$& 2323&.6(2)\ensuremath{^{q}}& $2^+$& 137&.80& & 2185&.6(6)\ensuremath{^{o}}& 0&.31(10)& 12&(4)& & && &\\
 & && $4^+$& 404&.18& & 1919&.8(4)\ensuremath{^{o}}& 0&.61(13)& 24&(5)& & && &\\
 & && $2^+$& 828&.66& & 1494&.5(5)\ensuremath{^{o}}& 0&.29(7)& 11&(3)& & && &\\
 & && $2^+$& 890&.50& & 1432&.8(2)\ensuremath{^{o}}& 1&.00(10)& 39&(4)& & && &\\
 & && $3^+$& 1022&.10& & 1301&.5(4)\ensuremath{^{o}}& 2&.58(14)& 100&(5)& & && &\\
 & && $4^+$& 1088&.28& & 1235&.3(2)\ensuremath{^{o}}& 0&.43(9)& 17&(3)& & && &\\
 & && $4^+$& 1168&.4& & 1155&.3(2)\ensuremath{^{o}}& 1&.26(9)& 49&(3)& & && &\\
 & && $3^-$& 1368&.53& & 955&.4(4)\ensuremath{^{o}}& 0&.19(4)& 7&.4(16)& & && &\\

 \colrule 
$(?)$& 2331&.7(3)\ensuremath{^{q}}& $3^+$& 1022&.10& & 1309&.7(4)\ensuremath{^{o}}& 0&.37(8)& 100&(22)& & && &\\
 & && $4^+$& 1168&.4& & 1163&.1(6)\ensuremath{^{o}}\ensuremath{^,}\ensuremath{^{p}}& 0&.10(5)& 27&(14)& & && &\\
 & && $5^+$& 1335&.51& & 996&.1(4)\ensuremath{^{o}}& 0&.14(5)& 37&(14)& & && &\\

 \colrule 
$(?)$& 2342&.6(3)\ensuremath{^{q}}& $4^+$& 1168&.4& & 1174&.2(2)\ensuremath{^{o}}& 0&.42(7)& 100&(17)& & && &\\

 \colrule 
$(?)$& 2372&.1(3)\ensuremath{^{q}}& $2^+$& 137&.80& & 2234&.2(4)\ensuremath{^{o}}& 1&.7(4)& 100&(23)& & && &\\
 & && $4^+$& 404&.18& & 1967&.9(3)\ensuremath{^{o}}& 0&.59(16)& 35&(9)& & && &\\

 \colrule 
$(?)$& 2385&.6(3)\ensuremath{^{q}}& $3^+$& 1022&.10& & 1363&.4(7)\ensuremath{^{o}}\ensuremath{^,}\ensuremath{^{p}}& 0&.08(3)& 32&(12)& & && &\\
 & && $4^+$& 1168&.4& & 1217&.2(3)\ensuremath{^{o}}& 0&.25(7)& 100&(28)& & && &\\
 & && $5^+$& 1335&.51& & 1050&.0(5)\ensuremath{^{o}}& 0&.11(3)& 44&(12)& & && &\\

 \colrule 
$(?)$& 2408&.7(4)\ensuremath{^{bb}}& $2^+$& 137&.80& & 2271&.0(2)\ensuremath{^{o}}& 0&.88(13)& 100&(15)& & && &\\
 & && $4^+$& 404&.18& & 2003&.7(7)\ensuremath{^{p}}& 0&.34(10)& 39&(11)& & 0&.24(5)& 36&(7)\\
 & && $2^+$& 828&.66& & 1580&.3(4)\ensuremath{^{o}}& 0&.11(3)& 13&(3)& & && &\\
 & && $2^+$& 890&.50& & 1518&.7(3)\ensuremath{^{o}}& 0&.25(7)& 28&(8)& & && &\\
 & && $3^+$& 1022&.10& & 1386&.3(2)& 0&.67(6)& 76&(7)& & 0&.67(8)& 100&(12)\\
 & && $4^+$& 1088&.28& & 1320&.3(15)& 0&.12(5)& 14&(6)& & 0&.31(5)& 46&(7)\\
 & && $4^+$& 1168&.4& & 1241&.3(12)\ensuremath{^{o}}\ensuremath{^,}\ensuremath{^{p}}& 0&.14(6)& 16&(7)& & && &\\
 & && $3^-$& 1368&.53& & 1040&.0(7)\ensuremath{^{o}}& 0&.11(4)& 13&(5)& & && &\\
 & && $$& && & [880&]\ensuremath{^{v}}& && && & 0&.66(8)& 99&(12)\\
 & && $(?)$& 2103&.3& & 304&.6(7)\ensuremath{^{o}}\ensuremath{^,}\ensuremath{^{p}}& 0&.10(3)& 11&(3)& & && &\\

 \colrule 
$(?)$& 2418&.9(6)\ensuremath{^{q}}& $4^+$& 404&.18& & 2014&.9(6)\ensuremath{^{o}}& 0&.42(10)& 100&(24)& & && &\\
 & && $6^+$& 770&.40& & 1648&.1(7)\ensuremath{^{o}}\ensuremath{^,}\ensuremath{^{p}}& 0&.19(6)& 45&(14)& & && &\\

 \colrule 
$(?)$& 2433&.8(2)\ensuremath{^{q}}& $4^+$& 404&.18& & 2029&.70(18)\ensuremath{^{o}}& 2&.17(16)& 100&(7)& & && &\\
 & && $6^+$& 770&.40& & 1663&.3(2)\ensuremath{^{o}}& 0&.52(10)& 24&(5)& & && &\\
 & && $4^+$& 1088&.28& & 1345&.6(3)\ensuremath{^{o}}& 0&.19(5)& 9&(2)& & && &\\
 & && $(5^-)$& 1526&.0& & 908&.0(10)\ensuremath{^{o}}\ensuremath{^,}\ensuremath{^{p}}& 0&.19(6)& 9&(3)& & && &\\

 \colrule 
$(?)$& 2439&.2(2)\ensuremath{^{q}}& $4^+$& 404&.18& & 2035&.0(2)\ensuremath{^{o}}& 1&.7(2)& 100&(12)& & && &\\
 & && $6^+$& 770&.40& & 1668&.7(2)\ensuremath{^{o}}& 0&.32(7)& 19&(4)& & && &\\
 & && $4^+$& 1088&.28& & 1351&.3(6)\ensuremath{^{o}}\ensuremath{^,}\ensuremath{^{p}}& 0&.10(4)& 6&(2)& & && &\\

 \colrule 
$(?)$& 2445&.2(3)\ensuremath{^{q}}& $2^+$& 137&.80& & 2307&.4(8)\ensuremath{^{o}}& 0&.27(11)& 40&(16)& & && &\\
 & && $3^+$& 1022&.10& & 1423&.0(2)\ensuremath{^{o}}& 0&.68(9)& 100&(13)& & && &\\
 & && $5^+$& 1335&.51& & 1110&.7(7)\ensuremath{^{o}}\ensuremath{^,}\ensuremath{^{y}}& 0&.29(6)& 43&(9)& & && &\\
 & && $(?)$& 1624&.6& & 820&.9(6)\ensuremath{^{o}}\ensuremath{^,}\ensuremath{^{p}}& 0&.08(2)& 12&(3)& & && &\\
 & && $(4)^+$& 1627&.5& & 818&.7(4)\ensuremath{^{o}}\ensuremath{^,}\ensuremath{^{p}}& 0&.19(5)& 28&(7)& & && &\\

 \colrule 
$(?)$& 2489&.5(5)\ensuremath{^{q}}& $4^+$& 404&.18& & 2085&.4(5)\ensuremath{^{o}}& 0&.49(10)& 100&(20)& & && &\\
 & && $3^+$& 1022&.10& & 1467&.1(8)\ensuremath{^{o}}& 0&.10(5)& 20&(10)& & && &\\
 & && $5^+$& 1335&.51& & 1154&.4(8)\ensuremath{^{o}}\ensuremath{^,}\ensuremath{^{p}}& 0&.14(6)& 29&(12)& & && &\\

 \colrule 
$(?)$& 2492&.0(3)\ensuremath{^{q}}& $2^+$& 137&.80& & 2354&.1(2)\ensuremath{^{o}}& 0&.90(8)& 100&(9)& & && &\\
 & && $4^+$& 404&.18& & 2088&.2(6)\ensuremath{^{o}}& 0&.37(15)& 41&(17)& & && &\\
 & && $3^+$& 1022&.10& & 1469&.9(5)\ensuremath{^{o}}& 0&.19(6)& 21&(7)& & && &\\
 & && $4^+$& 1168&.4& & 1323&.2(4)\ensuremath{^{o}}& 0&.17(5)& 19&(6)& & && &\\

 \colrule 
$(?)$& 2516&.6(7)\ensuremath{^{cc}}& $0^+$& 675&.6& & [1841&]& \lt0&.03& \lt17&& & 0&.14(3)& 19&(4)\\
 & && $2^+$& 828&.66& & 1688&.2(15)\ensuremath{^{p}}& 0&.07(5)& 32&(25)& & 0&.15(7)& 20&(9)\\
 & && $2^+$& 890&.50& & 1626&.8(6)\ensuremath{^{p}}& 0&.16(6)& 80&(30)& & 0&.18(3)& 24&(4)\\
 & && $3^+$& 1022&.10& & 1493&.8(10)& 0&.20(5)& 100&(25)& & 0&.74(7)& 99&(9)\\
 & && $4^+$& 1168&.4& & 1348&.9(5)& 0&.19(5)& 95&(25)& & 0&.10(8)& 13&(11)\\
 & && $$& && & [1297&]\ensuremath{^{v}}& && && & 0&.75(5)& 100&(7)\\
 & && $3^-$& 1368&.53& & [1148&]& \lt0&.15& \lt75&& & 0&.12(5)& 16&(7)\\
 & && $(3)^-$& 1609&.4& & [907&]& \lt0&.10& \lt50&& & 0&.29(5)& 39&(7)\\

 \colrule 
$(?)$& 2572&.0(5)\ensuremath{^{q}}& $4^+$& 404&.18& & 2168&.9(7)\ensuremath{^{o}}\ensuremath{^,}\ensuremath{^{p}}& 0&.23(8)& 100&(35)& & && &\\
 & && $(4)^+$& 1627&.5& & 944&.3(4)\ensuremath{^{o}}& 0&.15(3)& 65&(13)& & && &\\

 \colrule 
$(?)$& 2594&.3(4)\ensuremath{^{q}}& $3^+$& 1022&.10& & 1572&.0(5)\ensuremath{^{o}}& 0&.13(5)& 62&(25)& & && &\\
 & && $4^+$& 1168&.4& & 1425&.9(4)\ensuremath{^{o}}& 0&.20(5)& 100&(25)& & && &\\
 & && $5^+$& 1335&.51& & 1259&.1(7)\ensuremath{^{o}}& 0&.19(8)& 95&(40)& & && &\\

 \colrule 
$(?)$& 2642&.5(3)\ensuremath{^{q}}& $4^+$& 404&.18& & 2238&.3(2)\ensuremath{^{o}}& 0&.77(13)& 100&(17)& & && &\\

 \colrule 
$(?)$& 2653&.4(6)\ensuremath{^{q}}& $4^+$& 404&.18& & 2249&(2)\ensuremath{^{o}}\ensuremath{^,}\ensuremath{^{p}}& 0&.32(15)& 100&(47)& & && &\\
 & && $2^+$& 828&.66& & 1824&.7(6)\ensuremath{^{o}}& 0&.20(5)& 63&(16)& & && &\\

 \colrule 
$(?)$& 2757&.8(6)\ensuremath{^{q}}& $3^+$& 1022&.10& & 1735&.7(5)\ensuremath{^{o}}& 0&.18(5)& 100&(28)& & && &\\

 \colrule 
$(?)$& 2788&.1(9)\ensuremath{^{q}}& $8^+$& 1215&.6& & 1572&.5(8)\ensuremath{^{o}}& 0&.08(2)& 100&(25)& & && &\\

 \colrule 
$(?)$& 2810&.4(6)\ensuremath{^{q}}& $4^+$& 404&.18& & 2406&.2(7)\ensuremath{^{o}}& 0&.38(11)& 100&(29)& & && &\\
 & && $6^+$& 770&.40& & 2039&.9(10)\ensuremath{^{o}}\ensuremath{^,}\ensuremath{^{p}}& 0&.11(4)& 29&(11)& & && &\\

 \colrule 
$(?)$& 2818&.4(2)& $4^+$& 404&.18& & 2414&.2(2)\ensuremath{^{o}}& 1&.60(18)& 100&(11)& & && &\\
 & && $6^+$& 770&.40& & 2048&.0(2)\ensuremath{^{o}}& 0&.19(6)& 12&(4)& & && &\\
 & && $2^+$& 828&.66& & [1990&]& \lt0&.07& \lt4&& & 0&.21(9)& 11&(5)\\
 & && $3^+$& 1022&.10& & [1796&]& \lt0&.13& \lt8&& & 0&.46(4)& 25&.0(22)\\
 & && $4^+$& 1088&.28& & 1730&.1(2)& 0&.57(6)& 36&(4)& & 0&.49(9)& 27&(5)\\
 & && $4^+$& 1168&.4& & 1649&.7(2)& 1&.37(11)& 86&(7)& & 1&.84(4)& 100&.0(22)\\
 & && $$& && & [1525&]\ensuremath{^{v}}& && && & 0&.68(7)& 37&(4)\\
 & && $5^+$& 1335&.51& & 1482&.7(2)\ensuremath{^{o}}& 0&.30(5)& 19&(3)& & && &\\
 & && $3^-$& 1368&.53& & 1450&.0(8)\ensuremath{^{o}}\ensuremath{^,}\ensuremath{^{p}}& 0&.15(6)& 9&(4)& & && &\\
 & && $6^+$& 1437&.28& & 1380&.9(2)\ensuremath{^{o}}& 0&.65(6)& 41&(4)& & && &\\
 & && $6^+$& 1525&.3& & 1293&.4(15)\ensuremath{^{o}}& 0&.27(4)& 17&(3)& & && &\\
 & && $(5^-)$& 1526&.0& & 1292&.3(3)\ensuremath{^{o}}& 0&.87(11)& 54&(7)& & && &\\
 & && $(4)^+$& 1627&.5& & 1191&.1(5)& 0&.43(6)& 27&(4)& & 0&.20(7)& 11&(4)\\
 & && $(?)$& 1857&.82& & 960&.6(3)\ensuremath{^{o}}& 0&.69(7)& 43&(4)& & && &\\
 & && $(6)$& 1898&.5& & 919&.7(15)\ensuremath{^{o}}& 0&.13(5)& 8&(3)& & && &\\
 & && $(?)$& 1933&.6& & 884&.3(8)\ensuremath{^{o}}& 0&.11(5)& 7&(3)& & && &\\

 \colrule 
$(?)$& 2823&.3(2)\ensuremath{^{dd}}& $4^+$& 404&.18& & 2419&.2(2)& 3&.3(3)& 100&(9)& & 2&.93(6)& 100&.0(20)\\
 & && $6^+$& 770&.40& & 2052&.8(2)\ensuremath{^{o}}& 0&.69(11)& 21&(3)& & && &\\
 & && $2^+$& 828&.66& & [1994&]& \lt0&.07& \lt2&& & 0&.13(9)& 4&(3)\\
 & && $2^+$& 890&.50& & [1932&]& \lt0&.11& \lt3&& & 0&.26(4)& 8&.9(14)\\
 & && $4^+$& 1168&.4& & 1654&.0(11)\ensuremath{^{o}}\ensuremath{^,}\ensuremath{^{p}}& 0&.14(6)& 4&.2(18)& & && &\\
 & && $(5^-)$& 1526&.0& & 1297&.3(2)\ensuremath{^{o}}& 0&.33(8)& 10&(2)& & && &\\
 & && $(?)$& 1857&.82& & 965&.3(8)\ensuremath{^{o}}& 0&.10(5)& 3&.0(15)& & && &\\

 \colrule 
$(?)$& 2833&.6(4)\ensuremath{^{q}}& $4^+$& 404&.18& & 2429&.5(7)\ensuremath{^{o}}& 0&.63(9)& 100&(14)& & && &\\
 & && $6^+$& 770&.40& & 2063&.2(4)\ensuremath{^{o}}& 0&.23(4)& 37&(6)& & && &\\

 \colrule 
$(?)$& 2894&.9(4)\ensuremath{^{q}}& $4^+$& 404&.18& & 2490&.7(6)\ensuremath{^{o}}& 0&.21(7)& 100&(33)& & && &\\
 & && $2^+$& 890&.50& & 2004&.2(9)\ensuremath{^{o}}\ensuremath{^,}\ensuremath{^{p}}& 0&.10(4)& 48&(19)& & && &\\
 & && $3^+$& 1022&.10& & 1872&.9(4)\ensuremath{^{o}}& 0&.21(5)& 100&(24)& & && &\\

 \colrule 
$(?)$& 2981&.5(13)\ensuremath{^{q}}& $4^+$& 404&.18& & 2577&.3(13)\ensuremath{^{o}}& 0&.33(7)& 100&(21)& & && &\\

\end{longtable}
\begin{flushleft}
\footnotesize
\ensuremath{^{a}}Level spin assignments are nominal
assignments from the evaluation~\cite{ndsboth:156}, except as noted.  
\\ 
\ensuremath{^{b}}Literature values for absolute intensities are
from the evaluated $^{156}$Ho EC decay data of
Ref.~\cite{ndsboth:156}, which is based primarily upon
Ref.~\cite{gromov1976:156dy-beta}, except as noted. 
\\ 
\ensuremath{^{c}}Literature values for relative intensities are from the
adopted $\gamma$ radiations of the evaluation~\cite{ndsboth:156}, except as
noted.  Where the evaluation~\cite{ndsboth:156} gives more than one possible
adopted value, all are listed here.
\\ 
\ensuremath{^{d}}The literature $\gamma$-ray intensity is deduced from conversion
electron data only, using an assumed conversion coefficient from
Ref.~\cite{gromov1976:156dy-beta}.
\\ 
\ensuremath{^{e}}Listed in the evaluation~\cite{ndsboth:156} as 5.9, but the
intensity and conversion coefficient given in the original
literature~\cite{gromov1976:156dy-beta} actually yield 1.9.
\\ 
\ensuremath{^{f}}Listed in the evaluation~\cite{ndsboth:156} as 7.2, but the
intensity and conversion coefficient given in the original
literature~\cite{gromov1976:156dy-beta} actually yield 3.5.  The evaluation
also notes larger but ambiguous intensities reported in
$(\alpha,4n)$~\cite{deboer1977:156dy-a4np4n}.
\\ 
\ensuremath{^{g}}Transition was previously reported, but not in $\beta$
decay.
\\ 
\ensuremath{^{h}}Relative intensity from $(\alpha,4n)$ is listed in
the evaluation~\cite{ndsboth:156} as 57, but the intensities given in
the original literature~\cite{deboer1977:156dy-a4np4n} actually yield
45(14).  The uncertainties from the original
$(p,4n)$ literature~\cite{elmasri1976:156dy-p4n} are used here to
obtain the value 11(2).
\\ 
\ensuremath{^{i}}Relative intensity from $(\alpha,4n)$ is listed in
the evaluation~\cite{ndsboth:156} as 370, but the intensities given in
the original literature~\cite{deboer1977:156dy-a4np4n} actually yield
557(90).
\\ 
\ensuremath{^{j}}Relative intensity from $(\alpha,4n)$ is listed in
the evaluation~\cite{ndsboth:156} as 515.  The uncertainties, and
unrounded intensities, from the original
literature~\cite{deboer1977:156dy-a4np4n} are used here to obtain 530(60).
\\ 
\ensuremath{^{k}}This is a literature $I^\text{ce}$ value from conversion
electron data from Ref.~\cite{gromov1976:156dy-beta}.
Ref.~\cite{gromov1976:156dy-beta} makes no prediction for the corresponding
$\gamma$-ray intensity since an unknown portion of the electron
intensity may result from an $E0$ contribution.
\\ 
\ensuremath{^{l}}Level was previously reported, but not in $\beta$ decay
\cite{ndsboth:156}.
\\ 
\ensuremath{^{m}}Probable spin assignment for newly-identified level is given on the
basis of observed transitions to levels of known spin.
\\ 
\ensuremath{^{n}}The level at 1382.3(2)\,keV may be identified with the
adopted $(3^-)$ level at 1385(5)\,keV~\cite{ndsboth:156} reported in
$(p,t)$ scattering~\cite{kolata1977:156dy-pt}.
\\ 
\ensuremath{^{o}}Gamma-ray line was not previously reported or was not
reported in this placement.
\\ 
\ensuremath{^{p}}Identification of transition is tentative.
\\ 
\ensuremath{^{q}}Level was not previously reported.
\\ 
\ensuremath{^{r}}Energy and intensity are deduced from gated spectrum after
subtraction of contribution(s) from other placement(s).  \seeepaps
\\ 
\ensuremath{^{s}}Level is not identified as having been
populated in $\beta$ decay in the published
literature~\cite{gromov1976:156dy-beta} but is quoted as having been
populated in an unpublished $\beta$-decay study \cite{koldewijn:156dy-beta}.
\\ 
\ensuremath{^{t}}Transitions from the closely-spaced pair of levels at
1525.3(2) and 1526.0(2)\,keV are all potentially doublets.  Each
transition is assigned a primary placement as depopulating one of
these levels on the basis of transition energy as measured in gated
spectra but may contain a significant unresolved contribution
depopulating the other member of the pair.  (See text.)
\\ 
\ensuremath{^{u}}The placement here, reported in $(\alpha,4n)$
\cite{deboer1977:156dy-a4np4n}, is noted by the evaluation~\cite{ndsboth:156} to
be uncertain.
\\ 
\ensuremath{^{v}}A literature transition was reported to a level the
existence of which is not supported by the present data. (See text.)
\\ 
\ensuremath{^{w}}Limit obtained on any possible absolute intensity coincident
with the alleged 1310\,keV transition from
Ref.~\cite{deboer1977:156dy-a4np4n}, which is claimed as the only branch
from the literature $(2^+)_{1447}$ level.  (See text.)
\\ 
\ensuremath{^{x}}The intensity for the adopted 178.93(20)\,keV branch is
omitted in the evaluation~\cite{ndsboth:156}, with an indication that
the placement of the transition in the level scheme is uncertain.  The
relative intensity for this branch deduced from the original
literature~\cite{deboer1977:156dy-a4np4n} is 133(43).
\\ 
\ensuremath{^{y}}Peak in gated spectrum has abnormally large width.
\\ 
\ensuremath{^{z}}The evaluation~\cite{ndsboth:156} does not deduce
relative intensities for the 1039 and 594\,keV branches.  The 1039\,keV
branch was reported in $(\text{HI},xn)$, $(\alpha,4n)$, and
$(p,4n)$~\cite{riley1988:156dy-hixn,deboer1977:156dy-a4np4n,elmasri1976:156dy-p4n}.
The 594\,keV branch was only reported in $(\alpha,4n)$ and
$(p,4n)$~\cite{deboer1977:156dy-a4np4n}, and the line was noted by the
authors to contain $^{127}$I contamination.  
\\ 
\ensuremath{^{aa}}The nominal spin assignment for the adopted level at
1898.64(10)\,keV~\cite{ndsboth:156} is $(6,7^-)$.  (See text.)
\\ 
\ensuremath{^{bb}}The level at 2408.7(4)\,keV may be identified with the
adopted $(2^-)$ level at 2409.64(20)\,keV~\cite{ndsboth:156} reported
in $\beta$ decay~\cite{gromov1976:156dy-beta}.  (See text.)
\\ 
\ensuremath{^{cc}}The level at 2516.6(7)\,keV may be identified with the
adopted $(1)^-$ level at 2517.55(16)\,keV~\cite{ndsboth:156} reported
in $\beta$ decay~\cite{gromov1976:156dy-beta}.  (See text.)
\\ 
\ensuremath{^{dd}}The level at 2823.3(2)\,keV may be identified with the
adopted level at 2822.2(4)\,keV~\cite{ndsboth:156} reported in $\beta$
decay~\cite{gromov1976:156dy-beta}.
\\ 

\end{flushleft}

}


\addtocontents{toc}{\addvspace{2.25em}}

\makeatletter
\renewcommand{\toclevel@chapter}{-1}
\makeatother


\addtocontents{toc}{
\protect\pseudofootnotetext{Submitted to the Yale 
Graduate School April 5, 2003.  University Microfilms catalog number
3084266.  E-print \texttt{arXiv:nucl-ex/0502004}.  Dissertation
Abstracts International B
\textbf{64}, 1312 (2003).  Dissertation Abstracts Online AAT~3084266.
}
}

\end{document}